# NANOSCALE ELECTRIC PHENOMENA AT OXIDE SURFACES AND INTERFACES BY SCANNING PROBE MICROSCOPY

***Sergei V. Kalinin***

A DISSERTATION

in the

Materials Science and Engineering

Presented to the Faculties of the University of Pennsylvania in Partial Fulfillment of the Requirements for the Degree of Doctor of Philosophy

2002

# ACKNOWLEDGEMENTS

First and foremost, I would like to thank Dawn Bonnell for being an outstanding supervisor. She introduced me to the field of nanoscience and has guided my research from the first day at the AFM controls to the finer points of science. To her I owe my devotion to SPMs and all the things nano.

Many parts of this dissertation would be impossible if it were not for the collaboration with many people at Penn and elsewhere. I am deeply grateful to Prof. Michael Cohen and Prof. Vasek Vitek (Upenn), Prof. Mark Kachanov (Tufts) and Dr. A.E. Giannakopoulos and Prof. Subra Suresh (MIT) for enlightening discussions on electrostatic theory and continuum mechanics. Collaborations with Albina Borisevich, Matt Suchomel, and Prof. Peter Davies at U.Penn and G. Popov and Prof. Martha Greenblatt at Rutgers have allowed me to satisfy my curiosity and them to learn more about their favorite materials. The discussions with Dr. A. Gruverman (NCSU) were most useful to understand what's going on under the AFM tip. Samples from Steve Dunn (Cranfied) have helped us to jumpstart our ferroelectric lithography program. Marcus Freitag and Prof. Charlie Johnson (UPenn) have introduced me, albeit briefly, to the wondrous nanotubeland. This thesis would be incomplete without the strontium titanate samples from and discussions with Prof. Gerd Duscher (ORNL and NCSU) and Dr. Steve Pennycook (ORNL). The assistance from Dr. Marco Radosavljevic and Juraj Vavro (Upenn) in low temperature transport measurements is greatly appreciated. Last but not least, I would like to acknowledge D. Gorbachev for software support and advice during my four years at Penn and before. Finally, I am extremely grateful to the members of my dissertation committee – Prof. Takeshi Egami, Prof. I-Wei Chan and Prof. Charlie Johnson, who have undertaken the heavy burden of reading this manuscript.

During my time at Penn, I constantly felt the support and encouragement from the members of Bonnell group. There are several people in Bonnell group whom I want to thank in particular. Brian Huey and Nick Angert for introducing me to AFM and sharing its small tricks, which would otherwise take years to discover. Tony Alvarez, for providing constant support in most endeavors, both in the lab and outside, and sharing the experience on how to live in the USA. Rui Shao, for invaluable contribution in putting our nanoscience on the solid computerized basis and taking on the projects I held in the high esteem. I would also like to thank all other people at Bonnell group, – Jack Smith, Ed Peng, Prof. Jeff Sellar, Carolyn Johnson, Fabio Weibel, Cyrill Ruegg, Dr. Zonghai Hu and Dr. Xiaojun Lei – thank you all.

The research at Penn would be impossible without the assistance from the staff members. I am especially grateful to Vladimir Dominko, whose experience and personality rule over the microfabrications laboratory, for assistance in micropatterning and transport measurements; Dr. Jim Ferris and Dr. Doug Yates, for help in electron microscopy and numerous useful discussions; Bill Romanow, for always lending helping hand; Pat Overend and Irene Clements for making life so much easier.

Finally, I greatly acknowledge the financial support from NSF Grant DMR 00-79909 and NSF Grant DMR 00-80863 and DoE grant DE-FG02-00ER45813-A000.

# ABSTRACT

## NANOSCALE ELECTRIC PHENOMENA AT OXIDE SURFACES AND INTERFACES BY SCANNING PROBE MICROSCOPY

Sergei V. Kalinin
Dissertation Supervisor: Prof. Dawn A. Bonnell

Strong coupling between mechanical, electrical and magnetic properties in oxide materials, heterostructures and devices enable their widespread applications. Achieving the full potential of oxide electronics necessitates quantitative knowledge of material and device properties on the nanoscale level. In this thesis, Scanning Probe Microscopy is used to study and quantify the nanoscale electric phenomena in the two classes of oxide systems, namely transport at electroactive grain boundaries and surface behavior of ferroelectric materials.

The groundwork for the application of SPM for the determination of interface *I-V* characteristics avoiding contact and bulk resistivity effects is established. Scanning Impedance Microscopy (SIM) is developed to access ac transport properties. SIM allowed the interface capacitance and local *C-V* characteristic of the interface to be determined thus combining the spatial resolution of traditional SPMs with the precision of conventional electrical measurements. SPM of $SrTiO_3$ grain boundaries in conjunction with variable temperature impedance spectroscopy and *I-V* measurements allowed to find and theoretically justify the effect of field suppression of dielectric constant in the vicinity of the electroactive interfaces in strontium titanate. Similar approaches were used to study ferroelectric properties and ac and dc transport behavior in a number of polycrystalline oxides.

Polarization-related chemical properties of ferroelectric materials were investigated and quantified, leading to the discovery of the effects of potential retention above Curie temperature and temperature induced potential inversion. The origins of these phenomena were traced to the interplay between fast polarization and slow screening charge dynamics. Piezoresponse Force Microscopy (PFM) was used to study the polarization dynamics. An extensive description of contrast mechanisms in PFM conveniently represented in the form of "Contrast Mechanism Maps" was developed to relate experimental conditions such as tip radius and indentation force with the dominant tip-surface interactions. This topic was further developed to study the photochemical activity on ferroelectric surfaces as a function of domain orientation and use PFM to create predefined domain structures paving the way for photochemical assembly of metallic nanostructures on ferroelectrics.

# 1. INTRODUCTION

One of the most fascinating aspects of chemistry and physics of oxide materials is a wide variety of the properties they exhibit. While traditional semiconductor materials typically exhibit a single functionality and the coupling between mechanical, electrical and magnetic properties is relatively weak, this is not the case for the oxide materials. Century-old examples include strong electromechanical coupling in ferroelectric and piezoelectric materials that enable multiple applications as sensors, actuator and transducers.[1] More recent examples include perovskite manganites, in which the interplay between magnetic ordering and transport properties gives rise to the effect of colossal magnetoresistance and enables their potential applications for magnetic field sensors and magnetic heads.[2] Oxide systems allow novel paradigms for the electronic devices. The high degree of spin polarization in manganites provides one of the possible material bases for spintronics devices.[3] High temperature superconductors (HTSC) lend themselves for superconductive electronics; alternatively, mesoscopic quantum effects in Josephson junctions enable quantum-computing applications.[4] Switchable polarization in ferroelectric materials might enable non-volatile memory devices.[5]

This multitude of properties comes with the price. In traditional semiconductors, only few parameters must be controlled to achieve reliable device performance. Extensive knowledge has been accumulated on suitable microfabrication routes that do not degrade material properties and allow assembly of systems with complex device functionality. In comparison, the tendency to form oxygen vacancies and to develop concentration gradients due to dopant segregation in oxides significantly hinders the preparation of device-ready materials. Significant progress has been achieved in the last decade using Molecular Beam Epitaxy (MBE) and Pulsed Laser Deposition (PLD) techniques, which for the first time allowed preparation of epitaxial oxide films with extremely low defect density and enabled subatomic control of the composition, sparking a new interests for these applications.

It can be expected that oxide heterostructures and devices will significantly further existing technology. Indeed, in addition to the wide spectrum of physical properties of oxides *per se*, careful control and engineering of oxide interfaces presents multiple new opportunities. To mention a few, it was reported recently that multiplayer structures based on $SrTiO_3/BaTiO_3$ multilayer allow materials with advanced dielectric properties.[6] Other examples include interfacial magnetism in the $CaMnO_3/CaRuO_3$ system,[7] ferroelectric-semiconductor heterostructures,[8] and so on. An extremely important class of oxide interfaces is constituted by grain boundaries.[9] While the electronic applications described above are a nascent, albeit rapidly developing field, grain boundaries in bulk ceramic materials have been extensively studied and used for a better half of the century. In polycrystalline semiconductive oxides, the formation of electroactive interfaces due to dopant or vacancy segregation or the presence of interface states is a ubiquitous phenomena. Structure and topology of electroactive interfaces are known to influence greatly, and, in some cases, govern the properties of material. Grain boundary related transport properties of oxide materials provide the basis for applications such as varistors, thermistors, boundary layer capacitors, low field magnetoresistive devices, etc.[10] Alternatively, grain boundaries in HTSC materials act as weak links that limit the critical current density in these materials.[4]

One of the most fascinating aspects of oxide physics is related to the interplay between transport and non-linear dielectric properties. Operation of most electronic devices is associated with large potential gradients. In the incipient ferroelectrics high fields in the vicinity of the electroactive interfaces and grain boundaries can significantly affect the local dielectric properties. The immediate consequence of this effect is that traditional semiconductor characterization techniques such as *C-V* analysis, which are based on the assumption of constant dielectric properties, lead to the erroneous conclusions. An additional set of rich and complex phenomena stem from the presence of ferroelectric polarization. Indeed, it is well known that discontinuity in polarization gives rise to charge. Therefore, surfaces and interfaces in ferroelectrics in general are expected to be charged and this charge state can be controlled by external influences. In most perovskites, the polarization is sufficiently high to affect the electronic structure and chemical reactivity of surfaces and interfaces. In polycrystalline materials, spontaneous polarization partially compensates this interface charge below Curie temperature. Polarization decreases at higher temperatures, resulting in the increase of effective interface charge and grain boundary (GB) resistance.[10] On ferroelectric surfaces in the absence of surface states (which is typically the case on well-defined surfaces), polarization charge can induce surface band bending thus driving the surface in accumulation or inversion.[11] The opportunities presented by this approach are enormous – imagine the semiconductor structure that can be rendered *p*- or *n*- doped at will.

However, experimental opportunities for fundamental studies of these phenomena are limited. Ferroelectric domain size is usually small (tens of microns), precluding macroscopic studies of polarization dependent surface properties. The vast majority of transport measurements to date were performed on polycrystalline materials; thus, no relationship between the structure and properties of individual interface can be established. Moreover, transport measurements on lowly doped oxides suffer from poor contact quality that does not allow contact and GB phenomena to be separated unambiguously. Thus, the key to study these phenomena is the ability to perform local studies of electrical phenomena in oxides, most particularly transport and ferroelectric properties.

In this thesis, I have summarized the results of SPM studies of the local electric properties of oxide materials. The primary objects studied are $SrTiO_3$ (incipient ferroelectric) and $BaTiO_3$ (ferroelectric). These materials can be considered as models of ferroelectric perovskite with and without ferroelectric coupling. As such, they exhibit a fascinating interplay between ferroelectric, photoelectric and transport properties. The length scale of corresponding phenomena is small, thus necessitating spatially resolved studies, i.e. use of SPM techniques. Despite the tremendous progress in this field in the last decade, image formation mechanisms in most SPM techniques are not well understood.[12] Furthermore, some crucial techniques (e.g. imaging of ac transport properties) are lacking. Hence, the goal of this research is twofold. The first aim is to establish a reliable approach for the quantification of non-contact (EFM, SSPM) and contact (PFM) SPM imaging and extraction of materials properties from the experimental data. The second is to use this approach to study grain boundary behavior in $SrTiO_3$ as a model example of charged interfaces and domain structure and screening in $BaTiO_3$ as a model ferroelectric material. Local studies of phenomena at electroactive grain boundaries in piezoelectric oxides are included for extension of these results.

The general concept of SPM as applied to electric and ferroelectric phenomena is considered in Chapter 2 that summarizes current electrostatic SPM techniques.

In Chapter 3 the applicability of SPM for lateral transport measurements is considered. To rationalize existing SPM techniques, a classification scheme based on the measurements set-up functionality is developed. It is shown that Scanning Surface Potential Microscopy can be used in a manner similar to the 4 probe resistivity measurements employing the tip as a moving voltage probe. The development of Scanning Impedance Microscopy, which allows ac transport imaging, is described. Depending on the imaging frequency, both resistive and capacitive barriers at the interfaces can be visualized and quantified. Relevant theory and calibration methods are discussed. It is shown that combination of SSPM and SIM allows simultaneous acquisition of *I-V* and *C-V* data without the contribution of contacts and bulk resistance.

Chapter 4 presents the results of transport studies at the oxide interfaces. The physics of GB transport is briefly reviewed and it is shown that potential imaging can be used to detect the charged GBs. However, this technique is limited by the screening on the surface interface junction precluding the quantitative measurements of interface potential and depletion width. SSPM under lateral bias and SIM are applied to study dc and ac transport at the strontium titanate (STO) and the agreement between macroscopic and microscopic transport measurements is established thus excluding contact and grain bulk contributions. The results of variable temperature macroscopic I-V and impedance spectroscopy measurements are interpreted in terms of dielectric non-linearity at STO interface and corresponding theory is developed. Results of SPM imaging of transport behavior in polycrystalline materials and a number of examples are considered.

Properties of ferroelectric surfaces are considered in Chapter 5. The domain structure reconstruction from the combined topographic and electric measurements is presented. Origins of the electrostatic domain contrast are analyzed and it is shown that the polarization charge is almost completely screened. Variable temperature SSPM and PFM studies of surface potential on $BaTiO_3$ (100) are summarized and the relaxation times for polarization and screening charge are established. Finally, thermodynamic parameters of screening process are quantified.

The alternative approach to ferroelectric domain imaging is Piezoresponse Force Microscopy. In PFM, the image contrast can be attributed both to the electrostatic and electromechanical tip-surface interactions. A detailed analysis of tip-surface interactions in PFM is presented in Chapter 6 and the guidelines for quantitative imaging are developed.

The relationship between polarization and surface electronic properties is further considered in Chapter 7. It was recently discovered that photoelectric activity of BTO is strongly dependent on domain orientation. Here, the local domain structure and photochemical activity are correlated using PFM and AFM. A charged tip is used to control local polarization on the submicron scale, paving the way for the controlled assembly of metallic nanostructures. The applicability of e-beam patterning for selective poling of ferroelectric surface is investigated. Finally, the implications for nanotechnology are discussed on an example of the device fabrication using ferroelectric lithography.

[10] L.L. Hench and J.K. West, *Principles of Electronic Ceramics* (Wiley, New York, 1990).

[11] V.M. Fridkin, *Ferroelectric Semiconductors* (Consultants Bureau, New York, 1980).

[12] *Scanning Probe Microscopy and Spectroscopy: Theory, Techniques and Applications*, ed. D.A. Bonnell (Wiley VCH, New York, 2000).

# 2. FUNDAMENTALS OF ELECTRICAL SCANNING PROBE MICROSCOPIES

## 2.1. Introduction

Progress in modern science is impossible without reliable tools to characterize the structure and physical properties of materials and devices on micron and submicron length scales. While structural information in most cases can be obtained by such established techniques as scanning and transmission electron microscopy,[1,2] determination of local electronic structure, electric potential and field, and chemical functionality with high spatial resolution is a much more daunting problem. Local electronic properties became accessible after the invention of Scanning Tunneling Microscopy.[3] The rapid development of STM was due to the demonstrated atomic resolution and the ability to probe electronic structure (Scanning Tunneling Spectroscopy).[4] However, dc current feedback used in the vast majority of STMs limits this technique to conductive surfaces. On semiconductive surfaces, tip-induced band bending severely limits the resolution.

It was realized that an alternative to the current probe used in STM is the force probe comprised of the probe tip and force sensor. The concept of local probes interacting with the surface via a pliable cantilever, used for the detection of the probe-surface force was established in the seminal work by Binnig, Quate and Gerber.[5] Initially, Atomic Force Microscopy (AFM) was developed as a tool sensitive to strong short-range repulsive forces. It was almost immediately realized that AFM can be extended to map forces of a different nature such as magnetic and electrostatic forces or chemical interactions.[6,7,8,9,10,11,12,13,14] Since then, the number of available Scanning Probe Microscopy (SPM) techniques has greatly increased, not in the least due to availability and adaptability of commercial AFM devices. It wouldn't be an exaggeration to say that the rapid development of nanoscience and technology in the last decade is due largely to the availability of AFM and SPM that enable imaging and manipulation of submicron structures. The purpose of this chapter is to provide a general introduction to the Scanning Probe Microscopy techniques starting from topographic imaging to more complex issues of local property measurements and to summarize some of the unresolved problems in this field that will be addressed in the present thesis. The specific question of transport measurements by SPM is deferred to Chapter 3.

## 2.2. Origins of SPM and Topographic Imaging

Historically, Atomic Force Microscopy (AFM) was designed to measure the strong short-range repulsive forces between a tip and surface implemented as contact mode imaging. The probe (tip) is brought into contact with the surface (hence the name) and repulsive Van-der-Waals (VdW) forces result in the deflection of the cantilever, which is monitored by an STM, optical, heterodyne or capacitive system. A feedback loop keeps the deflection constant by adjusting the vertical position of the cantilever while scanning along a surface. The feedback signal provides the topographical profile of the surface. Operation in contact mode typically implies relatively large shear forces that can damage the tip and the surface, limiting the range of samples that can be imaged. Progress in topographic AFM led to the development of intermittent contact mode imaging.[15] This approach utilizes a mechanically driven cantilever-tip system. The mechanical oscillations are imposed by a piezoelectric actuator. The oscillation amplitude at a fixed driving voltage on the actuator is detected with a lock-in amplifier. While approaching the surface, the tip eventually comes into intermittent contact and the oscillation amplitude decreases. Similar to contact mode imaging, the feedback loop keeps the oscillation amplitude constant while scanning and the feedback signal provides the topographic profile of the surface. In this regime, the tip touches the surface at high incidence angle, precluding surface damage. AFM in an ultra high vacuum (UHV) environment is typically performed in the non-contact mode, in which the tip oscillates in the attractive region of the Van der Waals forces, and the resonant frequency shift due to the Van der Waals force gradient is used as the feedback signal. In this case, the tip doesn't contact the surface at all.

In principle, local property measurements can be performed in each of the topographic regimes. However, reliable detection of the electromagnetic interactions requires imaging at relatively large tip-surface separations, where they dominate over short–range Van der Waals forces thus excluding imaging in contact mode (Figure 2.1).

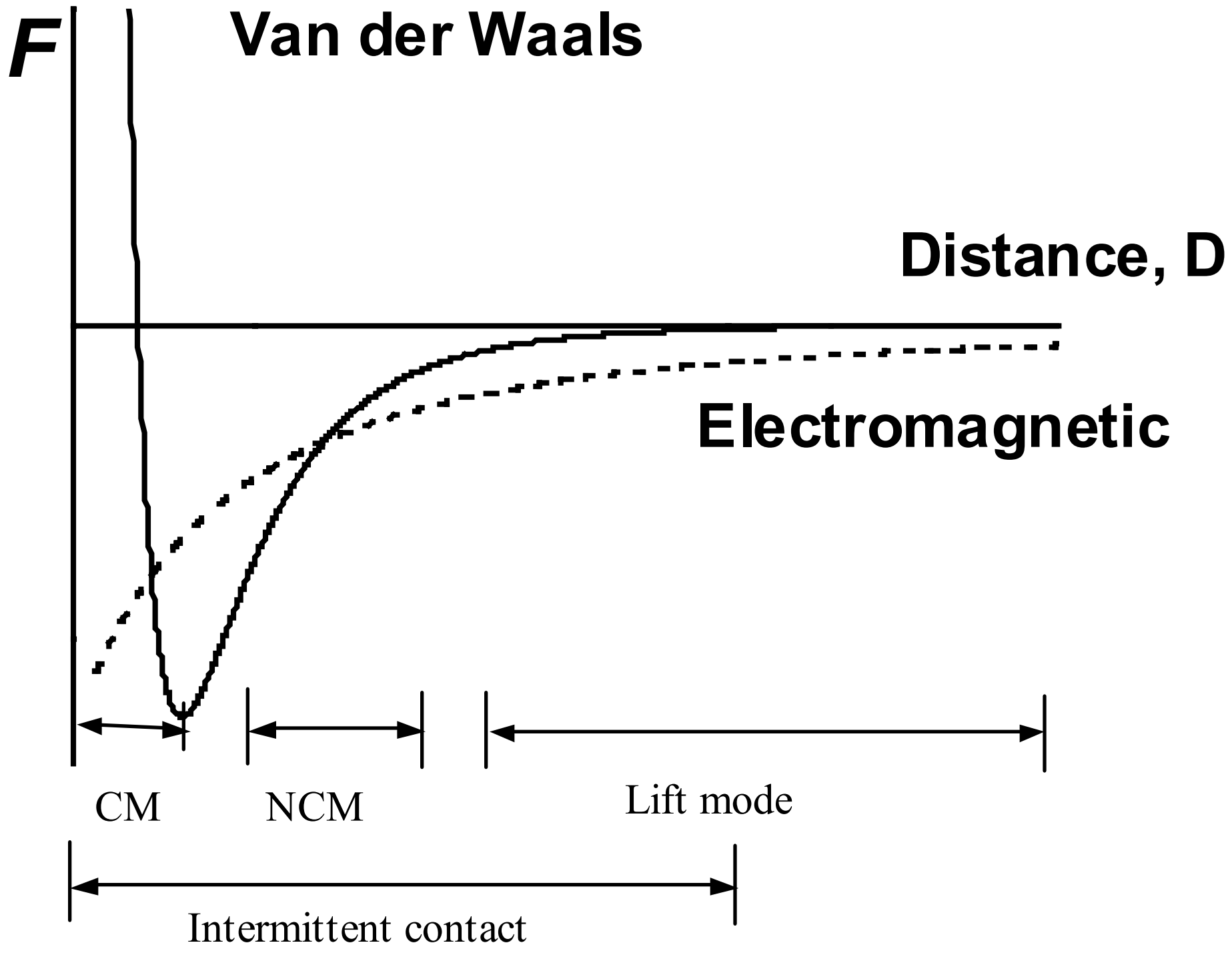

**Figure 2.1.** Distance dependence of Van Der Waals and electrostatic forces compared to the typical tip-surface separations in the contact mode (CM), non-contact mode (NCM), intermittent contact mode and lift mode. In the last case, the tip cannot acquire the topographic information and additional scan is necessary to position the tip at required separation from the surface.

In intermittent contact mode periodic tip-surface interactions, rather than long-range forces, provide the dominant contribution to the dynamic behavior of the cantilever. The major step in the electromagnetic force detection was achieved with the invention of a lift mode.[16] In this mode, the tip acquires topographic data near the surface, and then retraces the topographic profile at a predefined height above the surface to measure the force interactions. A similar approach can be implemented in point-by-point lift mode measurements.[17]

### 2.3. Electrical Interaction Based SPMs

The first applications of SPM for local potential and electromagnetic field measurements were based on its functionality as a force sensor. Since reliable measurement of small static deflections is difficult, most SPM techniques utilize the dynamic response of the tip to a periodic mechanical force (tapping or intermittent contact mode AFM), oscillating bias on the tip or on the sample (voltage modulation techniques) or an oscillating magnetic field. Periodic perturbation can result either in a static (DC) response of the tip, a response at the main frequency of perturbation (first harmonic signal) or a response at twice the main frequency of the perturbation (second harmonic signal). Lock-in techniques allow extraction of the amplitude and phase of the first or the second harmonic of the response for subsequent use as feedback or data signals. Major SPM techniques employ either mechanically or electrostatically driven cantilevers, even though other driving regimes are possible.[18] The techniques based on the detection of force-induced deflection of the cantilever are referred to as scanning force microscopies.

In addition to force detection, the ac or dc current through the probe tip can be monitored. The nature of this current strongly depends on the characteristic tip-surface separation and the experimental environment. For large (10-100 nm) separations, only ac displacement current due to capacitive tip-surface coupling is detected. In the tunneling range (0.1 – 10 nm), both displacement current (ac) and tunnel current (both ac and dc) can be measured.[19,20] In this regime, the microscope can operate as an AFM and STM probe simultaneously providing the complementry information on the surface structure and properties. Finally, in the contact regime the resistive tip-surface current can be monitored and used to quantify tip-surface junction properties and local resistivity of the surface.[21] The tip can also be used as a local capacitive sensor to measure tip-sample capacitance, giving rise to scanning capacitance microscopy.[22] The local probe techniques based on the detection of the current, optical or other signals in addition to force- or current based topography are referred to as scanning probe microscopies. Some of the more advanced experimental setups based on conventional electric SPMs are discussed at length in Chapter 3.

## 2.4. Force Detection: Non-contact SPMs

All electrostatic SFM techniques are based on the dynamic response of the probe to a mechanical or voltage modulation. During operation in the mechanically driven mode, the voltage on the actuator driving the cantilever, $V_{\text{piezo}}$, is

$$V_{piezo} = V_{acp}\sin(\omega_p t), \tag{2.1}$$

where $\omega_p$ is the driving frequency. The oscillating voltage on the actuator induces cantilever oscillations and the tip-surface separation is

$$d = d_0 + A(\omega_p)\sin(\omega_p t + \varphi_c), \tag{2.2}$$

where $A(\omega_p)$ is the frequency dependent oscillation amplitude and $\varphi_c$ is the phase shift between the driving voltage on the piezo and the cantilever oscillations. The driving frequency, $\omega_p$, is typically selected close to the resonant frequency of the cantilever to ensure a strong response. During operation, the conductive tip is either grounded or biased with a DC voltage.[23,24] The presence of an electrostatic force gradient near the surface results in a change of the resonance frequency of the cantilever that can be detected as a resonance frequency shift (frequency detection), while the feedback adjusts $\omega_p$ to keep $A(\omega_p)$ maximal, a shift of the oscillation amplitude at constant driving frequency (amplitude detection), or a phase shift (phase detection). In order to minimize the influence of surface topography on the dynamic properties of the cantilever, this technique is usually implemented in the lift mode. Mechanical contact between the tip and the surface strongly influences the dynamic properties of the cantilever and information on electrostatic interactions can no longer be unambiguously extracted from experimental data. Instead, mechanical tip-surface interactions can be quantified. Phase detection in the intermittent contact regime provides information on the elastic properties of the surface (phase imaging),[25] which is out of the scope of the present chapter.

An alternative approach involves voltage modulation techniques. The driving voltage at the actuator is set to zero (the tip is no longer mechanically driven) and a conductive tip is biased by an AC voltage. In this operational regime the tip potential is

$$V_{tip} = V_{dc} + V_{ac}\sin(\omega t). \tag{2.3}$$

Biasing the tip at large tip-surface separations above metallic or linear dielectric ($\kappa$ = constant) surfaces results in a static force, forces at the frequency of the tip voltage (first harmonic), and at twice the frequency of the tip voltage (second harmonic). All components contribute to the deflection of the cantilever, and tip-surface separation is

$$d = d_0 + A_0 + A_1 \sin(\omega t + \varphi_1) + A_2 \sin(2\omega t + \varphi_2), \qquad (2.4)$$

where $d_0$ is tip-surface separation when $V_{tip} = 0$, $A_0$ is static response and $A_1$, $A_2$, $\varphi_1$ and $\varphi_2$ are amplitudes and phase shifts of first and second harmonic responses. Magnitudes of $A_0$, $A_1$ and $A_2$ are relatively small and only the latter two components along with corresponding phase shifts can be determined by lock-in technique. Separation of the first and second harmonic responses allows quantification of different components of tip-surface force, in contrast to the determination of the total force gradient in case of the mechanically driven mode.

Biasing the tip in contact mode results in tip displacement owing to both electrostatic forces and electromechanical effects, such as the inverse piezoelectric effect and electrostriction. For linear piezoelectric materials tip deflection as a function of applied bias is similar to Eq.(2.4). However, this technique is most widely used for the characterization of ferroelectric materials, in which case the electromechanical response of the surface to applied voltage is considerably more complicated and will be discussed extensively in Chapter 6.

Voltage modulation and mechanical modulation can be combined, i.e., the tip can be driven both mechanically by the piezo at frequency $\omega_p$ and electrostatically at frequency $\omega$. Depending on whether these modulations are applied in the intermittent contact or non-contact mode, interpretation of the responses at the main frequency, $\omega$, and the second harmonic can be done along the lines discussed above. Obviously, quantification of electrostatic, electromechanical and elastic contributions to the signal is more challenging in this case and the first two modulation modes are far more widespread.

### 2.4.1. Tip-surface Forces and Contrast Transfer

Measurement of local materials properties by SPM requires quantitative understanding of the contrast formation mechanism. In SFM techniques, the key contributions to the contrast are tip-surface forces determined by material characteristics of interest and the tip properties and the force detection scheme. Hence, the discussion of electrostatic SFM would be incomplete without a consideration of the electrostatic forces and their implications of resolution and sensitivity of the techniques.

The electrostatic tip-surface interaction depends both on the electric and geometric properties of the tip (conductive/dielectric, tip shape) and the surface (conductor with a well-defined potential or an insulator with given volume or surface charge density, surface topography). For all practical purposes, however, conductive probes provide much better control over the tip properties since the voltage on the tip can be easily controlled. For such probes, tip potential is well defined and the tip shape can be either measured directly or approximated by appropriate geometric model.

For a conductive surface with a constant potential, the force between the tip and surface is:

$$F(z) = \frac{1}{2}\left(V_{tip} - V_{surf}\right)^2 \frac{\partial C(z)}{\partial z} , \tag{2.5}$$

where $F(z)$ is the force, $V_{tip}$ is tip potential, $V_{surf}$ is surface potential, $z$ is vertical tip-surface separation, $C(z)$ is a tip-surface capacitance. The surface potential is defined as $V_{surf} = V_{el} + \Delta CPD$, where $V_{el}$ is electrostatic potential with respect to the microscope ground and ΔCPD is contact potential difference between the tip and the surface (electrochemical potential).[26] Eq.(2.5) implies that the force is a function of the tip and surface geometry through the $C(z)$ term. It can be easily shown that $C(z)$, and consequently $F(z)$, are rapidly decaying functions of tip-surface separation, thus the measurable signal can be obtained only at relatively small tip-surface separations. The exact functional form of tip-surface capacitance $C(z)$ is very complicated even for flat surfaces and can be obtained only by numerical methods, such as finite element analysis (FEA). For the typical probe geometry the total capacitance $C(z)$ can be conveniently approximated as a sum of the contributions due to the tip apex, tip bulk and the cantilever that having the spherical, conical and plane geometry correspondingly:

$$C(z) = C_{apex}(z) + C_b(z) + C_c(z). \tag{2.6}$$

It is understood in Eq.(2.6) that the local part of the force that enables the high resolution of the SFM contribution is due to the tip apex, whereas the cantilever contribution is non-local. The tip bulk contribution is more difficult to interpret. However, comparing the relative magnitudes of $C_{apex}(z)$, $C_b(z)$ and $C_c(z)$ as a function of tip-surface separation, $z$, and relevant tip parameters, allows insight into the image formation mechanism of SFM.

A number of approximate models have been suggested to quantify the capacitive force between the tip and the surface. Some of these models use approximate geometric descriptions of the tip as a plate capacitor, a sphere,[27], a hyperboloid,[28,29,30] a cone[31] or a cone with spherical apex.[32] An alternative is to use an equivalent image charge such that the corresponding constant potential surface represents the actual tip.[33,34] Examples are point charge and line charge configurations.[35] Image charge distributions can be also found by numerical methods.[34] The advantage of the image charge approach is that it reduces a complicated boundary-value problem for potential to a much simpler problem of Coulombic charge-charge interaction, while preserving the characteristic features (distance and voltage dependence of the force, tip shape effects) of the original problem.

At small tip-surface separations ($z<R$, where $R$ is tip radius of curvature), the spherical tip apex provides the major contribution to the force. The bias and distance dependence of the force is best described by sphere or point charge models with the solution of the form $F = \gamma V^2 / z$, where $\gamma$ is a constant depending on the specific model. In the point-charge model the spherical part of the tip is represented by the point charge such that the curvature of the isopotential surface is equal to physical curvature of the tip. In this case

$$F_{cap} = \frac{3\pi}{4}\varepsilon_0 V^2 \frac{R}{z} \quad \text{and} \quad \frac{dF_{cap}}{dz} = -\frac{3\pi}{4}\varepsilon_0 V^2 \frac{R}{z^2}. \qquad (2.7a,b)$$

For larger tip-surface separations ($z>R$, where $R$ is tip radius of curvature) hyperboloid, cone or line charge approximations provide the best description. These models predict a logarithmic dependence of capacitive force on tip-surface separation of the general form $F = \eta V^2 \ln(D/z)$, where $\eta$ and $D$ are parameters related to the tip geometry. Particularly simple description of tip-surface interactions in this regime can be achieved using line charge model. Here, image charge distribution is approximated by a semi-infinite uniformly charged line with line charge density $\lambda$. The axially symmetric potential for the line is

$$V(x,z)=\frac{\lambda}{4\pi\varepsilon_0}\ln\left(\frac{d+z+\sqrt{(d+z)^2+x^2}}{d-z+\sqrt{(d+z)^2+x^2}}\right), \tag{2.8}$$

where $d$ is the distance from the lower end of the line to the surface. It can be shown that

$$d=h\alpha\,, \quad \text{where} \quad \alpha=\sqrt{1+\tan^2\theta}\,, \tag{2.9a,b}$$

and $h$ is the separation between the tip apex and surface and $\theta$ is the half-angle of the cone, which is the equipotential surface that represents the tip. The line charge density:

$$\lambda=\frac{4\pi\varepsilon_0 V}{\beta}, \quad \text{where} \quad \beta=\ln\left(\frac{1+\cos\theta}{1-\cos\theta}\right), \tag{2.10a,b}$$

depends on the equipotential surface geometry. The expression for the force is then:

$$F_{cap}=\frac{\lambda^2}{4\pi\varepsilon_0}\ln\left(\frac{(2d+L)^2}{4d(d+L)}\right), \tag{2.11}$$

where $L$ is the effective tip size. For $d<<L$ and small angles Eq.(2.11) reduces to:

$$F_{cap}=\frac{\lambda^2}{4\pi\varepsilon_0}\ln\left(\frac{L}{4z}\right) \quad \text{and} \quad \frac{\mathrm{d}F_{cap}}{\mathrm{d}z}=\frac{\lambda^2}{4\pi\varepsilon_0}\frac{1}{z}. \tag{2.12a,b}$$

This relation predicts the logarithmic dependence of capacitive force on tip-surface separation expected from simple considerations, but includes the effects of actual tip geometry. For a realistic tip shape including tip bulk and rounded tip apex, total force and force gradient acting on the tip can thus be approximated as

$$F_{cap}=V^2\left(\frac{\gamma}{z}+\eta\ln\left(\frac{D}{z}\right)\right) \quad \text{and} \quad \frac{\mathrm{d}F_{cap}}{\mathrm{d}z}=V^2\left(-\frac{\gamma}{z^2}+\frac{\eta}{z}\right), \tag{2.13a,b}$$

where $\gamma$ and $\eta$ are tip-shape dependent parameters that can be found experimentally from the analysis of force or force-gradient - distance curves and used to extract geometric parameters of the tip.

The cantilever contribution to total force and force gradient can be approximated by plane-plane capacitor model, in which case

$$F_{cap}=\frac{\varepsilon_0 V^2}{2}\frac{S}{(z+L)^2} \quad \text{and} \quad \frac{\mathrm{d}F_{cap}}{\mathrm{d}z}=-\varepsilon_0 V^2\frac{S}{(z+L)^3}, \tag{2.14a,b}$$

where $S$ is effective cantilever area and $L$ is tip length. For typical metal coated tip used in the EFM/SSPM measurements with $R \approx 30$nm, $\theta = 17°$, $L \approx 10$ μm, $S \approx 2 \cdot 10^3$ μm$^2$ contributions of tip apex, tip bulk and cantilever to overall force and force gradient are shown in Figure 2.2.

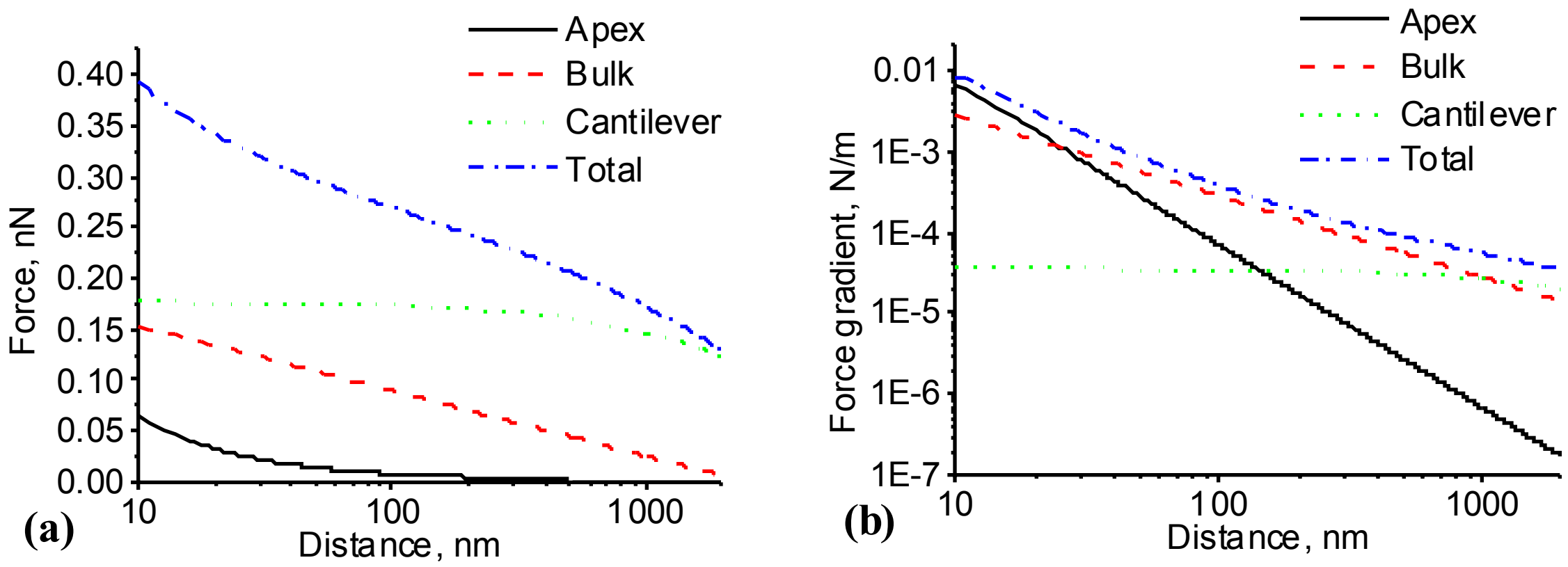

**Figure 2.2.** Relative contributions of tip apex, tip bulk and the cantilever to the total electrostatic force (a) and force gradient (b) for tip parameters defined in text.

As seen from Figure 2.2a, for intermediate and large tip-surface separation the cantilever provides the largest contribution to electrostatic force. At the same time, the major contribution to force gradient acting on the tip is due to the tip bulk, the cantilever providing distance-independent offset. It should be noted that the force acting on the cantilever is distributed, unlike the localized force acting on the tip and tip apex, and thus its influence on the dynamic and static properties of cantilever is different. In addition, actual cantilever configuration (i.e. tilted with respect to the surface) results in smaller force and larger force gradient; however, the estimates presented here are still valid.

Quantification of the electrostatic tip-surface interaction is significantly more complicated for the lossless dielectric materials. Using the analogy between the interaction of a point charge, $Q$, and a conductive or a dielectric surface, the electrostatic force can be written as that due to the interaction of charge with the corresponding image charge $Q_{im}$. For a conductive plane and a dielectric plane the forces are:

$$F_{cond}(z) = -\frac{1}{4\pi\varepsilon_0}\frac{Q^2}{4z^2}, \qquad F_{diel}(z) = -\frac{1}{4\pi\varepsilon_0}\left(\frac{\kappa-1}{\kappa+1}\right)\frac{Q^2}{4z^2}, \qquad (2.15a,b)$$

i.e. $Q_{im}$ = - $Q$ for the metal while $Q_{im} = -(\kappa-1)/(\kappa+1)Q$ for the dielectric plane. In both cases, the force is proportional to the square of the charge, resulting in a parabolic dependence on the bias voltage. Thus, the interaction force between a dielectric surface and a biased tip can be described in terms of an effective capacitance $C(z,\kappa)$, where $\kappa$ is the dielectric constant of the material. In the limit of high dielectric constant $C(z,\kappa) \approx C(z)$ and this assumption is widely used in quantification of EFM data.

The primary difference between conductors and dielectrics is that, in addition to tip-induced image charges responsible for the capacitive interaction, dielectric media can also sustain both surface charges and volume trapped charges, which contribute to the total electrostatic force. In fact, determination of surface- or volume charge density in dielectric is often the purpose of the EFM experiment. Analytical treatment of this problem is possible only for the simplest cases.[36,37] However, the important features of the electrostatic tip-surface interaction on dielectrics can be understood even within the framework of the point-charge model presented for the case of oscillatory tip bias.

Following Sarid,[18] for a charge $Q_s$ on a dielectric surface the electrostatic interaction between the biased tip and the sample includes three distinct components: effective surface charge on the sample, $Q_s$, constant dc bias applied between the tip and the sample, $V_{dc}$, and periodic ac bias applied between the tip and the sample, $V_{ac}\cos(\omega t)$. The charge on the tip can be approximated as

$$Q_t = Q'_s + Q_{dc} + Q_{ac}, \tag{2.16}$$

where $Q'_s$ is the image charge on the tip and

$$Q_{dc} = V_{dc}\, C, \tag{2.17}$$

$$Q_{ac} = C\, V_{ac} \cos \omega t, \tag{2.18}$$

and $C$ is the effective capacitance of the tip-sample system. Here the interaction between the tip and the surface is approximated by a corresponding capacitance. The induced charge on the tip can be found only for simple geometries. For localized charge $Q_s$ close to spherical tip apex ($z << R$, $R$ is the apex radius of curvature) induced charge $Q'_s \approx -Q_s$.

The force between the tip and the sample is a sum of two contributions; the first being due to the charge-induced charge interaction, the other due to the capacitance. The total expression for the force can be qualitatively described as:

$$F = \frac{Q_s Q_t}{4\pi\varepsilon_0 z^2} + \frac{1}{2} C'(V_{ac} + V_{dc})^2 . \qquad (2.19)$$

Substituting the expression for $Q_t$, Eq.(2.16) becomes the sum of three components:

$$F = F_{dc} + F_{1\omega} + F_{2\omega}, \qquad (2.20)$$

where

$$F_{dc} = \frac{Q_s^2}{4\pi\varepsilon_0 z^2} + \frac{Q_s V_{dc}}{4\pi\varepsilon_0} \frac{C}{z^2} + \frac{1}{2} C' \left( V_{dc}^2 + \frac{1}{2} V_{ac}^2 \right) \qquad (2.21)$$

is a static component of the force,

$$F_{1\omega} = \left( \frac{Q_s}{4\pi\varepsilon_0} \frac{C}{z^2} + C' V_{dc} \right) V_{ac} \qquad (2.22)$$

is the first harmonic component, and

$$F_{2\omega} = -\frac{1}{4} C' V_{ac}^2 \qquad (2.23)$$

is the second harmonic component. The expression for the corresponding force gradients can be easily obtained from Eq.(2.21-23).

Thus, the electrostatic force on the tip due to the oscillating voltage at frequency, $\omega$, applied between the tip and linear dielectric surface is subdivided into three categories:

1. A harmonic response with the same frequency as external bias due to surface and volume trapped charges and dipole charges of remanent (tip-bias independent) polarization.
2. A second harmonic response due to the interaction of the tip charges and bias-induced polarization image charges on the surface.
3. A static response due the surface charge- tip image charge interaction, tip charge - surface image charge interaction and static component of second harmonic response.

Therefore, static dielectric charges and tip-induced charges contribute to the harmonic components of the force differently and can be distinguished. In practice, however, the lack of the information on tip-surface capacitance and the charge state of the dielectric significantly complicate data analysis.

Further complications in analysis of tip-surface interactions arise for the lossy dielectric materials. In this case, tip-induced currents in the sample result in the increased damping that affects tip dynamics, i.e. phase and amplitude behavior. Preliminary results utilizing these phenomena were recently reported.[38]

### 2.4.2. Electrostatic Force Microscopy

Force gradient imaging with conductive cantilevers is one of the most common EFM techniques and is available on most commercial instruments. This technique is usually implemented in the lift mode. The grounded tip first acquires the surface topography using standard intermittent contact AFM. Electrostatic data are collected in the second scan, during which the tip retraces the topographic profile separated from the surface 50 to 100 nm, thereby maintaining a constant tip-sample separation. During operation in this mode, the cantilever is mechanically driven by the actuator.

The equation of motion for the tip in the point-mass approximation is

$$m\ddot{z} + \gamma\dot{z} + kz = F(z), \tag{2.24}$$

where $m$ is effective mass of the tip, $z$ is tip displacement, $\gamma$ is damping coefficient, $k$ is the cantilever spring constant and $F(z)$ is tip-surface force. Expanding $F(z)$ in Taylor series around equilibrium position of the tip $z_0 = 0$ as $F(z) = F_0 + \frac{dF}{dz}z$, Eq.(2.24) can be rewritten as

$$m\ddot{z} + \gamma\dot{z} + \left(k - \frac{dF}{dz}\right)z = F_0. \tag{2.25}$$

In other words, the presence of an electrostatic force gradient results in a change of the effective spring constant of the cantilever. Attractive tip-surface force renders the cantilever "softer", while repulsive force makes it "stiffer". The change of the spring constant changes the resonant frequency of the cantilever $\omega_r = \sqrt{k/m}$ and to first order the resonant frequency shift is

$$\Delta\omega = \frac{\omega_0}{2k}\frac{dF}{dz}, \tag{2.26}$$

where $\omega_0$ is the resonant frequency of free oscillating cantilever.

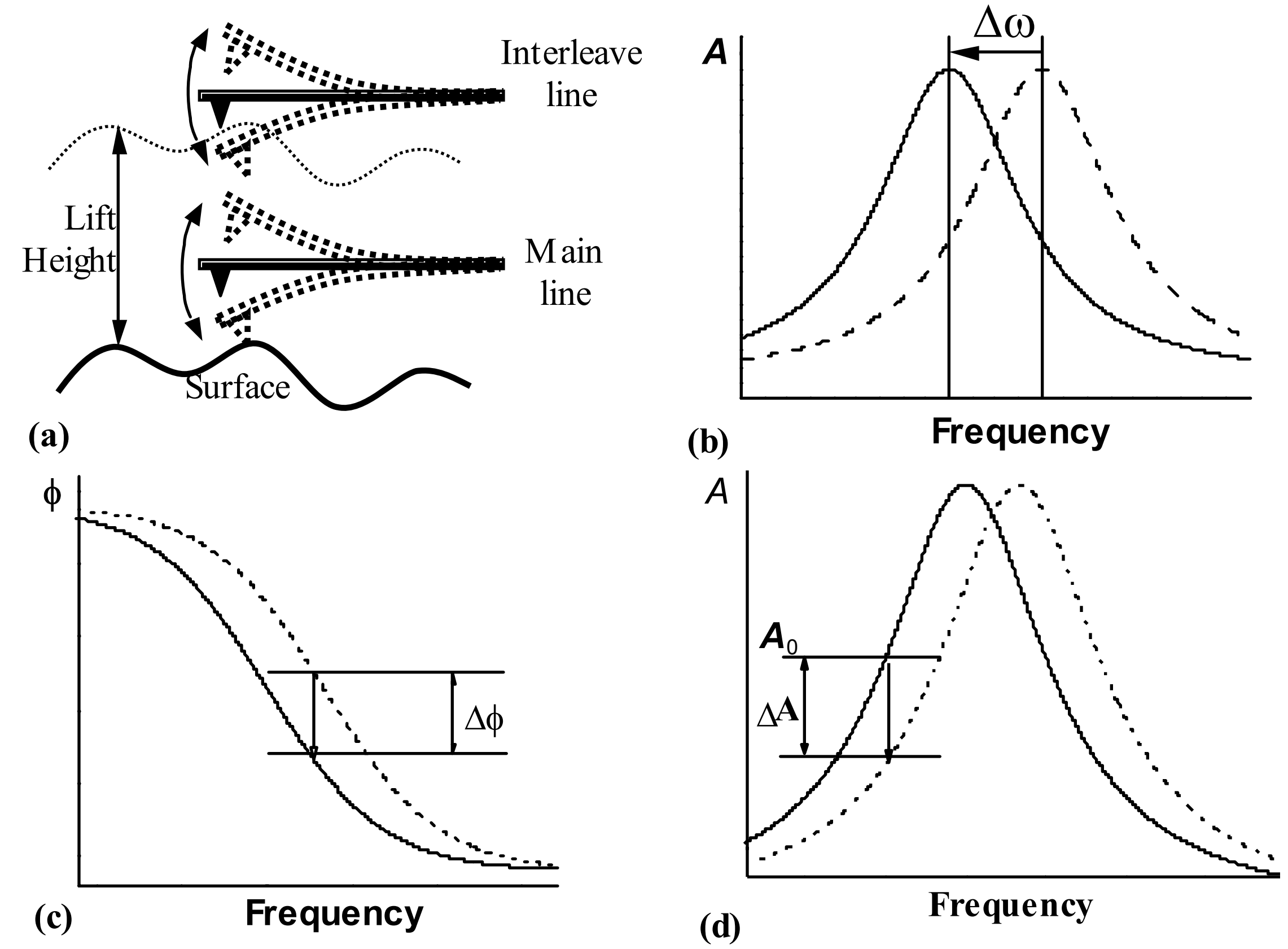

**Figure 2.3.** Schematic experimental set-up for Electrostatic Force Microscopy (a). Electrostatic force gradient shifts the resonance frequency of the cantilever which can be detected directly (b), through the change of oscillation phase at the resonant frequency of free cantilever (c) or through the amplitude change at the point of maximum slope in the amplitude-frequency curve (d).

As illustrated in Figure 2.3, resonance frequency shift due to the electrostatic force gradients results in the phase or amplitude shifts relative to the tip oscillating in the absence of the force gradient. Therefore, collecting phase and amplitude for mechanically modulated cantilever at a constant driving frequency as a function of tip bias allows imaging electrostatic properties of the surface. For small force gradients the phase shift at the resonance frequency is

$$\Delta\phi = \frac{Q}{k}\frac{dF}{dz}, \tag{2.27}$$

where $Q = m\omega_0/\gamma$ is the quality factor of the cantilever. Alternatively, the amplitude shift at the steepest part of amplitude-frequency curve is

$$\Delta A = \frac{2A_0}{3\sqrt{3}}\frac{Q}{k}\frac{dF}{dz}, \qquad (2.28)$$

where $A_0$ is the oscillation amplitude for a free cantilever.

For large force gradients Eqs.(2.27,28) are no longer valid since the resonance frequency shift due to the force gradient may exceed the width of the resonant peak; in addition, it amplitude and phase data are sensitive to topographic artifacts. An alternative approach for force gradient imaging includes the use of the Phase Locked Loop (PLL) to determine the resonant frequency of the cantilever and an additional feedback loop to adjust the driving frequency so that the cantilever is always at resonance. In this frequency-detection regime, the relationship between the resonant frequency shift and the force gradient is given by Eq.(2.26).

Further insight into the physics of tip-surface interaction can be obtained from the oscillation amplitude at constant excitation at the resonant frequency. The oscillation amplitude of the free cantilever is determined by the damping factor $\gamma$. Losses in the sample will increase $\gamma$ and decrease the oscillation amplitude. The latter, therefore, is related to the losses in materials below the tip and measuring the oscillation amplitude yields information on local conductivity. This approach was first developed for magnetic force imaging in Magnetic Dissipation Force Microscopy (MDFM).[39,40] The dissipation force EFM on semiconductor materials was first demonstrated using dithering topographic feedback.[41] The losses in this case arise due to the lateral motion of tip-induced depleted region and were shown to be related to the carrier concentration. It was shown recently that a similar technique can be implemented with conventional EFM as well and can be used to study single-electron charging dynamics in carbon nanotubes.[42,43]

### 2.4.3. Scanning Surface Potential Microscopy

As shown in the previous section, a tip can be made sensitive to electrostatic forces by biasing it with respect to the surface. However, for a DC biased tip the force gradient signal results from the combination of several possible interactions and individual components can be resolved only if the measurements are performed at different scan heights (thus varying $C(z)$) or different tip biases. In many cases, the sensitivity of the tip to the electrostatic forces and topographical artifacts doesn't allow high potential resolution. These disadvantages of force gradient EFM led to the development of an alternative voltage modulation SFM techniques, one of the most well-known examples of which is scanning surface potential microscopy (SSPM), also known as Kelvin Probe Microscopy (KPM).[44,45,46]

As in EFM, this technique is usually implemented in the lift mode. In the second scan, the actuator is disengaged and an oscillating bias is applied to the tip. Assuming that the surface is characterized by uniform (or, rather, slow varying) surface potential, $V_{surf}$, the capacitive force between the tip and the surface is given by Eq.(2.5). Modulation of the tip potential $V_{tip} = V_{dc} + V_{ac}\cos(\omega t)$ yields

$$F(z) = \frac{1}{2}\frac{\partial C(z)}{\partial z}\left[\left(V_{dc} - V_{surf}\right)^2 + \frac{1}{2}V_{ac}^2\left[1 - \sin(2\omega t)\right] + 2\left(V_{dc} - V_{surf}\right)V_{ac}\cos(\omega t)\right]. \quad (2.29)$$

Eq.(2.29) shows that the force between the tip and the surface due to the application of an ac bias on the tip has static component $F_{dc}$, first harmonic component $F_{1\omega}$ and second harmonic component $F_{2\omega}$, explicit expressions for which are given below. Note that the dc component of tip bias results in static and first harmonic components, while the ac component contributes to all three components.

$$F_{dc}(z) = \frac{1}{2}\frac{\partial C(z)}{\partial z}\left[\left(V_{dc} - V_{surf}\right)^2 + \frac{1}{2}V_{ac}^2\right], \quad (2.30)$$

$$F_{1\omega}(z) = \frac{\partial C(z)}{\partial z}\left(V_{dc} - V_{surf}\right)V_{ac}, \quad (2.31)$$

$$F_{2\omega}(z) = \frac{1}{4}\frac{\partial C(z)}{\partial z}V_{ac}^2. \quad (2.32)$$

The lock-in technique allows extraction of the first harmonic signal in the form of first harmonic of tip deflection proportional to $F_{1\omega}$. A feedback loop is employed to keep it equal to zero (hence the term nulling force approach) by adjusting $V_{dc}$ on the tip. Obviously, the condition $F_{1\omega} = 0$ is achieved when $V_{dc}$ is equal to $V_{surf}$ (Eq.(2.31)). Thus, the surface potential is directly measured by adjusting the potential offset on the tip and keeping the first harmonic response zeroed. It is noteworthy that the signal is independent of the geometric properties of tip-surface system (i.e. $C(z)$) and the modulation voltage. This technique allows very high (~mV) potential resolution.

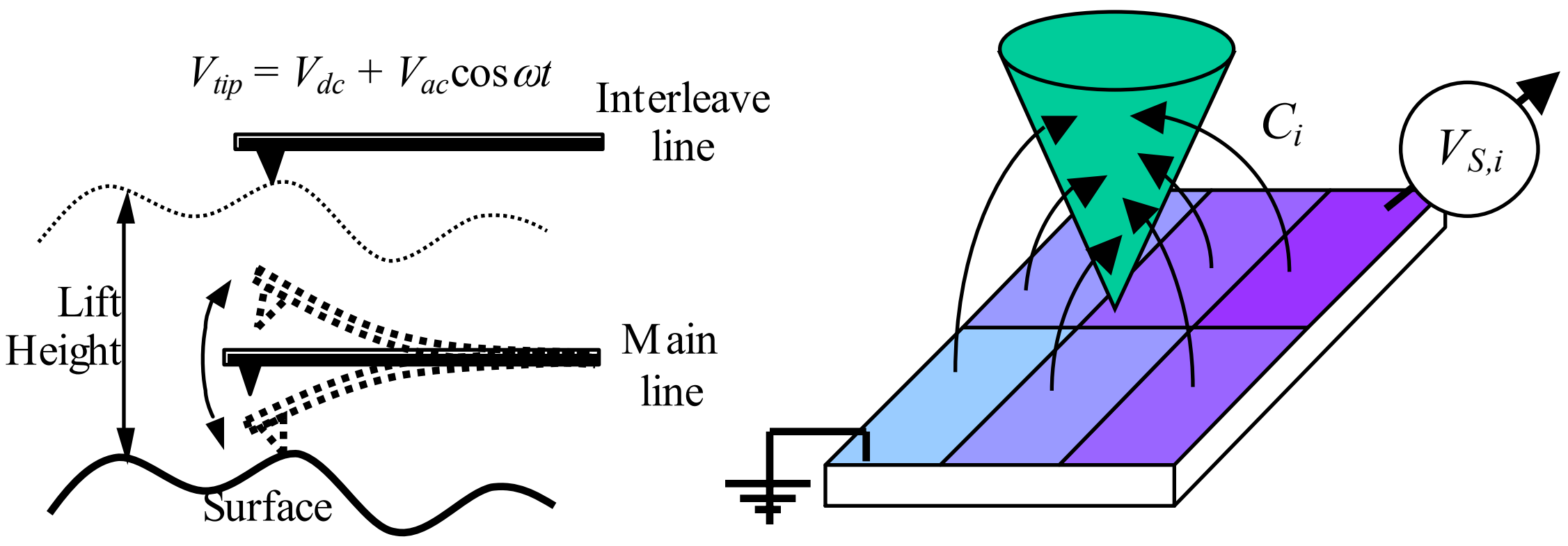

**(a)** **(b)**

**Figure 2.4.** Schematic experimental set-up for scanning capacitance microscopy (a) and diagram of how a heterogeneous potential distribution on a sample surface contributes to the capacitance measured by a tip (b).

For a realistic surface with a non-uniform surface potential distribution and non-uniform topography, Eq.(2.5) can be rewritten as the sum of partial capacitive interactions of the tip with the different regions on the surface (Figure 2.4.).[47] The force in this approximation is

$$F_{1\omega} = V_{ac} \sum_{i=1}^{n} \frac{\partial C_{eff,i}}{\partial z} \left( V_{dc} - V_{S,i} \right), \tag{2.33}$$

and effective surface potential determined by SSPM is

$$V_{dc} = \sum_{i=1}^{n} \frac{\partial C_{eff,i}}{\partial z} V_{S,i} \Big/ \sum_{i=1}^{n} \frac{\partial C_{eff,i}}{\partial z}, \tag{2.34}$$

where $C_{eff,i}(z)$ is partial capacitance between the tip and $i$-th region on the surface. For inhomogeneous surfaces local potential $V_{S,i}$ also reflects the difference in the surface work functions of dissimilar materials.[48] These equations can also be rewritten in the integral form as

$$V_{dc} = \frac{1}{C_z'} \int C_z'(x,y) V_{surf}(x,y) dS, \tag{2.35}$$

where $C_z'(x,y)$ is differential tip-surface capacitance. Spatial resolution of SSPM measurements is directly related to $C_z'(x,y)$ and experimental methods for the determination of this tip-surface transfer function (tip calibration) are of great interest. Additional contributions to the first harmonic of the force are due to the non-local cantilever contribution and feedback electronic effects, which determine the precision of SSPM voltage measurements. These effects will be addressed in Chapters 3 and 5.

### 2.4.4. Other Techniques

The examples considered above are EFM techniques that utilize either a mechanically driven (force gradient techniques) or an electrostatically driven (voltage modulation) cantilever. A combined approach simultaneously imposing both modulations is also possible. One such approach was developed by Terris *et al.*[27] and successfully applied to studies of contact electrification. In this technique, the voltage modulation frequency is higher than the feedback loop response frequency, but much lower than the oscillation frequency of the tip $\omega_t$. This range of frequencies is chosen so that the surface performs many oscillations while the cantilever has almost constant lateral position (which contributes to resolution), but there is no resonance between oscillations of the cantilever and the sample. The force equations in this case can be obtained from Eq.(2.21) for $V_{dc} = 0$. Thus

$$F = \frac{Q_s Q_t}{4\pi\varepsilon_0 z^2} + \frac{1}{2}\frac{\partial C}{\partial z} V_{ac}^2 \sin^2 \omega t \,. \tag{2.36}$$

Since $Q_t = -(Q_s + Q_{ac})$, $Q_{ac} = C(z)V_{ac}\sin(\omega t)$, the equation for the force gradient can be obtained from this expression as:

$$F' = \frac{1}{2} V_{ac}^2 \sin^2(\omega t) \frac{\partial^2 C}{\partial z^2} + \frac{Q_s V_{ac} \sin\omega t}{2\pi\varepsilon_0 z^2}\left(\frac{C}{z} - \frac{1}{2}\frac{\partial C}{\partial z}\right) + \frac{Q_s^2}{2\pi\varepsilon_0 z^3} \,. \tag{2.37}$$

For the uncharged surface, $Q_s = 0$, only the second harmonic term is non-zero and the force gradient oscillates at the frequency $2\omega$. This second harmonic term causes the tip oscillations at $\omega_t$ to be modulated at $2\omega$. For the charged surfaces, the harmonic term is also present and tip oscillations are modulated at frequency $\omega$, as well. This harmonic signal is detected from the output of feedback lock-in amplifier (reference frequency $\omega_t$) with a second lock-in (reference frequency $\omega$). The phase of the signal reflects the sign of the surface charge. The combined modulation technique was further developed by Ohgami *et al.*[49] who considered the effect of an additional constant bias applied to the tip, $V_{dc} \neq 0$, on the first harmonic signal. The general property of these techniques is that topographical and electrostatic information are collected in a single scan. The observed signal is the sum of electrostatic and electromechanical contributions and quantification of images in terms of surface charge, dielectric constant and electromechanical properties can be a very complicated problem.[50] Therefore, after the invention of the lift mode, combined modulation schemes are seldom used, albeit it is possible that they will be further development for studies of frequency mixing phenomena, etc. Additional possibilities are provided by combination of nulling scheme and force gradient detection that allows higher resolution of EFM and interpretability of SSPM in a single measurement.[51]

### 2.5. Current Detection: Contact SPMs

The SFM applications considered above utilized cantilever functionality as a force sensor measuring the electrostatic interactions between the localized tip and the surface. The complimentary approach for electric measurements by SPM is based on current detection. The most well known example of such technique is STM. Tip-surface current in the AFM experiment provides an additional degree of freedom. As discussed above, application of dc potential to the tip is expected to produce dc current only when the tip is relatively close to the surface, i.e. in the tunneling regime or in direct contact. In contrast, ac bias will always result in the ac current owing to the displacement current and conduction current. Finally, mechanical modulation of the tip in the vicinity of the surface is also expected to produce a displacement current due to the variations in the tip-surface capacitance. This approach is expected to provide a wealth of information on surface potential, charge and surface density of state.[52,53] Unfortunately, the relative magnitude of tip-surface capacitance limits the resolution of this technique to the micron range, even though the spectroscopic results are extremely promising.

#### 2.5.1. Scanning Spreading Resistance Microscopy

The SFM techniques considered so far detect the force acting on a tip. Possible charge transfer between the tip and the sample in these techniques is avoided either by maintaining finite tip-surface separation or ensuring the existence of dielectric layer on the tip (e.g. oxidized silicon tip) or on the surface. Detection of DC current through tip-surface junction under applied bias provides information on the resistivity of the sample. It can be easily shown that, due to the local nature of the probe, for good tip-surface contact the current is limited by the resistivity of the surface directly below the tip.[54,55,56] This spreading resistance, $R_{sp}$, is related to the conductivity of material, σ, as $R = 1/4\sigma a$, where $a$ is tip-surface contact radius. When applied to semiconductor materials, this technique is referred to as scanning spreading resistance microscopy (SSRM). Measuring the current for constant bias voltage allows local resistivity mapping that can be related to the chemical composition and carrier concentration. Successful SSRM imaging is possible only for high-quality tip-surface contact, such that contact resistance is smaller than the spreading resistance. This typically requires the compatibility of tip and surface material so that they don't form a Schottky pair. The spreading resistance is inversely proportional to contact area, which must be optimized during imaging. Practically it requires imaging at large indentation forces and use of mechanically stable tips (e.g. conductive diamond coated). A similar approach is used for probing leakage currents in thin dielectric films and is referred to as leakage current microscopy.[57,58]

In the last several years, the growing interest to the transport behavior of molecular systems led to AFM current measurements localized within single or several single molecules.[59] This approach, generally referred to as conductive AFM, employs relatively soft cantilevers that do not damage the sample surface. Sometimes the imaging is performed in fluid. The current magnitudes in this case are typically much smaller than that in SSRM (fA-pA vs. nA-mA).

Finally, tip surface resistance can be measured with ac current, as well. However, the additional contribution to the ac current is provided by the displacement current owing to tip-surface capacitance. While the localized part of this capacitance is small (tip apex-surface capacitance is in the fF range), the non-local capacitance between the cantilever and the surface and stray circuit capacitance is often in the nF range, resulting in large current offsets. AC conductive AFM imaging is possible if the tip-surface resistance is small and/or the modulation frequency is low; alternatively, when the capacitive part of tip surface impedance is smaller than contact resistance, $1/\omega C < R$, the capacitive coupling will dominate.

### 2.5.2. Scanning Capacitance Microscopy

The alternative paradigm in current based detection is tip-surface capacitance measurements. As discussed in Section 2.4, capacitance is a function of tip-surface separation, topography, tip shape, etc. The local part of tip-surface capacitance is significantly smaller that the non-local and stray capacitances and therefore cannot be measured directly. However, for semiconductor materials, local tip-surface capacitance depends on applied voltage, while stray capacitances are bias independent. Therefore, the voltage derivative of capacitance, $dC/dV$, can be collected as a local signal.

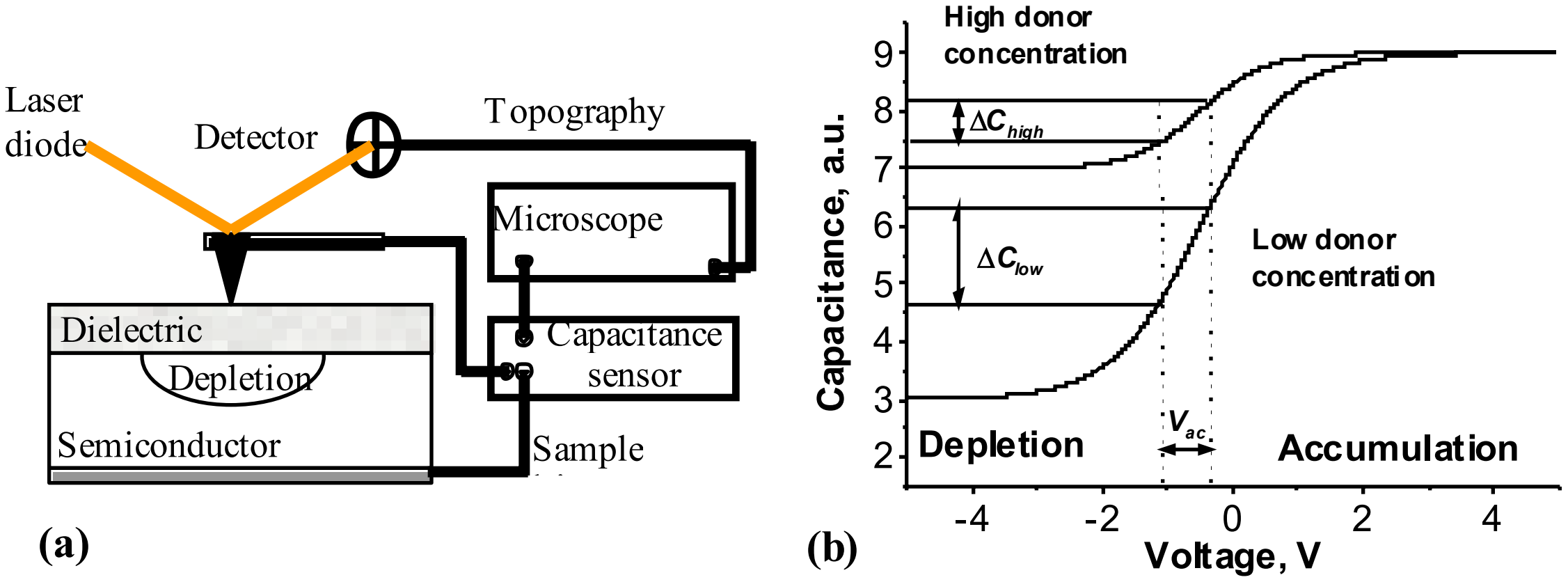

**Figure 2.5.** Schematic experimental set-up for scanning capacitance microscopy (a) and high frequency C-V curve for n-doped semiconductor with high (top) and low donor concentrations.

Image acquisition in SCM is performed on the semiconductor surface covered with an oxide film of known thickness in contact mode with conventional deflection feedback.[22,60] The advantage of this approach is that contact mode imaging is relatively insensitive to local dielectric constant, surface charge density or conductivity of the surface and thus provides reliable topographic data. For the SCM signal an ac bias in the kilohertz range is applied between the tip and the surface and the capacitance is detected with the ultra-high frequency (UHF) capacitance sensor (Figure 2.5a).[61] The variation of the capacitance $\Delta C$ due to depletion/accumulation below the tip is small compared to overall tip-surface capacitance and is measured with the lock-in technique. As seen from Figure 2.5b, the measured quantity is essentially the derivative of capacitance with respect to bias voltage $\Delta C/\Delta V$, or the slope of the *C-V* curve. In regions with higher carrier concentration the depletion depth is smaller, consequently the slope is small. In regions with lower carrier concentration the depletion depth and the slope are large.[62] Alternatively, for a uniformly doped sample the thickness of the surface oxide film is inversely related to the slope of the *C-V* curve. This imaging technique ($\Delta C$ mode) is easily realized; however, measurable change in the capacitance can often be obtained only if the driving amplitude is sufficiently high (several volts). This effectively smoothes *C-V* curve. For the samples with low carrier concentration, high driving voltages result in big depletion volumes and the loss of lateral resolution. An alternative SCM imaging technique utilizes additional feedback loops that keep $\Delta C$ constant by adjusting the driving voltage $\Delta V$. This technique has an advantage of constant depletion geometry, but the experimental setup is more complicated.[63] Interestingly enough, SCM cannot distinguish linear materials with different carrier concentration, e.g. metal and dielectric. Therefore, characterization of semiconductor device structures requires simultaneous SSRM and SCM imaging.

The exact shape of *C-V* curve depends on the applied bias, thickness of the oxide layer and doping level of the semiconductor. Provided the oxide thickness is known, a *C-V* curve can be used to determine the doping level of a semiconductor. Obviously the application of a very small capacitor (SPM probe) provides spatial localization of *C-V* curves and thus is used to map 2D dopant profile of semiconductor devices. [64] In many cases, information on doping can be obtained even without a detailed knowledge of the *C-V* curve. In addition, the *C-V* curve for a plane-plane arrangement necessarily differs from that for the SPM geometry, since in the latter case, fringe effects are significant and for large tip-semiconductor separation, will dominate the capacitance. In spite of these difficulties, both quantitative and qualitative SCM is proven to be one of the most versatile tools for dopant profiling in semiconductor devices.

Since the shape of the *C-V* curve strongly depends on the oxide thickness, SCM data can be obtained only from semiconductor surfaces with uniform oxide thickness. In addition, the surfaces must be very flat in order to avoid the geometrical effects in the capacitance. Interpretation of SCM data obtained even under ideal imaging conditions represents an extremely complicated problem due to the 3D geometry of the problem that precludes the analytical solution and requires application of numerical techniques.

### 2.6. Electromechanical Detection: Piezoresponse Force Microscopy

Some of the EFM techniques discussed above are based on the detection of an electrostatic field far from the surface, where the contribution of strong short-range Van-der-Waals forces is negligible. Alternatively, the tip is in contact and the current response of the surface to the applied bias is measured. However, for a broad class of materials, i.e. piezoelectrics, the application of a bias results in a significant mechanical response that can be detected by SPM. The set of techniques based on the detection of mechanical response to the applied bias[65,66,67] is generally referred to as piezoresponse force microscopy (PFM).

As in SSPM and SCM, a breakthrough in the detection of local electromechanical properties of materials was achieved by the application of voltage modulation techniques. The oscillating electric field applied between the tip and the surface results in the sample deformation. As in other voltage-modulation techniques, the applied bias results in a static response as well as first and second harmonic terms. The static signal is superimposed on the height data for the non-biased surface and usually cannot be detected. First and second harmonic signals can be extracted with the lock-in technique and used to map local piezoelectric and electrostrictive properties of the surface. The amplitude of the first harmonic signal provides information on the absolute value of the *z*-component of polarization vector. The phase of the piezoresponse signal depends on the sign of piezoelectric coefficient and allows determination of the polarization direction. This approach is limited to materials with sufficiently high piezoelectric coefficients; otherwise, the electric field necessary to produce the observable deformations will be greater than the coercive field and ferroelectric switching below the tip will complicate image interpretation.[68]

Superposition of a dc potential offset to the tip allows measurement of the local hysteresis loop for the ferroelectric material and investigation of fatigue effects in ferroelectric structures.[69,70,71] Further development of this technique based on simultaneous measurement of all components of displacement (rather than only the vertical component) allows determination of all components of polarization vector in each point of the surface and yields a reconstruction of surface crystallography.[72]

One of the primary difficulties in the interpretation of PFM is a variety of tip-surface and cantilever-surface interaction possible in contact mode. In addition, the relationship between local materials properties and PFM amplitude is extremely complex. These issues will be addressed in detail in Chapter 6.

### 2.7. Surface Modification by Electric SPMs.

In the last few years significant attention has been directed to the fabrication and patterning of nanoscale structures using AFM based techniques. Surface modification can be performed on very small length scales compared to the effective tip size and thus, in principle, can surpass the capabilities of photolithography. Currently, this patterning approach is serial rather than parallel and may not be sufficiently fast to be commercially viable. Nevertheless, the number and potential of AFM patterning techniques is enormous. Mechanical modification of the surface, i.e. nanoindentation has become a paradigm in the AFM and is supported by most commercial instruments.[73,74] Another approach for AFM patterning is based on chemically modified tips, e.g. dip-pen lithography.[75] To keep in line with the general direction of this thesis, discussed here are only electric based patterning techniques.

Application of electric potential through the tip in contact or intermittent contact mode can modify the surface. Depending on the nature of the surface, the material of the tip, the surrounding media, and applied bias, a wide spectrum of electrochemical or physical processes can occur. Most attention to date has been focused on the electrochemical oxidation of semiconductor surfaces. An example is the electrochemical oxidation of silicon, which allows formation of extremely thin $SiO_2$ lines.[76] The proposed mechanism includes formation of a water droplet between the surface and the tip; the latter serves as an electrode and causes the oxidation of the substrate. This method allows patterning of silicon surface with characteristic feature size of few tens of nanometers.[77] This patterning approach can be further combined with the standard photolithographic techniques, where selective etching of oxidized or pristine surface is used to fabricate complicated 3D structures. A related nanofabrication method is based on the local reduction of the surface or metal deposition using mixed ionic conductor probe. Application of bias in this case results in the deposition of corresponding metal on the surface.[78,79]

An alternative approach is based on charge deposition by the biased tip. This deposited charge is typically unstable and diffuses along the surface, limiting the minimal feature size to hundreds of nanometers.[80] In this technique, a deposited charge pattern can be developed by the colloid solution containing the particles with opposite charge.[81]

Another spectacular example of potential-induced surface modification is ferroelectric switching. One of the most prominent features of ferroelectrics is their ability to change the direction of polarization under the influence of electric field or mechanical stress and to retain the polarization direction after the field is switched off. This property makes ferroelectric materials and ferroelectric heterostructures promising materials for memory devices. Due to the small size of the probe tip in SFM, even a small applied field or force can cause local domain reversal and thus can be used for storing information on a ferroelectric substrate.[82] Feature resolution of 50 nm was achieved for epitaxial PZT(001)/Pt(111)/sapphire(0001) films by Hidaka *et al.*[83] Comparable resolution of less than 100 nm was achieved for PZT films on insulating $SrTiO_3$ or metallic Nb-doped $SrTiO_3$ substrates.[84] Similar experiments have been done recently on semi- or superconductor - ferroelectric heterostructures. In this case, ferroelectric switching in the ferroelectric layers results in electron or hole injection to the second component and can result in the metal-insulator or superconductor-dielectric phase transitions as demonstrated by Ahn *et al.*[85,86]

Recently, it was found recently that ferroelectric domain orientation affects the chemical reactivity of the surface in adsorption, catalytic and photochemical processes.[87,88,89] Chapter 7 presents a lithographic approach for the fabrication of metallic and semiconductive nanostructures based on the combination of ferroelectric domain writing and photodeposition.

## 2.8. Summary

The SPM techniques presented in this Chapter form a general background for the understanding of local electric properties of materials. These techniques provide a basis on which more complex techniques designed specifically to address transport phenomena in mesoscopic and nanoscale systems are developed as described in depth in the following Chapters.

[21] P. De Wolf, T. Clarysse, W. Vandervorst, L. Hellemans, Ph. Niedermann, and W. Hanni, J. Vac. Sci. Technol. **B 16**, 355 (1998).

[22] C.C. Williams, W.P. Hough, and S.A. Rishton, Appl. Phys. Lett. **55**, 203 (1989).

[23] Electric condition of dielectric or semiconductive tips is difficult to control. Contact of such tip with surface usually leads to contact electrification and resulting tip charge and electrostatic force are generally undefined. However, qualitative imaging with non-conductive tips is possible.

[24] *Electric Force Microscopy (EFM)*, Digital Instruments Support Note SN230, 1996.

[25] K.L. Babcock and C.B. Prater, Phase imaging: beyond the topography, Digital Instruments Support Note AN11.

[26] M. Fujihira, Annu. Rev. Mat. Sci. **29**, 353 (1999).

[27] B.D. Terris, J.E. Stern, D. Rugar, H. J. Mamin, *Phys. Rev. Lett.* **63**, 2669 (1989).

[28] N.M. Miskovsky, P.H. Cutler, T.E. Feuchtwang, *Int. Journ. Infrared And Millimeter Waves*, **2**, 739 (1981).

[29] L.-H. Pan, T.E. Sullivan, V.J. Peridier, P.H. Cutler, N.M. Miskovsky, *Appl. Phys. Lett.* **65**, 2151 (1994).

[30] J. He, P.H. Cutler, N.M. Miskovsky, T.E. Feuchtwang, Surf. Sci. 246, 348 (1991).

[31] H. Yokoyama, T. Inoue, J. Itoh, Appl. Phys. Lett. **65**, 3143 (1994).

[32] S. Hudlet, M. SaintJean, C. Guthmann, J. Berger, Eur. Phys. J. **B 2**, 5(1998).

[33] G. Mesa, E. Dobado-Fuentes, J.J. Saenz, J. Appl. Phys. **79**, 39 (1996).

[34] S. Belaidi, P. Girard, and G. Leveque, J. Appl. Phys. **81**, 1023 (1997).

[35] H.W. Hao, A.M. Baro, J.J. Saenz, J. Vac. Sci. Tech. **B 9**, 1323 (1991).

[36] J.D. Jackson, *Classical Electrodynamics* (John Wiley, New York, 1998).

[37] N.N. Lebedev, I.P. Skal'skaya, and Ya.S. Uflyand, *Problems in Mathematical Physics* (Pergamon Press, 1966).

[38] M. Bockrath, N. Markovic, A. Shepard, M. Tinkham, L. Gurevich, L.P. Kouwenhoven, M. Wu, L.L. Sohn, Nano Letters 2, 187 (2002).

[39] P. Grutter, Y. Liu, P. LeBlanc, U. Durig, Appl. Phys. Lett. **71**, 279 (1997).

[40] Y. Liu, B. Ellman, P. Grutter, Appl. Phys. Lett. **71**, 1418 (1997).

[41] T.D. Stowe, T.W. Kenny, D.J. Thomson, D. Rugar, Appl. Phys. Lett. **75**, 2785 (1999).

[42] M.T. Woodside, Ph.D. Thesis, UC Berkeley, 2001.

[43] M.T. Woodside, P.L. McEuen, Science **296**, 1098 (2002).

[44] J.M.R. Weaver, D.W. Abraham, J. Vac. Sci. Technol. **B 9**, 1559 (1991).

[45] M. Nonnenmacher, M.P. O'Boyle, H.K. Wickramasinghe, Appl. Phys. Lett. **58**, 2921 (1991).

[46] J. Janata, *100th anniversary of the Kelvin probe.* AVS Newsletter, May/June, 4 (1997).

[47] A.K. Henning and T. Hochwitz, Mater Sci Eng. **B 42**, 88 (1996).

[48] M.P. O'Boyle, T.T. Hwang, and H.K. Wickramasinghe, Appl. Phys. Lett. **74**, 2641 (1999).

[49] J. Ohgami, Y. Sugawara, S. Morita, E. Nakamura, and T. Ozaki, Jpn. J. Appl. Phys. **35**, 2734 (1996).

[50] K. Franke, H. Huelz, and M. Weihnacht, Surf. Sci. **41**, 178 (1998).

[51] J. Colchero, A. Gil, and A. M. Baro, Phys. Rev. **B 64**, 245403 (2001)

[52] Y. Oyama, Y. Majima, M. Iwamoto, J. Appl. Phys. **86**, 7087 (1999).

[53] Y. Majima, Y. Oyama, M. Iwamoto, Phys. Rev. B 62, 1971 (2000).

[54] P. De Wolf, J. Snauwaert, L. Hellemans, T. Clarysse, W. Vandervorst, M. D'Olieslaeger, D. Quaeyhaegens, J. Vac. Sci. Technol. **A 13**, 1699 (1995).

[55] A.C. Diebold, M. Kump, J.J. Kopanski, D.G. Seiler, Proc. - Electrochem. Soc. **94**, 78 (1994).

[56] P. De Wolf, M. Geva, T. Hantschel, W. Vandervorst, R.B. Bylsma, Appl. Phys. Lett. **73**, 2155 (1998).

[57] S. Landau, B.O. Kolbesen, R. Tillmann, R. Bruchhaus, A. Olbrich, E. Fritsch, C. Dehm, G. Schindler, W. Hartner, C. Mazure, Proc. - Electrochem. Soc. **98**, 789 (1998).

[58] Z. Xie, E.Z. Luo, H.B. Peng, B.R. Zhao, G.D. Hu, I.H. Wilson, J.B. Xu, L.H. Zhao, J. Non-Cryst. Solids **254**, 112 (1999).

[59] X.D. Cui, A. Primak, X. Zarate, J. Tomfohr, O.F. Sankey, A.L. Moore, T.A. Moore, D. Gust, G. Harris, S.M. Lindsay, Science 294, 571 (2001).

[60] C.C. Williams, J. Slinkman, W.P. Hough, and H.K. Wikramasinghe, Appl. Phys. Lett. **55**, 1662 (1989).

[61] C.C. Williams, Annu. Rev. Mater. Sci. **29**, 471 (1999).

[62] S.M. Sze, *Physics of Semiconductor Devices*, (John Wiley, New York, 1981).

[63] Y. Huang and C.C. Williams, J. Slinkman, Appl. Phys. Lett. **66**, 344 (1995).

[64] A.C. Diebold, M.R. Kump, J.J. Kopanski, and D.G. Seiler, J. Vac. Sci. Technol. **B 14**, 196 (1996).

[65] P. Guthner and K. Dransfeld, Appl. Phys. Lett. **61**, 1137 (1992).

[66] A. Gruveman, O. Auciello, and H. Tokumoto, J. Vac. Sci. Technol. **B 14**, 602 (1996).

[67] K. Takata, K. Kushida, K. Torii, and H. Miki, *Jpn. J. Appl. Phys.* **33**, 3193-3196 (1994).

[68] A. Gruverman, O. Auciello, and H. Tokumoto, Annu. Rev. Mater. Sci. **28**, 101 (1998).

[69] G. Zavala, J.H. Fendler, S. Trolier-McKinstry, J. Appl. Phys., **81**, 7480 (1997).

[70] A. Gruverman, O. Auciello, and H. Tokumoto, Appl. Phys. Lett. **69**, 3191-3193 (1996).

[71] E.L. Colla, S. Hong, D.V. Taylor, A.K. Tagantsev, N. Setter, and K. No, Appl. Phys. Lett. **72**, 2763 (1998).

[72] L.M. Eng, H.-J. Guntherodt, G.A. Schneider, U. Kopke, and J. Munoz Saldana, Appl. Phys. Lett. **74**, 233 (1999).

[73] B. Bhushan, *Nanomechanical properties of solid surfaces and thin films,* Handb. Micro/Nano Tribol.(2nd Ed.) 433 (1999).

[74] F.M. Serry, Y.E. Strausser, J. Elings, S. Magonov, J. Thornton, and L. Ge, Surf. Eng. **15**, 285 (1999).

[75] R.D. Piner, J. Zhu, F. Xu, S. Hong, and C.A. Mirkin, Science 283, 661 (1999).

[76] J.A. Dagata, *Science* **270**, 1625 (1995).

[77] J.A. Dagata, *NATO ASI Ser.* **E 264**(Nanolithography), 189 (1994).

[78] A. Spangenberg, J. Fleig, and J. Maier, Adv. Mat. **13**, 1466 (2001).

[79] K. Terabe, T. Nakayama, T. Hasegawa, and M. Aono, Appl. Phys. Lett. **80**, 4009 (2002).

[80] S.V. Kalinin and Dawn A. Bonnell, unpublished

[81] P. Mesquida and A. Stemmer, Adv. Mat. **13**, 1395 (2001).

[82] A. Gruverman, O. Auciello, R. Ramesh, H. Tokumoto, Nanotechnology **8**, A38 (1997).

[83] T. Hidaka, T. Maruyama, M. Saitoh, N. Mikoshiba, M. Shimizu, T. Shiosaki, L.A. Wills, R. Hiskes, S.A. Dicarolis, and J. Amano, Appl. Phys. Lett. **68**, 2358 (1996).

[84] T. Tybell, C.H. Ahn, and J.-M. Triscone, Appl. Phys. Lett. **72**, 1454 (1998).

[85] C.H. Ahn, S. Gariglio, P. Paruch, T. Tybell, L. Antognazza, and J.-M. Triscone, Science **284**, 1152 (1999).

[86] C.H. Ahn, T. Tybell, L. Antognazza, K. Char, R.H. Hammond, M.R. Beasley, O. Fischer, and J.-M. Triscone, Science **276**, 1100 (1997).

[87] J.L. Giocondi and G.S. Rohrer, Chem. Mater. **13**, 241 (2001).

[88] Sergei V. Kalinin, Dawn A. Bonnell, T. Alvarez, X. Lei, Z. Hu, J.H. Ferris, Q. Zhang, and S. Dunn, Nano Letters **2**, 589 (2002).

[89] Sergei V. Kalinin, C.Y. Johnson, and Dawn A. Bonnell, J. Appl. Phys. **91**, 3816 (2002).

# 3. LOCAL TRANSPORT MEASUREMENTS BY SPM

## 3.1. Introduction

One of the hallmarks of our times is a constant drive for the miniaturization of electronic devices. Conventional electronics have long passed the micron barrier and are now close to the 100-nanometer range.[1] At the same time, a significant scientific and industrial effort is focused on the "bottom-up" assembly of electronic devices from molecular and cluster sized building blocks, contributing to the rapidly developing fields of nanoscience and nanotechnology.[2] In addition to the potential for ultra high-density electronic devices, nanoscale systems often exhibit unusual physical properties that have attracted the attention of experimentalists and theorists alike.[3,4] The range of available methods for the fabrication of functional nanodevices has grown immensely in the last decade and encompasses a wide range of techniques from traditional photo- and e-beam lithography to self-assembly and nanopatterning. However, the successful implementation of nanodevices in large-scale manufacture also requires reliable techniques for device characterization and failure analysis. Since, by definition, most electronic devices are based on electronic phenomena and communicate with outside world by electric signals, transport measurements on the nanoscale present an important challenge for nanotechnology.

### 3.1.1. Qualitative Imaging of Transport Phenomena on the Nanoscale

Currently there exist multiple electric characterization techniques ranging from traditional *I-V*, *C-V* and Hall measurements to relatively complicated and less commonly used techniques such as deep level transient spectroscopy (DLTS).[5] A vast majority of these techniques are based on the detection of ac and dc current induced by a bias applied to the sample, which, in turn, requires the fabrication of relatively large-scale contacts to the device. The spatial resolution of these techniques is determined by contact separation, which in most cases is limited to > 10 micron.[6] In many cases, contact and spreading resistances of the material impose further restrictions on contact size. This limitation can be illustrated on such relatively large (> 1 micron length) objects as carbon nanotubes. It is known that depending on the growth technique, carbon nanotubes (CN) contain a number of defects, which can be imaged using techniques such as Scanning Gate Microscopy (SGM) as illustrated in Figure 3.1.[7]

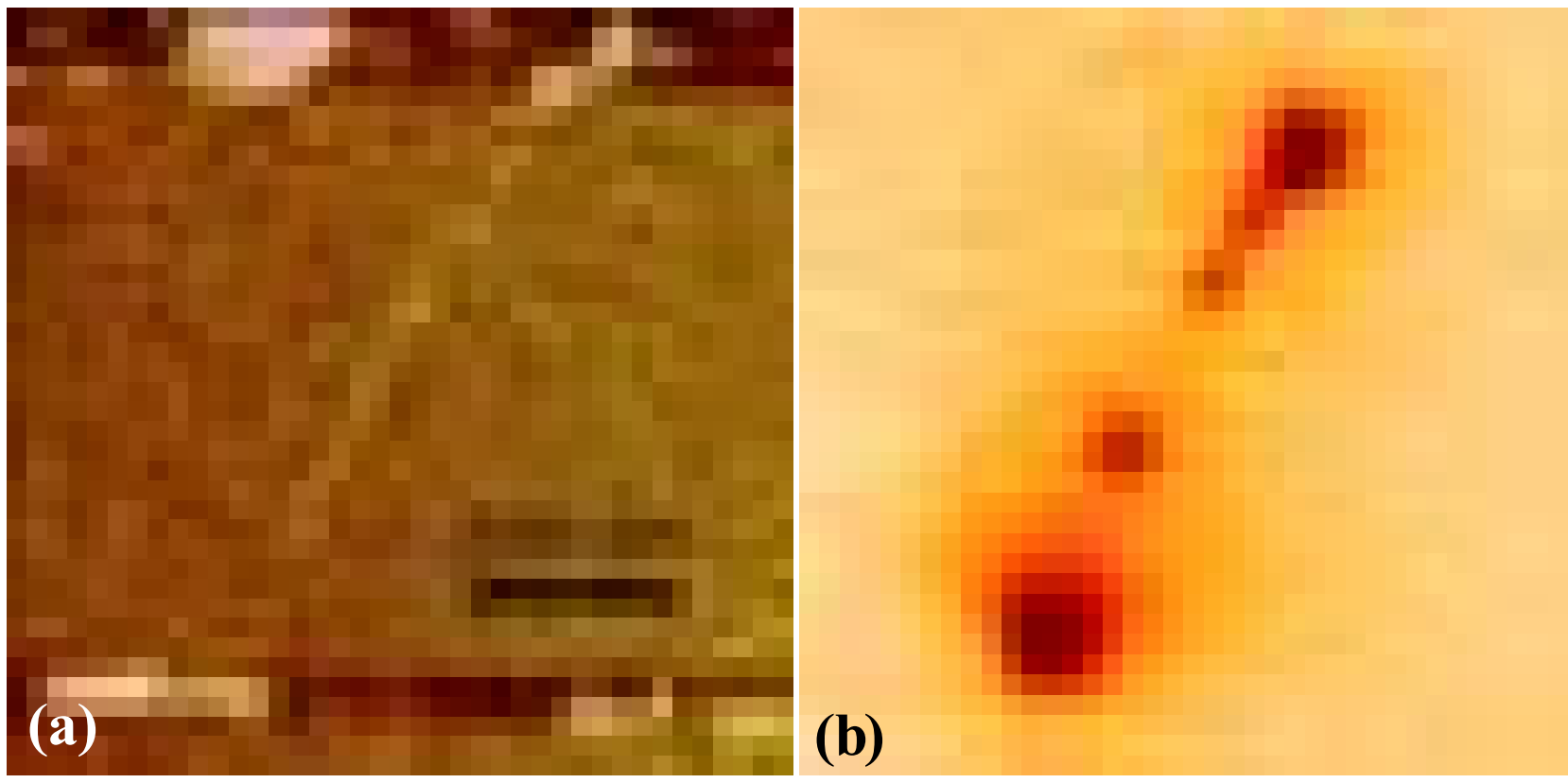

**Figure 3.1.** Surface topography (a) and scanning gate microscopy (b) images of carbon nanotube. A number of defects are clearly seen on SGM image.

However, differentiation of individual defect contributions to device properties using conventional current based semiconductor characterization techniques is impossible, since defect spacing is often in the 100-nanometer range and their location is *a priori* unknown, precluding contact fabrication to individual nanotube segments. Similar considerations can be applied to transport in polycrystalline materials, in which only average properties can be determined from macroscopic measurements and differentiation of contribution of individual microstructural elements to transport behavior is a complex problem.

Some of these problems can be circumvented by techniques such as impedance spectroscopy.[8,9,10] Through analysis of the frequency dependence of the amplitude and phase of a bias induced current the major relaxation processes in the solid can be determined and associated with microstructural elements. The typical application of impedance spectroscopy to polycrystalline materials differentiates grain boundary, grain interior and electrode impedances by fitting the impedance data to corresponding equivalent circuit models. However, even in this case no information is obtained about the properties of the individual elements. A number of approaches have been suggested to separate the impedance response of individual structural elements, such as microimpedance spectroscopy using patterned contact arrays.[11,12,13] However, the small contact area inevitably leads to high contact resistance and precludes quantitative measurements even in the four-probe configuration.[14,15]

In addition to problems with contact fabrication, fundamental limitations on transport measurements on the nanoscale arise from the unique physics of the mesoscopic systems. For example, for ballistic conductors such as metallic carbon nanotubes the resistance of the tube *per se* is zero; the overall device resistance is due to the resistance between the nanotube and the contacts.[3] In these and similar mesoscopic systems, the transport is no longer a local phenomenon that can be described by Ohm's law, but is determined by the geometry of the device as a whole.

To summarize, the progress in nanoscience and particularly nanoelectronics necessitates an understanding of the structure and transport properties on nanometer level. Due to both practical restrictions in contact fabrications and fundamental limitations arising from the mesoscopic nature of these systems traditional current based techniques do not allow device characterization on the sub micron scale. Ideally, one would be interested in quantitative real space imaging of transport phenomena, which suggests the use of scanning probe microscopy.[16, 17]

### 3.1.2. This Thesis: towards Spatially Resolved dc and ac Measurements

It is the purpose of this chapter to determine the limits of SPM to quantifying local dc and ac transport measurements. SPM is a large and rapidly developing field and a number of studies in the fields ranging from low-temperature physics to materials science have been reported. These studies vary significantly in the degree of technique sophistication and quantitative understanding of the reported results. In order to rationalize existing results, a classification scheme of SPM measurements based on the device functionality and SPM mode is developed in Section 3.2. It is noted that most SPM studies to date are limited to qualitative imaging of linear systems and the general framework for quantitative dc transport measurements in non-linear single and multiple interface systems is developed in Section 3.3. This section also presents a new calibration procedure. To date, the majority of experimental SPM studies of transport phenomena have been performed on dc biased devices, thus the information has been limited to dc transport properties. In other words, only resistive elements in the distributed equivalent circuit were determined. Section 3.4 describes Scanning Impedance Microscopy, the first technique that allows imaging of both resistive and capacitive transport behavior. The tip calibration standard for dc and ac transport imaging is described in Section 3.5. Finally, some considerations on imaging artifacts, systematic errors and degree of invasiveness are presented in Section 3.6. Conclusions and some further directions are summarized in Section 3.7.

### 3.2. General Framework for Transport Measurements with SPM

Progress in SPM is associated with the emergence of a large number of electrical characterization techniques enabled by several tip-surface interaction regimes (contact, non-contact, intermittent contact) and a number of modulation schemes for probing the local properties (mechanical modulation, voltage modulation, magnetic modulation, etc.) as summarized in Chapter 2. Further progress is facilitated by the classification of SPM techniques for electrical characterization in terms of implementation, measured property and imaging mechanism. Such classification will reduce the number of redundant reports on similar SPM techniques and facilitate the quantitative understanding of existing results.

The general framework suggested here is based on the measurement set-up functionality, as illustrated in Figure 3.2. The SPM tip can be considered as a moving terminal of the device. In single-terminal measurements, the properties of material directly below the tip are probed. In this case, the tip-surface impedance is dominated by the near-contact region. The rest of the sample provides the current sink (ground) and the detailed structure is not reflected in SPM data. In two-terminal measurements, the current path between the SPM probe and macroscopic electrode or two SPM probes is well-defined and the properties of material between the probes are studied. In three-terminal measurements, the current is applied across the system through macroscopic contacts, while the tip acts as a voltage probe or a local gate probe providing information on lateral transport properties. Depending on the experimental set-up, transport data can be acquired simultaneously with topography acquisition (main line), or using separate scans performed at fixed probe-surface separation (lift mode). In all of these cases, the tip can act as a current, capacitive or voltage probe depending on the imaging regime, as summarized in Table 3.I. In the three-terminal case, the tip can additionally act as a gate probe, influencing the overall transport through the system.

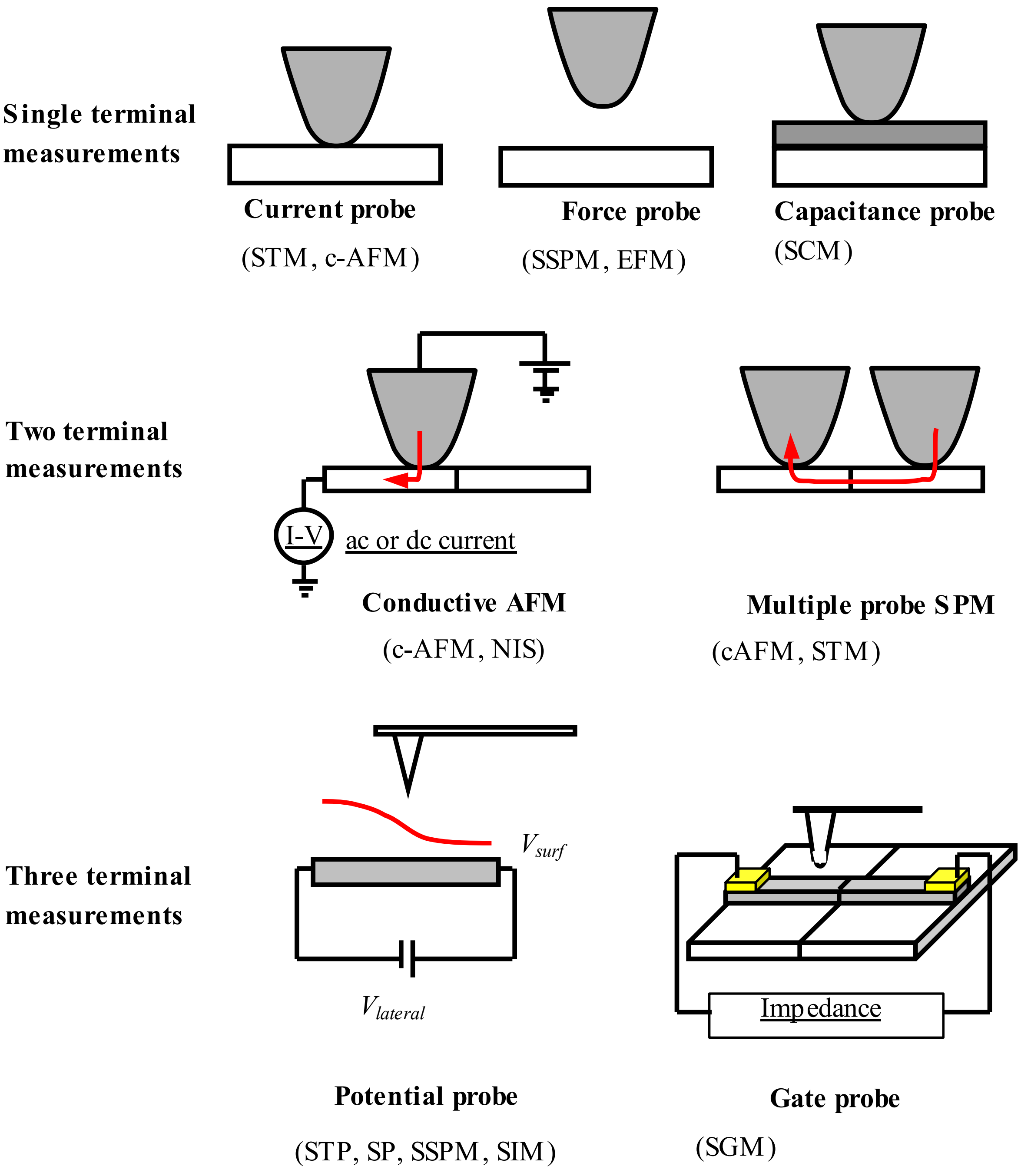

**Figure 3.2.** General framework for transport measurements by SPM

Table 3.I

*Local electric property measurements by SPM*

| **Technique (mode)** | **Topography feedback** | **Probe Principle** |
|---|---|---|
| SSPM (lift) | Contact, non-contact or intermittent contact | The first harmonic of tip-surface force is nulled by varying dc tip bias. Tip acts as a voltage probe. |
| EFM (lift) | Contact, non-contact or intermittent contact | Electrostatic force gradient between the dc biased tip and the surface. |
| Dissipation force EFM (lift) | Contact, non-contact or intermittent contact | Losses induced by mechanically modulated dc biased tip. |
| Conductive AFM (main) | Contact mode | Resistive or tunneling current |
| Tunneling AFM (main) | Intermittent contact | Tunneling current |
| Scanning Spreading Resistance Microscopy (main) | Contact mode | Resistive current |
| Scanning Capacitance Microscopy (main) | Contact mode | Tip-surface capacitance voltage derivative, d$C$/d$V$. |
| Scanning Tunneling Microscopy (main) | Current | Tunneling current |
| ac-STM (main) | Current | AC current probe |
| Piezoresponse Force Microscopy | Contact mode | Electromechanical response of the surface to ac tip bias |

Non-contact techniques such as Scanning Surface Potential Microscopy (SSPM) and Electrostatic Force Microscopy (EFM) can be used to measure local potential and work function, as discussed in Chapters 2 and 5.[18,19,20] In these techniques, the probe is capacitively coupled to the surface and capacitive force-induced changes in the dynamic behavior of the cantilever are recorded. The combination of SSPM with photoexcitation allows imaging of various photoelectric phenomena, such as diffusion length measurements of minority carriers.[21] Contact techniques such as Scanning Spreading Resistance Microscopy (SSRM)[22,23] and Scanning Capacitance Microscopy (SCM)[24] are based on the detection of tip-surface current. The dc tip-surface current measured in SSRM and conductive AFM provides information on the contact resistance and spreading resistance below the tip, from which local conductivity can be determined. These techniques require highly conductive tip-surface contact in order to be quantitative. In SCM, the voltage derivative of tip-surface capacitance is measured. Here, the tip-surface coupling is purely capacitive and leakage current can preclude the measurements. The combination of SCM and SSRM is used to determine local carrier concentration and doping level in semiconductors, e.g. to delineate *p-n* junctions or image more complex device structures.[25,26,27] Scanning Tunneling Microscopy (STM) and spectroscopy (STS) are used to determine local atomic structure and Fermi level.[28,29,30] The applicability of conventional STM is limited to conductive surfaces only; however, a number of approaches utilizing ac current feedback for STM imaging insulators (e.g. glass) were reported.[31] However, being single terminal measurements, all the techniques in Table 3.I are limited to the material properties in the near tip region.

The next level in SPM transport measurements is represented by the two-terminal measurements. Here, the position dependent impedance between the probe tip and fixed electrode is measured. Many conductive AFM measurements of non-uniform objects such as thin-films (bottom electrode used) nominally belong to this class. Some of the two-terminal SPM techniques are listed in Table 3.II according to the topography feedback (**T**) and bias modulation (**S**).

Table 3.II

*Two-terminal measurements by SPM*

| Technique | Feedbacks and modulation | Probe Principle |
|---|---|---|
| Conductive AFM | **T:** force<br>**S:** dc bias | dc current probe |
| Nanoimpedance Spectroscopy | **T:** force<br>**S:** ac bias | ac current probe |

By definition, these techniques are limited to current detection and, depending on the character of the bias (ac vs. dc), conductive AFM and nanoimpedance spectroscopy can be distinguished. It should be noted that the techniques listed in Table 3.II rely on force feedback for topography. The use of topographic current feedback requires a conductive substrate, imposing the limitations on the range of the systems that can be studied with this approach.

Finally, the three-terminal configuration allows several paradigms for SPM transport measurements. One type of conductivity measurements is based on using potential-sensitive SPM techniques such as scanning tunneling potentiometry (STP),[32,33,34,35,36,37] scanning surface potential microscopy (SSPM, or Kelvin Probe Force Microscopy, KPFM)[38,39,40,41] or Scanning Potentiometry[42] on laterally biased surfaces as shown in Figure 3.3.

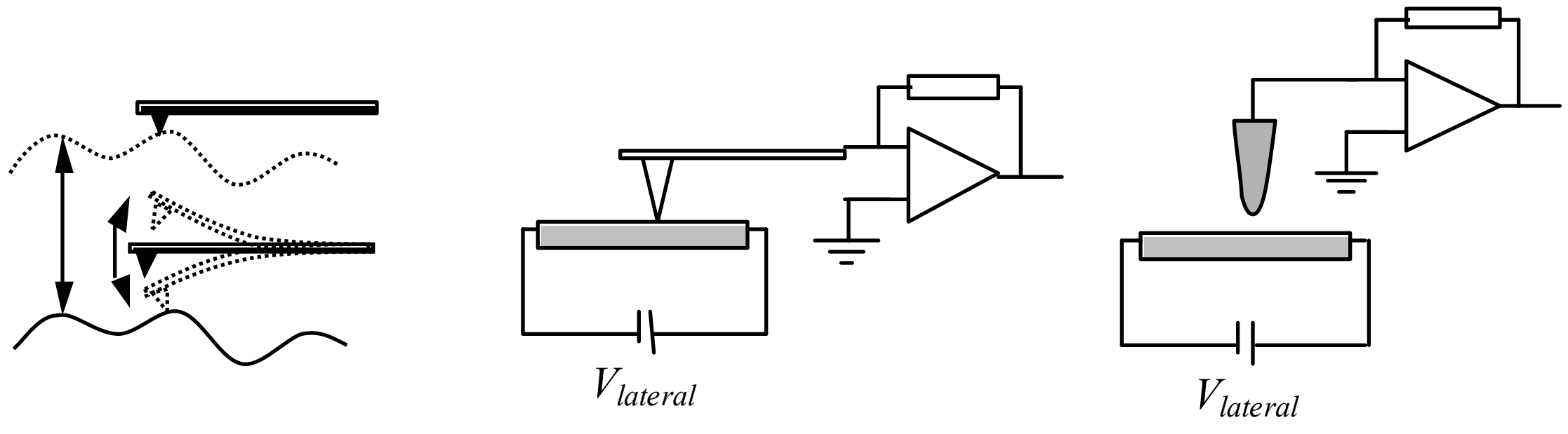

| **SSPM and SIM** | **Scanning Potentiometry** | **STP** |
|---|---|---|
| Topography: force<br>Potential: force | Topography: force<br>Potential: current | Topography: current<br>Potential: current |

**Figure 3.3.** Three-terminal potential probes for lateral transport measurements

This approach is very similar to 4-probe resistivity measurements, but instead of two fixed voltage electrodes, the SPM tip acts as a moving voltage probe providing the advantage of spatial resolution. The voltage sensitivity is enabled either by nulling the tunneling current (STP), capacitive force (SSPM) or direct connection to high input impedance electrometer (equivalent to nulling the spreading current, SP). In these measurements, the probe is often approximated as non-invasive, i.e. the presence of the AFM tip is assumed not to affect the potential and current distribution in the device. This assumption is completely justified in macroscopic systems with high conductivity or pinned surface Fermi level. This is not always true in the mesoscopic systems,[43] in which a biased SPM tip can introduce significant perturbation in the current and potential distribution. This effect is exploited in Scanning Gate Microscopy.[44,45,46,47,48] In this technique, a dc biased tip is scanned across the surface while a dc or ac bias is applied across a nanodevice, e.g. nanotube, nanowire or microfabricated Hall bridge. The global current vs. tip position constitutes the SGM image. In the framework of classical semiconductor theory, when the tip is positioned above the defect, tip-induced band bending results in depletion and an increase of local resistivity. When the tip is positioned above non-defect regions or far from the nanotube, the current is unaltered. The tip acts as a moving gate that allows the regions with different doping level in the nanostructure (e.g. carbon nanotube) to be distinguished. Interpretation of SGM images in quantum and mesoscopic systems (e.g. mesoscopic conductors in the fractional Hall effect regime) is more complex.[48,46]

Three-terminal SPM techniques are classified by the type of topography feedback (**T**), probe type (**P**) and sample modulation (**S**) and are summarized in Table 3.III.

Table 3.III

*Lateral transport measurements by SPM*

| Technique | Feedbacks and modulation | Probe Principle |
|---|---|---|
| Scanning Tunneling Potentiometry | **T:** current<br>**P:** current<br>**S:** dc bias | DC Voltage probe |
| Scanning Surface Potential Microscopy | **T:** force<br>**P:** force<br>**S:** dc bias | DC Voltage probe |
| Scanning Impedance micorscopy | **T:** force<br>**P:** force<br>**S:** ac bias | AC voltage probe |
| Scanning Potentiometry | **T:** force<br>**P:** dc current<br>**S:** dc bias | DC Voltage probe |
| SCM | **T:** force<br>**P:** ac current<br>**S:** dc bias | Capacitance probe |
| Scanning Gate Microscopy | **T:** force<br>**S:** dc bias | Gate probe |
| Ac-Scanning Gate Microscopy | **T:** force<br>**S:** ac bias | Gate probe |

As is obvious from Table 3.III, the variety of SPM techniques for transport measurements is immense. They differ by the spatial resolution, sensitivity, degree of invasiveness and applicability range. One of the limitations of all SPM techniques arises from the complicated geometry of the tip-surface system. In non-contact techniques based on the force (SSPM, SIM) or force-gradient (EFM) detection scheme, the first- or second derivative of tip-surface capacitance limits the spatial resolution. In the techniques based on current detection, the lateral resolution is limited either by contact area (~10 nm in conductive AFM, nanopotentiometry) or, in Scanning Tunneling Microscopy (STM) based techniques can potentially achieve atomic resolution. However, high contact resistance and poorly defined contact area hinder quantitative interpretation of SSRM and SCM results, whereas STP is applicable to conductive surfaces only. Very often tip properties define the image contrast (the probe is invasive), e.g. tip bias in SCM influences the apparent position of the junction.[49,50] Nevertheless, the opportunities provided by SPM for nanoscale imaging of semiconductor device operation are unparalleled by any conventional technique and a number of studies have been reported recently.[50,51,52,53,54,55,56]

To date, the lateral transport studies by SPM have been qualitative or semiquantitative at best, with a few exceptions.[34,39,41,57] The goal of this Chapter is to develop the quantitative guidelines for transport property measurements with SPM by comparing measurements on model systems with analytical solutions.

### 3.3. Quantifying DC transport by SSPM

Despite a number of SPM observations of potential barriers at interfaces in laterally biased electroceramics and semiconductor devices, few attempts have been made to characterize the transport properties of the interface directly from SPM data.[34,58] In all cases, the relative resistances of individual elements were determined, but no attempts were made to quantify local non-linear current-voltage behavior. Here, we develop the formalism to quantify the dc transport properties of a non-linear electroactive interface. Unlike conventional two or four probe resistivity measurements, SSPM is sensitive to variations in local potential, while (local) current is generally unknown. For single interfaces such as in bicrystals or metal-semiconductor junctions, the system can be represented by an equivalent circuit, which defines the current. The elements of the equivalent circuit are associated with microstructural features as observed in topographic image or by any other microscopic technique (e.g. Scanning Electron Microscopy or optical microscopy). Such analysis can be readily extended for multiple interface systems with 1D current flow.

### 3.3.1. Single Interface Systems

In a dc transport SPM experiment, a biased interface is connected to a voltage source in series with current limiting resistors to prevent accidental current flow to the tip. For such a circuit, the total resistivity of the sample $R_\Sigma$, is

$$R_\Sigma = 2R + R_d(V_d), \tag{3.1}$$

where $V_d$ is the potential across the interface, $R_d(V_d)$ is the voltage dependent resistivity of the interface and $R$ is the resistivity of the external current limiting resistors. The applied bias dependence of the potential drop at the interface is directly assessable by SPM and is referred to as the voltage characteristics of the interface. Since some of the circuit parameters (e.g. resistivity of current limiting resistors, $R$) are known and can be varied deliberately, variation of $R$ can be used to quantify the interface properties.

In the general case, when the functional form of the interface $I$-$V$ characteristics is unknown, it can be reconstructed directly from the interface potential drop provided the values of the current limiting resistors are known. In this case,

$$I(V_d) = (V - V_d)/2R, \tag{3.2}$$

where $I(V_d)$ is the current-voltage characteristics of the non-linear element.

As an alternative to varying the current limiting resistors, the current in the circuit can be measured directly. Such measurements can be conveniently done by applying a slow (approximately mHz range) triangular voltage ramp across the interface with the slow scan disengaged. To minimize the lateral bias variation during the single scan line, the stepwise increase in voltage can be synchronized with the line start TTL signal. The first image is then the SSPM image in which each line corresponds to different lateral bias conditions (i.e. potential profile across the interface, from which $V_d(V)$ is obtained), the second image stores the actual lateral bias ($V$) and the third image is current in the circuit measured by an $I$-$V$ converter ($I = I_d$). This approach can be extended to systems with multiple interfaces, such as *p-i-n* diodes, etc., providing $I$-$V$ characteristics for each interface. In all cases, the potential is defined as the difference between the potential under bias and the potential of the grounded surface, which accounts for the contact potential difference (CPD) variations across the interface between dissimilar materials.

This approach is somewhat limited for interfaces with highly asymmetric *I-V* characteristics, such as a metal-semiconductor interface. Here, the potential drop under forward bias is small compared to noise in potential measurements. For such systems, a different analysis is required. For example, for a metal-semiconductor junction the current is

$$I_d = I_0\left(\exp\left(\frac{qV_d}{nkT}\right) - 1\right), \tag{3.3}$$

where $V_d = V_2 - V_1$ is potential across the junction, $q = 1.6 \cdot 10^{-19}$ C is electron charge, $n$ is ideality factor, $k = 1.38 \cdot 10^{-23}$ J/K is the Boltzmann constant and $T$ is temperature. Saturation current, $I_0$, in the thermionic emission model is related to the Schottky potential barrier height $\phi_B$ as

$$I_0 = A^* T^2 \exp\left(-\frac{q\phi_B}{kT}\right), \tag{3.4}$$

where $A^* = 130$ A/(cm$^2$ K$^2$) is the Richardson constant.

For the ideal case of a diode in series with two current limiting resistors, the potential drop at the diode as a function of lateral dc bias $V$ is

$$V_d = \frac{R_d(V_d)}{R_\Sigma} V. \tag{3.5}$$

Eqs.(3.2,3,5) can be written for $V$ as a function of $V_d$:

$$V = 2RI_0\left(exp\left(\frac{qV_d}{kT}\right) - 1\right) + V_d, \tag{3.6}$$

which in the limit of a large forward bias simplifies to

$$V_d = \frac{kT}{q} \ln\left(\frac{V}{2RI_0}\right), \tag{3.7}$$

and a large reverse bias to

$$V_d = V + 2RI_0. \tag{3.8}$$

Therefore, for a forward biased diode the potential drop at the interface is expected to be small (of order of few mV) and hardly detectable by SSPM. For a large negative bias, however, the potential drop occurs primarily at the interface. The crossover between the two regimes is expected at a lateral bias $V = -2RI_0$. Incorporation of a leakage term, which is always present in real systems, for reverse bias, $I_l = \sigma V_d$ into Eq.(3.3) results in a deviation in the slope of the voltage characteristics under reverse bias from unity and Eq.(3.8) becomes

$$V_d = (V + 2RI_0)/(1 + 2R\sigma). \tag{3.9}$$

Eq.(3.9) implies that for finite conductivity in a reverse biased diode the potential drop occurs both at the metal-semiconductor interface and at the current limiting resistors. Therefore, experimental voltage characteristics of the interface can be used to obtain both the saturation and leakage current components of diode resistivity.

### 3.3.2. Multiple Interface System

The more general case of a polycrystalline material can be analyzed using a brick-layer type model.[59] In this model, the grain structure is approximated by a periodic arrangement of rectangular grains separated by the grain boundaries. Here, we neglect the conductance along the grain boundaries, i.e. grain boundaries are resistive.

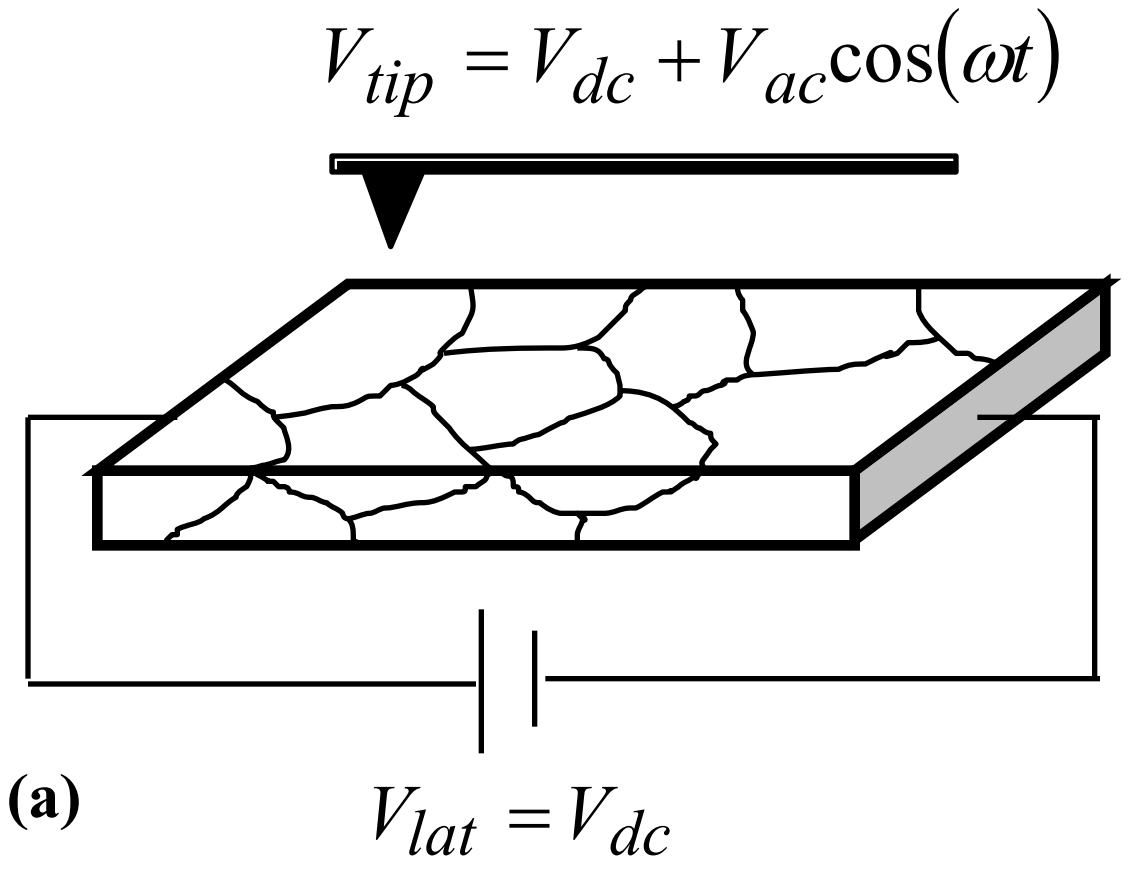

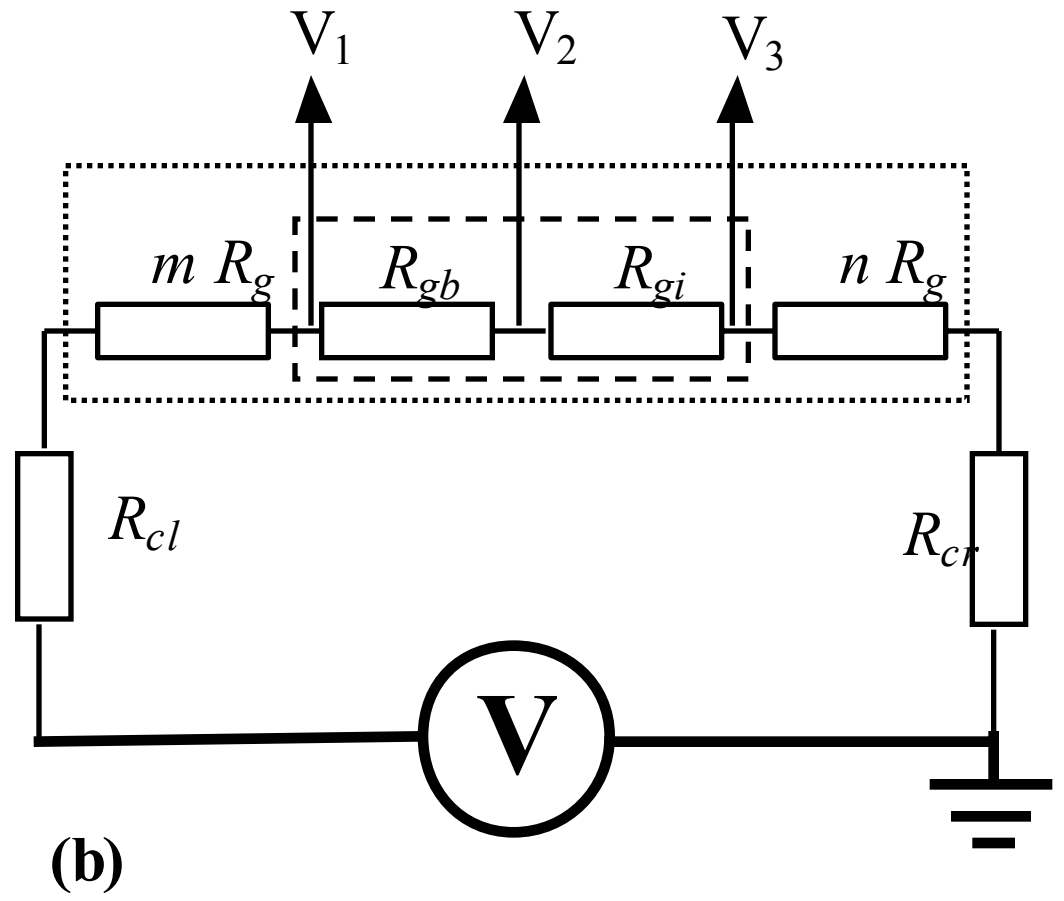

**Figure 3.4.** Experimental set-up for dc transport measurements by SSPM in multiple interface systems (a) and corresponding equivalent circuit (b).

Assuming a series arrangement of the grains and uniform current density, total resistance of the sample, $R_\Sigma$, can be written as (Figure 3.4)

$$R_\Sigma = R_{lc} + NR_{gi} + (N-1)R_{gb} + R_{rc}, \tag{3.10}$$

where $R_{lc}$ is the resistance of the left contact, $R_{rc}$ is the resistance of the right contact, $R_{gb}$ is the grain boundary resistance, $R_{gi}$ is the resistance of grain interior and $N = n+m+1$ is the number of the grains, $n$ and $m$ being the number of grains to the right and left of the investigated grain. In terms of the brick-layer model, $R_\Sigma$ is $R_{sample}\ S_{sample}/S_{grain}$, where $R_{sample}$ is sample resistance, $S_{grain}$ is average grain size and $S_{sample}$ is sample cross-section area. The potential drop at the individual grain boundary, $\Delta V_{gb} = V_1 - V_2$, is

$$\Delta V_{gb} = \frac{R_{gb}}{R_\Sigma} V, \tag{3.11}$$

where $V$ is the lateral dc bias. The potential drop within the grain, $\Delta V_{gi} = V_2 - V_3$, is

$$\Delta V_{gi} = \frac{R_{gi}}{R_\Sigma} V = \frac{dV}{dx} l, \tag{3.12}$$

where $\frac{dV}{dx}$ is the experimentally determined potential gradient along the grain and $l$ is the grain size. Therefore, the ratio of the potential drop at the grain boundary and in the grain interior, $\alpha$, is equal to the ratio of the grain boundary and grain interior resistivities

$$\frac{\Delta V_{gb}}{\Delta V_{gi}} = \frac{R_{gb}}{R_{gi}} = \alpha. \tag{3.13}$$

Provided that the electrode resistance is small, $R_{rc}+R_{lc} << N(R_{gb}+R_{gi})$, the total resistance can be measured directly and the grain boundary and grain interior resistances are

$$R_{gb} = \frac{1}{N}\frac{\alpha}{\alpha+1} R_\Sigma, \qquad R_{gi} = \frac{1}{N}\frac{1}{\alpha+1} R_\Sigma. \tag{3.14a,b}$$

This approach can also be extended to high electrode resistance. In this case, direct measurement of surface potential close to the right electrode and close to the left electrode allows potential drops at the electrodes to be distinguished from the potential drop in the bulk and all three components of dc resistivity can be determined.

In a realistic multiple interface system, the current flow is not uniform and not limited to a single direction. For 2D systems (e.g. thin films), the quantitative description of dc transport properties can be still be achieved by either numerical solution of the corresponding Kirchhoff equations for individual grain boundary resistivities or by finite element modeling (e.g. using models developed by Fleig *et. al.*[60]). In the 3D case (e.g.

bulk ceramics), the potential distribution *inside* the material is not accessed by SPM. In this case, the interface and bulk resistivity can be obtained from the Eq.(3.13); however, the properties of individual structural elements cannot be unambiguously determined.

### 3.3.3. Calibration

A model metal-semiconductor interface was used to calibrate the technique. The sample was prepared and characterized as described in Appendix 3.A. The metal and silicon are clearly identified from differences in topographic structure (Figure 3.5a). Surface potential under a slowly varying sample voltage is shown in Figure 3.5b. For a positively biased interface, no potential drop is observed (light contrast regions), while for the negatively biased device potential drops at the Schottky barrier in agreement with theoretical expectations. For small current limiting resistors (500 Ohm) the potential on the right hand side of the barrier *increased* for large negative biases. This behavior is attributed to photoinduced carrier generation in the junction region at a level equivalent to constant current element of opposite polarity in the circuit. The effect was suppressed by introducing large resistors in the circuit and disappeared for $R \geq 10$ kOhm.

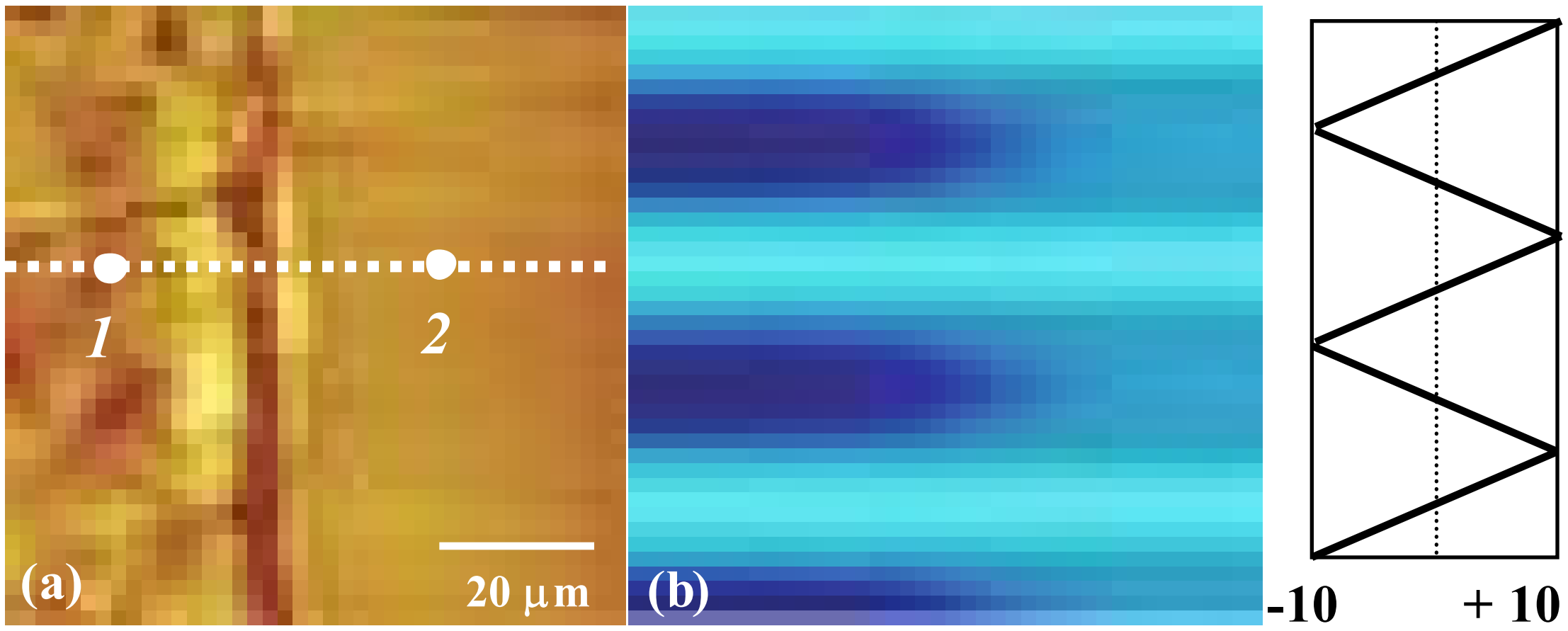

**Figure 3.5.** Surface topography (a) of the cross-sectioned diode. The potential profiles were acquired along the dotted line. The changes of potential, phase and amplitude were determined from positions 1 and 2. Surface potential (b) during a 0.002 Hz triangular voltage ramp to the sample for $R$ = 500 Ohm. The scale is 300 nm (a) and 10 V (b)

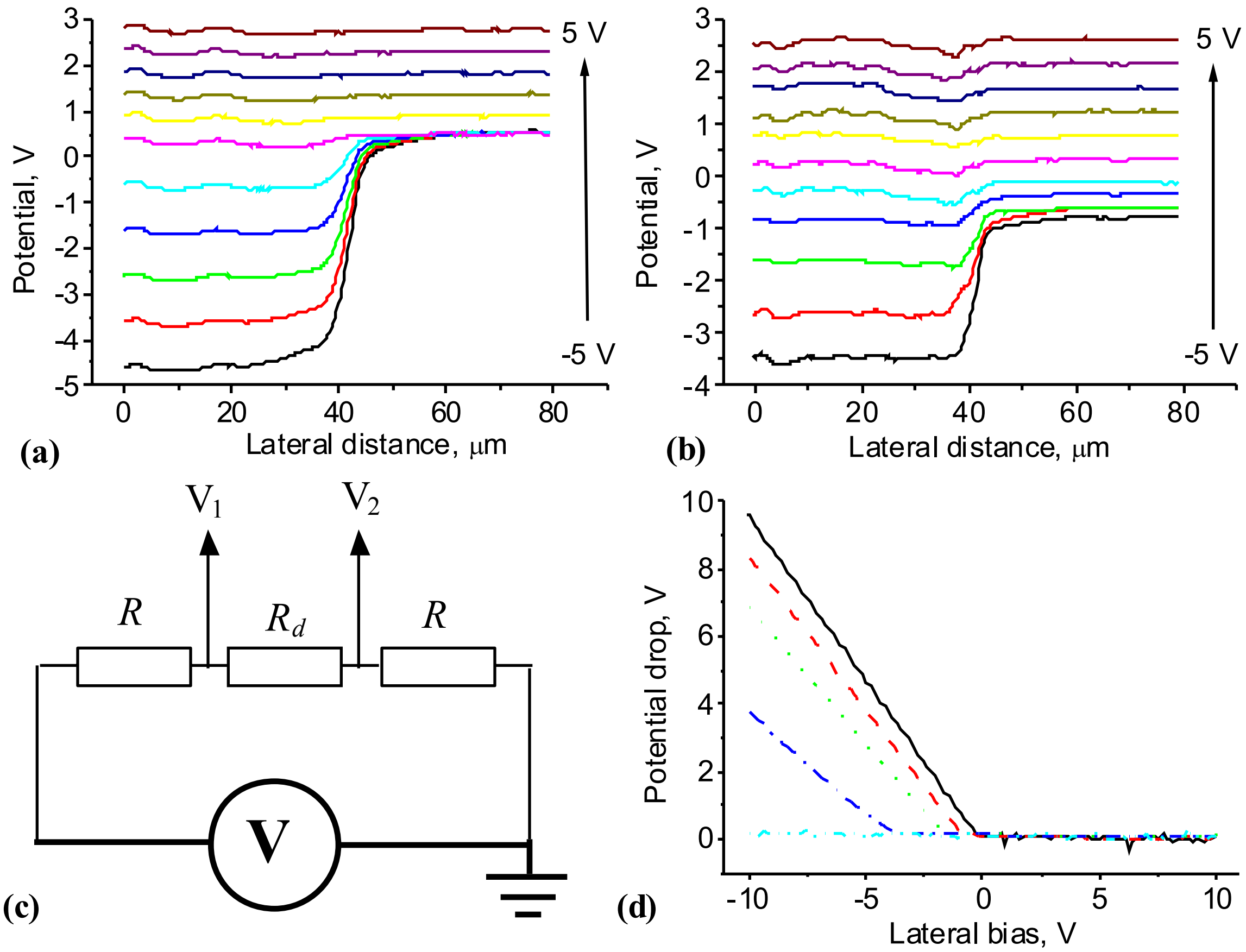

**Figure 3.6.** Potential profiles across the interface for different lateral biases for $R$ = 500 Ohm (a) and 100 kOhm (b). Profiles are acquired with 1 V interval. Note that the onset of the reverse bias condition is shifted for high resistivity circuit terminations. Equivalent circuit (c) for SSPM. Potential drop at the interface (d) as a function of lateral bias for circuit terminations 10 kOhm ( ), 47 kOhm ( ), 100 kOhm ( ), 220 kOhm ( ) and 1 MOhm ( ).

Potential distribution across the interface as a function of lateral bias for low-and high resistivity circuit termination is shown in Figure 3.6a,b. Note that for small $R$ the diode switches from forward to reverse bias at $V$ = 0, but the onset of this effect is delayed for large $R$, as suggested by Eq.(3.8).

The effect of lateral bias and $R$ on interface potential drop is displayed in Figure 3.6d as the voltage characteristics of the interface for different circuit terminations. Both the linear segment under reverse bias and almost zero potential drop under forward bias predicted by Eq.(3.9) are clearly seen. These data, approximated by the linear function $V_d = a+b\ V_{dc}$, are described by the fitting parameters summarized in Table 3.IV.

Table 3.IV.

*Interface voltage characteristics by SSPM*

| $R$, kOhm | $a$ | $b$ | $I_0$, $10^{-6}$ A | $R_l$, kOhm |
|---|---|---|---|---|
| 10 | -0.088 ± 0.005 | -0.971 ± 0.001 | 4.55 | 670 |
| 47 | -0.698 ± 0.004 | -0.909 ± 0.001 | 8.17 | 940 |
| 100 | -1.20 ± 0.01 | -0.807 ± 0.001 | 7.44 | 836 |
| 220 | -2.23 ± 0.02 | -0.596 ± 0.003 | 8.51 | 649 |

From Eq.(3.9), the intercept of voltage characteristic is proportional to saturation current, $I_0 = a/2Rb$. Saturation current calculated for different circuit termination resistors is shown in Table 3.IV. These values are in excellent agreement with saturation current obtained from the macroscopic *I-V* measurements, $I_0 = 7.83 \cdot 10^{-6}$ A. The deviation of slope from unity in the linear part of the curve results from a leakage current contribution to the diode conductivity. Associated resistance $R_l = 1/\sigma$ can be calculated from the voltage characteristics of the interface as $R_l = 2Rb/(1-b)$. Calculated values of leakage resistivity are summarized in Table 3.IV. These values are again very close to resistivities calculated from the macroscopic *I-V* curve under reverse bias conditions (~650 kOhm). These results demonstrate that the technique is quantitative and allows local transport imaging.

### 3.4. AC Transport: Scanning Impedance Microscopy – A Novel SPM Technique

In order to extend SPM transport measurements to ac regime, we have developed Scanning Impedance Microscopy (SIM), whose implementation, calibration and applications are discussed in this Section and in a number of related publications.[41,58,61]

Scanning Impedance Microscopy is based on dual pass imaging (lift mode). Electrostatic data are collected 50 to 100 nm above the surface as illustrated in Figure 3.7. The tip is held at constant bias $V_{dc}$ and a lateral bias $V_{lat} = V_{dc}+V_{ac}\cos(\omega t)$, is applied across the sample. This lateral bias induces an oscillation in surface potential

$$V_{surf} = V_s + V_{ac}(x)\cos(\omega t + \varphi(x)), \qquad (3.15)$$

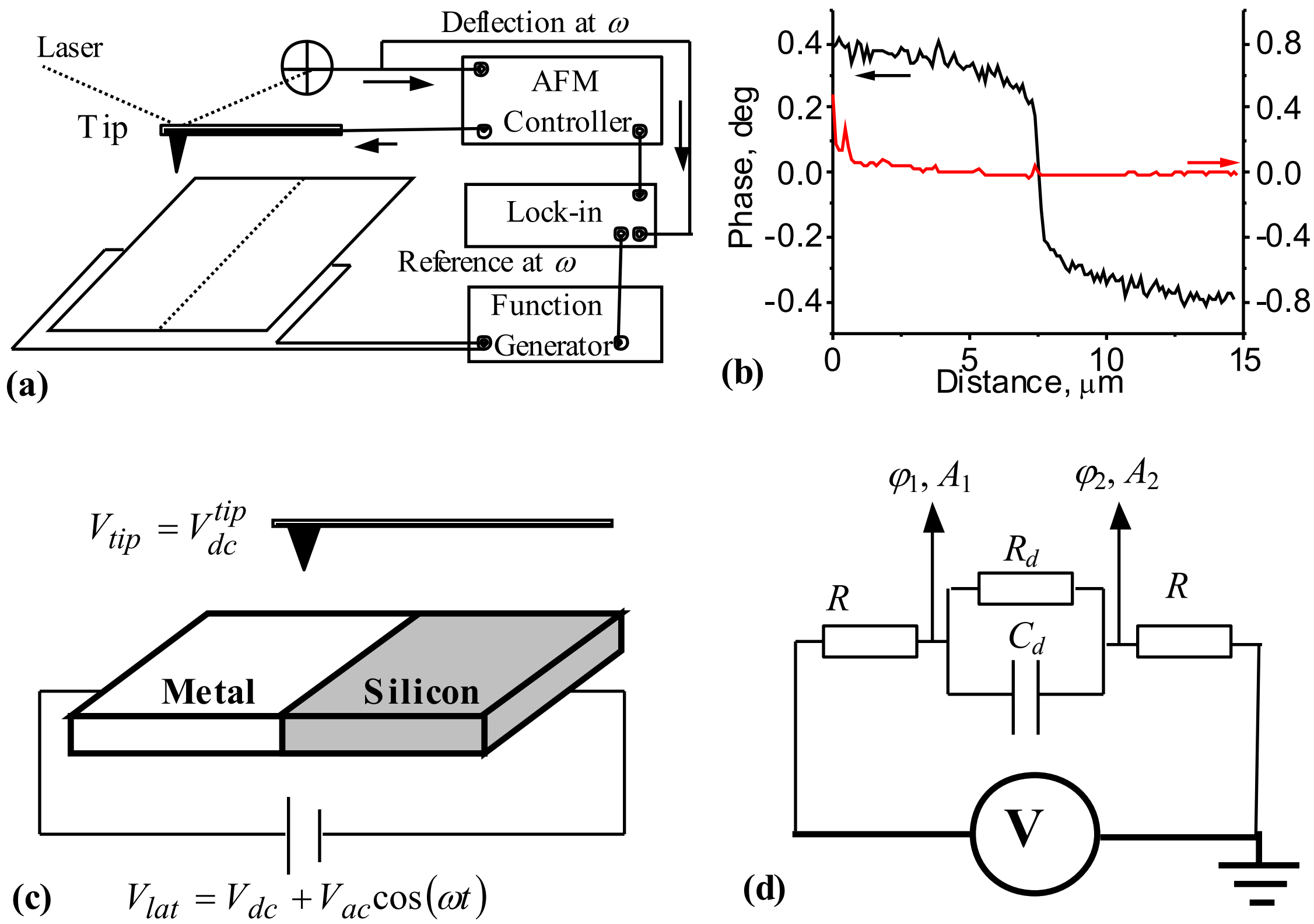

**Figure 3.7.** Experimental setup for scanning impedance microscopy (a) and typical phase and amplitude profiles across a grain boundary in $SrTiO_3$ bicrystal in the high frequency regime (b). Schematics of scanning impedance microscopy (c) and equivalent circuits for SIM of the single interface (d).

where $\varphi(x)$ and $V_{ac}(x)$ are the position dependent phase shift and voltage oscillation amplitude and $V_s$ is the dc surface potential.

The variation in surface potential results in a capacitive tip-surface force. The first harmonic of the force is

$$F_{1\omega}^{cap}(z) = C_z'(V_{dc} - V_s)V_{ac}, \tag{3.16}$$

where $C_z$ is tip-surface capacitance, $z$ is tip-surface separation and $V_s$ is surface potential. The magnitude, $A(\omega)$, and phase, $\varphi_c$, of the cantilever response to the periodic force induced by the voltage are:[62]

$$A(\omega) = \frac{F_{1\omega}}{m}\frac{1}{\sqrt{\left(\omega^2 - \omega_{rc}^2\right)^2 + \omega^2\gamma^2}} \quad \text{and} \quad \tan(\varphi_c) = \frac{\omega\gamma}{\omega^2 - \omega_{rc}^2}, \tag{3.17a,b}$$

where $m$ is the effective mass, $\gamma$ the damping coefficient and $\omega_{rc}$ the resonant frequency of the cantilever. Eqs.(3.15-17) imply that the local phase shift between the applied voltage and the cantilever oscillation is $\varphi(x)+\varphi_c$ and the oscillation amplitude $A(\omega)$ is proportional to the local voltage oscillation amplitude $V_{ac}(x)$. Therefore, variation in the phase shift (phase image) along the surface is equal to the variation of the true voltage phase shift with a constant offset due to the inertia between the sample and tip. The spatially resolved phase shift signal constitutes the SIM phase image of the device. The tip oscillation amplitude is proportional to the local voltage oscillation amplitude and constitutes the SIM amplitude image. This information is similar to that obtained by 4-probe impedance spectroscopy; therefore future references to the scanning probe technique based on cantilever phase detection induced by lateral ac bias applied to the surface will refer to Scanning Impedance Microscopy (SIM).

To determine the absolute value of local amplitude, $V_{ac}(x)$, from SIM data, the microscope is reconfigured to the open-loop SSPM mode, in which the feedback is disengaged, and tip oscillation in response to an ac bias applied to the tip is determined. The local voltage oscillation amplitude is then

$$V_{ac}(x)=\frac{V_{ac}A_{sim}(x)\left(V_{surf}(x)-V_{tip}^{sspm}\right)}{A_{sspm}(x)\left(V_{surf}(x)-V_{tip}^{sim}\right)}, \tag{3.18}$$

where $A$ is oscillation amplitude, $V_{tip}$ is the tip dc bias, $V_{ac}$ is the tip ac bias and *sim* and *sspm* refer to SIM and open-loop SSPM modes respectively. $V_{surf}(x)$ is the surface potential which varies with $x$ in the presence of a lateral bias and can be determined by independent SSPM measurement.

The major limitation of SIM is that the driving frequency must be selected far from the resonant frequency of the cantilever to minimize the variations of the phase lag between tip and surface due to electrostatic force gradients related to non-uniform surface potential.[63] In practice, however, SIM is used to obtain frequency dependent phase shift and amplitude; therefore, this limitation is not important. For frequencies above resonant frequency of the cantilever the amplitude of the response decreases rapidly with frequency [Eq.(3.17a)], therefore the phase error increases.[64] Noteworthy is that imaging is possible in the dc limit for harmonic oscillator response. Consequently, there is no fundamental limitation on imaging at the low frequencies; moreover, spectroscopic variants of this technique can be performed in all frequency ranges below cantilever resonant frequency $\omega_{rc}$. In the high frequency region, the cantilever response no longer follows harmonic oscillator type models and a number of cantilever resonances exist. Consequently, SIM can be extended to the high frequency regime except for the antiresonances where the cantilever frequency response is essentially zero.

### 3.4.1. Single Interface Systems

For a single interface system, the analysis of the SIM imaging mechanism is similar to that of SSPM. For the equivalent circuit in Figure 3.7d, the total impedance of the circuit, $Z_\Sigma$, is

$$Z_\Sigma = 2R + Z_d, \tag{3.19}$$

where $Z_d$ is the interface impedance. Under reverse biased conditions, the interface equivalent circuit is represented by a parallel *R-C* element and the impedance is:

$$Z_d = \frac{1}{1/R_d + i\omega C_d}, \tag{3.20}$$

where $R_d$ and $C_d$ are the voltage dependent interface resistance and capacitance, respectively. The voltage phase difference $\varphi_d = \varphi_2 - \varphi_1$ across the interface measured by SIM is calculated from the ratio of impedances on each side,

$$\beta = \frac{R}{Z_d + R}, \tag{3.21}$$

as (impedance divider effect):

$$\tan(\varphi_d) = \frac{\mathrm{Im}(\beta)}{\mathrm{Re}(\beta)}. \tag{3.22}$$

For the equivalent circuit in Figure 3.7d,

$$\tan(\varphi_d) = \frac{\omega C_d R_d^2}{(R + R_d) + R\omega^2 C_d^2 R_d^2}. \tag{3.23}$$

The voltage amplitude ratio, $A_1/A_2 = |\beta|^{-1}$, can be calculated from Eq.(3.21) as

$$\beta^{-2} = \frac{\left\{(R+R_d) + R\omega^2 C_d^2 R_d^2\right\}^2 + \omega^2 C_d^2 R_d^4}{R^2\left(1+\omega^2 C_d^2 R_d^2\right)^2}. \tag{3.24}$$

Depending on frequency, Eqs.(3.23,24) provide information on different aspects of interface conductance. In the low frequency limit, $\omega << \omega_0 = 1/(C_d R_d)$, the voltage phase shift across the interface is determined by interface resistance, interface capacitance and circuit termination resistance as

$$\tan(\varphi_d) = \omega C_d \frac{R_d^2}{(R+R_d)} . \tag{3.25}$$

In the same low frequency limit, the amplitude ratio is determined by the resistance ratio across the interface,

$$\frac{A_1}{A_2} = \frac{R+R_d}{R} . \tag{3.26}$$

In the high frequency limit, $\omega >> \omega_0$, Eq.(3.23) is

$$\tan(\varphi_d) = \frac{1}{\omega C_d R}, \tag{3.27}$$

while the amplitude ratio is equal to unity.

The crossover between low and high frequency limits occurs at the voltage relaxation frequency of the interface, $\omega_r$, defined as

$$\omega_r = \sqrt{\frac{R+R_d}{R}} \frac{1}{R_d C_d}, \tag{3.28}$$

at which the voltage phase angle attains it's maximal value

$$\tan(\varphi_d) = \frac{R_d}{2\sqrt{R(R+R_d)}} . \tag{3.29}$$

In the low frequency limit voltage phase shift at the interface is determined by interface capacitance and resistance, as well as by the resistance of the current limiting resistor [Eq.(3.25)]. In the high frequency limit, however, it is determined by interface capacitance and circuit termination only [Eg.(3.27)]. At the same time, the amplitude ratio is determined by interface and circuit termination resistance at low frequencies [Eq.(3.26)] and is unity in the high frequency regime. Therefore, SIM amplitude imaging at frequencies *below* the interface relaxation frequency visualizes <u>resistive</u> barriers at the interfaces and provides quantitative measure of interface resistance. SIM phase imaging at frequencies *above* the interface relaxation frequency visualizes capacitive barriers at the interfaces and provides a quantitative measure of interface capacitance. Variation of the circuit termination resistor, $R$, verifies the validity of Eqs.(3.26,27). For other circuit terminations, including both resistive and capacitive elements, a similar but more complex analysis is required.[65]

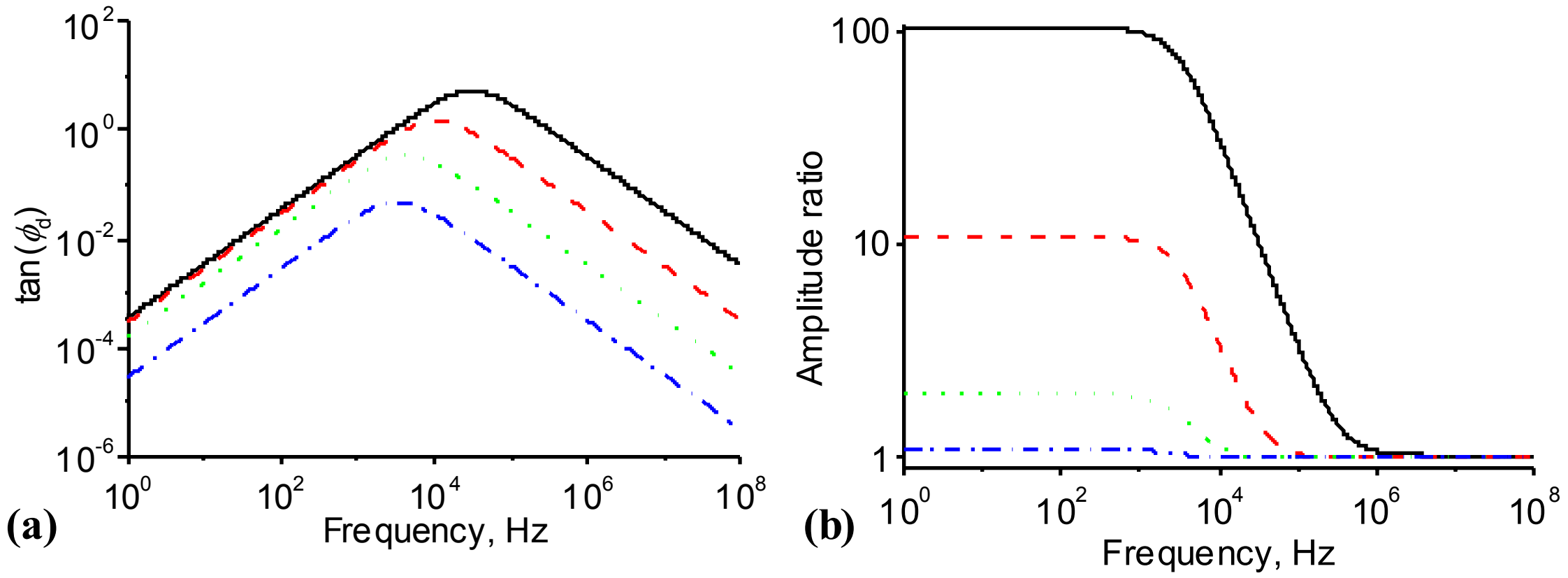

**Figure 3.8.** Calculated phase shift (a) and amplitude ratios across the interface for different circuit terminations. The interface resistance and capacitance are taken to be $C_d$ = 1.0 $10^{-10}$ F, $R_d$ = 500 kOhm for current limiting resistors of $R$ = 5 kOhm ( ), 50 kOhm ( ), 500 kOhm ( ) and 5 MOhm ( ).

It is illustrative to calculate the response for an ideal interface as shown in Figure 3.8. Both regimes for the frequency dependence of $\tan(\varphi_d)$ can be seen. An amplitude drop is expected at the interface for low frequencies, while no amplitude drop occurs at high frequencies. For a well-defined circuit termination the amplitude ratio and phase shift can be used to determine frequency dependent interface resistance, $R_d(\omega)$, and capacitance, $C_d(\omega)$, by solving Eqs.(3.23,24) for each frequency. In practice, such analysis is limited to relatively low frequency region where the amplitude ratio is far from unity. The alternative approach for SIM data interpretation is direct fitting of the frequency dependences of phase shift and amplitude ratio by Eqs.(3.23,24) that allows estimates of frequency independent interface resistance, $R_d$, and capacitance, $C_d$. This approach is applied to the analysis of transport properties of atomically abrupt grain boundary in $SrTiO_3$ in Chapter 4.

### 3.4.2. Multiple Interface Systems

SIM allows quantitative measurements of voltage amplitude and phase within the grain and at the grain boundaries as well as delineation of the resistive vs. capacitive behavior of individual microstructural elements in multiple interface systems. It should be noted here that unlike conventional impedance spectroscopy, in which the elements of equivalent circuit are associated with material microstructure in an averaged fashion, SIM allows direct correlation between impedance image and local microstructure.

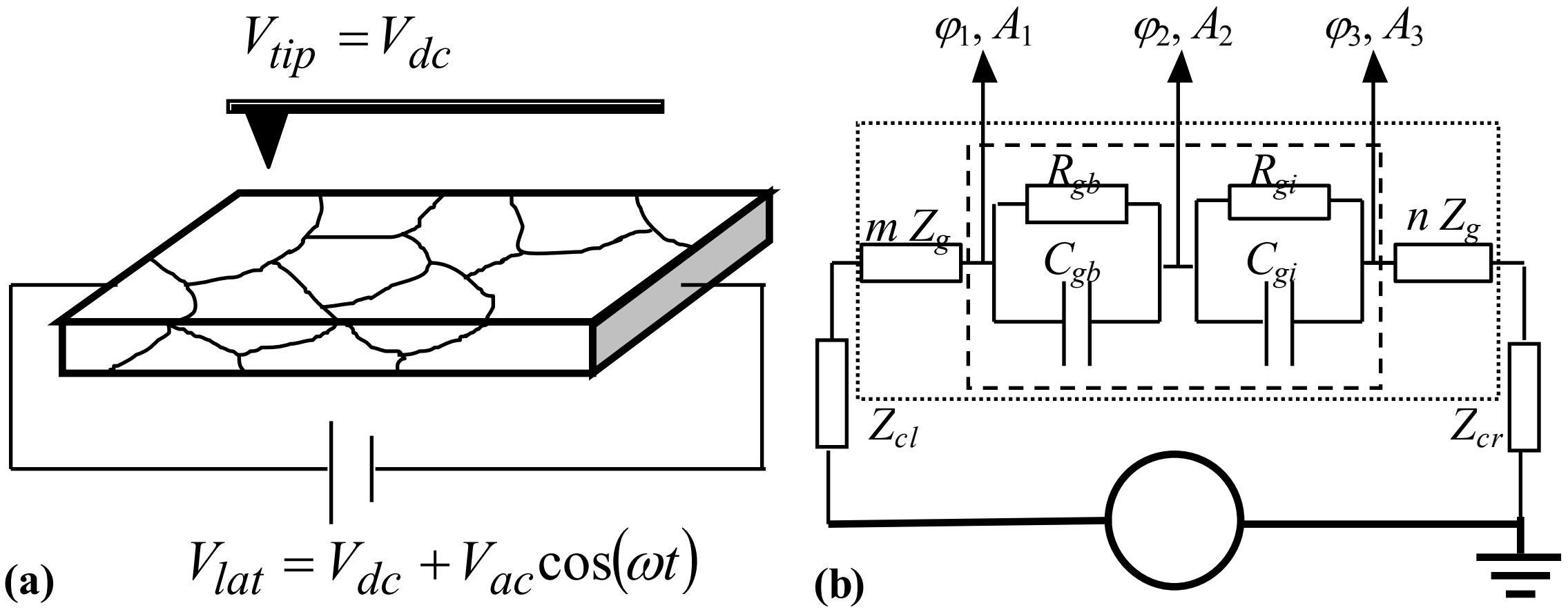

**Figure 3.9.** Experimental set-up for ac transport measurements by SIM in multiple interface systems (a) and corresponding equivalent circuit (b)

Assuming a series arrangement of the grains (Figure 3.9), the total impedance of the sample, $Z_\Sigma$, is:

$$Z_\Sigma = Z_{lc} + N\left(Z_{gb} + Z_{gi}\right) + Z_{rc}, \tag{3.30}$$

where $Z_{lc}$ is the impedance of the left contact, $Z_{rc}$ is the impedance of the right contact, $Z_{gb}$ is the grain boundary impedance, $Z_{gi}$ is the impedance of grain interior and $N = n+m+1$ is the number of the grains. Grain boundary and grain interior impedances are modeled by capacitive and resistive elements in parallel,

$$Z_{gb} = \frac{1}{1/R_{gb} + i\omega C_{gb}}, \qquad Z_{gi} = \frac{1}{1/R_{gi} + i\omega C_{gi}}, \tag{3.31a,b}$$

where $\omega$ is frequency, $R_{gb}$ and $C_{gb}$ are the grain boundary resistance and capacitance and $R_{gi}$ and $C_{gi}$ are the grain interior resistance and capacitance. As for the DC transport,

Eq.(3.31a,b) can be interpreted in terms of the brick-layer model, where measured grain boundary and bulk resistances and capacitances for the sample are scaled linearly and reciprocally by the number of grains in the cross-section of the sample. Eqs.(3.31a,b) can also be extended to alternative frequency dependent impedance models such as constant phase angle element (CPE).

The phase change at the grain boundary is calculated from the ratio of impedances between the region to the left and to the right of grain boundary and the ground,

$$\beta = \frac{nZ_{gb} + (n+1)Z_{gi} + Z_{rc}}{(n+1)(Z_{gb} + Z_{gi}) + Z_{rc}}, \tag{3.32}$$

as (impedance divider effect):

$$\tan(\varphi_{gb}) = \frac{\mathrm{Im}(\beta)}{\mathrm{Re}(\beta)}. \tag{3.33}$$

The ratio of voltage oscillation amplitudes on the left and on the right is

$$\frac{A_2}{A_1} = |\beta|. \tag{3.34}$$

For high tip biases during SIM measurement, this ratio is equal to the ratio of the tip oscillation signal (lock-in output) and is independent of the properties of the tip. Alternatively, Eq.(3.18) should be used to analyze the experimental data. Similar analysis for the grain interior and electrodes is straightforward.

It should be noted that Eqs.(3.30,31) are directly interpretable in terms of the brick-layer model. Indeed, the grain boundary and bulk impedances scale reciprocally with cross-section area; therefore, impedance ratios defined in Eqs.(3.32,33,34) do not depend on sample area.

It is illustrative to model the typical behavior of these values for a realistic material. Figure 3.10 shows impedance spectra and the SIM phase and amplitude characteristics of a grain boundary and grain interior calculated for the circuit in Figure 3.9b with $C_{gi}$ = 10 nF, $R_{gi}$ = 1 kOhm, $C_{gb}$ = 1 μF, $R_{gi}$ = 3 kOhm, $C_{lc}$ = 3 μF and $R_{lc}$ = 10 kOhm.[66] Calculated responses without a contact impedance contribution (generic termination) (1), including contact impedance (2), assuming only contact resistance (3) and for contact resistance $R_{lc}$ = 100 MOhm (4) are shown. The impedance spectra exhibit two well-defined half arcs corresponding to grain boundaries and grain interior and a partially overlapping electrode half arc. The phase shift is a maximum for the frequency between characteristic relaxation frequencies for the bulk and grain boundary. This frequency is further referred to as the voltage resonant frequency of the interface. The grain interior phase shift is equal to the grain boundary phase shift in magnitude and opposite in sign. The only exception is for a purely resistive circuit termination. In this case, there is an asymmetry between the grain boundary and grain interior phase shifts. It should be noted that the typical setup for SIM imaging includes current limiting resistors in the circuit and, therefore, resistive termination is usual. The grain boundary amplitude ratio is small below the resonant frequency and goes to unity above the resonant frequency. Therefore, amplitude changes are expected on the grain boundaries below the resonant frequency (dc limit), while for high frequencies grain boundaries are not associated with amplitude changes. In contrast, the grain interior amplitude ratio (i.e. local slope) in the dc limit is equal to $R_{gb}/(R_{gi}+R_{gb})$ and decreases for higher frequencies, but never achieves unity. Therefore, in the high frequency limit the SIM amplitude is expected to exhibit uniform decay along the sample surface and grain boundary barriers are not visible on the SIM amplitude image.

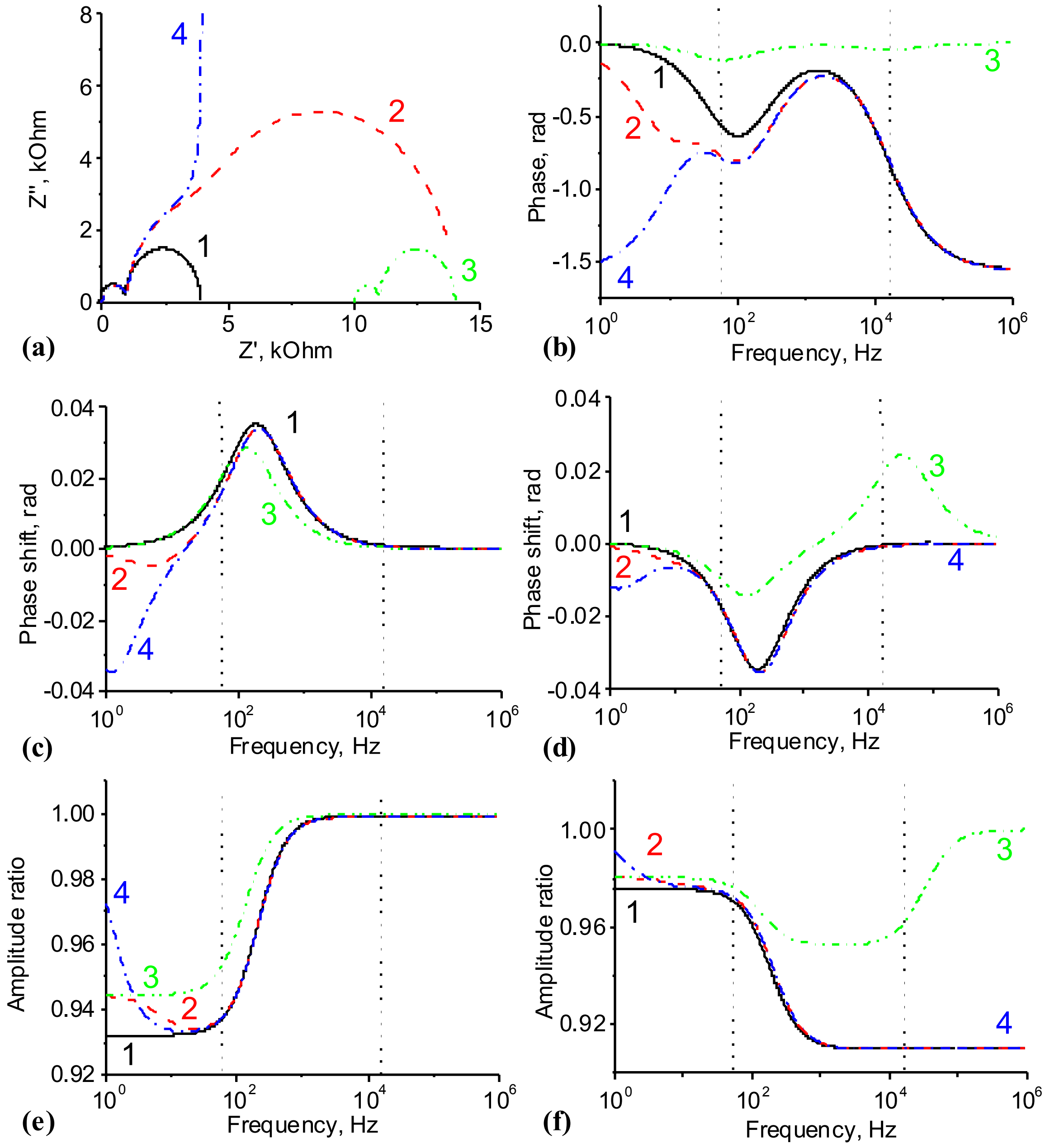

**Figure 3.10.** Calculated Cole-Cole plots (a), impedance phase angle vs. frequency (b), SIM phase shifts at the grain boundary (c) and within the grain (d) and SIM amplitude ratios at the grain boundary (e) and within the grain (f), for a circuit without a contact contribution (generic termination) (1), with a contact contribution (2), with purely resistive termination (3), and with high- resistance contacts (4).

This observation implies that the SIM imaging of polycrystalline materials will exhibit phase shifts on the interface and phase shifts of the opposite sign in the grain interior. The presence of the resistive circuit termination suppresses the latter for frequencies above the resonant frequency. Below the resonant frequency, the amplitude drops at the interfaces and exhibits uniform behavior within the grains similarly to the dc potential behavior. Above the resonant frequency, there is no amplitude change at the interfaces, while there is an amplitude drop within the grain that can be determined as a uniform slope.

### 3.4.3. Calibration

The analysis above was based on the assumption that interface phase shift and amplitude ratio are independent of probe properties and imaging conditions (tip-surface separation, tip bias, etc). Given that, interface resistance and capacitance can be determined from phase shift and amplitude ratio. In this section, the probe effect on imaging is determined and the interface properties measured on the calibration standard by SIM are compared to macroscopic impedance measurements.

Probe Effect

To determine the probe effects on imaging, phase and amplitude images across an atomically abrupt $SrTiO_3$ grain boundary were acquired under varying imaging conditions. The interface was located using SSPM on a grounded surface.[67,68] To quantify the experimental data the average amplitude and phase of the tip response were defined as the averages of unprocessed amplitude and phase images. To analyze the grain boundary phase shift averaged phase profiles were extracted and fitted by a Boltzman function $\varphi = \varphi_0 + \Delta\varphi_{gb}\left(1+exp\left(\left(x-x_0\right)/w\right)\right)^{-1}$, where $w$ is the width and $x_0$ is the center of phase profile. The driving frequency dependence of the average phase shift and amplitude are found to be in excellent agreement Eq.(3.17a,b). The amplitude was found to be linear in tip bias (Figure 3.11a), in agreement with Eq.(3.16,17b). The amplitude is nullified when the tip bias is $V_{dc}$ = 0.28 ± 0.02 V independent of tip-surface separation. Eq.(3.16) implies that this condition is achieved when $V_{dc} = V_s$, thus yielding the value of surface potential. The phase of the response changes by 180° between $V_{dc}$ = 0 and $V_{dc}$ = 1 (Figure 3.11c). Grain boundary phase shift is independent of tip bias (Figure 3.11e). A small variation in grain boundary phase shift occurs when tip potential, $V_{dc}$, is close to the surface potential, $V_s$, and the amplitude of the cantilever response is small. Note that the slopes of lines in Figure 3.11a are smaller for large tip-surface separations, indicative of a decrease in capacitive force, while the grain boundary phase shift does not depend on distance. The amplitude is linear in driving bias, $V_{ac}$, as shown in Figure 3.11b. Both the average and grain boundary phase shift are essentially driving amplitude independent (Fig. 3.11d,e).

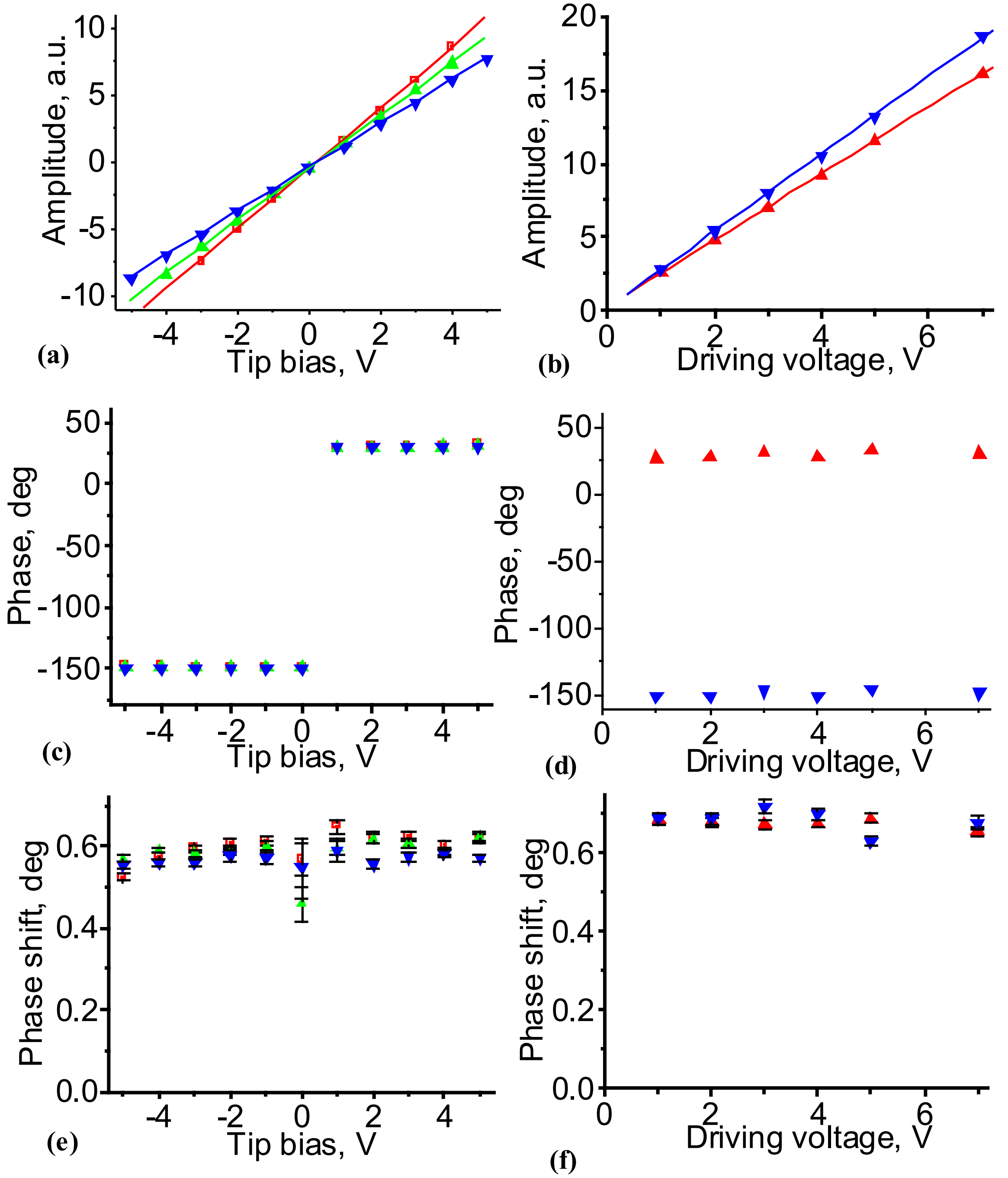

**Figure 3.11.** Cantilever oscillation amplitude (a,b), average phase shift (c,d) and interface phase shift (e,f) dependence on tip dc bias (a,c,e) and lateral ac bias (b,d,f). ■, ▲ and ▼ denote tip-surface separations of 50, 100 and 250 nm (a,c,e). ▲ and ▼ denote tip bias of $V_{dc}$ = 5 and -5 V (b,d,f).

Interface properties

To calibrate SIM and verify analytical solution developed in Section 3.4.1, we used the same calibration standard (Schottky diode) as for the dc transport measurements. SIM phase profiles across the metal-semiconductor interface are shown in Figure 3.12a,b.

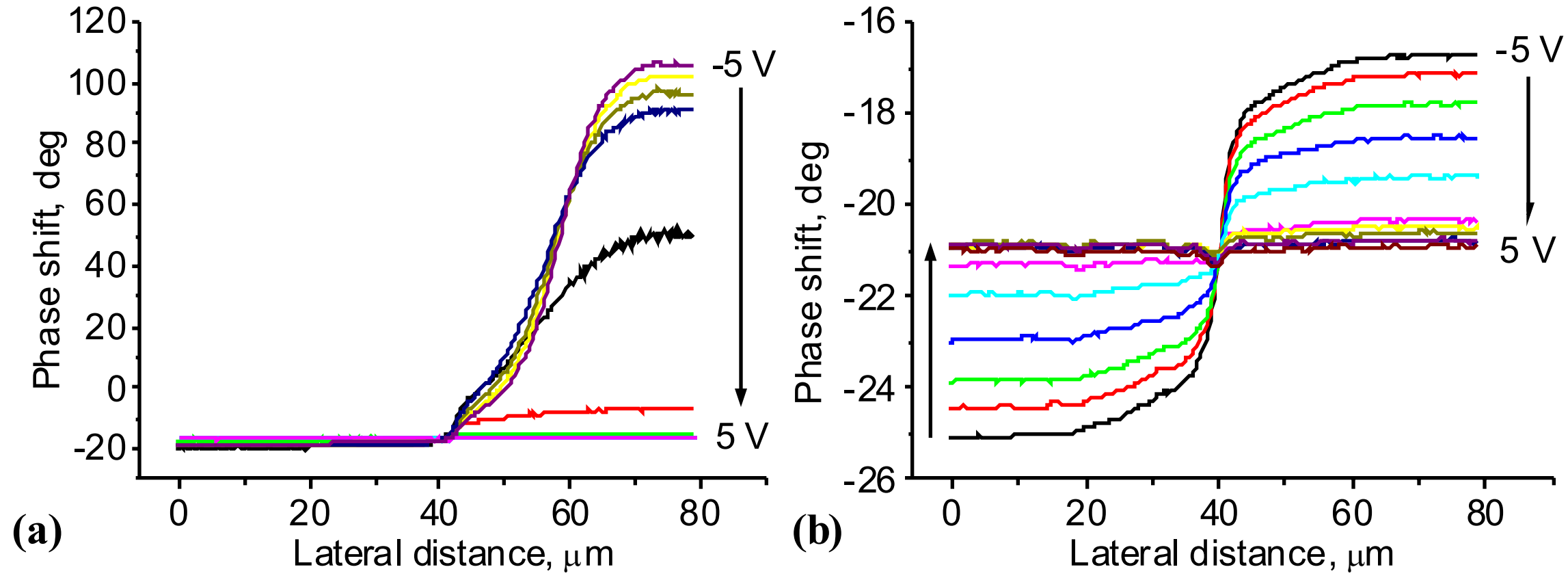

**Figure 3.12.** Scanning Impedance microscopy phase profiles across the interface for different lateral biases and $R$ = 500 Ohm (a) and 100 kOhm (b).

For small current limiting resistors the phase shift is anomalously large: ~172° at 3 kHz and 106° at 100 kHz. Phase shifts $\varphi_d > 90°$ imply that a *negative* bias to the device results in an *increase* of surface potential.[69] This behavior is analogous to the dc potential behavior observed in SSPM and is attributed to photoelectric carrier generation in the junction region. Again, this effect is completely suppressed by circuit termination with resistors $R \geq 10$ kOhm and phase shift at the interface for 100 kOhm termination is shown in Figure 3.12b. Note that for forward bias the phase shift on the left is voltage independent, while there is some residual phase shift on the right of the Schottky barrier. This phase shift is attributed to the diffusion capacitance of a forward biased junction.

Frequency dependence of tip phase and amplitude on the left and on the right of the junction is shown in Figure 3.13a,b. Tip dynamics are determined by a convolution of the harmonic response of the tip to the periodic bias [Eq.(3.17a,b)] and the frequency dependence of position dependent voltage phase and amplitude induced by the lateral bias. Nevertheless, the abrupt phase change by *ca.* 180° and tip oscillation amplitude maximum at $f$ = 72 kHz are indicative of a mechanical tip resonance. Detailed analysis of the frequency dependence of the amplitude has demonstrated that the resonant frequency on the left and right of the junction are shifted by ~1kHz due to the difference in surface potential and electrostatic force gradient.[63] As suggested by Eqs.(3.17a,b), the phase and amplitude of a harmonic oscillator are very steep close to the resonant frequency. Therefore, minute changes of the resonance frequency result in major errors in phase and amplitude in this frequency region. To minimize this effect, data were collected from 3kHz to 65 kHz and 75 kHz to 100kHz.

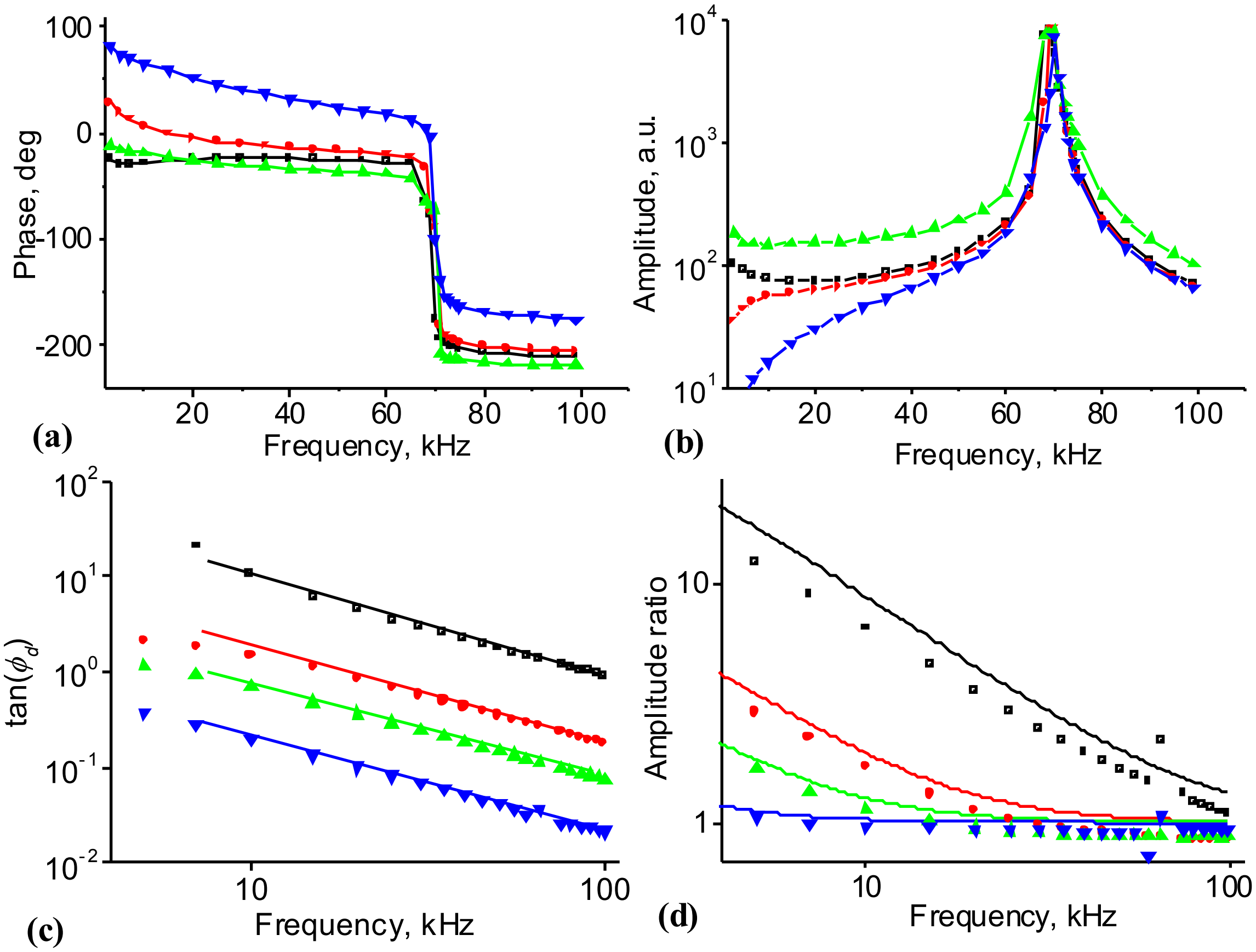

**Figure 3.13.** Frequency dependence of tip oscillation phase (a) and amplitude (b) on the left and on the right of the junction for $R$ = 10 kOhm (▲, ▼) and 100 kOhm (■,●). Frequency dependence of phase shift (c) and amplitude (d) for circuit terminations 10 kOhm (■), 47 kOhm (●), 100 kOhm (▲) and 220 kOhm (▼). Solid lines are linear fits (c) summarized in Table 3.V, and amplitude ratios calculated with interface capacitance from Table 3.V (d).

The frequency dependence of phase shift for different circuit terminations is shown in Figure 3.13c. From macroscopic impedance spectroscopy the relaxation frequency of the junction is estimated as 1.5 kHz at -5V reverse bias. Therefore, SIM measurements are performed in the high frequency region in which Eq.(3.27) is valid. In agreement with Eq.(3.27), $\tan(\varphi_d)$ is inversely proportional to frequency with a proportionality coefficient determined by the product of interface capacitance and circuit termination resistance. The experimental data are described by $\log(\tan(\varphi_d)) = a + b\log(f)$ and corresponding fitting parameters are listed in Table 3.V. Note that *b* is within experimental error of the theoretical value *b* = -1, in agreement with the parallel *R-C* model for the interface. As follows from Eq.(3.27), interface capacitance can be determined as $C_d = 10^{-a}/(2\pi R)$ and capacitances for different circuit terminations are listed in Table 3.V. Interface capacitance increases with the current limiting resistor and in all cases is larger than the capacitance obtained from macroscopic impedance spectroscopy, $C_d = 1.71 \cdot 10^{-10}$ F at -5 V. Amplitude ratios were calculated from Eq.(3.24) for interface capacitances in Table 3.V and $R_d$ = 603 kOhm and compared with experimental results [Eq.(3.18)] in Figure 3.13d. Note the excellent agreement between experimental and calculated values despite the absence of free parameters.

Table 3.V.

*Frequency dependence of SIM phase shift*

| R, kOhm | *a* | *b* | $C_d$, $10^{-10}$ F | $V_d$, V |
|---|---|---|---|---|
| 10 | 4.94 ± 0.02 | -0.99 ± 0.01 | 1.83 | 4.83 |
| 47 | 4.21 ± 0.01 | -0.98 ± 0.01 | 2.11 | 3.85 |
| 100 | 3.84 ± 0.01 | -0.98 ± 0.01 | 2.32 | 2.86 |
| 220 | 3.29 ± 0.04 | -0.98 ± 0.02 | 3.76 | 0.80 |

To quantify the microscopic *C-V* behavior, interface phase shift was measured as a function of lateral dc bias for different circuit terminations [Figure 3.14a]. Under reverse bias $\tan(\varphi_d)$ changes by almost two orders of magnitude from $\tan(\varphi_d)$ = 1.8 for *R* = 10 kOhm to $\tan(\varphi_d)$ = 0.042 for *R* = 220 kOhm. Interface capacitance can be calculated from the data in Figure 3.14a, while potential drop at the interface is directly accessible from SSPM measurements [Figure 3.6d]. Combination of the two determines the *C-V* characteristics of the interface, which is shown in Figure 3.14b. The resulting curve exhibits universal behavior independent of the current limiting resistance. This dependence is described as $1/C^2 = (4.0 \pm 0.6)\cdot 10^{18} + (6.1 \pm 0.3)\cdot 10^{18} V_d$. Using the relation for an ideal metal-semiconductor junction[70] yields the Schottky barrier height as $\phi_B = 0.6 \pm 0.1$ V. This value agrees with the barrier height obtained from macroscopic *I-V* measurements as $\phi_B = 0.55$ V. From the slope of the line the dopant concentration for the material is estimated as $N_B = 1.06\cdot 10^{24}$ m$^{-3}$.

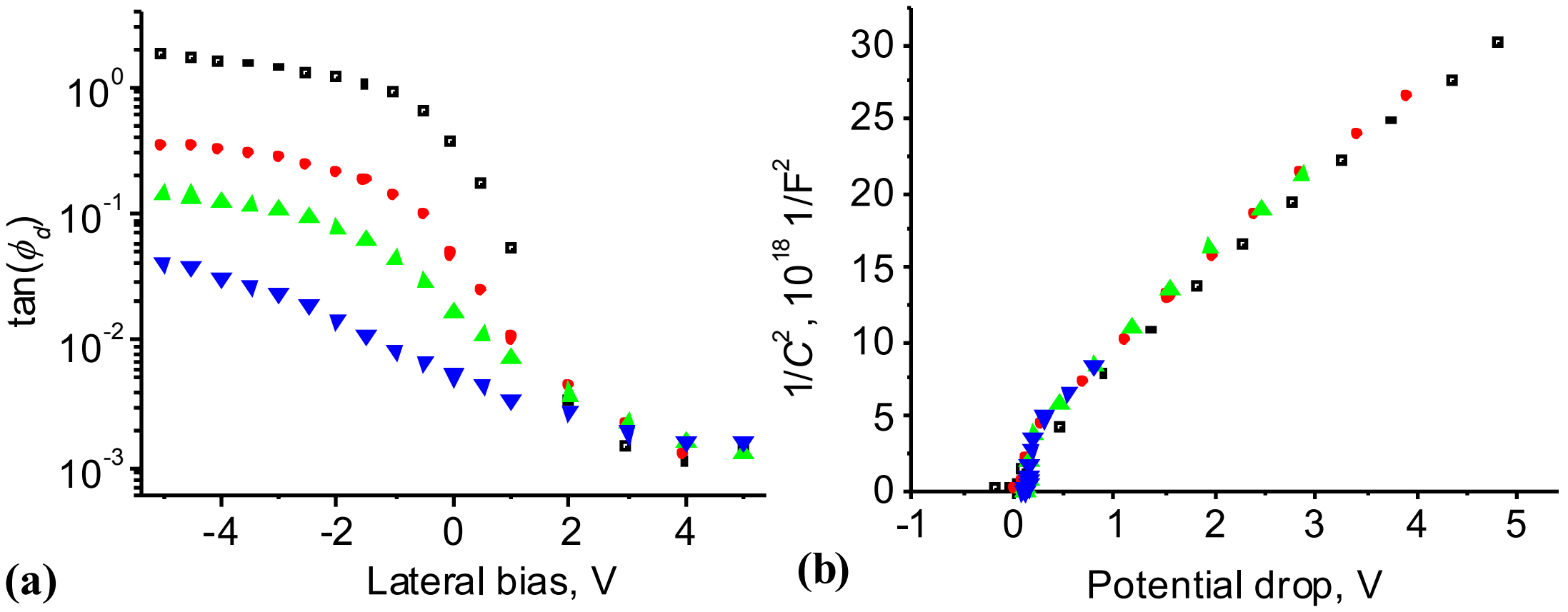

**Figure 3.14.** (a) Voltage phase angle tangent as a function of lateral bias for circuit terminations 10 kOhm (■), 47 kOhm (●), 100 kOhm (▲) and 220 kOhm (▼). (b) Calculated $1/C^2$ vs. $V_d$ for different circuit terminations. Though $\tan(\varphi_d)$ varies by 2 orders of magnitude, $1/C^2$ exhibits universal behavior.

These results demonstrate that local interface imaging of a metal-semiconductor interfaces yields junction properties completely consistent with properties determined by macroscopic techniques, thus verifying the quantitative nature of SIM.

### 3.4.4. Structure of Profiles

Further insight into the potential, field and current distribution in the junction region can be obtained from the analysis of SSPM and SIM profiles. Shown in Figure 3.15a is the lateral gradient of surface potential across the junction under reverse bias conditions, $V$ = -5V. The peak is asymmetric and can be deconvoluted to two Gaussian profiles of effective width ~2.1 μm corresponding to total effective width of ~ 2.8 μm. The half-width of SIM phase profile is 2.8 μm for $R$ = 10kOhm, 100kOhm [Figure 3.15b]. Therefore, the resolution of both techniques is comparable, as expected due to the similarities of tip-surface interactions employed in imaging mechanism.

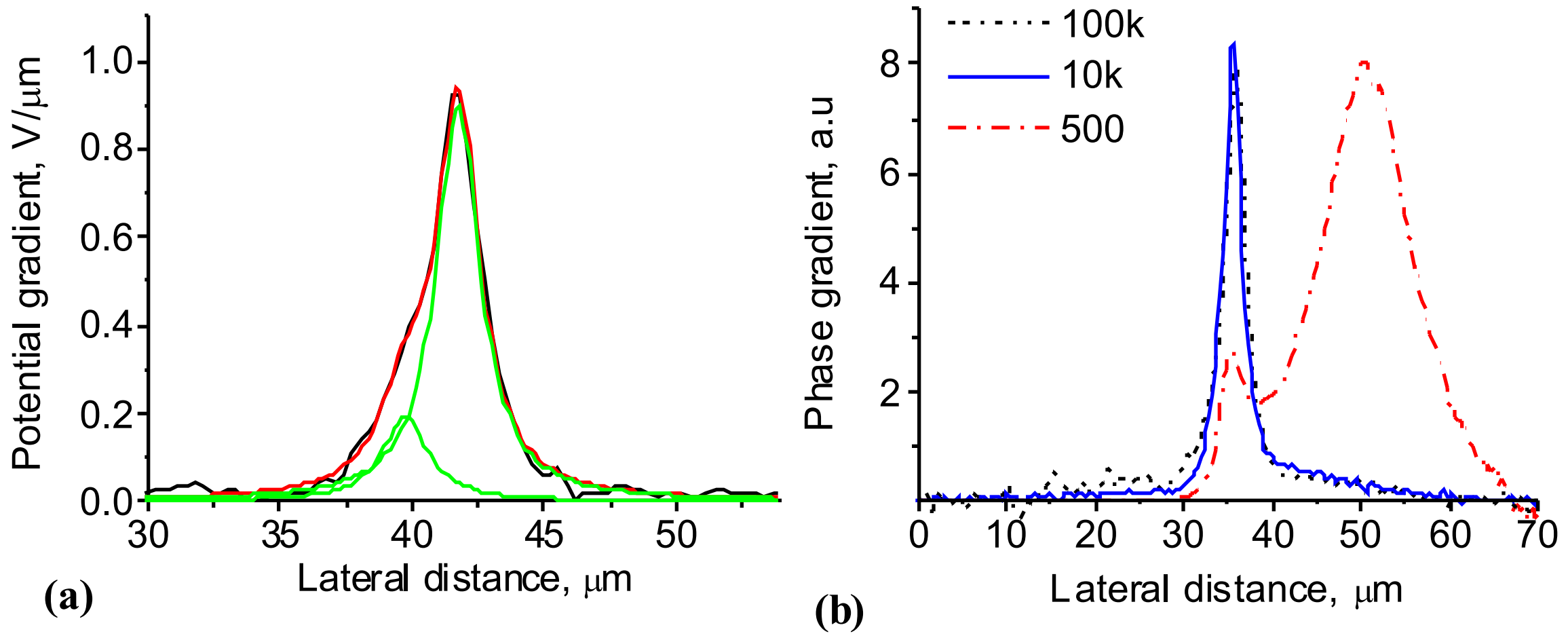

**Figure 3.15.** (a) Lateral potential gradient across the junction and (b) SIM phase angle gradient across the junction. Note that for small $R$ = 500 Ohm a large phase angle feature exists on the right of the interface. This effect is completely suppressed by high resistance circuit termination $R$ = 10kOhm and larger.

Noteworthy is that phase angle gradient across the surface for low resistance termination $R$ = 500 Ohm clearly exhibits both a junction peak with a half-width of 2 μm and a much broader peak with half-width of 11 μm. As mentioned throughout the text, this phenomenon is attributed to photoinduced carrier generation in the junction area. Analysis of SIM profile shape allows distinguishing the spatial localization of the two effects thus clarifying the origin of anomalous phase shift effect. It should also be noted that the exact shape of potential and SIM profiles is sensitive to surface topography and, in this case, relatively poor topographic structure of the metal-silicon interface precludes more quantitative studies.

### 3.4.5. DC and ac Transport in *p*-doped Polycrystalline Silicon

The question of extending this approach in a quantitative manner to more complex structures is illustrated with polycrystalline silicon. Topographic structure of polished *p*-doped silicon is compared to surface potential in is Figure 3.16a,b. Grain boundaries are associated with positive potential features of order of ~30 mV. This sign is expected for positively charged grain boundaries in a *p*-doped material. Application of 10 V lateral bias results in potential drops at the interfaces confirming the high resistivity of grain boundary region [Figure 3.16c]. The direction of the potential drop is inverted for -10 V lateral bias [Figure 3.16d]. SIM phase and amplitude images are shown in Figure 3.16e,f. Potential drop at the interface (~180 mV for 10 V bias and current termination resistance $R$ = 2.18 kOhm) yields the contribution of that interface as ~80 Ohm, which corresponds to ~18% of the total sample resistivity. Ramping the lateral bias from -10V to 10V shows that the interface is ohmic for grain boundary potentials between -200-200 mV. The SIM phase signal was linear with frequency in the frequency range 50-100kHz, suggesting that this region represents the low frequency limit of the interface. Indeed, the relaxation frequency of the interface was confirmed by macroscopic impedance spectroscopy to be 400kHz. The SIM amplitude image suggests that ac losses at the interface and in the bulk provide comparable contributions to overall sample resistance (IS has shown $R_{gb}$ = 400 Ohm, $R_{bulk}$ = 70 Ohm). The shape of the SIM phase signal is rather complex in comparison with the diode. Specifically, there appears to be a minimum in the vicinity of the grain boundary rather than a step [Figure 3.17c]. The experimental observations are rationalized if we assume that the application of an ac bias results in the generation of minority carriers in the interface region. Minority carrier dynamics are given by

$$\frac{\partial n}{\partial t} = D\frac{\partial^2 n}{\partial x^2} - \frac{n}{\tau}, \tag{3.34}$$

where $n$ is minority carrier concentration, $D$ is diffusion coefficient and $\tau$ is relaxation time. The solution of Eq.(3.34) for boundary conditions at the grain boundary $n(0) = n_0 e^{i\omega t}$ and in the bulk $n(\infty) = 0$ is

$$n = n_0\, exp\left(-x\sqrt{\left(i\omega + \tau^{-1}\right)/D}\right)e^{i\omega t}. \tag{3.35}$$

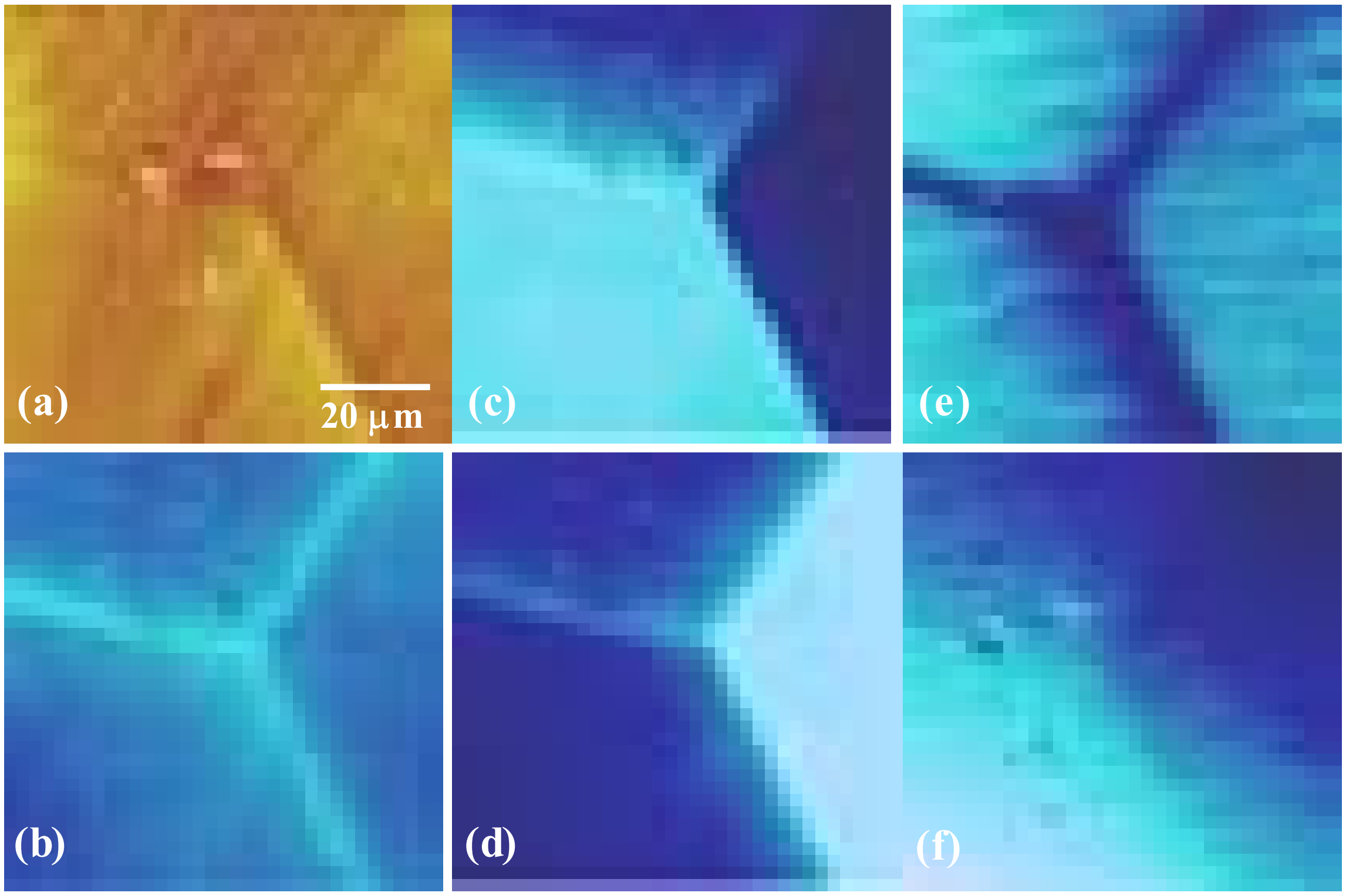

**Figure 3.16.** Surface topography (a) and surface potential of grounded (b), positively (c) and negatively (d) biased polycrystalline p-doped silicon surface. SIM phase (e) and amplitude (f) images of the same region at 90 kHz. Scale is 200 nm (a), 200 mV (b,c,d), 0.36 degree (e).

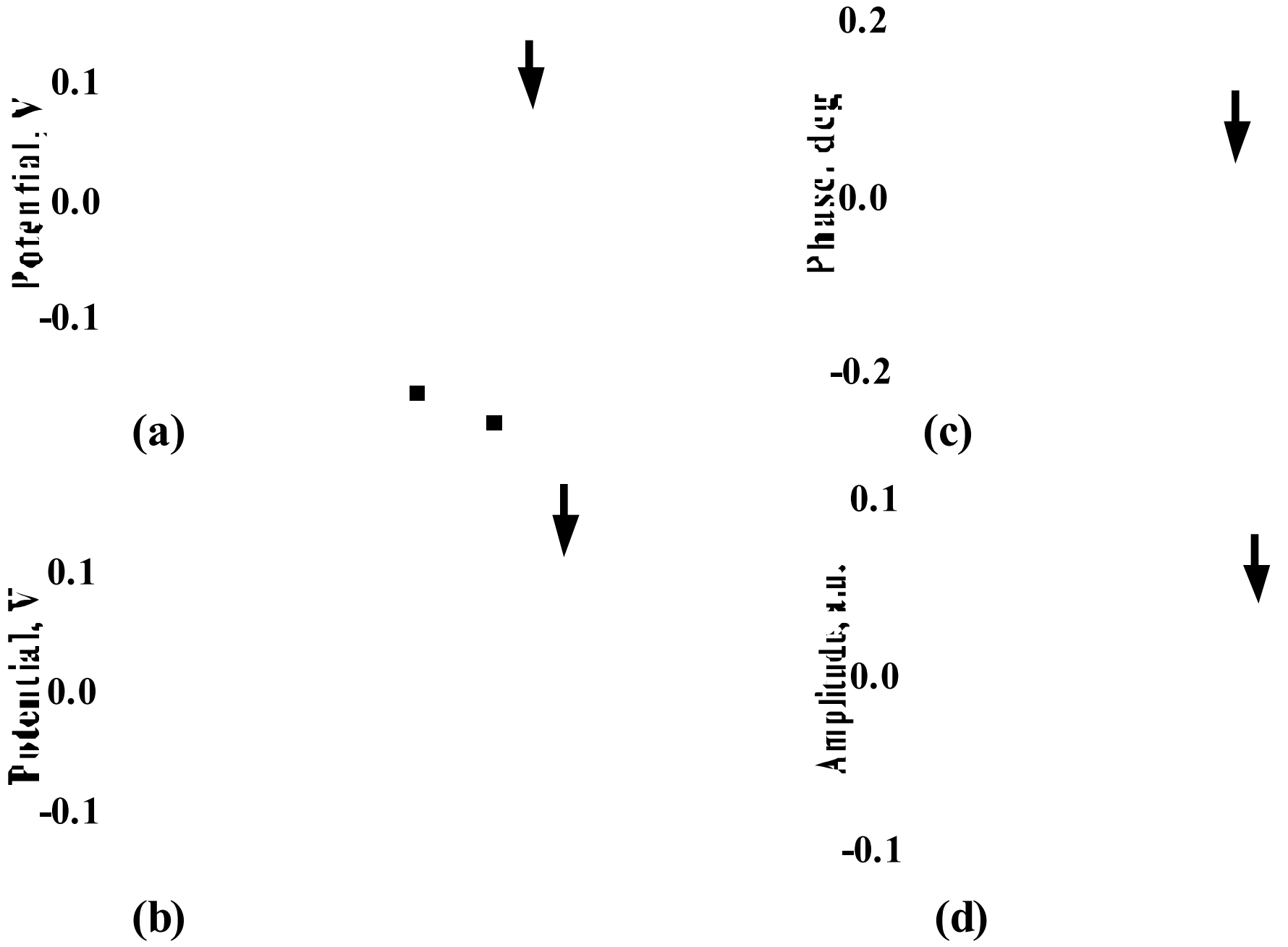

**Figure 3.17.** Potential profiles across the grain boundary for positive (a) and negative (b) lateral bias. Position of the grain boundary is indicated by the arrow. SIM phase (c) and amplitude (d) profiles across the same boundary.

In the low frequency limit, $\tau^{-1} >> \omega$, Eq.(3.35) is $n = n_0 e^{-x/L} e^{i\omega\tau/2} e^{i\omega t}$, where diffusion length $L = \sqrt{D\tau}$. Surface potential is related to the minority carrier concentration as $V_s = V_{s0} + \alpha n$, where $V_{s0}$ is potential of the surface far from the interface and $\alpha$ is a proportionality coefficient. Surface potential in the vicinity of the interface contains contributions from the applied lateral bias and the minority carrier current, $V = V_{s0} + V_{ac}e^{i\omega t} + \alpha n_0 e^{-x/L} e^{i\omega\tau/2} e^{i\omega t}$. Therefore, the SIM phase shift is

$$\tan(\varphi) = \frac{\alpha n}{V_{ac}} e^{-x/L} \frac{\omega\tau}{2}. \tag{3.36}$$

Eq.(3.36) predicts an exponential decay of SIM phase shift on the length scale comparable to the minority carrier diffusion length. Phase shift is expected to be linear in frequency in a good agreement with experimental observations.

### 3.5. Tip Calibration in Electrostatic SPMs

The resolution of force-based electrostatic SPMs such as SSPM and SIM for quantitative nanoscale imaging is limited by geometry of the tip.[71,72,73] For small tip-surface separations tip geometry can be accounted for using the spherical tip approximation and the corresponding geometric parameters can be obtained from electrostatic force- or force gradient distance and bias dependences.[74,75] Such a calibration process is often tedious and tip parameters tend to change with time due to mechanical tip instabilities.[76] Alternatively, properties can be quantified directly using an appropriate calibration method.[77] A tip-surface transfer function could be used to deconvolute the tip contribution from experimental data if it were known. Recently, well-defined metal-semiconductor interfaces have been considered as "potential step" standards.[78] However, the presence of surface states and mobile charges significantly affects potential distributions of even grounded surfaces. In addition, such a standard is expected to be sensitive to environmental conditions (humidity, temperature, etc).[79]

Here we propose a carbon nanotube based standard for tip calibration in electrostatic SPM. An ac voltage bias is applied across a nanotube resulting in the oscillation of the SPM tip due to the capacitive force (SIM amplitude imaging).[80,81] Since the nanotube is significantly smaller than the tip, it effectively probes the tip geometry. The force between the tip and the surface can be written as a function of capacitances as

$$2F_z = C'_{ts}(V_t - V_s)^2 + C'_{ns}(V_n - V_s)^2 + C'_{tn}(V_t - V_n)^2, \tag{3.37}$$

where $V_t$ is tip potential, $V_n$ is nanotube potential and $V_s$ is surface potential, $C_{ts}$ is tip-surface capacitance, $C_{ns}$ is nanotube-surface capacitance and $C_{tn}$ is tip-nanotube capacitance. $C'$ refers to derivative of capacitance with respect to the $z$ direction perpendicular to the surface. When an ac bias is applied across the nanotube, $V_n = V_0 + V_{ac}\cos(\omega t)$, and $V_s = V_0$. Therefore, the first harmonic of tip-surface force is:

$$F_{1\omega} = C'_{tn} V_{ac}(V_t - V_0). \tag{3.38}$$

In comparison, application of an ac bias to the tip, $V_t = V_{dc} + V_{ac}\cos(\omega t)$ yields

$$F_{1\omega} = C'_{tn} V_{ac}(V_{dc} - V_0) + C'_{ts} V_{ac}(V_{dc} - V_s). \tag{3.39}$$

Therefore, applying an ac bias directly to the carbon nanotube allows the tip-surface capacitance to be excluded from the overall force.

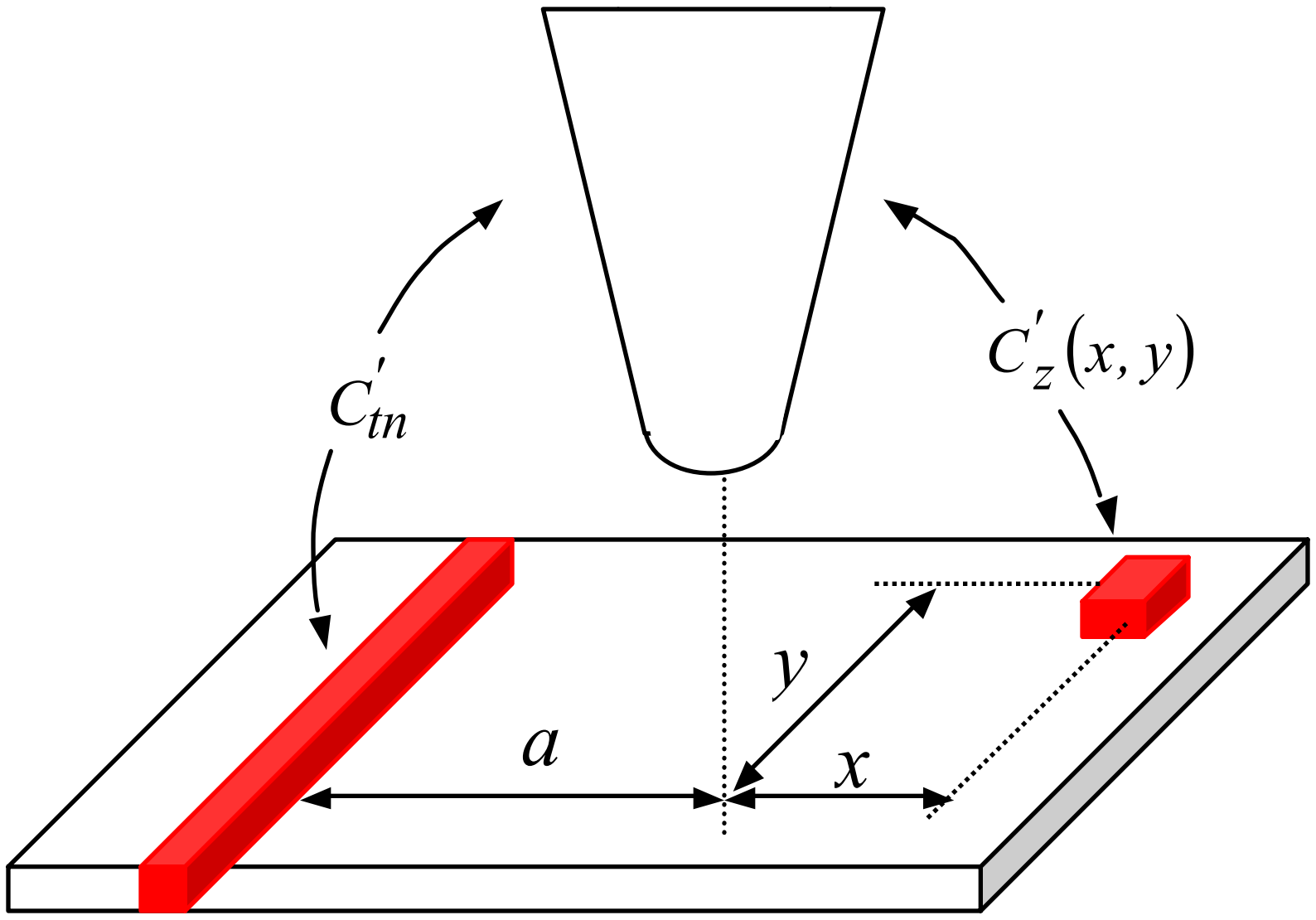

**Figure 3.18.** Tip-surface transfer function is defined as capacitance gradient, $C'_z(x,y)$, between the tip and the region *dxdy* located at position *x,y*. Experimentally determined is an integral of $C'_z(x,y)$ [Eq.(3.40)] as a function of distance from the nanotube, *a*.

Eq.(3.39) can be generalized in terms of the tip-surface transfer function $C'_z(x,y)$, defined as the capacitance gradient between the tip and a region, d*x*d*y*, on the surface (Figure 3.18)[78] as

$$F_{1\omega} = (V_t - V_0)\int C'_z(x,y) V_{ac}(x,y)\,dxdy. \tag{3.40}$$

Considering the nanotube as 1D, i.e. width of the nanotube, $w_0 << R$,

$$F_{1\omega}(a) = w_0 V_{ac}(V_t - V_0)\int C_z'(a, y)dy, \qquad (3.41)$$

where $a$ is the distance between the projection of the tip and the nanotube. Assuming a rotationally invariant tip, differential tip-surface capacitance is $C_z(x,y) = C_z(r)$, where $r = \sqrt{x^2 + y^2}$ and Eq.(3.41) can be rewritten as a function of a single variable, $a$. Therefore, the partial tip-surface capacitance gradient $C_z'(r)$ can be found by numerically solving Eq.(3.41) using experimental force profiles across the nanotube, $F_{1\omega}(a)$.

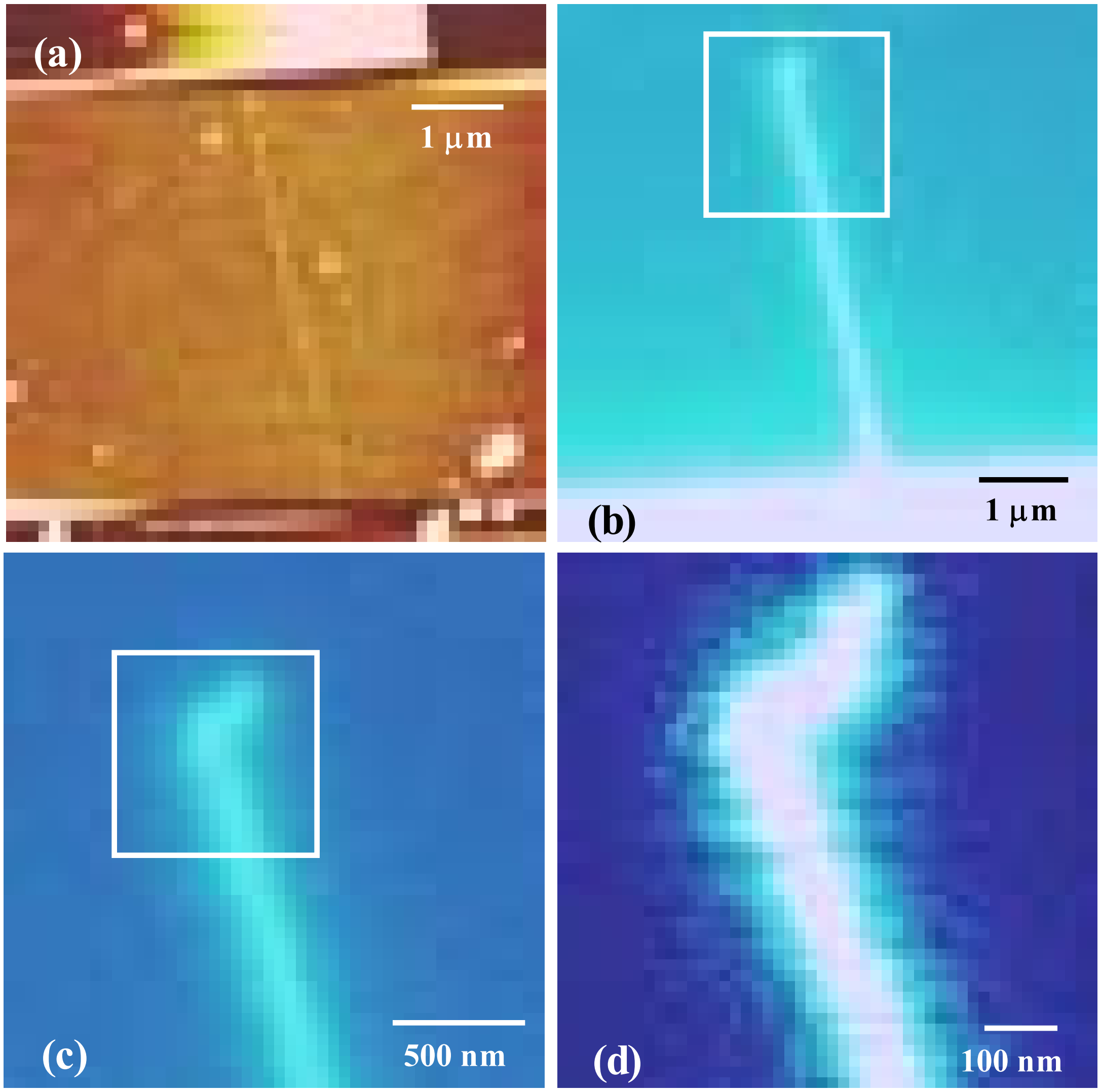

**Figure 3.19.** Surface topography (a) and SIM amplitude images (b,c,d) for a carbon nanotube circuit. The contrast is uniform along the tube. Scale is 10 nm (a).

The validity of this technique is illustrated in Figure 3.19. If the measurements are made sufficiently far (1-2 μm) from the biasing contact, the image background and potential distribution along the nanotube are uniform, confirming the absence of contact-probe interactions. Figure 3.20 shows topographic and amplitude profiles across a nanotube. The height of the nanotube is 2.7 nm, while the apparent width is ~40 nm due to the convolution with the tip shape. Simple geometric considerations yield a tip radius of curvature as $R \approx 75$ nm. Full width at half maximum (FWHM) of the amplitude profile can be as small as ~100 nm and increases with tip-surface separation. This profile is a direct measure of the tip-surface transfer function through Eq.(3.51).

To analyze the distance dependence and properties of $F_{1\omega}$, amplitude profiles were averaged over ~32 lines and fitted by the Lorentzian function,

$$y = y_0 + \frac{2A}{\pi} \frac{w}{4(x - x_c)^2 + w^2}, \tag{3.42}$$

where $y_0$ is an offset, $A$ is area below the peak, $w$ is peak width and $x_c$ is position of the peak. Note that Eq.(3.42) provides an extremely good description of the experimental data as illustrated in Figure 3.21c. The offset $y_0$ provides a direct measure of the non-local contribution to the SPM signal due to the cantilever and conical part of the tip.[82,83,84] The profile shape is tip dependent and profiles for two different tips (tip 1 and 2) are compared in Figure 3.21d. The distance dependence of peak height $h = 2A/\pi w$ is shown in Figure 3.21e. For large tip-surface separations $h \sim 1/d$. The distance dependence of width, $w$, is shown in Figure 3.21f and is almost linear in distance for $d > 100$ nm. Similar behavior was found for profile width for "potential step" type standards such as ferroelectric domain walls and biased interfaces.[85]

In the case of the amplitude profile given by Eq.(3.42), the local part of the differential tip-surface capacitance can be found solving Eq.(3.41) as

$$C'_z = \frac{2A}{\pi} \frac{w}{\left(4r^2 + w^2\right)^{3/2}}, \tag{3.43}$$

where $A$ and $w$ are $z$-dependent parameters determined in Eq.(3.42) and $r$ is radial distance.

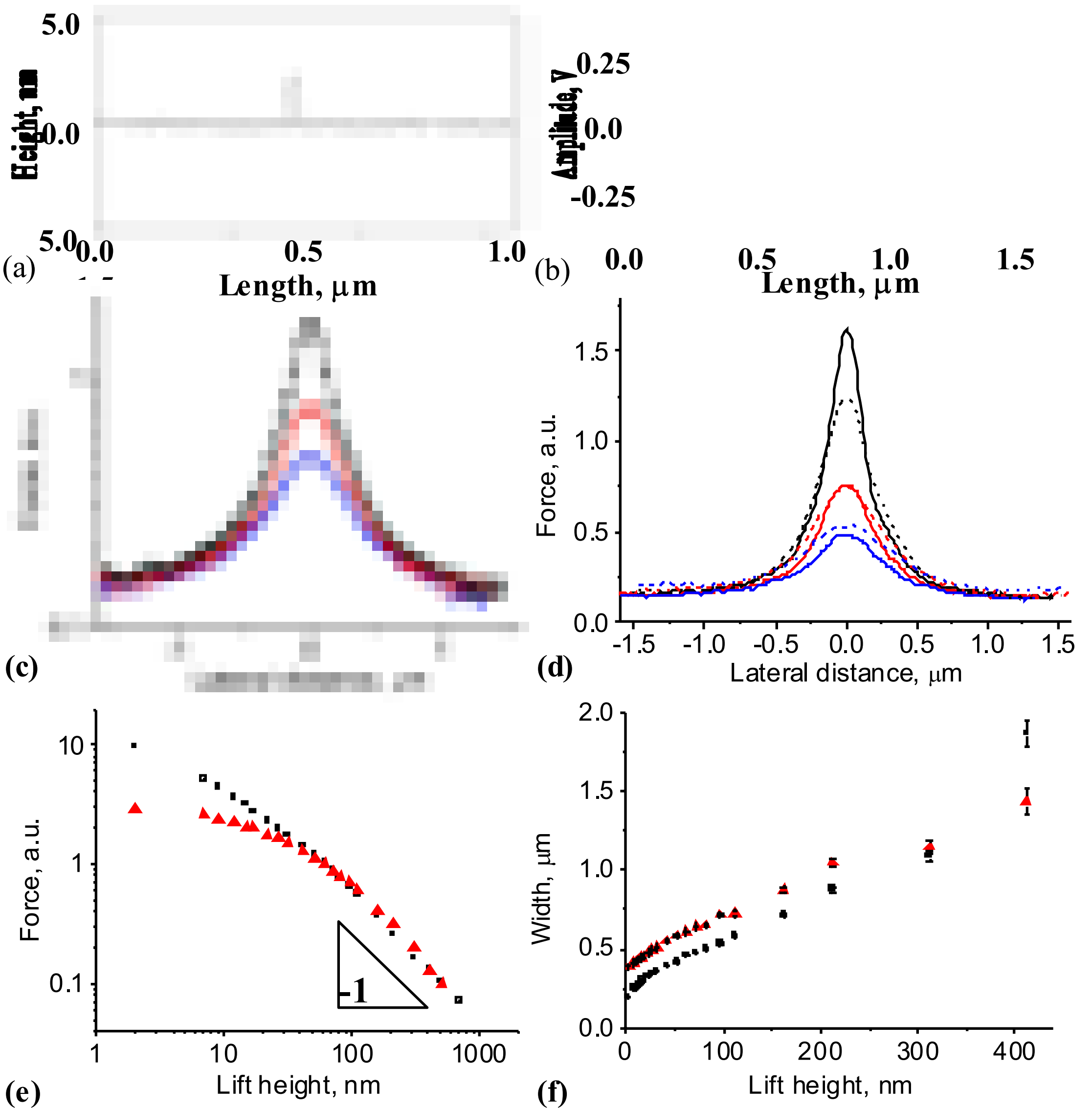

**Figure 3.20** Topographic profile (a) and SIM amplitude profile (b) across a carbon nanotube. (c) Force profiles at lift height of 10 nm (■), 30 nm (▲) and 100 nm (▼) and corresponding Lorentzian fits. (d) Force profiles at lift height of 10 nm, 30nm and 100nm for tip 1 (solid line) and tip 2 (dash line). Peak height (e) and width (f) as a function of tip-surface separation for tip 1 (■) and tip 2 (▲).

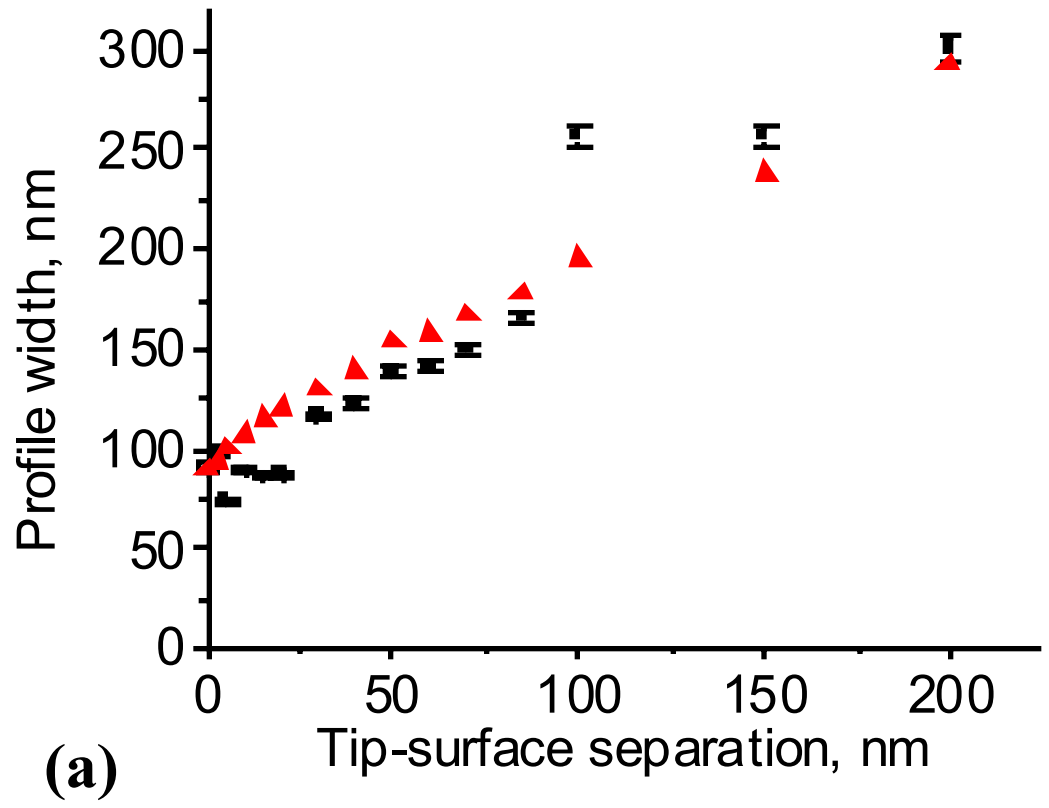

(a)

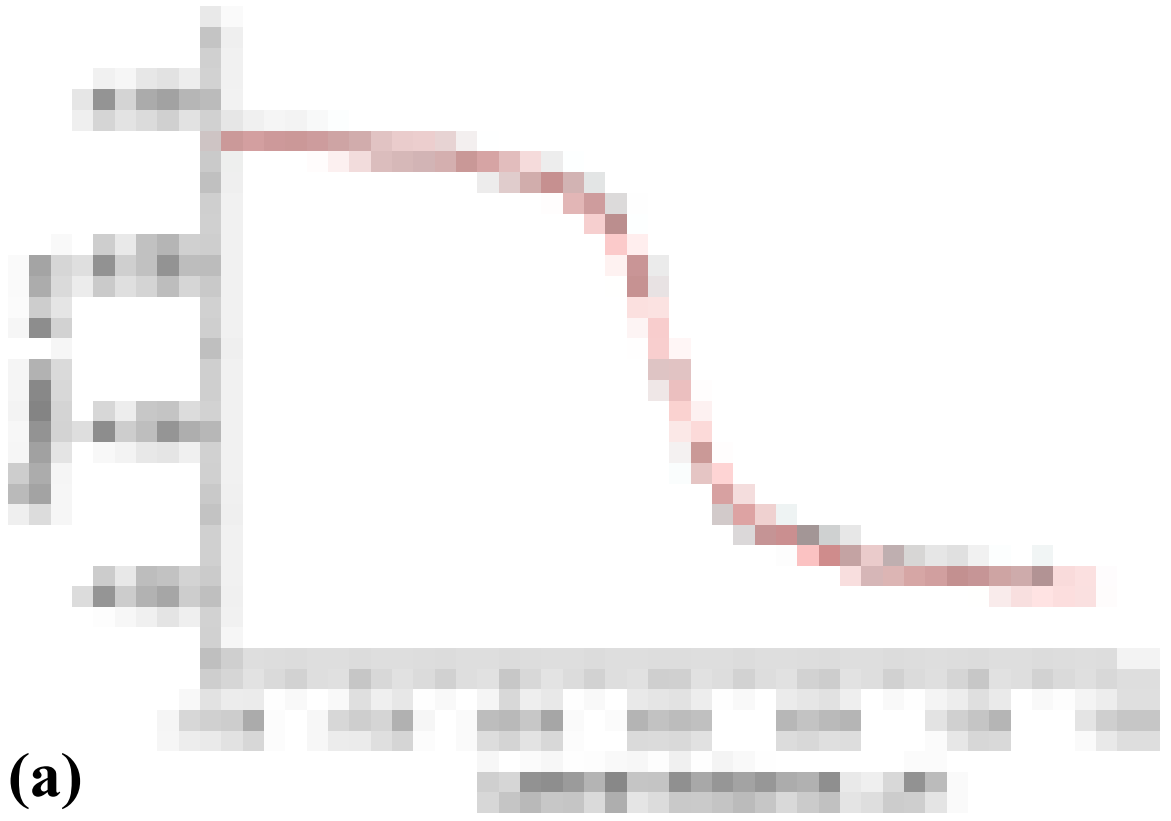

(a)

**Figure 3.21.** Profile width for carbon nanotube standard (▲) and SIM phase image of the $SrTiO_3$ grain boundary (■) as a function of lift height (a). Measured (■) and simulated (line) phase profiles (b).

Eq (3.43) can be used to determine the tip shape contribution to electrostatic SPM measurements in systems with arbitrary surface potential distributions. For a stepwise surface potential distribution, $V_{surf} = V_1 + (V_2 - V_1)\theta(x)$, where θ(x) is a Heaviside step function, the measured potential profile is $V_{eff} = V_1 + V_2 \arctan(2x/w)/\pi$, provided that the cantilever contribution to the measured potential is small. A similar phenomenological expression is expected to describe phase and amplitude profiles in open-loop SSPM and Scanning Impedance Microscopy (SIM).[86] Figure 3.22 shows the phase profile across a grain boundary in a Nb-doped $SrTiO_3$ bicrystal. From independent measurements the double Schottky barrier width is <20 nm, i.e. well below the SPM resolution. Note the excellent agreement between the measured and simulated profile shape. The distance dependence of profile width for the nanotube standard and SIM phase image of grain boundary are compared in Figure 3.22b, demonstrating excellent agreement. The profile width determined from SSPM measurements is significantly larger indicating feedback and mobile surface charge contribution to the profile width.[87]

To summarize, we have developed a carbon nanotube based standard for the calibration of SPM tips in voltage-modulated SPM. The nanotube standard provides a simultaneous measure of topographic and electrostatic resolution, as well as the convolution function for electrostatic SPM. In contrast to traditional SPM measurements (tip is ac biased) in which the tip interacts both with the dc biased nanotube and the substrate, the latter interaction is effectively excluded. Moreover, surface and oxide trapped charges contribute to the signal for ac tip biasing.[88] Mobile surface charges redistribute under the dc bias, resulting in "smearing" of the potential or electrostatic profile. The characteristic relaxation times for surface charges in air are relatively high and are of order of seconds;[89,90,91] therefore, surface charge dynamics do not contribute to measurements at high (~10-100 kHz) frequencies.

### 3.6. Imaging Artifacts and Some Considerations on Invasiveness

Quantitative and sometimes even qualitative SPM studies of fundamental physical phenomena on micron and nanometer level are often hindered by SPM imaging artifacts. In fact, the wide availability of advanced SPM techniques and lack of universal standards and reference handbooks in the field have contributed to a significant number of papers with dubious and even obviously erroneous interpretation of SPM data. Here we briefly discuss the major sources of artifacts in electrostatic measurements by SPM. The origins of the artifacts in the surface potential measurements can be traced either to the non-idealities of microscope electronics and feedback effects or to tip effects. The latter include the effects of tip geometry and non-local cantilever effects and effects of surface topography.

#### 3.6.1. Current Leakage

One possibility for erroneous potential measurements in SSPM is dc and ac voltage drops in the circuit. DC voltage drops in the electronics can be important if the tip-surface resistance is very small. One of the weakest links in the measurement set-up is the contact between the tip holder and tip substrate that can have resistances on the order of several kOhm and larger depending on the type of tip coating (e.g. for diamond coated cantilevers). However, during SSPM imaging tip-surface separation is usually large (> 10 nm), therefore resistance is also large (>>GOhm) and dc voltage drops in the electronics are negligible (unless there are problems with insulation, etc.). If the tip accidentally touches the surface so that current flows, the effects are obvious: for a nonconductive or contaminated metal surface a charged patch forms from contact electrification. Overall, dc leakage is negligible with a possible exception for extremely small tip-surface separations, in which case tip-surface charge transfer is possible. This is not true for ac leakage. Modest capacitive coupling between the ac biased channel and the rest of the microscope on the order of ~nF at the typical frequencies of 10-100 kHz is equivalent to a leakage resistance on the order of 10-100 kOhm. For cantilevers with semiconductive coatings and large probe-tip holder contact resistances, this leakage resistance can be (but usually is not) comparable to the contact resistance between tip holder and the substrate, resulting in significant attenuation of driving voltage. This effect can be compensated by improving probe holder contact, e.g. by placing a drop of silver print or a piece of indium on the contact area.[92]

### 3.6.2. Feedback and Non-local Cantilever Effects

Other sources of error are feedback effects and non-local cantilever effects. Due to feedback non-ideality, the first harmonic of the cantilever response is not exactly nullified as suggested by Eq.(3.21); rather it is reduced to some small, but finite value. In addition, simple calculations suggest that at typical tip surface separations (10 – 100 nm), tip-surface and cantilever-surface capacitive gradients are comparable (Chapter 2). Therefore, both contributions are important and the signal can be subdivided into the local tip part and non-local cantilever part. In this case, the nulling condition in SSPM corresponds to

$$F_{loc}(V_{dc} - V_s) + F_{nl}(V_{dc} - V_{av}) = \delta / V_{ac} , \tag{3.44}$$

where $F_{loc}$ is the local part of the tip surface capacitance gradient, $F_{nl}$ is the non-local and cantilever part, $V_s$ is the local potential below the tip, $V_{av}$ is the surface potential averaged over the cantilever length, and $\delta$ is the feedback constant (which, of course, depends on the gain values for feedback loop). The nulling voltage $V_{dc}$ is then

$$V_{dc} = V_s \frac{F_{loc}}{F_{loc} + F_{nl}} + V_{av} \frac{F_{nl}}{F_{loc} + F_{nl}} + \frac{\delta}{V_{ac}(F_{loc} + F_{nl})} . \tag{3.45}$$

Reliable measurement of local surface potential is possible if, and only if, $F_{loc} >> F_{nl}$, i.e. the tip is close to the surface, and the third term in Eq.(3.45) is small. This is when the feedback error is minimized. Fortunately, relative potential variations across the surface are independent of feedback effects, i.e. the measured potential variation between two points relates to the true potential difference as:

$$\left(V_1 - V_2\right)_{measured} = \left(V_1 - V_2\right)\frac{F_{loc}}{F_{loc} + F_{nl}}. \tag{3.46}$$

The absolute surface potential value, however, depends on feedback effects:

$$V_{dc} = V_{av} + \frac{\delta}{V_{ac}\left(F_{loc} + F_{nl}\right)}. \tag{3.47}$$

The implications of Eqs.(3.46,47) are two fold. Measurements of absolute surface potential (or CPD) by SSPM are subject to errors due to the feedback effect. The reciprocal dependence of absolute surface potential on driving voltage can be easily confirmed experimentally by variation of the driving amplitude.[93] Doing so at different tip-surface separations allows the effect of $F_{loc}$ and $F_{nl}$ to be determined and true surface potential can be obtained from the analysis of the data.[94] Potential variations across the surface do not depend on $V_{ac}$ (except that small driving amplitudes result in the increased noise level), but the effective potential difference between two spots on the surface (say, 10-40 micron apart) decays logarithmically with lift height. For very large tip-surface separations, the potential contrast along the surface disappears.[93,95]

SSPM metrology of laterally biased devices is limited by a significant cantilever contribution to the measured potential, minimization of which requires imaging at small tip-surface separations. Under optimal conditions, the potential drop measured at the interface (i.e., ± 500 nm from the interface) is approximately 90% of its true value. The rest decays at the lateral distances of order of approximately 10 microns from the interface, i.e., comparable to the cantilever widths (similar considerations apply to SIM profiles, e.g. Figure 3.12b). Therefore, by measuring potential distribution in ceramics with grain sizes of order of 10 to 20 micrometers, grain boundary conductivity can be determined reliably, whereas grain bulk conductivity can not. Similar problems exist for carbon nanotube circuits. Due to the fact that interaction area of SSPM (30 to 100 nm) is much larger than diameter of a nanotube, the measured potential is a weighted average of nanotube potential and back gate potential.[96]

Of course, more complex artifacts are possible. For example, capacitive crosstalk between an tip bias channel and photodiode detector channel will result in an error signal proportional to the driving voltage and this effect can be minimized only by decreasing the tip surface separation, but not by increasing driving amplitude.

### 3.6.3. Topographic Artifacts

Another source of artifacts in surface potential measurements is geometric inhomogeneities of the surface. The cross talk between potential and topographic images is well-known[97] to result in imaging artifacts in both EFM and SSPM images [Figure 3.18]. For EFM in the frequency detection mode, the signal is proportional to the second derivative of the tip surface capacitance rather than the first, as in SSPM. Therefore, the signal is more localized and simple numerical estimates suggest that the cantilever effect can be ignored (Chapter 2).

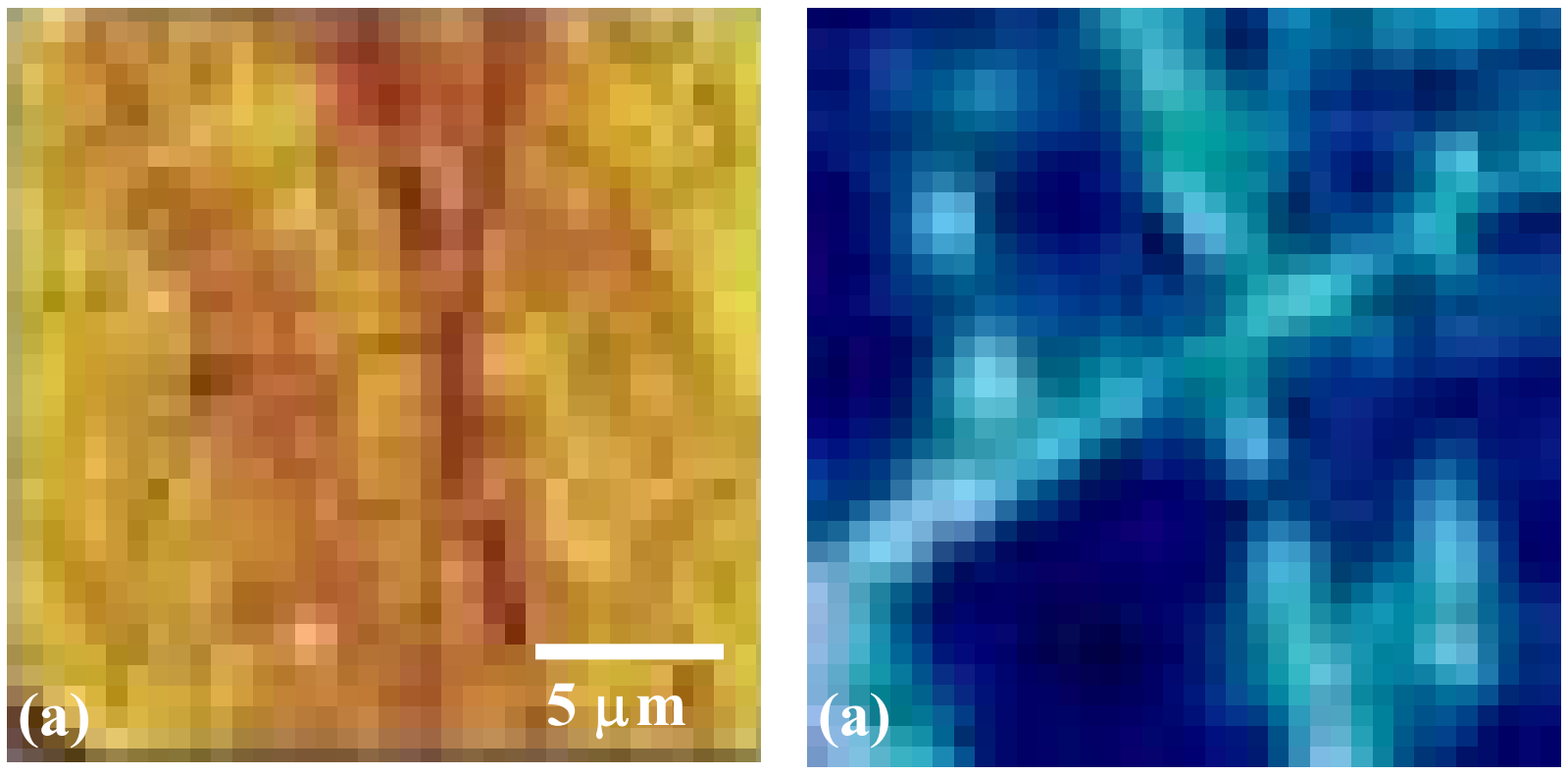

**Figure 3.22.** Surface topography (a) and surface potential of polished $SrTiO_3$ surface. Notice close correspondence between topographic and potential images.

However, the capacitive force gradient between the tip and the surface measured in EFM renders this technique extremely sensitive to the topographic structure of the surface. The topographic artifacts in the EFM images can be distinguished from the bias dependence of effective image contrast. Indeed, the presence of topographic inhomogeneities results in a variation of tip-surface capacitance and therefore effective EFM contrast is

$$\Delta\omega \sim \frac{\mathrm{d}F_1^{cap}(z)}{\mathrm{d}z} - \frac{\mathrm{d}F_2^{cap}(z)}{\mathrm{d}z} = \frac{1}{2}\left(V_{tip} - V_s\right)^2 \left(\frac{\partial^2 C_1(z)}{\partial z^2} - \frac{\partial^2 C_2(z)}{\partial z^2}\right), \qquad (3.48)$$

where $F_1^{cap}(z)$, $F_2^{cap}(z)$, $C_1(z)$and $C_2(z)$ are electrostatic forces and tip-surface capacitances over regions 1 and 2 correspondingly. Eq.(3.48) shows that features on EFM image are quadratic in tip bias and do not change sign when the bias is changed from positive to negative. On the other hand, variation of local potential results in the variation of EFM signal as

$$\Delta\omega \sim \left\{ \left(V_{tip} - V_1\right)^2 - \left(V_{tip} - V_2\right)^2 \right\} \frac{\partial^2 C(z)}{\partial z^2} = \left\{ -2V_{tip}\left(V_1 - V_2\right) + \left(V_1^2 - V_2^2\right) \right\} \frac{\partial^2 C(z)}{\partial z^2}, \quad (3.49)$$

i.e. the EFM signal is linear in tip bias and therefore EFM features reverse sign when the bias is changed from positive to negative. To measure local potential by EFM a number of images must be collected for different tip biases. The local potential can be determined from the apex of parabolic dependence between frequency shifts and tip bias. The direct measurement of surface potential can be performed by SSPM, but topographic artifacts there are more complex. Under optimal imaging conditions, protruding topographic features are associated with negative potential artifacts. This implies that most commercial systems err in overcompensating the topographic artifacts. A number of algorithms to reduce this effect have been undertaken;[98] however, currently this problem is not solved.

### 3.6.4. Invasiveness

A very interesting, and almost unstudied issue is the tip perturbation on surface properties.[99] Voltage or mechanical modulation of the biased tip induces an ac current in the region directly below the tip proportional to $V\frac{dC}{dt} + C\frac{dV}{dt}$, where the first term originates from oscillation in tip position and the second from voltage oscillations. For a voltage modulated tip with a driving amplitude of 10 V, driving frequency of 100 kHz and a (heavily underestimated) tip surface capacitance of 10 aF, the displacement current amplitude is ~100 pA. Assuming the lateral size of the biased region to be ~100 nm, variations in surface potential due to the displacement current become significant for resistivities higher than $10^{-2}$ Ohm·m. For well conducting surfaces or semiconductor surfaces with pinned Fermi levels, changes in surface potential due to displacement current are negligible. However, this might not be the case for surfaces with unpinned Fermi levels, where tip-induced band bending and associated variations in surface potential can be important. Another interesting case is imaging of nanoscale objects, for which injection of even several electrons can severely affect properties. This analysis has a number of interesting implications for lateral transport measurements on biased nanoscale devices as discussed in Section 3.2 (Scanning Gate Microscopy).

### 4.7. Summary

The general framework for SPM transport measurements is formulated distinguishing single, two and three terminal cases. The procedure to quantify dc transport properties from potential data is established by appropriate equivalent circuit model and calibration standards. A novel scanning probe microscopy technique, referred to as Scanning Impedance Microscopy, is developed for the quantitative imaging of ac transport properties. Excellent agreement between interface capacitance of a model metal-semiconductor interface from spatially resolved SIM measurements and macroscopic impedance spectroscopy is demonstrated.

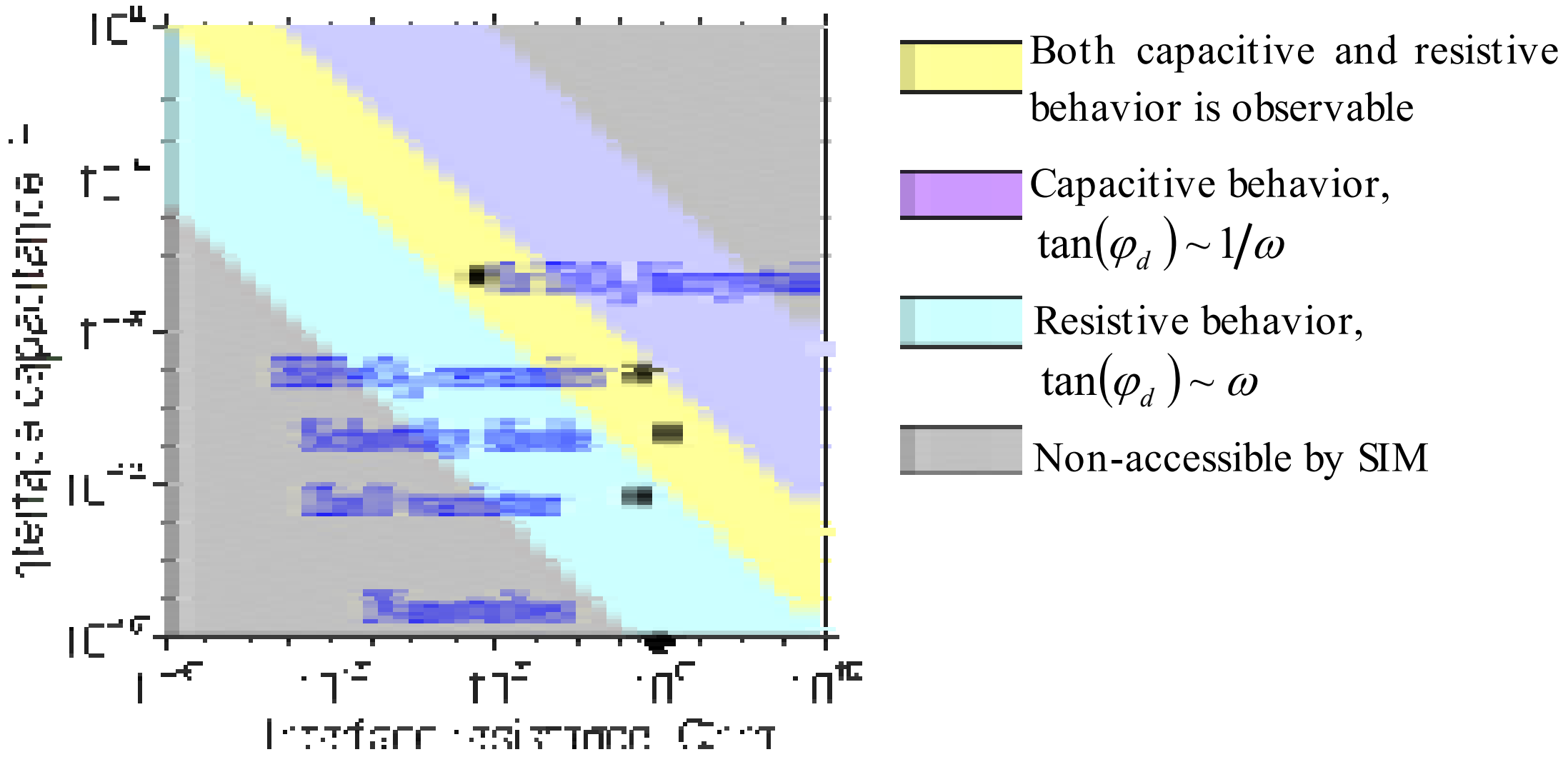

**Figure 3.23.** Systems accessible by SIM in the 100 Hz-100kHz frequency range.

The range of the interfaces amenable to SIM studies and those studied in the present thesis is summarized in Figure 3.23. Variation of the dc component of lateral bias in SIM measurements allows reconstruction of *C-V* characteristics of the interface. SIM and SSPM yield local voltage and *I-V* characteristics of the interfaces. The combination of SSPM and SIM is established as a quantitative tool for the characterization of local dc and ac transport properties in semiconductor devices on micron and submicron level.

This approach allows quantitative local imaging of electronic properties of the interfaces including contact potential barriers, frequency dependent capacitance and interface trap states, etc. As was shown by Wiesendanger et al.[100] force based SPM techniques can provide electrostatic information at the atomic level. Therefore, SSPM and SIM are expected to be applicable for quantitative characterization of ac and dc transport properties of nano- and molecular electronic devices, etc.

## Appendix 3.A.

Experimental procedure

The AFM and SSPM measurements were performed on a commercial instrument (Digital Instruments Dimension 3000 NS-IIIA) using metal coated tips (l ≈ 225 μm, resonant frequency ~ 72 kHz, $k$ ≈ 1-5 N/m). For SIM, the AFM was additionally equipped with a function generator and lock-in amplifier (DS340, SRS 830, Stanford Research Systems). The lift height for the interleave scans in the SSPM and SIM was usually 100nm.[101] The scan rate varied from 0.5 Hz for large scans (~80 μm) to 1 Hz for smaller scans (~10 μm). Driving voltage $V_{ac}$ in the interleave scan was 5 V for the SSPM and 1 V for the SIM. To reduce the effect of drift the images were acquired with the interface oriented along the slow scan axis. This approach also minimizes non-local cantilever contributions to the measured potential, SIM phase and amplitude. To minimize photoelectric effects, the laser beam was focused *ca.* 120 μm from the cantilever end and the optical microscope and room illumination were switched off. To quantify the transport properties of the interface, the tip was repeatedly scanned along the same line across the surface and a slow (~mHz) triangular voltage ramp was applied across the boundary. The resulting image represents potential profiles at different lateral biases, from which voltage characteristics of the interface can be determined. To quantify the voltage dependence of interface capacitance, numerous SIM images were collected for varying lateral biases. The driving frequency $f$ = 50 kHz in this case was selected to be sufficiently far from both the resonant frequency of the cantilever (~72 kHz) and the relaxation frequency of the interface (~1.5 kHz). Finally, circuit termination resistors in SSPM and SIM was varied from 500Ohm to 1 MOhm.

Model system

The calibration sample was prepared by cross-sectioning commercial Schottky barrier diode. The top of the diode was removed by polishing with diamond media down to 1 μm grit size. Further polishing was precluded by selective polishing of the interconnect material, resulting in large topographical variations from the metal to silicon.

To establish the nature of the Schottky barrier, energy dispersive spectroscopy (EDS) was performed (JEOL JSM-6400) and the interface was identified as Au-Si.

To demonstrate applicability of this technique to the general case of polycrystalline semiconductors, a sample was prepared by polishing a silicon solar cell on diamond media to 1 μm grit size and alumina media to 0.05 μm size. To reveal grain structure, samples were etched in 0.1 M NaOH at 80 °C for ~ 1 min. Macroscopic indium contacts were soldered on the surface.

<u>Diode properties</u>

Two-point dc transport properties were measured both for initial and cross-sectioned device by *I-V* measurements (HP4145B Semiconductor Parameter Analyzer). AC transport properties were measured by impedance spectroscopy (HP4276A LCZ meter) in the frequency range 0.2-20 kHz under lateral dc bias.

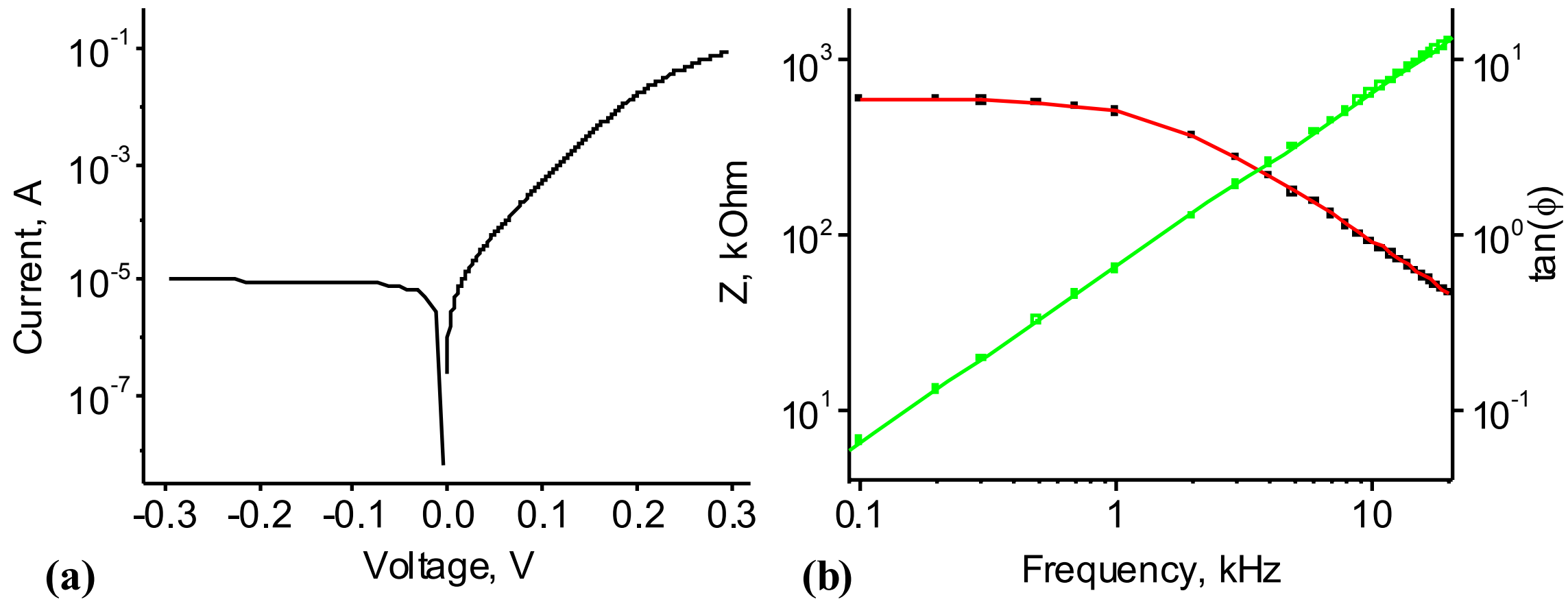

**Figure 3.24.** (a) Current-voltage characteristic for crossectioned diode and (b) impedance magnitude and phase angle vs. frequency dependence. Solid lines are fits by Eqs.(3.A.3,4).

Prior to SPM studies, the properties of crossectioned and non-crossectioned devices were studied. The *I-V* characteristic of the crossectioned sample is shown on Figure 3.24a. In the forward bias region, the current voltage characteristic is given by

$$\log(I) = (-5.106 \pm 0.005) + (17.01 \pm 0.05)\ V, \quad (3.A.1)$$

Using Eq.(3.3) and estimating the junction area of crossectioned device as $S = 1.4\ 10^{-3}\ cm^2$, the saturation current density is $I_0 = 5.5\ 10^{-3}\ A/cm^2$. From Eq.(3.4) Schottky potential barrier height is therefore $\phi_B = 0.55$ V.

The slope of the forward *I-V* curve is very close to theoretical value $0.434q/kT = 16.78$ at 300 K. For reverse-biased device the leakage current change by less then factor of 2 from 1.3 $10^{-5}$ A at -2 V to 2.5 $10^{-5}$ A at -10 V. Therefore, the crossectioned sample demonstrated almost ideal diode behavior in the bias range from -5 to 0.3 V. No deviations between the I-V characteristics for crossectioned and initial device ware found in the forward bias region, while in the reverse bias region deviation in current between pristine and crossectioned device didn't exceed 5%.

The ac transport properties for forward and reverse bias regimes were analyzed by impedance spectroscopy. The frequency dependence of the impedance modulus, $|Z|$, and impedance phase angle, $\theta$, under -5V reverse bias is shown on Figure 3.20b. For ideal parallel *R-C* element the impedance phase angle is

$$\tan(\theta) = \omega R_d C_d\ , \tag{3.A.2}$$

and impedance modulus is

$$|Z| = \frac{R_d}{\sqrt{1 + (\omega R_d C_d)^2}}\ . \tag{3.A.3}$$

Experimentally tangent of phase angle is found to be linear in frequency verifying the selected equivalent circuit and frequency dependence of the phase angle is

$$\tan(\theta) = (0.646 \pm 0.002) \cdot 10^{-3}\ \omega/2\pi\ . \tag{3.A.4}$$

The product of junction resistivity and capacitance is therefore estimated as $R_d C_d = (0.103 \pm 0.001)10^{-3}$ s. Fitting Eq.(3.A.4) to impedance data allows the junction resistance to be estimated as $R_d = 603 \pm 1$ KOhm at -5 V. Therefore, junction capacitance is calculated as $(1.71 \pm 0.01)\ 10^{-10}$ F at -5 V. It was also found that application of large reverse biases (V < -10V - -12 V) results in rapid decrease of impedance and phase angle for crossectioned device, while this behavior is not observed in the pristine device. This behavior is attributed to the formation of surface region with higher conductivity due to near-surface space charge layer on the crossectioned device. Further evidence supporting this assumption is obtained during scanning probe measurements.

Nanotube calibration standard

Nanotubes are grown by catalytic chemical vapor deposition (CVD)[102,103] directly on a $SiO_2$/Si wafer. Fe/Mo particles on porous alumina act as the catalyst. The nanotubes are grown in an Ar/$H_2$/Ethylene atmosphere at 820°C. This process yields predominantly single wall carbon nanotubes (SWNT) with a small fraction of multiwall nanotubes with a few shells. SWNTs can be distinguished based on the apparent height of 3 nm or less as measured by AFM. The substrate has an oxide layer with a thickness of 225 nm. The degenerately doped silicon acts as a back gate and is grounded. Leads are patterned by e-beam lithography and thermal evaporation of Cr and Au so that the nanotube is a molecular size element in a circuit.

The standard is based on the detection of the amplitude of cantilever oscillation induced by an ac voltage bias ($V_{pp}$ = 200 mV) applied to the carbon nanotube. The tip acquires surface topography in the intermittent contact mode and then retraces the surface profile maintaining constant tip-surface separation. Measurements were performed using CoCr coated tips (Metal coated etched silicon probe, Digital Instruments, $l \approx 225$ μm, resonant frequency ~ 62 kHz) and Pt coated tips (NCSC-12 F, Micromasch, $l \approx 250$ μm, resonant frequency ~ 41 kHz), further referred to as tip 1 and tip 2. A lock-in amplifier is used to determine the magnitude and phase of cantilever response. The output amplitude, $R$, and phase shift, $\theta$, are recorded by the AFM electronics (Nanoscope-IIIA, Digital Instruments). To avoid cross-talk between the sample modulation signal and topographic imaging, the frequency of ac voltage applied to the nanotube (50 kHz) was selected to be far from the cantilever resonant frequency.

[19] A.K. Henning, T. Hochwitz, J. Slinkman, J. Never, S. Hoffman, P. Kaszuba, and C. Daghlian, J. Appl. Phys. **77**, 1888 (1995).

[20] Q. Xu and J. W. P. Hsu , J. Appl. Phys. **85**, 2465 (1999).

[21] R. Shikler, N. Fried, T. Meoded, and Y. Rosenwaks, Phys. Rev. **B 61**, 11041 (2000).

[22] P. De Wolf, R. Stephenson, T. Trenkler, T. Clarysse, T. Hantschel, and W. Vandervorst, J. Vac. Sci. Techol. **B 18**, 361 (2000).

[23] P. De Wolf, J. Snauwaert, L. Hellemans, T. Clarysse, W. Vandervorst, M. D'Olieslaeger, and D. Quaeyhaegens, J. Vac. Sci. Techol. **A 13**, 1699 (1995).

[24] C.C. Williams, Annu. Rev. Mater. Sci. **29**, 471 (1999).

[25] M.L. O'Malley, G.L. Timp, S.V. Moccio, J.P. Garno, and R.N. Kleiman, Appl. Phys. Lett. **74**, 272 (1999).

[26] J.N. Nxumalo, D.T. Shimizu, and D.J. Thomson, J. Vac. Sci. Techol. **B 14**, 386 (1996).

[27] H. Edwards, V.A. Ukraintsev, R. San Martin, F.S. Johnson, P. Menz, S. Walsh, S. Ashburn, K.S. Wills, K. Harvey, and M.-C. Chang, J. Appl. Phys. **87**, 1485 (2000).

[28] E.T. Yu, K.Barmak, P. Rosenheim, M.B. Johnson, and J.-M. Halbout, J. Appl. Phys. **79**, 2115 (1996).

[29] R.M. Silver, J.A. Dagata, and W. Tseng, J. Vac. Sci. Techol. **A 13**, 1705 (1995).

[30] W. Wu, S.L. Skala, J.R. Tucker, J.W. Lyding, A. Seabaugh, E.A. Beam III, and D. Jovanovic, J. Vac. Sci. Techol. **A 13**, 602 (1995).

[31] S.J. Stranick and P.S. Weiss, Rev. Sci. Instrum. **64**, 1232 (1993).

[32] P. Muralt and D.W. Pohl, Appl. Phys. Lett. **48**, 515 (1986).

[33] J.R. Kirtley, S. Washburn, and M.J. Brady, Phys. Rev. Lett. **60**, 1546 (1988).

[34] B. Grevin, I. Maggio-Aprile, A. Bentzen, L. Ranno, A. Llobet, and O. Fischer, Phys. Rev. **B 62**, 8596 (2000).

[35] G. Ramaswami and A.K. Raychaudhyri, Appl. Phys. Lett. **75**, 1982 (1999).

[36] D.G. Cahill and R.J. Hamers, J. Vac. Sci. Technol. **B 9**, 564 (1991).

[37] M. Tanimoto and K. Arai, J. Vac. Sci. Technol. **B 12**, 2125 (1994).

[38] O. Vatel and M. Tanimoto, J. Appl. Phys. **77**, 2358 ~1995!.

[39] L. Bürgi, H. Sirringhaus, and R. H. Friend, Appl. Phys. Lett. **80**, 2913 (2002)

[57] D.A. Bonnell and S.V. Kalinin, in *Polycrystalline Semiconductors VI -Bulk Materials, Thin Films and Devices*, ed. O. Bonnaud, T. Mohammed-Brahim, H.P. Strunk, and J.H. Werner, in series "Solid State Phenomena" (Scitech Publ, Uettikon am See, Switzerland, 33-46, 2001).

[58] S.V. Kalinin and D.A. Bonnell, Appl. Phys. Lett. **78**, 1306 (2001).

[59] *Principles of electronic ceramics*, ed. by L.L. Hench and J.K West (Wiley Interscience, New York, 1990).

[60] J. Fleig, S. Rodewald, and J. Maier, J. Appl. Phys. **87**, 2372 (2000).

[61] D.A. Bonnell, R.A. Alvarez, and S.V. Kalinin, patent pending

[62] P.M. Chaikin and T.C. Lubensky, *Principles of Condensed Matter Physics* (Cambridge University Press, New York, 1997).

[63] D. Sarid, *Scanning Force Microscopy*, (Oxford University Press, New York, 1991).

[64] It should be noted that this decrease of the response is predicted by harmonic oscillator type model, applicability of which to the description of cantilever dynamics is limited. It is expected that SIM imaging is possible near the high-order cantilever resonances as well. Experimentally, we have found that potential SSPM imaging is possible under such conditions, albeit it is sometimes unstable. Similar behavior can be expected for SIM.

[65] Sergei V. Kalinin, M. Suchomel, P.K. Davies, and Dawn A. Bonnell, J. Am. Cer. Soc., in print

[66] Resistance and capacitance of contacts on the left does not influence the voltage phase angle shift and voltage oscillation amplitude ratio across the interface.

[67] S.V. Kalinin and D.A. Bonnell, Phys. Rev. **B 62**, 10419 (2000).

[68] B.D. Huey and D.A. Bonnell, Appl. Phys. Lett. **76**, 1012 (2000).

[69] Note that for small resistivity circuit termination the amplitude of voltage oscillations on the right of the junction is very small, thus introducing large error in the lock-in output. This explanation, however, does not account for observed dc behavior.

[70] S.M. Sze, *Physics of Semiconductor Devices* (Wiley, New York, 1981).

[71] Z.Y. Li, B.Y. Gu, and G.Z. Yang, Phys. Rev. **B 57**, 9225 (1998).

[72] S. Lanyi, J. Torok, and P. Rehurek, J. Vac. Sci. Technol. **B 14**, 892 (1996).

[73] A. Eimov and S.R. Cohen, J. Vac. Sci. Technol. **A 18**, 1051 (2000)

[74] S. Belaidi, P. Girard, and G. leveque, J. Appl. Phys. **81**, 1023 (1997).

[75] L. Olsson, N. Lin, V. Yakimov, and R. Erlandsson, J. Appl. Phys. **84**, 4060 (1998).

[76] H.O. Jacobs, H.F. Knapp, and A. Stemmer, Rev. Sci. Instr. **70**, 1756 (1999)

[77] F. Robin, H. Jacobs, O. Homan, A. Stemmer, and W. Bächtold, Appl. Phys. Lett. **76**, 2907 (2000)

[78] H. O. Jacobs, P. Leuchtmann, O. J. Homan, and A. Stemmer, J. Appl. Phys. **84**, 1168 (1998).

[79] H. Sugimura, Y. Ishida, K. Hayashi, O. Takai, and N. Nakagiri, Appl. Phys. Lett. **80**, 1459 (2002)

[80] A. Bachtold, M. S. Fuhrer, S. Plyasunov, M. Forero, E.H. Anderson, A. Zettl, and P.L. McEuen, Phys. Rev. Lett. **84**, 6082 (2000).

[81] Sergei V. Kalinin and Dawn A. Bonnell, Appl. Phys. Lett. **78**, 1306 (2001).

[82] A.K. Henning, T. Hochwitz, J. Slinkman, J. Never, S. Hoffmann, P. Kaszuba, C. Daghlian, J. Appl. Phys. **77**, 1888 (1995).

[83] Sergei V. Kalinin and Dawn A. Bonnell, Phys. Rev. B **63**, 125411 (2001).

[84] G. Koley, M. G. Spencer, and H. R. Bhangale, Appl. Phys. Lett. **79**, 545 (2001)

[85] Sergei V. Kalinin and Dawn A. Bonnell, unpublished

[86] This result is valid for SPM techniques based on force detection. The resolution in force gradient based techniques such as EFM is better because it is defined by the second derivative of tip-surface capacitance.

[87] Sergei V. Kalinin and Dawn A. Bonnell, Phys. Rev. B **62**, 10419 (2000).

[88] M. Freitag, A.T. Johnson, unpublished.

[89] S. Cunningham, I.A. Larkin, and J.H. Davis, Appl. Phys. Lett. **73**, 123 (1998).

[90] G.H. Buh, H.J. Chung, and Y. Kuk, Appl. Phys. Lett. **79**, 2010 (2001).

[91] T. Tybell, C.H. Ahn, and J.-M. Triscone, Appl. Phys. Lett. **75**, 856 (1999).

[92] Digital Instruments listserver, spm@di.com

[93] Sergei V. Kalinin and Dawn A. Bonnell, Phys. Rev. B **63**, 125411 (2001).

[94] In ambience, mobile surface charges and adsorbtion all contribute to measured potential. However, here we consider only *measurement* artifacts.

[95] G. Koley, M. G. Spencer, and H. R. Bhangale, Appl. Phys. Lett. **79**, 545 (2001).

[96] S.V. Kalinin, M. Freitag, A.T. Johnson, and D.A. Bonnell, Appl. Phys. Lett. **81**, 754 (2002).

[97] It can be easily demonstrated by acquiring topographic and potential images on the clean rough gold surface: topographic protrusions are associated with depressions on potential images.

[98] A. Efimov and S.R. Cohen, J. Vac. Sci. Technol. **A 18**, 1051 (2000).

[99] This is well known problem for magnetic force microscopy: tip magnetization can distort surface magnetization in the magnetically soft samples. For potential imaging, the only relevant reference I am aware of is: Y. Leng, C. C. Williams, L. C. Su, and G. B. Stringfellow, Appl. Phys. Lett. **66**, 1264 (1995).

[100] A. Schwarz, W. Allers, U.D. Schwarz, and R. Wiesendanger, Phys. Rev. **B 32**, 13617 (2000).

[101] Scanning Impedance Microscopy is possible in the normal intermittent contact as well if lateral ac bias frequency is far from the resonant frequency of the cantilever. This approach allows reducing the image acquisition time by a factor of two and minimizing sample drift effects. However, the non-linear character of tip-surface interaction in this case alters observed values of amplitude and phase shift (Typically by 30-50%).

[102] J. H. Hafner, C. L. Cheung, and C. M. Lieber, J. Am. Chem. Soc. **121**, 9750 (1999).

[103] M. Freitag, M. Radosavljevic, Y. Zhou, A. T. Johnson, and W. F. Smith, Appl. Phys. Lett. **79**, 3326 (2001)

# 4. EFFECT OF GRAIN BOUNDARIES ON ELECTRONIC TRANSPORT IN SEMICONDUCTING TRANSITION METAL OXIDES

## 4.1. Introduction

### 4.1.1. Grain Boundaries in Oxide Materials

Oxide materials are widely used for many commercial applications such as sensors, electroceramics, actuators, etc. The interplay between structure, charge, orbital and spin degrees of freedom gives rise to semiconducting,[1] dielectric,[2] ferroelectric,[3,4] superconducting,[5] and magnetoresistive[6,7] properties. Currently, the vast majority of experimental studies and most applications are based on polycrystalline bulk or thin film materials characterized by the presence of the large number of grain boundaries between regions with dissimilar crystallographic orientation. Depending on the properties and the fabrication route, the grain boundaries are associated with various degrees of structural and compositional disorder. Even coherent grain boundaries result in the deviation of local symmetry of the structure. The matching between adjacent lattices defines the intrinsic grain boundary states and interface charge. The enthalpy of defect formation at the grain boundaries is typically smaller than in the bulk; grain boundaries therefore act as sinks for dopant atoms and oxygen vacancies, resulting in the large composition and carrier concentration gradients.[8] In extreme cases, grain boundaries can be associated with the impurity phase wetting layers. Interestingly, the deviations of symmetry from cubic results in several orientation variants and twin grain boundaries occur even in the nominally single crystal non-cubic perovskite materials. In many cases, grain boundaries are associated with interface potential barriers, local lattice distortions or magnetic disorder, thus significantly affecting the macroscopic transport properties.

Often grain boundaries enable useful behavior, such as low-field magnetoresistance,[9,10,11] grain boundary Josephson junctions,[12] positive temperature coefficient of resistivity (PTCR)[13,14] and varistor behavior.[15] In other cases, grain boundaries limit the performance of the material, e.g. critical current density in polycrystalline YBCO HTSC materials. Recent developments in epitaxial oxide growth enabled oxide heterostructures with novel functionalities. Properties of these systems are critically dependent on interface structure. Therefore, knowledge of structure-property relations and control of grain boundaries and interfaces in oxide systems is crucial for progress in this field. Fundamental insights come from structure-property relations at well-defined coherent interfaces, for which the structure is unique (there is no amorphous second phase layers, the structure is reproducible, etc.) and structural and transport measurements can be correlated with theoretical studies. In most cases, grain boundary transport phenomena are governed by the interface charge, i.e. the electronic properties of the interface, even though more subtle effects associated with magnetic disorder and strain are possible.

### 4.1.2. Structure and Properties of Grain Boundaries in $SrTiO_3$

The perovskite $SrTiO_3$ is a prototype of oxides in which the presence of interface charge results in grain boundary potential barriers. Consequent nonlinear transport properties constitute the basis for numerous applications, e.g. varistors and boundary layer capacitors.[16,17] In ferroelectric perovskites such as $Sr_xBa_{1-x}TiO_3$ polarization induced compensation of charged grain boundaries and the associated reduction of the Schottky barrier below the Curie temperature give rise to PTCR behavior.[18]

The local properties of grain boundaries in $SrTiO_3$ on atomic level have recently been studied by High Resolution Transmission Electron Microscopy (HRTEM), Electron Energy Loss Spectroscopy (EELS), electron holography,[19,20,21,22] and are the subject of intensive theoretical studies.[23,24] HRTEM studies by Oak-Ridge, Max-Plank and Northwestern groups suggest that the atomic structure of well-defined grain boundaries (GBs) in $SrTiO_3$ viewed in (100) plane is formed by pentagonal blocks comprised of 2 or 3 cation columns.[25] For example, the $\Sigma 5$ (every fifth atomic plane is coincident) tilt boundary consists of two alternating pentagonal oxygen structural units containing two Sr columns and two Ti columns respectively, as illustrated in Figure 4.1. In both units, the cation positions are half occupied, forming a grain boundary reconstruction.[26,27] This model predicts existence of undercoordinated titanium atoms. If the requisite number of oxygen atoms resides at the grain boundary, it is stoichiometric and neutral in terms of formal charge. The exact occupancy of the titanium columns cannot be determined by HRTEM; however, even small variations in oxygen and cation stoichiometry will result in significant interface charge. Calculations predict small oxygen vacancy formation

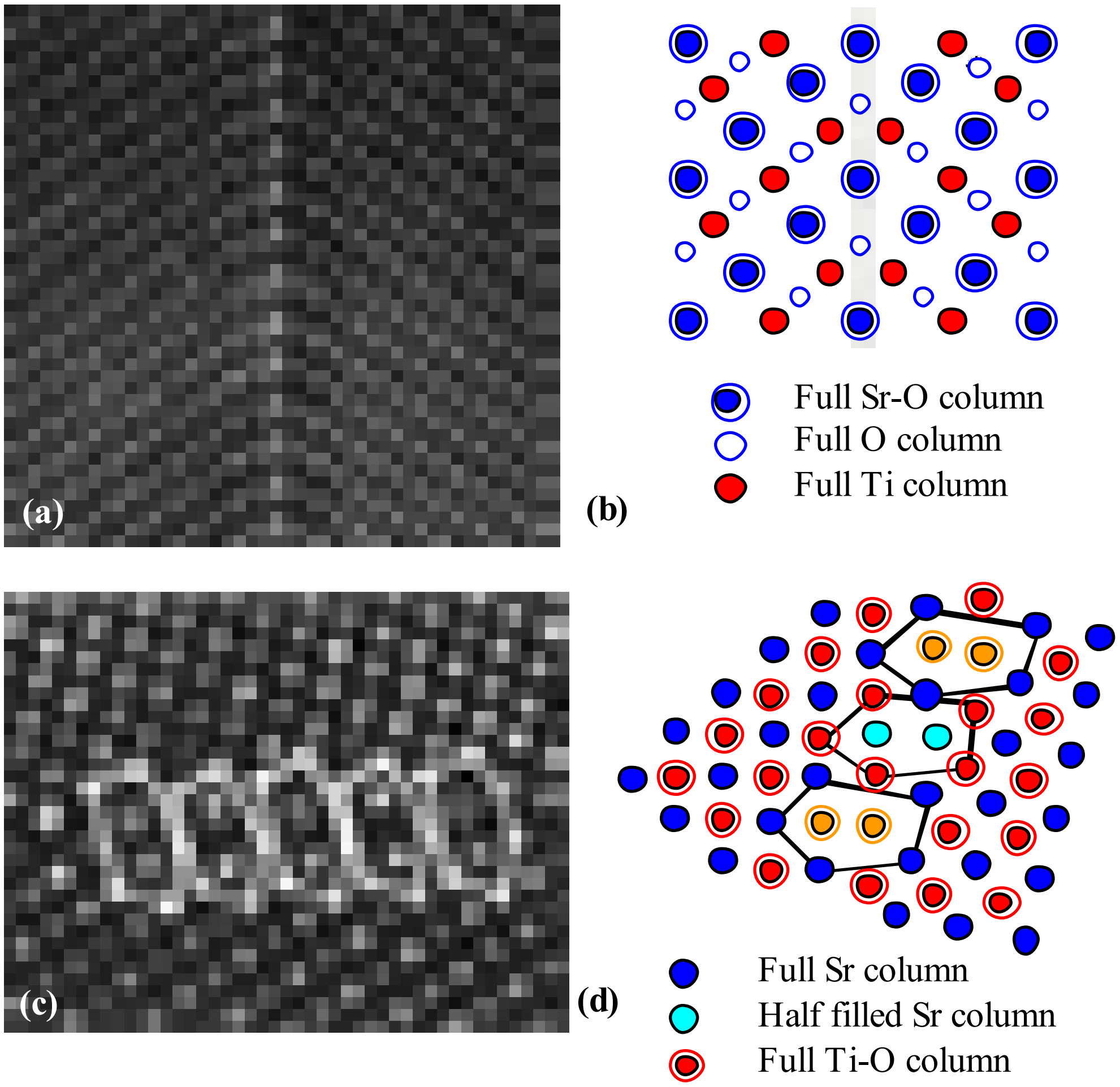

**Figure 4.1.** *Z*-contrast image (a,c) and schematic structure (b,d) of Σ3 (a,b) and Σ5 (c,d) grain boundary in $SrTiO_3$. [Images courtesy of G. Duscher, ORNL]

energy on the GB, making vacancy segregation on the GB possible. If the grain boundary is oxygen deficient, then the Ti will be partially reduced and the interface carries formal charge compensated by free carriers in an adjacent depletion regions. Controversy persists regarding the exact origins and properties of potential barriers at $SrTiO_3$ grain boundaries.[28,29,30,31]

### 4.1.3. This Thesis: Dielectric Non-linearities and Transport across $SrTiO_3$ Interfaces

Currently, a wealth of information is available on the atomic structure of grain boundaries in $SrTiO_3$. However, little is known on the structure of the potential barrier at the interface and transport properties of these grain boundaries. In order to determine the formal interface charge and the structure of space-charge layer, structural investigations must be connected to transport measurements such as *I-V* measurements, capacitance and impedance spectroscopy, etc. Complementary to HRTEM information can be obtained from spatially resolved SPM imaging of local properties.

The purpose of this Chapter is to investigate the transport mechanisms in coherent $SrTiO_3$ grain boundaries and establish the applicability of SPM for local probe measurements in this and other systems. The physics of electroactive interfaces is briefly reviewed in Section 4.2 along with possible transport mechanisms in heavily doped $SrTiO_3$ bicrystals and possible effects of non-linear dielectric properties on interface potential barrier. SPM imaging of static and transport properties of $\Sigma 5$ grain boundary in $SrTiO_3$ bicrystal is discussed in detail in Sections 4.3.1 and 4.3.2. The results of variable temperature *I-V* and impedance measurements and the interpretation in terms of non-linear dielectric behavior at $SrTiO_3$ GBs are summarized in Section 4.3.3 and 4.3.4. Finally, application of SPM for imaging transport behavior in polycrystalline oxides is presented in Section 4.4.

## 4.2. Physics of Electroactive Interfaces

### 4.2.1. Grain Boundary Potential Barriers

In polycrystalline oxide materials, aliovalent dopant or oxygen vacancy segregation or the presence of intrinsic states at the interfaces result in interface charge. If the sign of the interface charge is the same as that of the majority carriers, it results in adjacent depletion regions with low carrier concentration. The potential distribution in the vicinity of the interface can be estimated using the abrupt junction approximation. The grain boundary is characterized by an interface charge density, $\sigma$, and an adjacent space charge layer with width $d$.[32] Charge neutrality requires that $\sigma = -2\, d\, q\, N_d$, where $N_d$ is the ionized donor concentration. For a linear dielectric material with constant donor density, the potential distribution is found from the solution of one-dimensional Poisson equation

$$\frac{d^2\varphi}{dz^2} = \frac{qN_d}{\kappa\varepsilon_0}, \tag{4.1}$$

where $q$ = 1.602·10$^{-19}$ C is electron charge, $\kappa$ is dielectric constant of material and $\varepsilon_0$ = 8.854·10$^{-12}$ F/m, as

$$\varphi(z) = \varphi_{gb}(1 - z/d)^2, \tag{4.2}$$

where potential at the grain boundary is

$$\varphi_{gb} = \frac{\sigma^2}{8q\kappa\varepsilon_0 N_d}. \tag{4.3}$$

The grain boundary potential barrier changes the capacitive behavior of the material. Indeed, a resistive grain boundary region between the conductive grains acts as a plane-plane capacitor. The interface capacitance is related to the depletion width as $C_{gb} = \kappa\varepsilon_0/2d$. For future discussion, we introduce effective interface charge $\sigma_{eff} = 8C_{gb}\varphi_{gb}$. In the abrupt junction approximation, for linear dielectric material the effective interface charge and interface charge are equal, $\sigma_{eff} = \sigma$. Non-linear polarization behavior near the interface alters the effective interface charge.

One of the shortcomings of this model is that it does not clarify the origin of the interface (or surface) charge. When the charge is extrinsic, i.e. due to the impurity or vacancy segregation, the grain boundary charge and equilibrium impurity distribution can be determined from the minimization of the free energy functional if the enthalpy of segregation is known.[33,34] An alternative description of the interface charge is based on the interface state model. These interface states can be due to the impurity atoms (e.g. Bi at ZnO GBs); alternatively, interface states can exist even in the pure material without any concentration gradients due to the lattice discontinuity at the interface. These intrinsic states depend sensitively on the grain boundary structure. The formation of the grain boundary potential barrier can be described using the model originally suggested by Pike and Seager.[35] In this model, the grain boundary is formed by bringing into contact the "grain boundary phase" and "bulk phase" as illustrated in Figure 4.2. The Fermi level in the GB phase differs from the bulk due to the presence of the interface states. It is assumed that the positions of the valence and conduction band edges in both phases coincide. This assumption is reasonable given that the nature of atomic orbitals responsible for the bottom of conduction band and top of valence band is similar in both phases. For $SrTiO_3$, these are oxygen *p*-orbitals and titanium *d*-orbitals respectively. In contact, the Fermi levels in both phases equilibrate due to charge transfer from bulk to the interface and the transferred charge is

$$\sigma = \int_{E_v}^{E_c} N_{ss}(E) f(E, E_F) dE - \int_{E_{vc}}^{E_c} N_{ss}(E) f(E, E_{FB}) dE, \qquad (4.4)$$

where $N_{ss}(E)$ is interface density of states (IDOS), $E_v$ is the top of the valence band, $E_c$ is the bottom of the conduction band, $f(E,E_F) = 1/(1+\exp\{(E-E_F)/kT\})$ is the Fermi function, $E_F$ is the Fermi energy in the bulk and $E_{FB}$ is the Fermi energy in the grain boundary phase before the contact, $T$ is temperature, and $k = 1.38 \cdot 10^{-23}$ J/K is Boltzmann constant. In the low temperature approximation the Fermi function can be approximated by the Heaviside step function, $f(E,E_F) = \theta(E-E_F)$ and the interface charge is

$$\sigma = \int_{E_{FB}}^{E_F} N_{ss}(E) dE. \qquad (4.5)$$

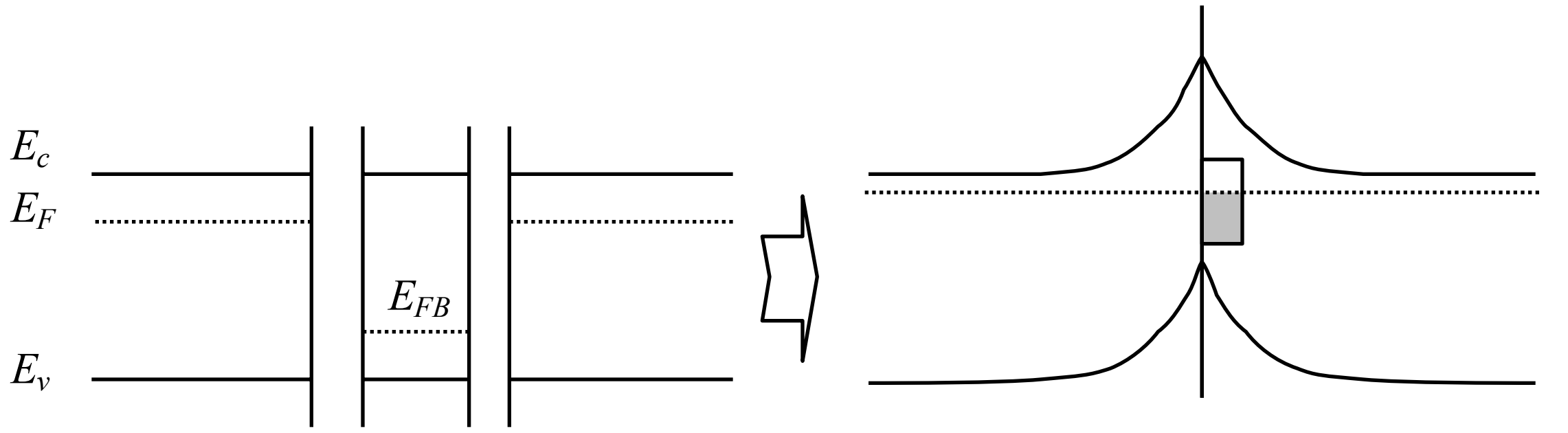

**Figure 4.2.** Formation mechanism of Schottky double barriers (DSB) at the interfaces.

The significance of this formalism is that it can be readily extended to the analysis of grain boundary transport phenomena. Application of a lateral bias across the interface results in the shift of the bulk and interface Fermi levels and occupation of the empty interface states. For most semiconductors, a detailed analysis of electron injection and emission at the grain boundary yields the Fermi level at the interface within $kT$ of Fermi level of the negatively biased grain.[36,37] The interface charge is then

$$\sigma = \left(\frac{\varphi_B}{4\gamma}\right)^{1/2} + \left(\frac{\varphi_B + qV}{4\gamma}\right)^{1/2}, \tag{4.6}$$

where $\varphi_B$ is bias-dependent Schottky barrier and $\gamma = q^2/8\kappa\varepsilon_0 N_d$. Simultaneous solution of Eq.(4.4) and (4.6) allows the bias dependence of the Schottky barrier height to be obtained. For the interfaces with zero IDOS, the bias dependence of the Schottky barrier height is particularly simple[38]

$$\varphi_B = \varphi_{gb}\left(1 - \frac{qV}{4\varphi_{gb}}\right)^2. \tag{4.7}$$

When the applied bias achieves a critical breakdown value, $V_{break}$, the grain boundary potential barrier vanishes. For the interfaces with zero IDOS the breakdown bias is related to the zero bias barrier height as $V_{break} = 4\ \varphi_{gb}/q$, while the presence of interface states increases breakdown voltage.

One of the major difficulties in the analysis of transport phenomena using Eqs.(4.4,6) is that the voltage dependent current density across the interface, rather than bias dependent Schottky barrier height, is experimentally accessible. In order to relate the two, knowledge of the transport mechanism across the interface is required. Alternatively, the depletion width and, therefore, interface charge can be obtained from *C-V* measurements. However, the charge dynamics of the interface states can be rather complex and capacitance can exhibit significant frequency dispersion, preventing direct utilization of this technique in many important cases (e.g. ZnO, Si). For such materials, detailed analysis of *I-V* curves and frequency dependence of capacitance is required.[39,40]

### 4.2.2. Transport across grain boundaries

Electronic transport across the space charge regions and thin insulating layers has been extensively studied with respect to transport on metal-semiconductor interfaces, *p-n* junctions, and, to a smaller extent, grain boundaries. A number of primary transport mechanisms can be distinguished.[41] Depending on interface properties, the major transport mechanisms across semiconductor grain boundaries are thermionic emission, diffusion, tunneling and space charge current as illustrated in Figure 4.3.

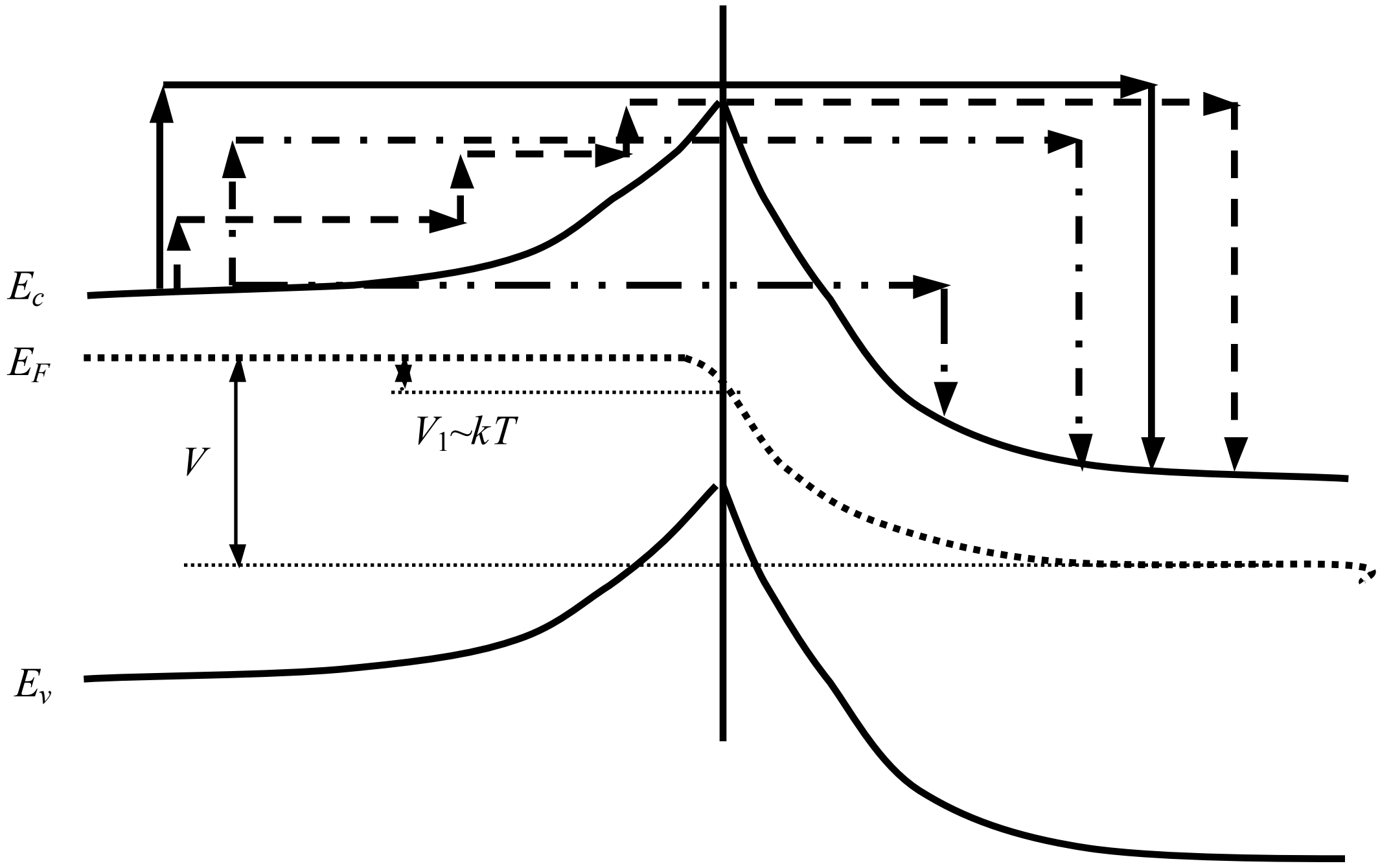

**Figure 4.3.** Transport mechanisms across electroactive grain boundaries. Schematically shown is thermionic emission ( ——— ), diffusion ( – – – ), thermionic field emission ( – · – · · ) and field emission ( – · · – ).

Thermionic emission

Thermionic emission theory is based on the assumptions that the barrier height is large compared to $kT$, electron collisions within the depletion region are neglected and the image force effects are neglected. The current-voltage dependence is given by

$$J_{TE} = A^* T^2 \exp\left(-\frac{q\varphi_{gb}(V)}{kT}\right)\left(1-\exp\left(-\frac{qV}{kT}\right)\right), \tag{4.8}$$

where $\varphi_{gb}(V)$ is voltage dependent potential barrier at the interface, $A^*$ is Richardson constant. In the small bias limit, the interface resistance is

$$R_{TE} = \frac{k}{A^* Tq}\exp\left(\frac{q\varphi_{gb}}{kT}\right). \tag{4.9}$$

Diffusion.

Diffusion theory is based on the assumptions that the barrier height is large compared to $kT$, electron collisions within the depletion region are included and the carrier concentration on the edges of depletion region are unaffected by current flow. The current-voltage dependence is given by

$$J_D = qN_c \mu E_0 \exp\left(-\frac{q\varphi_{gb}(V)}{kT}\right)\left(1-\exp\left(-\frac{qV}{kT}\right)\right), \qquad (4.10)$$

where $N_c$ is the effective density of states in the conduction band, $\mu$ is the mobility and $E_0$ is the field at the interface. In the small bias limit, the interface resistance is

$$R_D = \frac{kT}{q^2 N_c \mu E_0} \exp\left(\frac{q\varphi_{gb}}{kT}\right). \qquad (4.11)$$

The crossover between thermionic emission and diffusive transport occurs when $v_r = \mu E_0$, where thermal velocity $v_r = A^* T^2 / qN_c$. The similarity in the assumptions used in the derivation of Eqs.(4.8,10) implies that they can be combined in the thermionic emission-diffusion theory.

From the discussion above, it is clear that for high-mobility materials the dominant transport mechanism is thermionic emission, while for low mobility materials diffusion through space charge layer limits the transport.

<u>Tunneling and thermionic field emission.</u>

For very thin space charge layers, an additional transport mechanism across the potential barrier is thermionic field emission (TFE) and field emission (FE). In the former case, a thermally activated electron tunnels through the top of the potential barrier, while in the tunneling case the electron tunnels directly through the potential barrier. The low bias resistances for TFE and FE mechanisms are[42]

$$R_{TFE} = C_1 \frac{k}{qA^* T} \exp\left(\frac{q\varphi_{gb}}{E_0}\right), \qquad (4.12)$$

and

$$R_{FE} = C_2 \frac{k}{qA^* T} \exp\left(\frac{q\varphi_{gb}}{E_{00}}\right), \qquad (4.13)$$

where $C_1$ and $C_2$ are functions dependent on $N_d$, $T$ and $\phi_B$ and

$$E_0 = E_{00} \coth\left(\frac{E_{00}}{kT}\right). \tag{4.14}$$

The characteristic energy $E_{00}$ is given by

$$E_{00} = \frac{qh}{4\pi}\sqrt{\frac{N_d}{\varepsilon\varepsilon_0 m^*}}. \tag{4.15}$$

TE, TFE and FE regimes can be differentiated by considering the ratio between $E_{00}$ and $kT$. TE dominates for $kT >> E_{00}$, TFE corresponds to $kT \approx E_{00}$ and FE corresponds to $kT << E_{00}$. From Eqs.(4,9,12,13,14) TE dominates for current transport in the low doped semiconductors (large depletion width) and FE dominates to highly doped semiconductors (small depletion width), while TFE corresponds to the intermediate doping levels.

Space Charge Limited Current

A common feature of the mechanisms described above is the exponential dependence on potential barrier height. Unlike metal-semiconductor junctions, for grain boundaries the potential barrier rapidly decreases with bias (e.g. Eq.(4.7)) and for high enough bias the potential barrier disappears. Nevertheless, even under the breakdown conditions, a depletion region exists near the interface. It is natural to assume that transport across the interface in this case is limited by the drift, rather than diffusion, through the depletion region. This transport mechanism is somewhat similar to the space charge limiting current in the *n-i-n* structures.[43] The field in the depletion region is related to the carrier density, $\rho$, as $dE/dz = \rho/\kappa\varepsilon_0$, while the drift current is related to electric field as $J = -\mu\rho E$. Combination of these equations yields the field profile in the insulating layer as $E^2 = 2Jx/\kappa\varepsilon_0\mu$ and integration yields

$$J_{SC} = \frac{8\kappa\varepsilon_0\mu V^2}{9d^3}. \tag{4.16}$$

This equation is valid for $V > kT$, while for lower biases the more rigorous theory by Luryi must be used.[44] Obviously, space charge limited current in the grain boundary depletion region differs from that in the *n-i-n* structure; however, Eq.(4.16) is expected to provide a reasonable estimate for the drift limit of grain boundary resistance.

Transport mechanism and the I-V curve.

Provided that the transport mechanism across the interface is known, Eqs.(4.8,10,12,13) can be used to determine the bias dependence of interface potential barrier and calculate interface density of states using Eq.(4.4). Indeed, for conventional semiconductors such as Si or GaAs with high mobilities and relatively large GB depletion regions the dominant transport mechanism is thermionic emission and a number of *I-V* reconstructions of IDOS have been reported.[45,46] Similar analysis was performed for diffusion transport.[47] However, for the vast majority of oxide materials the dominant transport mechanism and associated materials parameters are sensitively dependent on the preparation route and, in the most cases, are unknown. The *C-V* measurements are hindered by the frequency dispersion of capacitance that precludes unambiguous determination of depletion width.[40,48]

In addition to the theoretical problems, transport measurements on oxides often suffer from large series resistance effects, i.e. grain boundary resistance is comparable to grain resistance or contact resistance at high biases and the reliable separation of these contributions is impossible unless a special 6 probe configuration on a well-defined bicrystal sample is used.[8] Therefore, local SPM transport probes as described in Chapter 3 are expected to be particularly useful for grain boundary characterization, since they allow unambiguous separation of grain boundary, bulk and contact responses for both bicrystal and polycrystalline materials alike.

### 4.2.3. Possible Grain Boundary Transport Mechanisms in $SrTiO_3$

The transport in $SrTiO_3$ based materials has been a subject of extensive experimental and theoretical research. The majority of experimental efforts to date were focused on studies of the transport processes in single crystals[49,50,51,52,53] or, alternatively, in polycrystalline materials.[54,55] Most grain boundary studies were limited to high temperature properties. The experimental effort has been primarily focused on acceptor-doped materials with relatively low dopant concentration, in which diffusion was shown to be the dominant transport mechanism. Only a few measurements on well-defined bicrystal samples have been reported and they were limited to room temperature.[56,57,58]

To the best of my knowledge, no detailed analysis of *I-V* behavior at such interfaces has been attempted. In order to fill the gap, the *I-V* and impedance properties of well-defined interfaces in a $SrTiO_3$ bicrystal in the low temperature range are studied.

Transport mechanism at model $SrTiO_3$ grain boundary

Prior to experimental study of transport behavior at grain boundaries of $SrTiO_3$, the possible transport mechanisms are estimated using the models considered above [Eqs.(4.8,10,15 16)] and the available material parameters. The effective density of states in the conduction band is calculated as $N_c = 2\left(2\pi m^* kT/h^2\right)^{3/2} M_c$, where $M_c$ is a number of equivalent conduction band minima and $m^*$ = 1.32 $m_0$ for $SrTiO_3$.[59] This estimate yields $N_c$ = 3.8·10$^{25}$ m$^{-3}$ at room temperature. The Richardson constant for $SrTiO_3$ is $A^* = 4\pi q m^* k^2/h^3$ = 158 A/cm$^2$ K$^2$. The temperature dependence of mobility is taken as $\mu = 0.83(\exp(600/T)-1)$ from Ref.[52], the corresponding room temperature value is $\mu$ = 5.3 cm$^2$/Vs (compare to 1400 cm$^2$/Vs for Si and 200 cm$^2$/Vs for ZnO).[42] The dielectric constant in $SrTiO_3$ is high, $\kappa$ = 300 at 300 K and strongly temperature dependent (compare to $\kappa$ =11.7 for Si and 8.5 for ZnO). To estimate the temperature dependence of the dielectric constant of $SrTiO_3$, the Curie Weiss law, $\kappa = 83700/(T-28)$, valid for $T$ > 100 K, is used.[59] In these estimates, the field dependence of the dielectric constant is neglected. To calculate the dominant transport mechanism as a function of temperature and bias, the bias dependence of the interface potential barrier is described by Eq.(4.7), i.e. the interface state density is assumed to be zero. The grain boundary depletion width and potential at room temperature were taken as $d$ = 14 nm and $\varphi_{gb}$ = 0.55 eV (these values are close to the experimentally measured parameters for 1 at. % Nb doped $SrTiO_3$ bicrystals studied). These values correspond to the carrier concentration $N_d$ = 3.75·10$^{25}$ m$^{-3}$. From the resistivity $\rho$ = 0.017 Ohm·cm [Ref. 60] and mobility $\mu$ = 5.3 cm$^2$/Vs the carrier concentration can be estimated as $N_d$ = 6.94 10$^{25}$ m$^{-3}$, i.e. close to this value.

To estimate the possible contribution of the tunneling processes, $E_{00}$ for the material is calculated from Eq.(4.15) as $E_{00}$ = 5.75 meV. In this case, $E_{00} < kT$ = 26 meV at 300 K. At lower temperatures, $E_{00}$ decreases due to the increase of dielectric constant and at 100 K the corresponding value is $E_{00}$ = 3.3 meV, still smaller than $kT$ = 8.6 meV. From these estimates, the tunneling contribution to the transport through the grain boundary with parameters specified above can be neglected.

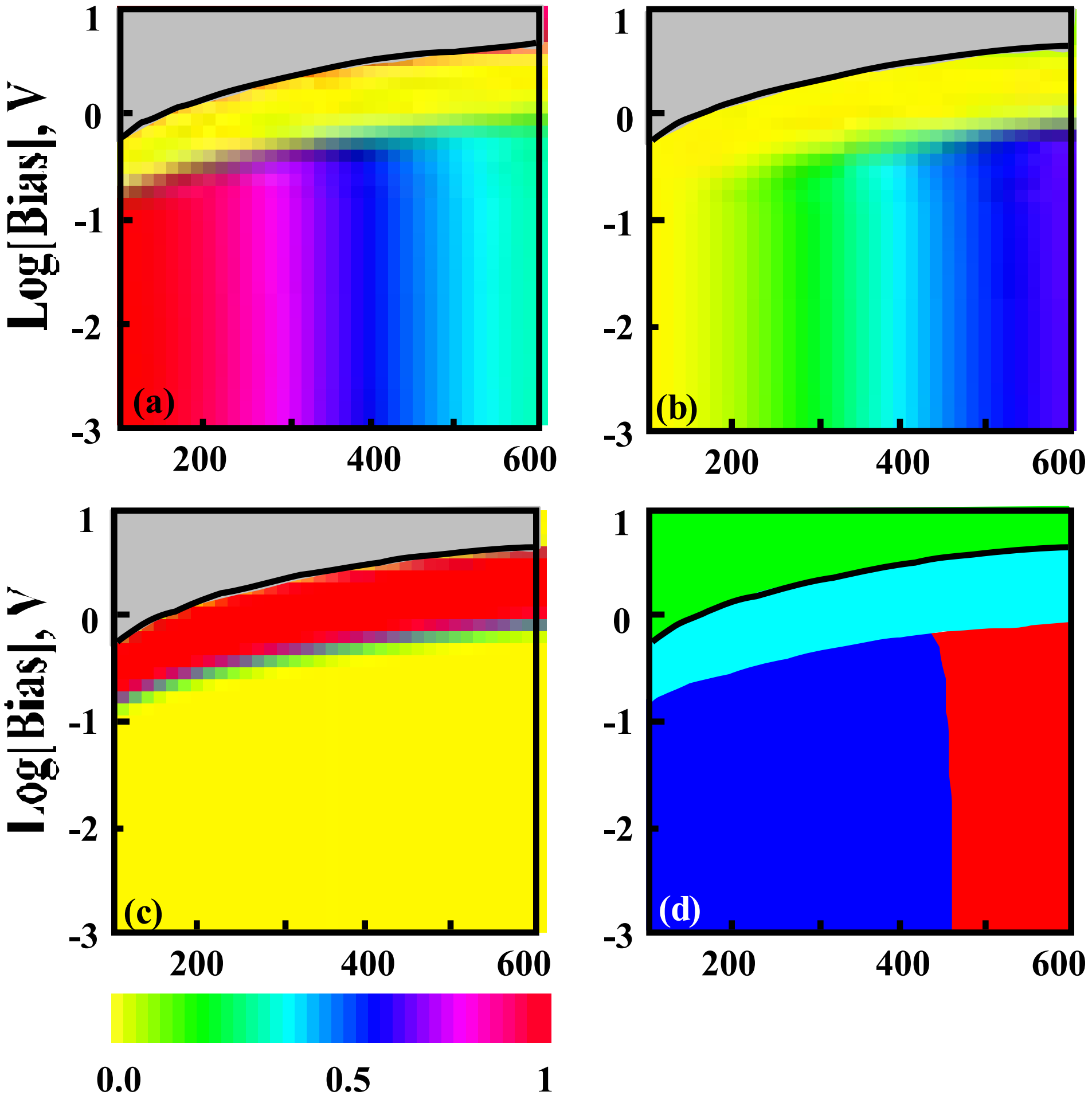

**Figure 4.4.** Temperature and bias dependence of thermionic (a), diffusion (b) and space charge (c) currents in a strontium titanate grain boundary (parameters in text). Conductance phase diagram (d). TE – thermionic emission, D – diffusion, SC – space charge, B – breakdown.

From consideration of the transport mechanisms in Section 4.2.2, the dominant mechanisms for $SrTiO_3$ are expected to be thermionic emission or diffusion. The latter mechanism is favored by the low carrier mobility in $SrTiO_3$, while relatively small depletion widths at GBs favors the former. As discussed above, neither mechanism is expected to apply to transport under high biases, when drift through the depletion layer, rather than emission or diffusion over the potential barrier, limits the current. To estimate the relative contribution of these mechanisms, the fraction of current carried by each was estimated as

$$f(J_i) = \frac{1/J_i}{1/J_{TE} + 1/J_D + 1/J_{SC}}, \tag{4.17}$$

where $i$ corresponds to thermionic emission (TE), diffusion (D) or space charge limited current (SC). The resulting phase diagrams for dominant conduction mechanism through $SrTiO_3$ grain boundary are shown in Figure 4.4.

Depending on bias and temperature, the dominant transport mechanism is thermionic emission for low temperatures and biases, diffusion for high temperatures and low biases and space charge limited current at high biases. It must be noted that the complexity of the phase diagram in Figure 4.4 suggests that transport in this case does not belong to any pure scenario and combined thermionic emission-diffusion or thermionic field emission-diffusion models must be used.[61] However, such an approach is less illustrative and in such complex cases, numerical modeling of grain boundary transport may be required. Similar phase diagrams can be obtained for arbitrary interface parameters, providing valuable predictive tool for description of transport processes.

Non-linear dielectric phenomena at $SrTiO_3$ interfaces

One of the key assumptions in the abrupt junction model considered in Section 4.2.1 is the constant value of dielectric constant, $\kappa$, in the vicinity of the interface. While this assumption is undoubtedly correct for such materials as Si, GaAs or ZnO, it may not hold for $SrTiO_3$, especially at low temperatures. Indeed, the dielectric constant of incipient ferroelectrics such as $SrTiO_3$ is field dependent, enabling its application in frequency-agile materials. A deviation of dielectric constant from the bulk value has been reported for $SrTiO_3$ thin films for $T < 80$ K at fields $\sim 10^7$ V/m [Ref.62] or $T < 50$ K for field $\sim 10^6$ V/m.[63] Studies at higher biases, for which the deviations from bulk behavior can be expected at higher temperatures, were precluded by the breakdown strength of the $SrTiO_3$.

It can be conjectured that high electric fields ($10^7$ - $10^8$ V/m) in the vicinity of grain boundaries can significantly affect the dielectric properties of material. In fact, the maximal electric field at the interface with $d$ = 10 nm and $\varphi_{gb}$ = 0.5 eV is $E = 2\cdot10^8$ V/m, i.e. at least an order of magnitude higher than the maximum field achievable for thin films. Therefore, deviation from the bulk dielectric behavior can be expected for $T > 100$ K, contributing to measured *I-V* and *C-V* data. At lower temperatures, field induced transition to a ferroelectric phase can occur.[64] The evidence for the formation of a polar phase near the grain boundaries in polycrystalline undoped $SrTiO_3$ for $T < 132$ K from Raman measurements was recently reported.[65] An additional origin of unusual dielectric phenomena at these interfaces is strain, which can induce ferroelectric phase transition in $SrTiO_3$.[66] A ferroelectric or piezoelectric phase at the interface and the associated polarization discontinuity will contribute to the interface charge.

Despite these reports, the evidence for non-linear dielectric effects in transport at $SrTiO_3$ interfaces is scarce. A number of theoretical and experimental studies of current transport in PTCR materials above the corresponding Curie temperature (i.e. in the nominally cubic phase of $BaTiO_3$) were reported; however, the vast majority of the experimental studies were limited to polycrystalline materials. Combined with the lack of reliable information on the dominant transport mechanism and large dopant and vacancy concentration gradients generic in ceramic materials, this precludes unambiguous identification of non-linear dielectric effects. The effect of dielectric non-linearity was recently reported at contacts to $SrTiO_3$, namely at Au-$SrTiO_3$ interface.[67] However, until now non-linear dielectric behavior at the $SrTiO_3$ grain boundaries have not been reported.

### 4.3. Transport in $SrTiO_3$ Bicrystals: The Model System

In this Section, we analyze the applicability of SPM based techniques for the direct spatially resolved imaging of interface properties on the grounded and biased surfaces. To complement the SPM data, the transport in the Nb-doped Σ5 $SrTiO_3$ bicrystal was studied by variable temperature transport measurements.

#### 4.3.1. Grain Boundary Potential Barrier Imaging and Surface Screening

In order to relate the grain boundary properties to atomic configuration, an interface with a known structure was used. Nb-doped Σ5 $SrTiO_3$ bicrystals (0.5 wt%) were produced by diffusion bonding. Numerous high-resolution transmission electron microscopy studies on similar bicrystals have shown that the interfaces are atomically abrupt, with no impurity segregation.[68] A 10x10x0.5 mm crystal, dark-blue due to the donor doping, is sectioned such that the grain boundary is perpendicular to the (100) surface. The grain boundary can be easily detected by means of transmission optical microscopy as a dark blue (almost black) line on the lighter background perpendicular to the edge of the crystal. Topographic features were used as markers to determine the position of grain boundaries in the EFM and SSPM measurements; very often, a wedge-like divot is associated with the grain boundary-crystal edge junction. Prior to imaging the crystal was repeatedly washed with ethanol, acetone and distilled water.

The AFM, EFM and SSPM measurements were performed on a commercial instrument (Digital Instruments Dimension 3000 NS-III) with Co/Cr coated tips ($l \approx 225$ μm, resonant frequency ~ 70 kHz). SSPM measurements were performed from ~10 nm to 1.5 μm above the surface; EFM measurements were performed from 20 to 200 nm above the surface. The scan rate varied from 0.2 Hz to 0.5 Hz. The driving voltage dependence of surface potential observed by SSPM saturates at driving voltage ~ 1-2 V for the lift heights of interest; therefore, $V_{ac}$, was chosen to be > 5 V. To reduce the effect of drift the images were acquired with the grain boundary oriented along the slow scan axis. Topographic and EFM images were processed by line flattening.[69] SSPM images were processed only by a constant background subtraction. Force gradient and potential profiles were obtained by averaging the flattened EFM and unprocessed SSPM images along the slow scan axis. A generic feature of SSPM is fewer imaging artifacts due to topography (Figure 4.5). The use of low dc bias voltages contributes to higher image stability. In addition, detection in SSPM implies much smaller tip oscillation amplitude than that in topographic or EFM imaging. From a direct comparison of the signal from the photodiode in the main scan line and non-contact scan line the characteristic oscillation amplitude during potential detection is < 1 nm depending on feedback circuit parameters.[70] Thus, for flat surfaces imaging is possible at very small tip-surface separation. To perform SIM measurements, the microscope was augmented by external electronics as described in Chapter 3.

The surface topography, surface potential, and force gradients of the grain boundary-surface intersection are compared in Figure 4.5. The surface is extremely flat with RMS roughness less then 1 nm and a number of spots due to contaminates. Pores with diameters of ~100-200 nm are distributed non-uniformly along the interface. This observation suggests that the pores exist in the bulk as well and contribute to the optical contrast of the grain boundary. Similar observations of pores at $SrTiO_3$ bicrystal interfaces are reported by other groups.[71] The surface potential measured 50 nm above the surface exhibits a sharp protrusion associated with the grain boundary, Figure 4.5b. The half-width of the broad potential feature is ~700 nm. Superimposed on this contrast is a much narrower and larger amplitude feature of half-width ~200 nm localized at the grain boundary. The force gradient images acquired with +5 V and -5V tip bias are shown in Figure 4.5 c,d. The grain boundary is again associated with a feature of half-width ~700 nm. The contrast in the force gradient image reverses with tip polarity, indicative of a dominant Coulombic contribution to the interaction.

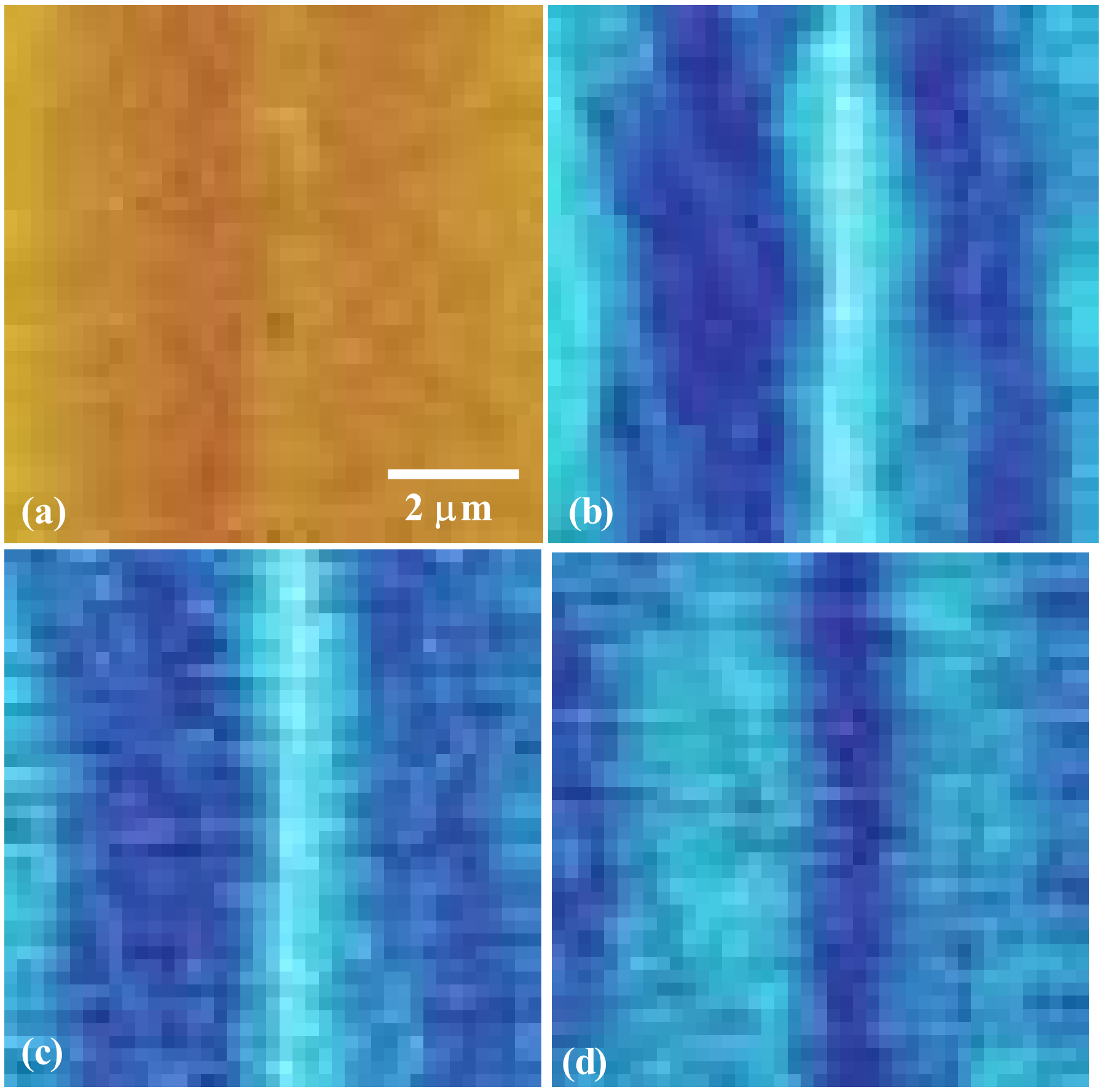

**Figure 4.5.** Topography of Nb-doped 36.8° $SrTiO_3$ bicrystal in the vicinity of grain boundary (a). SSPM image of the same region (b) and EFM (force gradient) images at tip bias $V_{tip}$ = 5V (c) and $V_{tip}$ = -5V. Range is 5 nm (a), 20 mV (b) and 2 Hz (c,d).

Quantification of surface properties from the EFM and SSPM data requires the solution of several independent problems, solved here for the case of an interface intersecting a surface. First, these techniques are ultimately sensitive to the force gradient (EFM) or the force (SSPM) between the tip and the surface. Second, the measurements are performed above the grain boundary-surface junction rather than at the grain boundary itself. Therefore, the relationship between the properties of a grain boundary in the bulk and the potential distribution above the surface for ideal surface termination is considered in Section A. The grain boundary contribution to force gradient (EFM data) is considered in Section B and corresponding experimental results are presented in section C. The influence of a charged grain boundary on the effective potential determined by SSPM and the influence of imaging conditions on SSPM data is discussed in Section D and E. Finally, the possibility for grain boundary screening are presented in section F.

A. Potential at grain-boundary-surface junction

In order to quantify the properties of a grain boundary a relationship between the bulk properties and potential above a grain boundary-surface junction is required. Since the potential above the junction is relatively insensitive to the details of charge distribution in the space charge regions, the abrupt junction approximation is used as discussed in Section 4.2. In order to find the potential above the grain boundary-interface junction the image charge method for a dielectric half-plane is applied:

$$\varphi(\mathbf{r}) = \frac{1}{2\pi\varepsilon_0(\kappa+1)}\left(\int d\mathbf{r}' \frac{\rho(\mathbf{r}')}{|\mathbf{r}-\mathbf{r}'|} + \int d\mathbf{s} \frac{\sigma(\mathbf{r}'')}{|\mathbf{r}-\mathbf{r}''|}\right), \qquad (4.18)$$

where the volume and surface integrals are taken within space charge layer and at the interface, respectively. This integral is solved in a closed form, however the resulting expression has a complicated functional form that is not useful for image analysis and is not shown here. Instead, the maximal value of potential and the Lorentzian width of the peak are used to describe the potential behavior far from the surface. As expected, the potential achieves a maximum above the grain boundary-surface junction and the calculated potential both within the crystal and above the surface for $\kappa = 4$ is shown in Figure 4.6b.

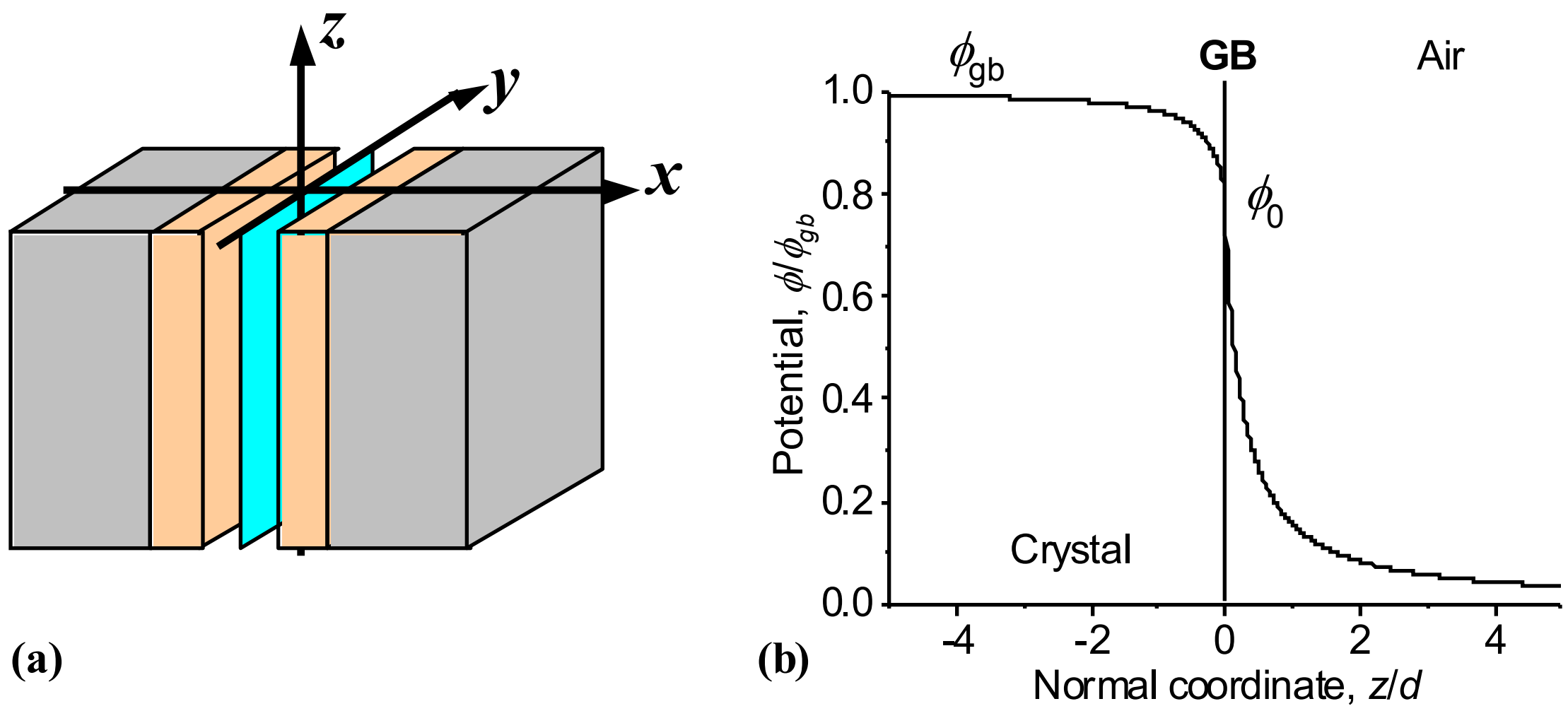

**Figure 4.6.** Simplified charge distribution near the grain boundary (a) and potential near the grain boundary-surface junction for $\kappa = 4$ (b).

Within the crystal the potential is almost constant, decreasing only to $\kappa/(\kappa+1)$ of its bulk value at the boundary. Above the surface, the potential decays rapidly. Potential profiles at different heights above the surface are shown in Figure 4.7a. For tip-surface separations $z > 0.1\,d$ the potential profile above the junction is well fitted by the Lorentzian function $y = (2A/\pi)\ w/(4x^2+w^2)$ and the dependence of the width, $w$, on separation is shown in Figure 4.7b. The line fit is $w/d = 0.4 + 1.9\ z/d$ for $0.1<z/d<10$, i.e. within the experimentally accessible region for a typical $d$ of order 100 nm for low-doped semiconductors. Therefore, the measured potential profile width as a function of tip-surface separation can be used to determine the size of space charge layer.

The potential above the surface-interface junction for $x = 0$, i.e. the height of potential profile is described by:

$$\varphi(z) = \frac{qN_d}{\pi\varepsilon_0(\kappa+1)}\left[\left(d^2 - z^2\right)\arctan\left(\frac{d}{z}\right) + d_{sc}z\left(1 + 2\ln(z) - \ln\left(d^2 + z^2\right)\right)\right], \quad (4.19)$$

which yields $\varphi_0 = \varphi(0) = \dfrac{qN_d d^2}{2\varepsilon_0(\kappa+1)}$ for the potential at the grain boundary-surface junction. The distance dependence of the potential is shown in Figure 4.7b. The relationship between the potential at the grain boundary in the bulk and the potential at the junction has the following form:

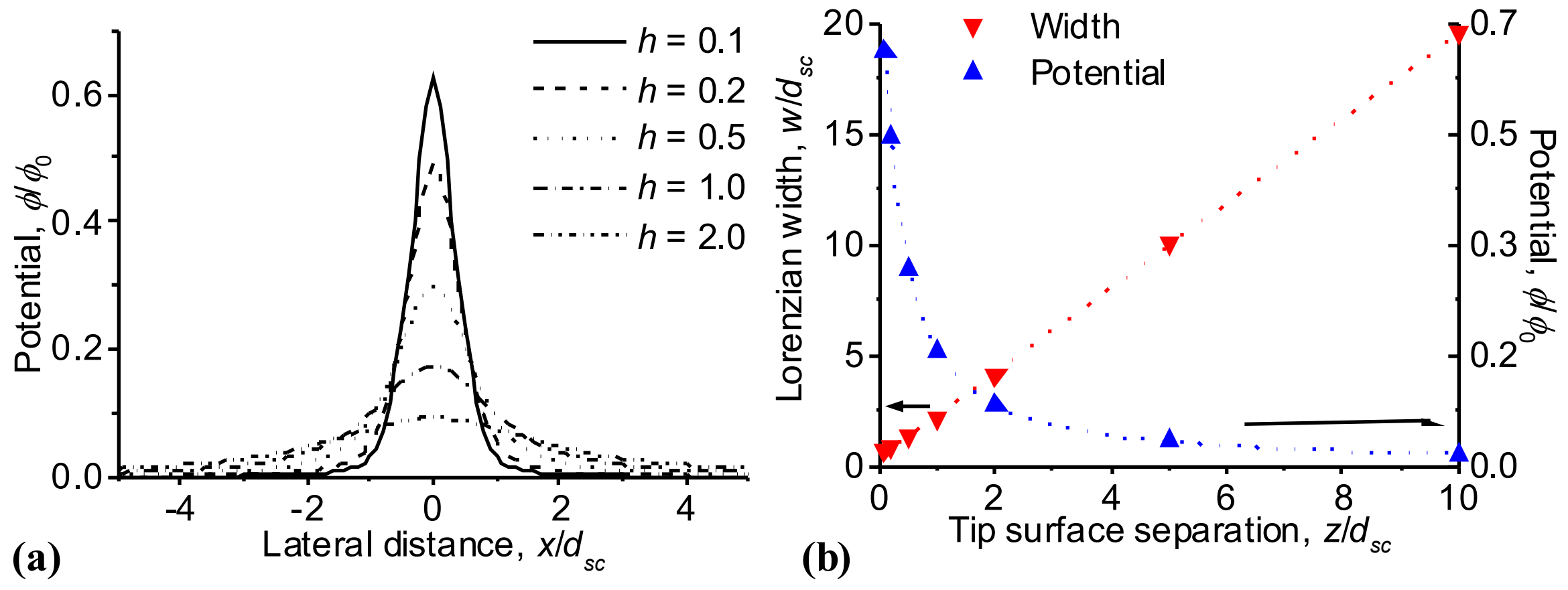

**Figure 4.7.** Potential profiles at different heights above grain boundary-surface junction (a) and grain boundary potential and profile width for different heights (b).

$$\varphi_{gb} = \frac{\kappa + 1}{\kappa}\varphi_0 . \tag{4.20}$$

This equation implies that for systems with high dielectric constants the potential at the grain boundary-surface junction accessible by scanning probe techniques is almost equal to that at the grain boundary in the bulk, provided that no charge accumulation takes place at surface-interface junction. Unfortunately, EFM and SSPM imaging at small tip-surface separations is difficult due to topographic artifacts and imaging instabilities, while for larger tip-surface separations Eq.(4.19) predicts rapid decay of potential. Nevertheless, if the potential is a known function of tip-surface separation, Eq.(4.19) can be used to extrapolate the lift height-potential dependence to obtain the potential at the junction and Eq.(4.20) can then be used to determine the grain boundary potential in the bulk. Rewriting Eq.(4.19) in terms of $\varphi_0$ and $d$ yields:

$$\varphi(z) = \varphi_0 \frac{2}{\pi d^2}\left[\left(d^2 - z^2\right)\arctan\left(\frac{d}{z}\right) + dz\left(1 + 2\ln(z) - \ln\left(d^2 + z^2\right)\right)\right]. \tag{4.21}$$

Noteworthy is that the potential above the grain boundary-surface junction is reduced to a function of only two variables; the potential at the grain boundary-surface junction and the width of the space-charge layer. As expected, Eq.(4.21) is homogeneous in $z/d$, since it is the only length scale in the problem. Fitting the experimental data by Eq.(4.21) yields $\varphi_0$ and $d$. Both EFM and SSPM provide information about the electrostatic forces acting on the tip rather than about the actual potential. Therefore, in the next two sections the forces acting on the tip and their relationship with EFM and SSPM signals are considered.

B. EFM imaging of grain boundary

For large separations (>10nm) electrostatic forces between the tip and surface dominate short-range Van-der-Waals forces[72] and for conductive materials are capacitive in nature, i.e. the distance dependence is that of the derivative of tip-surface capacitance. In the limit of high dielectric constant $C(z,\varepsilon) \approx C(z)$. Since the dielectric constant of $SrTiO_3$ is high ($\varepsilon \approx 300$ at 293K),[59] in subsequent analyses the effective capacitance $C(z,\varepsilon)$ is replaced by true capacitance $C(z)$.

To quantify the EFM data the grain boundary contribution to the force gradient must be included in the overall interaction. The total force between the biased tip and the surface can be written as

$$F(z) = \frac{\mathrm{d}C(z)}{\mathrm{d}z}\Delta V^2 + \int \frac{\partial \varphi_{gb}}{\partial \mathbf{n}}\left(\sigma_{tip} + \sigma_{ind}\right)\mathrm{d}\mathbf{S}_{tip}, \tag{4.22}$$

where the first term is the capacitive force, $F_{cap}(z)$, discussed in Chapter 2 and the second term is a contribution due to the Coulombic interaction of grain boundary charges with the metallic tip, $F_q(z)$. $\sigma_{tip}$ is the surface charge density of the tip in the absence of a grain boundary, $\sigma_{ind}$ is the image charge density induced by grain boundary charges and $\mathbf{n}$ is the normal vector to the tip surface. Assuming that the second term in Eq.(4.22) is much smaller than the first, i.e. the grain boundary contribution to the EFM signal is small, $\sigma_{ind} << \sigma_{tip}$, the charge state of the tip is not influenced by the grain boundary. This justifies the use of the line charge model to describe the Coulombic interaction between the grain boundary and the tip and the second term in Eq.(4.22) becomes:

$$\int \frac{\partial \varphi_{gb}}{\partial \mathbf{n}}\sigma_{tip}\mathrm{d}\mathbf{S}_{tip} = \int_d^L \lambda_{tip}\varphi'_{gb}\mathrm{d}z \approx \lambda_{tip}\varphi_{gb}(d), \tag{4.23}$$

since $\varphi_{gb}(z)$ rapidly decays with tip-surface separation. From Eqs.(2.10,11) the capacitive, $F_{cap}(z)$, and Coulombic, $F_q(z)$, components of the force are

$$F_{cap}(h) = \frac{4\pi\varepsilon_0 V^2}{\beta^2} \ln\left(\frac{L}{4h}\right), \quad \text{and} \quad F_q(h) = \frac{4\pi\varepsilon_0 V}{\beta} \varphi_{gb}(\alpha h). \tag{4.24a,b}$$

The force gradients proportional to the frequency shift measured in EFM are:

$$\frac{\mathrm{d}F_{cap}}{\mathrm{d}z} = \frac{4\pi\varepsilon_0 V^2}{\beta^2} \frac{1}{h}, \qquad \frac{\mathrm{d}F_q}{\mathrm{d}z} = \frac{4\pi\varepsilon_0 V\alpha}{\beta} \varphi'_{gb}(\alpha h). \tag{4.25a,b}$$

The cantilever spring constant, $k$, that relates the experimentally determined frequency shift and force gradient is usually unknown and depends strongly on cantilever properties. However, the ratio of frequency shift at the grain boundary to that of the surface is independent of cantilever properties and is equal to:

$$\frac{F'_q}{F'_{cap}} = \frac{\alpha\beta}{V} h\varphi'_{gb}(\alpha h). \tag{4.26}$$

From Eq.(4.21) the gradient of grain boundary potential can be found as

$$\varphi'_{gb}(z) = \varphi_0 \frac{2}{\pi d^2}\left[-4z \arctan\left(\frac{d}{z}\right) + 2d\left(2 + 2\ln(z) - \ln\left(d^2 + z^2\right)\right)\right]. \tag{4.27}$$

By fitting the measured ratio with the function

$$\frac{F'_q}{F'_{cap}} = \frac{A}{\pi d^2} z\left[4\alpha z \arctan\left(\frac{d}{\alpha z}\right) - 2d\left(2 + 2\ln(\alpha z) - \ln\left(d^2 + \alpha^2 z^2\right)\right)\right], \tag{4.28}$$

where $A$ and $d$ are now fitting parameters and $\alpha$ is given in Eq.(2.9b), both the space charge layer width, $d$, and potential at the grain boundary-surface junction $\varphi_0$ can be extracted as

$$\varphi_0 = \frac{AV}{\sqrt{1+\tan^2\theta}}\left\{\ln\left(\frac{1+\cos\theta}{1-\cos\theta}\right)\right\}^{-1}, \tag{4.29}$$

where $\theta$ is tip half-angle. The width of the EFM profile provides an independent measurement of $d$.

C. Bias and distance dependence of EFM image

To quantify the bias dependence of capacitive force the average frequency shift due to the surface, defined as the image average of the unprocessed image, is analyzed by Eqs. (2.12). The difference between the maximum frequency shift at the grain boundary and the frequency shift far from the grain boundary is referred to as the grain boundary frequency shift. In order to improve the estimates for large separations, force gradient profiles were fitted by a Lorentzian function. The bias dependence of the average frequency shift and grain boundary frequency shift are compared in Figure 4.8a,b. As expected, the average frequency shift is a parabolic function of bias voltage, $\Delta\omega_{im} = \Delta\omega_0 + a(V - V_s)^2$, where $\Delta\omega_{im}$ is the total frequency shift, $\Delta\omega_0$ is the frequency offset, $V_s$ is the potential offset and $a$ is a proportionality constant (Table 4.I).

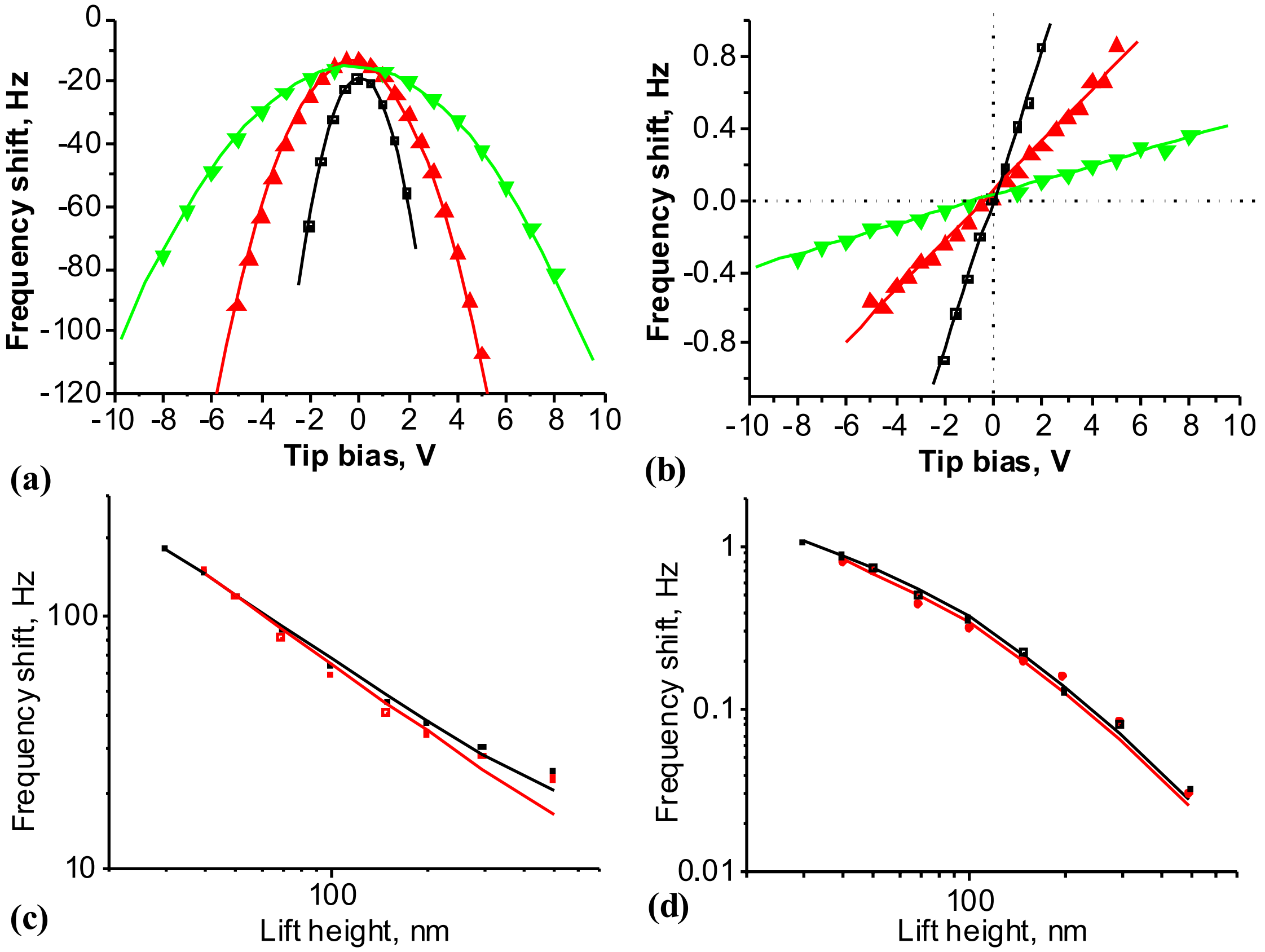

**Figure 4.8.** Total frequency shift (image average) (a) and frequency shift due to grain boundary (b) as a function of tip bias for lift heights 20 nm (■), 50nm (▲), and 120 nm (▼). Fitting the average frequency shift (c) and grain boundary frequency shift (d) as a function of tip-surface separation for tip biases 5 V (■) and -5V (●).

Table 4.I

*EFM Bias-dependence fitting parameters*

| Lift height, nm | $\Delta\omega_0$, Hz | $V_s$, mV | $a$, Hz/V$^2$ | Slope, Hz/V |
|---|---|---|---|---|
| 26 | -19.6 ± 0.2 | 117 ± 5 | -10.4 ± 0.1 | 0.42 ± 0.01 |
| 56 | -13.9 ± 0.1 | -216 ± 3 | -3.44 ± 0.01 | 0.140 ± 0.003 |
| 126 | -15.7 ± 0.1 | -188 ± 5 | -0.986 ± 0.002 | 0.041 ± 0.001 |

The potential offset $V_s$ is related to the existence of a uniform double layer due to intrinsic surface states or adsorption as well as to the work function difference between the tip and the surface. The frequency offset $\Delta\omega_0 \approx 10$Hz is due to drift in the oscillating characteristics of the cantilever after calibration. The dependence of the grain boundary frequency shift on bias voltage is shown in Figure 4.8b. This dependence is linear and intersects the origin. These results indicate that the average force interaction originates from a capacitive tip-surface interaction, thus the quadratic bias dependence. The grain boundary component to the EFM contrast is linear in tip bias and therefore can be attributed to Coulombic interactions of charges at the grain boundary with the biased tip. The magnitude of the former effect is much larger than that of the latter, in agreement with the assumptions done in the derivation of Eq.(4.23).

The distance dependencies of the average frequency shift and the grain boundary frequency shift are shown in Figure 4.8c,d for tip biases of 5 and -5 V. For small tip-surface separations the dependence in log-log coordinates is almost linear with a slope close to unity, in agreement with Eq. (4.24a). In order to take into the account frequency offset $\Delta\omega_0$, the following fitting function is used:

$$\Delta\omega_{im} = b + \frac{c}{z}, \tag{4.30}$$

where $\Delta\omega_{im}$ is total frequency shift, $b$ and $c$ are fitting parameters. For tip biases of 5 V and -5 V, $b$ = 7.8 ± 1.7 Hz, $c$ = 6200 ± 115 nm/s and $b$ = 3.2 ± 3.5 Hz, $c$ = 6431 ± 295 nm/s, respectively. The frequency shift can be found from Eq.(2.12) and (4.25) as

$$\Delta\omega_{im} = \frac{\omega_0}{2k}\frac{4\pi\varepsilon_0 V^2}{\beta^2}\frac{1}{h} \tag{4.31}$$

Substituting the resonant frequency of the cantilever $\omega_0$ = 76.6 kHz, a typical spring constant for the tip $k$ = 1-5 N/m and a typical tip half-angle $\theta \approx 17°$, the frequency shift according to Eq.(4.31) yields coefficient $c$ equal to 6400 - 1290 nm/s, which is in excellent agreement with our experimental results. From this the cantilever spring constant for the probe used is close to 1 N/m.

The lift height dependence for the grain boundary frequency shift shown in Figure 4.8d was also quantified. The force gradient maxima were fitted by Eq.(4.28) yielding $A$ = 152 ± 3, $d$ = 229 ± 15 nm for V= 5 V and $A$ = 142 ± 4, $d$ = 217 ± 32 nm for V= -5 V.

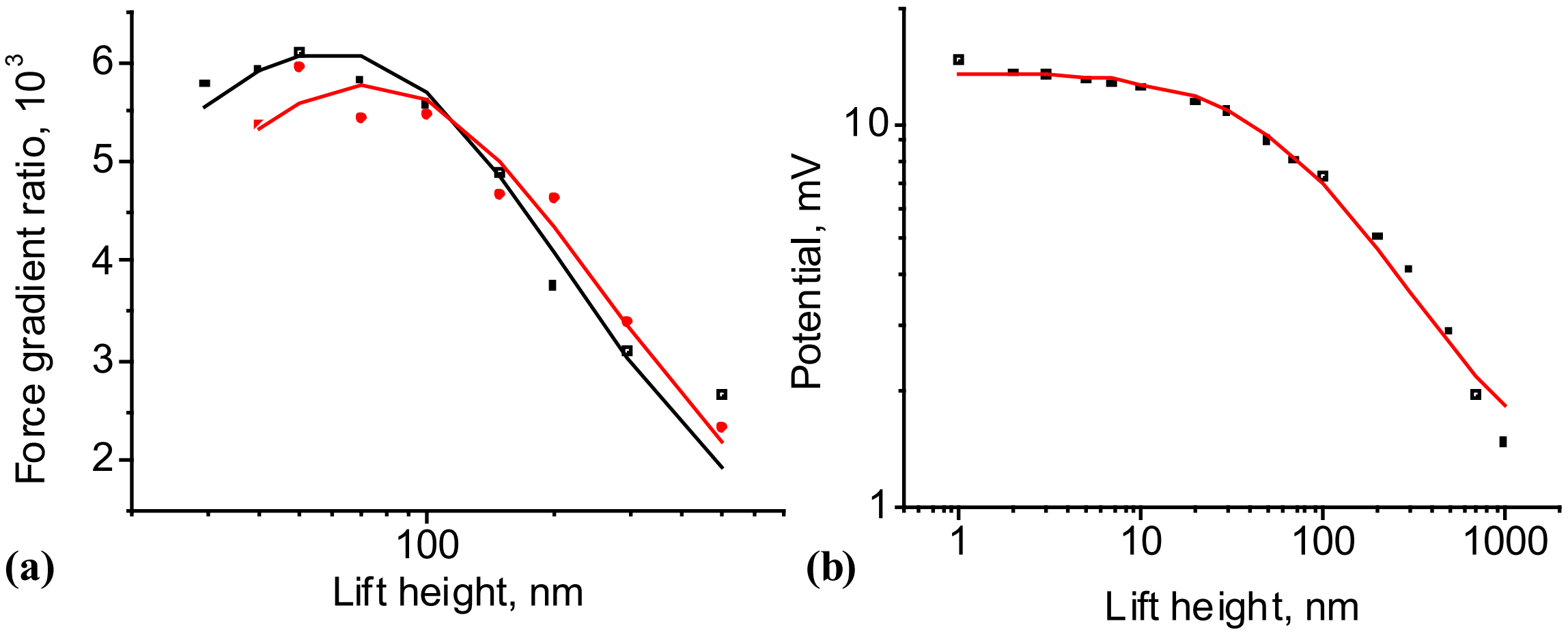

**Figure 4.9.** Fit of the ratio of frequency shift due to grain boundary to total frequency shift as a function of tip-surface separation (a) for tip biases 5 V (■) and -5V (●). (b) Fit of potential barrier height determined by SSPM as a function of lift height.

From Eqs.(2.12, 4.25) the frequency shift for the grain boundary is:

$$\Delta\omega_{gb} = \frac{\omega_0}{2k}\frac{4\pi\varepsilon_0 V\alpha}{\beta}\varphi'_{gb}(\alpha h). \tag{4.32}$$

Using the same tip parameters, these results yield the potential at the grain boundary-surface junction as 30-140 mV. Using the spring constant derived from the total frequency shift - distance analysis, the potential at the grain boundary-surface junction is calculated as ~ 30mV.

Finally, the cantilever-dependent constant in Eq.(2.12) can be eliminated by taking the *ratio* of experimentally observed grain boundary frequency shift and total frequency shift. The ratio and the fit by Eq.(4.28) are shown in Figure 4.9a. Using tip parameters $\theta \approx 17°$, $\alpha \approx 1.04$, $\beta \approx 4.1$, the depletion width is 240 nm and the potential at grain boundary-surface junction is 29 mV.

D. SSPM imaging of grain boundaries

Quantification of SSPM contrast is treated similarly to that for EFM contrast, using the line charge description of the tip. The electrostatic force between the tip and the surface is described by Eqs.(2.12,31) and the first harmonic of capacitive force is:

$$F_{1\omega}^{cap} = \frac{4\pi\varepsilon_0 V_{ac}\left(V_{dc} - V_s\right)}{\beta^2} \ln\left(\frac{L}{4h}\right), \tag{4.33}$$

where $V_s$ is the effective surface potential originating from intrinsic surface states or adsorbates. The first harmonic of Coulombic force is:

$$F_{1\omega}^{q} = 4\pi\varepsilon_0 V_{ac}\left\{\ln\left(\frac{1+\cos\theta}{1-\cos\theta}\right)\right\}^{-1} \varphi_{gb}\left(h\sqrt{1+\tan^2\theta}\right). \tag{4.34}$$

Using the analysis presented in Chapter 3, the nulling bias measured in SSPM is

$$V_{dc} = V_s - \beta\frac{\varphi_{gb}(\alpha h)}{\ln(L/4h)} + \frac{\delta}{V_{ac}\ln(L/4h)}. \tag{4.35}$$

The average image potential far from the boundary is

$$V_{dc} = V_s + \frac{\delta}{V_{ac}\ln(L/4h)}. \tag{4.36}$$

In ideal SSPM, imaging the nulling voltage is equal to surface potential and does not depend on tip-surface capacitance and driving voltage. This corresponds to $\delta = 0$ in Eq.(4.36), i.e. ideal feedback. For a realistic system, however, Eq.(4.36) predicts SSPM signal dependence on tip-surface separation and driving amplitude. The potential contrast due to the grain boundary is then the difference of $V_{dc}$ above the boundary and far from the boundary:

$$\Delta V_{dc} = \beta\frac{\varphi_{gb}(\alpha h)}{\ln(L/4h)}. \tag{4.37}$$

Provided that experimental SSPM contrast follows Eq.(3.38), i.e. feedback error is described by single parameter, $\delta$, and grain boundary potential is independent of driving voltage as suggested by Eq.(4.37), fitting the experimentally determined distance dependence of potential amplitude at the grain boundary by the function

$$\Psi(z) = \frac{\beta\varphi_0}{c - \ln(z)} \frac{2}{\pi d^2}\left[\left(d^2 - \alpha^2 z^2\right)\arctan\left(\frac{d}{\alpha z}\right) + \alpha d z\left(1 + 2\ln(\alpha z) - \ln\left(d^2 + \alpha^2 z^2\right)\right)\right], \quad (4.38)$$

where $c$, $d$ and $\varphi_0$ are fitting parameters and $\alpha$ are $\beta$ parameters of line charge model, can be used to extract the space charge width, $d$, and the potential at the grain boundary-surface junction, $\varphi_0$. The parameter $c = \ln(L/4)$ is related to the tip and cantilever geometry.

## E. Driving amplitude and distance dependence of SSPM image

Quantification of SSPM data is done similarly to that of EFM data, i.e. average image potential and potential difference between the grain boundary and the rest of the image were determined. Both driving voltage and tip surface separation dependencies were measured. The non-ideality of the feedback loop results in $1/V_{ac}$ dependence of measured average surface potential on driving amplitude as predicted by Eq.(3.40). Thus, the average image potential $V_{av}$ is fit by $V_{av} = V_s + B/V_{ac}$, where $V_s$ is surface potential and $B$ is a fitting parameter. The fitting results are summarized in Table 4.II and illustrated in Figure 4.10a.

Table 4.II

*Driving voltage dependence of SSPM image average*

| Lift height, nm | $V_s$, mV | $B = \delta\ln(L/4h)$ |
|---|---|---|
| 10 | -89 ± 2 | 85 ± 1 |
| 30 | -147 ± 3 | 103 ± 2 |
| 50 | -117 ± 3 | 135 ± 3 |
| 0 | 323 ± 12 | 122 ± 3 |

Noteworthy is that the measured potential amplitude due to the grain boundary, $\Delta V_{gb}$, is virtually $V_{ac}$ independent above 2 V as seen in Figure 4.10b. At low driving voltages, there is considerable noise and possibly a small increase in potential. However, this effect does not exceed ~2-4 mV, while the dependence of the average image potential (See Figure 4.10c) indicates a strong driving voltage dependence. This observation implies that grain boundary potential amplitudes obtained by SSPM are relatively insensitive to imaging conditions and Eq.(4.37) can be used to describe potential-distance relation. This demonstrates that system parameters that strongly influence the absolute value of measured surface potential do not alter measured potential variations.

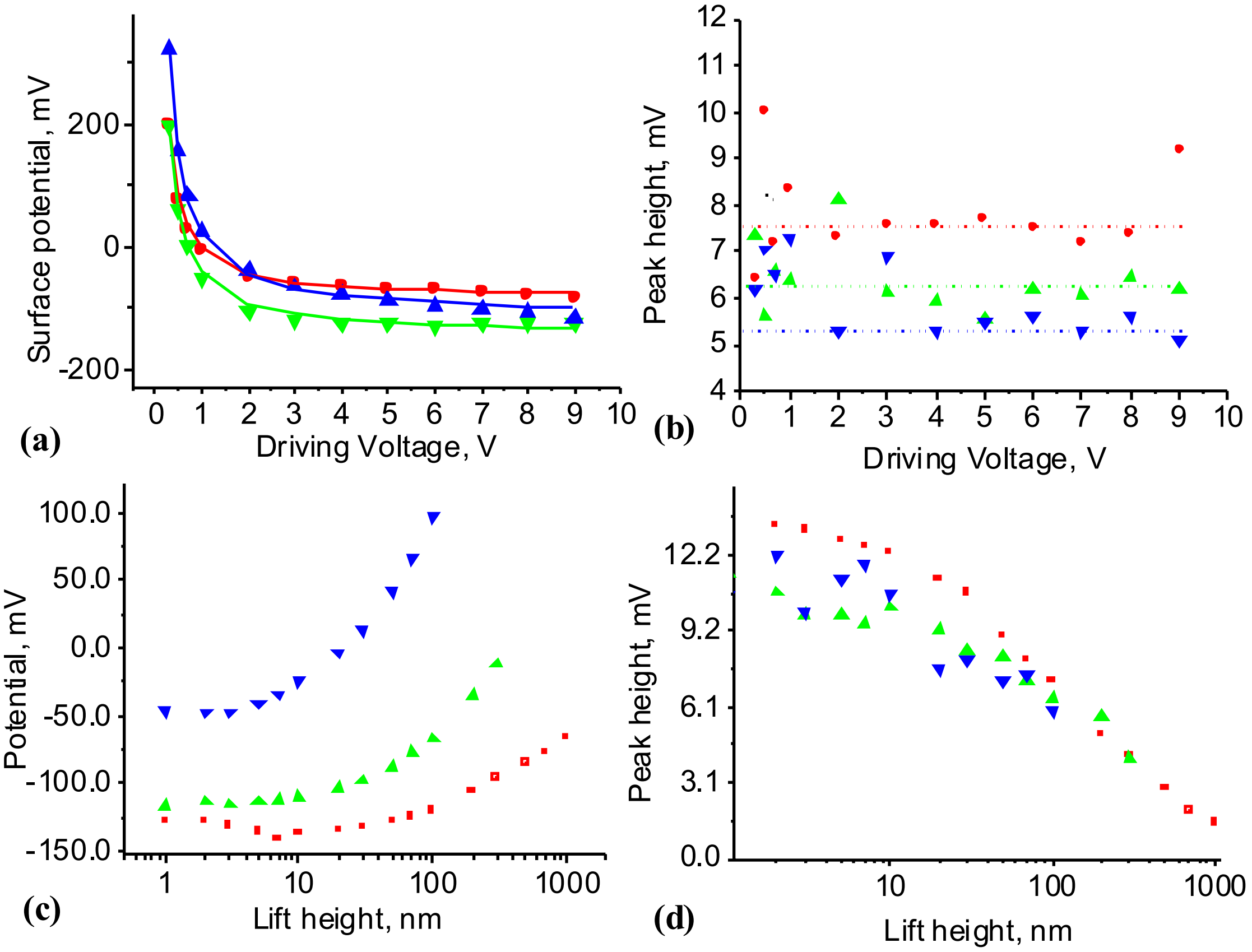

**Figure 4.10.** Average surface potential (a) and grain boundary potential (b) as a function of driving voltage for lift heights 10 nm (●), 30 nm(▲), and 50 nm (▼). Dependence of average surface potentials (c) and grain boundary potential barrier (d) as a function of lift height for driving voltages 5 V (■), 1V (▲), and 300 mV (▼).

The distance dependence for different driving voltages is shown in Figure 4.10c. These results are in a qualitative agreement with Eq.(4.36), i.e. $V_{dc}$ is higher for small driving voltages. In contrast, the distance dependence of grain boundary potential is independent of driving voltage as shown in Figure 4.10d. The noise is significantly higher for low driving voltages. The grain boundary potential-distance dependence fitted by Eq.(4.38) is shown in Figure 4.9b. Depletion width and potential at the grain boundary-surface junction determined from these data are 147 ± 21 nm and 28 ± 4 mV, respectively. Note the close agreement with the properties obtained from the quantification of EFM measurements despite the difference in imaging mechanism and analysis. From the fit we also estimate that the parameter $c \approx 8.8 \pm 0.7$ is close to ideal value $c \approx 8$ for the real tips. The larger value of $c$ in the measurement is consistent with a contribution of cantilever to the electrostatic force.

F. Summary of static measurements and screening at the surface interface junction

Quantification of both the EFM and SSPM results lead to a depletion width of $d_{sc} \approx 200$ nm and a potential of $\phi_{gb} \approx 30$ mV for the grain boundary. This potential value is significantly smaller than expected for typical $SrTiO_3$ interface (~ 0.5 V). The width of the observed grain boundary contrast (~ 700 nm) is larger than the total depletion width obtained from distance dependencies (~ 400 nm). The potential distribution is non-uniform within the broad feature (Figure 4.5), exhibiting a narrow peak with the "correct" shape superimposed on a wide asymmetric region. Most strikingly, the sign of the grain boundary potential feature as observed by SSPM is positive. In the *n*-doped material, this corresponds to the accumulation type grain boundary, which can account for the small value of grain boundary potential, which is limited by the separation between the donor levels and the bottom of conductions band. However, using imaging under applied bias (Section 4.3.2) the grain boundary is unambiguously associated with potential barrier and, therefore, is depletion type. In order to rationalize these observations, we introduce a screening model for the surface interface junction as shown in Figure 4.11a,b. In this model, accumulation of charged adsorbates at the surface-interface junction results in the widening of the grain boundary potential feature and, most notably, in the sign inversion. To verify this assumption, we attempted to remove the screening charges. In the first case, application of lateral bias across the interface results in the high lateral field in the interface region (~$10^7$ V/m). The electrostatic forces induced by the field swipes the screening charges from the surface-interface junction area. After the bias is switched off, the true sign of the grain boundary is observed, as illustrated in Figure 4.11c. This potential distribution is metastable and the accumulation of screening charges reduces the magnitude of the negative feature with subsequent sign reversal. Corresponding relaxation times are large (30 min - several hours) and strongly depend on the surface treatment prior to the experiment. It can be argued that this effect can be attributed to the charge trapping at the interface; however, characteristic retention time is much larger than can be expected for typical interface states. Moreover, such relaxation process would be observed in the impedance spectroscopy data, contrary to the results presented in Section 4.4.

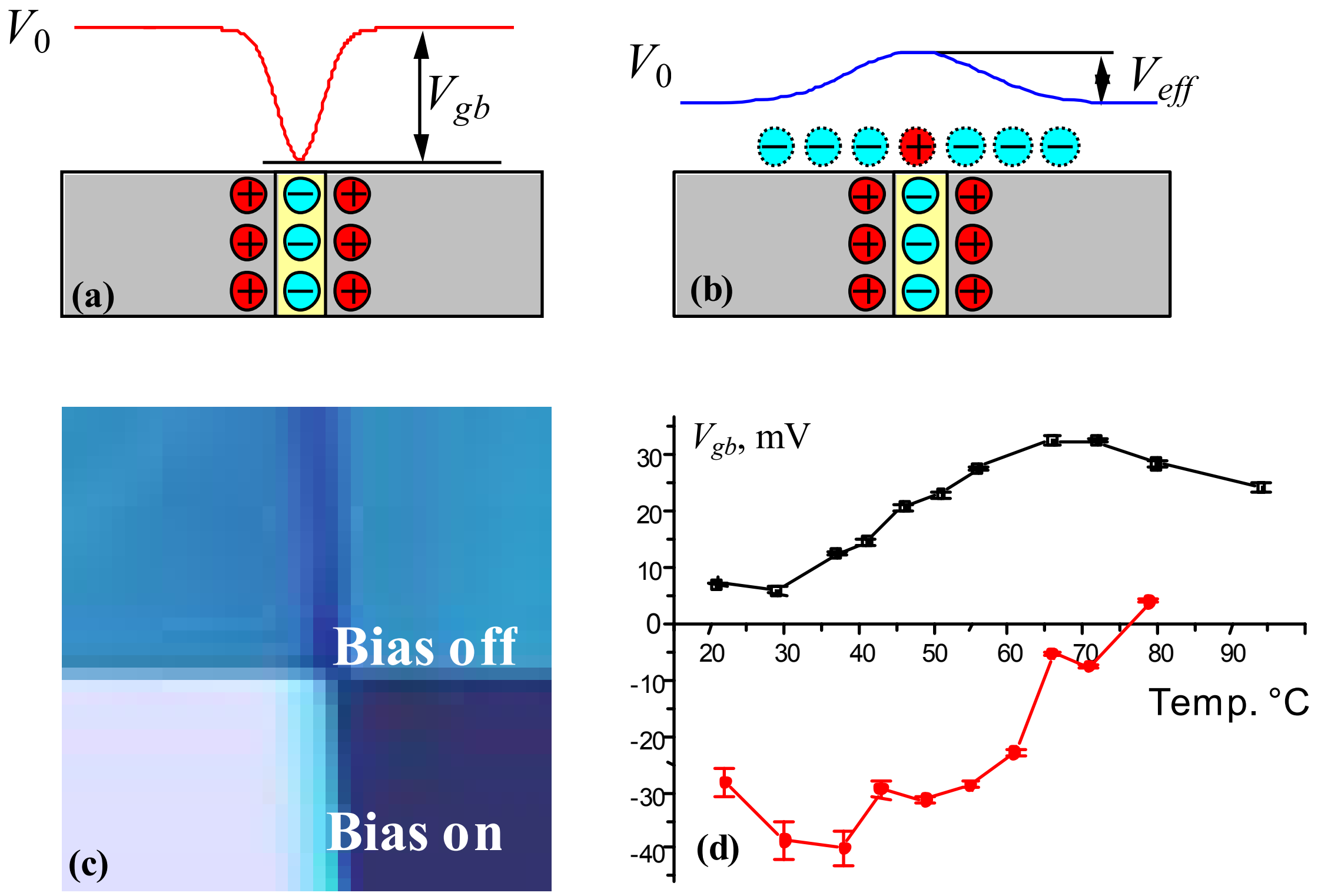

**Figure 4.11.** Potential distribution at pristine (a) and screened (b) grain boundary-surface junction. (c) Relaxation of surface potential at $SrTiO_3$ bicrystal grain boundaryin the turn-off experiment. (d) Temperature dependence of grain boundary potential in polycrystalline barium-strontium titanate (in collaboration with F. Weibel).

An additional approach to studying the screening equilibrium includes temperature variation. In this case, increasing the temperature results in an increase of the apparent interface potential in polycrystalline BST as illustrated in Figure 4.11d. On decreasing the temperature, the sign of the grain boundary potential feature is inverted; the relaxation time to the equilibrium positive value is ~ 30 min.

These results illustrate that in ambient the screening charges preclude reliable measurements of the grain boundary potential barrier and depletion width. Even though the potential on the surface-interface junction, $\varphi_0$, can be determined reliably, the relationship Eq.(4.20) between the potential in the bulk and on the surface does not hold. In fact, even the sign of the potential can be determined erroneously. The depletion width measured by SPM in this case corresponds to the Debye length of the screening charges on the surface and the observed potential profile width (~700 nm) roughly corresponds to the measured surface depletion width ($2d \approx 400$ nm) with the tip smearing effect (~200 nm)[73] taken into account. Despite the fact that potential at the surface-interface junction and grain boundary potential in the bulk are not simply related in air, it can be expected that the analysis procedure developed in this section will be valuable for the quantification of the interface potential data obtained under UHV conditions, under which external screening is negligible.

### 4.3.2. AC and DC Transport across the Grain Boundary

DC transport in STO

To study the dc transport in $SrTiO_3$ grain boundary, Nb-doped $\Sigma 5$ $SrTiO_3$ bicrystals (1 at.%) was soldered by indium to copper contact pads. Surface topography, potential on the grounded and biased surface is illustrated in Figure 4.12. Note that on applying the bias the potential drop develops at the grain boundary, clearly illustrating its resistive nature. To quantitatively study interface transport properties, a slow triangular ramp (0.002 Hz) is applied across the bicrystal during SSPM imaging. Collected was the SSPM image in which each line corresponds to different lateral bias conditions, i.e. potential profile across the interface; the second image stores the actual lateral bias. The voltage characteristics of the interface for different current limiting resistors are shown in Figure 4.13a. The voltage drop across the interface is almost linear for small lateral biases and then saturates, illustrating the decrease of the interface resistance with bias. The maximum observed potential drop across the interface is ~ 1 V; application of higher biases or the use of smaller circuit termination resistors resulted in the current flow to the tip and the destruction of the latter. Using the formalism developed in Chapter 3, the *I-V* curve reconstructed from the potential data is shown in Figure 4.13b. The *I-V* curves for different circuit terminations coincide with each other and coincide with the *I-V* curve obtained by direct two probe measurements between the contacts. These results indicate that the bulk and contact contribution to the material resistance is negligibly small compared to the grain boundary resistance, which is thus the dominant resistive element of the circuit. Therefore, in the subsequent analysis of the ac and dc transport properties of grain boundary bulk and resistive contributions can be neglected.

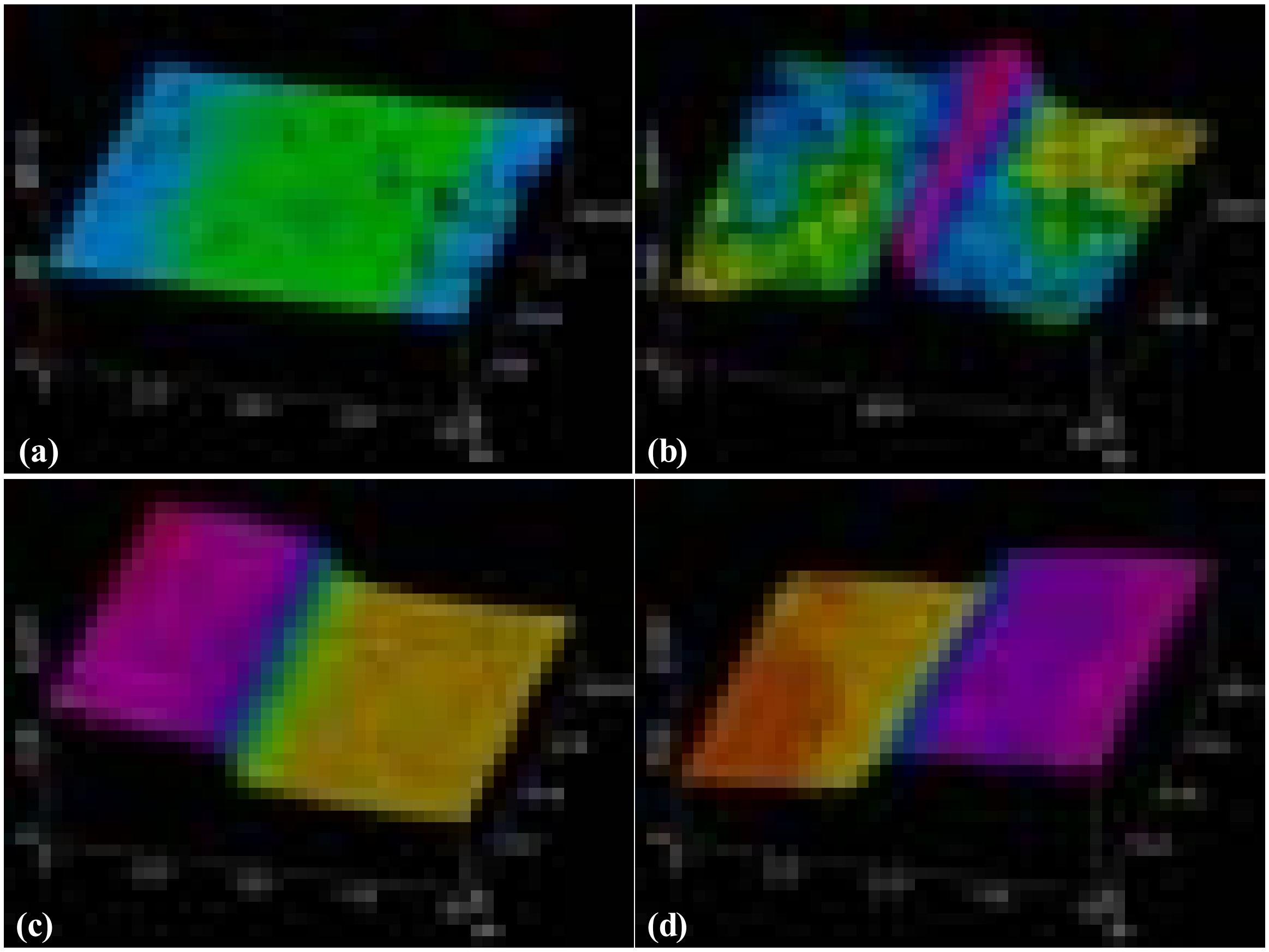

**Figure 4.12.** Surface topography (a), surface potential of the grounded surface (b) (large scan size) and surface potential for forward (c) and reverse (d) bias. Note the difference in scales for (b) and (c), (d).

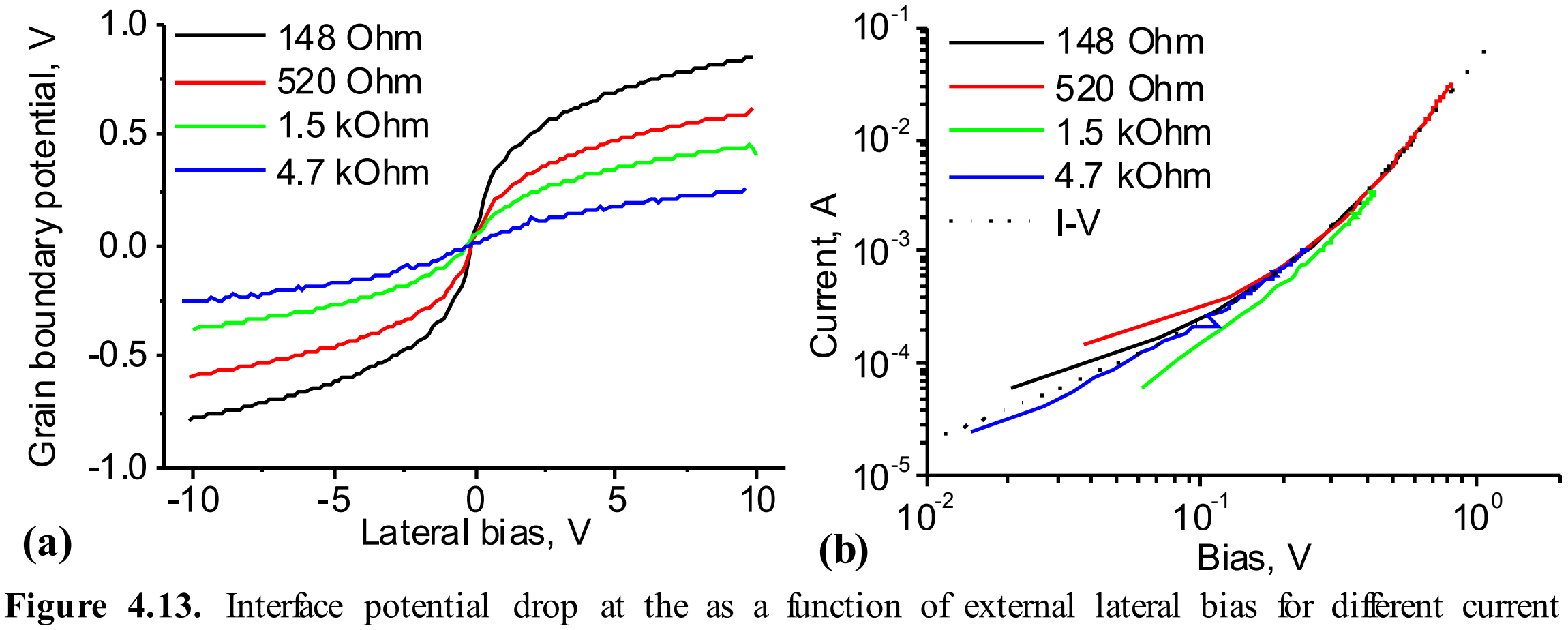

**Figure 4.13.** Interface potential drop at the as a function of external lateral bias for different current limiting resistors (a) and I-V curve reconstruction from SSPM data (b).

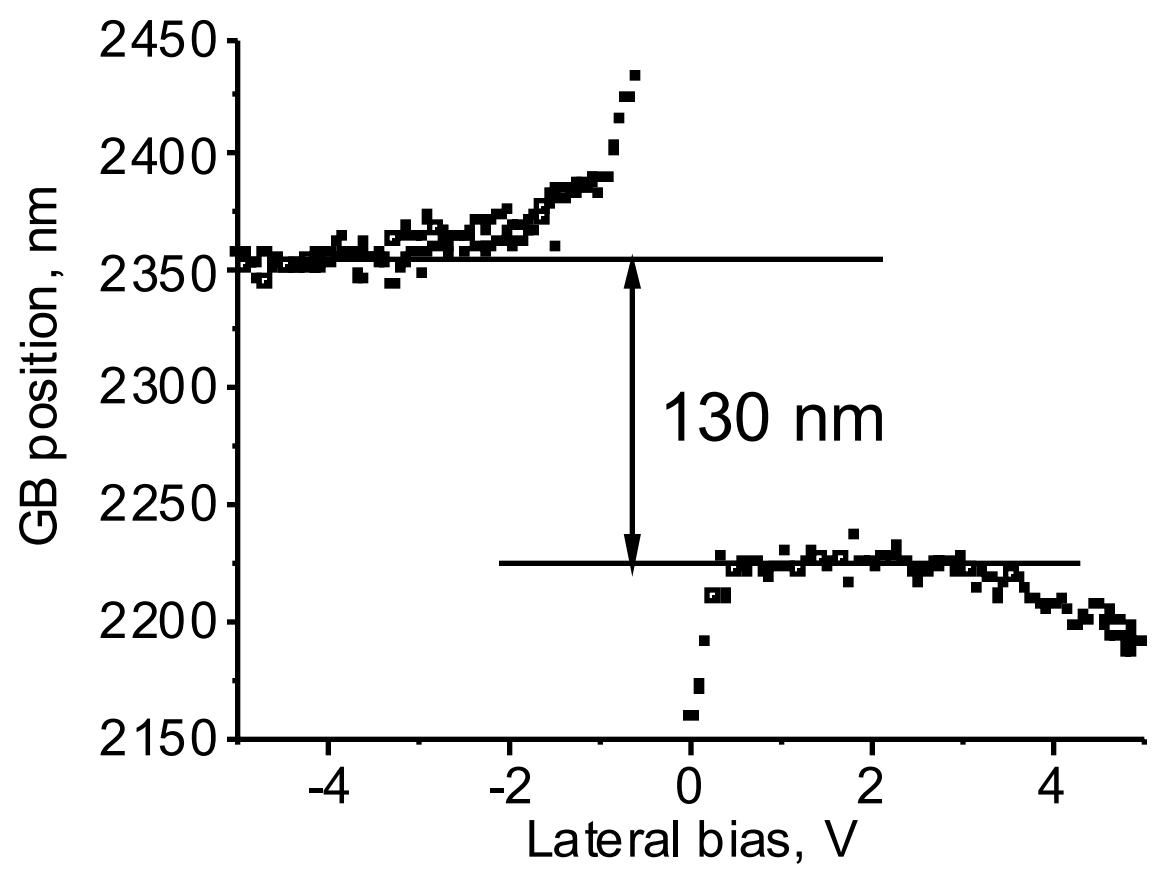

**Figure 4.14.** Bias dependence of grain boundary position.

Additional information on grain boundary properties can be obtained from the structure of potential profile under applied bias. Biasing the grain boundary is accompanied by the displacement of the center of mass of the depletion region. This displacement from negative to positive breakdown voltage is equal to depletion width. To analyze the grain boundary position, 256 potential profiles at different biases were extracted and fitted by the Boltzmann function. Bias sweep was performed several times to ensure the absence of the drift. Shown in Figure 4.14 is the position of the potential profile as a function of external bias and from these data the displacement is estimated as ~130 nm. This value is also very similar to the depletion width determined from the force gradient-distance and force - distance analysis. However, based on data on the surface screening, this value pertains to the characteristic Debye length of the screening charges rather than intrinsic depletion width.

AC transport in $SrTiO_3$

The ac transport properties of the interface were studied using the Scanning Impedance Microscopy. Shown in Figure 4.15a is the frequency dependence of the interface phase angle for different circuit terminations.

At the first step of analysis, phase data was fitted by Eq.(3.23) for frequency independent $R_{gb}$, $C_{gb}$ (Model 1) and the fitting results are summarized in Table 4.III.

Table 4.III.

*Interface properties by Model 1*

| $R$, Ohm | $R_{gb}$, Ohm | $C_{gb}$, $10^{-7}$ F |
|---|---|---|
| 148 | 243.7 ± 3.5 | 2.14 ± 0.04 |
| 520 | 387.8 ± 4.5 | 2.15 ± 0.04 |
| 1480 | 510.1 ± 3.0 | 2.25 ± 0.03 |
| 4700 | 666.3 ± 7.0 | 2.21 ± 0.04 |

Note that the interface capacitance is virtually independent on circuit termination resistance, while interface resistance is smaller for small circuit termination resistances. This behavior is ascribed to the large driving amplitude used in this experiment (1 $V_{pp}$), which results in the decrease of effective interface resistance due to the non-linear *I-V* characteristic of the grain boundary. The effective oscillation amplitude is larger for small $R$, resulting in smaller $R_{gb}$. In the future, care must be taken to ensure imaging in the small signal regime using cantilevers with higher sensitivity. The amplitude ratios calculated using Eq.(3.24) and data in Table 4.II are shown in Figure 4.15b. Note the excellent agreement between the measured and calculated values despite the absence of free parameters.

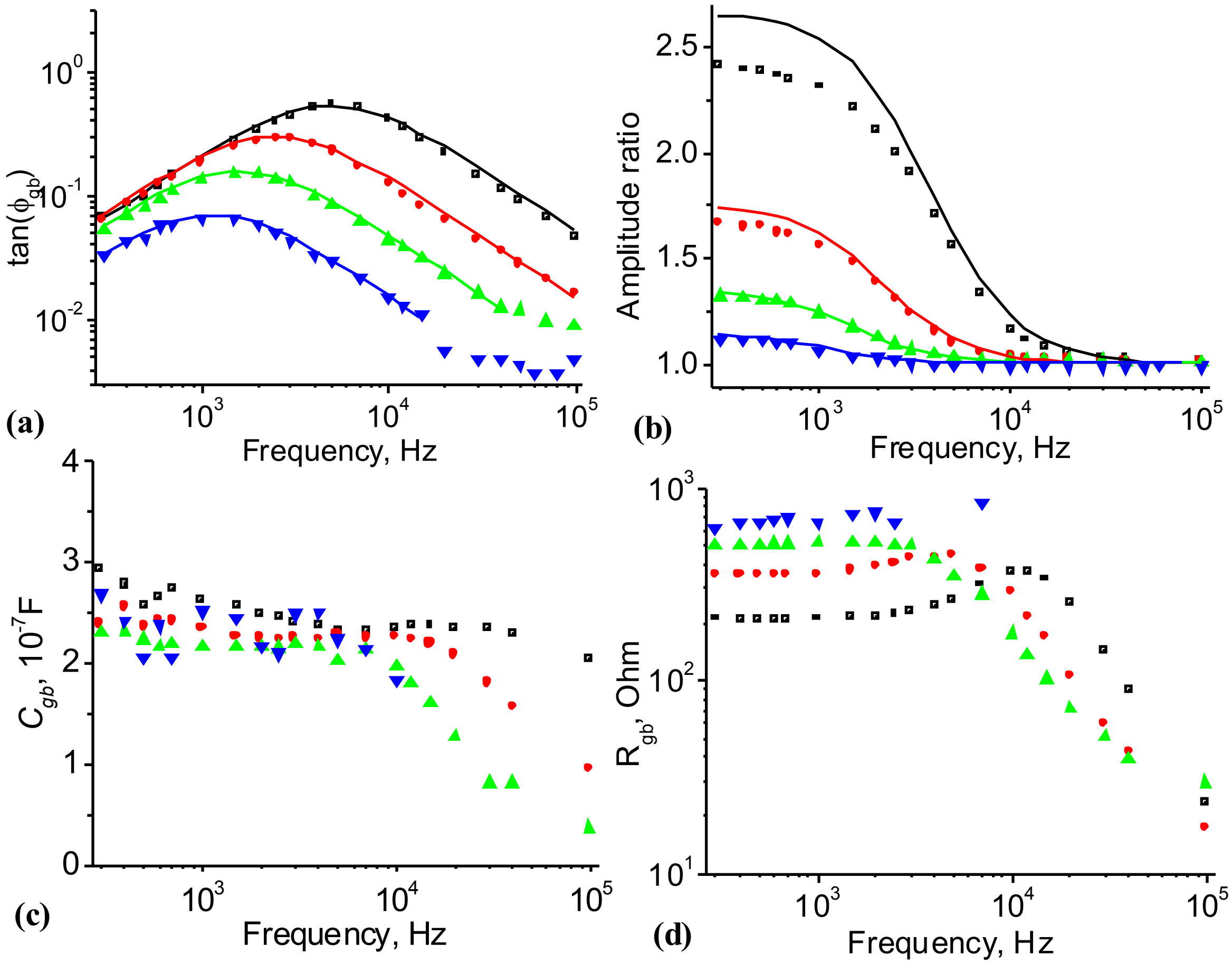

**Figure 4.15.** Frequency dependence of interface phase angle (a) and amplitude ratio (b) in $SrTiO_3$ bicrystal. Solid lines on (a) are fits by Eq.(3.23), on (b) – calculations by Eq.(3.24) using data from Table 4.III. Frequency dependent interface capacitance (c) and resistance (d) calculated from Eqs.(3.23,24). Data are shown for circuit terminations 148 Ohm (■), 520 Ohm (●), 1.48 kOhm (▲), and 4.8 kOhm (▼).

At the second step, we studied the frequency dependence of interface resistance and capacitance. In this case, Eqs.(3.23,24) are solved at each frequency for $R_{gb}$, $C_{gb}$ and the resulting values are plotted in Figure 4.15c,d. Thus determined capacitance values are relatively frequency independent, while the interface resistance rapidly decreases in the high frequency region. This behavior is because the amplitude ratio is close to unity for high frequencies and cannot be reliably determined from SPM data.

The SIM capacitance-voltage curve is illustrated in Figure 4.16. Here, the phase and amplitude data are measured as a function of tip bias at 5 kHz, i.e. in the region where the reliable determination of $R_{gb}$, $C_{gb}$ is possible. The correction for tip bias and surface bias variation is introduced according to Eq.(3.18). The symmetric shape of the amplitude-bias curve in Figure 4.16b with respect to tip bias is indicative of the adequate correction. The grain boundary resistance and capacitance calculated from these data for different circuit terminations are shown in Figure 4.16c,d. Note that the bias dependence of the interface resistance is not sensitive to circuit termination, but is well below the corresponding value determined from the *I-V* curve due to the large driving amplitude effect. The interface capacitance exhibits weak bias dependence, which is attributed to the aforementioned errors in the experimentally measured amplitude ratio. In comparison, shown in Figure 4.16e is the bias dependence of interface capacitance at different frequencies obtained using conventional *C-V* measurements using the same oscillation amplitude. SIM data for $R$ = 520 Ohm and 5 kHz (matched resistances) and *C-V* data for 10 kHz are compared in Figure 4.16f, illustrating the fair agreement between the two.

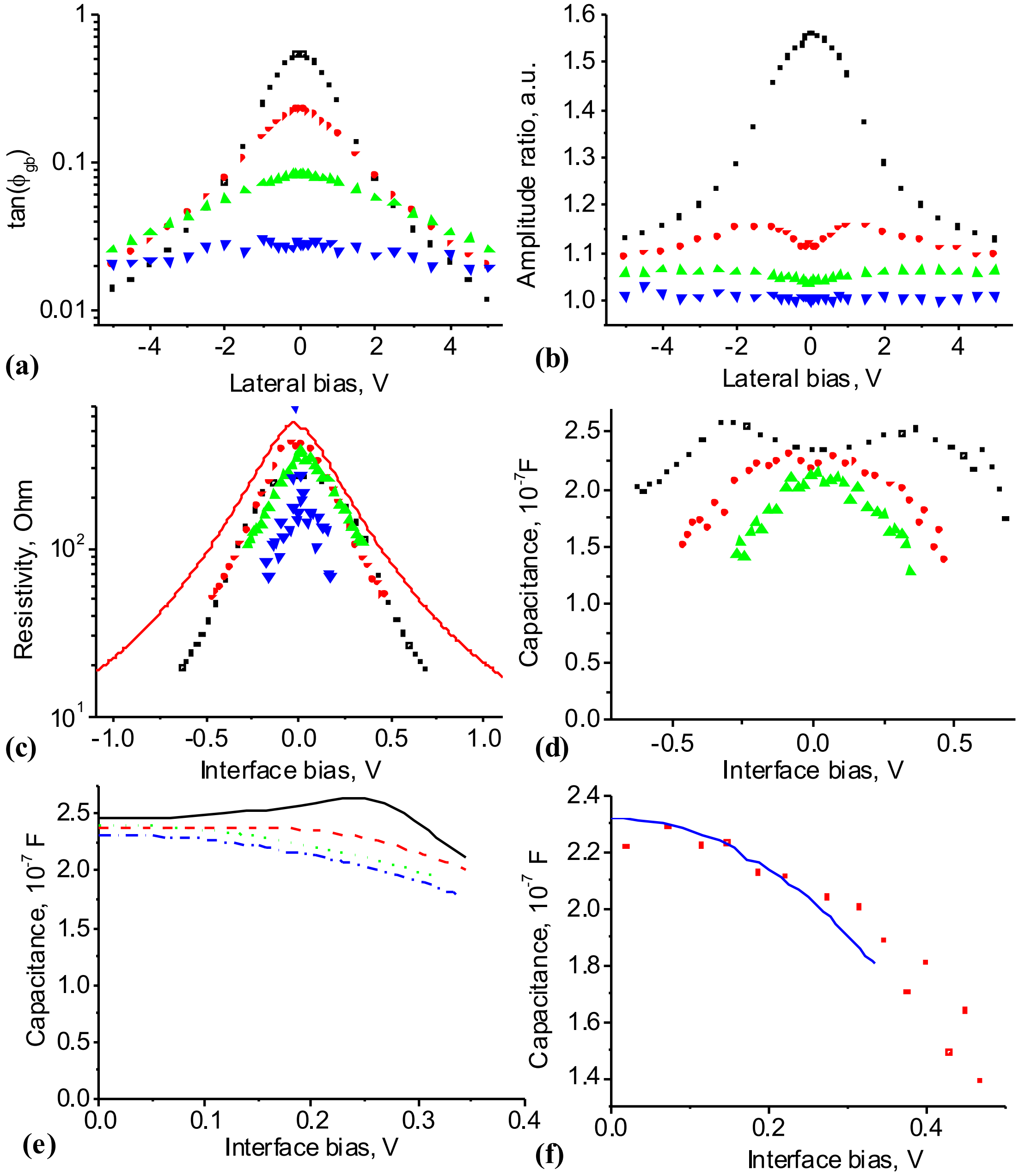

**Figure 4.16.** Lateral bias dependence of interface phase angle (a) and amplitude ratio at 5 kHz. Grain boundary bias dependence of GB resistance (c) and capacitance (d). Data are shown for circuit terminations 148 Ohm (■), 520 Ohm (●), 1.48 kOhm (▲), and 4.8 kOhm (▼). (e) Bias dependence of interface capacitance at 1 kHz ( ———— ), 3 kHz ( – – – ), 10 kHz ( ······ ), and 20 kHz ( – · – · ). (f) Bias dependence of interface capacitance from *C-V* measurements (10 kHz) and SIM measurements (520 Ohm).

<u>Conductive AFM studies of $SrTiO_3$ interface</u>

As illustrated above, the presence of the screening charges at the surface-interface junction limits the applicability of non-contact SPM techniques for analysis of the interface properties. To obtain more quantitative information on the interface properties, we used the variants of conductive AFM as described in Chapter 3. To perform the single-terminal measurements, the microscope was equipped with Ithaco current amplifier (for single-terminal measurements) and additional custom-built current amplifier[74] (for two-terminal measurements) as illustrated in Figure 4.17a,b. The single-terminal current image (both contacts are grounded) along the surface and corresponding current profile are shown in Figure 4.17c,d. The grain boundary region clearly has lower conductivity than the bulk of the crystal. The width of the profile is determined by the depletion width of the grain boundary and imaging conditions such as the rise time of the current amplifier and tip-surface contact area. The former effect can be minimized by scanning at smaller areas; however, for yet unclear reasons (most likely, contamination build-up or tip oxidation) the signal was lost for small scan sizes and small tip velocities. The average tip-surface current far from the grain boundary is strongly tip and bias polarity dependent and for the probe used was 1.34 mA at tip bias of 1 V. Using the resistivity value $\rho$ = 0.017 Ohm·cm, the contact radius can be estimated as 18 nm. The width of the grain boundary feature is ~ 100 nm, the conductivity is suppressed by ~18%.

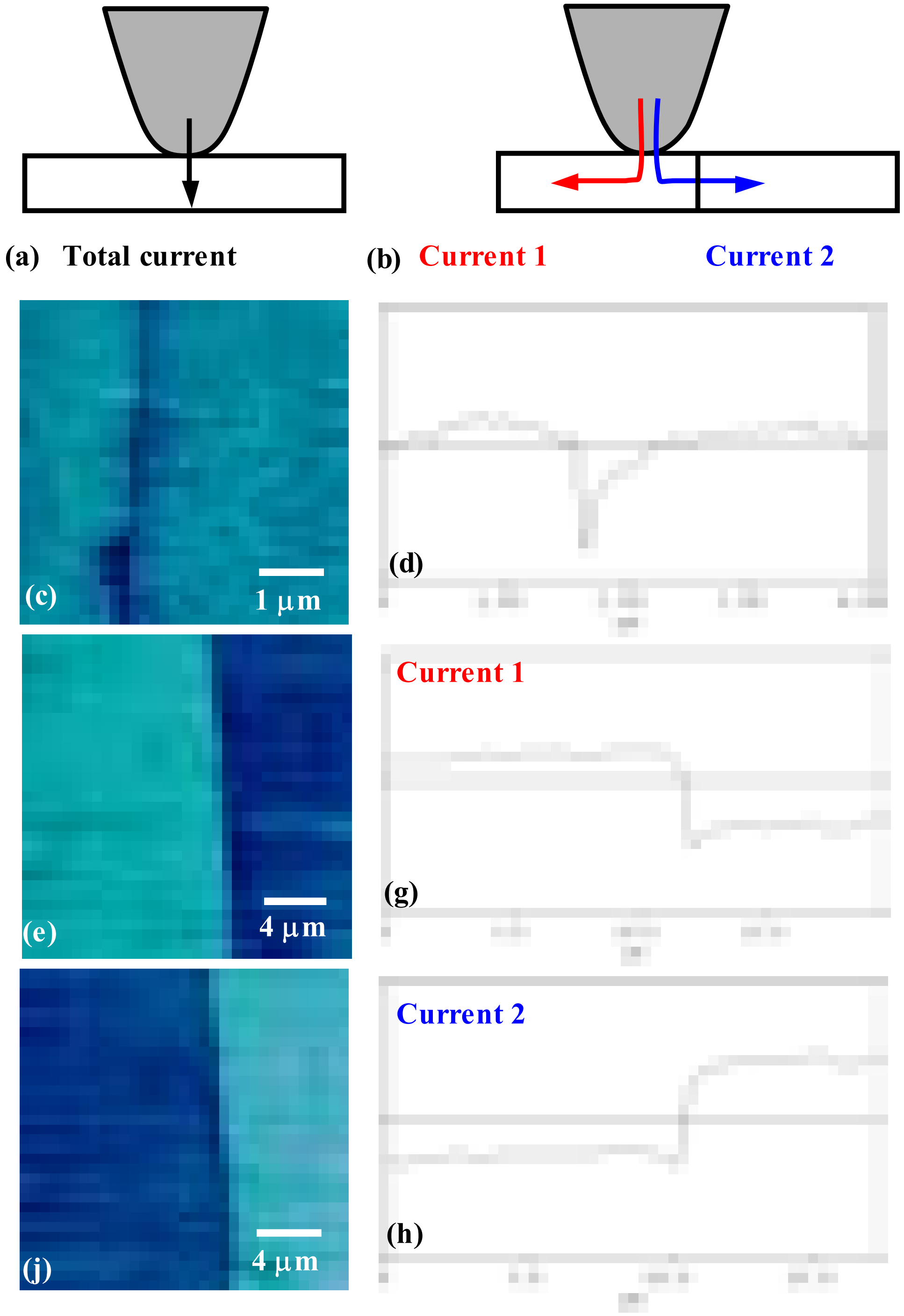

**Figure 4.17.** Schematic of single-terminal (a) and two-terminal (b) current measurements on $SrTiO_3$ bicrystal. Current images and current profiles across the grain boundary in the single terminal (c,d) and two-terminal set-up (e-h).

The observed interface width and magnitude are weighted average of interface and the bulk due to the instrumental broadening. If the conductance in the GB region is much smaller than in the bulk and the interface current is zero, the real depletion width can be estimated as 18 nm. This is close to the value estimated from capacitance measurements ($d$ = 22 nm).

Illustrated in Figure 4.17e-h are the results of two terminal measurements of the same interface. Note the formation of the sharp current step when the tip traverses the grain boundary. The magnitude of the potential drop is determined by the voltage divider ratio formed by the tip-surface contact resistance, grain boundary resistance and circuit termination resistance, $\frac{I_{right}}{I_{left}} = \frac{R_{ts} + R}{R_{ts} + R_{gb} + R}$, where $R_{ts}$ = 750 Ohm is tip-surface resistance, $R_{gb}$ is grain boundary resistance and $R$ = 10 Ohm is circuit termination resistance. The relative current drop at the interface agrees well with that expected from the ratio of tip-surface contact resistance and the total resistance (0.81 for current 1, 0.73 for current 2, 0.57 expected for $R_{gb}$ = 600 Ohm).

These results illustrate huge potential of the c-AFM for the interface characterization. Here, the surface charge effect on the measurements is minimal; therefore, interface properties can be characterized reliably. However, these measurements are complicated by the nature of the tip-surface contact, which, until now, limited the number of successful experiments. Further research in this direction using both dc and ac probes is under way.

### 4.3.3. Temperature Dependence of Interface Transport Properties

The crystal was soldered by indium to copper contact pads; additional indium contacts were deposited to perform 4 probe *I-V* measurements. AC transport properties were measured by impedance spectroscopy (HP4282A LCR meter) in the frequency range 20 Hz-1 MHz and modulation signal of 20 mV. Measurements were performed in the temperature range 123-373 K (Delta temperature chamber). Four probe dc *I-V* measurements were performed in the temperature range 78-300 K using home built cryogenic system.

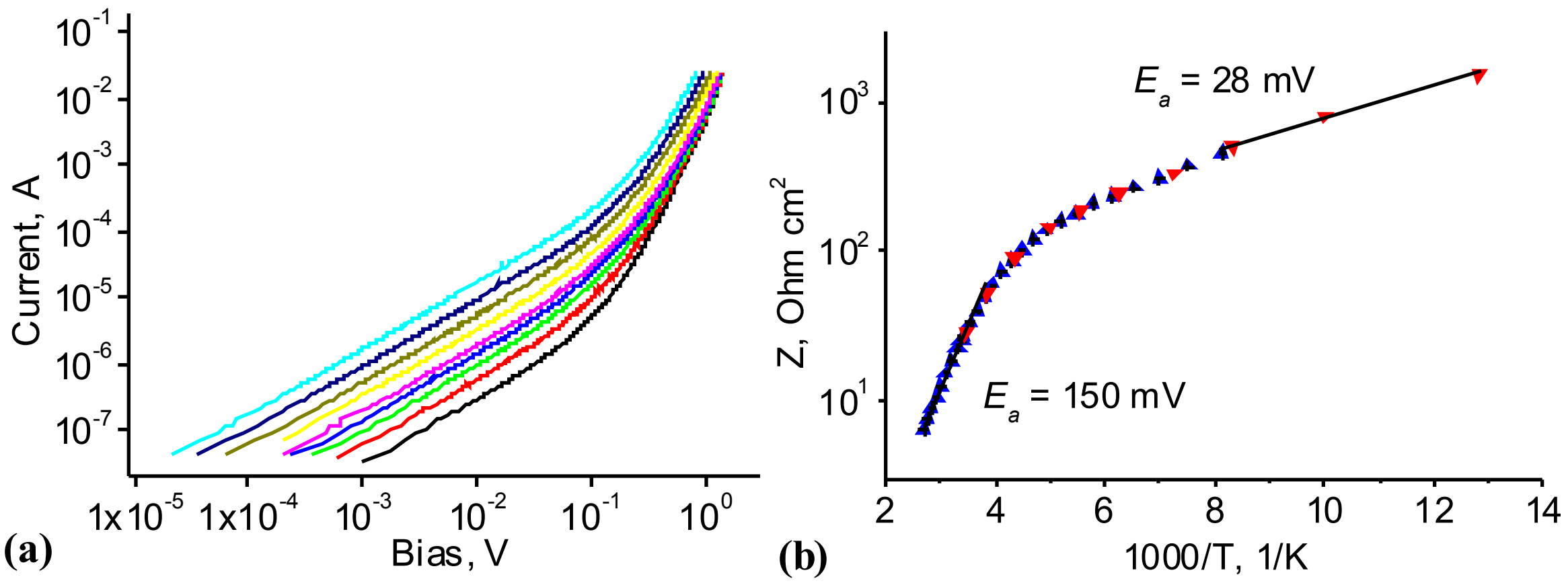

**Figure 4.18.** Temperature dependent *I-V* characteristics of Σ5 grain boundary in $SrTiO_3$ bicrystal (a) and temperature dependence of zero-bias resistance (b).

Shown in Figure 4.18a are the *I-V* curves obtained at different temperatures (lower curve corresponds to 78 K, upper curve – to 300 K). Note that for low biases the current is linear in bias, while for higher biases (> 50-100 mV) the onset of non-linear behavior is observed. The non-linearity coefficient $\alpha$ in $I \sim V^{\alpha}$ is relatively small, $\alpha$ = 2.3-3.5 depending on temperature. This value is typical for $SrTiO_3$, but is significantly smaller than for oxide varistor materials such as ZnO, for which $\alpha$ can be as high as 40. Interestingly, this small non-linearity value suggests significant contribution of space-charge current ($\alpha$ = 2), since thermionic emission and diffusion models predict significantly higher values of nonlinearity coefficient.

The attempts to deconvolute the interface density of states from the variable temperature *I-V* data using the formalism developed by Pike and Seager[35] were unsuccessful. As suggested by the theoretical arguments developed in Section 4.2, the transport at the $SrTiO_3$ interfaces is limited by the low carrier mobility and does not belong to the pure thermionic emission or diffusion case at high biases. Therefore, *I-V* curve reconstruction is impossible without more detailed numerical analysis.

The low-bias temperature dependence of resistance for bicrystal interface determined from the linear part of the *I-V* curves and impedance data is shown in Figure 4.18b. Note that the activation energy $E_a = k/q\, d\ln R_{gb} / d(1/T)$ for conductivity has two well defined linear regimes $E_a$ = 28 mV for $T < 220$ K and $E_a$ = 150 mV for $T > 220$ K. In both cases, $E_a$ is much smaller than expected potential barrier height $\varphi_{gb}$ ~0.5-1 eV. This is because activation energy for conduction incorporates the temperature dependence for potential barrier, $E_a \approx T\, d\varphi_{gb}/dT - \varphi_{gb}$. Since $\varphi_{gb} \sim 1/\varepsilon \sim T - \theta$, and $\theta$ = 28 K for $SrTiO_3$, $E_a << \varphi_{gb}$ for $T > \theta$. Based on the theoretical estimates, it is conjectured that the change in activation energy corresponds to the crossover between thermionic and diffusive transport. To verify this assumption, the interface parameters were estimated in the linear dielectric approximation. From room temperature impedance spectroscopy the interface capacitance is $C_{gb} = 5.7\cdot10^{-2}$ F/m$^2$, from which depletion width is estimated as 23 nm for $\kappa$ = 300. From the interface resistance the potential barrier is calculated as ~570 mV for both thermionic [Eq.(4.9)] and diffusion [Eq.(4.11)] models. The thermal velocity is $23.5\cdot10^3$ m/s and the mobility at room temperature is 5.3 cm$^2$/Vs. The effective field at the interface is $E_0 = 4.96\cdot10^7$ V/m. Therefore, $\mu E_0 = 26.3\cdot10^3$ m/s is close to thermal velocity at room temperature and both diffusion and thermionic mechanisms are active. For lower temperatures thermal velocity decreases, while mobility increases resulting in the onset of purely thermionic behavior. Therefore, it is feasible that the low temperature conductivity region corresponds to purely thermionic case, while for $T > 220$ K diffusion dominates.

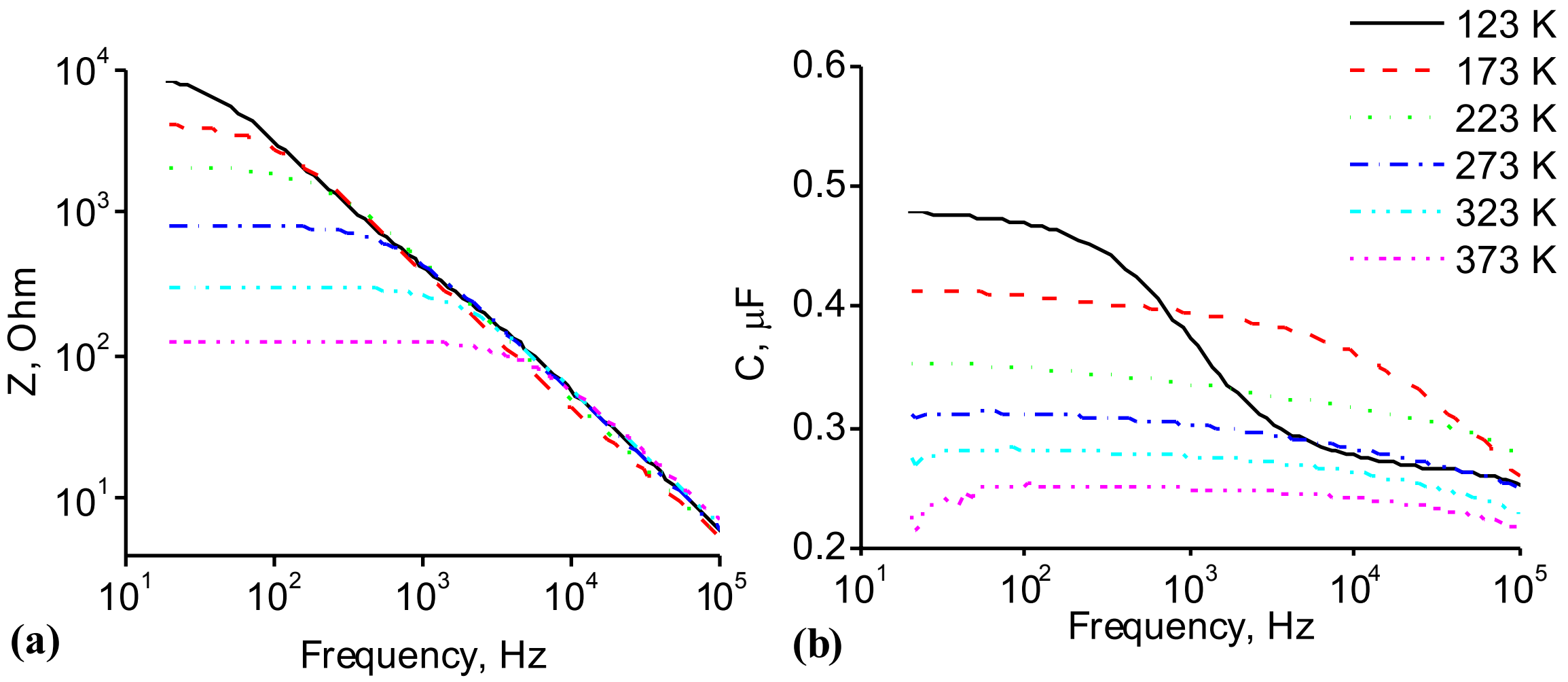

**Figure 4.19.** Frequency dependence of interface impedance (a) and interface capacitance (b) at different temperatures.

Variable temperature impedance spectra were interpreted in terms of single *R-C* element corresponding to grain boundary for $T > 150$ K (single undistorted arc on the Cole-Cole plot). The impedance amplitude was fitted by $|Z| = R_{gb} \Big/ \sqrt{1 + \left(\omega R_{gb} C_{gb}\right)^2}$, yielding the interface resistance and capacitance as shown in Figure 4.19a. Alternatively, frequency dependence of interface capacitance was calculated as $C_{gb}(\omega) = -\sin(\theta)/\omega|Z|$, where $\theta$ is impedance phase angle. For lower temperature, an additional relaxation process is observed as seen in Figure 4.19b. It is tempting to associate this relaxation process with the formation of non-cubic phase below 132 K as reported by Waser,[65] however, more extensive studies of the low temperature transport (both *I-V* and *C-V*) are clearly required. The resistivity for crystal bulk was < 1 Ohm.

Temperature dependence of interface potential $\varphi_{gb}$ was calculated from thermionic emission model. Shown in Figure 4.20a is interface capacitance and grain boundary potential from impedance spectroscopy. Noteworthy is that temperature dependence of capacitance is significantly weaker than that of a bulk dielectric constant, $\kappa = C/(T - 28)$, suggesting that dielectric constant is reduced in the vicinity of the interface compared to the bulk. At the same time, $\varphi_{gb}$ is almost linear in temperature and relatively insensitive to the details of the calculations. It might be argued that mechanism for reduction of interface capacitance relative to the bulk dielectric constant for lower temperatures the band bending at the interface decreases, resulting in the filling of the empty interface states and increase of the depletion width. However, the effective interface charge also decreases with the temperature as illustrated in Figure 4.20b, while reverse would be expected for the interface state filling.

### 4.3.4. Non-linear Dielectric Behavior at STO Grain Boundaries

One of the key assumptions in the conventional semiconductor theory is field-independent dielectric constant. As suggested by the theoretical arguments in Section 4.2 and the experimental results, this is not the case for $SrTiO_3$. In the material with field dependent dielectric constant, the usual Poisson equation is inapplicable for the calculation of field and potential in the grain boundary region. Rather, the solution of Maxwell equations is required. For 1D non-linear dielectric, one writes

$$\partial D/\partial z = qN_d, \tag{4.39}$$

where $D$ is the displacement vector. The displacement to the right of grain boundary, $0 < z < d$, is

$$D = \sigma/2(1 - z/d), \tag{4.40}$$

and $D = 0$ for $z > d$. By definition, $D = \varepsilon_0 E + P$, where $E$ is electric field, $E = \partial\varphi/\partial z$, and $P$ is polarization. In the non-linear dielectric the relationship between field and polarization is

$$E = AP + BP^3, \tag{4.41}$$

where $A = A_0(T - \theta)$, $A_0$ and $B$ are temperature independent constants, $T$ is temperature and $\theta$ is Curie temperature. For linear dielectric material, $A = 1/\kappa\varepsilon_0$ and $B = 0$. From Eqs.(4.40,41) polarization and electric field are calculated as a function of distance from the interface. Interface potential is

$$\varphi_{gb} = \int_0^d E(z)dz. \tag{4.42}$$

Local dielectric constant is

$$\kappa(E) = 1 + \frac{1}{\varepsilon_0}\frac{\partial P}{\partial E}, \tag{4.43}$$

and interface capacitance is

$$\varepsilon_0 C_{gb}^{-1} = \int_0^d \kappa^{-1}(z)dz. \tag{4.44}$$

In the general case, Eqs.(4.42-44) must be solved numerically. However, for weakly non-linear dielectric material such as $SrTiO_3$ at room temperature, approximate solutions can be found using asymptotic expansion for polarization,[75]

$$P = P_0 + \varepsilon_0 B P_1 + \dots, \qquad (4.45)$$

where $P_0$ is polarization for linear dielectric and $P_1$ is first order correction due to non-linearity. Using $1+\varepsilon_0 A \approx 1$, polarization distribution in the vicinity of the interface is $P = D - \varepsilon_0 B D^3$. The electric field is $E = AD + BD^3$ and interface potential is

$$\varphi_{gb} = \frac{A\sigma d}{4} + \frac{B\sigma^3 d}{32}. \qquad (4.46)$$

The reciprocal interface capacitance from Eq.(4.44) is $C_{gb}^{-1} = 2Ad + B\sigma^2 d/2$ and interface capacitance is

$$C_{gb} = \frac{1}{2Ad} - \frac{B\sigma^2}{8A^2 d}. \qquad (4.47)$$

The effective interface charge is

$$\sigma_{eff} = \sigma - \frac{B\sigma^3}{8A}. \qquad (4.48)$$

As follows from Eqs.(4.46-48), the dielectric non-linearity results in the suppression of dielectric constant in the vicinity of the interface, increase of potential barrier height and reduction of interface capacitance. The effective interface charge is reduced compared to linear case. Since coefficient $A = A_0(T-\theta)$ is linear in temperature, deviations from linear dielectric behavior become more pronounced for low temperatures.

Eqs.(4.46-48) predict that interface charge density, $\sigma$, depletion width, $d$, and non-linearity coefficient $B$ can be determined from the temperature dependence of effective interface charge, $\sigma_{eff} = 8C_{gb}\varphi_{gb}$, and interface potential. Fitting the temperature dependence of effective interface charge by $y = a + b/(T-28)$ yields $\sigma$= 0.29 C/m$^2$ and $B$ = 4.37·10$^9$ Vm$^5$/C$^3$. The reciprocal interface capacitance was fitted by linear function $y = a + bx$. The slope yields depletion width as $d = b/2A_0$, $d$ = 13.6 nm. The intersect,

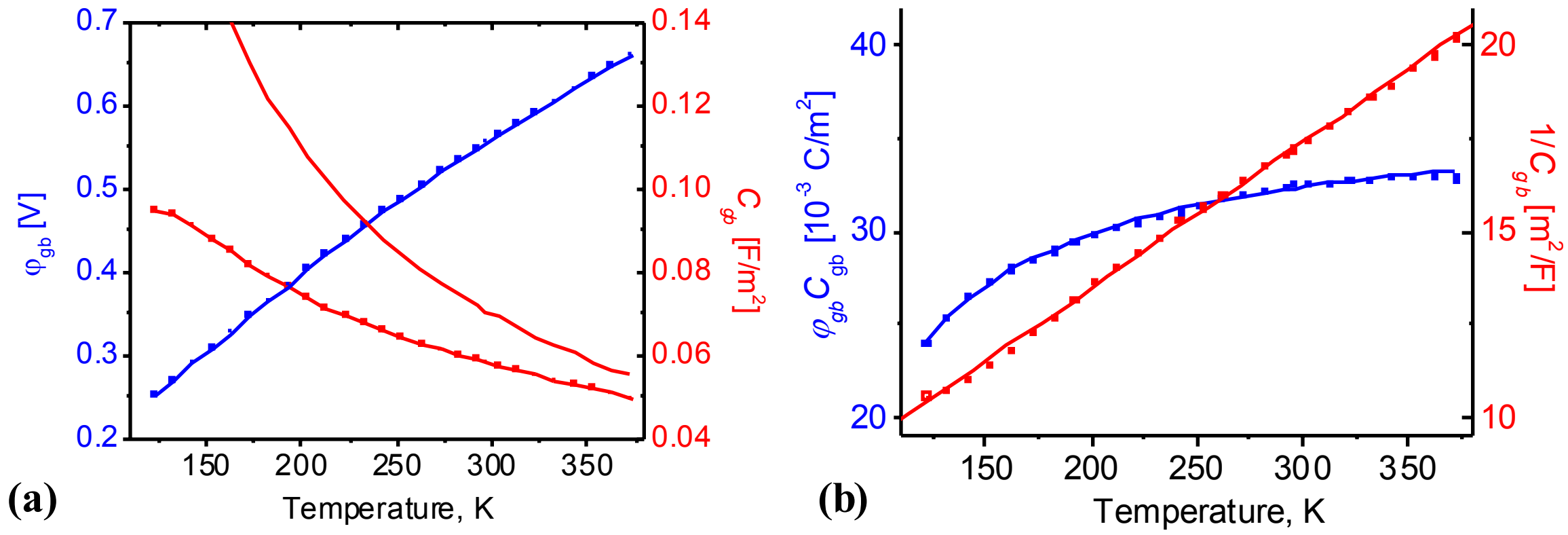

**Figure 4.20.** Temperature dependence of interface potential barrier and interface capacitance compared to the temperature dependence of dielectric constant (a). Temperature dependence of interface charge and reciprocal capacitance. Solid lines in (b) are fits by Eqs.(4.46-48).

$a = B\sigma^2 d/2 - A_0\theta$, yields $B$ = 11.3·10$^9$ Vm$^5$/C$^3$. Finally, fitting the temperature dependence of interface potential yields $d$ = 19.4 nm and $B$ = 4.85·10$^9$ Vm$^5$/C$^3$. In comparison, dielectric measurements at low temperatures yield $B$ = 4-8 Vm$^5$/C$^3$ [Ref.62]. The fitting results are summarized in Table 4.IV.

Table 4.IV

*Interface properties of non-linear $SrTiO_3$ interface*

| Interface properties | $\sigma$, C/m2 | $d$, nm | $B$, 10$^9$ Vm$^5$/C$^3$ |
|---|---|---|---|
| $\sigma_{eff}$ | 0.29 | | 4.37 |
| $1/C_{gb}$ | | 13.6 | 11.3 |
| $\varphi_{gb}$ | | 19.4 | 4.85 |
| Reference | | | ~4 - 8 |

From these values, the donor concentration $N_d = \sigma/2qd$ is $N_d$ = 6.54·10$^{25}$ m$^{-3}$ for $d$ = 14 nm. This value is very close to carrier concentration for Nb doped $SrTiO_3$ with resistivity ρ = 0.017 Ohm·cm [Ref.60], and mobility $\mu$ = 5.3 cm$^2$/Vs, $N_d$ = 6.94·10$^{25}$ m$^{-3}$. Noteworthy is that activation energy in thermionic emission regime $E_a = \theta\sigma^2/8q\varepsilon_0 N_d C$ is estimated as 41 meV in a close agreement with experimental value $E_a$ = 28 meV.

### 4.4. Transport in Polycrystalline Oxide Materials

After an understanding of grain boundary phenomena in $SrTiO_3$ bicrystal samples by SPM was achieved, this approach was extended to polycrystalline ceramic materials. Depending on the nature of the material, local potential and transport studies can be complemented by high-resolution magnetic (in $La_xSr_{1-x}MnO_3$ material with colossal magnetoresistance)[76] and ferroelectric domain imaging (described in more details in Chapter 6). Summarized below are the results of the SPM studies of several polycrystalline oxide systems.

#### 4.4.1. Grain Boundary Limited Transport in ZnO

Polycrystalline ZnO is widely used as a prominent electroceramic material due to non-linear current voltage characteristics, which enable its application in varistors and surge protectors.[1,2] This material was extensively studied by conventional impedance, *I-V* and *C-V* techniques on the bulk samples and microimpedance measurements using fabricated contact arrays on sample surface. However, the former access only averaged interface properties in the polycrystalline samples, while the latter are typically associated with the non-uniform current distribution in the sample, thus hindering the quantitative interpretation of the *I-V* and impedance data. Here, we illustrate the applicability and limitations of SPM transport measurements for the analysis of transport phenomena in such materials. In the three-terminal set-up, SPM tip acts a moving voltage probe, while the current is induced by the macroscopic external electrodes. Therefore, current in the sample is macroscopically uniform precluding current crowding effects, while potential drop at each interface can be quantified.

Shown in Figure 4.21 are surface topography and surface potential on a polycrystalline ZnO surface under different bias conditions. The topographic image exhibits a number of spots due to contaminants and depressions due to inter- and intragranular pores. The surface potential of the grounded ZnO surface is essentially uniform over ZnO grains and exhibits well-defined contrast due to the chemical inhomogeneity of the surface. Small potential depressions in the center of the image are associated with second phase (Bi-based spinel phase) inclusions that can be clearly observed under the optical microscopy. On application of 5 V lateral bias potential drops at the grain boundaries become evident (Figure 4.21c). The contrast inverts on application of bias of opposite polarity (Figure 4.21d). Note that the potential depression on the second phase inclusion is independent of applied bias. Some grain boundaries demonstrated rectifying behavior. Shown in Figure 4.22 is surface topography and surface potential of the central part of image in Figure 4.21. Grain boundaries are now seen as depressions on topographic image, probably due to the selective polishing. The surface potential of the unbiased surface exhibits potential depressions at grain boundaries (Figure 4.22b). Unlike $SrTiO_3$ bicrystal, the origins of the interface potential variations are traced to the presence of a second Bi-rich phase as shown by SEM imaging in the backscattering regime. Application of a direct bias clearly delineates three large grains within the image (Figure 4.22c), and reversal of the bias indicates that at least two of the grain boundaries clearly exhibit rectifying behavior (Figure 4.22d).

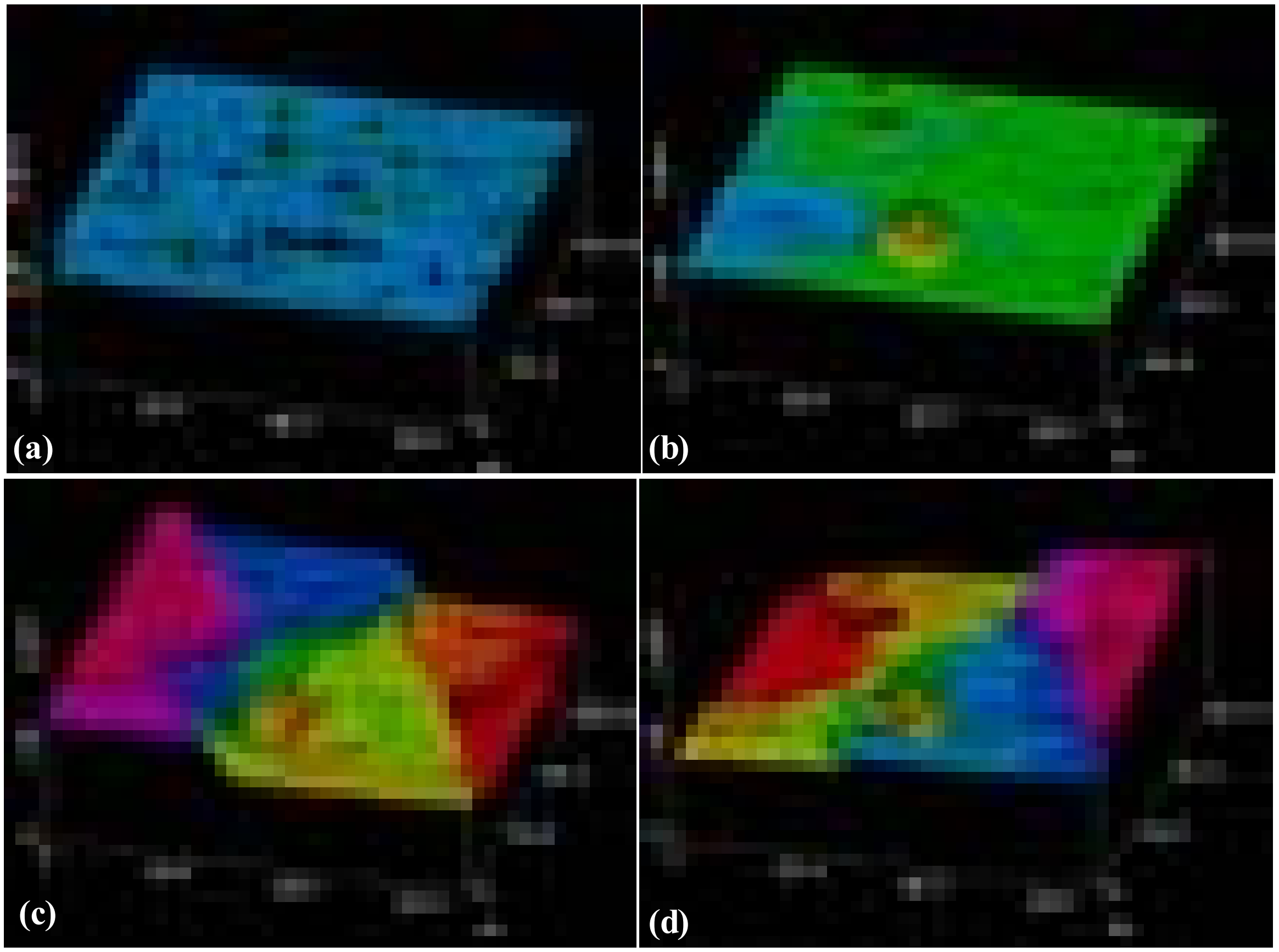

**Figure 4.21.** Surface topography (a) and surface potential for grounded (b), direct (c) and reverse (d) biased ZnO varistor surface.

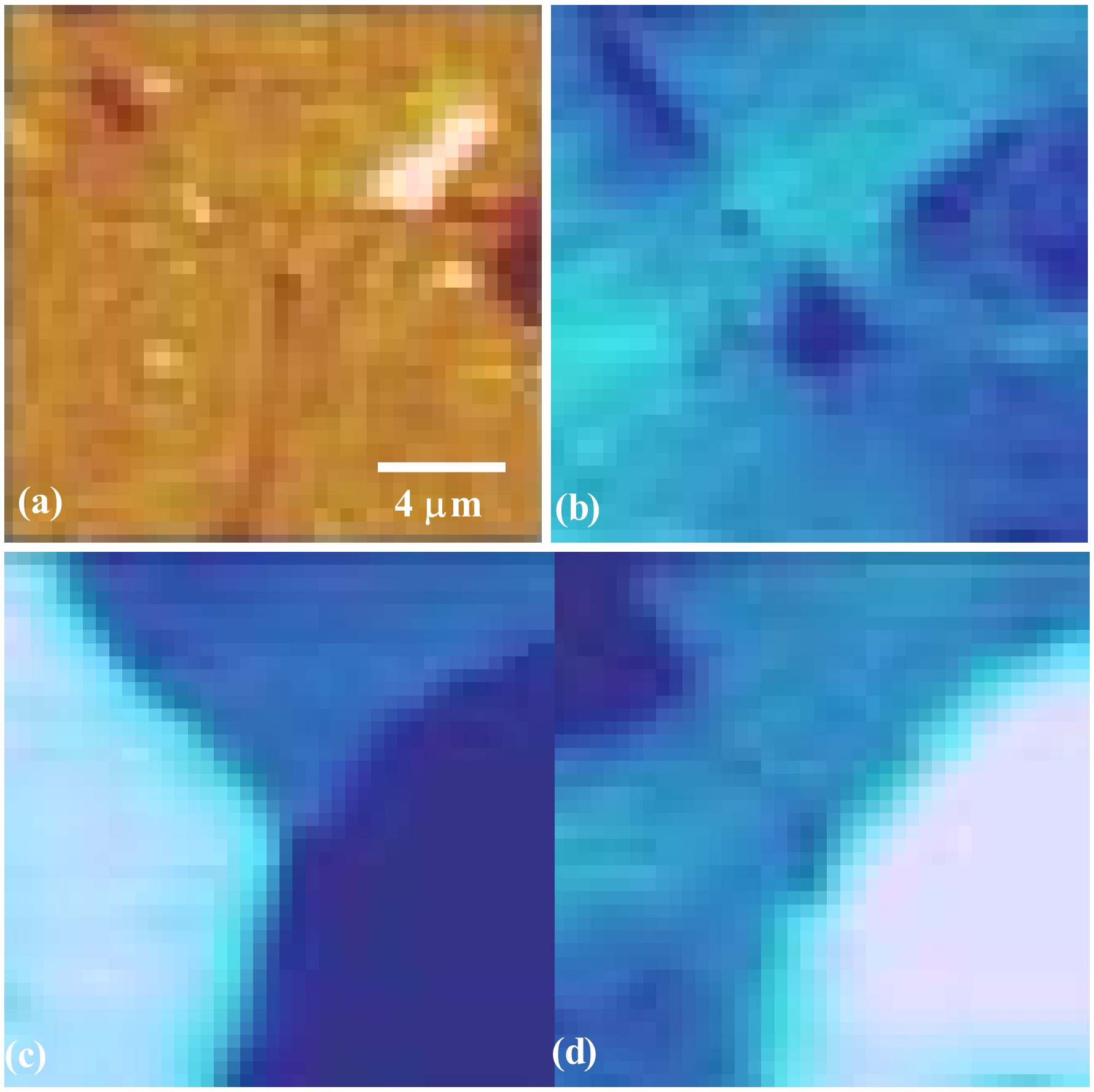

**Figure 4.22.** Surface topography (a) and surface potential for grounded (b), forward (c) and reverse (d) biased ZnO varistor surface.

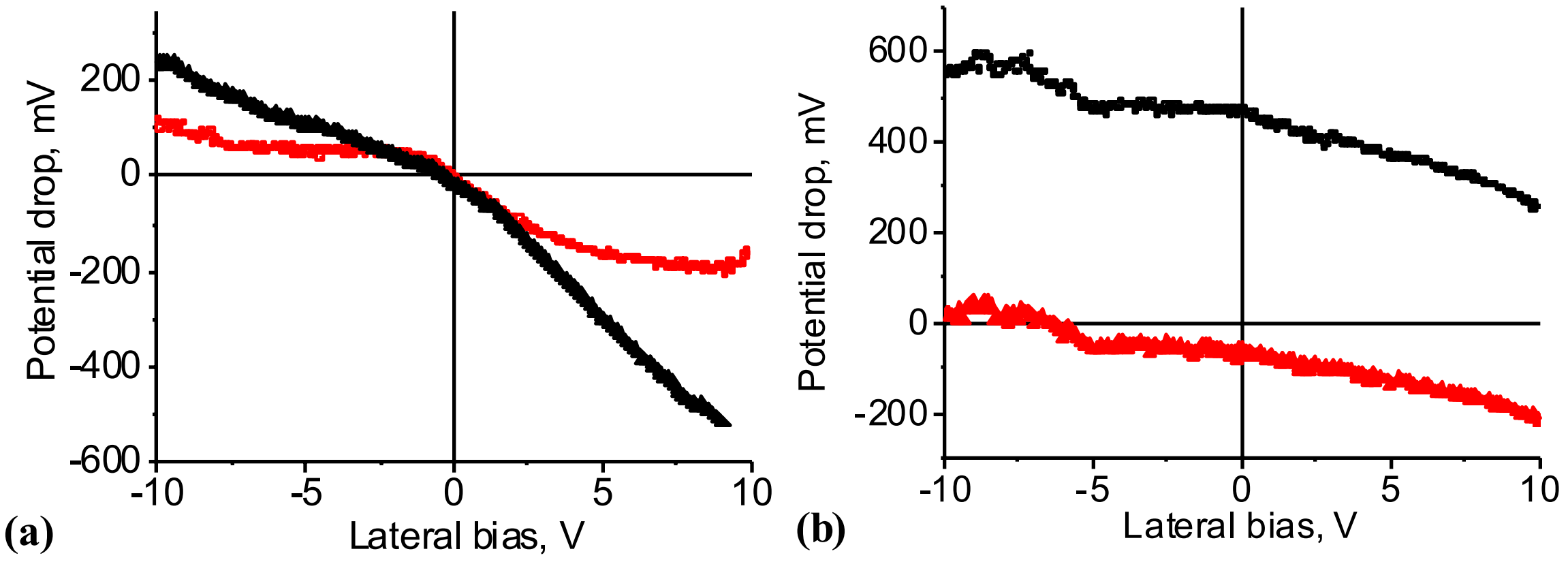

**Figure 4.23.** Voltage characteristics of two individual grain boundaries in a ZnO varistor (a) exhibiting asymmetric ohmic and non-ohmic behavior (a). Voltage characteristic of second-phase inclusion (b). Shown are potential drops on the left (black) and the right (red) boundaries.

Further indication of this behavior is obtained from the potential drop-bias voltage behavior, as shown in Figure 4.23a. Grain boundaries are non-uniform and exhibit both ohmic and non-ohmic behavior for forward and reverse biases. In both case, the potential drop at the interface is zero for a grounded surface, while the potential drop at a ZnO-second phase interfaces is non-zero and depends only slightly on external bias (Figure 4.23b).

SSPM of polycrystalline samples can be further extended to reconstruct the current distribution through a complex microstructure. The ohmic behavior of the grain gives rise to a potential gradient in the current direction. Hence, a general pattern of current distribution can be obtained by differentiating the surface potential map along *x*- and *y*-directions and averaging the derivative maps over the individual grains. Corresponding derivative patterns are shown in Figure 4.24. Average *x*- and *y*-derivatives of potential are proportional to the *x*- and *y*- components of the current in the grain and hence the direction of the current can be determined. The reconstructed potential distribution is shown in Figure 4.25. It should be noted that the magnitude of the current couldn't be obtained solely from the SPM image; additional information on the total current through the sample, specific conductivity of the grains or multiple SPM measurements with different shunting resistors would be required.

Scanning Impedance Microscopy of polycrystalline ZnO has shown that even for the highest experimentally accessible frequencies (100 kHz as limited by the lock-in amplifier used) the interfaces are in the low-frequency regime. Both amplitude and phase images exhibit abrupt changes of signal at the interfaces (not shown) similar to the dc potential. Further progress can be achieved by higher-frequency SIM measurements; in particular, such measurements can establish the origins of the frequency dispersion of interface capacitance (i.e. whether it is generic for individual grain boundaries or collective phenomena).

Conductive AFM imaging in single-terminal and two-terminal configurations on ZnO is illustrated in Figure 4.26. The current image on the etched surface closely resembles the sample topography, probably due to the topographic artifacts (tip does not penetrate the grooves between the protrusions on the surface). In the two-terminal configuration, the grain contrast is determined by the separation from the macroscopic electrodes, illustrating the potential for the spatially resolved characterization.

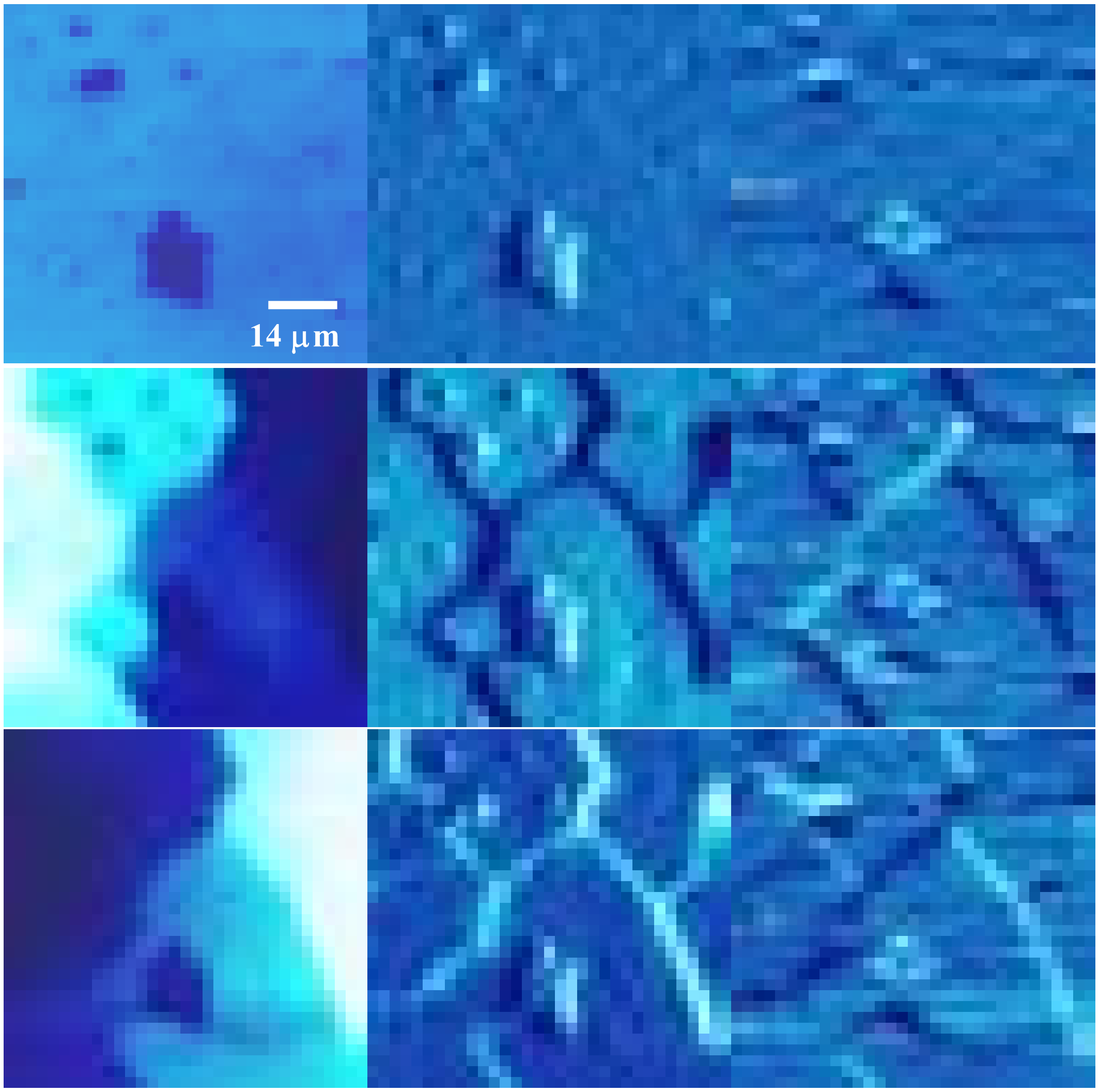

**Figure 4.24.** Surface potential (left column), *x* derivative (central column) and *y* derivative (right column) for grounded (1st row), forward (2nd row) and reverse (3d row)-ZnO varistor.

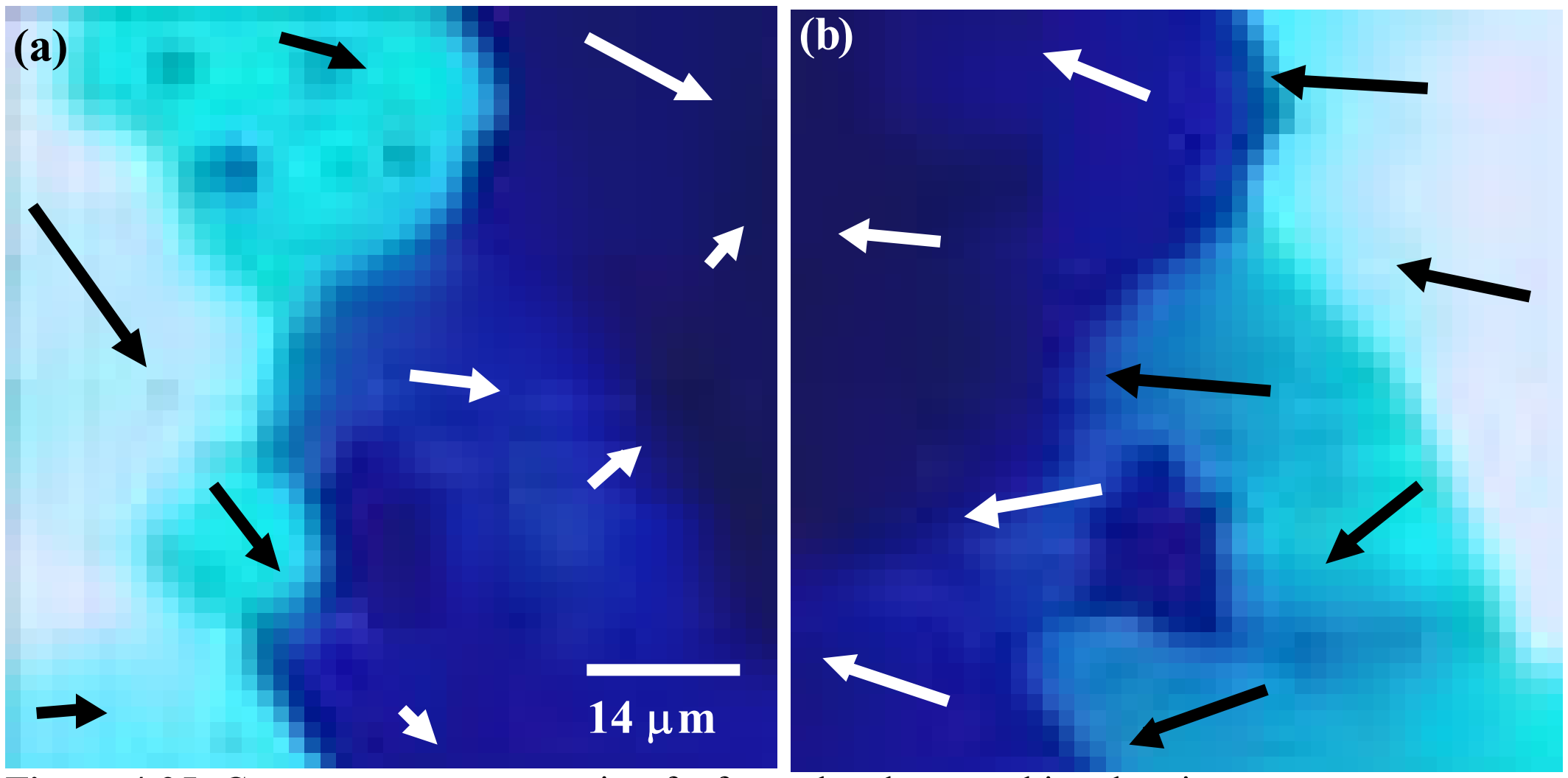

**Figure 4.25.** Current map reconstruction for forward and reverse biased varistor.

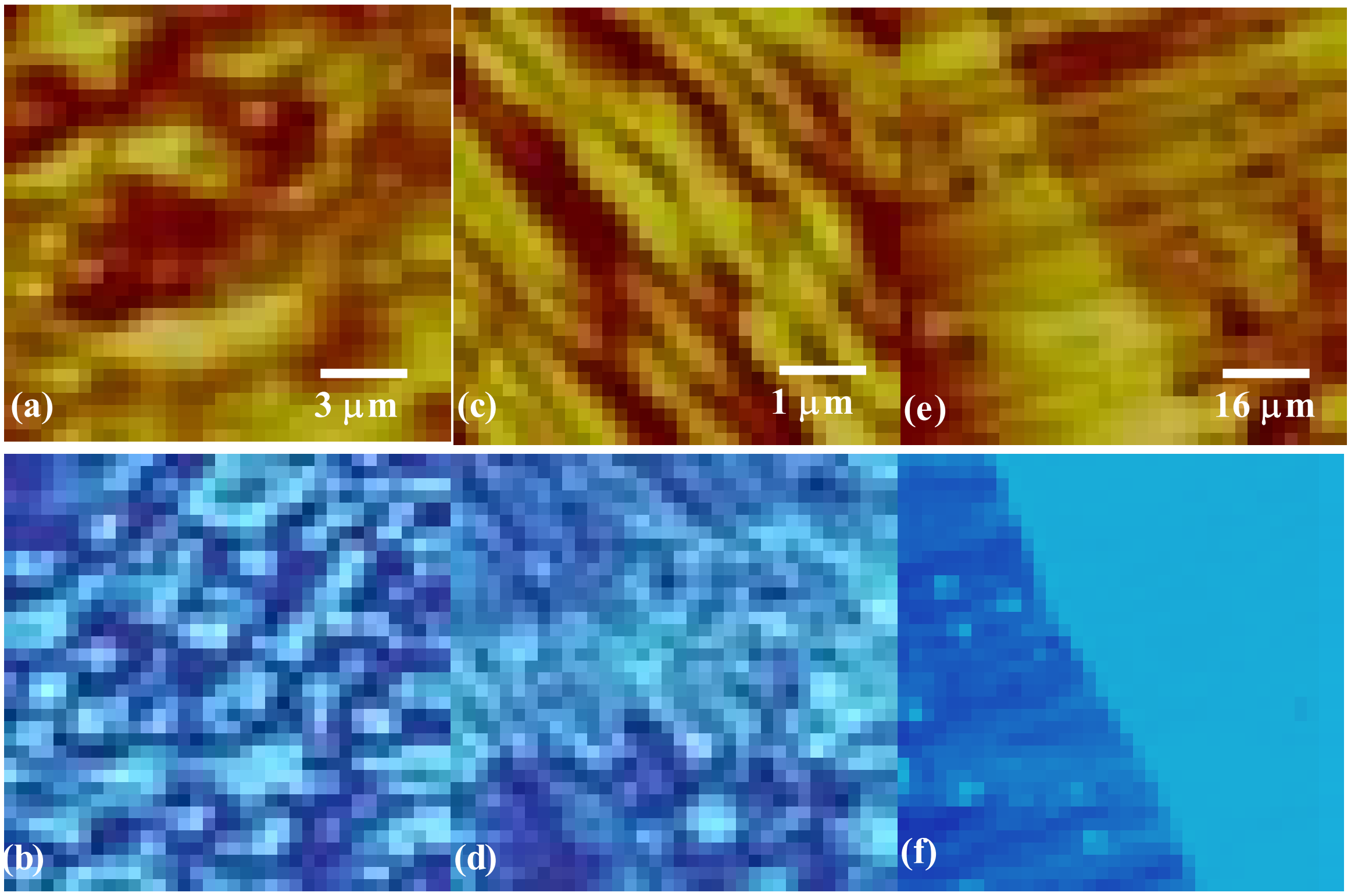

**Figure 4.26.** Surface topography (a,c,e) and current (b,d,f) images of etched (a-c) and non-etched ZnO surface. Imaging is made is the single terminal (a-c) and two-terminal (e,f) configuration.

To summarize, in polycrystalline ZnO surface potential and two-terminal cAFM measurements illustrate the potential for local resistivity mapping. The individual grain boundaries typically exhibit asymmetric *I-V* characteristics and a number of rectifying grain boundaries was observed. This behavior was attributed recently to the piezoelectric effect on grain boundary potential barriers;[77] however, attempts to perform simultaneous piezoresponse and potential imaging in ZnO were unsuccessful. Further progress in this field is expected if the local AFM studies are combined numerical modeling similar to Fleig and Maier.[78,79]

#### 4.4.2. Variable Temperature Transport and Piezoelectricity in $BaTiO_3$

An interesting example for SPM studies is polycrystalline semiconducting $BaTiO_3$ with positive temperature coefficient of resistance (PTCR). In this material, below ferroelectric Curie temperature the interface charge is compensated by the spontaneous polarization. On increasing the temperature, polarization decreases and the interface potential barrier develop, resulting in the drastic increase of interface resistance. Illustrated in Figure 4.27 is surface potential on the grounded, forward and reverse biased PTCR surface room temperature. The grain boundaries are clearly conductive. On increasing the temperature, potential barriers at grain boundaries become visible.

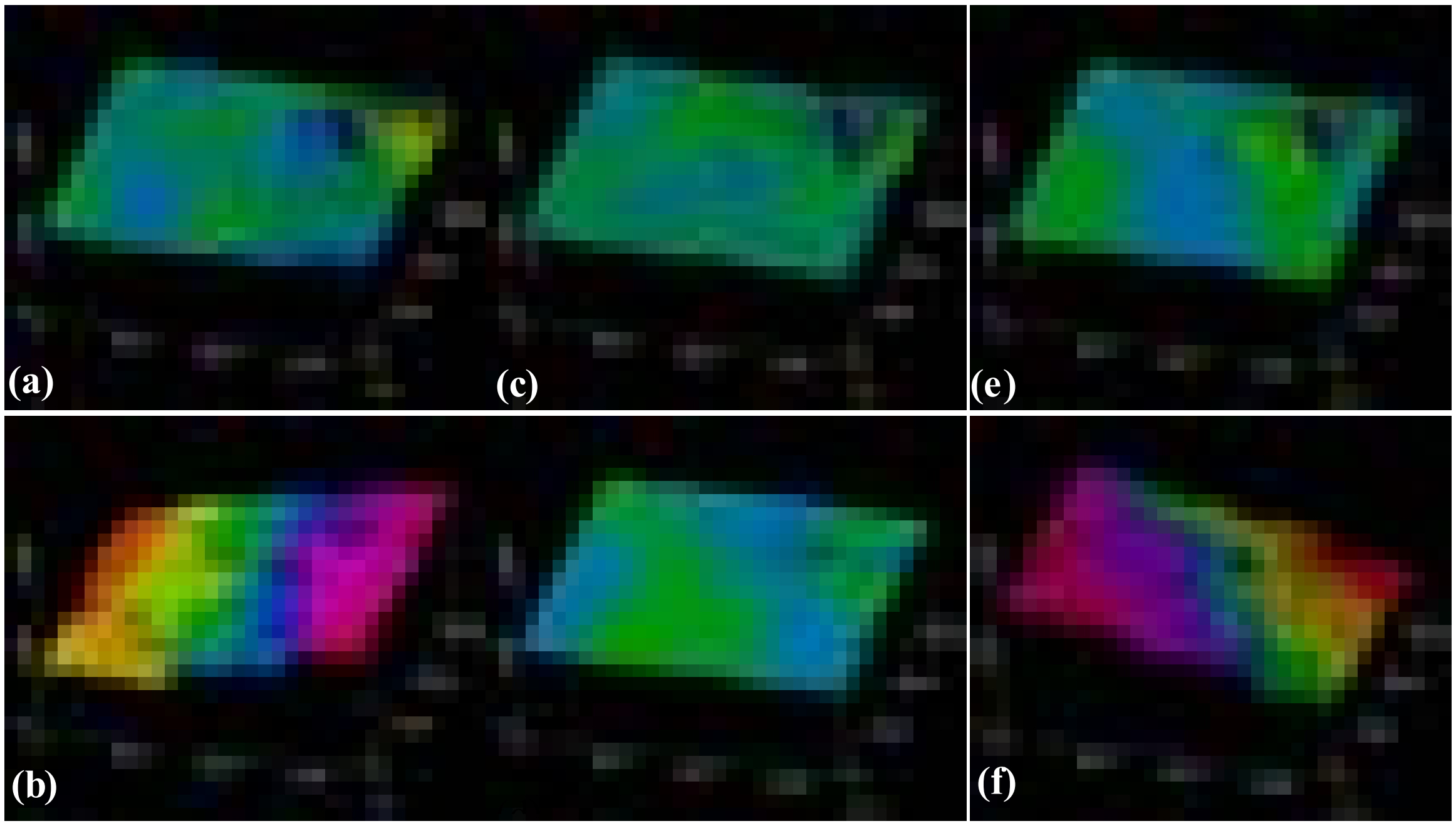

**Figure 4.27.** Surface potential of forward biased (a,b), grounded (c,d) and reverse biased (e,f) PTCR $BaTiO_3$ surface at room temperature (a,c,e) and at 35°C (b,d,f).

Ferroelectric activity of $BaTiO_3$ can be accessed locally by piezoresponse force microscopy (PFM). Shown in Figure 4.28 are piezoresponse and surface potential images of PTCR surface at room temperature. The fact that individual grains consist of single domain with an absence of surface potential variations on the grounded surface (not shown) is consistent with a high carrier concentration in semiconducting $BaTiO_3$ that screens spontaneous polarization and stabilizes a single domain structure. From PFM image, the individual grains are clearly in the single domain state. On increasing the temperature, the piezoelectric activity decreases as illustrated in Figure 4.28b. Resistive barriers do not exist at the grain boundaries at room temperature; however, SSPM indicates the development of resistive barriers below the nominal transition temperature

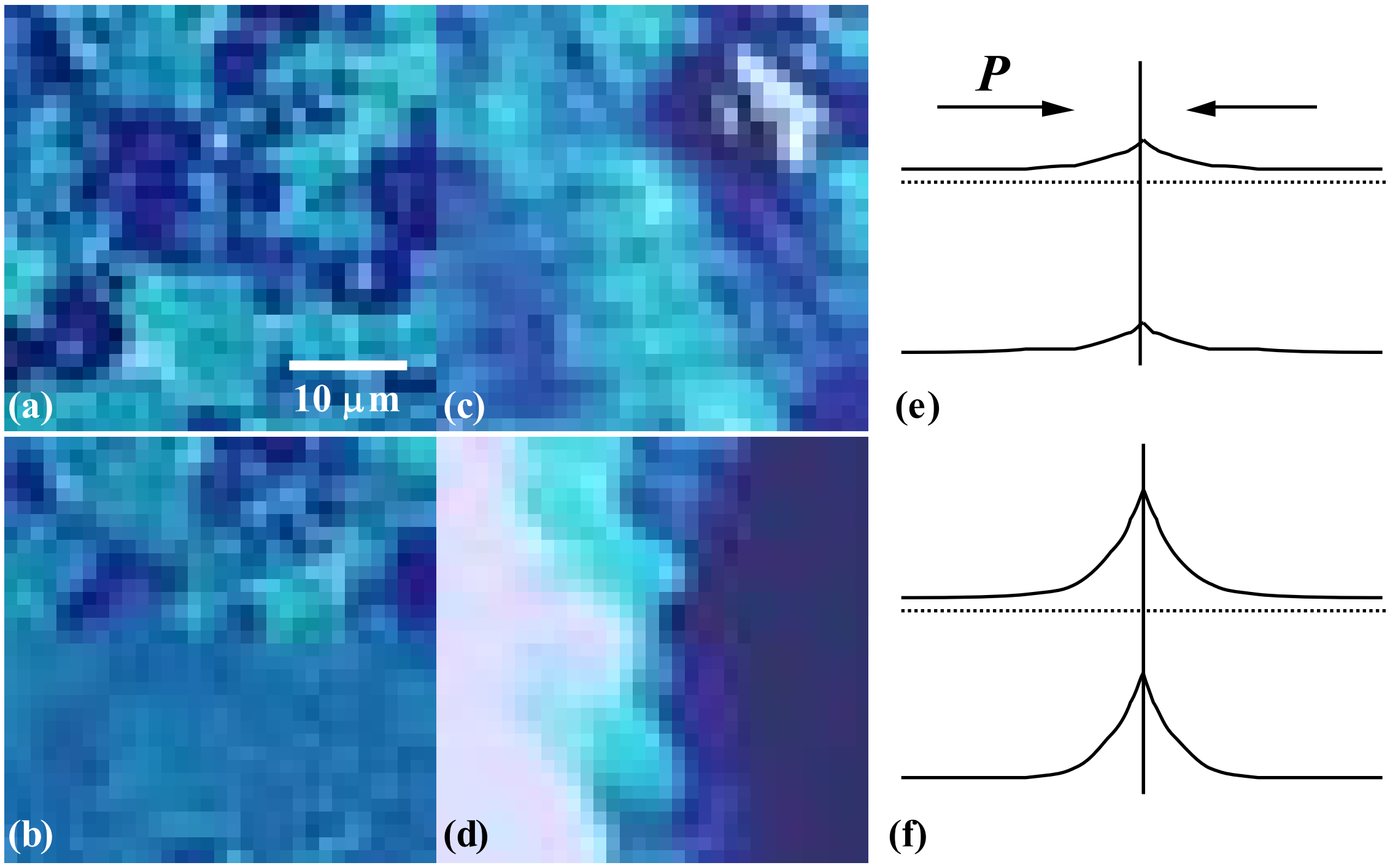

**Figure 4.28.** Piezoresponse (a,b) and surface potential (c,d) images of laterally biased PTCR $BaTiO_3$ surface at room temperature (a,c) and at 35°C (b,d). Schematic of band structure of the interface below (e) and above phase transition temperature (f).

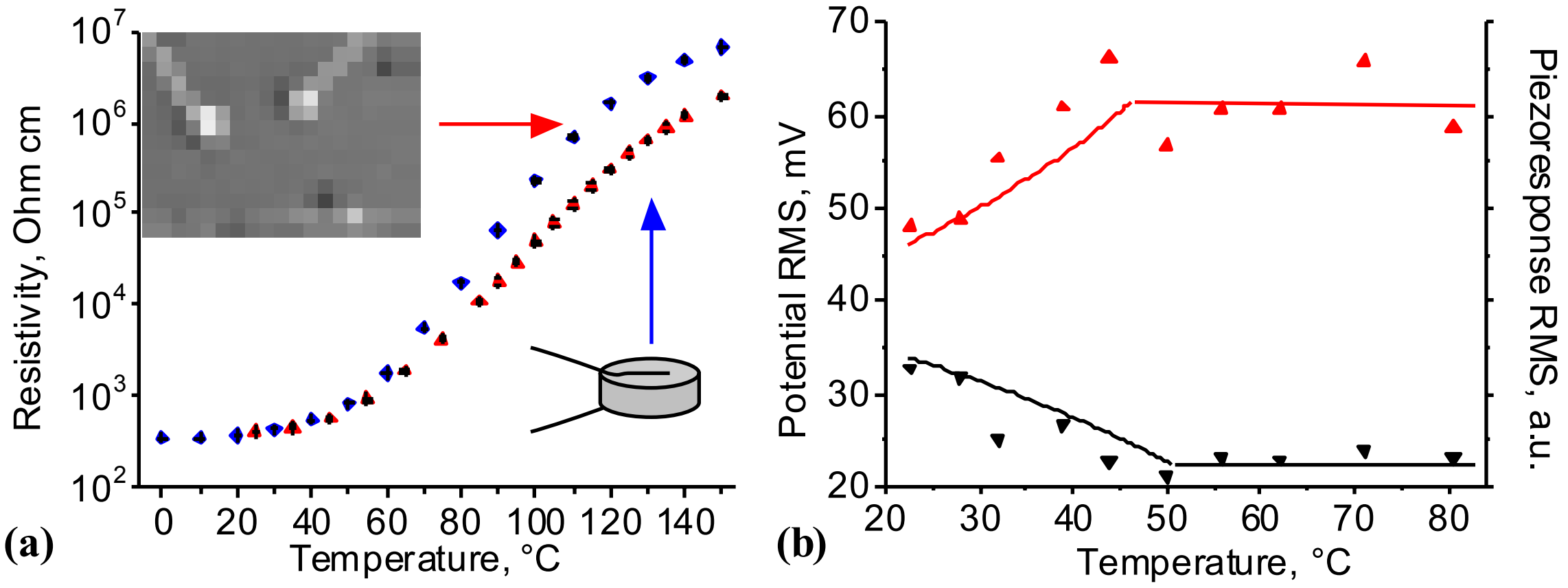

**Figure 4.29.** (a) Temperature dependence of resistance measured for macroscopic device (◆) and between the micropatterned contact pads (▲). The discrepancy between the curves is due to the varistor effect. (b) Locally measured temperature dependence of electrical activity of the interfaces (▲) and averaged piezoelectric activity of the surface (▼).

(50 °C). Simultaneous piezoresponse imaging confirms the localization of potential drop at the grain boundaries. PFM contrast does not disappear well above the onset of PTCR behavior in agreement with broad transition observed in the impedance spectroscopy measurements performed both on the macroscopic device and between micropatterned electrodes (Figure 4.29a). The variation of PFM and SSPM signals within the image provide quantitative measures of grain boundary resistivity and piezoresponse activity and indicate the concurrent increase in resistivity and decrease of piezoresponse with temperature as illustrated in Figure 4.29b.

To summarize, a combination of variable temperature SSPM under external lateral bias and piezoresponse imaging allows real space imaging of grain boundary PTCR behavior. At room temperature, most of the grains are in the single domain state, consistent with high conductivity of the material. Piezoresponse activity decreases with temperature along with the increase of the resistivity of grain boundary regions. The formation of resistive grain boundary barriers begins below the nominal transition temperature, while piezoresponse activity is observed in the PTCR region. These results indicate the gradual nature of the transition, which is a direct consequence of large dispersion of grain boundary properties. Further progress in this direction is expected for PTCR samples with larger grain sizes facilitating the quantitative observations of interface potential behavior under variable temperature conditions.

### 4.4.3. Grain Boundary and Domain Wall Mediated Transport in $BiFeO_3$

Bismuth ferrite $BiFeO_3$ simultaneously exhibits both ferroelectric ($T_C$ = 830 $^oC$) and long range antiferromagnetic G-type ordering ($T_N$ = 370 $^oC$).[80] Because of this magnetoelectric coupling, it has been proposed that $BiFeO_3$ ceramics systems could be used to develop novel memory device applications. Extensive structural, magnetic, and electric studies of various $BiFeO_3$ solid solutions systems have been reported.[81,82,83] The electric and dielectric properties of $BiFeO_3$, which could be strongly affected by small amounts of impurities and ferroelectric behavior, have been inadequately investigated. It was suggested that the impurity segregation on grain boundaries could lead to complex impedance behavior and grain boundary barrier layer (GBBL) dielectric effects. Here, electrical SPM is used to study piezoresponse and microscopic ac and dc transport in polycrystalline $BiFeO_3$, thus facilitating the understanding of the macroscopic impedance and dielectric properties.

Piezoresponse Force Microscopy of $BiFeO_3$

It was long known that $BiFeO_3$ exhibits ferroelectric properties. However, experimental measurements of electromechanical properties are hindered by relatively high conductivity of this material. Due to the lack of transparency in the visible range, optical observation of domain structure is also impossible. Here we attempted local studies of piezoelectric activity of $BiFeO_3$ by piezoresponse force microscopy. In these experiments, the modulating bias is applied to the sample, while variation of tip bias allowed to record local hysteresis loops and study local switching behavior.

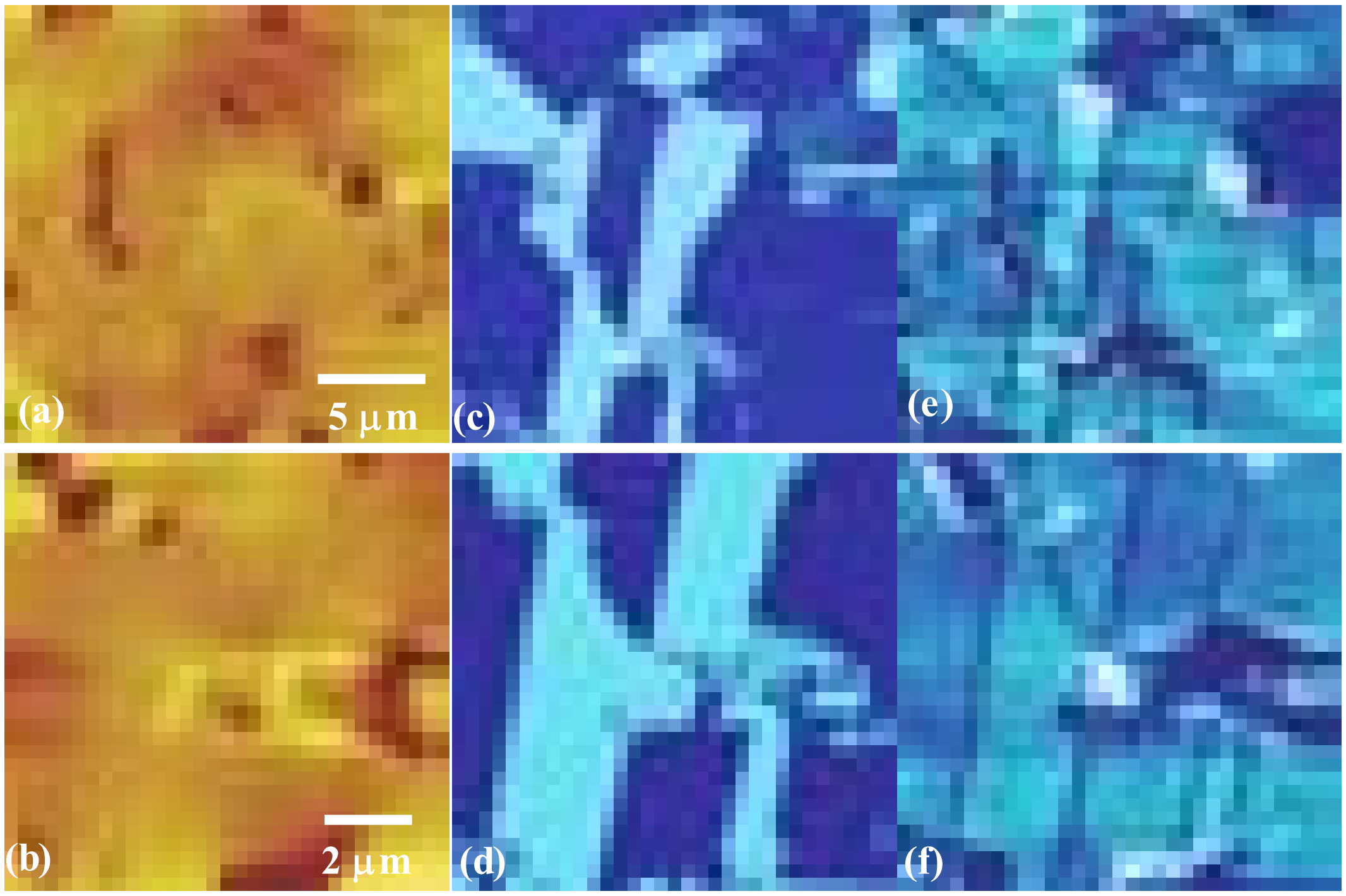

**Figure 4.30.** Surface topography (a,b), piezoresponse phase (c,d) and amplitude (e,f) at different magnifications. Note that extremely clear PFM contrast is observable despite relatively high (~100 kOhm) conductivity of the sample.

Shown in Figure 4.30 are surface topography and piezoresponse phase and amplitude images of polished $BiFeO_3$ surface. PFM images exhibits clear domain structure, in which the amplitude is constant for the antiparallel domain and the phase changes by 180°. The maximum response amplitude depends on grain orientation and some grains are characterized by virtually zero amplitude. A number of such grains are located at the junctions and can be interpreted as impurity inclusions, similarly to previous analysis of phase distribution in $Li_2O$-$Nb_2O_5$-$TiO_2$ system.[84]

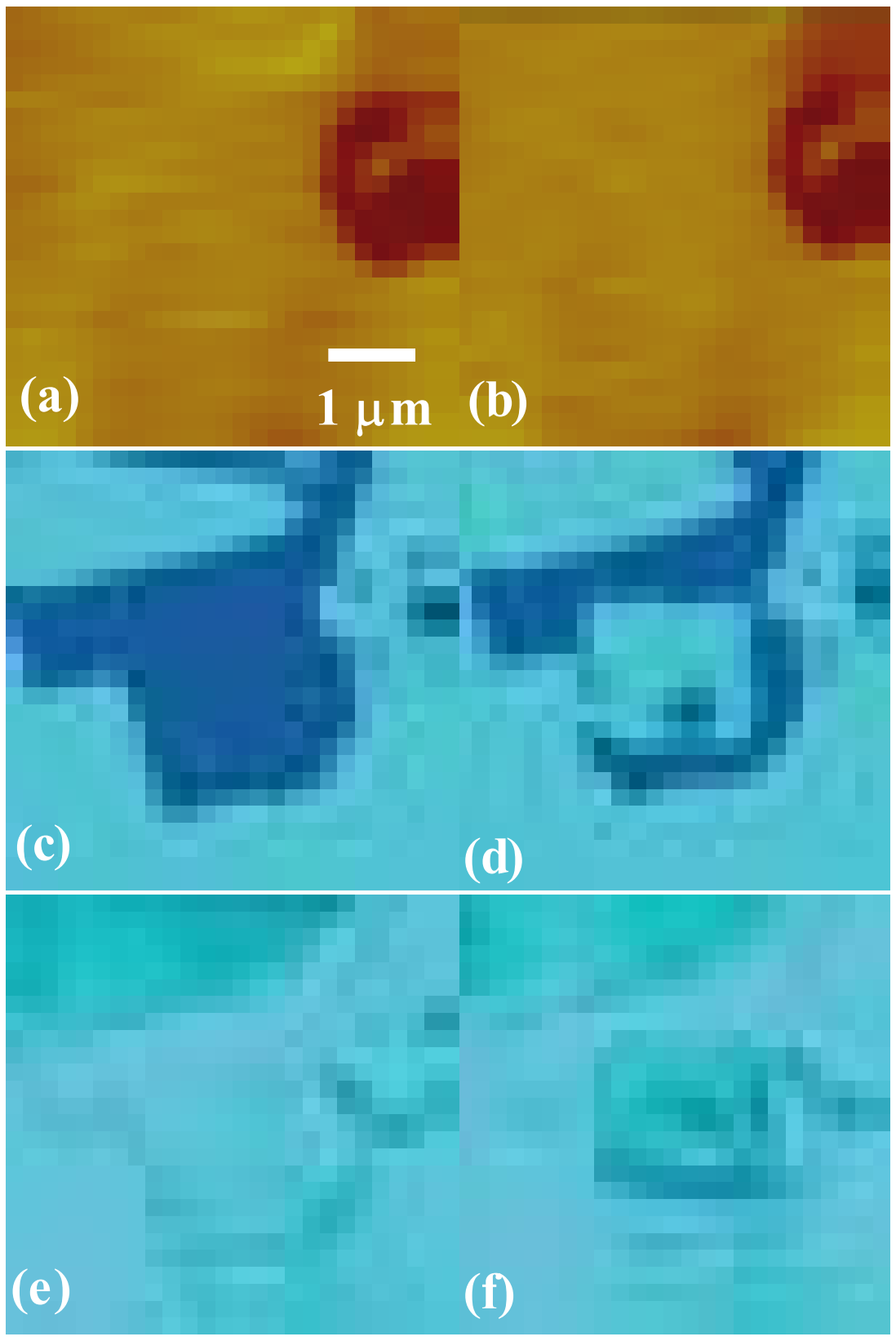

**Figure 4.31.** Surface topography (a,b), PFM phase (c,d) and amplitude (e,f) of $BiFeO_3$ surface before (a,c,e) and after (b,d,f) single scan switching by –10 V.

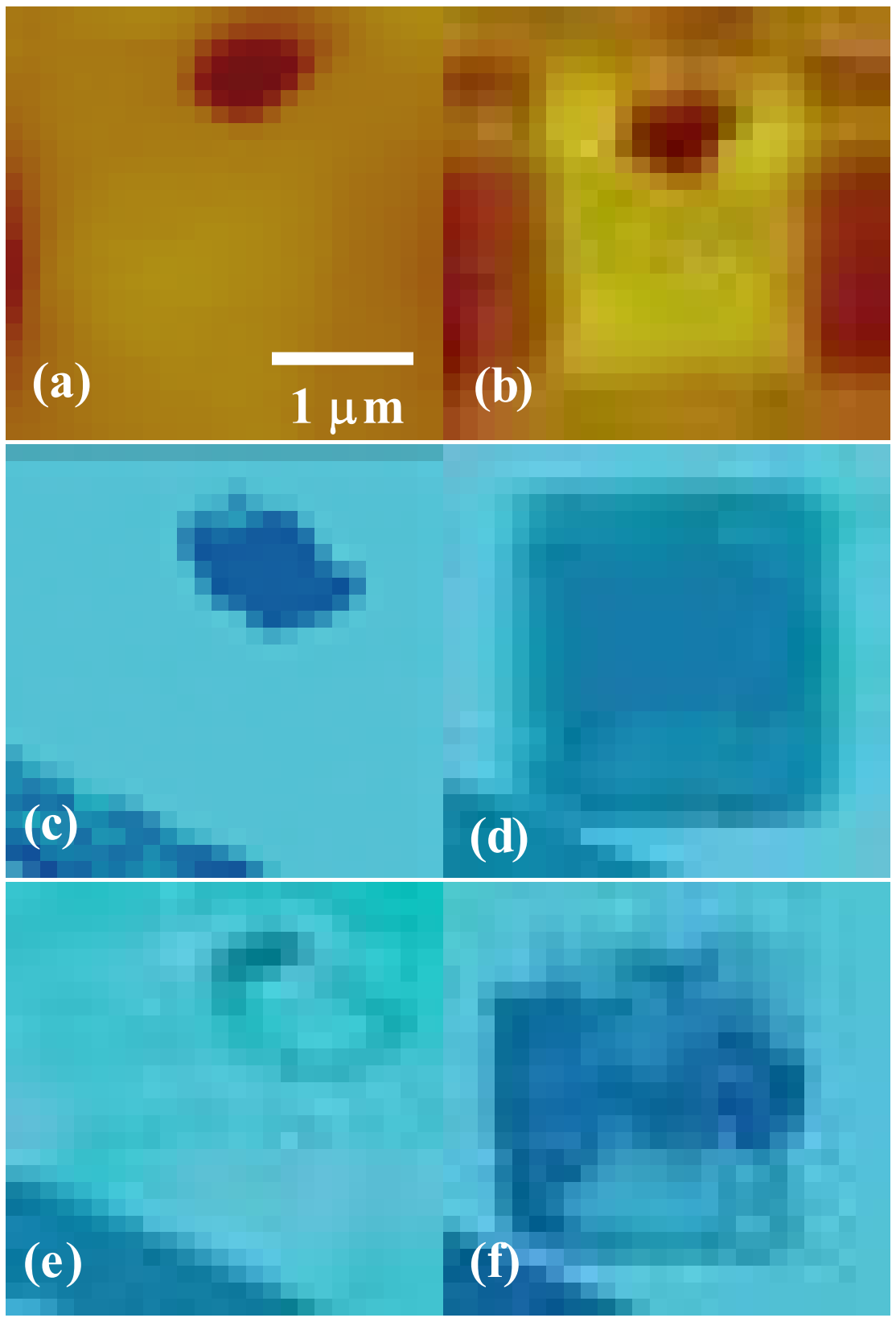

**Figure 4.32.** Surface topography (a,b), PFM phase (c,d) and amplitude (e,f) of $BiFeO_3$ surface before (a,c,e) and after (b,d,f) ~20 scan switching by –10 V.

Local polarization can be easily switched in $BiFeO_3$. Shown in Figure 4.31 is the sequence of PFM images before and after poling by SPM tip. Noteworthy is that the response amplitude is reduced in the switched area, which is not the case for the pristine domains. Similarly, PFM electromechanical hysteresis loops (not shown) indicate significant asymmetry with respect to bias. To account for this observation, we suggest tip-induced oxidation of the surface. To verify this assumption, we have performed ~20 continuous scans by the tip over the same area. The evolution of the piezoresponse image was recorded and shown in Figure 4.32 is the surface topography and piezoresponse image after switching. Note that the amplitude is zero in the switched region and surface topography exhibit protrusion. It might be speculated that the continuous exposure of the $BiFeO_3$ surface to the negatively biased tip results in the electrochemical reaction of the type

$$2BiFeO_3 + (x+2)\ H_2O + 2e^- \rightarrow Bi_2O_3{\cdot}xH_2O + 2Fe(OH)_2 + 2OH^- \qquad (4.46)$$

that leads to the non-ferroelectric products. Further confirmation of the reaction mechanism is clearly required; however, these studies demonstrate the unique potential for ferroelectric and chemical modification of $BiFeO_3$. In conjunction with perspective thin film applications, this provides unique avenues for the local surface patterning required for the device fabrication.

SSPM under lateral bias

The surface topography and surface potential at a $BiFeO_3$ surface under different bias conditions are shown in Figure 4.33. The topographic image exhibits a number of spots due to contaminants and depressions due to inter- and intragranular pores. Grain boundaries can be seen due to selective polishing of grains with different orientations. The surface potential of the grounded $BiFeO_3$ surface exhibits large-scale potential variations due to ferroelectric domains and surface contaminants. On application of a 10 V lateral bias, the potential drops at the grain boundaries become evident (Figure 4.33c). The contrast inverts on application of a bias of opposite polarity (Figure 4.33d). Note that the potential features related to ferroelectric polarization are independent of the applied bias. Ramping the dc bias across the sample has shown that the potential drop at the interface is linear in external bias and the grain boundaries exhibit ohmic behavior for small biases ($\Delta V_{gb} < 50$ mV)).

AC transport by SIM

The surface topography, SIM phase images at 20 and 70 kHz and the SIM amplitude image at 70 kHz of the same region are shown in Figure 4.34. Note that the phase images exhibit well-defined phase shifts at the grain boundaries, while the amplitude image shows a uniform decrease of amplitude across the surface. Positive phase shifts at the grain boundary and a negative phase shift in the bulk are clearly observed in agreement with theoretical arguments. For higher frequencies phase shifts in the grain interior are not observed due to the resistive component in the experimental circuit. At the same time, the amplitude decreases linearly in the direction of current flow indicating that the experimental frequency range (10-100kHz) is above the resonant frequency of the interface. To quantify the frequency dependence of the grain boundary phase shift, the latter was determined for a series of images collected at 10kHz steps. The analysis in the vicinity of the resonant frequency of the cantilever (60 kHz) is complex due to a force-gradient induced resonant frequency shift and associated non-linear phase behavior. To relate the SIM phase shift to the material properties, the latter were independently determined by impedance spectroscopy and the corresponding spectra are shown in Figure 4.35a. From the impedance spectroscopy data, the average grain boundary resistivity and capacitance are estimated as $R_{gb}$ = 116 kOhm cm and $C_{gb}$ = 7.6 nF/cm, while the grain interior resistivity and capacitance are $R_{gi}$ = 812 Ohm cm and $C_{gi}$ = 7 pF/cm. It should be noted that two RC elements provide a relatively poor description of the high frequency region of the experimental impedance spectra; the properties of the grain boundary component are well defined, whereas bulk properties can be determined only approximately. Figure 4.35b shows the calculated grain boundary phase shift vs. frequency dependence as compared to experimental SIM data. The only free parameter in the calculations is the effective grain number. The best fit is obtained for $n = 210$ grains, which is comparable with grain number N ~ 70 estimated from the grain size (~ 20-30 μm) and the distance between measurement point and left contact (~ 1-2 mm). The discrepancy between the two is due to the uncertainty in the bulk resistance and variation in grain boundary properties and orientation. Note the excellent agreement between phase angle frequency dependences obtained from local measurements and impedance spectroscopy.

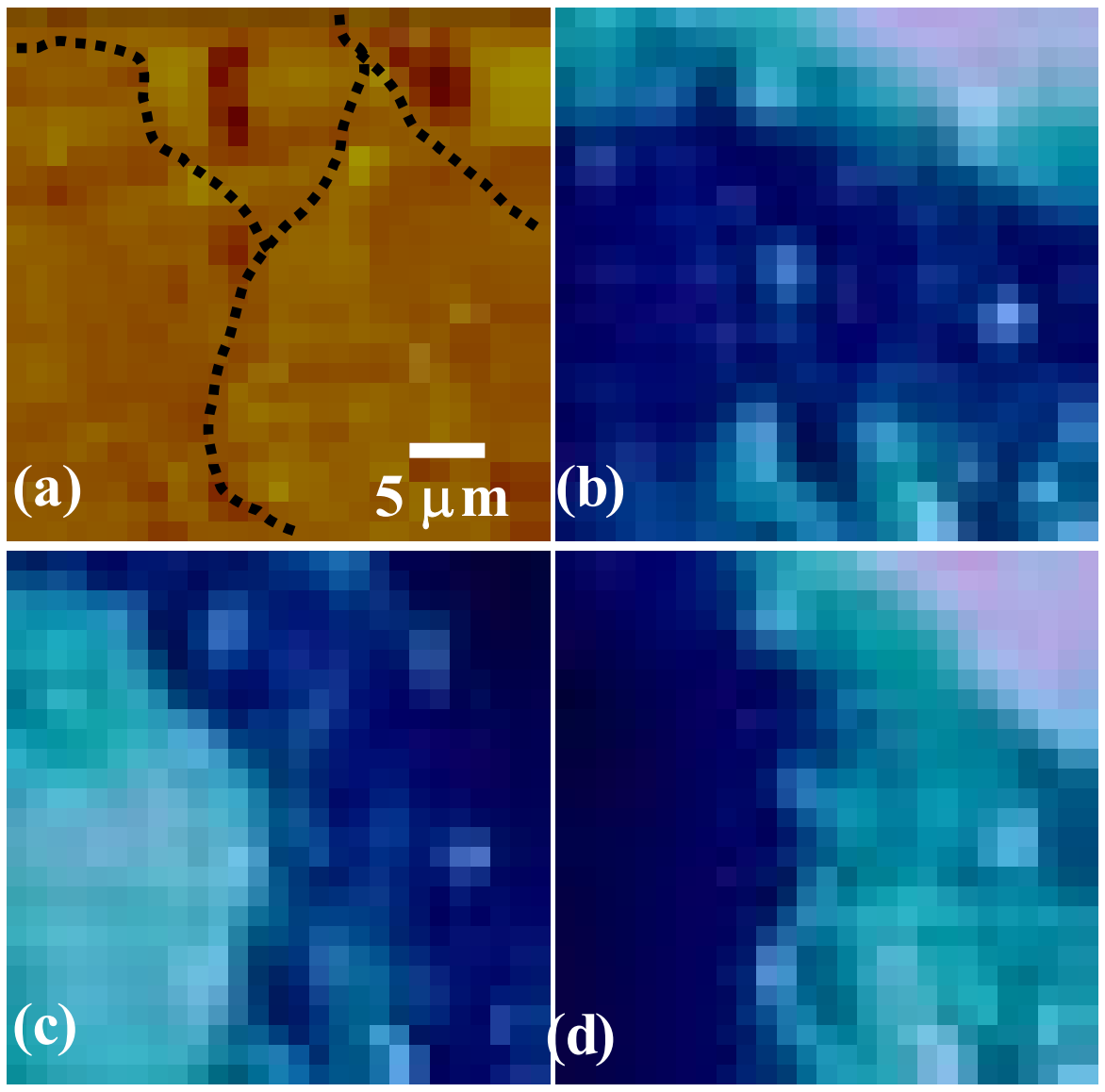

**Figure 4.33.** Surface topography (a), surface potential of the grounded surface (b), and surface under lateral bias of 10 V (c) and –10 V (d). Scale is 200 nm (a), 50 mV (b,c,d).

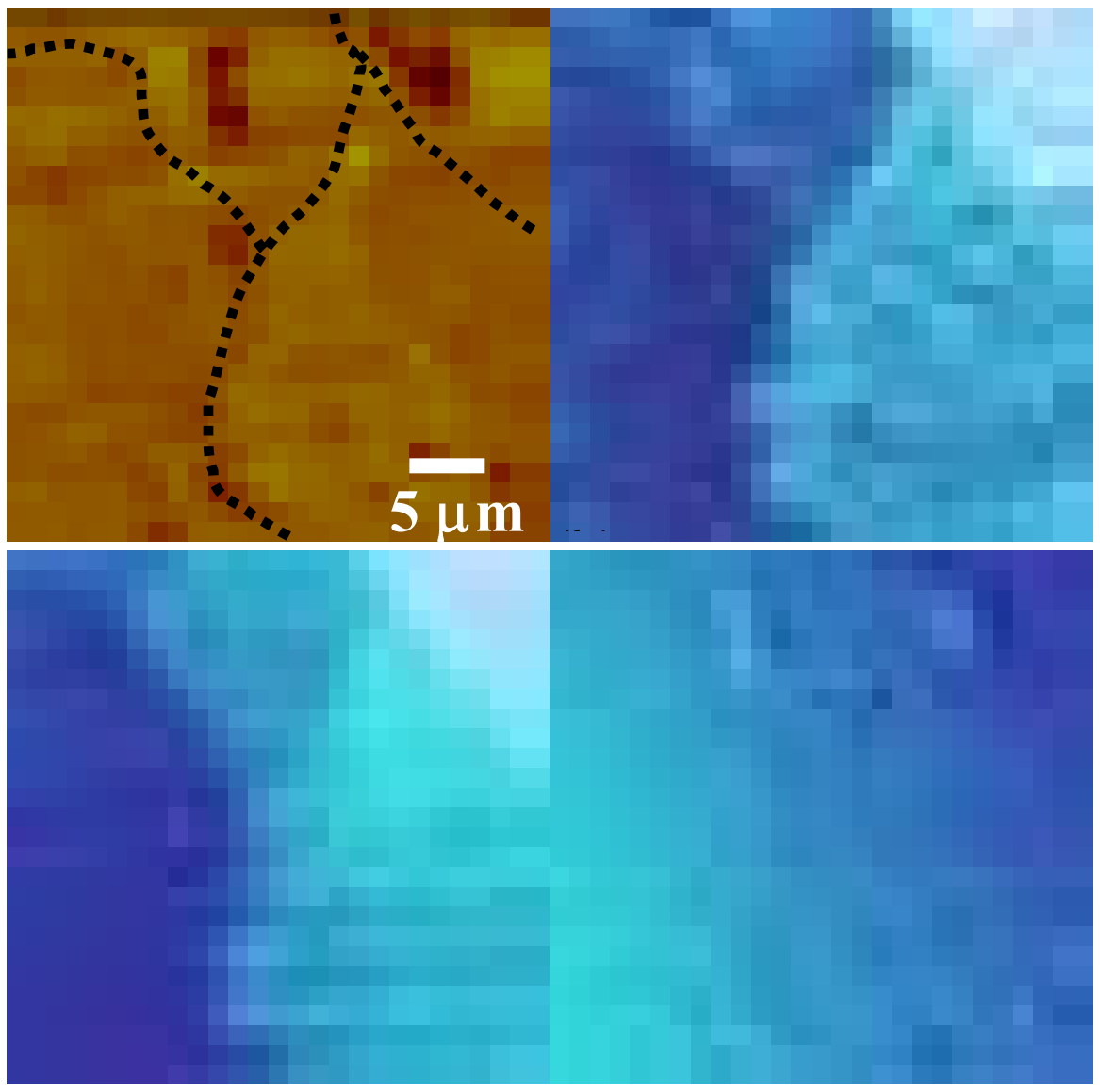

**Figure 4.34.** Surface topography (a), SIM phase image at 30 kHz (b) and 70 kHz (c) and SIM amplitude image at 70 kHz (d) of the same region. Scale is 200 nm (a), 0.2 degree (b,c).

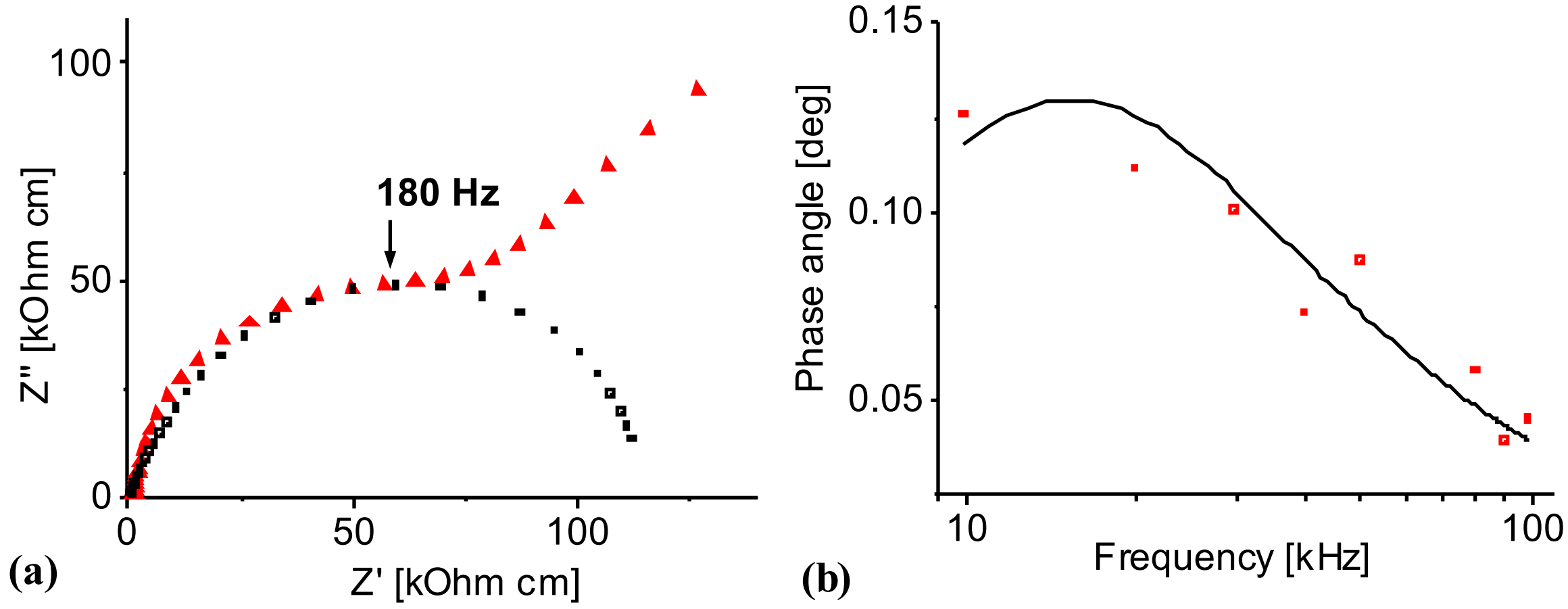

**Figure 4.35.** (a) Cole-Cole plots of as prepared $BiFeO_3$ pellets (■) and the rectangular sample (▲) used for scanning probe microscopy studies. (b) Experimental SIM phase shift across the interface and theoretical curve calculated from the impedance data.

To summarize, ferroelectric domain structures in $BiFeO_3$ can be observed and local hysteresis loops can be obtained by PFM despite the high conductivity of the material. Both ferroelectric domain structure and local chemical composition can be modified by PFM. The fingerprints of the domain structure can be observed in surface potential images as well. Grain boundaries are shown to be associated with resistive barriers by SSPM. It is shown that ferroelectric domain boundaries do not contribute to the SIM image, thus allowing unambiguous correlation of impedance spectra with electroactive grain boundaries. For $BiFeO_3$ ceramics excellent agreement between local SIM measurements and impedance spectroscopy data was found.

## 4.5. Conclusions

In this Chapter, electric SPM techniques developed in Chapter 3 were used for the characterization of a number of single- and multiple interface oxide systems. A numerical procedure for the extraction of potential at a surface-interface junction from the SSPM and EFM data is developed. It is shown that in air this potential is significantly smaller and in some cases opposite in sign to the grain boundary potential in the bulk, due to screening by mobile adsorbates. DC and ac current measurements were shown to be insensitive to the presence of the screening charge. Transport properties of $\Sigma 5$ grain boundary in Nb-doped $SrTiO_3$ bicrystal were characterized and compared to the conventional measurements. Based on the temperature dependence of interface capacitance, it was conjectured that the dielectric constant is suppressed at the interface. We developed a description of grain boundary properties in non-linear dielectric material. We have shown the grain boundary transport in metallic $SrTiO_3$ to be thermionic emission below RT and diffusion above RT. Temperature dependence of interface potential and capacitance was used to determine grain boundary parameters and non-linear dielectric properties of $SrTiO_3$. The values obtained are in a good agreement with permittivity measurements.

Polarization-mediated local transport behavior was studied in a number of piezoelectric and ferroelectric polycrystalline oxides. In polycrystalline ZnO a number of grain boundaries with asymmetric I-V characteristics were observed; however, attempts to perform simultaneous piezoresponse and potential imaging were unsuccessful. In polycrystalline $BaTiO_3$, much higher electromechanical activity allowed combination of variable temperature SSPM and piezoresponse imaging of grain boundary PTCR behavior. The formation of resistive grain boundary barriers was observed below the nominal transition temperature, while piezoresponse activity was observed in the PTCR region. These results indicate the gradual nature of the transition, which is a direct consequence of large dispersion of grain boundary properties. In polycrystalline $BiFeO_3$, ferroelectric domain structure was observed and local hysteresis loops were obtained by PFM, unambiguously proving it's ferroelectric nature. Grain boundaries, rather than ferroelectric domain walls, are shown to be responsible for low-frequency dielectric behavior. Excellent agreement between SIM and impedance spectroscopy was found.

---

[70] H.O. Jacobs, H.F. Knapp, A. Stemmer, Rev. Sci. Instr. **70**, 1756 (1999).

[71] E.B. McDaniel, S.C. McClain, J.W. P. Hsu, Appl. Optics **37**, 84 (1998).

[72] M. Saint Jean, S. Hudlet, C. Guthmann, J. Berger, J. Appl. Phys. **86**, 5245 (1999).

[73] S.V. Kalinin, M. Freitag, A.T. Johnson, and D.A. Bonnell, Appl. Phys. Lett. **81**, 754 (2002).

[74] Current Designs, Inc

[75] A. Nayfeh, *Introduction to Perturbation Techniques* (John Wiley, New York, 1981).

[76] G. Popov, S.V. Kalinin, T. Alvarez, T.J. Emge, M. Greenblatt, and D.A. Bonnell, Phys. Rev. **B 65**, 064426 (2002).

[77] P.M. Verghese and D.R. Clarke, J. Appl. Phys. **87**, 4430 (2000).

[78] J. Fleig, Solid State Ionics **150**, 181 (2002).

[79] S. Rodewald, J. Fleig, and J. Maier, J. Eur. Ceram. Soc. **21**, 1749 (2001).

[80] P. Fischer, M. Polomska, I. Sosnowska, and M. Szymanski, J. Phys. **C 13,** 1931 (1980).

[81] M. Mahesh Kumar, A. Srinivas. S.V. Suryanarayana, and T. Bhimasankaram, Phys. Stat. Sol. **A 165,** 317 (1998).

[82] M. Polomska, W. Kaczmarek, Z. Pajak, Phys. Stat. Sol. **A 23,** 567 (1974).

[83] J.B. MacChesney, J.J. Jetzt, J.F. Potter, H.J. Williams, and R.C. Sherwood, J. Am. Ceram. Soc. **49**, 644 (1966).

[84] A.Y. Borisevich, S.V. Kalinin, D.A. Bonnell, and P.K. Davies, J. Mater. Res. **16**, 329 (2001).

# 5. POLARIZATION AND CHARGE DYNAMICS ON FERROELECTRIC SURFACES

## 5.1. Introduction

### 5.1.1. Physics and Applications of Ferroelectric Materials

Since the discovery of the ferroelectric properties of Rochelle salt approximately 80 years ago,[1] the ability of ferroelectric materials to sustain spontaneous polarization below the Curie temperature remains one of the most fascinating materials phenomena and constitutes the basis for their wide technological applicability.[2,3,4] For more than 20 years after their discovery, Rochelle salt was the only known ferroelectric material and was universally considered as a curiosity rather than a potentially useful material. Consequently, only one other ferroelectric material ($KH_2PO_4$, KDP) was found over two decades after the discovery of ferroelectricity in Rochelle salt. The complicated crystallographic structure of these compounds (112 atoms per unit cell for Rochelle salt, 16 atoms per unit cell for KDP) made ferroelectricity extremely difficult for theoretical interpretation. The fact that these compounds contained hydrogen bonds led to the erroneous assumption that the existence of hydrogen bonding is a precondition for ferroelectricity and therefore the search for new ferroelectric materials performed over these two decades was limited to hydrogen-containing compounds.

The situation changed in the early forties when the search for materials with high dielectric constants as a substitute for natural mica in capacitor applications led to the discovery of ferroelectricity in the perovskite $BaTiO_3$ simultaneously in the USA, Russia and Japan. Immediately $BaTiO_3$ and related ferroelectric perovskites were recognized as promising materials for the submarine sonar arrays, heralding the beginning of intensive research in the field. These materials rapidly became extremely widespread due to high chemical stability, good mechanical properties and ease of preparation. In addition, the relatively simple perovskite structure made ferroelectric perovskites more amenable for theoretical treatment originating a number of models ranging from the original "rattling atom" model to soft-mode based description of ferroelectricity and thermodynamic Ginzburg-Devonshire type models.[2,5] After the discovery of the strong electromechanical coupling in ferroelectrics, numerous applications as sensors, actuators, transducers, etc.

emerged. Major applications of ferroelectric ceramics can be divided into several distinct areas that can be derived from the different combinations of their properties:

1. The high dielectric permittivity and wide frequency range of response enables the creation of compact capacitors in the form of multilayers, thick or thin films.
2. The piezoelectric and electrostrictive responses in poled and unpoled ferroelectrics are employed in transducers for converting electrical to mechanical response and vice versa. Interestingly, it is the piezoelectric effect that enables very high precision position control in STM and AFM devices.
3. The strong temperature sensitivity of polarization (pyroelectric effect) makes these materials applicable for wide range of imaging systems and thermal-medical diagnostics. In polycrystalline materials, interplay between polarization and grain boundary phenomena gives rise to the positive temperature coefficient of resistance (PTCR) phenomenon,[6] which is employed in thermal sensors.
4. The high quadratic and linear electro-optic coefficients are used in modulators, guided wave structures, light valves and electrooptical devices.

In the last decade, the developments of deposition techniques for epitaxial ferroelectric thin films and advanced ceramic fabrication have resulted in numerous novel applications such as those in microelectromechanical systems (MEMS).[7,8,9] The ability of ferroelectric materials to exist in two or more polarized states, conserve polarization for a finite period, and change the polarization in a field allows their consideration for non-volatile computer memory devices (FeRAM).[10,11,12]

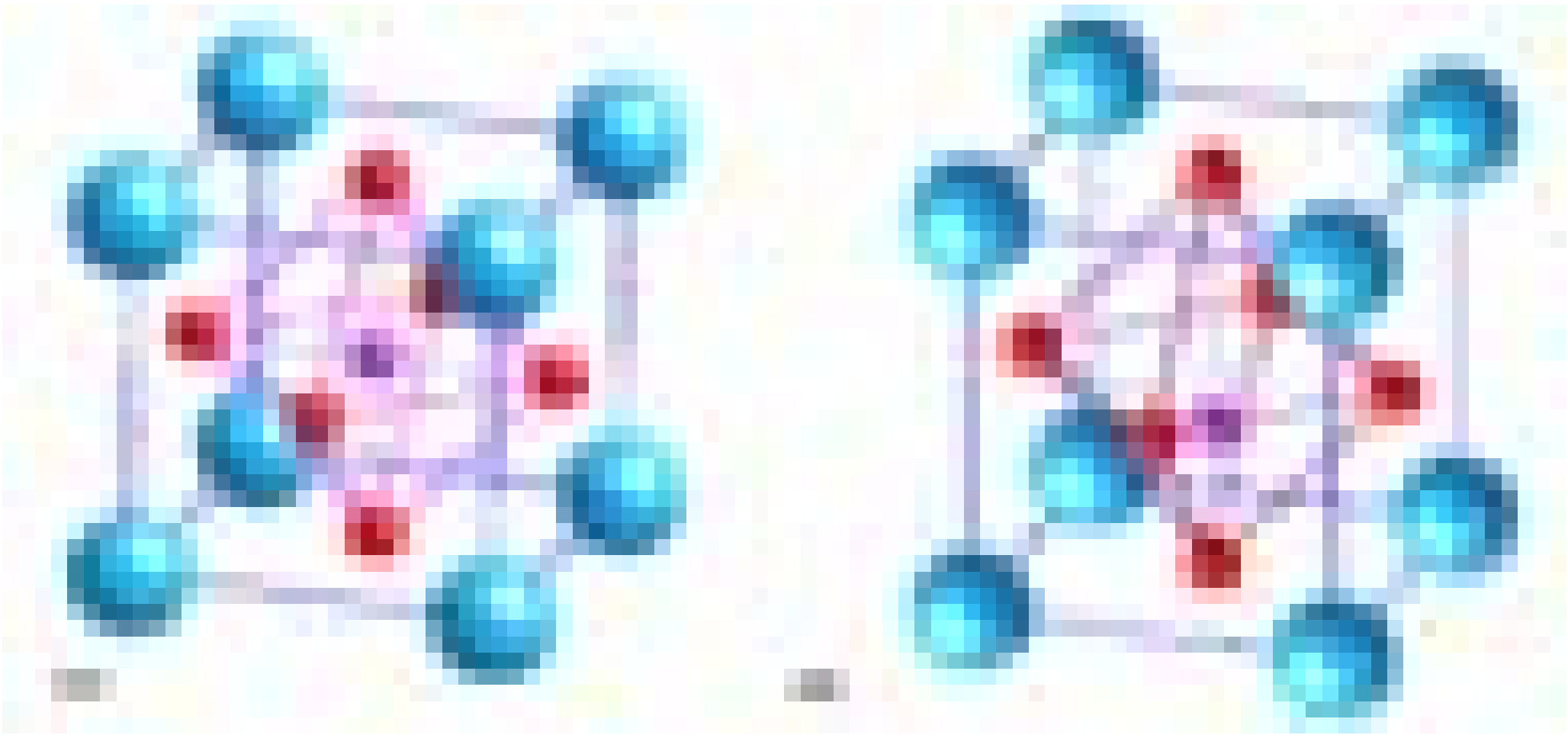

**Figure 5.1.** Crystallographic structure of $BaTiO_3$ above (a) and below (b) ferroelectric Curie temperature.

The origins of the useful properties exhibited by ferroelectric materials can be traced to the crystallographic structure illustrated here on the example of ferroelectric $BaTiO_3$. Above the Curie temperature, $T_c$ = 130°C, $BaTiO_3$ exists in cubic paraelectric phase with space group *Pm*3*m* (Figure 5.1a). On decreasing the temperature, the titanium atom shifts along one of the (100) directions, resulting in the change of the symmetry of the unit cell from cubic to tetragonal (space group *P*4*mm*) and simultaneous development of a dipole moment. Below 5°C and -90°C it transforms into more complicated orthorhombic (*Amm*2) and rhombohedral (*R*3*m*) phases. In the tetragonal phase of $BaTiO_3$, the spontaneous polarization is parallel to the *c*-axis of the tetragonal unit cell and can point in one of the six (100) directions. Dipole interactions result in ferroelectric domains. The depolarization energy limits the maximum domain size, giving rise to complex domain patterns. Different orientations of polarization vectors in adjacent domains result in 180° and 90° domain walls. Several possible domain configurations on $BaTiO_3$ (100) surface are illustrated in Figure 5.2.

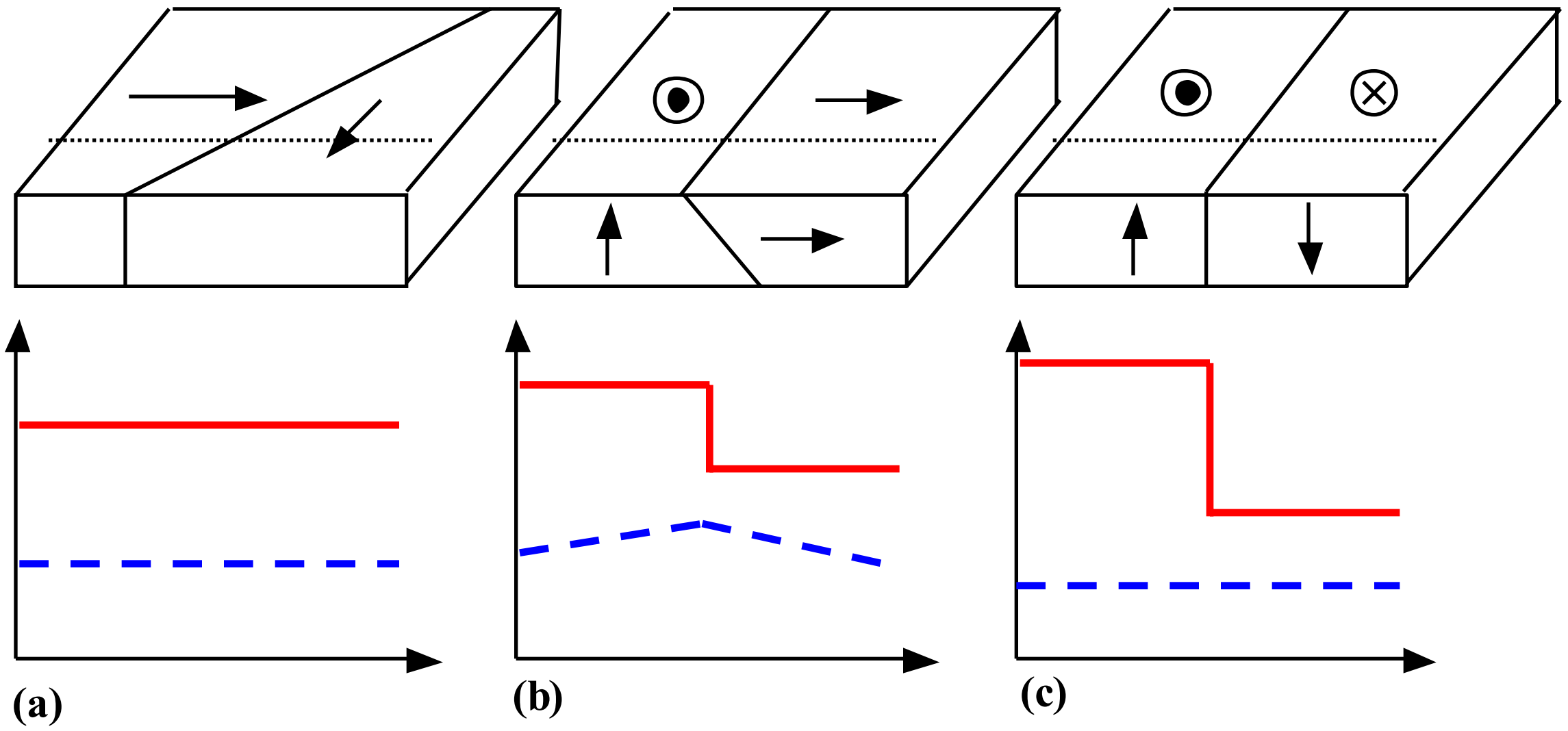

**Figure 5.2.** Domain arrangements on (100) surface of tetragonal $BaTiO_3$. Arrows represent the orientations of polarization vectors. (a) 90° *a*1-*a*2 boundary, (b) 90° $c^+$-*a*1 boundary, (c) 180° $c^+$-$c^-$ boundary. Shown below is surface potential (solid) and surface topography (dashed) expected along the dotted line.

The immediate implication of domains is that experimentally accessible properties of a ferroelectric crystal, and especially ceramics, are averaged over multiple domains. Domain structure significantly influences many physical properties, such as piezoelectricity, electrooptical properties, hysteresis and switching behavior. High correlation energy (compared to ferromagnetic materials) and multiple electromechanical couplings in ferroelectrics imply that domain structure is strongly influenced by the stress fields in material. These effects are especially pronounced in the epitaxial films, in which coupling to the substrate can stabilize certain types of domain structures and change thermodynamic parameters (e.g. $T_c$) of ferroelectric.[13,14,15,16] The high surface energy of ferroelectric domain walls imply that domain growth is nucleation-limited and makes polarization switching processes sensitive to the local defect chemistry, since atomic scale defects (e.g. oxygen vacancies) can serve as pinning centers, especially for charged walls. Therefore, complete description of ferroelectric structure requires not only morphological information on domain orientation, but also the knowledge of local stress and electric fields. While statistically averaged descriptions of ferroelectric microstructure can be used when the device size is much larger than the domain/grain size, current miniaturization of ferroelectric-based electronic devices including MEMS, FeRAMs and DRAMs requires quantitative understanding of domain structure – properties relationship.

### 5.1.2. Spatially Resolved Studies of Ferroelectric Phenomena

The formation and static properties of domains in bulk crystalline ferroelectrics have been extensively studied by such techniques as polarizing optical microscopy, etching, surface decoration, etc.[17,18,19] However, these methods provide relatively low spatial resolution, of the order of 1 μm, limited by optical diffraction. Techniques such as etching damage the surface of material and, therefore, cannot be used for the investigations of dynamic behavior of domains under applied fields and with temperature. Higher spatial resolution can be achieved by electron-beam based probes such as scanning and transmission electron microscopies. However, sample charging, local heating and beam damage cause significant problems. As a result, only a relatively small number of *in-situ* experiments on domain wall motion under applied lateral bias or ferroelectric phase transition by optical microscopy,[20,21] scanning[22,23] and transmission[24,25,26,27,28,29,30] electron microscopy have been reported. Significant progress in ferroelectric imaging was achieved after the invention of SPM based techniques. Contact and intermittent mode atomic force microscopy along with lateral force microscopy has been widely used to characterize domain-related topographic features.[31,32,33,34,35,36,37,38,39] Electrostatic scanning probe techniques such as Electrostatic Force Microscopy (EFM) and Scanning Surface Potential Microscopy (SSPM) have been used to image electric fields associated with polarization charge on ferroelectric surfaces.[40,41,42,43,44] Piezoresponse Force Microscopy (PFM)[45,46,47,48,49] based on the local electromechanical response of the surface to tip bias was established as a prominent technique for domain imaging with sub-10 nm resolution. These and other SPM studies of ferroelectric surfaces are discussed in detail in Chapter 6. However, most SPM techniques such as PFM, SCM, Scanning Near Field Optical Microscopy probe the finite volume of material directly below the probe; alternatively, non-contact SPM imaging was limited to the qualitative information on domain morphology and no quantitative information on polarization-related surface properties was obtained.

### 5.1.3. Polarization Related Chemical Properties of Ferroelectric Surfaces

The electronic and chemical properties of ferroelectric surfaces and interfaces are significantly affected by the polarization charge. If the polarization vector in the vicinity of the surface has an out-of-plane component, the polarization discontinuity will be associated with polarization charge. For $BaTiO_3$ polarization charge is ~0.26 $C/m^2$ corresponding to ¼ of an electron per unit cell. This surface charge is sufficient to induce accumulation or strong inversion, affecting photoelectric and catalytic activity of the surface. A detailed discussion of surface space charge phenomena in ferroelectric semiconductors is given by Fridkin.[50] Domain structure (and hence the polarization charge distribution) can be controlled and, in fact, engineered. It was shown by Ahn[51,52] that a ferroelectric field effect in the ferroelectric/semiconductor heterostructures could result in a metal/insulator transition dependent on the local polarization orientation, introducing new paradigms in oxide electronics. While extensive research in this field is currently under way in a number of groups worldwide (in particular, RTW Aachen,

University of Maryland, Yale University, Oak Ridge National Laboratory, IBM), relatively little is known about the polarization-related properties of ferroelectric surfaces.

The purpose of this chapter is to establish the applicability of local SPM techniques for spatially resolved quantitative studies of physical phenomena on ferroelectric surfaces. To achieve this, electrostatic SPM techniques are used to determine the thermal and temporal evolution of electrostatic fields above the surface as related to the polarization screening processes. Single crystal of $BaTiO_3$ is chosen as model surface, due to the fact that barium titanate is one of the most important and thus extensively studied of ferroelectric materials. Sample preparation and experimental procedure is described in Section 5.2. In Section 5.3 qualitative domain imaging on $BaTiO_3$ (100) surface and domain structure reconstruction from the combined topographic and property measurements are presented. In Section 5.4, the origins of domain contrast in EFM and SPM are discussed and based on the magnitude of observed domain potential contrast the surface polarization charge is shown to be screened in ambience. Dynamic studies of thermal and temporal evolution of domain-related surface potential on $BaTiO_3$ surface are presented in Section 5.5. A thermodynamic model for screening process is developed in Section 5.6 and based on the experimental observations the relevant thermodynamic and kinetic parameters as well as possible mechanism of screening are established.

## 5.2. Experimental Procedures

The AFM and SSPM measurements were performed on commercial instrument (Digital Instruments Dimension 3000 NS-III). Conventional silicon tips (TESP, $l \approx 125$ µm, resonant frequency ~ 270 kHz) and metal-coated tips (MESP, $l \approx 225$ µm, resonant frequency ~ 60 kHz) were used. The lift height for the interleave scan in the SSPM was usually 100nm. The scan rate varied from 0.2 Hz for large scans (~40 µm) to 1 Hz for smaller scans (~10 µm). Our studies indicated that surface potential observed by SSPM signal saturates at driving voltage ~ 1-2 V for the lift heights used and thus driving voltage $V_{ac}$ in the interleave scan was taken to be 5 V. Variable temperature measurements were performed on the home-built heating stage. In order to reduce the noise during the imaging, a low noise power supply was used. During measurements, the temperature was increased in steps of ~10°C and the system was kept at the selected temperature for ~0.5 h in order to achieve the thermal equilibrium. The cantilever was re-tuned at each step in order to stay in the vicinity of the resonance frequency. Thermal drift was corrected by adjusting lateral offsets to position of domain-unrelated topographical features. The lateral drift of the tip with respect to the surface were usually 2-3 μm per 10°C except in the vicinity of Curie temperature, where the ferroelectric phase transition was accompanied by significant (~10 μm) lateral displacements of the surface.

In order to obtain reliable values of surface corrugation angle, the images were acquired so that the *c-a* domain walls were oriented along the slow scan axis. The profiles were averaged along the direction of the domains (*y*-axis). The errors for the corrugation angle values at different temperatures were estimated from the averaged values obtained for different domains, rather than from the averaging along the single domain. To improve the statistics, usually two images were acquired at each temperature. It should also be noted that due to the relatively large lateral size (~5-20 μm) of surface potential features as compared to image size (20-40 μm) special care is taken during the processing of SSPM images to avoid artifacts. For example, zero order flattening (i.e. the offset line is subtracted from each scan line) removes the potential variations due to potential features oriented perpendicular to the slow scan axis. All room temperature SSPM images are shown without any image processing or with a subtracted offset plane of zero tilt. Zero order flattening was applied for variable-temperature images to improve observed potential contrast. This image processing doesn't alter relative height of features in *x*-direction. Usual image processing ($1^{st}$ order flattening) was applied for all topographic images to compensate the tilt of the sample surface and the effect of thermal fluctuations during the image acquisition.

Barium titanate single crystals (5x5x1 mm, $T_c$ = 130°C, Superconductive Components, Inc) were used for these studies. As evidenced by surface morphology, these crystals were polished in the cubic phase. The roughness of the (100) face did not exceed 15 Å. Prior to further analysis the crystal was repeatedly washed in acetone and deionized water. In order to obtain a reproducible well-developed domain structure the crystal was heated above Curie temperature, kept at 140°C for ~0.5 h and cooled down on metallic surface. The domain structure of the crystal was then characterized by polarized-light optical microscopy.

### 5.3. Domain Structure Reconstruction from SSPM

The tetragonal symmetry of $BaTiO_3$ unit cell results in characteristic surface corrugations at 90° *a*-*c* domain walls (provided that the crystal was polished in cubic phase) as illustrated on Figure 5.2. The corrugation angle is $\theta = \pi/2 - 2\arctan(a/c)$, where $a$ and $c$ are the parameters of the tetragonal unit cell. Topographic imaging distinguishes *a*-*c* walls only. At the same time, the difference in electric properties of the surface, loosely referred to as "potential" in Figure 5.2, distinguishes *c*-domains of opposite polarity. Potential and topographic information alone does not distinguish antiparallel *a* domains (invisible in optical microscope) and *a*1-*a*2 domain walls (visible in optical microscope). In these cases, lateral PFM on NSOM measurements are required. Additional domain variants are presented by charged domain walls (e.g. head to head or tail to tail), which can be identified in PFM from polarization orientation and electrostatic SPM due to the presence of large domain wall charge. All $BaTiO_3$ crystals examined here the dominant elements of the domain structure were uncharged domain walls shown on Figure 5.2 and complementary information in topography and surface potential allowed the reconstruction of the surface domain structure.

Polarized light microscopy, AFM and SSPM allowed the following types of domain structures to be characterized. After a typical poling process, the central part of the crystal consists of large lamellar domains oriented at 45° to the edges of the crystal. The absence of significant topographic and potential variations allows this domain structure to be ascribed to *a*1-*a*2 domain arrangements. Close to the edge of the crystal, regions with *a*-*c* orientation are present. If the size of the *c*-domains is relatively small, then 180° walls perpendicular to 90° domain boundaries between *a* and *c* domains (Figure 5.3a,c,e) are formed. Similar domain arrangements have been reported elsewhere.[53] This domain pattern can be ascribed to *c* domain wedges in the crystal with dominating *a* domain structure. The formation of 180° walls within the wedge minimizes the depolarization energy. If *c* domain regions are large (Figure 5.3b,d), irregular 180° walls separating $c^+$-$c^-$ domains exist. These walls are continuous through *a* domain regions, indicating the presence of *a* wedge domains in preferentially *c* domain material (Figure 5.3f). More complex domain structures can also be observed. Figure 5.4 shows the boundary between regions with *a*1-*a*2 (left side) and $c^+$ - $c^-$ (right side) domain arrangements. The optical micrograph clearly indicates the presence of *a*1-*a*2 boundaries (left). Minor lines (right) can be observed only for small focus depths indicating a near-surface character. Large scale AFM imaging indicates that large surface corrugations (Figure 5.4a) are associated with the presence of 90° domain walls. The measured corrugation angle $\theta \approx 0.62°$ is very close to calculated value ($\theta = 0.629°$). The surface potential indicates that the left region of the image is not associated with significant potential features, while clear $c^+$-$c^-$ domain regions are present on the right side.

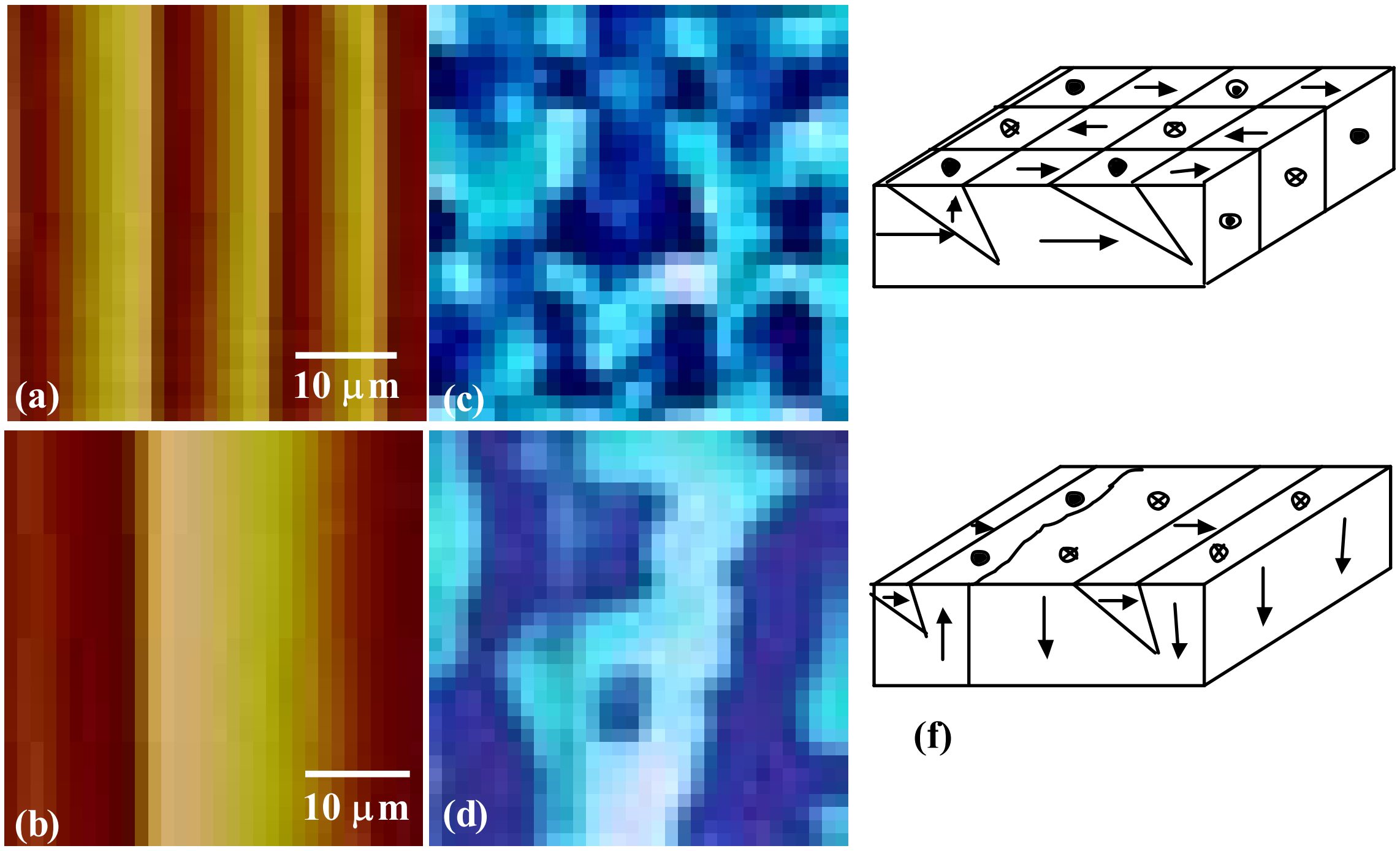

**Figure 5.3.** Surface topography (a,b), surface potential (c,d) and schematics of domain structure (e,f) in *a*-domain region with *c*-domain wedges (a,c,e) and in *c*-domain region with *a*-domain wedges (b,d,f).

Noteworthy is that small horizontal potential features are also observed on the SSPM image. Figure 5.4d,e shows the expanded scan of the right region. Surface corrugations corresponding to the 90° domain walls are now clearly seen (note the difference in vertical scales between Figure 5.4a and d). The surface potential image from the same region (Figure 5.4e) shows both potential features corresponding to surface *a*-*c* domain structure and bulk $c^+$-$c^-$ domain arrangement. In this case, domain contrast originates from surface polarization as well as from bound screening charges on charged domain boundaries. Formation of this domain structure is ascribed to the strain in the near-surface layer associated with macroscopic 90° domain wall between with *a*1-*a*2 and $c^+$-$c^-$ domain regions. Similar near-surface domain structure reconstruction was observed for polycrystalline $BaTiO_3$ as well, albeit in this case the built-in stresses inherent in ceramic material are expected to dominate the domain reconstruction process.

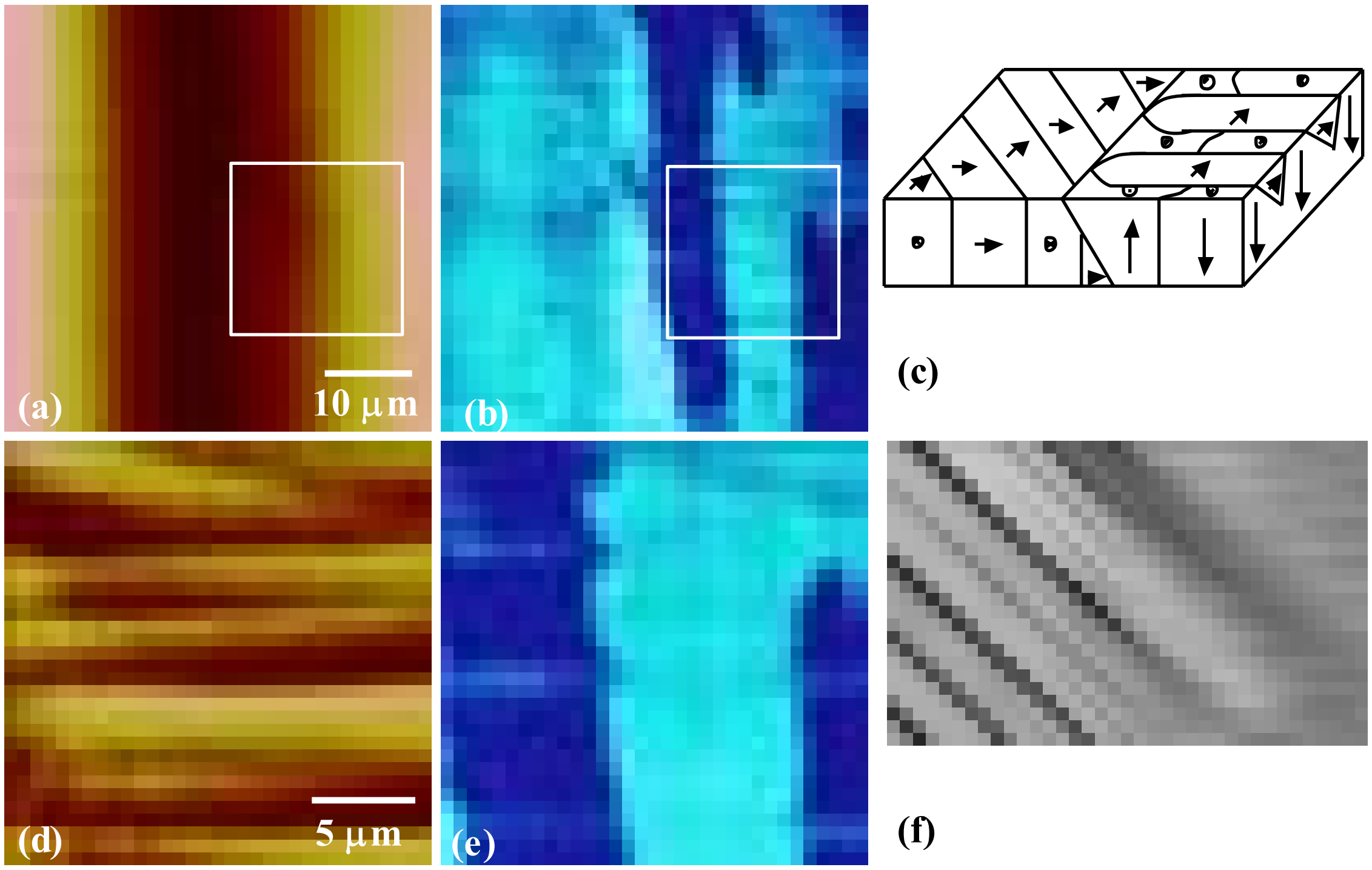

**Figure 5.4.** Surface topography (a,d), surface potential (b,e), domain structure reconstruction (c) and polarized light optical micrograph (f) in the region with complex domain arrangement. Scale is 100nm (a), 10 nm (d), 0.2 V (b,e).

SSPM imaging yields potential difference between $c^+$ and $c^-$ domains as ~150 mV; between *a* and *c* domains as ~75 mV. Surface potential is virtually uniform within the domains. The obvious question is what is the physical meaning of these numbers and how do they relate to the polarization charge and the chemistry of this surface. The less obvious question is what is the qualitative relationship between domain potential and the polarization direction is, i.e. whether domains positive on potential image are $c^+$ or $c^-$.

## 5.4. Origins of Domain Contrast in EFM and SSPM

In quantification of electrostatic SPM on ferroelectric materials the vast majority of authors assume that a ferroelectric surface is characterized by an unscreened polarization charge density $\sigma = \mathbf{P} \cdot \mathbf{n}$, where **P** is the polarization vector and **n** is the unit normal to the surface[42,54,55] It is well known, however, that polarization is always screened on ferroelectric surfaces.[50] Hence, in the present Section image formation in several SPM techniques is analyzed to establish the presence of the screening charges and analyze the effect on the surface properties.

To quantitatively address electrostatic properties of ferroelectric surfaces the surface layer is represented with polarization charge $\sigma_{pol} = \mathbf{P} \cdot \mathbf{n}$ and screening charge equivalent to surface charge density, $\sigma_S$, of the opposite polarity. The following cases can be distinguished:

1. Completely unscreened, $\sigma_S = 0$,
2. Partially screened, $\sigma_{pol} > -\sigma_S$,
3. Completely screened, $\sigma_{pol} = -\sigma_S$,
4. Overscreened, $\sigma_{pol} < -\sigma_S$.

A completely unscreened surface is extremely unfavorable from an energetic point of view due to the large depolarization energy. An overscreened surface is likely to occur during bias-induced domain switching and indeed has been observed.[56,57] Partially or completely screened surfaces are likely to be the usual state of ferroelectric surfaces in air. The charge distribution on a ferroelectric surface can be represented in terms of a double layer of width, *h*, dipole moment density $h \cdot min\lfloor \sigma_{pol}, \sigma_S \rfloor$ and an uncompensated charge component, $\delta\sigma = \sigma_{pol} - \sigma_S$. For future discussion, it is convenient to introduce degree of screening $\alpha = -\sigma_s / \sigma_{pol}$. Here it is assumed that the screening is symmetric, i.e. the degree of screening for $c^+$ and $c^-$ domains is the same. This assumption is supported by the experimentally observed near-equality of potential differences between $a$-$c^+$ and $a$-$c^-$ domains. Depending on the relative spatial localization of the polarization and screening charges (e.g. on the polarity of dipole layer), surface potential in the completely screened case can have the same sign as $\sigma_{pol}$, or be of the opposite sign.

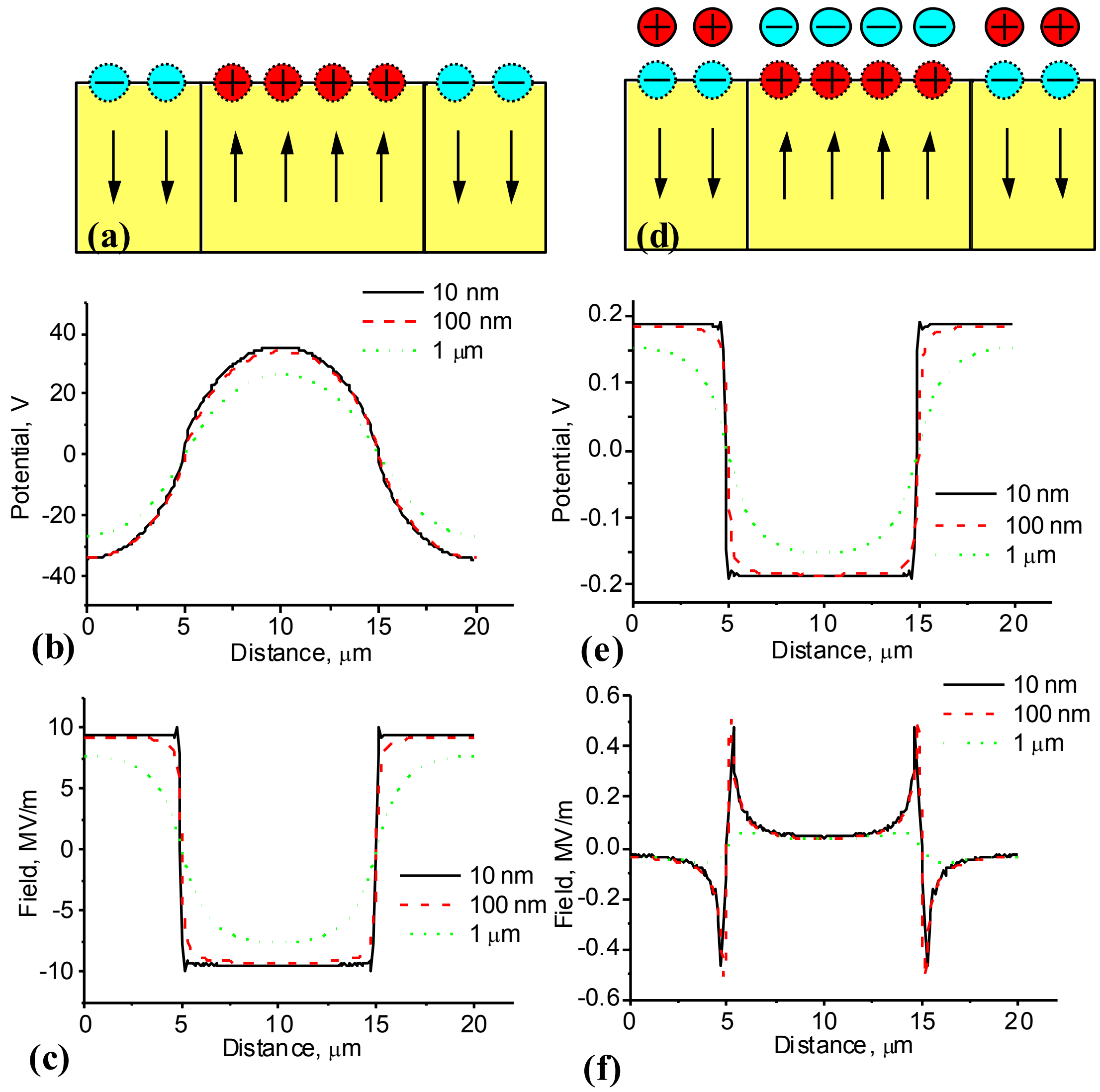

**Figure 5.5.** Simplified surface charge distribution (a,d), potential (b,e) and the field (c,f) in the vicinity of ferroelectric surface for unscreened (a,b,c) and completely screened (d,e,f) cases. Surface charge density is 0.25 C/m$^2$, domain size 10 μm, width of the double layer 2 nm.

To analyze the origins of image contrast in EFM and SSPM, it is instructive to calculate the potential and field distributions above ferroelectric surface in the completely screened and completely unscreened cases. Usually non-contact measurements are performed at tip-surface separations of 10-100 nm, which are much smaller then typical domain sizes (~1-10μm). Typical values of potential and field are calculated in Appendix 5.A and shown in Figure 5.5. For the partially screened surfaces, the potential and the field are a linear superposition of profiles for completely screened and unscreened surfaces. Simple arguments predict that the surface potential above the unscreened surfaces and electric field above the completely screened surfaces scale linearly and reciprocally with domain size, while electric field over the unscreened surfaces and potential over the screened surfaces are virtually domain size-independent.

Thus, experimentally observed uniform image contrast within the domain can be attributed either to the potential variation above the surface and corresponding change of the capacitive interaction, or to the variation in the surface charge density and normal electric field that results in additional Coulombic interaction between the tip and the surface. Hence, image contrast alone is insufficient to distinguish these contributions and detailed analysis of force gradient-distance (EFM) and force-distance (SSPM) data is required. Such detailed analysis of tip-surface interactions in EFM and SSPM of ferroelectric surfaces is presented in the next two sections.

### 5.4.1. Electrostatic Force Microscopy of the $BaTiO_3$ (100) Surface

As discussed in Chapter 2, the force gradient acting on the probe at intermediate tip-surface separations is governed by tip bulk and cantilever contributions. Assuming that domain size is comparable or larger than the tip size (which is usually true), but much smaller than the cantilever size, the tip interacts with a single domain, and the cantilever detects the average surface potential. In the following discussion the polarization charges are assumed to be almost completely screened by surface adsorbates and/or free carriers, equivalent to the presence of a double layer, characterized by potential $V_s$. The additional contribution due to the Coulombic interaction, $F_{coul}$, associated with the uncompensated surface charge density, $\delta\sigma$, is omitted in this description, but can be easily incorporated.

The capacitive electrostatic force between the tip and the surface is

$$F(z) = (V_{tip} - V_s)^2 F_t(z) + (V_{tip} - V_{av})^2 F_c(z), \tag{5.1}$$

where $F_t(z)$ is the tip contribution and $F_c(z)$ is a non-local cantilever contribution to the probe-surface capacitance gradient.

For a tip shape including the tip bulk and rounded tip apex the capacitive force and force gradient are related to tip parameters as:

$$F_{cap} = V^2\left(\frac{\gamma}{z} + \eta \ln\left(\frac{D}{z}\right)\right), \text{ and } \frac{dF_{cap}}{dz} = V^2\left(-\frac{\gamma}{z^2} + \frac{\eta}{z}\right), \tag{5.2a,b}$$

where $\gamma$, $\eta$ and $D$ are tip-shape dependent parameters which can be found experimentally from force or force-gradient - distance dependencies. The cantilever force and force gradient can be approximated by simple plane-plane capacitor model [Eqs.(2.14a,b)].

The force gradient can be derived from Eq.(5.1) and after grouping

$$F'(z) = V_{tip}^2\{F_t' + F_c'\} + V_{tip}\{-2V_s F_t' - 2V_{av}F_c'\} + V_s^2 F_t' + V_{av}^2 F_c'. \tag{5.3}$$

The average force gradient determined experimentally as the average of all image points is:

$$F_{av}'(z) = V_{tip}^2\{F_t' + F_c'\} - 2V_{tip}V_{av}\{F_t' + F_c'\} + V_{av}^2(F_t' + F_c'), \tag{5.4}$$

or

$$F_{av}'(z) = A_2 V_{tip}^2 + A_1 V_{tip} + A_0, \tag{5.5}$$

provided that the image size is large compared to the domain size. It should be noted here that the frequency shift proportional to the measured force gradient in EFM often has an additive constant due to the slow drift of the oscillation characteristics of the cantilever, but quadratic and linear coefficients in tip bias can be easily extracted.

The force gradient difference between domains of different polarities with surface potentials $V_1$ and $V_2$ is:

$$F_d'(z) = -2V_{tip}(V_1 - V_2)F_t' + (V_1^2 - V_2^2)F_t', \tag{5.6}$$

or

$$F_d'(z) = B_1 V_{tip} + B_0. \tag{5.7}$$

If the experimentally determined average force gradient and the difference in force gradients above domains with different polarity are quadratic and linear in voltage respectively, the constants $A_2$, $A_1$ and $B_1$, $B_0$ can be extracted. Our previous estimates (Chapter 2) suggest that $F_c'$ can be neglected compared to $F_t'$ for intermediate tip-surface separations. In this case, in the absence of a Coulombic contribution from unscreened charges the coefficients in Eqs.(5.5,7) yield the following universal ratios:

$$\frac{B_1}{A_2} = -2(V_1 - V_2), \tag{5.8a}$$

$$\frac{B_0}{B_1} = \frac{V_1 + V_2}{-2}, \tag{5.8b}$$

$$\frac{A_1}{A_2} = -2V_{av}, \tag{5.8c}$$

Note that these ratios are independent of the probe properties and are distance-independent. Conversely, if these ratios are distance independent, then the observed contrast between domains of different polarity can be attributed to the double layer contrast without a free charge contribution, since the distance dependencies of the two are different. By fitting the distance dependence of $A_2$ and $B_1$ to Eq.(5.2) and using the results summarized in Chapter 2, the relative contributions of the tip apex and the tip bulk to the overall force gradient can be estimated.

The Coulombic contribution to the tip-surface force and force gradient related to the unscreened charge can be estimated using a line charge model. The total force between the biased tip and the surface can be written as

$$F(z) = \frac{\mathrm{d}C(z)}{\mathrm{d}z}\Delta \mathrm{V}^2 + \int \frac{\partial \varphi_{sc}}{\partial \mathbf{n}}(\sigma_{tip} + \sigma_{ind})\mathrm{d}\mathbf{S}_{tip}, \tag{5.9}$$

where the first term is the capacitive force, $F_{cap}(z)$, and the second term is a contribution due to the Coulombic interaction of uncompensated charges with the metallic tip, $F_{coul}(z)$. $\sigma_{tip}$ is surface charge density of the tip without uncompensated charges, $\sigma_{ind}$ is the image charge density induced by uncompensated charge, and **n** is the normal vector to the tip surface. Assuming that the second term in Eq.(5.9) is much smaller than the first, $\sigma_{ind} << \sigma_{tip}$, the second term in Eq.(5.9) becomes:

$$\int \frac{\partial \varphi_{sc}}{\partial \mathbf{n}} \sigma_{tip}\, \mathrm{d}\mathbf{S}_{tip} = \int_{d}^{L+d} \lambda_{tip} \varphi'_{sc}\, \mathrm{d}z \approx \lambda_{tip} \varphi_{sc}(d), \tag{5.10}$$

since $\varphi_{sc}(z)$ rapidly decays with tip-surface separation. The decay length for the electric field is in this case is comparable with the domain size. For uniformly charged surfaces this assumption is no longer valid; however, the electric field can be assumed to be uniform in this case and the Coulombic force is then $F_{coul} = \lambda L E_n$, hence the SPM contrast for Coulombic and capacitive interactions is similar. Eq.(5.10) implies that for a dominant Coulombic interaction the tip-surface force is proportional to potential, while force gradient is proportional to electrostatic field. Hence, domain contrast in force sensitive (SSPM) and force gradient sensitive (EFM) SPMs can be expected to differ, unlike the completely screened scenario in which EFM and SSPM profiles are similar.

### 5.4.2. Bias and Height Dependence of Force Gradient

Surface topography, surface potential and force gradient images of a similar region are compared in Figure 5.6. Note that for positive tip bias (Figure 5.6c) the EFM image is similar to the SSPM image. For negative tip bias, the EFM image is inverted, as expected (Chapter 2). For zero tip bias the EFM image has the same sign as for a negatively biased tip, indicative of positive average surface potential. For large negative biases, the EFM image is unstable as seen in Figure 5.6f. It is unclear whether this effect should be attributed to a feedback loop instability, tip-induced desorption or charge transfer in the surface layer or dielectric constant (and hence capacitive force) difference between *a* and *c* domains. To minimize this effect on surface properties, quantitative measurements were performed well inside the linear region.

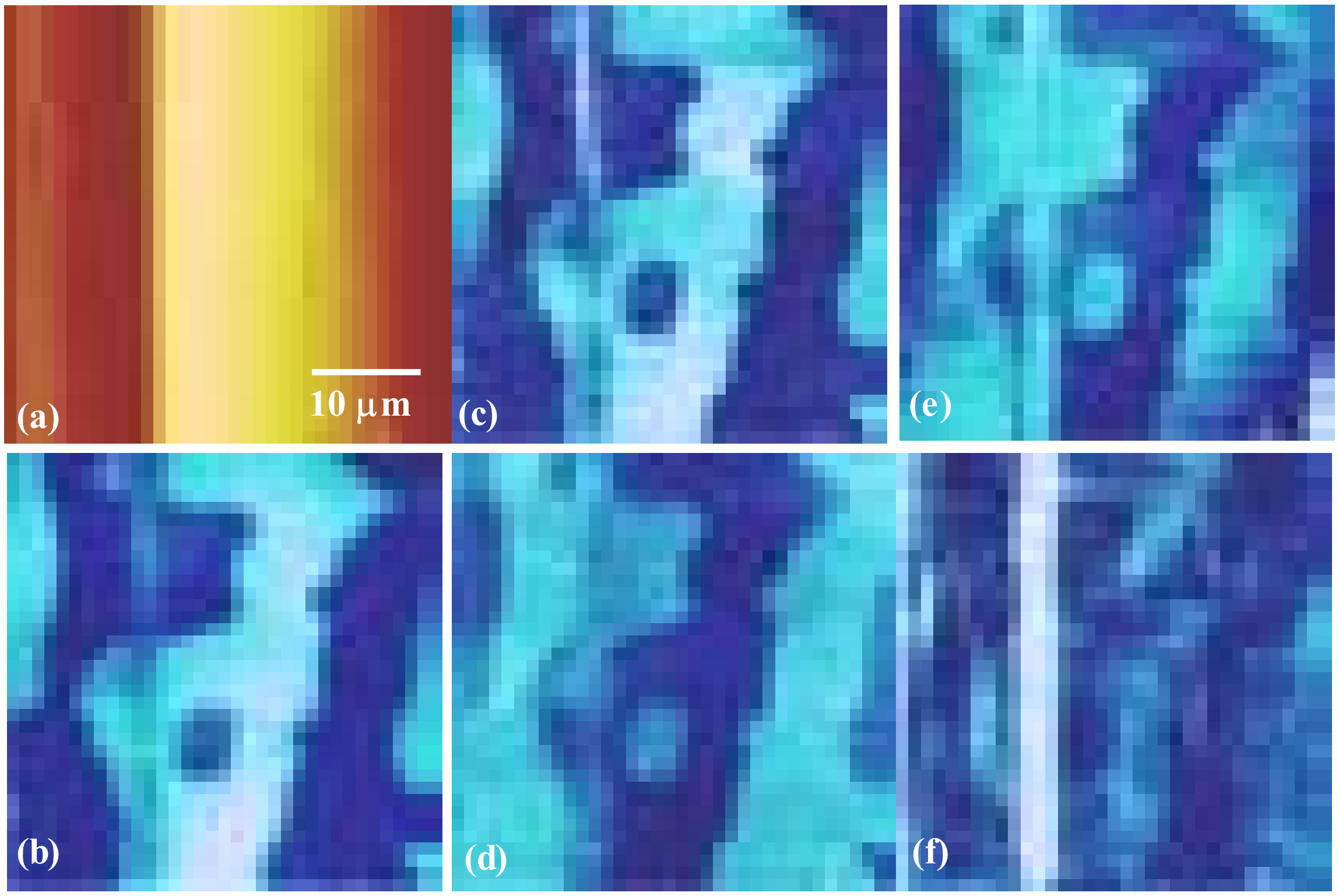

**Figure 5.6.** Surface topography (a), surface potential (b) and EFM images of $BaTiO_3$ (100) surface at tip bias of 5 V (c), 0 V (d), -2 V (e) and -5 V (f). Note the inversion of domain contrast with tip bias and abnormal image at large negative bias.

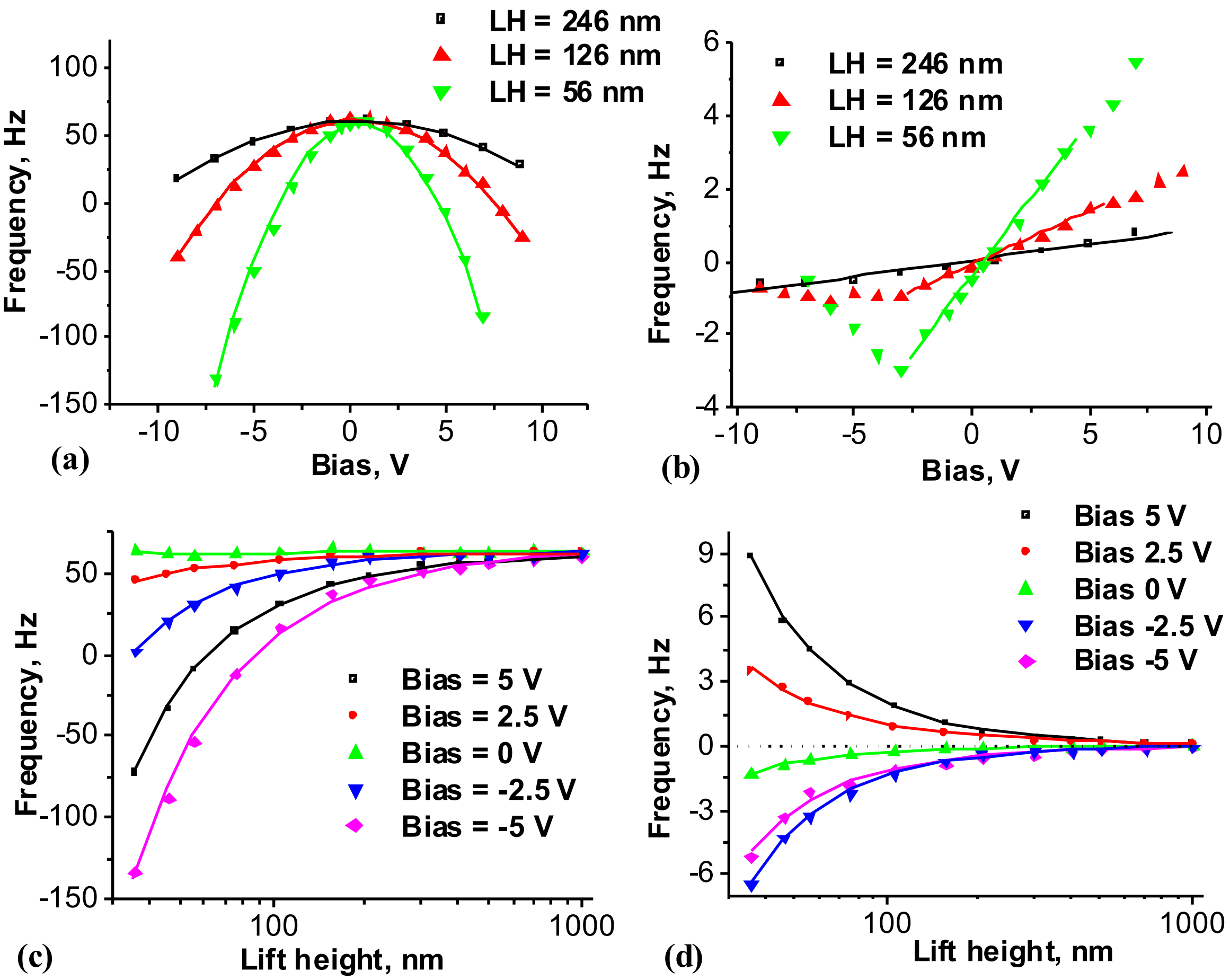

**Figure 5.7.** Bias dependence of average frequency shift (a) and domain frequency shift (b) in force gradient (EFM) images. Distance dependence of average frequency shift (c) and domain frequency shift (d) in force gradient (EFM) images for different tip biases.

The bias dependence of the average force gradient and the domain force gradient are compared in Figure 5.7a,b. As expected, the average force gradient is a parabolic function of the bias voltage; the data are described by Eq.(5.5). The zero order term $A_0 \approx$ 60Hz includes a frequency offset due to drift in the oscillating characteristics of the cantilever after calibration and depends on the tip. The domain force gradient dependence is linear and is approximated by Eq.(5.7). Large biases result in non-linear behavior of the domain force gradient even though the average force gradient follows Eq.(5.5) well. The distance dependence for the average force gradient and domain force gradient is shown in Figure 5.7c,d for several tip biases along with corresponding fits by Eq.(5.2b). It is clearly seen that a non-linear response in domain force gradient exists for all tip-surface separation studied, suggesting that this behavior should be attributed to the microscope instabilities rather than materials properties.

In order to quantify the distance dependence of EFM data, coefficients $A_2$ and $B_1$ for two tips were determined as a function of tip-surface separation (Figure 5.8a,b). These dependencies can be linearized in log-log coordinates and corresponding effective slopes are summarized in Table 5.I. The effective slopes are larger than expected for the line charge model (-1) and smaller than expected for the sphere model (-2), in agreement with previous studies on different systems.[58] As expected, the effective slope is smaller for a sharp tip, since the relative contribution of the tip bulk (i.e. line charge contribution) is larger in this case. In fact, the effective slope of the average force gradient for a sharp tip is almost equal to unity, implying that the line charge model can be used to describe the capacitive interaction in this case. To quantify the relative apex and bulk contributions to $A_2$ and $B_1$, experimental dependencies were fitted by Eq. 5.2b and fitting parameters are listed in Table 5.I.

Table 5.I.

*Distance dependence of average (A) and domain (D) frequency shifts.*

| Tip | Effective slope | $\gamma$, N nm$^2$/V$^2$ | $\eta$, N nm/V$^2$ |
|---|---|---|---|
| Dull A | -1.17 ± 0.04 | 4000 ± 300 | 136 ± 7 |
| Sharp A | -1.02 ± 0.05 | 860 ± 150 | 60 ± 4 |
| Dull D | -1.41 ± 0.02 | 1600 ± 70 | 28 ± 1.5 |
| Sharp D | -1.11 ± 0.01 | 144 ± 23 | 20.5 ± 0.6 |

The frequency shift due to force gradient can be found as

$$\Delta\omega_{im} = \frac{\omega_0}{2k}\frac{4\pi\varepsilon_0 V^2}{\beta^2}\frac{1}{h}. \tag{5.11}$$

Substituting the resonant frequency of the "dull" cantilever $\omega_0$ = 68.14 kHz, a typical spring constant for the cantilever $k$ = 1-5 N/m and a typical tip half-angle $\theta \approx 17°$, the frequency shift according to Eq.(5.11) yields proportionality coefficient equal to 235 - 47 nm/s $V^2$, which is in excellent agreement with the experimental results. The spring constant for the cantilever is therefore estimated as $k \approx 1.75$ N/m.

As shown above, the distance dependence of fitting coefficient ratios can be used to determine the relative contributions to imaging contrast. The distance dependence of ratios $B_1/A_2$ and $B_0/B_1$ for sharp and dull tips is compared in Figure 5.8c. It is clearly seen that for small tip surface separations ($z$ < 100 nm) the ratios are almost distance independent. For larger tip-surface separations the measured values of domain force gradient and variations of average force gradient are small compared to typical noise levels (~0.1-1Hz), consequently errors in fitting coefficients are large in this region. Average potential determined from $A_1/A_2$ (Eq.(5.8c)) is shown in Figure 5.8d and summarized in Table 5.II.

Table 5.II.

*Fitting coefficient ratios for EFM imaging of ferroelectric domains*

| Tip | $2\,(V_1-V_2) = -B_1/A_2$ | $(V_1+V_2)/2 = -B_0/B_1$ | $V_{av} = -A_1/2A_2$ |
|---|---|---|---|
| Dull | 0.27 ± 0.03 | 0.60 ± 0.08 | 0.53 ± 0.05 |
| Sharp *c-c* | 0.31 ± 0.04 | 0.55 ± 0.09 | 0.60 ± 0.07 |
| Sharp *a-c* | 0.17 ± 0.02 | 0.63 ± 0.09 | 0.60 ± 0.07 |

The absolute potential difference between adjacent domains is calculated as 668 mV - 533 mV = 135 mV (dull) and 628 mV - 473 mV = 155 mV (sharp). Therefore, the potential difference between $c^+$ and $c^-$ domains is $\Delta V_{c-c} \approx$ 135 - 155 mV. Noteworthy is that the average image potential, $V_{av}$, is approximately equal to $(V_1+V_2)/2$, i.e. effective surface areas of $c^+$ and $c^-$ domain regions are equal, as expected from energy considerations. The potential difference between $a$ and $c^+$ domains was similarly found to be 85 mV, i.e. approximately equal to the expected value $\Delta V_{a\text{-}c} \approx \Delta V_{c\text{-}c}/2$. Domain potentials $V_1$ and $V_2$, and average image potential $V_{av}$ are combinations of four independent parameters $A_1$, $A_2$, $B_0$ and $B_1$ and thus are independent.

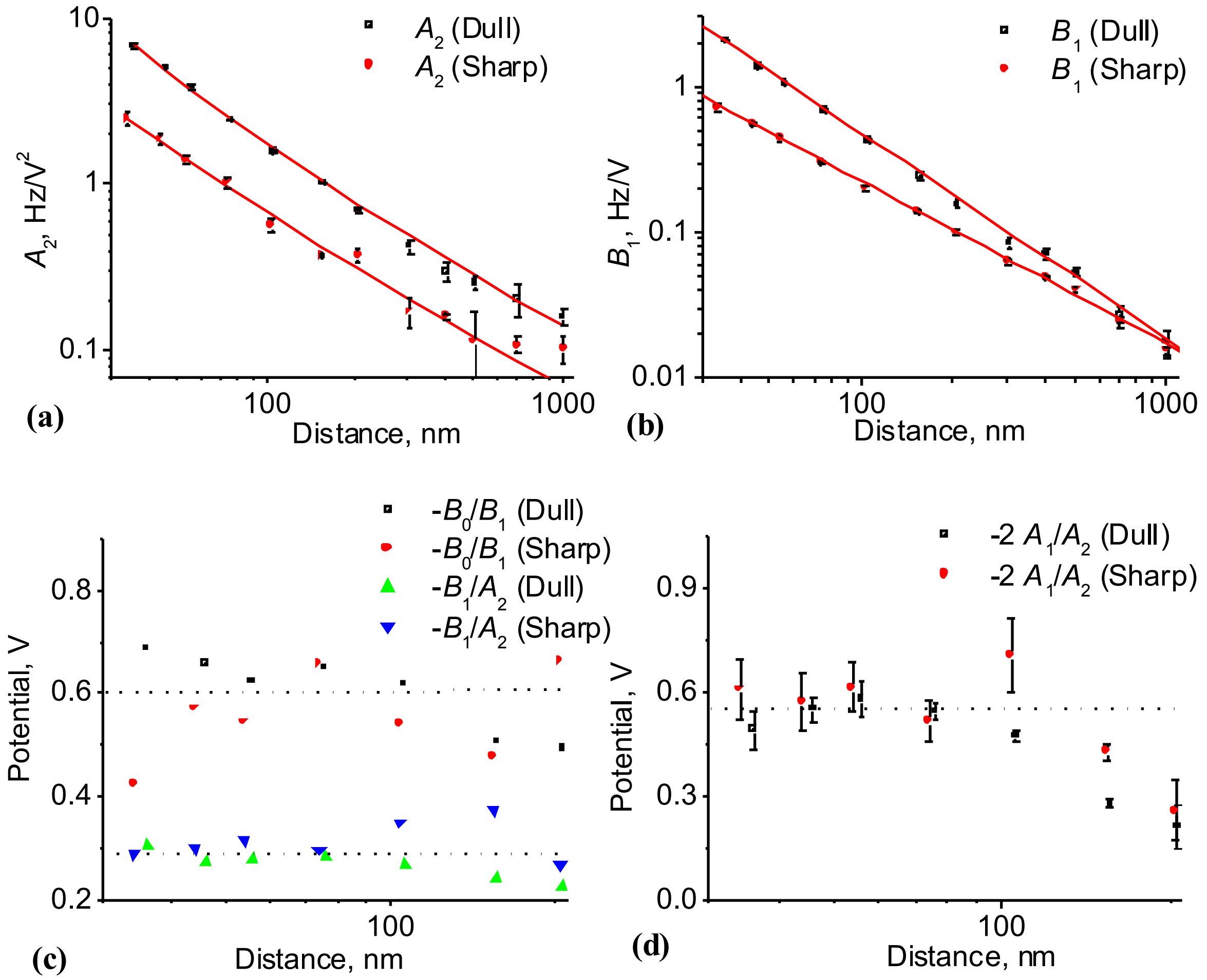

**Figure 5.8.** Coefficients $A_2$ (a) and $B_1$ (b) as a function of tip-surface separation for blunt and sharp tip. Distance dependence of universal fitting coefficient ratios $B_0/B_1$ and $B_1/A_2$ (c) and -2 $A_1/A_2$ (d) for sharp and dull tips. Note that the ratios are independent on distance, while the coefficient *per se* decrease by more than an order of magnitude.

### 5.4.3. Scanning Surface Potential Microscopy of $BaTiO_3$ (100) Surface

In order to quantify the SSPM contrast of ferroelectric surfaces both the cantilever contribution and the non-ideality of feedback loop must be taken into account.[59,60] Using the results in Chapter 2, the first harmonic of the electrostatic force between the tip and a complete screened ferroelectric surface is

$$F_{1\omega}(z) = V_{ac}(V_{dc} - V_s)F_t + V_{ac}(V_{dc} - V_{av})F_c. \tag{5.12}$$

The operation of SSPM implies that the measured surface potential is

$$V_{dc} = \frac{V_s F_t + V_{av} F_c}{F_t + F_c} + \frac{\delta}{V_{ac}(F_t + F_c)}, \tag{5.13}$$

where $\delta$ is constant dependent on feedback loop parameters. Similar to EFM image analysis, the average image potential, $V_{dc}^{av}$, and the potential difference between domains of different polarity, $\Delta V_{dc}$, are defined as:

$$V_{dc}^{av} = V_{av} + \frac{\delta}{V_{ac}(F_t + F_c)}, \tag{5.14}$$

and

$$\Delta V_{dc} = (V_1 - V_2)\frac{F_t}{F_t + F_c}. \tag{5.15}$$

If Eq.(5.14) hold, the domain potential difference is independent of feedback operation. Taking expressions for the distance dependence of tip-surface forces [Eq.(2.12)] and cantilever-surface forces [Eq.(2.14)] and taking into account that the cantilever contribution to the force dominates as shown in Figure 2.2a, the distance dependence of measured domain potential contrast $\Delta V_{dc}$ is

$$\Delta V_{dc} \approx (V_1 - V_2)\frac{F_t}{F_c} \approx (V_1 - V_2)\frac{4\pi S}{L^2 \beta^2}\left(\ln(L/4) - \ln(z)\right). \tag{5.16}$$

Thus, experimentally measured potential difference between domains decreases logarithmically with tip-surface separation. Figure 2.2a suggests that saturation occurs only for very small tip-surface separations, when the contribution of the tip apex to the force is dominant. In this case, however, the tip-induced field is very large and can induce polarization switching or screening charge redistribution below the tip, while electrostatic force can also result in the contact between the tip and the surface.

### 5.4.4. Bias and Height Dependence of Surface Potential

Quantification of the SSPM data was done similar to that of the EFM data, i.e. average image potential and potential difference across the domain boundary were determined. Both driving voltage and tip surface separation dependencies were measured. According to Eq.(2.31), surface potential measured by SSPM is independent of bias voltage. In practice, however, the non-ideality of the feedback loop results in $1/V_{ac}$ dependence on driving amplitude, as predicted by Eq.(3.40) and shown in Figure 5.9. Thus, the average image potential $V_{av}$ is fit by $V_{av} = V_s + B/V_{ac}$, where $V_s$ is surface potential and $B$ is fitting parameter (Figure 5.9). Average surface potential is virtually distance independent, $V_s = 600 \pm 20$mV and coincides with the average surface potential determined by EFM. The coefficient $B$ increases for large tip-surface separations as predicted by Eq.(5.13) and summarized in Table 5.III.

Table 5.III

*Driving voltage dependence of SSPM images*

| Lift height, nm | $V_s$, mV | $B$, mV$^2$ |
|---|---|---|
| 6 | 576 ± 13 | 211 ± 4 |
| 16 | 602 ± 18 | 245 ± 5 |
| 56 | 627 ± 14 | 366 ± 4 |
| 206 | 607 ± 10 | 533 ± 3 |

The domain potential difference, $\Delta V_{dc}$, is virtually $V_{ac}$ independent above 2 V, in agreement with Eq.(5.14). At low driving voltages, there is considerable noise and possibly a small increase in measured potential. However, this effect does not exceed ~10-20 mV, while the dependence of the average image potential (Figure 5.9c) indicates a strong driving voltage dependence. This observation implies that domain boundary potential differences obtained by SSPM are insensitive to feedback parameters and Eq.(5.16) can be used to describe potential-distance relations. This also demonstrates that feedback parameters that strongly influence the absolute value of measured surface potential do not affect measured potential variations. The domain potential-distance dependence is shown in Figure 5.9d. In agreement with the previous discussion, these values are almost independent of driving voltage, and in fact are almost linear in semilogarithmic coordinates in a good agreement with Eq.(5.16). The distance dependence of domain potential differences was fitted by $y = a + b\ln(x)$. From Eq.(5.16) the ratio $a/b = \ln(L/4)$ and yield the effective tip length as $L \approx 14$ μm for all tips used, i.e. very close to expected tip length ($L = 15$ μm). The distance dependence of domain potential difference does not saturate in the tip-surface separation range studied, i.e. SSPM doesn't determine "true" potential difference between the domains because of the significant cantilever contribution to the measurements. Imaging at even smaller tip-surface separations suffers from imaging instabilities and the possibility for tip-induced polarization switching and charge transfer.

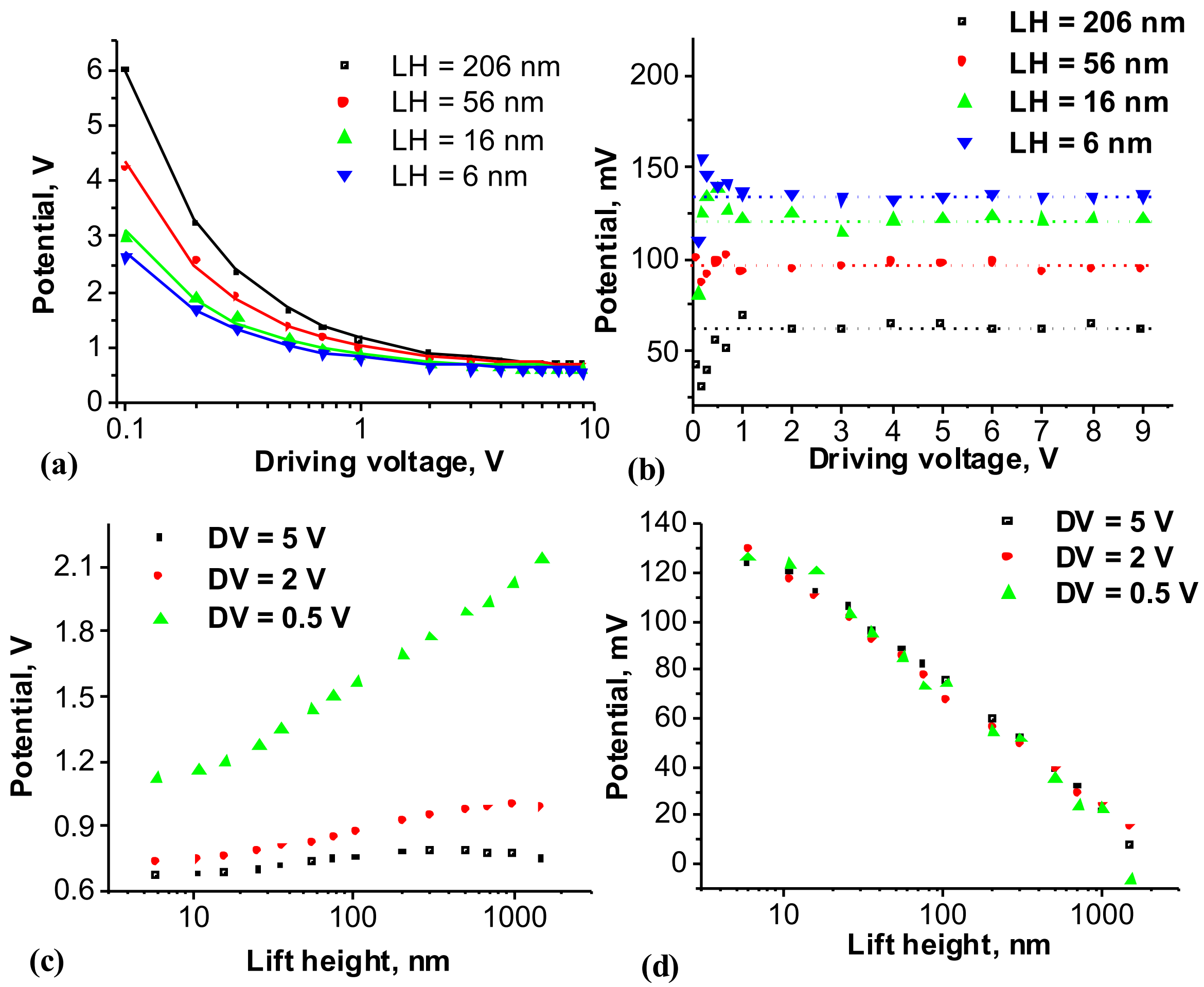

**Figure 5.9.** Driving voltage dependence of average image potential (a) and domain potential difference (b). Distance dependence of average image potential (c) and domain potential difference (d).

To summarize, EFM provides the true values of domain potential through universal ratios defined in Eq.(5.8a-c) obtained by multiple scans at different tip biases, while SSPM, though being experimentally simpler, is unable to provide the correct value of domain potential difference. Nevertheless, the potential difference from SSPM (~130 mV) measured directly in a single experiment, is remarkably close to the domain potential difference obtained from complex and time consuming analysis of EFM data (~150 mV), justifying the application of the former technique for ferroelectric materials characterization.

### 5.4.5. Polarization Screening

From qualitative observations, both EFM and SSPM contrast is found to be uniform within the domains with rapid variation at the domain boundaries. The magnitude of potential and force gradient features are virtually domain-size independent. From these observations, the contrast can be attributed either to electrostatic field for an unscreened surface (Figure 5.5c) or surface potential on a completely screened surface (Figure 5.5e). Both EFM and SSPM yield potential difference between $c^+$ and $c^-$ domains as $\Delta V_{c\text{-}c} \approx$ 150mV and between $a$ and $c$ domains as $\Delta V_{a\text{-}c} \approx \Delta V_{c\text{-}c}/2$. This value is much smaller than that expected for an unscreened surface, suggesting that polarization charge is largely screened. This is further verified by the distance dependence of the universal coefficient ratios (Figure 5.8). Therefore, the state of $BaTiO_3$ (100) surface under ambient conditions corresponds to almost complete screening of polarization bound charges.

As discussed, in ferroelectric semiconductors the screening can be attributed both to the adsorption and to redistribution of charge carriers in the material. Eq.(5.A.13) suggests that a potential difference of 0.140V is equivalent to a 0.20 nm double layer of a dielectric constant $\varepsilon_1$ = 80 ($H_2O$) on a ferroelectric substrate (external screening) or a 9.5 nm depletion layer in a ferroelectric with a dielectric constant $\varepsilon_2$ = 3000 (intrinsic screening). While the former estimate is reasonable for a molecular adsorbate layer or occupation/depletion of surface states, the latter is unreasonably small for a depletion layer width in a semiconductor with a low charge carrier concentration (~1 μm).

Moreover, potential differences between $c^+$-$a$ and $c^-$-$a$ domains are almost equal, suggesting that the screening is symmetric. This is not the case if the screening is due to the free carriers in materials with a predominant electron or hole conduction, in which the width of accumulation layer for the polarization charge opposite to the majority carrier charge and width of depletion layer for the polarization charge similar to the majority carrier charge are vastly different. Thus, surface adsorption or intrinsic surface states are the dominant mechanism for polarization screening on a ferroelectric surface in ambient conditions, though a minor contribution from intrinsic screening can not be excluded. Noteworthy is that the average surface potential is approximately equal to average domain potential between $c^+$ and $c^-$ domains, $V_{av} \approx (V_1+V_2)/2$. This observation implies that surface areas occupied by $c^+$ and $c^-$ domains are equal, as expected from considerations of electrostatic energy minimization.

### 5.4.6. Ferroelectric Domain Wall Widths

The additional feature of ferroelectric domain structure is domain wall width. Intrinsic ferroelectric domain wall width is widely debated in the literature, but based on HRTEM and piezoresponse measurements it can be estimated as < 10nm. Domain wall widths measured by SSPM are well above this value and can be as high as several hundred nanometers. The major contributions to the measured widths are related to SSPM resolution, screening charge redistribution in the vicinity of domain wall (the origin of this effect can be traced to the field discontinuities on Figure 5.5f) and lateral spreading of stray field with tip-surface distance and include:

1. Intrinsic domain wall width in the bulk
2. Lateral spreading of the field in air
3. Tip shape effect (SSPM resolution)
4. Feedback effects (SSPM artifacts)
5. Surface charge redistribution

Lateral spreading of the field can be calculated analytically as illustrated below. The resolution in SSPM experiment strongly depends on the properties of the probe used, however, using the appropriate calibration standard can be shown to be comparable or smaller than the observed domain wall widths. Under optimal imaging conditions, feedback effects are unlikely to cause significant deterioration in resolution. Surface charge redistribution can result in significant widening of potential features.

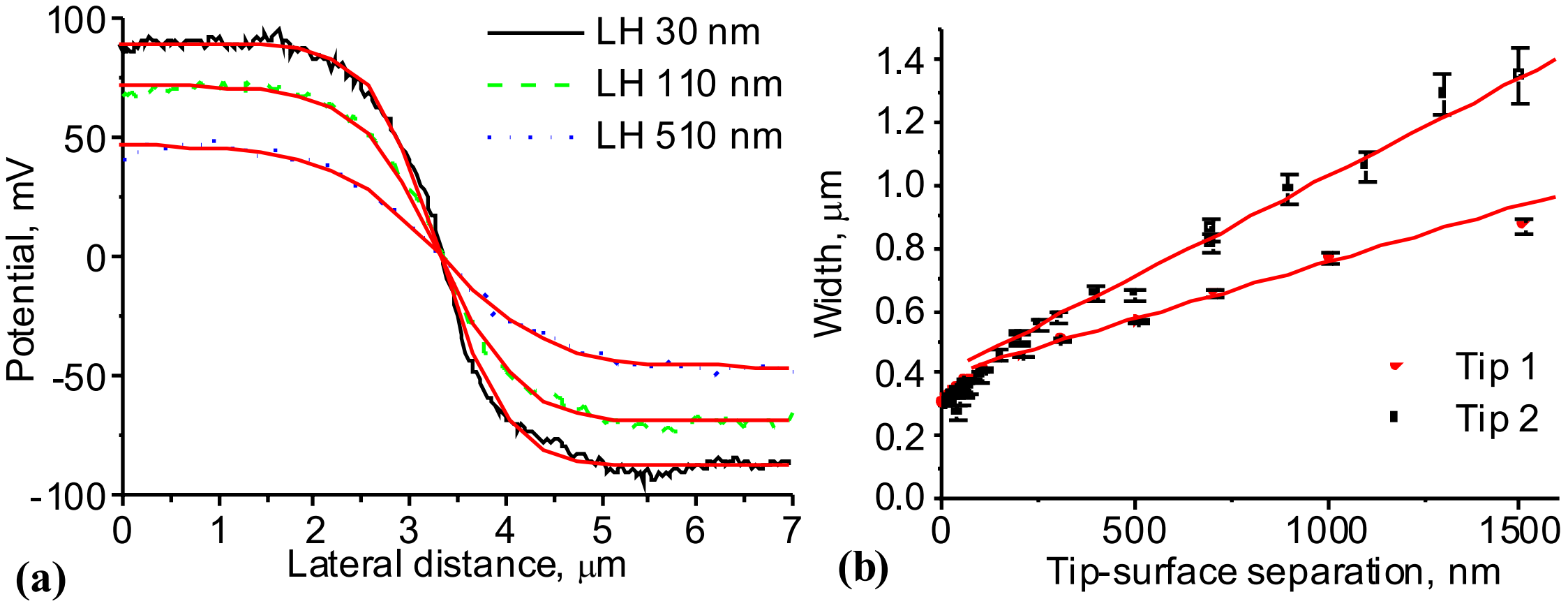

**Figure 5.10.** Averaged potential profiles at different tip-surface separations and corresponding sigmoidal fits (a) domain wall width as a function of tip surface separation for different tips (b).

To establish the limitations on measured domain wall in SSPM measurements, experimental results were compared to the modeling of the lateral spreading of field above atomically sharp domain wall. To theoretically estimate distance dependence of measured domain wall width, we define it as

$$w = \left(\varphi_{c^+} - \varphi_{c^-}\right) \Big/ \left(\frac{\partial \varphi}{\partial x}\right)_{dw}, \tag{5.17}$$

where the derivative is taken along the normal to the wall. From the solution for the completely screened case (Appendix 5.A) distance dependence of $w$ can be estimated as

$$w = \pi z - \frac{2\pi z^2}{L}, \tag{5.18}$$

i.e. domain wall width increases linearly with slope close to $\pi$. The slope obtained from the fit of the calculated potential profile is summarized in Table 5.IV. The width, $dx$, of the profile determined by fitting experimental data by Boltzmann function

$$y = \frac{A1 - A2}{1 + \exp\left(\left(x - x_0\right)/dx\right)} + A1, \tag{5.19}$$

where $A1$, $A2$ are offsets and $x_0$ is the position of the domain wall. The profile width is related to $w$ as $w = 4\,dx$.

The presence of surface charges results in spreading of potential on the surface, and in the first approximation this effect can be taken into account by writing

$$w \approx w_0 + w_s \approx w_0 + \pi z, \tag{5.20}$$

where $w_0$ is proportional to the Debye width of the free charges on the surface.

The average potential profiles across $c^+$-$c^-$ domain walls are shown in Figure 5.10a. The measurements were performed with a sharp tip (tip 1) and an uncharacterized tip (tip 2). Profiles were fitted by Boltzmann function and dependence of characteristic wall width on tip-surface separation is shown in Fig. 5.10b and summarized in Table 5.IV.

Table 5.IV

*Distance dependence of ferroelectric domain wall width*

| | Intersect, μm | Slope | Width |
|---|---|---|---|
| Theory | | 0.78 | |
| Modeling | 0.12 ± 0.02 | 0.61 ± 0.03 | 2.42 |
| Tip 1 | 0.38 ± 0.02 | 0.66 ± 0.03 | 2.64 |
| Tip 2 | 0.41 ± 0.02 | 0.32 ± 0.03 | 1.28 |

These data indicate that for large tip-surface separation domain wall width grows linearly with separation consistent with tip shape effects and generic spreading of field in air. The numerical value of the slope is very close to the value obtained from simulations and theoretical arguments for field spreading, indicating the latter is the major mechanism for the distance dependence of domain wall width. Note that for small separations width tends to non-zero value of ~300 nm. This value is consistent with surface adsorbate screening of polarization bound charges, in which case the domain wall width at the surface is comparable to the Debye length. Based on the tip calibration studies using carbon nanotube standards, profile widening due to the tip shape effect is somewhat smaller and the observations strongly suggest that there is a significant contribution of screening charges to the observed domain wall width. Our results indicate that this value (~300 nm) is generic for lateral potential screening on $BaTiO_3$ and $SrTiO_3$ surfaces.[61,62]

### 5.5. Polarization Dynamics of the $BaTiO_3$ (100) Surface

To study the dynamic behavior of ferroelectric domain structure (i.e. domain wall motion and phase transition) and to determine the nature and properties (e.g. mobility and relaxation times) of surface charges we perform variable temperature (VT) SSPM and PFM imaging of ferroelectric phase transition.

#### 5.5.1. Phase transition and Potential Dynamics by VT SSPM

With some understanding of image contrast and domain structure, it is possible to examine the ferroelectric phase transition. Figure 5.11 shows the temperature dependence of topographic structure. The topography of the surface consists of four large corrugations oriented in *y*-direction. A number of small spots due to contaminates are also evident. The overall domain structure (i.e. number and relative size of domains) doesn't change below the transition temperature; however, the surface corrugation angle, which is directly related to *c/a* ratio in the tetragonal unit cell, changes with temperature. In order to obtain reliable measurement of corrugation angle, it is averaged over *y*-direction and over four domain walls for each image. Figure 5.12 shows the corrugation angle as a function of temperature in which the angle decreases with temperature and drops to zero at Curie temperature $T_c$ = 130°C. Note the agreement between experimentally measured corrugation angle and the value calculated from the temperature dependence of *a/c* ratio in $BaTiO_3$.[63]

SSPM measurements show that the surface potential distribution doesn't significantly change during heating, i.e., no domain wall motion is observed. The surface potential image represented on Figure 5.11b shows the large vertical features aligned in the same directions as domains. Two zones of reverse contrast – band-shaped and edge shaped can also be seen. The dark spot indicated by arrow represents the surface contamination that significantly (~ 70 mV) depresses the surface potential. On increasing the temperature, the contaminates were used to adjust for thermal drift. A stepwise increase in temperature results in an increase of domain potential contrast (Figure 5.12b);

keeping the sample at constant temperature for ~30 min results in the decay of potential contrast. The result is a saw-tooth like potential - temperature dependence. The amplitude of potential oscillations increases with temperature.

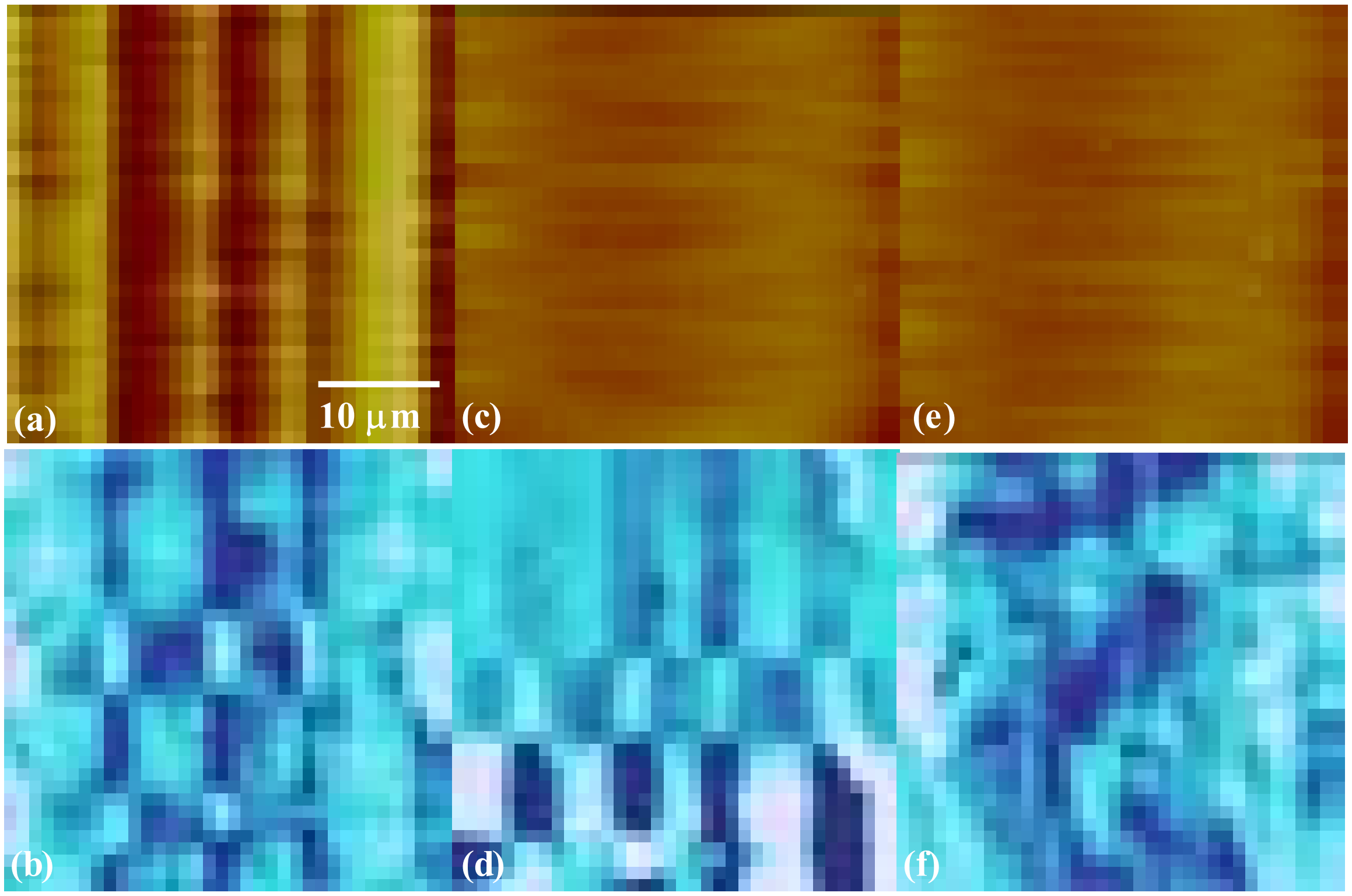

**Figure 5.11**. Surface topography and potential distribution at $BaTiO_3$ (100) surface before ferroelectric phase transition at 125°C (a,b), 4 min after transition (c,d) and after 2.5 h annealing at 140°C (e,f). Scale is 0.1 V (b), 0.5 V (d) and 0.05 V (f).

Above the Curie temperature, surface polarization disappears as indicated by the absence of characteristic surface corrugations. Unexpectedly, this is not the case for potential. The morphology of the potential features remains essentially the same (comp. Figure 5.11b,d), however, at the transition the potential amplitudes grow by almost 2 orders of magnitude. As can be seen from Figure 5.11d (the image was acquired from bottom to top 4 min after the transition, total acquisition time – 11 min) the potential amplitude decays with time. Surface potential distribution after remaining at 140°C for 2.5 h is shown in Figure 5.11f. The surface potential amplitude is now very small (~2-5 mV) and the potential distribution is almost random, though some resemblance to surface potential distribution below $T_c$ still exists. Note that the magnitude of domain unrelated potential features remains almost the same ($\phi_{hole}$ in Figure 5.13b).

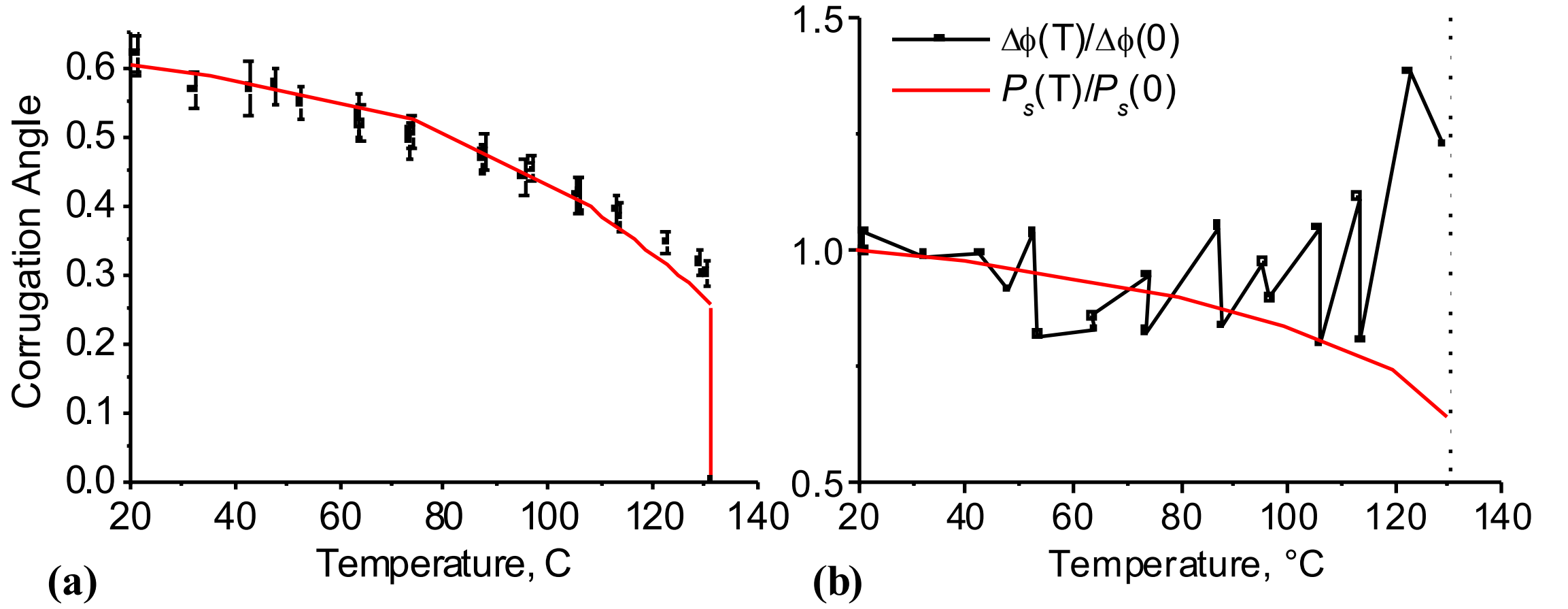

**Figure 5.12**. Temperature dependence of surface corrugation angle compared to the calculated value (solid line) (a) and domain potential contrast in SSPM measurements below Curie temperature(b).

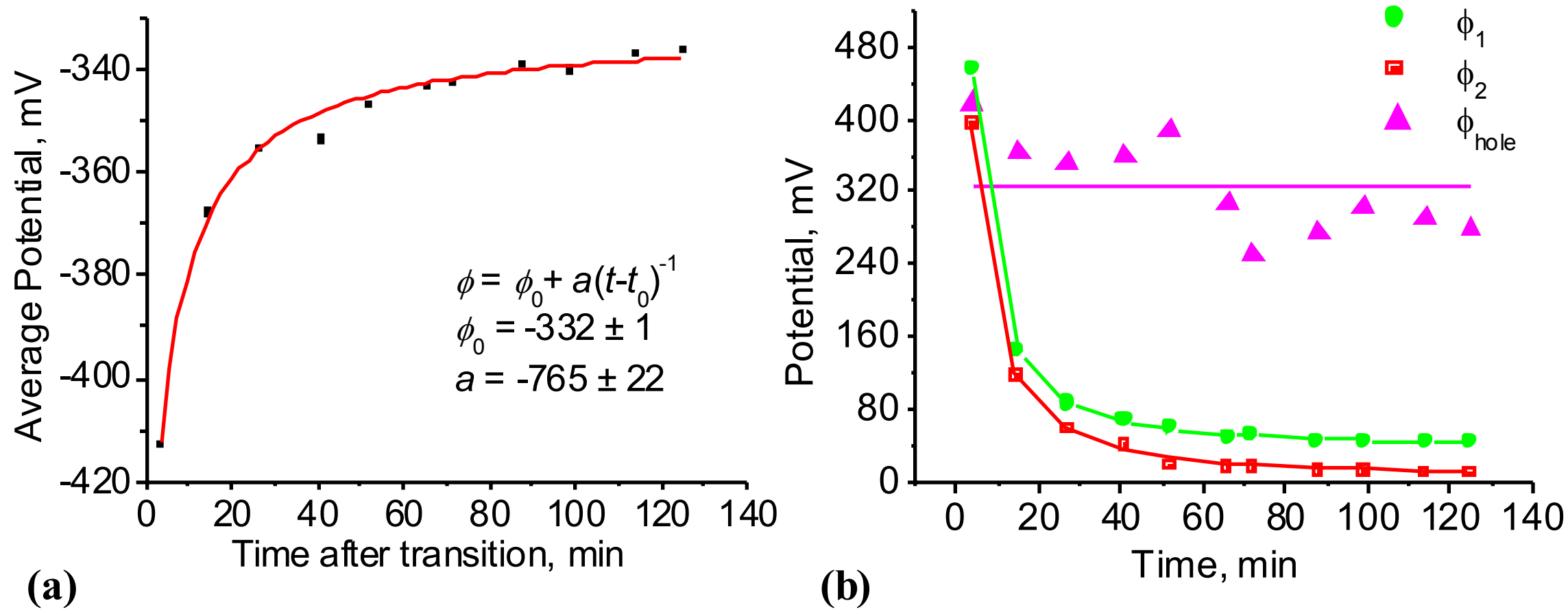

**Figure 5.13**. Time dependence of average surface potential (a) and domain potential contrast in SSPM measurements above Curie temperature (b).

In order to get preliminary insight into the kinetics of the relaxation process, the averaged surface potential and domain potential contrast were determined as a function of relaxation time (Figure 5.13). The kinetics of these relaxation processes were fitted by a power law of the form:

$$\phi = A + B(t + t_0)^d, \tag{5.21}$$

where $\phi$ is the surface potential amplitude, $t$ is the time of image acquisition, $t_0$ is the time when the region of interest within the image was acquired, $A$, $B$ and $d$ are fitting parameters. The power $d$ was ~ -1.5 for most potential amplitudes studied. Fits by other functions (e.g. exponential decay) were of much lower quality. The relaxation time was ~15 min both for average potential relaxation and for domain contrast relaxation, thus suggesting a similar origin for these relaxation processes (as opposed to two simultaneously occurring processes, e.g. charge relaxation on the surface and changes in the tip-surface contact potential difference tip due to the water desorption). The residual domain potential contrast for different spots varied from 3 to 10 mV after 2.5 h. The magnitude of domain-unrelated potential feature, however, remains almost constant.

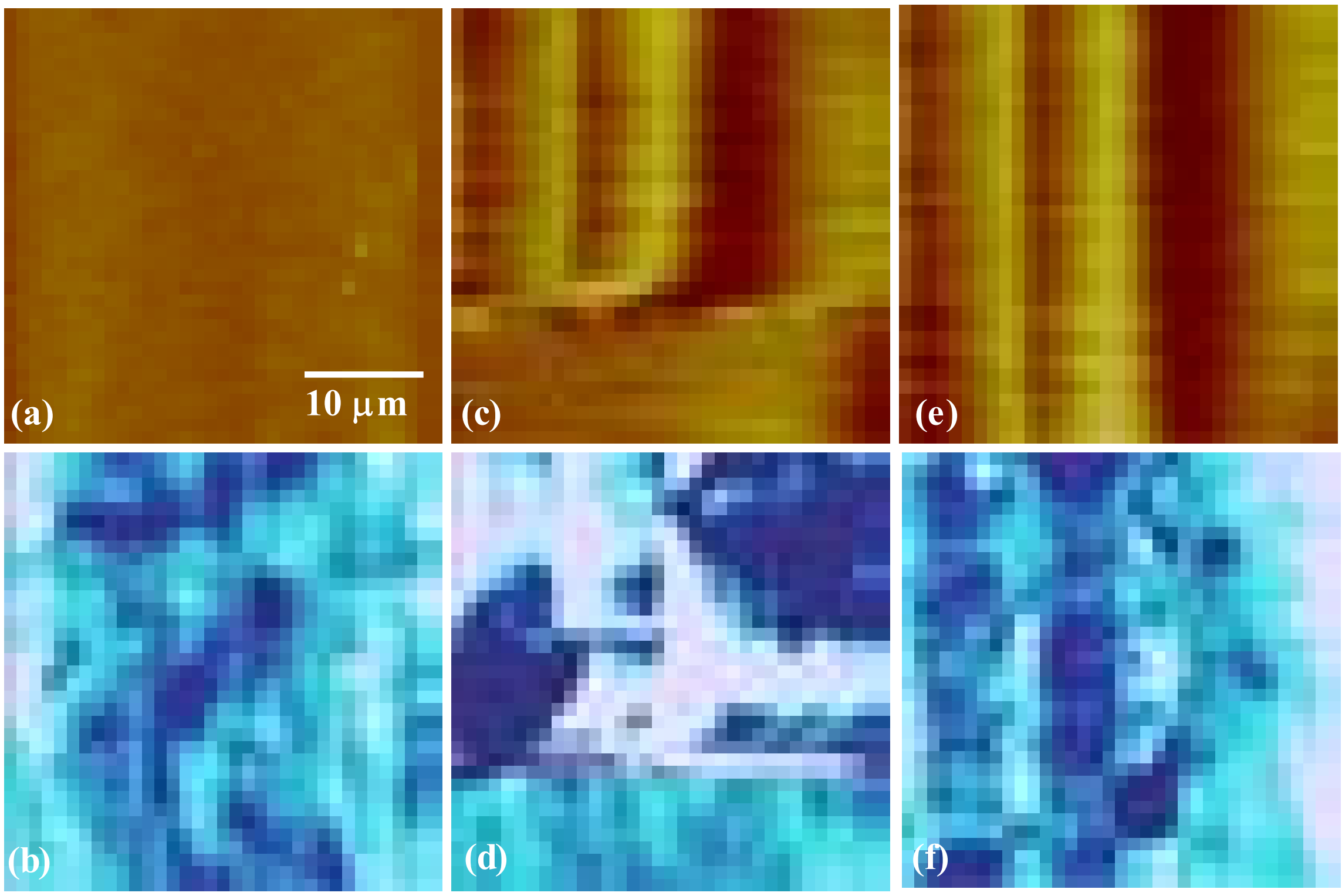

**Figure 5.14**. Surface topography (a,c,e) and surface potential distribution (b,d,f) on $BaTiO_3$ (100) surface above the Curie temperature (a,b), during the transition (c,d) and 1 h after the transition (e,f). Images are acquired from bottom to top. Scale is 30 nm (a,c,e), 0.05 V (b), 0.1 V (d,f).

The sequence of events on the reverse transition is similar to those in the forward transition. Figure 5.14a-f shows surface topography and surface potential distribution above the transition temperature, during the transition, and below the transition. It can be seen that during the transition the apparent topography is very volatile for ~30 s, then the new domain structure forms. New domains are oriented in the same direction as before the first transition; however, the size of the domains differs. At the transition surface potential exhibits large unstable potential amplitudes that may be attributed to depolarization currents associated with the formation of domain structure. After a relaxation period, the surface potential stabilizes. As clearly seen from Figure 5.14e,f, surface potential is again closely related to the new domain structure. Relaxation of newly formed potential features occurs much slower than on transition to the paraelectric phase.

The temperature dependence of potential can be rationalized only in terms of complete screening of polarization bound charge, i.e. domain potential in SSPM measurements should be attributed to the potential due to the double layer formed by polarization and screening charges. Domain potential has the sign of the screening charges and is reverse to that expected from polarization orientation, i.e. $c^+$ domains are negative and $c^-$ domains are positive on the SSPM image. A similar conclusion is reached from the comparison of SSPM and PFM images as reported by us and other groups.[64,65]

### 5.5.2. Temperature Induced Domain Potential Inversion

To verify the model considered above and determine the relevant kinetic and thermodynamic parameters of the screening process, isothermal kinetic studies were performed. The temperature was changed rapidly by ~10°C and ~8-9 SSPM images were obtained at constant temperature at 8 min intervals. On increasing the temperature, the initial increase in domain potential contrast decays with time to a stable and lower value (Figure 5.15). An unusual behavior is observed on *decreasing* the temperature (Figure 5.16). After a temperature decrease from 70°C to 50°C the domain contrast inverts (Figure 5.16a,b), i.e. a positive *c* domain becomes negative. The potential difference between the domains decreases with time, passing through an isopotential point corresponding to zero domain potential contrast (Figure 5.16c), and finally establishing an equilibrium value (Figure 5.16d).

This phenomenon, which will be referred to as temperature induced domain potential inversion, is consistent with the previously proposed explanation of screening on ferroelectric surfaces. In the case of complete screening, the surface potential has the sign of the screening charges and is reverse to that expected from polarization orientation, i.e. $c^+$ domains are negative and $c^-$ domains are positive on the SSPM image. Increasing the temperature results in a decrease of polarization bound charge leaving some of the screening charge uncompensated, thus increasing the effective surface potential. On decreasing the temperature spontaneous polarization increases and, for a short period, the sign of domain potential is determined by the polarization charge. Under isothermal conditions, polarization and screening charges equilibrate and the potential establishes an equilibrium value. Typical relaxation time for domain-related potential contrast is ~ 15 min, sufficiently slow to allow the kinetics of the screening process to be studied *in situ*. At the same time, after moderate (~1 h) annealing at a constant temperature domain contrast stabilizes and the temperature dependence of equilibrium domain potential difference can be determined. The latter can be related to the degree of screening and quantitative analysis of thermodynamics and kinetics of screening process is presented in Section 5.6.

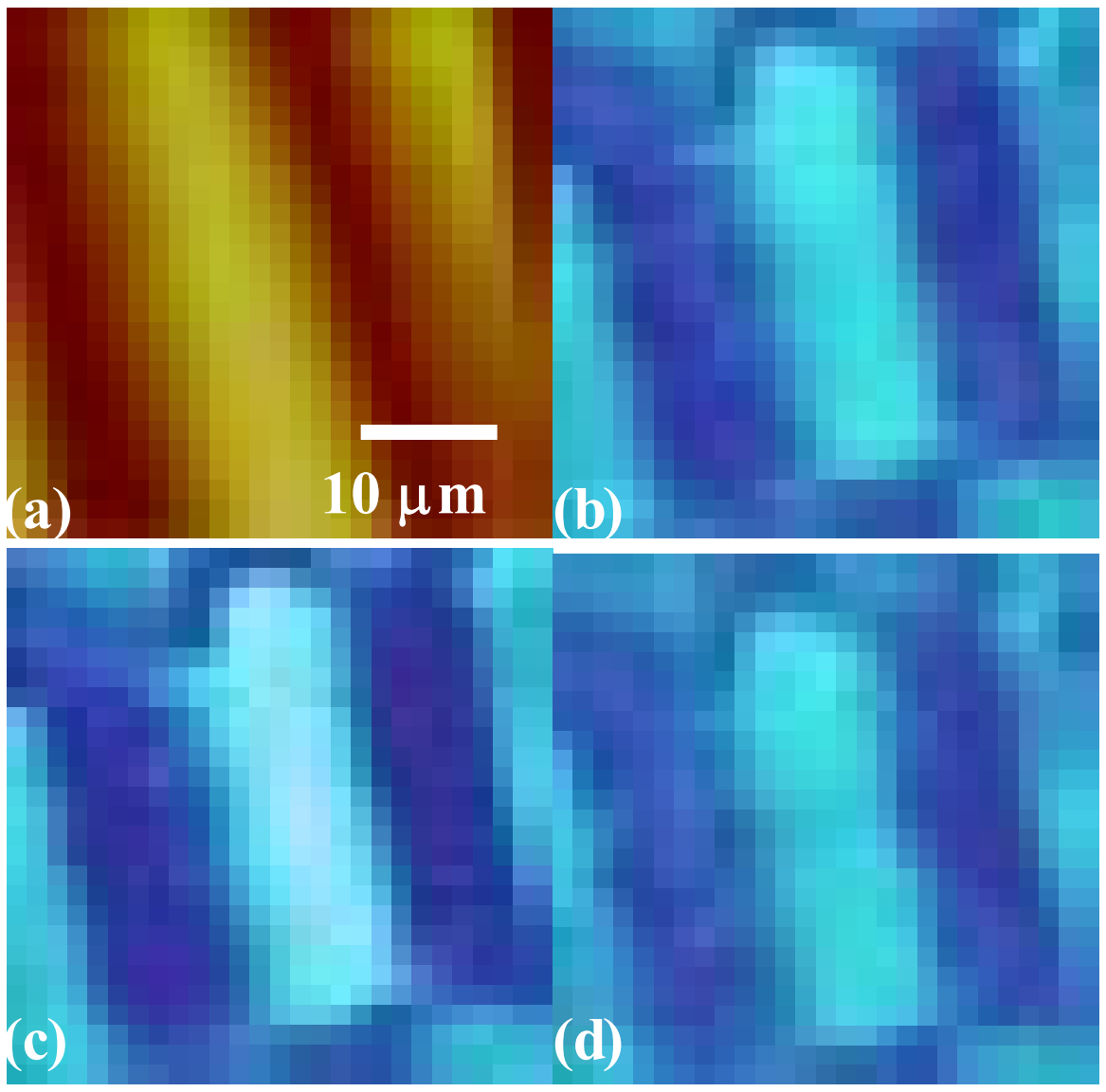

**Figure 5.15.** Surface topography (a) and surface potential (b) of the ferroelectric domain structure on a $BaTiO_3$ (100) surface at $T$ = 50°C. Surface potential after heating from 50°C to 70°C (c) and after annealing at 70°C for 50 min (d).

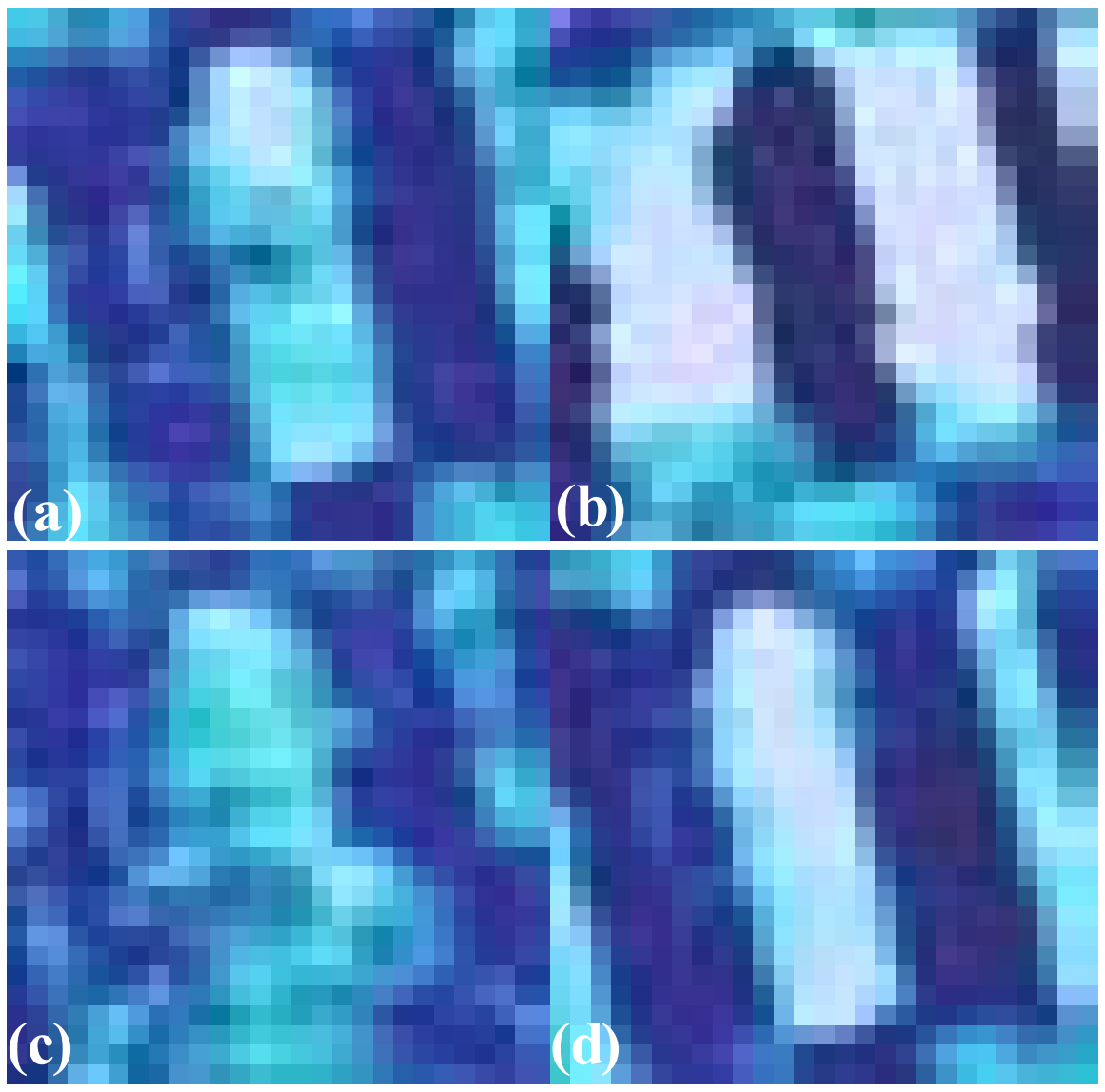

**Figure 5.16.** Surface potential (a) of ferroelectric domain structure on $BaTiO_3$ (100) surface at T = 90°C. Surface potential during cooling from 90°C to 70°C (b), at 70°C (c) and after annealing at 70°C for 50 min (d).

### 5.5.3. Screening Charge Relaxation during Domain Wall Motion

The relationship between polarization orientation and surface potential can also be established from the observation of domain wall motion. Fig. 5.17 shows SSPM images of $c^+$ - $c^-$ domain structures obtained at a 12 h interval. Shrinking of the negative domain results in a dark rim in the direction of domain wall motion (shrinking occurred spontaneously rather than under applied tip bias or lateral bias). The rim is ascribed to the slow relaxation of screening charges after the displacement of domain wall. Simple considerations (Figure 5.17c,f) imply that a negative rim in the direction of wall motion is possible only if domain related potential features are determined by the screening charges.

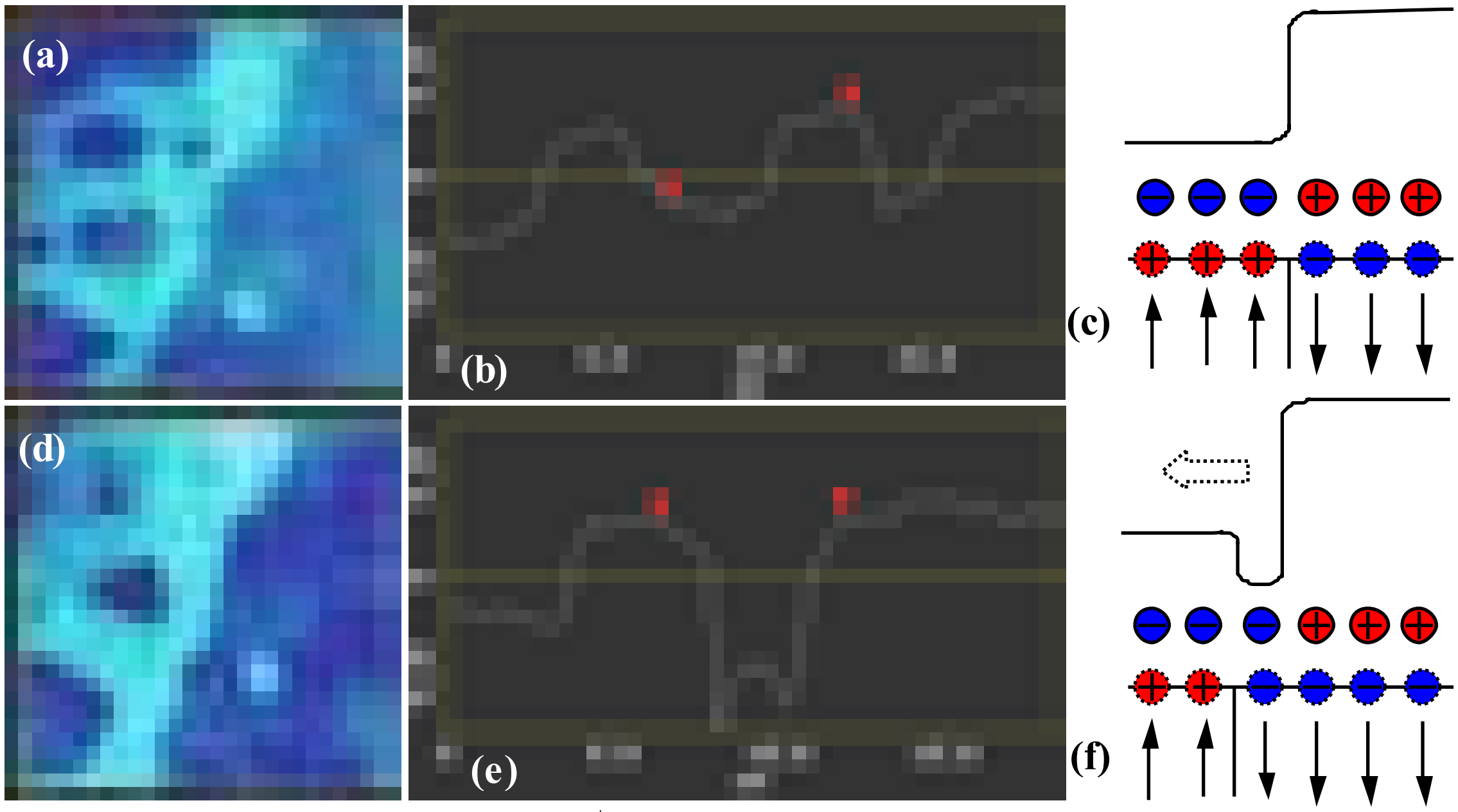

**Figure 5.17**. Surface potential images of $c^+$-$c^-$ domain region $BaTiO_3$ (100) acquired at 12 h interval (a,d), corresponding average profiles along the boxes (b,e) and the scheme of surface charge distribution (c,f).

### 5.5.4. Phase Transition and Polarization Dynamics by VT PFM

To distinguish the atomic polarization from surface potential, the phase transition was studied by PFM. The surface topography and piezoresponse at various temperatures are displayed in Figure 5.18. Surface corrugations indicate the presence of 90° *a-c* domain boundaries. The piezoresponse image reveals 180° domain walls separating regions of opposite polarity within *c*-domains. On heating from room temperature to 125°C the overall domain structure remains constant, however, small nuclei of domains of inverse polarity (Figure 5.18b,d) grow with temperature. On transition to the paraelectric state, both the surface corrugations and the piezoresponse contrast almost disappear. It should be noted that extremely weak inverted piezoresponse contrast could be observed after the transition (Figure 5.18f). This phenomenon is ascribed to the weak electrostatic interaction between the screening charges and SPM probe as discussed in Chapter 6. A similar observation is reported for PFM of the ferroelectric phase transition in TGS.[66] On reverse transition, domain-related corrugations form very quickly (Figure 5.19c). Piezoresponse variation during the transition is complex, but clear piezoresponse contrast develops after the transition (Figure 5.19d) and after equilibration below $T_c$, a new well-defined *a-c* domain structure is established (Figure 5.19e,f).

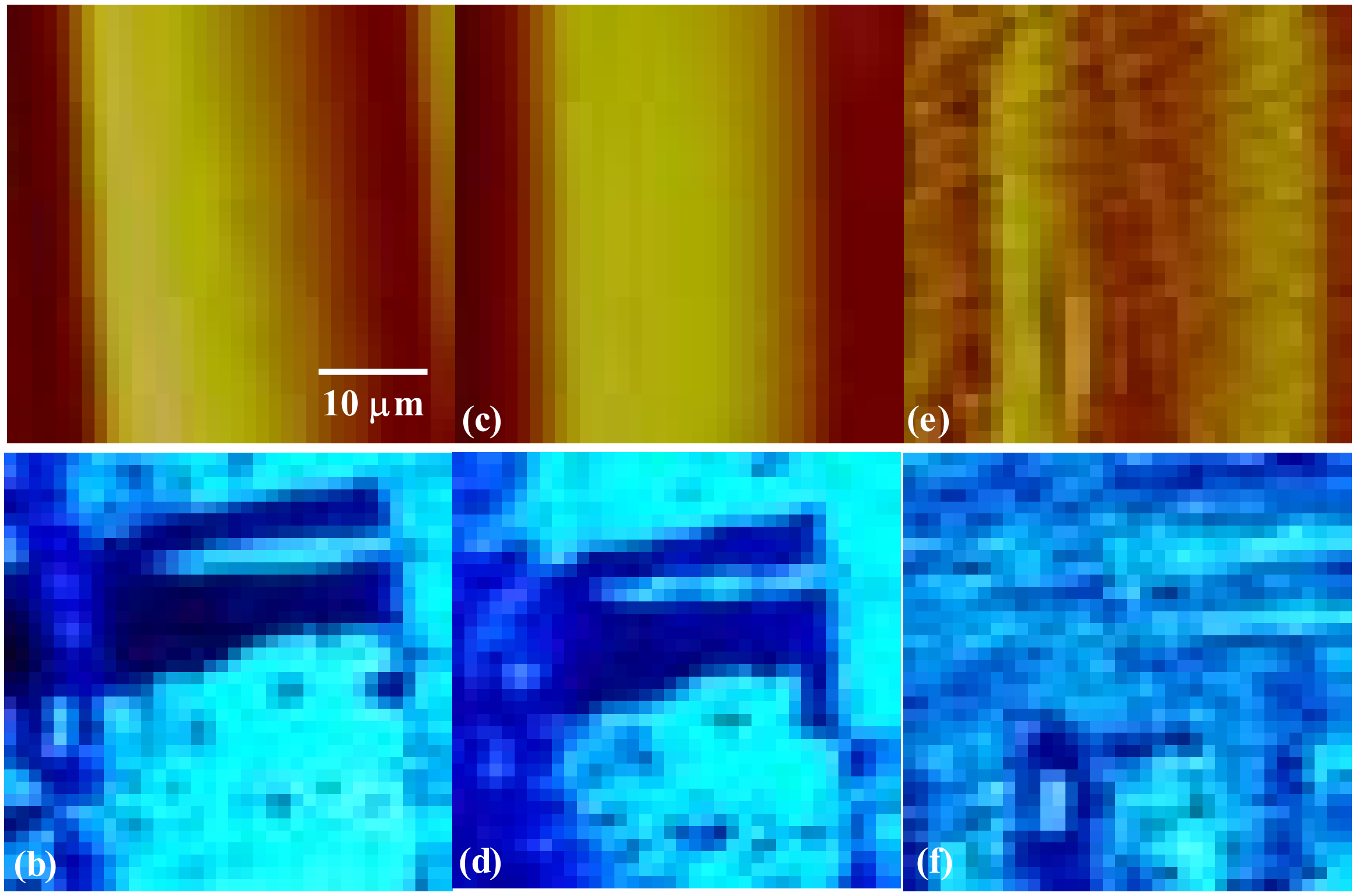

**Figure 5.18.** Surface topography (top) and piezoresponse (bottom) of $BaTiO_3$ (100) surface before ferroelectric phase transition at 20°C (a,b), at 125°C (c,d) and 4 min after transition at 140°C (e,f).

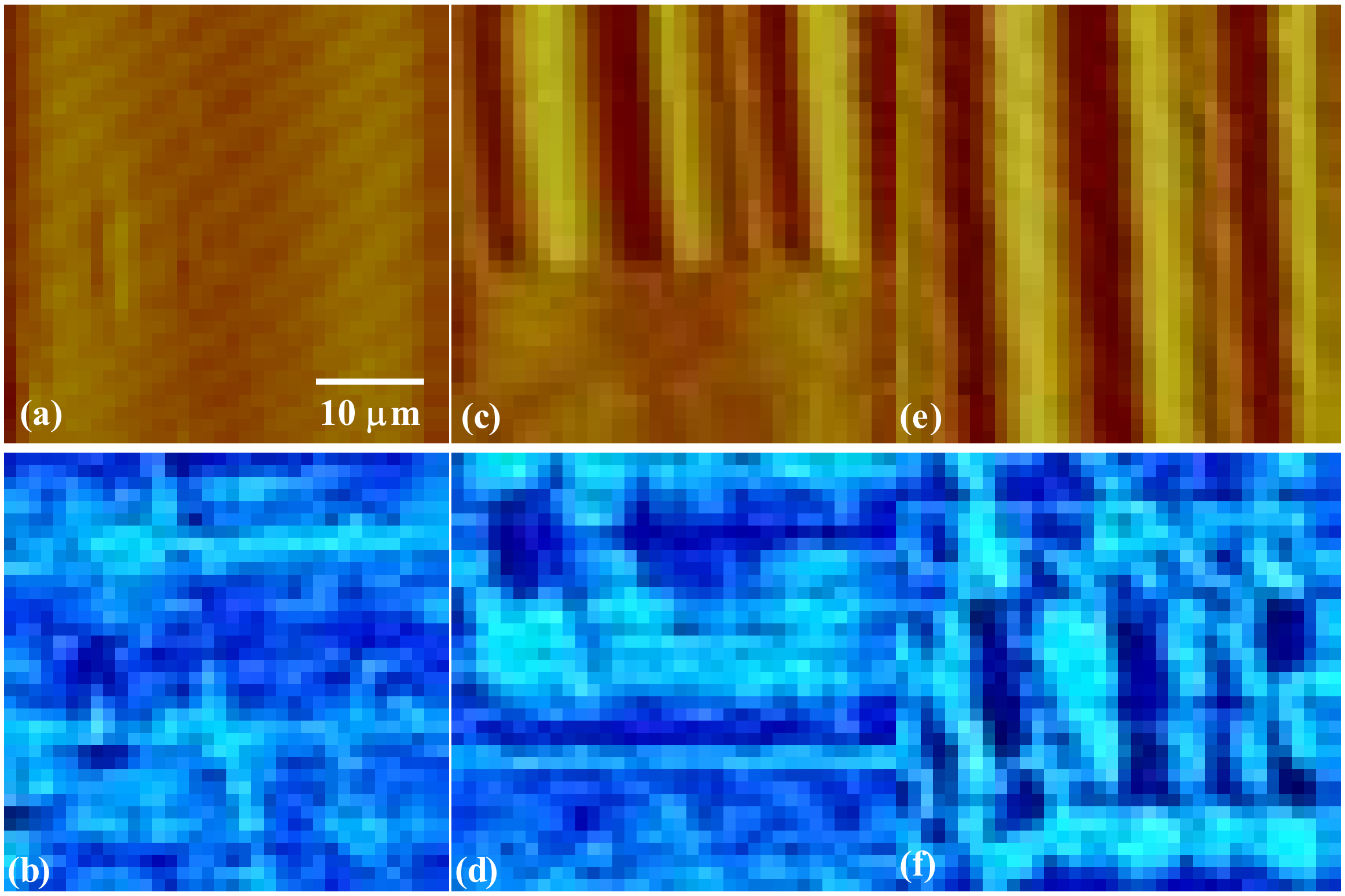

**Figure 5.19.** Surface topography (top) and piezoresponse (bottom) of $BaTiO_3$ (100) surface above Curie temperature at 140°C (a,b), during the reverse ferroelectric phase transition at 130°C (c,d) and after 30 min annealing at 120°C (e,f). Scale is 30 nm (a,c,e).

### 5.6. Screening and Thermodynamics of Adsorption on $BaTiO_3$ (100) Surfaces

Since the measured surface potential contains contributions both from capacitive and Coulombic interactions due to screened and unscreened polarization, quantification of the temperature dependence could yield thermodynamic and kinetic parameters as well as spatial localization of screening. To accomplish this a relationship must be established between measured potential and screening phenomena.

#### 5.6.1. Thermodynamics of Partially Screened Ferroelectric Surface

The surface of a ferroelectric material is characterized by a polarization charge density $\sigma = \mathbf{P} \cdot \mathbf{n}$, where **P** is the polarization vector and **n** is the unit normal to the surface. However, the unscreened state is unstable and extrinsic surface adsorption and/or intrinsic charge redistribution result in polarization screening at ferroelectric surfaces or interfaces. In the case when charge compensation is due to adsorption, the free energy for screening process is:

$$E(\alpha,T) = E_{el}(\alpha,T) + \alpha \frac{P}{qN_a} \Delta H_{\mathrm{ads}} - \alpha \frac{P}{qN_a} T\Delta S_{\mathrm{ads}} + E_{dw}, \tag{5.22}$$

where $q = 1.602 \cdot 10^{-19}$ C is electron charge, $P$ is spontaneous polarization, $N_a = 6.022 \cdot 10^{23}$ mol$^{-1}$ is Avogadro number, $\alpha$ is the degree of screening and $T$ is the temperature. Experimentally, the degree of screening is very close to unity, $\alpha \approx 1$,[67] therefore, in the subsequent description we introduce the fraction of the unscreened charge, $\gamma = 1 - \alpha$. The enthalpy and entropy of adsorption are denoted $\Delta H_{\mathrm{ads}}$ and $\Delta S_{\mathrm{ads}}$, respectively. The domain wall area is constant during the measurement and the corresponding free energy, $E_{dw}$, is assumed independent on the degree of screening. The electrostatic contribution to the free energy in Eq.(5.22), $E_{el}(\gamma, T)$, is derived in Appendix 5.A as:

$$E_{el}(\gamma,T)=\frac{P^2}{\varepsilon_0+\sqrt{\varepsilon_x\varepsilon_z}}\left\{\gamma^2 L\frac{7\zeta(3)}{2\pi^3}+(1-\gamma)h\frac{\varepsilon_0}{\varepsilon_2}+(1-\gamma)^2 h\frac{\sqrt{\varepsilon_x\varepsilon_z}}{\varepsilon_2}\right\}, \tag{5.23}$$

where $L$ is the domain size, $h$ is the screening layer width, $\varepsilon_2$ is the dielectric constant of the screening layer, $\varepsilon_x$ and $\varepsilon_z$ are the dielectric constants of the ferroelectric and $\varepsilon_0 = 8.854 \cdot 10^{-12}$ F/m is the dielectric constant of vacuum (Figure 5.20a). Eq.(5.23) reduces to Eq.(5.A.16) for $\gamma = 0$ (complete screening) and Eq.(5.A.21) for $\gamma = 1$ (unscreened surface).

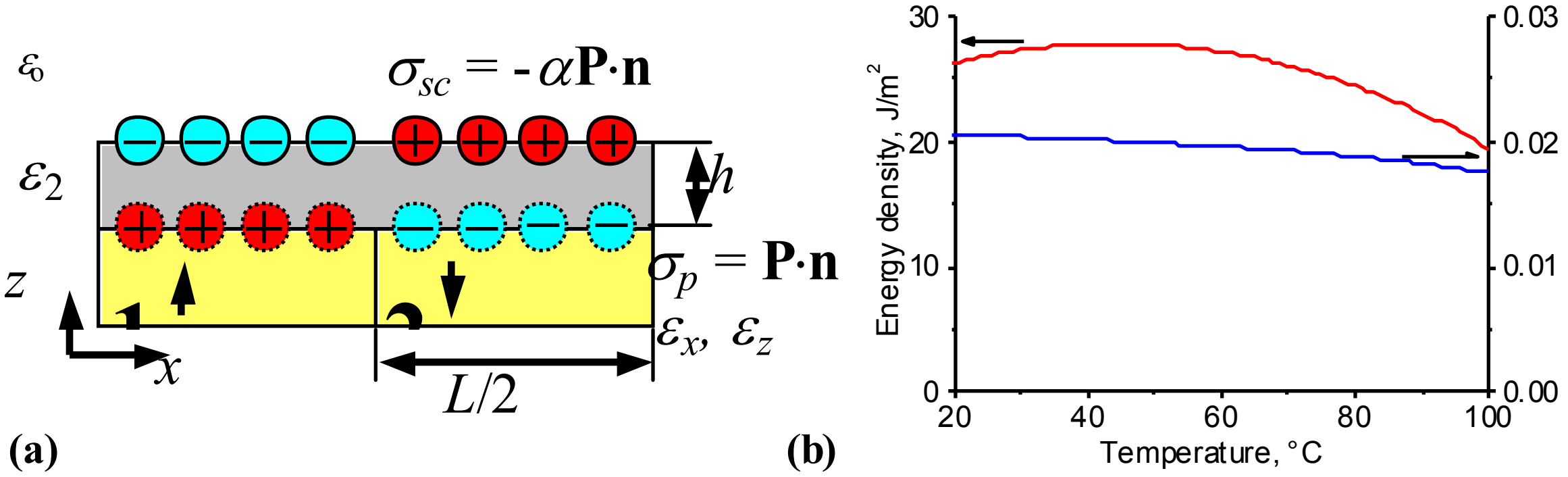

**Figure 5.20.** Charge distribution on the partially screened anisotropic ferroelectric surface (a) and temperature dependence of material constants.

The temperature dependence of the equilibrium screening can be obtained from the condition of the minimum of free energy

$$\frac{\partial E(\gamma,T)}{\partial\gamma}=0. \tag{5.24}$$

Since $E_{el}(\gamma, T)$ is a quadratic function of γ [Eq.(5.23)], this condition can be written as

$$\frac{\partial E_{el}}{\partial\gamma}=-b_1(T)\gamma+b_2(T), \tag{5.25}$$

where

$$b_1(T)=\frac{P^2}{\varepsilon_0+\sqrt{\varepsilon_x\varepsilon_z}}\left(2L\frac{7\zeta(3)}{2\pi^3}+2h\frac{\sqrt{\varepsilon_x\varepsilon_z}}{\varepsilon_2}\right), \tag{5.26}$$

and

$$b_2(T) = \frac{P^2}{\varepsilon_0 + \sqrt{\varepsilon_x \varepsilon_z}} \left( h\frac{\varepsilon_0}{\varepsilon_2} + 2h\frac{\sqrt{\varepsilon_x \varepsilon_z}}{\varepsilon_2} \right). \tag{5.27}$$

Thus, from Eq. (5.22) the equilibrium degree of screening is defined by

$$\gamma(T) = -\frac{P}{qN_a}\frac{\Delta H_{ads}}{b_1(T)} + \frac{b_2(T)}{b_1(T)} + T\frac{P}{qN_a}\frac{\Delta S_{ads}}{b_1(T)}, \tag{5.28}$$

where $b_1(T)$ and $b_2(T)$ are temperature dependent coefficients defined by domain structure and material properties.

The temperature dependencies of $b_1(T)$, $b_2(T)$ are calculated within the framework of the Ginzburg-Devonshire[18,68,69] theory. Spontaneous polarization, $P$, and susceptibilities, $\chi_{11}$ and $\chi_{33}$, are determined as

$$P^2 = \left( -\alpha_{11} + \sqrt{\alpha_{11}^2 - 3\alpha_1\alpha_{111}} \right) / 3\alpha_{111}, \tag{5.29}$$

$$\chi_{11}^{-1} = \left( 2\alpha_1 + 2\alpha_{12}P^2 + 2\alpha_{112}P^4 \right)\varepsilon_0, \tag{5.30}$$

$$\chi_{33}^{-1} = \left( 2\alpha_1 + 12\alpha_{11}P^2 + 30\alpha_{111}P^4 \right)\varepsilon_0. \tag{5.31}$$

The numerical values of coefficients $\alpha_1$, $\alpha_{11}$, $\alpha_{12}$, $\alpha_{111}$ and $\alpha_{112}$ are taken from Ref. [70].

Material related coefficients $b_1(T)$ and $b_2(T)$ in Eq.(5.25) can be loosely interpreted as free energy density related to the unscreened component of polarization charge and the free energy density related to the screened charge. Calculated temperature dependencies of $b_1$ and $b_2$ for domain size $L$ = 10 μm, $\varepsilon_2$ = 80 (water) and $h$ = 0.1 nm are shown on Figure 5.20b. It is clearly seen that $b_1$ and $b_2$ are only weakly temperature dependent. The physical origin of this behavior is that the product $\varepsilon_x\varepsilon_y$ is only weakly temperature dependent. Consequently, $b_1$ and $b_2$ can be approximated by their room temperature values. At $T$ = 25°C for $h$ = 0.1 nm $b_1$= 26.67 J/m$^2$ and $b_2$ = 0.02034 J/m$^2$. Eq.(5.28) suggests that the degree of screening is a linear function of temperature.

### 5.6.2. SSPM Contrast vs. Degree of Screening

The effective potential measured by SSPM contains contribution both from screened and unscreened components of polarization charge. The force acting on the biased tip above a partially screened ferroelectric surface is written as a sum of capacitive and Coulombic components:

$$F = \frac{dC}{dz}\left(V_{tip} - V_s\right)^2 + CV_{tip}E_z, \tag{5.32}$$

where $C = C(z)$ is the distance dependent tip-surface capacitance, $V_{tip}$ is the tip bias, $V_s$ is the surface potential due to electric double layer, and $E_z$ is the normal component of electric field due to unscreened polarization charge. In SSPM the tip bias is $V_{tip} = V_{dc}+V_{ac}\cos(\omega t)$, where $V_{ac}$ is driving voltage and the first harmonic of the force is:

$$F_{1\omega} = \frac{dC}{dz}\left(V_{tip} - V_s\right)V_{ac} + CV_{ac}E_z. \tag{5.33}$$

To quantify the capacitive and Coulombic components of the tip-surface interactions, a line charge model is used as described in Chapter 2.4. The image charge distribution within the tip is approximated by a uniformly charged line with line charge density $\lambda$ located at a distance $d = z\sqrt{1+\tan^2\theta}$ from the surface, where $z$ is separation between the tip apex and surface and $\theta$ is half-angle of the conical equipotential surface that represents the tip. The line charge density is

$$\lambda = \frac{4\pi\varepsilon_0 V}{\beta}, \quad \text{where} \quad \beta = \ln\left(\frac{1+\cos\theta}{1-\cos\theta}\right). \tag{5.34a,b}$$

The capacitive and Coulombic contributions to tip-surface force in this model are proportional to derivative of capacitance $C_z'$ and capacitance $C$, respectively:

$$\frac{dC}{dz} = \frac{4\pi\varepsilon_0}{\beta^2}\ln\left(\frac{H}{4d}\right), \qquad C = \frac{4\pi\varepsilon_0 H}{\beta}, \tag{5.35a,b}$$

where $H$ is tip length (~10-15 μm). Eq.(5.35b) for the Coulombic contribution is valid only if the decay length of the field, which is proportional to the domain size $L$, is larger than the tip size, $H$. In the realistic situation when $L$ and $H$ are comparable, Eq.(5.35b) provides the upper estimate for Coulombic force.

Using the line charge model, Eq.(5.33) is rewritten as

$$F_{1\omega} = \frac{4\pi\varepsilon_0}{\beta^2}\left(V_{tip} - V_s\right)V_{ac}\ln\left(\frac{H}{4d}\right) + \frac{4\pi\varepsilon_0}{\beta}HV_{ac}E_z. \tag{5.36}$$

The nulling condition for the first harmonic of the force is achieved when the dc offset of the tip bias collected as effective surface potential is

$$V_{dc} = V_s - E_z \beta H \Big/ \ln\left(\frac{H}{4d}\right). \tag{5.37}$$

The potential contrast between the domains is

$$\Delta V_{dc} = (V_1 - V_2) - (E_1 - E_2)\beta H \ln\left(\frac{H}{4d}\right)^{-1}. \tag{5.38}$$

As shown in Section 5.4, domain contrast is independent of feedback effects and is not susceptible to the temperature-induced shifts of cantilever resonant frequency, while this is not true for effective surface potential *per se*. For a typical metal coated tip used in the SSPM measurements with $\theta$ = 17°, $H \approx$ 10 μm and tip-surface separation $z$ = 50-100 nm Eq.(5.38) can be approximated as

$$\Delta V_{dc} = (V_1 - V_2) - 1.18H(E_1 - E_2), \tag{5.39}$$

since the logarithmic term is only weakly dependent on the tip length. Here the cantilever contribution to the measured potential that leads to a significant height dependence of measured potential contrast is neglected. Under the experimental conditions (lift height 100 nm), the deviation between the true and measured domain potential difference does not exceed ~30 % and the uncertainties in the other parameters (tip shape model, materials properties) are expected to be comparable.

Using the representation of a partially screened ferroelectric surface as a superposition of completely unscreened and completely screened regions, the potential difference between domains of opposite polarity is

$$\Delta\varphi_s = (1-\gamma)\frac{2h\sqrt{\varepsilon_x \varepsilon_z}P}{\varepsilon_2\left(\varepsilon_0 + \sqrt{\varepsilon_x \varepsilon_z}\right)}, \tag{5.40}$$

while the difference in the normal component of the electric field is

$$\Delta E_u = \gamma \frac{P}{\varepsilon_0 + \sqrt{\varepsilon_x \varepsilon_z}}. \tag{5.41}$$

It follows from Eqs.(5.38,40,41) that the measured potential difference between the domains is

$$\Delta V_{dc} = (1-\gamma)\frac{2h\sqrt{\varepsilon_x \varepsilon_z} P}{\varepsilon_2\left(\varepsilon_0 + \sqrt{\varepsilon_x \varepsilon_z}\right)} - \beta H \gamma \frac{P}{\varepsilon_0 + \sqrt{\varepsilon_x \varepsilon_z}} \ln\left(\frac{H}{4d}\right)^{-1}, \tag{5.42}$$

i.e. domain potential contrast is a linear function of degree of screening. Combination of Eq.(5.28) and Eq.(5.42) suggests that the surface potential difference between the domains as measured by SSPM is a linear function of temperature, as well. Therefore, the temperature dependence of domain potential contrast can be used to estimate the temperature dependence of the degree of screening and determine thermodynamic parameters associated with the screening process.

### 5.6.3. Kinetics of Polarization Screening

As shown above, domain potential contrast dynamics on heating and cooling are complex. The time dependence of domain potential contrast on heating and cooling is shown in Figure 5.21a,b. To quantify the kinetics, the time dependence of domain potential contrast, $\Delta\varphi$, was approximated by an exponential function

$$\Delta\varphi = \Delta\varphi_0 + A\exp(-t/\tau), \tag{5.43}$$

where $\tau$ is relaxation time and $A$ is a prefactor. Due to the finite heating and cooling rates, the domain potential contrast immediately after the temperature change cannot be reliably established; therefore, Eq.(5.43) describes the late stages of potential relaxation. Spectroscopic (single point) topographic measurements are limited by the response time of the cantilever while the experimental response time for surface potential measurements is ~ 10-100 ms and strongly depends on feedback parameters. Spatially resolved measurements are limited either by line acquisition time (0.5 - 5 sec) or image acquisition time (4 - 20 min).

The temperature dependence of the potential redistribution time is shown in Figure 5.21c. The redistribution time is almost temperature independent with an associated relaxation energy of ~ 4 kJ/mole. A possible explanation of this low value of activation energy is that the kinetics of relaxation process is limited by the transport of charged species to the surface. The characteristic redistribution time is ~ 20 min and is close to the relaxation time for domain potential contrast above $T_c$ (15 min).[71] The redistribution processes both on heating and cooling result in the same equilibrium value of domain potential contrast, $\Delta\varphi_0$. The temperature dependence of domain potential contrast, shown in Figure 5.21d, is almost linear, with the zero potential difference corresponding to temperature ~110°C well below the Curie temperature of $BaTiO_3$ ($T_c$ = 130°C). For higher temperatures, the degree of screening is smaller and the Coulombic contribution to the effective SSPM potential increases. Since polarization charge and screening charge contributions to the effective surface potential are of opposite sign, the decrease of the degree of screening results in the *decrease* of domain potential contrast. The thermodynamics of this process are expected to be strongly temperature dependent and to dominate over relatively weak variation of spontaneous polarization in this temperature range ($P$ = 0.26 C/m$^2$ at 25°C and 0.20 C/m$^2$ at 100°C).

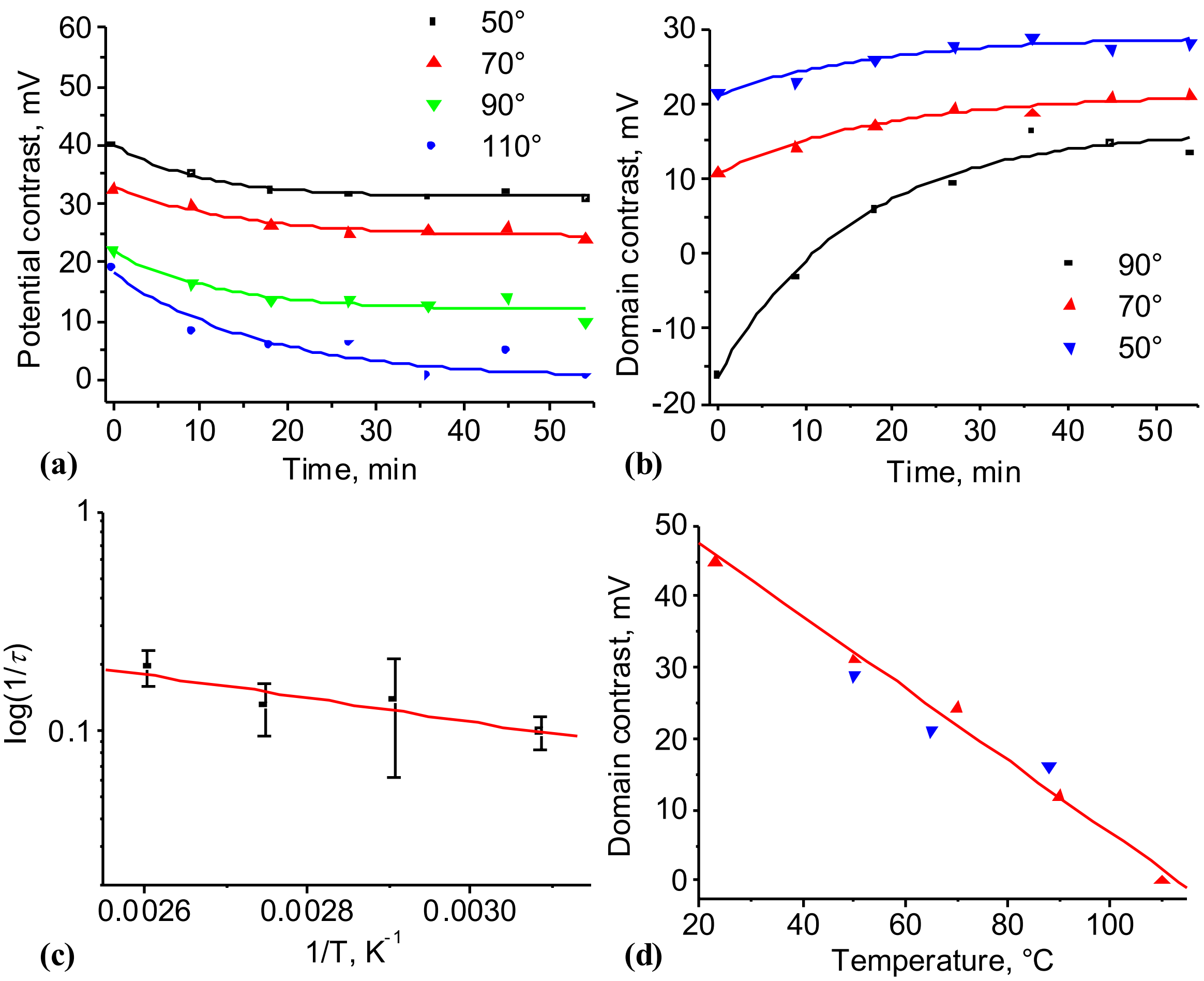

**Figure 5.21.** Time dependence of domain potential contrast on heating (a) and cooling (b). Solid lines are fits by Eq. (5.43). (c) Time constant for relaxation process on heating in Arrhenius coordinates and (d) temperature dependence of equilibrium domain potential contrast on heating (▲) and cooling (▼) and fit by Eq. (5.44) (solid line).

It is interesting to analyze the possibility of tip bias effects on charge dynamics. It is well known that while in the contact regime a large dc tip-surface potential difference results in charge transfer between the tip and the surface and local polarization switching, this is not the case in SSPM. Indeed, during SSPM imaging tip is located at ~100 nm from the surface. The nulling potential scheme insures that dc tip-surface potential difference is zeroed. The characteristic frequency of ac field (70kHz) is many orders of magnitude higher than the relaxation frequency of screening charges as observed here (~1 mHz) and elsewhere.[56,72,73,74,75] Using the plane-plane capacitor approximation, the magnitude of ac potential drop in the screening layer can be estimated as $\Delta V_{sl} = V_{ac} / (1 + z\varepsilon_2 / h\varepsilon_0)$, where $z$ is tip-surface separation, $h$ and $\varepsilon_2$ is width and dielectric constant of screening layer. For $z$ = 100 nm, $h$ = 0.2 nm, $\varepsilon_2 = 80\varepsilon_0$ and $V_{ac}$ = 5 V the potential drop in the screening layer is ~ 0.1 mV. Therefore, the ac field is not expected to affect screening charge dynamics. These considerations notwithstanding, the tip effect on charge dynamics can be experimentally detected as the dependence of effective domain potentials on driving voltage and, as shown in Section 5.4, no such effect is observed during SSPM imaging of domains in $BaTiO_3$.

#### 5.6.4. Thermodynamics of Polarization Screening

As discussed in Section 5.6.1, the domain potential contrast can be related to the degree of screening of spontaneous polarization. As shown in Figure 5.21d, the temperature dependence of equilibrium domain potential contrast in the temperature interval 30°C < $T$ < 100°C is linear and the domain potential contrast is the same on heating and cooling, i.e. equilibrium is achieved. This dependence can be represented by the linear function

$$\Delta V_{dc} = 0.059 - 5.3 \cdot 10^{-4} T, \tag{5.44}$$

where $T$ is temperature in Celsius degrees. The width of the screening layer must be estimated, and is taken to be $h$ = 0.1 nm such that the calculated potential difference between $a$ and $c$ domains at the room temperature is close to the measured value at 100 nm (50 mV) for almost complete screening (compare to 0.25 nm for $c^+$ and $c^-$ domains). The contribution of electric field to the room temperature potential and cantilever effects are ignored in this estimate. However, the lack of reliable information regarding the thickness and properties of the screening layer precludes better estimates.

The combination of Eq.(5.42) and Eq.(5.44) for a tip length of 10 μm yields the temperature dependence of equilibrium degree of screening

$$\gamma = 1.627 \cdot 10^{-5} + 1.23 \cdot 10^{-6} T \,. \tag{5.45}$$

A comparison of Eq.(5.28) and Eq.(5.45) allows the enthalpy, $\Delta H_{ads}$, and entropy, $\Delta S_{ads}$, of adsorption to be determined as $\Delta H_{ads}$ = 164.6 kJ/mole, $\Delta S_{ads}$ = -126.6 J/mole K. The enthalpy and entropy of adsorption thus obtained are within expected values in spite of the approximations inherent in this approach. Moreover, from Eq.(5.42) the Coulombic contribution to the effective potential can be estimated as < 10-20 % thus validating our previous conclusion that the surface is completely screened at room temperature.

### 5.6.5. The Origins of Polarization Screening on Ferroelectric Oxide Surfaces

The nature of the screening charges can not be determined from these experiments; however, these results are consistent with the well known fact that water and hydroxyl groups, -OH, adsorb on oxide surfaces in air[76,77,78] The formation of water layers on hydrophilic surfaces and it's implications for force measurements with AFM are well documented in literature.[79] On ferroelectric surfaces, the formation of water layers was reported recently to affect the domain nucleation in agreement with the results reported here.[80,81] Obviously, adsorbed water can provide the charge required to screen the polarization bound charge, since corresponding polarization charge densities are of order of 0.25 $C/m^2$ corresponding to $2.6 \cdot 10^{-6}$ $mole/m^2$. For a typical metal oxide surface with characteristic unit cell size of ~ 4 Å this corresponds to the coverage of order of 0.25 mL. Dissociative adsorption of water as a dominant screening mechanism on $BaTiO_3$ surface in air was verified using temperature programmed desorption experiments on poled $BaTiO_3$ crystals.[82] As shown by Vanderbilt,[83] perovskite (100) surfaces are free from the midgap surface states; therefore, the screening on $BaTiO_3$ (100) surface cannot be attributed to the surface states filling.

## 5.7. Conclusions

To summarize, in this Chapter non-contact electrostatic SPM was used to study the surface chemistry and physics of ferroelectric surface. On surfaces with known crystallographic orientation, the domain structure can unambiguously be determined by SPM. Surface topography in ferroelastic materials is directly related to the misorientation angle between domains with different polarization directions, e.g. for tetragonal perovskites the corrugation angle, $\theta$, associated with 90° *a-c* domain walls is $\theta = \pi/2 - 2\arctan(a/c)$, where *a* and *c* are the parameters of the tetragonal unit cell. Complimentary information obtained by non-contact (SSPM, EFM) or contact (PFM) SPM allows the orientation of the polarization vectors (e.g. $c^+$ - $c^-$ domains) to be distinguished. Both EFM and SSPM yield potential difference between $c^+$ and $c^-$ domains as $\Delta V_{c\text{-}c} \approx 150$mV and between *a* and *c* domains as $\Delta V_{a\text{-}c} \approx \Delta V_{c\text{-}c}/2$. The small potential variations between domains indicate that $BaTiO_3$ surface is screened in air. The absolute value of domain potential contrast and equality of potential differences between $c^+$-*a* and $c^-$-*a* domains suggest that the screening is due to the surface adsorbates. The average surface potential is approximately equal to average domain potential between $c^+$ and $c^-$ domains, $V_{av} \approx (V_1+V_2)/2$, implying that surface areas occupied by $c^+$ and $c^-$ domains are equal, as expected. At room temperature the surface potential has the sign of the screening charges and is reverse to that expected from polarization orientation, i.e. $c^+$ domains are negative and $c^-$ domains are positive on the SSPM image. Increasing the temperature results in a decrease of polarization bound charge leaving the screening charges uncompensated, thus increasing the effective surface potential. On decreasing the temperature spontaneous polarization increases and for a short period of time the sign of domain potential is determined by the polarization charge, thus giving rise to the effect of temperature induced domain potential inversion. Under isothermal conditions, polarization and screening charges equilibrate and the potential achieves an equilibrium value. Temperature and time dependent behavior of surface potential is shown to be governed by the rapid polarization dynamics and slow screening charge dynamics. The relaxation kinetics were found to be weakly dependent on temperature with activation energy $E_a \sim 4$ kJ/mole. Equilibrium domain potential difference was found to be linearly dependent on temperature; the zero potential contrast between the domains is observed at ~110°C. This behavior is ascribed to the reduction in the degree of screening with temperature and increase of electric field contribution to tip-surface force. A thermodynamic description of adsorbate screening of ferroelectric surfaces based on Ginzburg-Devonshire theory was developed and enthalpy and entropy of adsorption were obtained as $\Delta H_{ads}$ = 164.6 kJ/mole, $\Delta S_{ads}$ = -126.6 J/mole K. These values are well within the range expected for adsorption from the gas phase. Based on the existing models for the oxide surface behavior in ambient, the screening is attributed to the dissociative adsorption of water.

## Appendix 5.A

Assume an anisotropic ferroelectric with dielectric constants $\varepsilon_{xx} = \varepsilon_{yy} = \varepsilon_x$, $\varepsilon_{zz} = \varepsilon_z$ and polarization $\mathbf{P}(x) = P\cdot\mathbf{z}$ , $(0<x<L/2)$, $-P\cdot\mathbf{z}$ , $(L/2<x<L)$, where $L$ is characteristic domain size, is covered by a charged adsorbate layer of thickness $h$ and dielectric constant $\varepsilon_2$. Domain structure is uniform in $y$ direction. Adsorbate charge density is $\sigma(x) = -\sigma$ $(0<x<L/2)$ and $\sigma$ $(L/2<x<L)$. Potential in the air, in the adsorbate layer and in the ferroelectric are denoted as $\varphi_1$, $\varphi_2$ and $\varphi_3$ respectively. These potentials can be found as solutions of Laplace equations[84]

$$\nabla^2\varphi_1 = 0,\ z > \mathrm{h}, \tag{5.A.1}$$

$$\nabla^2\varphi_2 = 0,\ h>z>0, \tag{5.A.2}$$

$$\varepsilon_x \frac{\partial^2\varphi_3}{\partial x^2} + \varepsilon_z \frac{\partial^2\varphi_3}{\partial z^2} = 0,\ z < 0. \tag{5.A.3}$$

Corresponding boundary conditions are the continuity of potential (or, equivalently, tangential component of the electric field) at the interfaces

$$\varphi_1(h) = \varphi_2(h), \tag{5.A.4}$$

$$\varphi_2(0) = \varphi_3(0), \tag{5.A.5}$$

and the normal component of the displacement vectors

$$\varepsilon_0 \frac{\partial\varphi_1}{\partial z} - \varepsilon_2 \frac{\partial\varphi_2}{\partial z} = \sigma(x) \quad \text{for } z = h, \tag{5.A.6}$$

$$\varepsilon_2 \frac{\partial\varphi_2}{\partial z} - \varepsilon_z \frac{\partial\varphi_3}{\partial z} = P(x) \quad \text{for } z = 0. \tag{5.A.7}$$

The solution to Eqs. (5.A.1-3) can be found in the form of a Fourier series

$$\varphi_1 = \sum_{n=0} A_n \sin\left(\frac{2\pi(2n+1)x}{L}\right) exp\left(-\frac{2\pi(2n+1)z}{L}\right), \qquad z > \mathrm{h} \tag{5.A.8}$$

$$\varphi_2 = \sum_{n=0}\left(B_n \exp\left(-\frac{2\pi(2n+1)z}{L}\right) + C_n exp\left(\frac{2\pi(2n+1)z}{L}\right)\right)\sin\left(\frac{2\pi(2n+1)x}{L}\right), \tag{5.A.9}$$

$$\varphi_3 = \sum_{n=0} D_n \sin\left(\frac{2\pi(2n+1)x}{L}\right) exp\left(\frac{2\pi(2n+1)z}{L}\sqrt{\frac{\varepsilon_x}{\varepsilon_z}}\right), \qquad z < 0 \tag{5.A.10}$$

Of interest are potentials in the completely screened case, $\sigma = -P$, in which case

$$\varphi_1^s = -\frac{4h\sqrt{\varepsilon_x\varepsilon_z}\,\sigma}{\pi\varepsilon_2\left(\varepsilon_0 + \sqrt{\varepsilon_x\varepsilon_z}\right)}\sum_{n=0}\frac{1}{2\mathrm{n}+1}\sin\left(\frac{2\pi(2n+1)x}{L}\right)exp\left(-\frac{2\pi(2n+1)z}{L}\right), \quad (5.A.11)$$

$$\varphi_3^s = \frac{4h\varepsilon_0\sigma}{\pi\varepsilon_2\left(\varepsilon_0 + \sqrt{\varepsilon_x\varepsilon_z}\right)}\sum_{n=0}\frac{1}{2\mathrm{n}+1}\sin\left(\frac{2\pi(2n+1)x}{L}\right)exp\left(\frac{2\pi(2n+1)z}{L}\sqrt{\frac{\varepsilon_x}{\varepsilon_z}}\right), \quad (5.A.12)$$

The potential difference between the domains of opposite polarity in completely screened case is therefore:

$$\Delta\varphi_s = \frac{2h\sqrt{\varepsilon_x\varepsilon_z}\,\sigma}{\varepsilon_2\left(\varepsilon_0 + \sqrt{\varepsilon_x\varepsilon_z}\right)} \approx \frac{2h\sigma}{\varepsilon_2}, \quad (5.A.13)$$

while the field variation between the centers of the domains is

$$\Delta E^{\mathrm{s}} = \frac{2h\sigma\sqrt{\varepsilon_x\varepsilon_z}}{\varepsilon_2\left(\varepsilon_0 + \sqrt{\varepsilon_x\varepsilon_z}\right)L} \approx \frac{2h\sigma}{\varepsilon_2 L}. \quad (5.A.14)$$

Depolarization energy density for the screened case is:

$$E_{el}^s = \frac{1}{L}\int_0^L\left(-\varphi_1^s + \varphi_3^s\right)Pdx, \quad (5.A.15)$$

or

$$E_{el}^s = \frac{hP^2}{\varepsilon_2}. \quad (5.A.16)$$

In the unscreened case ($h$ = 0, $\sigma$= 0), so

$$\varphi_1^u = \frac{2LP}{\pi^2\left(\varepsilon_0 + \sqrt{\varepsilon_x\varepsilon_z}\right)}\sum_{n=0}\frac{1}{(2\mathrm{n}+1)^2}\sin\left(\frac{2\pi(2n+1)x}{L}\right)exp\left(-\frac{2\pi(2n+1)z}{L}\right), \quad (5.A.17)$$

$$\varphi_3^u = \frac{2LP}{\pi^2\left(\varepsilon_0 + \sqrt{\varepsilon_x\varepsilon_z}\right)}\sum_{n=0}\frac{1}{(2\mathrm{n}+1)^2}\sin\left(\frac{2\pi(2n+1)x}{L}\right)exp\left(\frac{2\pi(2n+1)z}{L}\sqrt{\frac{\varepsilon_x}{\varepsilon_z}}\right). \quad (5.A.18)$$

In the unscreened case potential and electric field difference between the centers of $c^+$ and $c^-$ domains are

$$\Delta\varphi_u = \frac{4CL\sigma}{\left(\varepsilon_0 + \sqrt{\varepsilon_x\varepsilon_z}\right)\pi^2}, \quad (5.A.19)$$

where $C \approx 0.916$ is the Catalan constant and

$$\Delta E^{u} = \frac{\sigma}{\varepsilon_0 + \sqrt{\varepsilon_x \varepsilon_z}} . \tag{5.A.20}$$

Depolarization energy density in the unscreened case is

$$E_{el}^{u} = \frac{P^2 L}{\varepsilon_0 + \sqrt{\varepsilon_x \varepsilon_z}} \frac{7\zeta(3)}{8\pi^3}, \tag{5.A.21}$$

where $\zeta(3) \approx 1.202$ is the zeta function.

Potential in the partially screened case is represented as the sum of the potential in the completely screened case with a surface charge density $\sigma$ and in the unscreened case with a polarization, $P$-$\sigma$. Therefore, depolarization energy is

$$E_{el} = \frac{1}{L}\int_0^L \left\{P\left(\varphi_3^u + \varphi_3^s\right) + \sigma\left(\varphi_1^u + \varphi_1^s\right)\right\} dx , \tag{5.A.22}$$

where $\varphi_1^u$ and $\varphi_3^u$ are given by Eqs. (5.A.13,19) with $P$-$\sigma$ rather $P$.

Introducing the degree of screening, $\alpha$, such that $\sigma = -\alpha P$, depolarization energy density is calculated as

$$E_{el} = \frac{P^2}{\varepsilon_0 + \sqrt{\varepsilon_x \varepsilon_z}} \left\{(1-\alpha)^2 L \frac{7\zeta(3)}{2\pi^3} + \alpha h \frac{\varepsilon_0}{\varepsilon_2} + \alpha^2 h \frac{\sqrt{\varepsilon_x \varepsilon_z}}{\varepsilon_2}\right\}. \tag{5.A.23}$$

[24] I.M. Reaney, Ferroelectrics **172**, 115 (1995).

[25] L.A. Bursill and P.J. Lin, Ferroelectrics **70**, 191 (1986).

[26] E. Snoeck, L. Normand, A. Thorel, and C. Roucau, Phase Transitions **46**, 77 (1994).

[27] Z. Xu, D. Viehland, P. Yang, and D.A. Payne, J. Appl. Phys. **74**, 3406 (1993).

[28] O.O. Popoola and W.M. Kriven, Philos. Mag. Lett. **75**, 1 (1997).

[29] E. Snoeck, C. Roucau, P. Baules, M.J. Casanove, M. Fagot, B. Astie, and J. Degauque, Microsc., Microanal., Microstruct. **4**, 249 (1993).

[30] N. Yamamoto, K. Yagi, and G. Honjo, Phys. Status Solidi **A 62**, 657 (1980).

[31] Y.G. Wang, J. Dec, and W. Kleemann, J. Appl. Phys. **84**, 6795 (1998).

[32] A.L. Gruverman, J. Hatano, and H. Tokumoto, Jpn. J. Appl. Phys. **A 36**, 2207 (1997).

[33] M. Takashige, S.-I. Hamazaki, N. Fukurai, F. Shimizu, and S. Kojima, Jpn. J. Appl. Phys. **B 35**, 5181 (1996).

[34] M. Takashige, S.-I. Hamazaki, F. Shimizu, and S. Kojima, Ferroelectrics **196**, 211 (1997).

[35] S. Balakumar, J.B. Xu, J.X. Ma, S. Ganesamoorthy, and I.H. Wilson, Jpn. J. Appl. Phys. **A 36**, 5566 (1997).

[36] G.K.H. Pang and K.Z. Baba-Kishi, J. Phys. **D 31**, 2846 (1998).

[37] L.M. Eng, M. Friedrich, J. Fousek, and P. Gunter, J. Vac. Sci. Technol. **B 14**, 1191 (1996).

[38] H. Bluhm, U.D. Schwartz, and R. Wiesendanger, Phys. Rev. **B 57**, 161 (1998).

[39] A. Correia, J. Massanell, N. Garcia, A.P. Levanyuk, A. Zlatkin, and J. Przeslawski, Appl. Phys. Lett. **68**, 2796 (1996).

[40] R. Luthi, H. Haefke, K.-P. Meyer, E. Meyer, L. Howald, and H.-J. Guntherodt, J. Appl. Phys. **74**, 7461 (1993).

[41] R. Luthi, H. Haefke, W. Gutmannsbauer, E. Meyer, L. Howald, and H.-J. Guntherodt, J. Vac. Sci. Technol. **B 12**, 2451 (1996).

[42] F. Saurenbach and B.D. Terris, Appl. Phys. Lett. **56**, 1703 (1990).

[43] J. Ohgami, Y. Sugawara, S. Morita, E. Nakamura and T. Ozaki, Jpn. J. Appl. Phys., **A 35**, 2734 (1996).

[44] L.M. Eng, J. Fousek and P. Gunter, Ferroelectrics, **191**, 211 (1997).

216

# 6. IMAGING MECHANISM OF PIEZORESPONSE FORCE MICROSCOPY ON FERROELECTRIC SURFACES

## 1. Introduction: SPM imaging of Ferroelectric Surfaces

In the recent years, scanning probe microscopy (SPM) based techniques have been successfully employed for the characterization of ferroelectric surfaces on the micron and nanometer level.[1,2,3,4] The nature of the probe and contrast formation mechanism in these techniques is vastly different; therefore, SPM image reflects different properties of ferroelectric surface. Table 6.I summarizes some of the common SPM imaging techniques used for the characterization of ferroelectric materials and briefly presents the information obtained.

Table 6.I

***Scanning Probe Techniques for Ferroelectric Imaging***

| Technique | Measured property | Materials contrast |
|---|---|---|
| Non-contact: EFM SSPM | Electrostatic force gradient (EFM) Effective surface potential (SSPM) | Stray fields above ferroelectric surface, which are related to local domain structure are measured. Only out-of–plane component can be determined. |
| Contact: PFM | Vertical (v-PFM) and lateral (l-PFM) electromechanical response of the surface. | Electromechanical response of the surface closely related to domain orientation is measured. Observation of vertical and lateral components of piezoresponse allows imaging both in-plane and out-of–plane components of polarization. |
| SCM | Voltage derivative of tip-surface capacitance | Only out-of–plane component can be determined. |
| NSOM | Optical properties of the surface | Optical indicatrix of the surface. Both in-plane and out-of-plane polarization components can be measured. |
| Friction force | Friction forces | Polarization-induced field dependent |
| microscopy | | friction force. |
| SNDM | Non-linear dielectric permittivity | Both in-plane and out-of-plane polarization components can be measured. |

The polarization related surface properties are uniform within the domain and change abruptly at domain walls, providing readily interpretable contrast in SPM measurements. In addition, most SPM techniques allow local poling of ferroelectric material with subsequent imaging of induced changes. These two factors contribute to the general interest of SPM community to ferroelectric materials. From the materials scientist point of view, the morphological information on domain structure and orientation obtained from SPM images is sufficient for many applications, and numerous observations of local domain dynamics as related to polarization switching processes,[5,6,7] ferroelectric fatigue,[8,9,10,11] phase transitions,[12,13,14,15] mechanical stresses,[16] etc. have been made.

The majority of SPM techniques listed in Table 6.I allow spectroscopic measurements, in which the local response is measured as a function of external parameter. The most widely used is voltage spectroscopy, i.e. local hysteresis loop measurements. However, unlike imaging applications, interpretation of spectroscopic measurements presents a significant challenge. The image formation mechanism in SPM techniques is usually quite complex and depends sensitively on the details of probe-surface interactions. Very often, the interaction volume of the SPM probe is small and minute contamination or damage of the surface precludes imaging. Non-local contribution to the signal is typically large (comparable to property variations between the domains), although it can usually be ignored in imaging. However, the spectroscopic analysis of local ferroelectric properties by SPM including hysteresis measurements,[17] stress effects in thin films,[18] size dependence of ferroelectric properties,[19,20] etc. requires quantitative interpretation of the SPM interaction.

Among the techniques in for ferroelectric surface imaging listed in Table 6.I the most widely used currently is Piezoresponse Force Microscopy, due to the ease of implementation, high resolution and relative insensitivity to topography and the state of the surface. It is not an exaggeration to say that PFM is rapidly becoming one of the primary imaging tools in the ferroelectric thin film research that routinely allows high resolution (~ 10 nm) domain imaging. In contrast to X-ray techniques, which are limited to averaged analysis of domain structure, PFM yields spatially resolved information on domain size, correlations, domain behavior near the inhomogeneities and grain boundaries. PFM can be used for imaging static domain structure in thin film, single crystals and polycrystalline materials, selective poling of specified regions on ferroelectric surface, studies of temporal and thermal evolution of domain structures, quantitative measurements of thermal phenomena and local hysteresis measurements. The information provided by PFM is summarized in Figure 6.1.

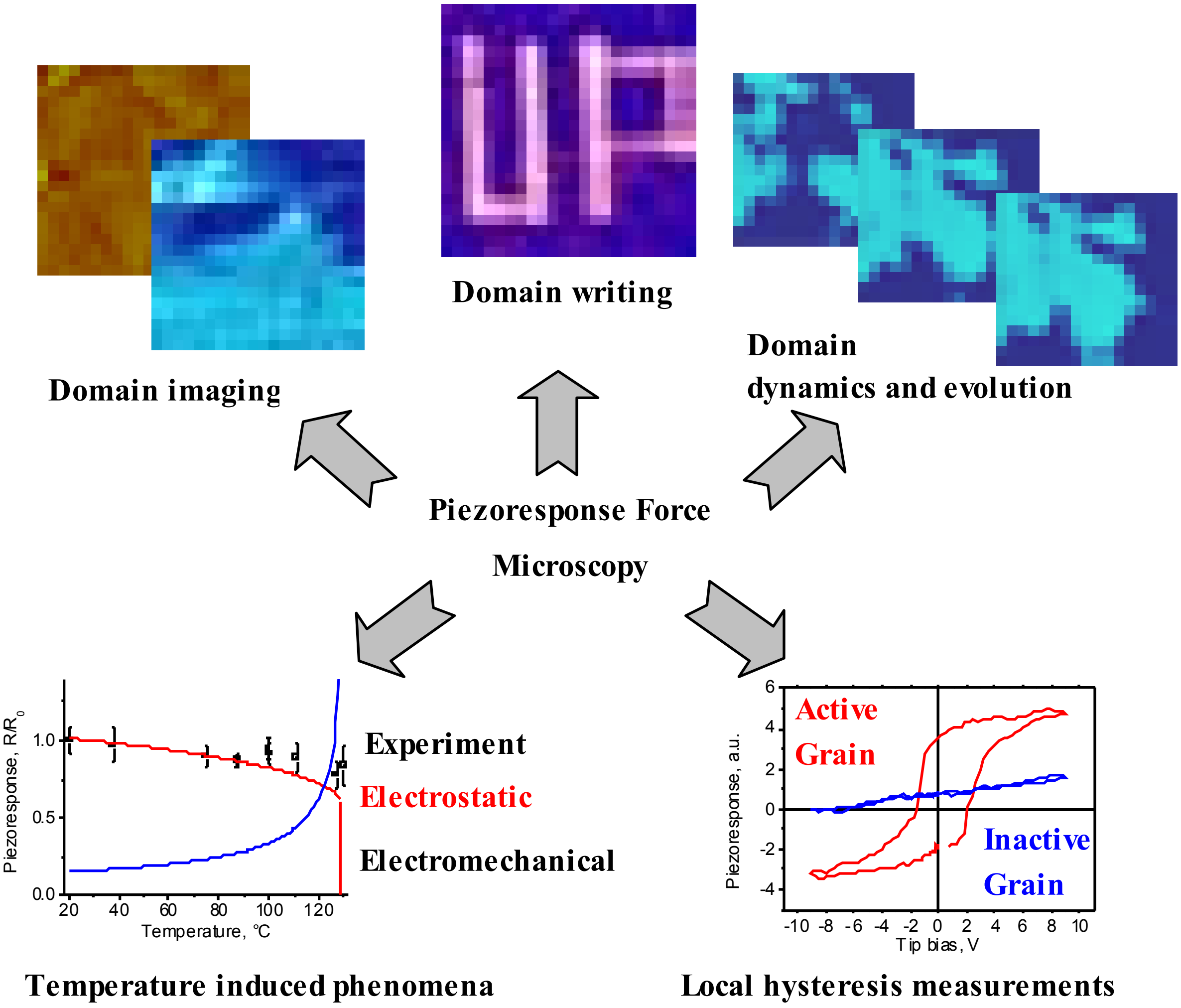

**Figure 6.1.** Applications of Piezoresponse Force Microscopy

Despite the wide range of applications, contrast formation mechanism in PFM is still under debate.[21,22,23,24] The early treatments assumed that the measured electromechanical response is equal or proportional to the piezoelectric constant $d_{33}$ of the material, with some deviations due to the clamping by surrounding material. However, Luo *et al.*[13] have found that the temperature dependence of piezoresponse contrast is similar to that of the spontaneous polarization. This behavior was attributed to the dominance of electrostatic interactions due to the presence of polarization bound charge,[25] since the electromechanical response based on the piezoelectric coefficient, $d_{33}$, would diverge in the vicinity of the Curie temperature. Similar observations were reported by other groups.[15,26,27] The presence of electrostatic forces hypothesis is also supported by measurements on the nonpiezoelectric surfaces.[28] In contrast, the existence of a lateral PFM signal,[29,30,31] the absence of relaxation behavior in PFM contrast as opposed to SSPM contrast, as well as numerous observations using both EFM/SSPM and PFM[32,33] clearly point to significant electromechanical contribution to PFM contrast. Significant progress in the understanding of PFM was achieved recently by Max-Plank Institute (Halle) group,[34] Ecole Federal Politechnique de Lozanne (Switzerland) group,[26] and University of Maryland group.[35] However, a comprehensive description of the Piezoresponse Force Microscopy contrast formation mechanism including electrostatic, electromechanical, and non-local cantilever interactions as well as tip-surface contact issues has not been achieved.

In this Chapter, contrast formation mechanism is analyzed and relative magnitudes of electrostatic vs. electromechanical contributions to PFM interaction for the model case of $c^+$, $c^-$ domains in tetragonal perovskite ferroelectrics are determined. The principles of PFM are summarized in Section 6.2. Electrostatic tip-surface interactions are analyzed in Section 6.3 using the image charge model for conductive sphere-anisotropic dielectric half plane. Electromechanical contrast is analyzed in detail in Section 6.4 and limiting cases for electromechanical response are obtained. It is shown that both electrostatic and electromechanical interactions can contribute to the PFM image. The relative contributions of these interactions depend on the experimental conditions. Contrast Mechanism Maps were constructed to delineate the regions with dominant electrostatic and electromechanical interactions in Section 6.5. Under some conditions, i.e. those corresponding to a relatively large indentation force and tip radius, the real piezoelectric coefficient can be determined. This analysis reconciles existing discrepancies in the interpretation of PFM imaging contrast. The non-local contribution to PFM signal arises due to the buckling oscillations of the cantilever and corresponding analysis is performed in Section 6.6. Temperature dependence of PFM contrast for $BaTiO_3$ is analyzed in Section 6.7. The description of any SPM technique would be incomplete without the analysis of imaging artifacts as described in Section 6.8. Finally, an approach to simultaneous acquisition of PFM and potential images is presented in Section 6.9.

### 6.2. Principles of Piezoresponse Force Microscopy

Piezoresponse Force Microscopy is based on the detection of bias-induced surface deformation. The tip is brought into contact with the surface and the piezoelectric response of the surface is detected as the first harmonic component of bias-induced tip deflection $d = d_0 + A\cos(\omega t + \varphi)$, as shown in Figure 6.2.

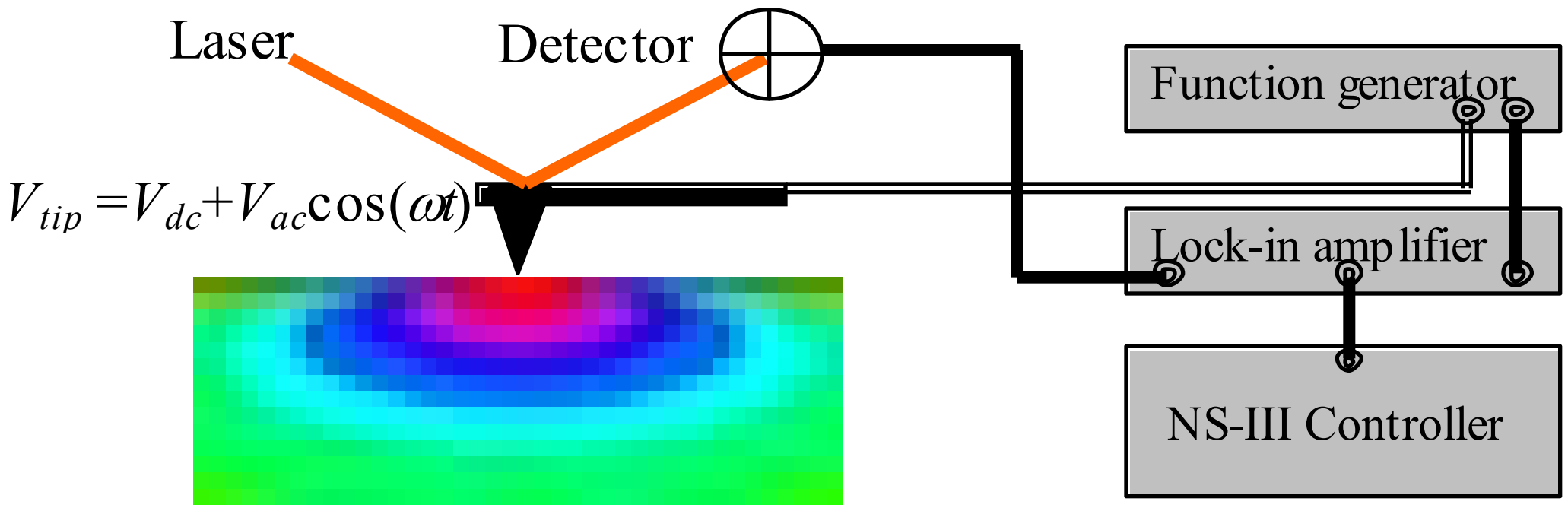

**Figure 6.2.** Schematics of PFM. Conductive tip is brought into contact with ferroelectric surface. The tip is ac biased by function generator. The first harmonic of tip oscillations is measured by lock-in amplifier and the phase and amplitude of the response are stored as the PFM image. Often, *x*-signal of the lock-in is acquired.

The phase of the electromechanical response of the surface, $\varphi$, yields information on the polarization direction below the tip. For $c^-$ domains (polarization vector pointing downward) the application of a positive tip bias results in the expansion of the sample and surface oscillations are in phase with the tip voltage, $\varphi = 0$. For $c^+$ domains $\varphi = 180°$. The piezoresponse amplitude, $PR = A/V_{ac}$, defines the local electromechanical activity of the surface. For a purely electromechanical response, piezoresponse amplitude is equal for $c^+$ and $c^-$ domains, while it is zero at the domain boundary. The width of the amplitude depression at the domain boundary is an experimental measure of the spatial resolution of the technique. The width of the phase profile cannot serve as the definition of the resolution; rather it represents the noise level and/or time constant of the lock-in amplifier.

As discussed in the introduction, both long range electrostatic forces and the electromechanical response of the surface contribute to the PFM signal so that the experimentally measured piezoresponse amplitude is $A = A_{el} + A_{piezo} + A_{nl}$, where $A_{el}$ is electrostatic contribution, $A_{piezo}$ is electromechanical contribution and $A_{nl}$ is non-local contribution due to capacitive cantilever-surface interactions.[25,29,36] Quantitative PFM imaging requires $A_{piezo}$ to be maximized to achieve predominantly electromechanical contrast. Alternatively, for $A_{el} >> A_{piezo}$, electrostatic properties of the surface will be imaged. Often local polarization provides the dominant contribution to both electromechanical and electrostatic properties of the surface; qualitative imaging of domain structures is thus possible in both electromechanical and electrostatic cases. Cantilever size is usually significantly larger than domain size; therefore, a non-local cantilever contribution is usually present in the form of additive offset on PFM *x*-image. It can however lead to the erroneous interpretation of phase and amplitude images as shown in the Figure 6.3.

It is illustrative to estimate the effect of these interactions on PFM images using formalism developed by Hong *et. al.*[37] Assuming that $A_{el} = F_{loc} V_{ac} (V_{tip} - V_{loc})$, $A_{piezo} = d_{eff} V_{ac}$ and $A_{nl} = F_{nl} V_{ac} (V_{tip} - V_{av})$, where $V_{tip}$ is tip potential, $V_{loc}$ is local potential below the tip apex, $d_{eff}$ is effective electromechanical response of the surface, $V_{av}$ is average surface potential below the cantilever, $F_{loc}$ and $F_{nl}$ are proportionality coefficients determined by tip-surface and cantilever surface capacitance gradients, tip-surface contact stiffness and spring constant of the cantilever. In the simplest approximation (the field is uniform) $d_{eff}$ is equal to $d_{33}$; taking into account second order effects, $d_{eff} = d_{33} + Q(V_{tip} - V_{loc})$, where $Q$ is corresponding electrostrictive coefficient.

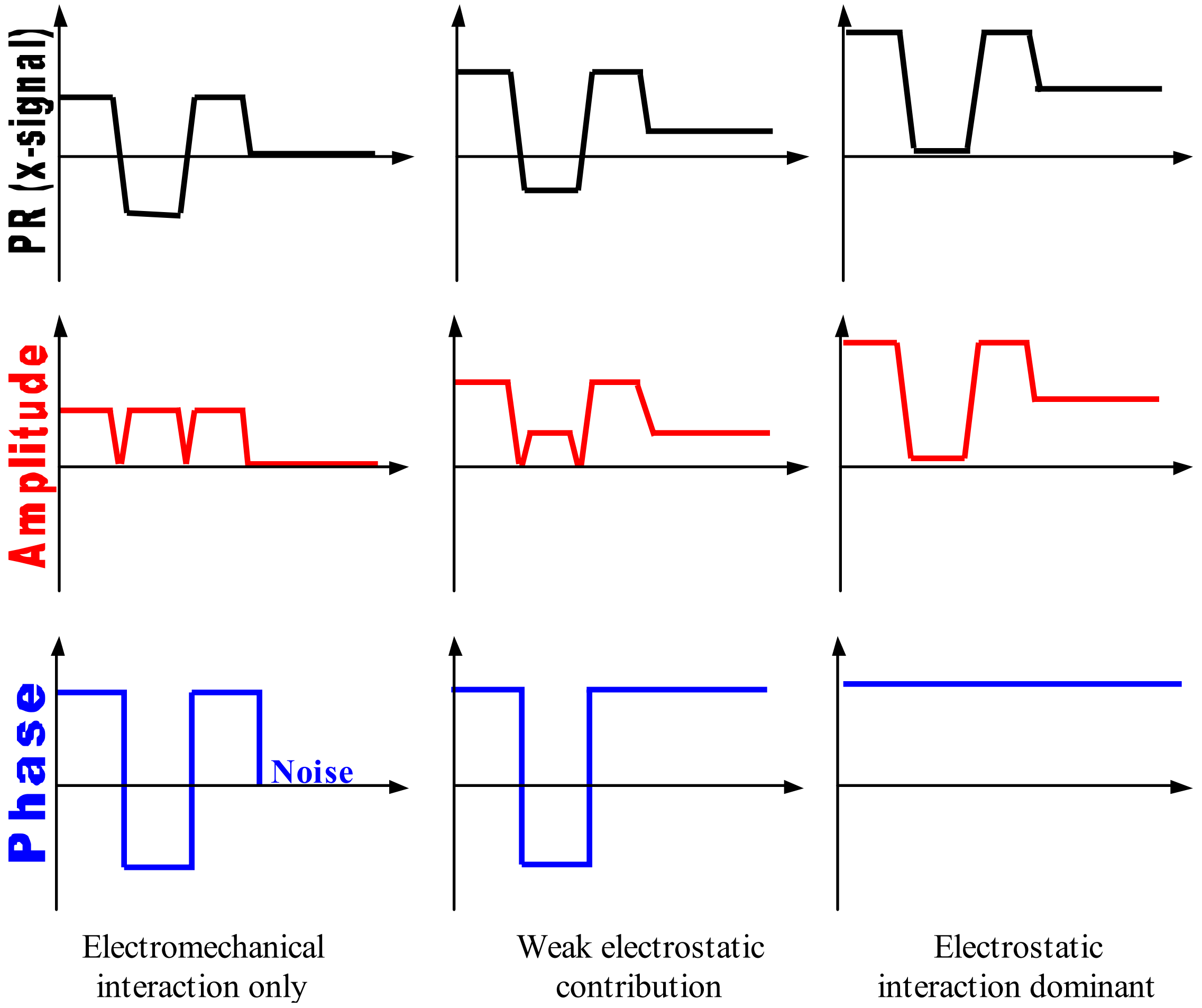

**Figure 6.3.** Schematic of signals in PFM of $c^+/c^-/c^+/a$ domain structure. Shown are *x*-signal, amplitude and phase signal for purely electromechanical case, with weak electrostatic contribution and strong electrostatic contribution.

In the purely electromechanical case, $F_{loc}$ and $F_{nl}$ are identically zero. In this case, the response amplitudes are equal in $c^+$ and $c^-$ domain regions, while phase changes by 180° between the domains. For domains with arbitrary orientation, the absolute value of the amplitude signal provides a measure of the piezoelectric activity of the domain; in-plane domains or non-ferroelectric regions are seen as the regions with zero response amplitude. PFM spectroscopy represents ideal electromechanical hysteretic behavior from which materials properties such as piezoelectric and electrostriction coefficients can be obtained. The relationship between the classical and electromechanical hysteresis loops is illustrated in Figure 6.4.

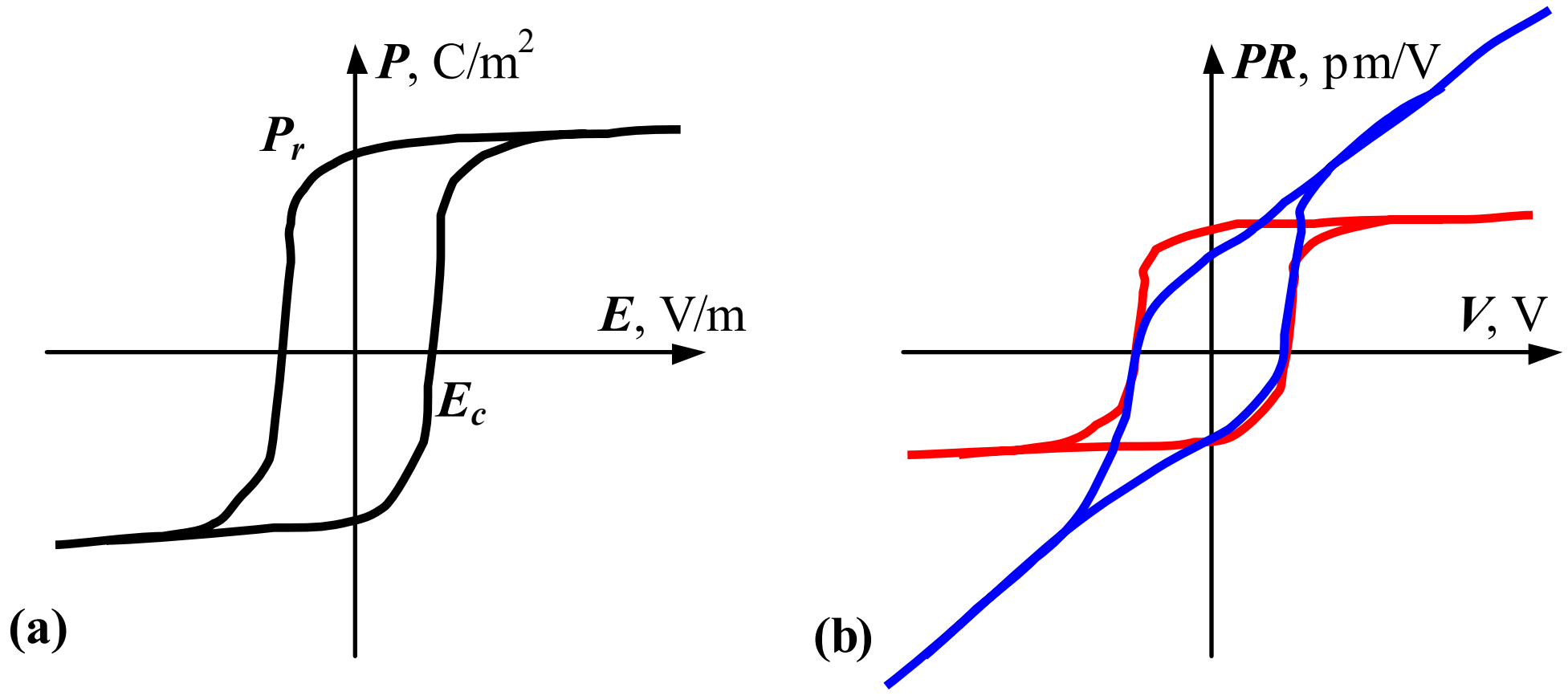

**Figure 6.4.** Standard hysteresis loop (a) and electromechanical hysteresis loop measured in piezoresponse force microscopy (b)

In the more realistic case, both electrostatic and electromechanical interactions contribute and the PFM *x*-signal over $c^+$ and $c^-$ domains can be written as:

$$PR_+ = d_{33} + Q(V_{tip} - V_1) + F_{loc}(V_{tip} - V_1) + F_{nl}(V_{tip} - V_{av}), \tag{6.1a}$$

$$PR_- = -d_{33} + Q(V_{tip} - V_2) + F_{loc}(V_{tip} - V_2) + F_{nl}(V_{tip} - V_{av}). \tag{6.1b}$$

For a small non-local electrostatic contribution, $d_{33} >> F_{nl}(V_{tip} - V_{av})$, the phase still changes by 180° between the domains (Figure 6.3); however, the response amplitudes are no longer equal in $c^+$ and $c^-$ domain regions. Non-ferroelectric regions are seen as the region with finite response amplitude. Similar behavior is expected for non-zero local electrostatic contribution, in which case the piezoresponse is a sum of electromechanical and electrostatic contributions and depends linearly on tip bias, $PR_\pm = d_{33} \pm F_{loc}(V_{tip} - V_{loc})$. The immediate implication of Eqs.(6.1a,b) is that second order electrostriction coefficients can be determined by PFM if and only if imaging is purely electromechanical; otherwise, electrostatic response is measured as discussed in Section 6.6. For a large non-local contribution, $F_{nl}(V_{tip} - V_{av}) >> d_{33}$, the phase is determined by electrostatic force only and does not change between the domains. The amplitude signal is strongly asymmetric and maximal response corresponds to either $c^+$ or $c^-$ domain, while minimal signal is observed to the domain with opposite polarity. In-plane domains or non-ferroelectric regions are seen as regions with intermediate contrast on amplitude image. Obviously, even qualitative analysis of PFM image in this case is difficult: for example, a sample comprised of $a$-$c^+$ and $c^-$-$c^+$ domains can not be distinguished based on a vertical PFM image only.

This analysis can be developed by considering two experimental parameters readily accessible in any SPM experiment. The signal average taken over $x$-image of the sample with random domain orientation (unpoled), provided that $V_{tip}$ is smaller than the switching bias and the image size is significantly larger than the domain size, can be written as:

$$PR_{av} = (Q + F_l + F_{nl})(V_{tip} - V_{av}), \tag{6.2}$$

while contrast between domains is

$$\Delta PR = 2d_{33} + (Q + F_l)(V_1 - V_2). \tag{6.3}$$

The averaged image signal provides a measure of electrostatic tip-surface interactions, while domain contrast comprises both electrostatic and electromechanical contributions. For imaging in ambience, the potential difference between domains of opposite polarity, $V_1$-$V_2$, is small and typically does not exceed 200 mV, while tip dc bias can be changed in the wide region of ~ -30- 30 volts. Therefore, the conditions for dominant electromechanical contrast in imaging are significantly less stringent that in spectroscopy. On a high quality ferroelectric surface, the spatial variations of both electromechanical and electrostatic properties are strongly related to ferroelectric domain structure. This implies that qualitative domain imaging is possible even if the electrostatic contribution dominates, $F_{loc}(V_1 - V_2) >> d_{33}$, provided that local polarization provides the dominant contribution to surface potential. Piezoresponse in this case is $PR_{\pm} = F_{loc}(V_{tip} - V_{loc})$ and changes linearly with tip bias. In fact, some of the existing references on resonance-enhanced PFM seem to belong to this category.[38,39] It should be noted that the application of high bias to the tip typically results in the significant charge transfer between the tip and the surface, thus altering the surface potential distribution.[33,40,41,42,43] This consideration is expected to limit imaging in this regime to small tip biases and indeed no hysteresis loop measurements in the "electrostatic" PFM has been reported.

As follows from Eqs.(6.1a,b), quantitative spectroscopic piezoresponse measurements require that the electrostatic and non-local components of the response be minimized. The first step in developing systematic approach is the reliable calculation of the magnitudes of electrostatic, electromechanical and non-local responses as a function of tip radius of curvature, indentation force and cantilever spring constant.

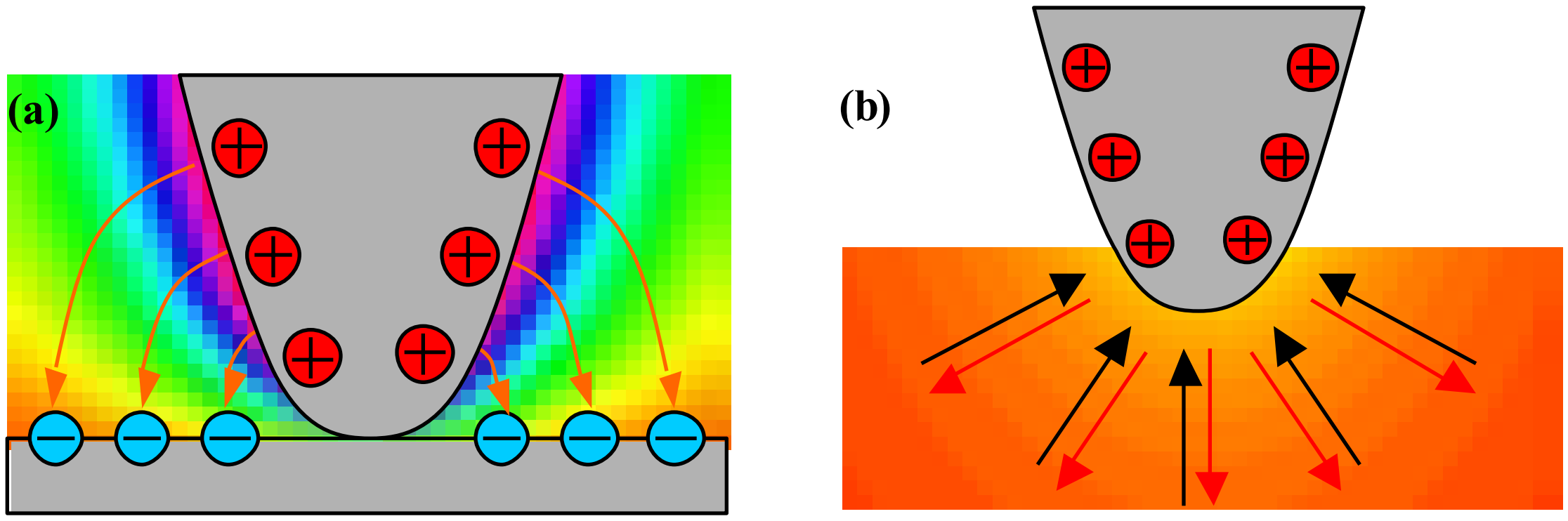

**Figure 6.5.** Electrostatic (a) and electromechanical (b) contrast in PFM.

One of the difficulties in a comparison of the relative magnitudes of electromechanical and electrostatic responses is the difference in the contrast transfer mechanism. In the electromechanical case, the surface displacement due to inverse piezoelectric effect is determined as a function of applied voltage (Figure 6.5). Tip deflection in this case is equal to surface displacement since the contact stiffness of the tip–surface junction is usually much larger than cantilever spring constant. In the electrostatic case, the force containing both local and non-local components is defined.

Here, the influence of imaging conditions on PFM contrast is analyzed. The magnitudes of $A_{el}$ and $A_{piezo}$ are determined as a function of indentation force and tip radius of curvature. The case when the measured response amplitude owes to electrostatic tip-surface interaction only, $A = A_{el}$, is further referred to as electrostatic regime. The case when the electromechanical contribution dominates, $A = A_{piezo}$, is referred to as electromechanical regime. The non-local contribution, $A_{nl}$, is governed by the cantilever spring constant rather than indentation force and is considered separately in Section 6.6.

## 6.3. Electrostatic Contrast

In the electrostatic regime of piezoresponse force microscopy the capacitive and Coulombic tip-surface interactions result in an attractive force between the tip and the surface, that causes surface indentation below the tip (Figure 6.5). The early attempts to quantify tip-surface interactions in this regime approximated tip-surface junction by a plane-plane capacitor. Obviously, this is inapplicable in contact because a capacitive force in planar geometry does not cause tip deflection. A correct description of the electrostatic tip-surface interaction must take into account the tip shape.

### 6.3.1. Potential Distribution in the Tip-surface Junction

The potential distribution in the tip-surface junction in non-contact imaging is often analyzed in the metallization limit for the surface.[44] In this limit, the tip-surface capacitance $C_d(z,\kappa)$, where $z$ is the tip-surface separation and $\kappa$ is the dielectric constant for the sample is approximated as $C_d(z,\kappa) \approx C_c(z)$, where $C_c(z)$ is the tip-surface capacitance for a conductive tip and conductive surface. This approximation breaks down for small tip-surface separations when the effect of field penetration in the dielectric sample is non-negligible. For ferroelectric surfaces, the effective dielectric constant is high, $\kappa \approx$ 100-1000, favoring the metallization limit. However, in contact tip-surface separation $z \approx 0$ leads to a divergence in the capacitance, $C_c(z)$, and the corresponding force. To avoid this difficulty and, more importantly, take into account the anisotropy of the ferroelectric medium, we calculate the tip-surface force using the image charge method for spherical tip geometry. This approach is applicable when the tip-surface separation is small, $z << R$, where $R$ is radius of curvature of the tip.

Table 6.II.

*Image charges for conductive, dielectric and anisotropic dielectric planes*

| | conductive | isotropic dielectric | anisotropic dielectric |
|---|---|---|---|
| $Q'$ | $-Q$ | $-\frac{\kappa-1}{k+1}Q$ | $-\frac{\sqrt{\kappa_z\kappa_x}-1}{\sqrt{\kappa_z\kappa_x}+1}Q$ |
| $d'$ | $-d$ | $-d$ | $-d$ |
| $Q''$ | 0 | $\frac{2\kappa}{k+1}Q$ | $\frac{2\sqrt{\kappa_z\kappa_x}}{\sqrt{\kappa_z\kappa_x}+1}Q$ |
| $d''$ | | $d$ | $d\sqrt{\kappa_z/\kappa_x}$ |

The potential in air produced by charge $Q$ at a distance $d$ above a conductive or dielectric plane located at $z = 0$ can be represented as a superposition of potentials produced by the original charge and the corresponding image charge $Q'$ located at position $z = d'$ below the plane. The potential in a dielectric material is equal to that produced by a different image charge $Q''$ located at $z = d''$.[45,46,47] Values of $Q'$, $Q''$, $d'$ and $d''$ for metal and isotropic or anisotropic dielectric materials are summarized in Table 6.II. Note that the potential in air above an anisotropic dielectric material is similar to the isotropic case with an effective dielectric constant $\kappa_{eff} = \sqrt{\kappa_x \kappa_z}$, where $\kappa_x$, $\kappa_z$ are the principal values of the dielectric constant tensor. This simple image solution can be used only for the dielectric for which one of the principal axes of dielectric constant tensor coincides with surface normal and in-plane dielectric constants are equal, $\kappa_x = \kappa_y$. For systems without in-plane isotropy, field distribution is no longer rotationally invariant and does not allow representation with single image charge; more complex solutions using distributed image charges are required.[48]

To address tip-surface interactions and taking the effect of the dielectric medium into account, the image charge distribution in the tip can be represented by charges $Q_i$ located at distances $r_i$ from the center of the sphere such that:

$$Q_{i+1} = \frac{\kappa - 1}{\kappa + 1} \frac{R}{2(R + d) - r_i} Q_i, \tag{6.4a}$$

$$r_{i+1} = \frac{R^2}{2(R + d) - r_i}, \tag{6.4b}$$

where $R$ is tip radius, $d$ is tip-surface separation, $Q_0 = 4\pi\varepsilon_0 RV$, $r_0 = 0$ and $V$ is tip bias. Tip-surface capacitance is

$$C_d(d, \kappa) = \frac{1}{V} \sum_{i=0}^{\infty} Q_i , \tag{6.5}$$

from which the force can be found. The rotationally invariant potential distribution in air can be found from Eqs.(6.4a,b). One of the important parameters for the description of tip-surface junction is potential on the surface directly below the tip, which defines the potential attenuation in the tip-surface gap. Specifically, for sphere plane model potential on the surface directly below the tip is

$$V(0,0)=\frac{1}{4\pi\varepsilon_0}\frac{2}{\kappa+1}\sum_{i=0}^{\infty}\frac{Q_i}{R+d-r_i}. \tag{6.6}$$

In the conductive surface limit, $\kappa=\infty$ and Eq.(6.5) is simplified to[49]

$$C_c=4\pi\varepsilon_0 R\sinh\beta_0\sum_{n=1}(\sinh n\beta_0)^{-1}, \tag{6.7}$$

where $\beta_0=\text{arccosh}((R+d)/R)$. Surface potential in this case is $V(0,0)\equiv 0$. For the conductive tip-dielectric surface

$$C_d=4\pi\varepsilon_0 R\sinh\beta_0\sum_{n=1}\left(\frac{\kappa-1}{\kappa+1}\right)^{n-1}(\sinh n\beta_0)^{-1}. \tag{6.8}$$

While in the limit of small tip-surface separation $C_c$ diverges logarithmically, $C_d$ converges to the universal "dielectric" limit[50]

$$C_d(\kappa)_{z=0}=4\pi\varepsilon_0 R\frac{\kappa-1}{\kappa+1}\ln\left(\frac{\kappa+1}{2}\right). \tag{6.9}$$

The distance dependence of tip-surface capacitance and surface potential directly below the tip are shown in Figure 6.6a,b.

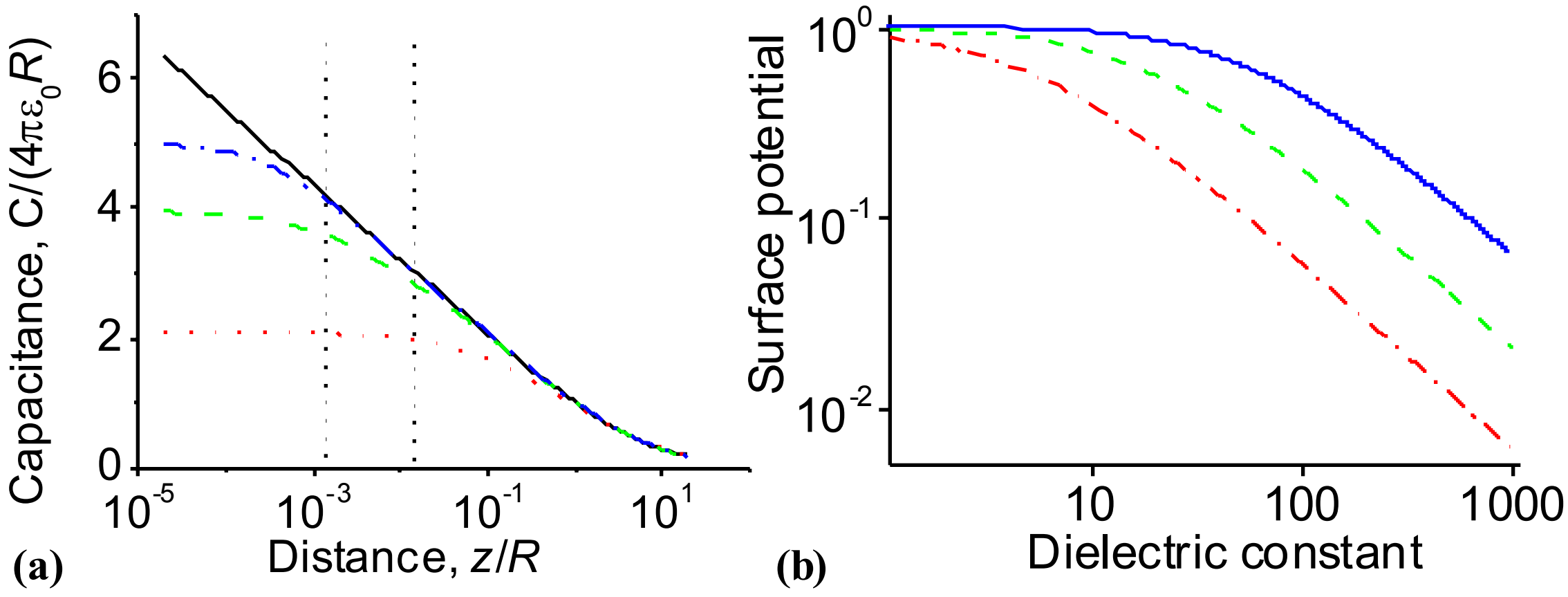

**Figure 6.6.** (a) Tip-dielectric surface capacitance for κ = 10 ( ), κ = 100 ( ) and κ = 1000 ( ), compared to the metallic limit ( ). Vertical lines delineate the region of characteristic tip-surface separations (0.1-1nm) in contact mode for tip radius $R$ = 50nm. (b) Surface potential below the tip for tip-surface separations $z$ = 0.1 $R$ ( ), $z$ = 0.01 $R$ ( ) and $z$ = 0.001 $R$ ( ) as a function of the dielectric constant of the surface.

For relatively large tip-surface separations $C_d(z,\kappa) \approx C_c(z)$, which is the usual assumption in non-contact SPM imaging. The most prominent feature of this solution is that, while for low-$\kappa$ dielectric materials the tip-surface capacitance achieves the dielectric limit in contact and hence surface potential is equal to the tip potential, this is not the case for high-$\kappa$ materials. Tip-surface capacitance, capacitive force and electric field can be significantly smaller than in the dielectric limit. The surface potential below the tip is smaller than the tip potential and is inversely proportional to dielectric constant (Figure 6.6b). This is equivalent to the presence of an apparent dielectric gap between the tip and the surface that attenuates the potential. Even though under typical experimental conditions contact area between the tip and the surface is non-zero and the sphere-plane model is not rigorous, this analysis is still valid. For perfect ferroelectric surface, the gap owes to intrinsic phenomena such as finite Thomas-Fermi length of the tip material (~0.5 A) and/or non-uniform polarization distribution in surface layer of ferroelectric. For small contact area and large dielectric constant of the material potential drop in such intrinsic gap can be significant. In many practical cases, surface damage, contamination or loss of volatile components during fabrication results in the extrinsic non-ferroelectric dead layer. In either case, Eq.(6.6) and Figure 6.6 illustrate the implications of a dielectric gap on tip-induced surface potential since the minimal tip-surface separation, $d$, is limited by dead layer width.

### 6.3.2. Tip-surface Forces and Indentation Depth

The electrostatic contribution, $A_{el}$, to piezoresponse amplitude is calculated assuming that the total force acting on the tip is comprised of the elastic contribution due to the cantilever, $F_0 = kd$, and a capacitive force, $F_{el} = C'\left(V_{tip} - V_{surf}\right)^2$, where $k$ is the cantilever spring constant, $d$ the setpoint deflection, $C'$ is the tip-surface capacitance gradient and $V$ is the potential. From Eqs.(6.5,7,8), the magnitudes of capacitive and Coulombic forces between the cantilever-tip assembly and the surface can be estimated. The capacitive force is:

$$2F_{cap} = C'_{loc}\left(V_{tip} - V_{loc}\right)^2 + C'_{nl}\left(V_{tip} - V_{av}\right)^2, \qquad (6.10)$$

where $V_{tip}$ is the tip potential, $V_{loc}$ is the domain-related local potential directly below the tip, $V_{av}$ is the surface potential averaged over the distance comparable to the cantilever length, $C'_{loc}$ is the local part of tip-surface capacitance gradient and $C'_{nl}$ is the non-local part due to the cantilever. Typically, the cantilever length is significantly larger than the characteristic size of ferroelectric domains; therefore, the non-local part results in a constant background on the image that does not preclude qualitative domain imaging. The non-local capacitance gradient can be estimated using plane-plane geometry as $C'_{nl} = \varepsilon_0 S(z+L)^{-2}$, where $S$ is the effective cantilever area and $L$ is the tip length. For a typical tip with $L \approx 10$ μm and $S \approx 2 \cdot 10^3$ μm$^2$, the non-local contribution is $C'_{nl} \approx 1.8 \cdot 10^{-10}$ F/m and is independent of tip radius. The force for a tip-surface potential difference of 1 V is $F_{nl} \approx 0.9 \cdot 10^{-10}$ N. Here, it is assumed that the force acting on the cantilever results in surface indentation only and the cantilever geometry does not change, i.e. measured is tip displacement. Practically, optical detection scheme used in most modern AFMs implies that measured is cantilever bending angle; buckling oscillations of the cantilever, which are not associated with vertical tip motion tip result in the effective displacement as analyzed in Section 6.6.

The local capacitive contribution due to the tip apex is $F_{loc} = 1.4 \cdot 10^{-8}$ N for $z$ = 0.1 nm, $R$ = 50 nm, i.e. two orders of magnitude larger. However, $C'_{loc}$ scales linearly with tip radius and, therefore, for the sharp tips capable of high-resolution non-local contributions to the signal increase. Similar behavior is found for non-contact SPMs.[51] The Coulombic tip-surface interaction due to polarization charge can be estimated using the expression for the electric field above a partially screened ferroelectric surface, $E^u = (1-\alpha)P\varepsilon_0^{-1}\left(1+\sqrt{\kappa_x \kappa_z}\right)^{-1}$, where $\alpha$ is the degree of screening and $P$ is spontaneous polarization ($P$ = 0.26 C/m$^2$ for $BaTiO_3$). For unscreened surfaces $\alpha$ = 0 so this Coulombic contribution in the limit $F_{coul} << F_{cap}$ is $F_{coul} = C_{loc}\left(V_{tip} - V_{loc}\right)E^u$ and for the same tip parameters as above $F_{coul} = 2.2 \cdot 10^{-9}$ N. However, polarization charge is almost completely screened in air (as discussed in Chapter 5), typically $1-\alpha << 10^{-3}$, and under these conditions the Coulombic contribution can be excluded from the electrostatic tip-surface interaction for isothermal experiments.

Capacitive force results in an indentation of the surface. In the Hertzian approximation the relationship between the indentation depth, $h$, tip radius of curvature, $R$, and load, $P$, is[52]

$$h = \left( \frac{3P}{4E^*} \right)^{\frac{2}{3}} R^{-\frac{1}{3}}, \tag{6.11}$$

where $E^*$ is the effective Young's modulus of the tip-surface system defined as

$$\frac{1}{E^*} = \frac{1-\nu_1^2}{E_1} + \frac{1-\nu_2^2}{E_2}. \tag{6.12}$$

$E_1$, $E_2$ and $\nu_1$, $\nu_2$ are Young's moduli and Poisson ratios of tip and surface materials [Figure 6.7]. For ferroelectric perovskites the Young's modulus is of the order of $E^* \approx$ 100GPa. The contact radius, $a$, is related to the indentation depth as $a = \sqrt{hR}$ . Hertzian contact does not account for adhesion and capillary forces in a tip-surface junction and a number of more complex models for nanoindentation processes are known.[53]

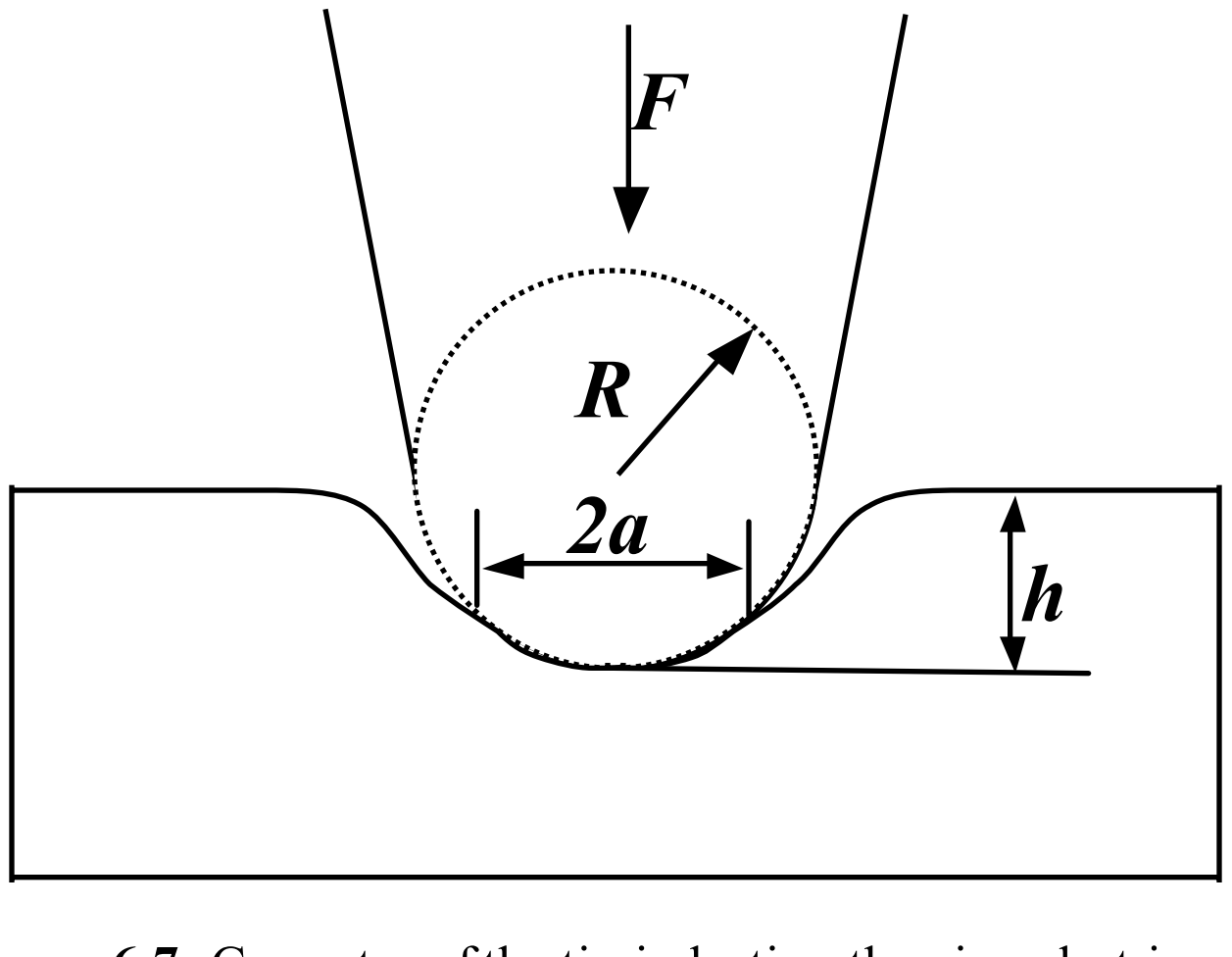

**Figure 6.7.** Geometry of the tip indenting the piezoelectric surface

Under typical PFM operating conditions the total force acting on the tip is $F = F_0 + F_{el}$, where $F_0 = k\ d_0$ is elastic force exerted by the cantilever of spring constant $k$ at setpoint deflection $d_0$ and $F_{el}$ is the electrostatic force. Since the electrostatic force is modulated, $V_{tip} = V_{dc} + V_{ac} \cos(\omega t)$, the first harmonic of tip deflection is

$$h_{1\omega} = \frac{\chi}{2\pi\omega}\int \left(F_0 + C'_{loc}\left(V_{dc} + V_{ac}\cos(\omega t) - V_{loc}\right)^2\right)^{\frac{2}{3}}\cos(\omega t)dt, \qquad (6.13)$$

where $\chi = \left(3/4E^*\right)^{\frac{2}{3}} R^{-\frac{1}{3}}$. In the limit when the indentation force is much larger than electrostatic force, $F_{el} << F_0$, the effective spring constant of the tip-surface junction is $k_{eff} = \partial P/\partial h$ and the first harmonic of cantilever response is $h_{1\omega} = F_{1\omega} / k_{eff}$. For a Hertzian indentation the response is:

$$h_{1\omega} = \frac{2}{3}\left(\frac{3}{4E^*}\right)^{2/3} R^{-1/3} F_0^{-1/3} F_{1\omega}. \qquad (6.14)$$

This equation can be also obtained directly from an expansion of the integrand in Eq.(6.13). For typical PFM imaging conditions the setpoint deflection is ~100 nm and the spring constant of the cantilever $k$ varies from ~0.01 to ~100 N/m. Consequently, imaging can be done under a range of loads spanning at least 4 orders of magnitude from 1 nN to 10 μN. For $F_0$ = 100 nN, $E^*$ = $10^{11}$ Pa and potential difference between the domains $\Delta V$ = 150 mV, PFM contrast between the domains of opposite polarities is $\Delta h_{1\omega} = 6.02 \cdot 10^{-12}$ m/V. It should be noted that the potential difference between ferroelectric domains in ambient is determined by the properties of the adsorbate layer that screens spontaneous polarization.[54] Under UHV conditions where the intrinsic screening by charge carriers[55] dominates the potential difference would be larger and can achieve the limiting value of $\Delta V$ = 3 V comparable to band gap. In this case, the electrostatic PFM contrast between the domains of opposite polarities can be as large as $\Delta h_{1\omega} = 1.2 \cdot 10^{-10}$ m/V.

It is useful to consider the effect of cantilever stiffness on the electrostatic contrast. For soft cantilevers, the indentation depth can be extremely small. The electrostatic tip-surface and even cantilever-surface interaction can dominate over the elastic load, especially for the large potential difference between the tip and surface typical during hysteresis measurements or polarization switching. In this case, the linear approximation of Eq.(6.14) is no longer valid. In the small signal approximation, $V_{ac} \rightarrow 0$, the response amplitude can still be obtained from Eq.(6.13) where the effective load is now $F_0 = kd_0 + C_{loc}^{'}(V_{dc} - V_{loc})^2$, predicting a decrease of response with bias. Interestingly, the integral in Eq.(6.13) is nullified for zero tip-surface potential difference, $V_{dc}$-$V_{loc}$ = 0. Therefore, the imaging mechanism bears close similarity to that of non-contact open-loop SSPM and feedback can be employed to obtain nulling potential map on any surface. On piezoelectric surfaces electromechanical contribution is non-zero and nulling condition does not correspond to equilibrium surface potential. For a small indentation force the cantilever dynamics are expected to be significantly more complex; the tip can lose contact with the surface in the upper part of the trajectory, the cantilever vibration can be significant, etc. Some of these effects are discussed in Sections 6.6 and 6.8.

## 6.4. Electromechanical Contrast

The analysis of electrostatic interactions above is applicable to any dielectric surface; however, for ferroelectric and, more generally, piezoelectric materials an additional bias-induced effect is a linear electromechanical response of the surface. A rigorous mathematical description of the problem is extremely complex and involves the solution of coupled electromechanical mixed boundary value problem for the anisotropic medium. Fortunately, the geometry of the tip-surface junction in PFM is remarkably similar to the well-studied piezoelectric indentation problem.[56,57,58,59,60] In the classical limit, the coupled electromechanical problem is solved for mixed value boundary conditions; $V_s = V_{tip}$ in the contact area and the normal component of the electric field $E_z$ = 0 elsewhere. It is shown in Section 6.3 that surface potential can be significantly attenuated due to the dielectric tip-surface gap. The gap can originate both from the intrinsic properties of tip-surface junction and due to the presence of "dead layers" on the ferroelectric surface.[61] To account for this effect, we introduce semiclassical contact limited strong indentation, in which $V_s = \gamma V_{tip}$ in the contact area, where $\gamma$ is the attenuation factor. SPM experiment can also be performed under the conditions when the contact area is negligibly small. Piezoelectric deformation occurs even when the tip is not in contact due to the tip-induced non-uniform electric field. In this case, the zero field approximation outside of the contact area is invalid; instead, the contact area itself can be neglected and the surface deformation can be ascribed solely to field effects. Therefore, we distinguish two limits for the PFM electromechanical regime:

1. Strong (classical) indentation: $V = V_{\text{tip}}$ in the contact area, $E_z = 0$ elsewhere

2. Weak (field induced) indentation: contact area is negligible, $E_z \neq 0$

In practice, both mechanisms might operate and the dominant contribution depends on imaging parameters. Interestingly, by this definition, response calculated in the weak indentation (WI) model allows to estimate the error associated with $E_z = 0$ assumption in the strong indentation (SI) model.

### 6.4.1. Strong Indentation Limit

A complete description of the strong indentation limit is given by Suresh and Giannakopoulos,[62,58] who extended Hertzian contact mechanics to piezoelectric materials. The relationship between the load, $P$, indentor potential, $V$, and indentation depths, $h$, is

$$h = \frac{a^2}{R} + \frac{2\beta}{3\alpha} V, \tag{6.15a}$$

$$P = \alpha \frac{a^3}{R} - \beta a V \tag{6.15b}$$

where $\alpha$ and $\beta$ are materials dependent constants and $a$ is contact radius (Appendix C). Solving Eqs.(6.15a,b) for indentation depth as a function of indentor bias relevant for PFM yields surface deformation as illustrated in Figure 6.8a. For small modulation amplitudes, the PFM contrast is $h_{1\omega} \approx h'(F, V_{dc}) V_{ac}$, where the functional form of $h(F, V_{dc})$ is given by Eq.(6.15a,b). The bias dependence of the piezoresponse coefficient is given by the local slope, $k = h'(F, V_{dc})$ shown in Figure 6.8b. For $V_{dc}$ = 0 the asymptotic analysis of Eqs.(6.15a,b) for the $c^+$ orientation yields $k_0 = 4/3\ \beta/\alpha$, while for $V_{dc} \to +\infty$ and $V_{dc} \to -\infty$ the respective limits are $k_{+\infty} = 5/3\ \beta/\alpha$ and $k_{-\infty} = 2/3\ \beta/\alpha$ (Figure 6.8b) and are independent of tip radius and contact force. The response amplitude in the strong indentation limit is comparable to the corresponding $d_{33}$ value (Table 6.III).

The applicability of the strong indentation regime to PFM contrast is limited. A high dielectric constant leads to a significant potential drop between the tip and the surface, $V_s << V_{tip}$; therefore, for an infinitely stiff tip and surface the basic assumption of the strong indentation limit, $V_s = V_{tip}$ in the contact area, is not fulfilled. Even for finite contact the potential on the surface below the tip is lower than the tip potential and differs from that assumed in the strong indentation limit. It is useful to consider the effect of contact radius on this assertion. A simple approximation for the surface potential below the tip is $V_s = \gamma V_{tip}$ in the contact area, where $\gamma$ is the attenuation factor (Figure 6.9a-d). Such behavior is referred to as contact limited strong indentation (CSI). Using a spherical approximation for contact region, attenuation factor is estimated as $\gamma = \left(1 + w\kappa_{eff}/a\kappa_d\right)^{-1}$, where $w$ is the thickness of the "apparent" dielectric gap ($w > 0.1$ nm), $\kappa_d$ is the dielectric constant in the gap ($\kappa_d$ = 1-100), $a$ is the contact radius and $\kappa_{eff}$ is the effective dielectric constant of the ferroelectric material. For planar geometry (i.e. $R >> a >> w$), $\kappa_{eff}$ is close to $\kappa_z$ for a ferroelectric material. For the spherical case, $\kappa_{eff}$ is close to $\sqrt{\kappa_x\kappa_z}$ , imposing an upper and lower limit on $\kappa_{eff}$. For a metallic tip, the gap effect is expected to be minimal; for a perfect conductor the lower limit on the effective tip-surface separation is set by the Thomas-Fermi length of the metal. For doped silicon tips $w$ will be comparable to the depletion width of the tip material. Even for thin dielectric layers (0.1 nm-1 nm) the effective surface potential can be attenuated by as much as a factor of 100 due to a large difference between dielectric constants of dielectric and ferroelectric. For imperfect contact, the magnitude of the piezoresponse in the strong indentation limit can become comparable to that of the electrostatic mechanism. The deviation of the tip shape from spherical (e.g. flattening due to wear, etc) reduces the electrostatic response due to a higher contact stiffness and increases electromechanical response. The resolution in the strong indentation limit is limited by indentation radius $a$.

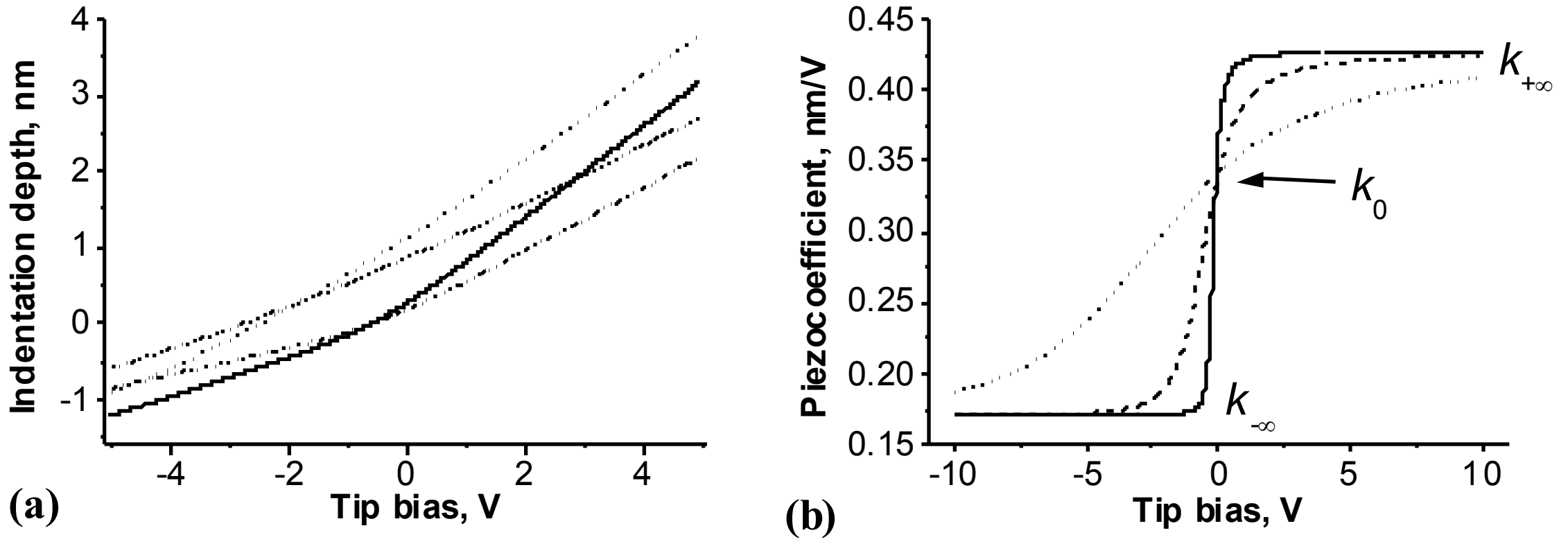

**Figure 6.8.** Indentation depth as a function of tip bias for different compositions and loads in the strong indentation limit (a) [PZT4 100 nN (———— ), PZT4 1000 nN (·············· ), BTC 10 nN (· — · · ) and BTC 100 nN ( · - - · - - ) and piezoresponse coefficient of $BaTiO_3$ as a function of tip bias for different loads (b) [10 nN (———— ), 100 nN (- - - ), 1000 nN (·········· )].

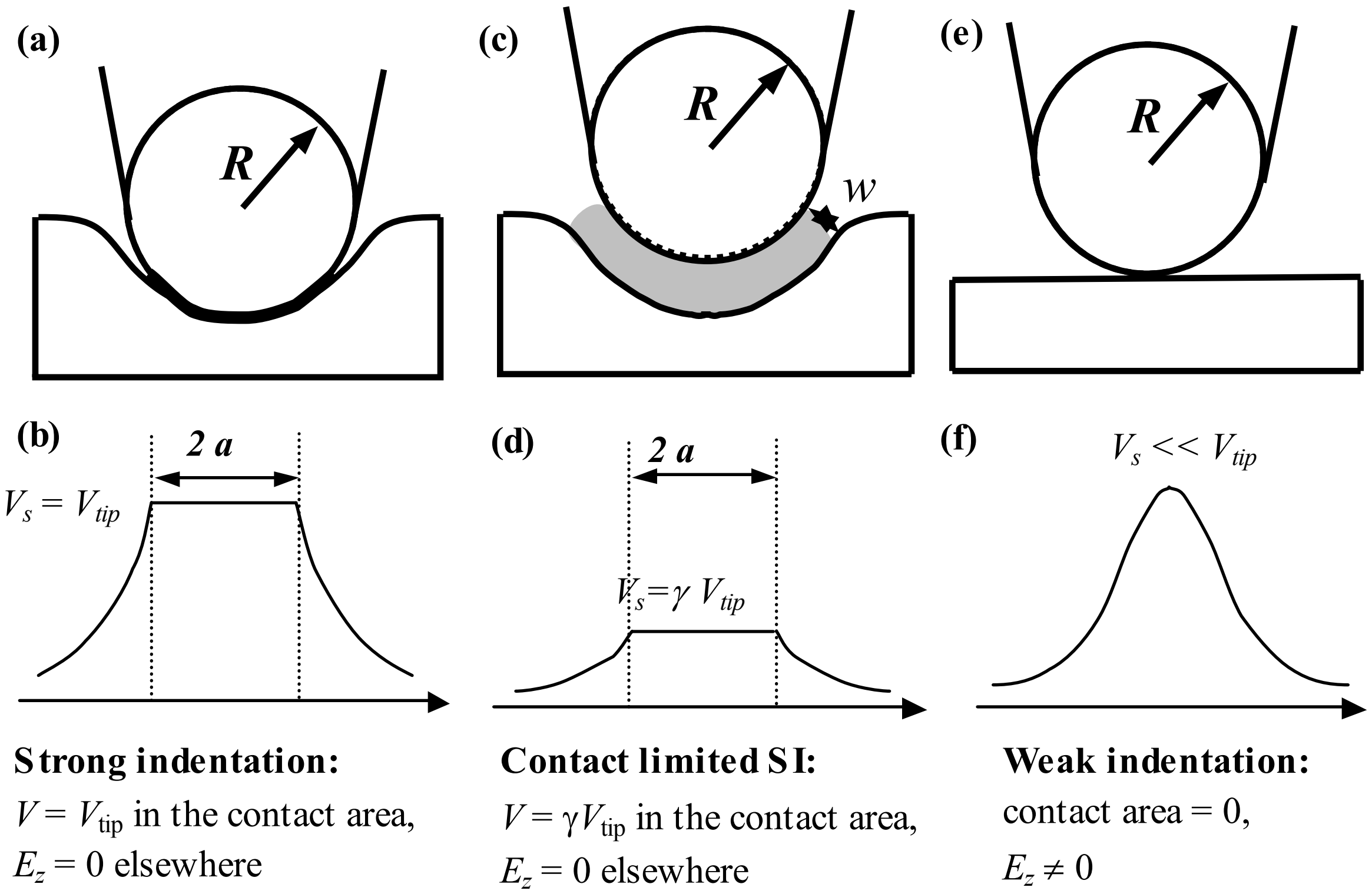

**Figure 6.9.** Limiting cases for the electromechanical interactions in the PFM. Tip-surface junction (a,c) and surface potential (b,d) in the strong indentation limit with and without the apparent gap effect and tip-surface junction (e) and surface potential (f) in the weak indentation limit.

### 6.4.2. Weak Indentation Limit

Weak indentation defines the other limiting case in the PFM experiment when the indentation load and contact area are small. In this limit, the contribution of the contact area to the total electromechanical response of the surface can be neglected (Figure 6.9e,f). The potential distribution in the tip-surface junction is calculated ignoring the electromechanical coupling (rigid electrostatic limit) as shown in Section 6.3 since the dielectric constant of material is sufficiently high and field penetration into ferroelectric is minimal. The electromechanical response of the surface is calculated using the Green's function for point force/charge obtained by Karapetian *et al.*[60]

$$h(r) = f\frac{A}{r} + q\frac{L(s_{ij}, e_{ij}, \varepsilon_{xx}, \varepsilon_{zz})}{r}, \tag{6.16}$$

where $h$ is vertical displacement, $r$ is the radial coordinate, $f$ is the point force, $q$ is the point charge, $A$ and $L$ are materials dependent constants and $r$ is the distance from the indentation point. For distributed charge, the surface deflection is:

$$h(\mathbf{r}) = L\int \frac{\sigma(\mathbf{r}_0)}{|\mathbf{r} - \mathbf{r}_0|}\mathrm{dS}, \quad \text{where} \quad \sigma(\mathbf{r}_0) = \varepsilon_0 E_z(\mathbf{r}_0) \tag{6.17}$$

The materials properties affect the PFM contrast through the coefficient $L$, while the geometric properties are described by the (materials independent) integral. This treatment implicitly assumes that the field penetration into the material is small.

For spherical tip geometry, the electromechanical surface response in the weak indentation limit can be evaluated using the image charge method developed in Section 6.3. The surface charge density induced by point charge $Q$ at distance $l$ from a conductive or high-$\kappa$ dielectric surface is $\sigma_0 = \frac{Q}{4\pi}\frac{2d}{\left(l^2 + r^2\right)^{3/2}}$. From Eq.(6.17), charge induced piezoelectric deformation of the surface is $h = QL/l$. Using the image charge series developed in Section 6.3, total tip-induced surface deformation is

$$h = L\sum_{i=0}^{\infty}\frac{Q_i}{R + d - r_i} = LG(R, d). \tag{6.18}$$

Note that this expression is remarkably similar to that of the tip-induced surface potential [Eq.(6.6)]. Thus, piezoresponse in the weak indentation limit can be related to tip-induced surface potential $V_s$ as $h = 2\pi\varepsilon_0 L(\kappa + 1)V_s$. Specifically, the surface deformation is linear in surface potential, $h = d_{eff}V_s$, where the effective piezoresponse constant, $d_{eff}$, in the weak indentation limit is $d_{eff} = 2\pi\varepsilon_0 L\left(\sqrt{\kappa_x\kappa_z} + 1\right)$ [Table 6.III].

Table 6.III.

*Piezoresponse coefficients for different materials.*

| Composition | $d_{33}$, m/V | SI, $k_0$, m/V | WI, $L$, $m^2$/C | WI, $d_{eff}$, m/V |
|---|---|---|---|---|
| $BaTiO_3$ | $1.91 \cdot 10^{-10}$ | $3.40 \cdot 10^{-10}$ | $1.54 \cdot 10^{-3}$ | $1.10 \cdot 10^{-10}$ |
| PZT4 | $2.91 \cdot 10^{-10}$ | $4.96 \cdot 10^{-10}$ | $2.41 \cdot 10^{-3}$ | $1.71 \cdot 10^{-10}$ |
| PZT5a | $3.73 \cdot 10^{-10}$ | $6.04 \cdot 10^{-10}$ | $2.66 \cdot 10^{-3}$ | $2.05 \cdot 10^{-10}$ |

For $R$ = 50 nm, $d$ = 0.1 nm and a typical value of $L \approx 2.5 \cdot 10^{-3}\ \mathrm{m}^2/\mathrm{C}$ the characteristic piezoresponse amplitude in the weak indentation limit is $h \approx 6.54 \cdot 10^{-12}$ m/V. The distance and tip radius dependence of the response is $h \sim (R/d)^{0.5}$, in agreement with a previously used point charge approximation.[63] The effective piezoelectric constant $d_{eff}$ for weak indentation limit is remarkably similar to $k_0$ for the strong indentation limit as shown in Table 6.III. The difference between the limits arises from the disparate ways the dielectric gap is taken into account (Figure 6.9). The weak indentation limit accounts for the effect of the gap directly in the functional form of coefficient $L$, which incorporates the dielectric properties of the surface. In the strong indentation limit, the effective dielectric gap must be introduced through the attenuation factor $\gamma$. The resolution in the weak indentation limit is determined by the tip radius of curvature and effective tip-surface separation and is proportional to $\sqrt{Rh}$ .

### 4.3. Further Development: Alternative Treatments of PFM

From the descriptions above, it is clear that strong and weak indentation regimes represent two limiting cases for the field penetration in the material, as governed by contact properties. For small electromechanical coupling, piezoresponse for arbitrary probe-surface geometry can be found using the Greens functions for the anisotropic piezoelectric medium if the solution for the purely electrostatic problem is known.

For the first time, such approach was developed by Ganpule *et. al.*,[5] who calculated piezoresponse in the non-uniform material as

$$PR = \int_0^{\infty} E_z(z) d_{33} \mathrm{d}z , \tag{6.19}$$

where $E_z(z)$ is the electric field exerted by the tip at depth $z$ within ferroelectric and $d_{33}$ is the piezoelectric constant. Here, the response is calculated using the field distribution in the material obtained from the solution of uncoupled electrostatic problem. It was shown that this model provides an adequate description of the PFM contrast in the PZT thin films with complex *a-c* domain structures. This approach can also be extended for analysis of contact effects as demonstrated by Kalinin and Bonnell[15] and for analysis of spatial resolution as shown Durkan *et al.*[21] The limitation of these treatments is that the solutions obtained pertain only to specific simplified cases.

Currently the approach using isotropic Green functions is being developed by G. Schneider and J. Munos-Saldana (University of Technology Hamburg).[64] For most piezoelectric materials, the electrostatic field distribution can be calculated ignoring the electromechanical coupling. In this case, coupled electromechanical problem for displacement and potential is reduced to much simpler 3D Laplace's equation for the potential in the anisotropic medium. Using calculated field distribution and Green functions for isotropic or anisotropic elastic medium, surface displacement as a function from the distance from tip apex can be calculated, providing both the quantitative estimate for the piezoresponse and the measure for the lateral resolution of the technique. In fact, calculation in Section 6.4 will illustrate that the response is only weakly dependent on the elastic properties of the material, thus justifying the use of isotropic Green's functions.

It is imperative to note that the response both in the strong and weak indentation cases and in the more general case is shown to be proportional to the surface potential directly below the tip with the proportionality coefficient comparable to the $d_{33}$ of the material. Response is only weakly dependent on the details of surface potential distribution and the model used. Therefore, the crucial step in the PFM modeling and interpretation is the analysis of this tip-surface potential transfer. The same is true in the experimental PFM imaging – unless good tip-surface contact is achieved, the information obtained from PFM image is highly unreliable. The contact quality and potential drop in the dielectric gap cannot be established from PFM experiment only and additional measurements are required.[65] Contact quality is particularly important for quantitative spectroscopic measurements. However, adequate description of spatial resolution and probed volume in PFM as well as quantitative interpretation of contrast behavior in the vicinity of domain walls and grain boundaries require advanced 3D models as presented in this Section.

### 6.4. Effect of Materials Properties on Piezoresponse

The solutions discussed in Section 6.3 and 6.4 present a mathematically rigorous description of PFM contrast mechanism. These solutions clearly illustrate that complete analysis of the electromechanical response of the surface in terms of materials properties is difficult. Even in the ideal case of known geometry, both strong and weak indentation limits lead to complex expressions that include 10 electroelastic constants for a transversally isotropic medium. For the systems with lower symmetry (e.g. ferroelectric grain with random orientation) the analytical treatment of the problem is even more complex. Therefore, the understanding of PFM contrast can be greatly facilitated if the simplified relationship between piezoresponse and material properties can be established. As illustrated in Table 6.III, in many cases the effective response calculated from rigorous models is comparable to the $d_{33}$ of the material. Given the difference between the geometries of the problems ($d_{33}$ defines the electromechanical response in $z$-direction to the uniform field applied in the $z$-direction, effective piezoresponse defines the electromechanical response in the $z$-direction to the highly non-uniform field below the tip), these results are quite surprising. In order to rationalize this observation, here we investigate the contribution of various electromechanical constants to piezoresponse in the strong and weak indentation limits.

Electroelastic properties of the solid can be described either in terms of elastic compliances $s_{ij}$ [m$^2$/N], piezoelectric constants $d_{ij}$ [C/N or m/V] and dielectric permittivities $\varepsilon_{ij}$ [F/m], or by elastic stiffness constant $c_{ij}$ [N/m$^2$], piezoelectric constants $e_{ij}$ [C/m$^2$ or Vm/N] and dielectric permittivities $\varepsilon_{ij}$ [F/m].[66] These sets of constants are related by tensorial relations $d_{nj} = e_{ni} s_{ij}$, $e_{nj} = d_{ni} c_{ij}$, $s_{ij} = c_{ij}^{-1}$ and $c_{ij} = s_{ij}^{-1}$. In order to clarify the relative contributions of different electroelastic constants to PFM, responses both in the strong and weak indentation limits are calculated for a variety of ferroelectric materials[66,67,68] A sensitivity function of piezoresponse is defined as the functional derivative of piezoresponse with respect to material parameter, $S(f_{ij}) = \delta PR(f_{ij})/\delta f_{ij}$. Here, the sensitivity is calculated numerically as

$$S(f_{ij}) = \frac{PR(f_{ij} = 1.1 f_{ij}^0) - PR(f_{ij} = 0.9 f_{ij}^0)}{0.2 PR(f_{ij} = f_{ij}^0)}, \tag{6.20}$$

where $f_{ij}$ is a selected electroelastic constant and $f_{ij}^0$ is a reference value for that constant. A positive value of $S(f_{ij})$ implies that a higher constant favors piezoresponse, while a negative value of $S(f_{ij})$ suggests that the response decreases with this constant. $S(f_{ij}) \approx 0$ indicates that the response is independent of that property.

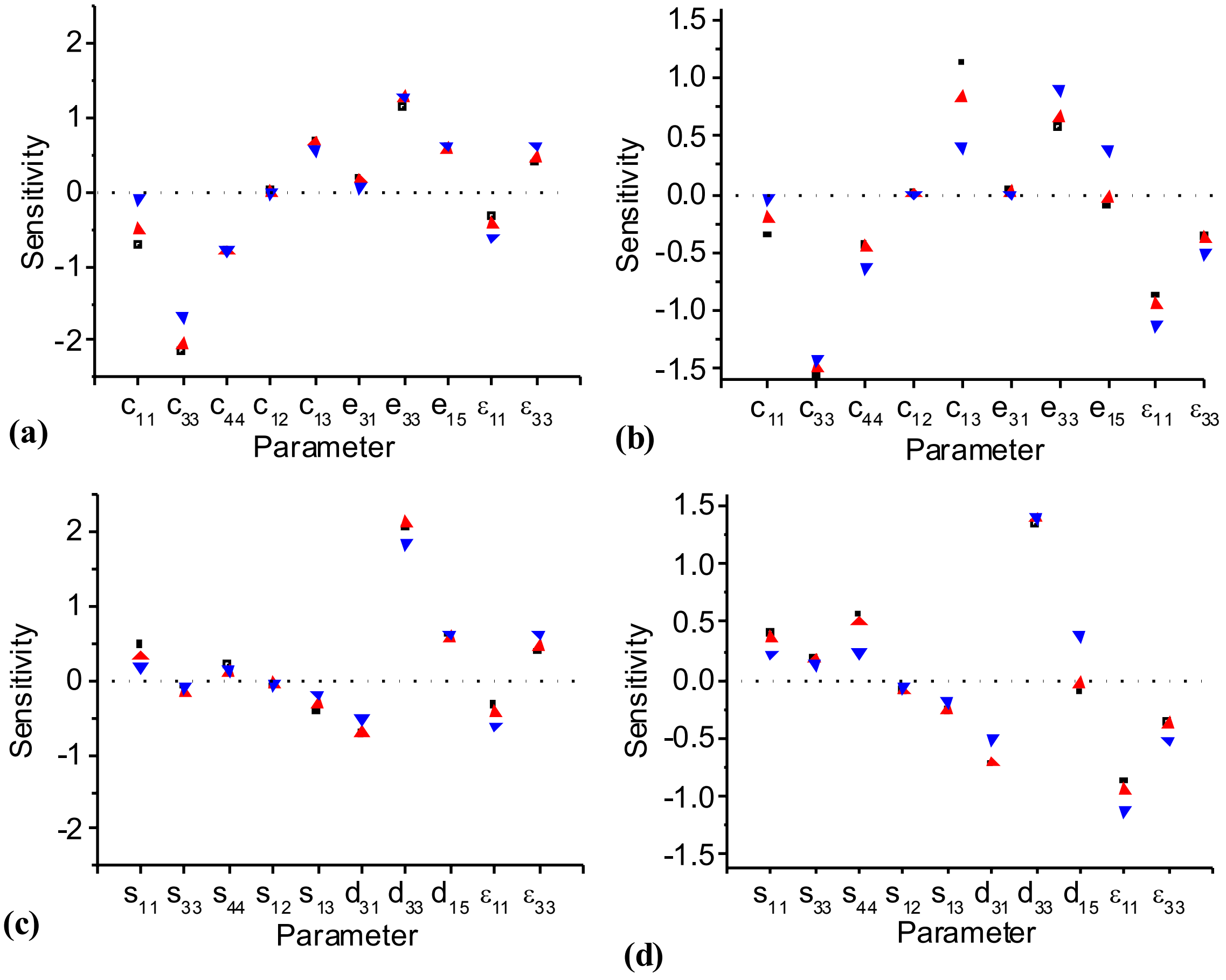

**Figure 6.10.** Sensitivity in the strong (a) and weak (b) indentation limits in the in the ($c_{ij}$, $e_{ij}$, $\varepsilon_{ij}$) representation. Shown are sensitivities for $BaTiO_3$ (■), PZT4 (▲) and PZT5a (▼). Sensitivity in the strong (c) and weak (d) indentation limits in the in the ($s_{ij}$, $d_{ij}$, $\varepsilon_{ij}$) representation. Shown are sensitivities for $BaTiO_3$ (■), PZT4 (▲) and PZT5a (▼).

Shown in Figure 6.10a,b is sensitivity of *PR* in strong indentation limit to electroelastic constants in the ($c_{ij}$, $e_{ij}$, $\varepsilon_{ij}$) representation. Note that response decreases for larger elastic stiffnesses, and the largest contribution originates from $c_{33}$. As expected, response increases for larger piezoelectric constants $e_{ij}$. The contributions of various electroelastic constants in this representation to piezoresponse are comparable. Sensitivity of piezoresponse for several ferroelectric materials in the ($s_{ij}$, $d_{ij}$, $\varepsilon_{ij}$) representation is shown in Figure 6.10c,d. Piezoresponse in the strong indentation limit is clearly dominated by the $d_{33}$ of the material, while other electroelastic constants provide minor contributions (Figure 6.10c). This observation implicitly justifies well-known assumption of measured piezoresponse being equal to $d_{33}$. In the weak indentation limit, both $d_{33}$ and $\varepsilon_{11}$ strongly influence the response, significant contributions being provided by $d_{31}$ and $\varepsilon_{33}$ as well (Figure 6.10d). The response increases with $d_{33}$ and decreases with $\varepsilon_{11}$ as expected. The response in both limits does not depend on elastic stiffness $c_{12}$ (Figure 6.10c,d).

The goal is to determine under what conditions a correlation exists between the measured piezoresponse and $d_{33}$ of the material. Earliest treatments of piezoresponse image contrast explicitly assumed that the response is proportional or equal to $d_{33}$. To test this assertion, the calculated piezoresponse coefficient is compared to piezoelectric constant for a series of ferroelectric materials. An almost linear correlation exists between the response in strong indentation limit and $d_{33}$, $PR \sim 1.5\ d_{33}$ (Figure 6.11a). In contrast, no such correlation is observed between $L$ and $d_{33}$ for the weak indentation limit (Figure 2.11b). The physical origin of this behavior is that $L$ defines the response of the surface to charge and therefore depends on ratios of the type $d_{ij}/\varepsilon_{ij}$. According to the Ginzburg-Devonshire theory these ratios are proportional to the corresponding second order electrostriction coefficients, $d_{ij}/\varepsilon_{ij} \sim Q_{ij}P$. Therefore, the effect of the electromechanical coupling coefficient and dielectric constants counteract each other. On the other hand, the effective piezoelectric constant in the weak indentation limit, $d_{eff}$, exhibits a good correlation with $d_{33}$, $d_{eff} \sim 0.5\ d_{33}$ (Figure 6.11c), since the dielectric constant effect is already accounted for. The effective piezoelectric response constants in the weak and strong indentation limits exhibit almost perfect linear dependence, $d_{eff} = 0.33\ k_0$ (Figure 6.11d). This similarity is due to the fact that in the first approximation the piezoresponse is defined by surface potential directly below the tip; the minor differences in the proportionality coefficient between piezoresponse and $d_{33}$ are due to the differences in surface potential profile as shown in Figure 6.9. The difference in the mechanisms is that in the strong indentation limit the *potential* in the contact area is assumed to be known and equal to the tip potential; the exact tip geometry is therefore not essential as long as contact is good. In the weak indentation limit the surface deflection is defined by tip-induced *charge distribution* on the surface, which strongly depends on tip geometry. If "true" PFM is the ability to quantify piezoelectric coefficient directly from the measurements, it can be achieved directly only in the strong indentation region. In the weak indentation regime, the electromechanical properties of the surface can be obtained indirectly provided that tip-surface geometry is known. However, experimental determination of relevant geometric parameters is usually very difficult; therefore, quantitative electromechanical information in the PFM can be obtained only in the strong indentation limit.

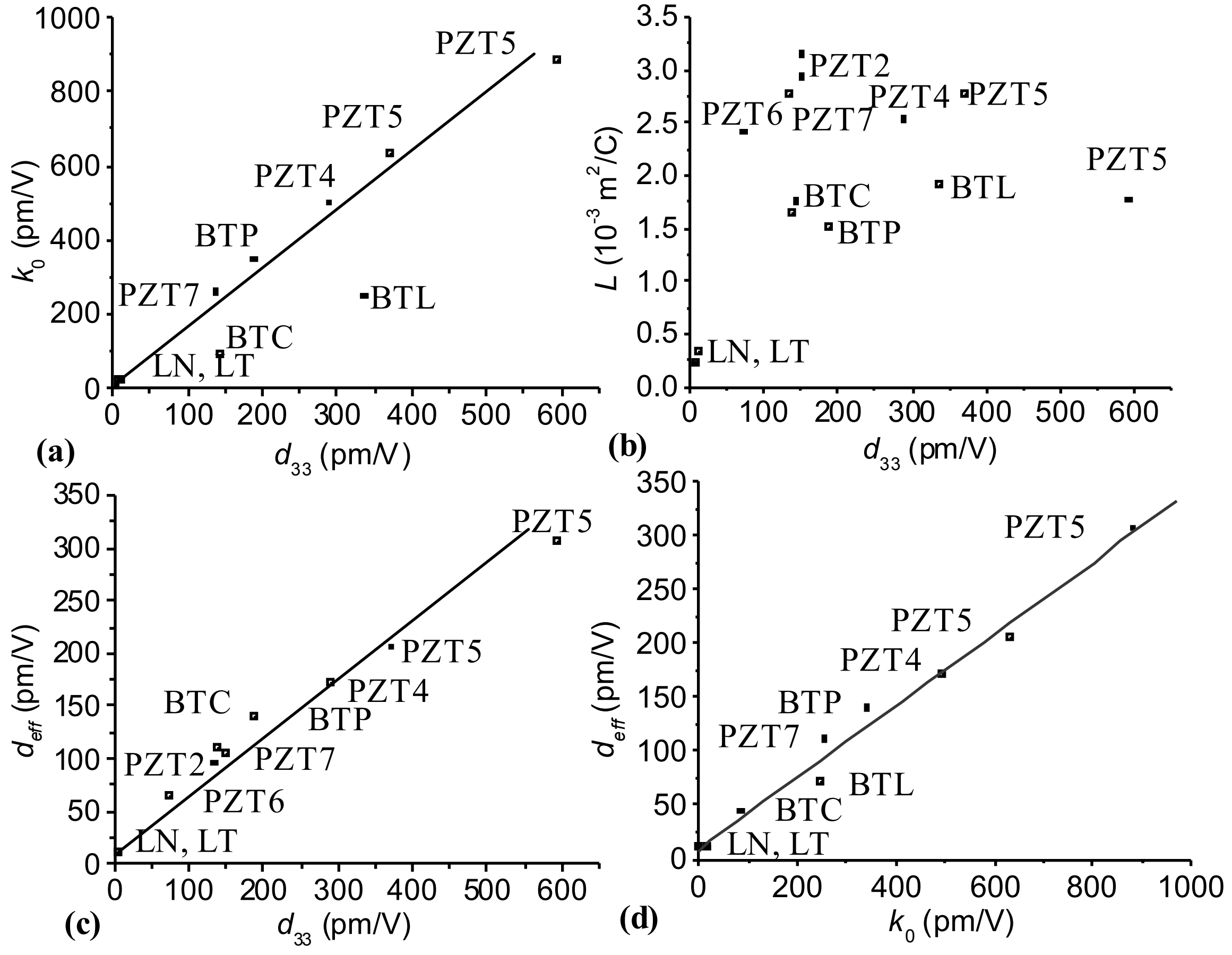

**Figure 6.11.** Correlation between piezoresponse and $d_{33}$ in the strong (a) and weak (b) indentation limits for some polycrystal and single crystal materials. Correlation between effective piezoelectric constant $d_{eff}$ and $d_{33}$ in the weak indentation limit (c) and correlation between $d_{eff}$ and piezoresponse in the strong indentation limit (d). PZT denotes different types of commercial lead zirconate-titanate ceramics, LN and LT are $LiNbO_3$ and $LiTiO_3$, BTC is 95%$BaTiO_3$/5%$CaTiO_3$ (ceramic B), BTP and BTL are $BaTiO_3$ polycrystals.

### 6.5. PFM Contrast Mechanism Maps

In the PFM measurement, the contrast mechanism will depend on details of the experimental conditions. Depending on tip radius and indentation force, both linear and non-linear electrostatic interactions and strong and weak indentation regimes can occur. In order to relate PFM imaging mechanisms to experimental conditions, Contrast Mechanism Maps were constructed as shown in Figure 6.12.[69,70] To delineate the regions with dominant interactions, surface deformation in the electrostatic case was estimated using the distance dependence of tip-surface capacitance as $F_{1\omega} = 2.7 \cdot 10^{-8} (R/50)(0.1/d)(V_{tip} - V_s)V_{ac}\ N$, where both $R$ and $d$ are in nanometers. The surface deformation, $h_{1\omega}^{el}$, was calculated from Eq.(6.14). The boundaries of the non-local regions are established by a comparison of tip apex-surface capacitance and cantilever-surface capacitance.[71] Surface deformation in the electromechanical regime was calculated including the "apparent dielectric gap" effect as $h_{1\omega}^{em} = d_{eff} / (1 + w\kappa_{eff} / a\kappa_d)$, where contact radius, $a$, is given by the Hertzian model and $\kappa_{eff}/\kappa_d = 30$. The boundary between strong and weak indentation regimes is given by attenuation factor of 0.3. It should be noted that the estimate of the attenuation factor is the major source of uncertainty in this treatment; the Green function based description as described in Section 6.4.3 is expected to overcome this difficulty. The boundary between the electromechanical and electrostatic regions is given by the condition $h_{1\omega}^{em} = h_{1\omega}^{el}$. For small indentation forces, a non-linear dynamic behavior of the cantilever is expected and the corresponding condition is $F_{el} = F_0$. For very large indentation forces, the load in the contact area can be sufficient to induce plastic deformation of the surface or the tip. The onset of this behavior is expected when $F_0 / \pi a^2 = E^*$. High pressures in the contact area can significantly affect the ferroelectric properties of material and induce local polarization switching, etc.[72,73,74] at a strain $P/d_{33} \sim 3 \cdot 10^9$ N/m$^2$ for a typical ferroelectric material. The effect of tip-surface potential difference and driving amplitude on imaging can be analyzed using formalism presented in Section 6.3.2 and 6.4.1.

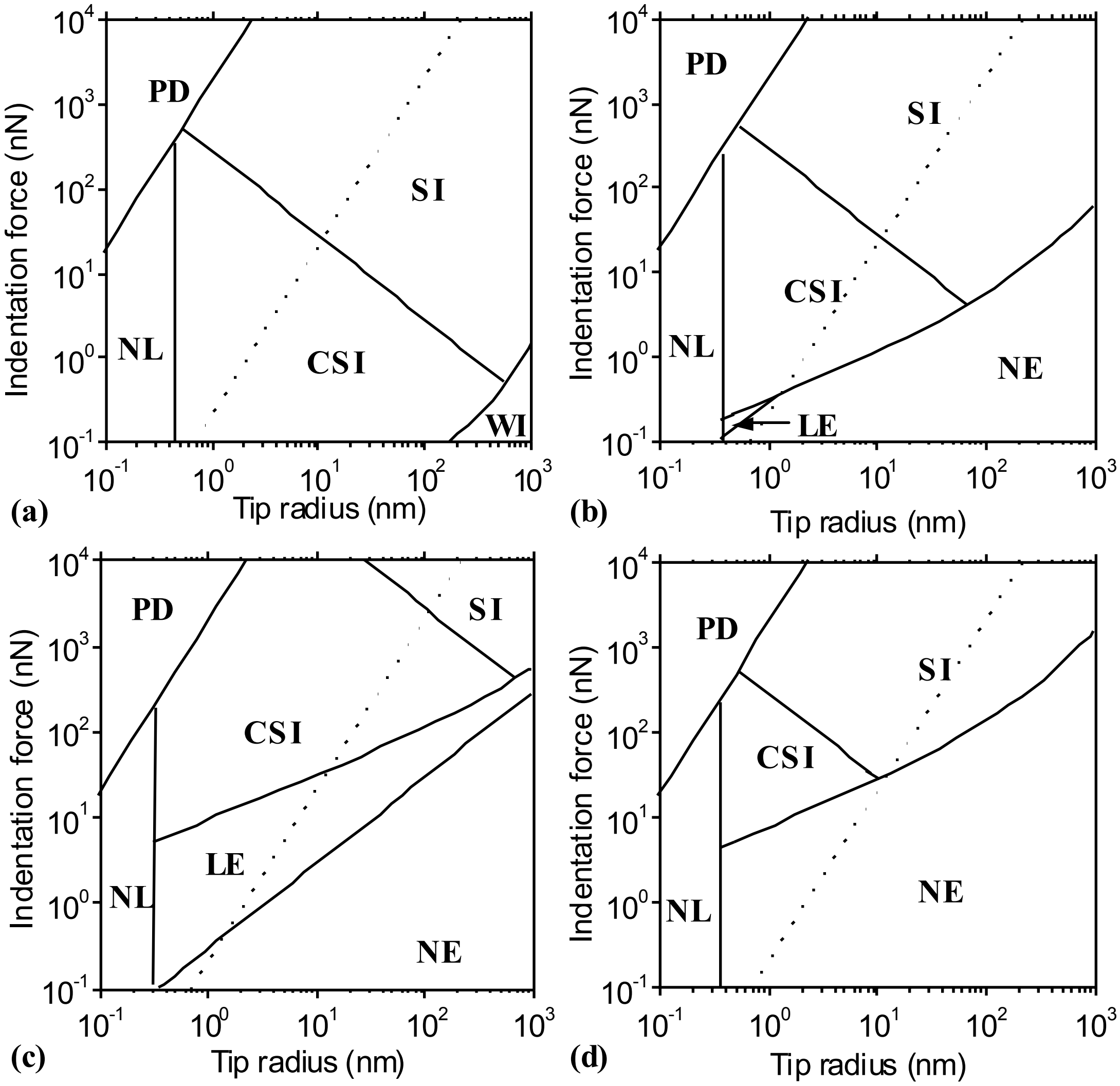

**Figure 6.12.** Contrast Mechanism Maps of piezoresponse force microscopy. SI is strong indentation regime, CSI - contact limited strong indentation, WI - weak indentation regime, LE - linear electrostatic regime, NE - nonlinear electrostatic regime, NL - non-local interactions, PD - plastic deformation. The dotted line delineates the region where stress-induced switching is possible. (a) $w$ = 0.1 nm, $\Delta V = V_{tip} - V_s$ = 0 V, (b) $w$ = 0.1 nm, ΔV = 1 V, (c) $w$ = 1 nm, ΔV = 1 V, (d) $w$ = 0.1 nm, ΔV = 5 V.

The Contrast Mechanism Map in Figure 6.12a corresponds to imaging under good tip-surface contact ($w$ = 0.1 nm) and zero tip-surface potential difference. The crossover from contact limited strong indentation to strong indentation limit depends on the choice of the attenuation factor. Pure weak indention behavior is observable only for large tip radii and small indentation forces. Typically, the ferroelectric domains are associated with surface potential variations and tip potential is not equal to surface potential. The Contrast Mechanism Map in Figure 6.12b corresponds to imaging under good tip-surface contact ($w$ = 0.1 nm) and moderate tip-surface potential difference ($V_{tip}$-$V_{loc}$ = 1 V). Less perfect contact that results from oxidized tips or poorly conductive coating, as well as the presence of contaminants will expand the weak indentation and linear electrostatic regions, primarily at the expense of the strong indentation region (comp. Figure 6.12b,c). Increasing the tip-surface potential difference increases the electrostatic contribution (Figure 6.12d). Consequently, the non-linear electrostatic region expands and can even eliminate the linear electrostatic region. However, above a certain tip-surface potential difference or driving voltage the linear approximation Eq.(6.14) is no longer valid and Eq.(6.13) must be used. The effect of high driving voltages and tip-surface potential difference is an increase of indentation force $F = F_0 + C'_{loc}\left(V_{tip} - V_{loc}\right)^2$, expanding the electromechanical region. Piezoelectric coefficient can be quantified directly from the measurements only in the strong indentation region. As shown in Figure 6.11, $k_0$ correlates linearly with $d_{33}$ in the strong indentation regime. In the weak indentation regime and contact limited strong indentation regime, the properties of the surface can still be obtained indirectly as discussed in Section 6.4.2. Finally, in the electrostatic regime, PFM image is dominated by long-range electrostatic interactions and piezoelectric properties of material are inaccessible. In certain cases surface charge distribution is directly correlated with ferroelectric domain structure; therefore, qualitative information on domain topology can still be obtained. These results allow multiple controversies in the interpretation of PFM contrast to be reconciled by elucidating experimental conditions under which electrostatic vs. electromechanical mechanisms dominate. Acquisition of quantitative information requires blunt tips and intermediate indentation forces to avoid pressure-induced polarization switching, i.e. operation regimes to the right of the dotted line in Figure 6.12. The use of top metallic electrode as proposed by Christman[75] or liquid electrode as proposed by Ganpule[35] is the limiting case of this consideration.

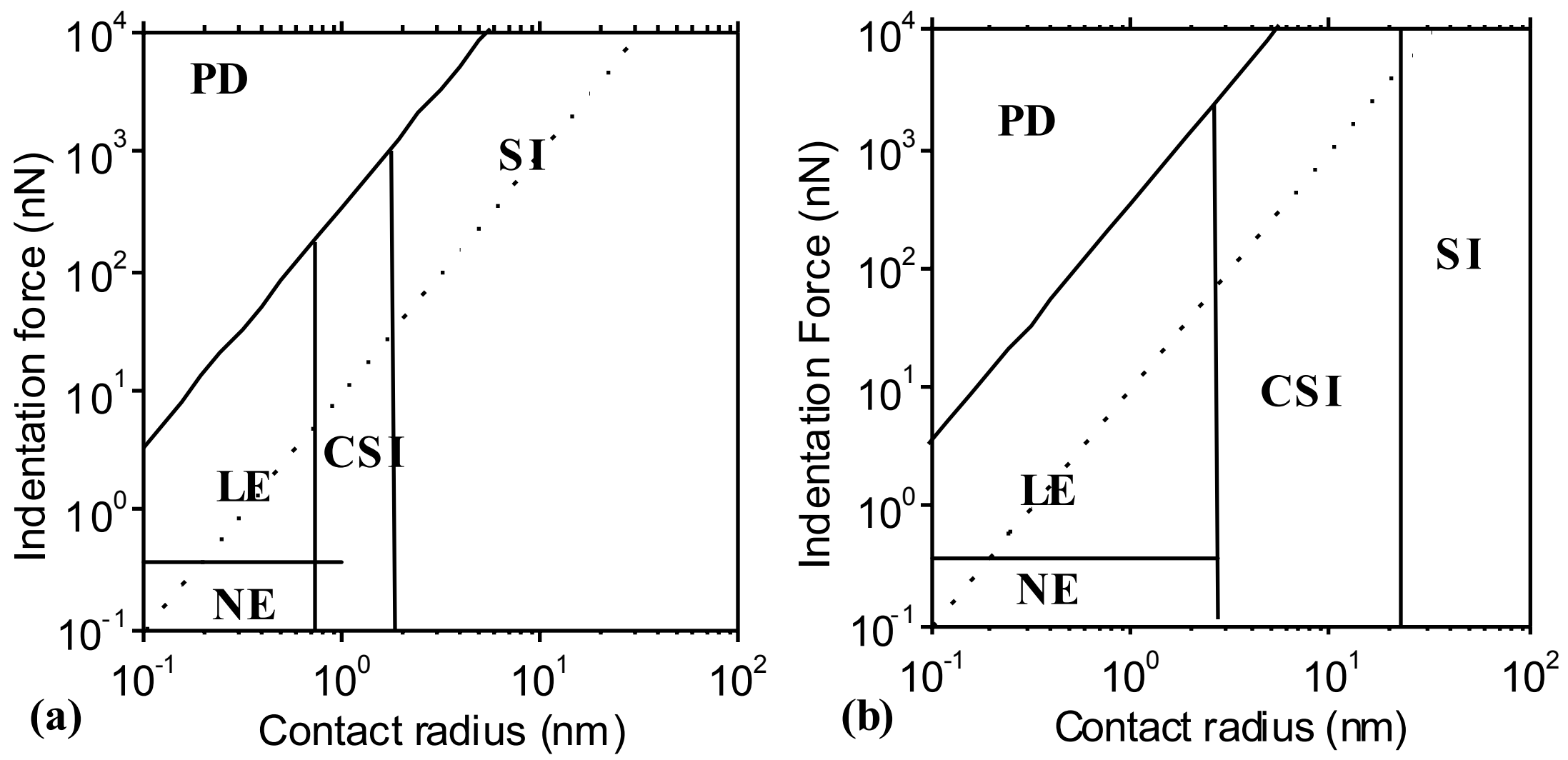

**Figure 6.13.** Contrast Mechanism Maps of piezoresponse force microscopy as a function of contact radius and indentation force. SI corresponds to strong indentation regime, CSI - contact limited strong indentation, WI - weak indentation regime, LE - linear electrostatic regime, NE - nonlinear electrostatic regime, PD - plastic deformation. Dotted line delineates the region where stress-induced switching is possible. The maps are constructed for good tip-surface contact ($w$ = 0.1 nm) and bad contact ($w$ = 1 nm).

The Contrast Mechanism Maps in Figure 6.12 are semiquantitative for a spherical tip; however, gradual tip wear during the imaging is inevitable and can be easily detected using appropriate calibration standards. The influence of tip flattening on PFM contrast mechanisms is shown in Figure 6.13a,b. The response was calculated as a function of contact radius for fixed electrostatic force corresponding to $R$ = 100 nm. In contrast to the spherical case, the contact stiffness for a flat indentor does not depend on the indentation force; hence, the crossover from the electrostatic to electromechanical regime occurs at some critical contact radius. Since the sphere/plane model is less accurate for this case, the degree of approximation associated with it results in the more qualitative nature of the contrast map. It should be noted, however, that electrostatic force can be measured directly[76] and used for the construction of the map for individual tip.

### 6.6. Non-local Effects: Cantilever Contribution to PFM

The non-local contribution to PFM, $A_{nl}$, arises due to the buckling oscillations of the cantilever[77] induced by capacitive cantilever-surface interactions as illustrated in Figure 6.14.[37] Typically, the cantilever length is significantly larger than the size of ferroelectric domains; therefore, the non-local interaction results in a constant background that does not preclude quantitative domain imaging but heavily contributes to local hysteresis measurements. In order to calculate the effective displacements due to the buckling oscillations, the simple harmonic-oscillator type models are no longer applicable; instead, realistic cantilever geometry must be taken into account.

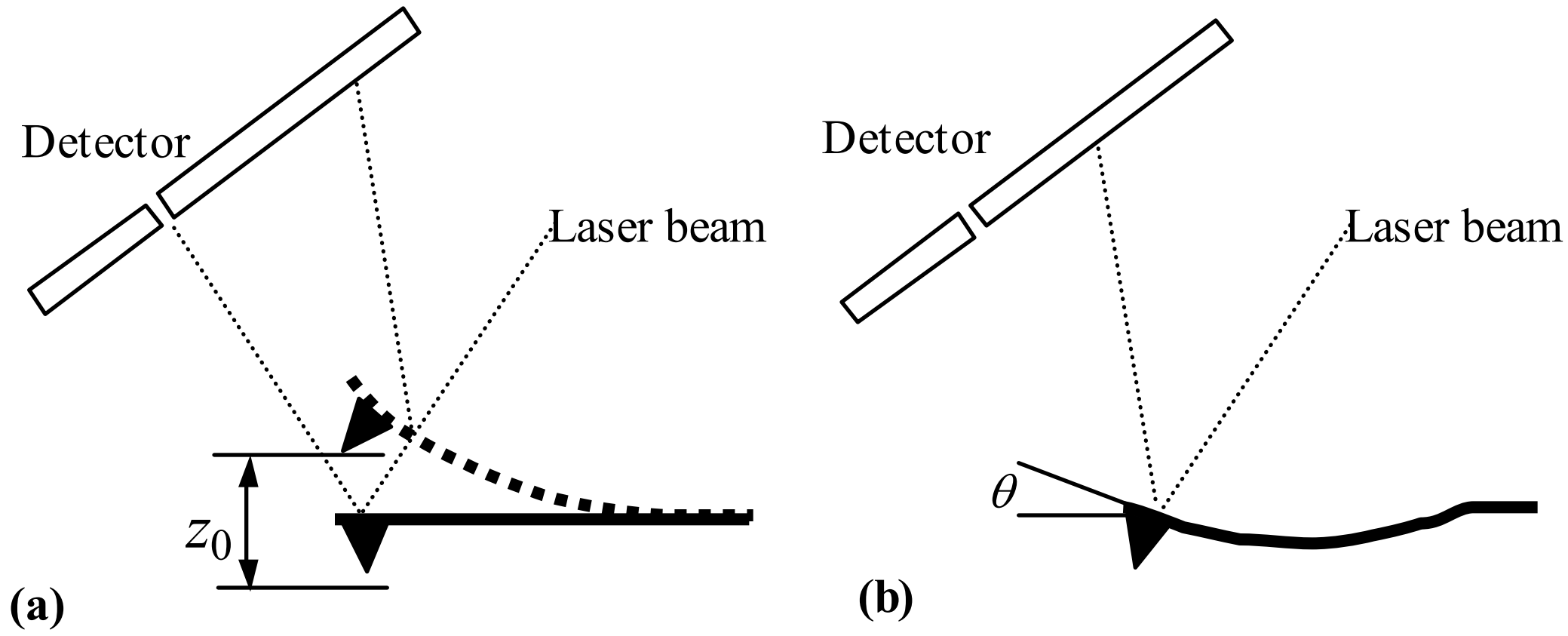

**Figure 6.14.** Schematic diagram of non-local cantilever effect in PFM. Displacement of laser beam induced by cantilever deflection (a) is equivalent to that due to cantilever buckling induced by uniformly distributed load (b).

Cantilever oscillations can be described by the beam equation[50]

$$\frac{d^4u}{dx^4} + \frac{\rho A}{EI}\frac{d^2u}{dt^2} = \frac{q(x,t)}{EI}, \tag{6.21}$$

where $E$ is the Young's modulus of cantilever material, $I$ is the moment of inertia of the crossection, $\rho$ is density, $A$ is cross-section area, and $q(x,t)$ is the distributed force acting on the cantilever. For a rectangular cantilever $A = wt$ and $I = wt^3/12$, where $w$ is cantilever width and $t$ is thickness. For a uniform periodic force Eq.(6.21) is solved by introducing $u(x,t) = u_0(x)e^{i\omega t}$, $q(x,t) = q_0 e^{i\omega t}$, where $u_0$ is displacement amplitude, $q_0$ is load per unit length, $t$ is time and $\omega$ is modulation frequency. After substitution Eq.(6.21) is:

$$\frac{d^4u_0}{dx^4} = k^4 u_0 + \tilde{q}, \tag{6.22}$$

where $k^4 = \omega^2 \rho A / EI$, $\tilde{q} = q_0 / EI$. The boundary conditions for Eq.(6.22) are $u_0(0) = 0$ and $u_0'(0) = 0$ on the clamped end and $u_0(L) = 0$, $u_0''(L) = 0$ on the supported end, where $L$ is cantilever length. In comparison, boundary conditions for a free oscillating cantilever are $u_0''(L) = 0$, $u_0'''(L) = 0$ on the free end. Eq.(6.22) is solved in the usual fashion. Of interest is the deflection angle $\theta$ at $x = L$, which is related to the local slope as $\theta = \operatorname{atan}(u_0'(L)) \approx u_0'(L)$. From Eq.(6.22) the frequency response of effective deflection for the buckling oscillations of supported cantilever is

$$\theta = \frac{\tilde{q}(\cos(kL) - \cosh(kL) + \sin(kL)\sinh(kL))}{k^3(\cosh(kL)\sin(kL) - \cos(kL)\sinh(kL))}. \tag{6.23}$$

Resonant frequencies for cantilever oscillations are found as a solution of

$$\cosh(kL)\sin(kL) = \cos(kL)\sinh(kL). \tag{6.24}$$

The lowest order solutions of Eq.(6.24) are $\beta_n = k\ L$ = 3.927, 7.067, 10.21. Corresponding eigenfrequencies are $\omega_n^2 = EI\beta_n^4 / \rho S L^4 = Et^2 \beta_n^4 / 12 \rho L^4$. In comparison, for the free oscillating cantilever the frequency response is given by

$$\theta = \frac{\tilde{q}(\sinh(kL) - \sin(kL))}{k^3(1 + \cosh(kL)\cos(kL))}, \tag{6.25}$$

and the resonance occurs for

$$\cos(kL)\cosh(kL) + 1 = 0, \tag{6.26}$$

Several lowest order solutions of Eq.(6.26) are $\alpha_n = k\ L$ = 1.875, 4.694, 7.855. Therefore, the first cantilever buckling resonance in contact mode occurs at ~ 4.4. times higher frequency than the resonance of the free cantilever.

In the low frequency limit Eq.(6.23) is simplified by $\theta = -L^3 q_0 / 48EI = -L^3 q_0 / 4wt^3 E$ and the effective oscillation amplitude detected by an optical detector is $2\theta L/3$. For the free oscillating cantilever in the low frequency limit, the response is larger by the factor of 8.

The capacitive cantilever-surface force is $F_{cap} = \varepsilon_0 S (V_{tip} - V_{surf})^2 / 2L^2$, where S is the cantilever area $S = Lw$. Therefore, the first harmonic of the load is $q_0 = \varepsilon_0 S V_{ac} \Delta V / 2LH^2$, where $H$ is tip height equal to cantilever-surface separation and $\Delta V$ = $V_{dc}$-$V_{surf}$. The non-local contribution to PFM signal is conveniently rewritten in terms of the spring constant of the free oscillating cantilever, $k_{eff} = Ewt^3 / 4L^3$ as $A_{nl} = -Lw\varepsilon_0 \mathrm{V_{ac}} \Delta V / 48 k_{eff} H^2$. Therefore, the piezoresponse signal in local hysteresis loop measurements comprising both electromechanical and non-local electrostatic parts is

$$PR = d_{eff} + \frac{Lw\varepsilon_0 \left(V_{dc} - V_{surf}\right)}{48 k_{eff} H^2}. \tag{6.27}$$

As discussed above, PFM imaging and quantitative piezoresponse spectroscopy requires electromechanical interaction to be much stronger than that of the non-local electrostatic interaction. From Eq.(6.27) the non-local contribution is inversely proportional to the cantilever spring constant, while the electromechanical contribution is spring constant independent. This condition can be written as $k_{eff} >> k^* = Lw\varepsilon_0 \Delta V / 48 d_{eff} H^2$, where $k^*$ is the critical cantilever spring constant corresponding to the equality of non-local cantilever-surface and electromechanical tip-surface interactions. Taking an estimate $d_{eff}$ = 50 pm/V, $\Delta V$ = 5 V, $L$ = 225 μm, $w$ = 30 μm, $H$ = 15 μm, the condition on the spring constant is $k_{eff}$ > 0.55 N/m. This condition can be easily modified for cantilevers with different geometric properties and can be rewritten as a condition for tip-surface potential difference. Note that while for $\Delta V$ = 0 non-local interactions are formally absent, this condition is hardly achieved experimentally unless a top-electrode set-up is used. Even though for the cantilever with high spring constants ($k_{eff}$ = 50 N/m) the electrostatic contribution is ~1% of electromechanical, it will hinder the determination of electrostriction coefficient from the saturated part of hysteresis loop as illustrated in Figure 6.4.

The frequency dependence of non-local contribution is given by $A_{eff} = A_{nl} f(\omega)$, where dimensionless $f(\omega)$ is

$$f(\omega) = \frac{48(\cos(x) - \cosh(x) + \sin(x)\sinh(x))}{x^3 (\cosh(x)\sin(x) - \cos(x)\sinh(x))}, \tag{6.28}$$

and $x^4 = \omega^2 L^4 A\rho / EI$ . Frequency dependence of oscillation amplitude and phase is shown in Figure 6.15a. Damping is expected to reduce the oscillation amplitude in the vicinity of the resonances. The presence of additional frequency-independent terms (e.g. due to electromechanical response) as well as additional components due to tip-surface interactions will severely modify this frequency dependence.

The ratio between the first resonance frequency of free and clamped cantilevers combined with the limited frequency range of current lock-ins (< 100 kHz) implies that this frequency dependence is not important if the spring constant of the cantilever is higher that ~30 kHz. However, experimental frequency dependence of PFM signal is extremely complex and a number of features corresponding to tip holder resonances, etc. are observed, complicating the interpretation of observed frequency spectra.

Nevertheless, Figure 6.15 provides a useful guideline to determine the conditions under which non-local electrostatic interactions become important. For cantilevers with spring constant $k_{eff}$ = 10 $k*$ response is purely electromechanical almost in the entire frequency spectrum, while for $k_{eff}$ = $k*$ local-and non-local responses are comparable. For $k_{eff}$ = 0.1 $k*$ the response is predominantly electrostatic. Figure 6.15 can also be interpreted in terms of tip-surface potential difference, since $k* \sim \Delta V$.

To provide a simple experimental criterion for the determination of non-local effects, we propose using average sensitivity function

$$\xi(V) = \int_{\omega_1}^{\omega 2} A(\omega, V) d\omega , \qquad (6.29)$$

where $\omega_1$, $\omega_2$ are the boundaries of the frequency window of interest and $A(\omega,V)$ is tip oscillation amplitude. Linear behavior of $\xi(V)$ implies electrostatic interaction, while constant or slowly varying $\xi$(V) is indicative of dominant electromechanical behavior.

To illustrate the cantilever effect on PFM measurements, we used cantilevers with different spring constants: Cantilever 1 (NSC-12 Pt type C, Micromasch, $\omega_r$ = 150 kHz, $k_{eff}$ = 4.5 N/m), Cantilever 2 (NSC-12 Pt type D, Micromasch, $\omega_r$ = 82.34 kHz, $k_{eff}$ = 0.35 N/m), Cantilever 3 (triangular, CSC21 type A, Micromasch, $\omega_r$ = 42.56 kHz, $k_{eff}$ = 0.12N/m), Cantilever 4 (DDESP, Digital instruments, $\omega_r$ = 363.5 kHz, $k_{eff}$ = 40 N/m). Frequency spectra were acquired from 10 to 100 kHz. It was found that while for stiff cantilevers (cantilever 1 and 4) the response amplitude is essentially frequency independent, soft cantilevers (cantilevers 2 and 3) show strong bias and frequency dependence of amplitude (Figure 6.15b,c). Note the difference in frequency dependence for cantilever 2 for $\Delta V = 0$ (no non-local effect) and $\Delta V = 1$ (non-local effect present). Average sensitivity function calculated from 10 to 100 kHz is shown in Figure 6.15d. Linear behavior for soft cantilevers and constant behavior for stiff cantilevers is clearly seen. The effect of cantilever stiffness on local hysteresis loops is illustrated in Figure 6.16. In comparison, cantilever parameters and corresponding critical stiffness $k^*$ for $\Delta V = 1$V are listed in Table 6.IV. For cantilevers 1 and 4, $k_{eff}$ is significantly higher than $k^*$. For cantilever 3, $k^*$ is larger than $k_{eff}$ and the crossover from local to non-local behavior is expected at $\Delta V$ as small as 200 mV. Finally, for cantilever 2 the two are comparable and non-local behavior is expected for $\Delta V$ > 2 V in a good agreement with experimental observations.

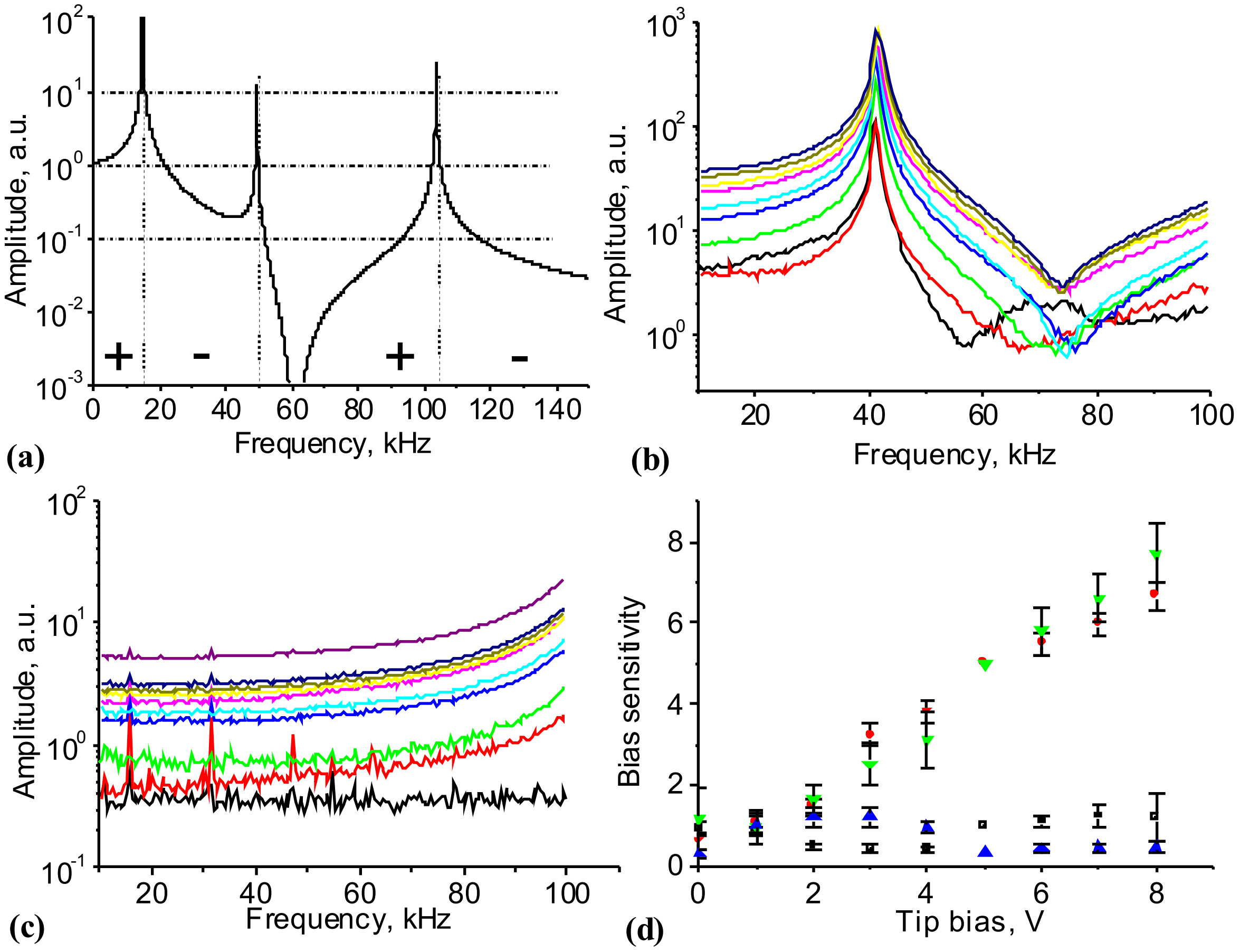

**Figure 6.15.** (a) Frequency dependence of amplitude for buckling cantilever oscillations. Vertical lines divide the regions with in-phase and out-of-phase oscillations. Experimental frequency spectra for cantilevers 2 (b) and 3 (c) for biases 1-10 V. Experimental sensitivity function for cantilevers 1 (■), 2 (●), 3 (▼), and 4 (▲).

Table 6.IV

*Cantilever parameters and critical stiffnesses*

| Cantilever | $L$, μm | $w$, μm | $H$, μm | $k^*$ for ΔV=1 V, N/m | $k_{eff}$, N/m |
|---|---|---|---|---|---|
| 1 | 130 | 35 | 15 | 0.075 | 4.5 |
| 2 | 300 | 35 | 15 | 0.17 | 0.35 |
| 3* | 410 | 80 | 15 | 0.53 | 0.12 |
| 4 | 125? | 30? | 15 | 0.061 | 40 |

*effective values

The non-local contribution to PFM is illustrated in Figure 6.16, which compares local hysteresis loops obtained using cantilevers with large ($k$ = 5 N/m) and small spring constants ($k$ = 0.1 N/m).

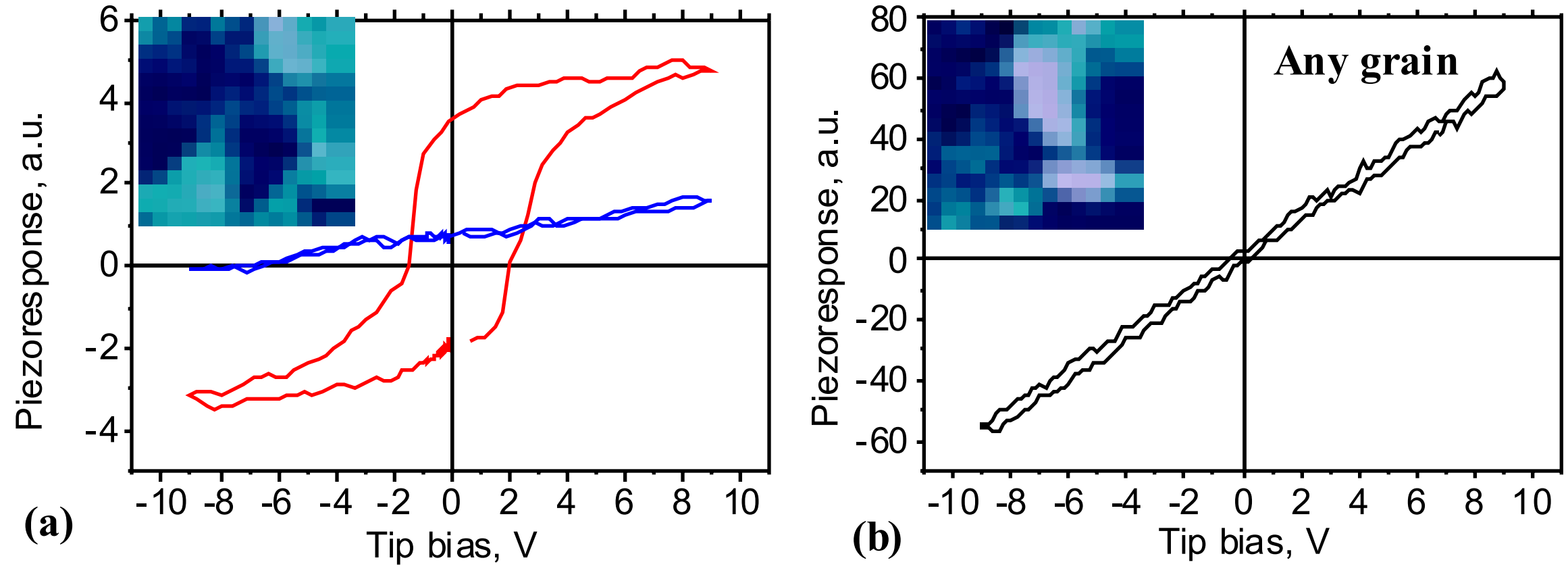

**Figure 6.16.** Piezoresponse hysteresis loops for stiff (a) and soft (b) cantilevers. Upper insets show 1 μm scans of the surface verifying that imaging is possible in both cases.

Both cantilevers allow successful PFM imaging since relative domain contrast in not influenced by the non-local contribution. However, only the stiff cantilever yields a well-defined local hysteresis loop. The soft cantilever exhibits a response linear in voltage due to the dominance of capacitive cantilever-surface force and cantilever buckling. Still, the contribution of electrostatic interactions is non-negligible for the first cantilever, as well, and can be detected on non-ferroelectric grains (Grain II). Note that the stiffness of the cantilever cannot be increased indefinitely: for a very stiff cantilever and a large indentation force materials properties (e.g. pressure induced polarization reversal or mechanical surface stability) limit the imaging as illustrated in the Figure 6.12. For small indentation force the spring constant of tip-surface junction [Eg.(6.14)] will become smaller than cantilever spring constant; hence tip deflection will be much smaller than the surface deflection.

### 6.7. Temperature Dependence of PFM Contrast

It was mentioned in the introduction that one of the origins of the existing ambiguity between electrostatic and electromechanical response mechanisms is the weak temperature dependence of experimentally measured piezoresponse. Here we apply the analytical solutions developed in Sections 6.3 and 6.4 to rationalize the temperature dependence of the piezoresponse of $BaTiO_3$. Experimentally measured temperature dependence of piezoresponse contrast is illustrated in Figure 6.17. This temperature dependence of PFM contrast is reminiscent of that of polarization and, indeed, capacitive interaction between the conductive tip and polarization charge has been used to describe the piezoresponse imaging mechanism. However, SSPM imaging suggests that polarization bound charge is completely screened on the surface as discussed in Chapter 5. Potential dynamics is extremely complex and exhibits relaxation behavior, which is not observed for the PFM signal. In order to explain the observed phenomena, we calculate the temperature dependence of PFM signal using the models developed in Section 6.3.

The temperature dependence of PFM contrast is calculated according to Karapetian *et. al.*[60] for weak indentation limit. The temperature dependence of the electroelastic constants for $BaTiO_3$ was calculated by Ginzburg-Devonshire theory[78,79] and the temperature dependence for $L(T)$ is compared to experimental measurements in Figure 6.18. In contrast to the strong indentation limit, no divergence occurs in the temperature dependence of the weak indentation and contact limited strong indentation limits, consistent with experimental behavior. The physical origin of this behavior is that not only the piezoelectric constant, but also the dielectric constant increases with temperature.

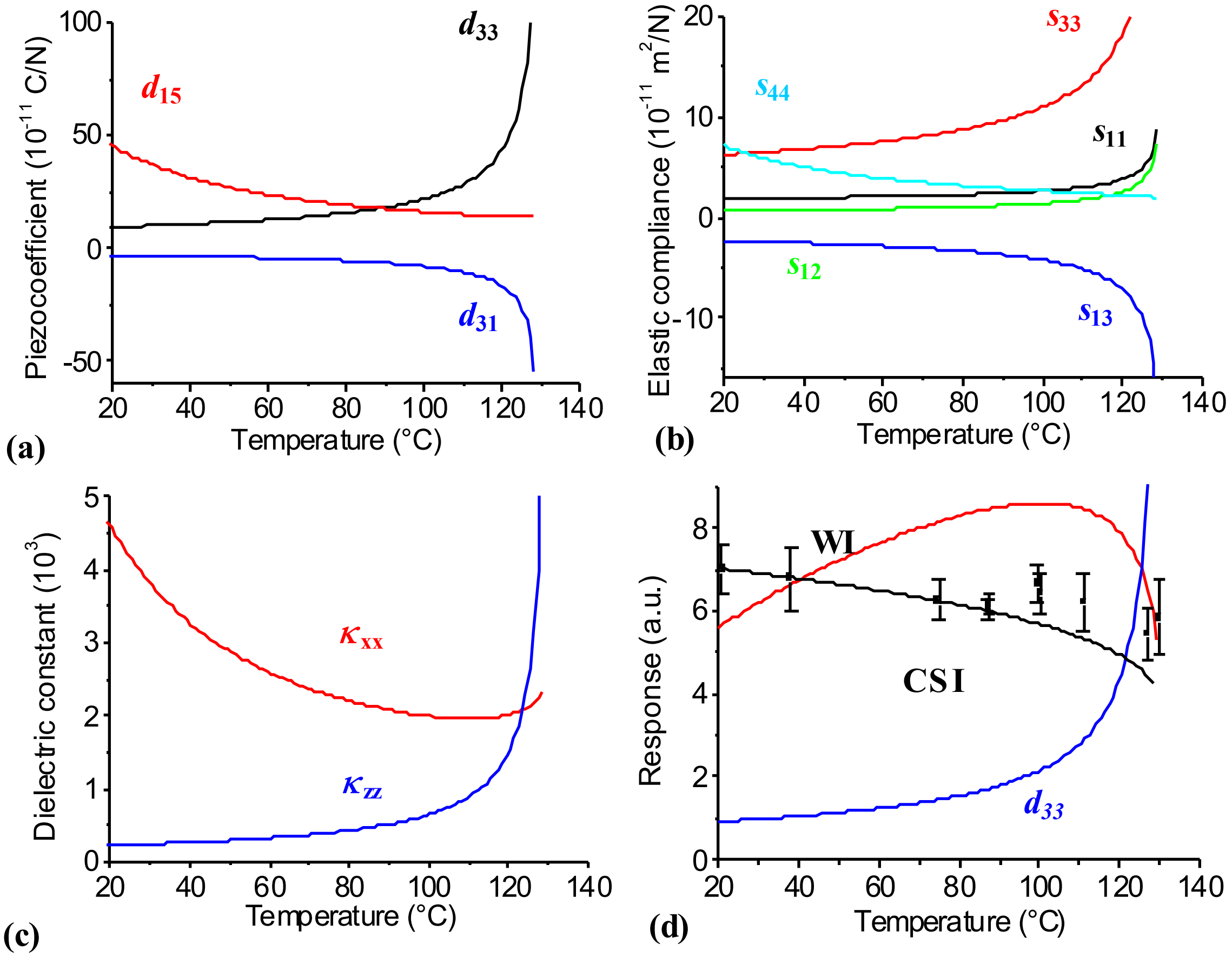

**Figure 6.17.** Temperature dependence of elastic constants (a), piezoelectric constants (b) and dielectric constants (c) for $BaTiO_3$ calculated from Ginzburg-Devonshire theory and temperature dependence of piezoresponse coefficient in the WI and CSI limits (d).

The simplified model for this behavior can be developed assuming that $PR \sim V_{eff} d_{33}$, where $V_{eff}$ is the potential on the ferroelectric surface. Using the attenuation factor developed in Section 6.3 for a spherical geometry for the tip, dielectric layer and the surface, the approximate relationship between $V_{tip}$ and $V_{eff}$ is

$$V_{eff} = \left(1 + \frac{\varepsilon_f}{\varepsilon_d} \frac{d}{a}\right)^{-1} V_{tip} = (1+\alpha)^{-1} V_{tip}, \tag{6.30}$$

where $\varepsilon_f$ is the dielectric constant of ferroelectric, $\varepsilon_d$ is the dielectric constant of surface layer, $d$ is the thickness of surface layer and $a$ is the effective tip radius. The dielectric constant increases with temperature and diverges at $T_c$; hence, at higher temperatures the potential drop through the dielectric layer increases. Thus, the effective potential on the dielectric-ferroelectric boundary decreases. At room temperature where $\varepsilon_{33} \approx 1000$, $\varepsilon_d \approx 100$, $d \approx 1$-10 nm and $a \approx 10$-100 nm, the attenuation factor, $\alpha$, is of order of unity and $V_{eff} \approx V_{tip}$. At higher temperatures and high $\varepsilon_{33}$ increases and Eq.(6.30) is reduced to:

$$V_{eff} \approx \frac{\varepsilon_d}{\varepsilon_{33}} \frac{a}{d} V_{tip} . \tag{6.31}$$

It should be noted that while Eq.(6.30) is correct only for the selected geometry, the reciprocal relationship between the effective potential and the dielectric constant of the media in Eq.(6.31) is universal for any tip-surface geometry (Figure 6.6). To estimate the temperature dependence of the piezoresponse both $d_{33}$ and $\varepsilon_{33}$ were calculated within the framework of the Ginzburg-Devonshire[80] theory, in which case the temperature dependence of spontaneous polarization, $P$, and susceptibility, $\chi_{33}$, is given by

$$P^2 = \left( -\alpha_{11} + \sqrt{\alpha_{11}^2 - 3\alpha_1 \alpha_{111}} \right) / 3\alpha_{111} , \tag{6.32}$$

$$\chi_{33}^{-1} = \left( 2\alpha_1 + 12\alpha_{11} P^2 + 30\alpha_{111} P^4 \right) \varepsilon_0 , \tag{6.33}$$

where the numerical values of coefficients $\alpha_1$, $\alpha_{11}$ and $\alpha_{111}$ are listed in Table 6.V and $\varepsilon_0 = 8.314 \cdot 10^{-12}$ F/m is the dielectric constant of vacuum. The piezoelectric constant, $d_{33}$, is calculated as

$$d_{33} = 2\varepsilon_0 \varepsilon_{33} Q_{11} P , \tag{6.34}$$

where the electrostrictive coefficient $Q_{11} = 1.11 \cdot 10^{-1}$ m$^4$/C$^2$. From Eqs.(6.31) and (6.34) the temperature dependence of piezoresponse can be calculated as

$$R \sim 2\varepsilon_0 \varepsilon_d Q_{11} P \frac{a}{d} V_{tip} \sim P , \tag{6.35}$$

i.e. the temperature dependence of piezoresponse is that of spontaneous polarization. Noteworthy the predicted temperature dependence of piezoresponse using simplified model [Eq.(6.35)] and rigorous calculation in the WI limit results in very similar temperature dependences. This is due to the fact that in both cases the response is determined by (almost) temperature independent ratios of the type $d_{ij}/\varepsilon_{ij} \sim PQ_{ij}$, rather than strongly temperature dependent piezoelectric coefficients. However, in the model [Eq.(6.35)] only $d_{33}/\varepsilon_{33}$ ratio is considered, while in the WI description all relevant parameters are incorporated into the model. In the WI, however, the physical origins of this weak temperature dependence are less obvious.

Table 6.V.

*Temperature dependence of free energy expansion coefficients*[81]

| Coefficient | Formula | Dimensionality |
|---|---|---|
| $\alpha_1$ | $(T-\theta)/2\varepsilon_0 C_{cw}$ | m/F |
| $\alpha_{11}$ | $[0.448(T-T_c)-21.5]\cdot 10^7$ | V m$^5$/C$^3$ |
| $\alpha_{111}$ | $[27.7-0.534(T-T_c)]\cdot 10^8$ | V m$^9$/C$^5$ |
| $C_{cw}$ | $1.7\cdot 10^5$ | |
| $\theta$ | 393 | K |
| $T_c$ | 403 | K |

Thus, the temperature dependence of experimental PFM contrast suggests that under the experimental conditions ($F_0 \approx$ 200 nN, nominal radius $R \approx$ 30 nm, tip is not blunted) the imaging mechanism of PFM is governed by the dielectric gap effect. The major contribution to piezoresponse is an electromechanical response of the surface to the tip bias, however, the properties of tip-surface contact change with temperature. From Figure 6.16 the width of the "apparent gap" in these measurements can be estimated as > 1 nm. This conclusion is verified by small experimental piezoresponse coefficients (~ 4 pm/V) [26, 34,82,83] as compared to the calculated value for $BaTiO_3$ (~ 50 - 100 pm/V).

## 8. Imaging Artifacts in PFM

One of the inherent problems of any SPM technique is the inevitable cross talk between the property and topographic images. Therefore, the analysis of any SPM imaging technique would be incomplete without the brief description of the measurement artifacts. In this section, we briefly analyze topographic artifacts related to the morphological inhomogeneity of the surface and certain phenomena related to the frequency dependence of PFM contrast.

### 8.1. Topographic Artifacts

In the electrostatic regime, topographic features on the surface result in the cross talk between topographic and PFM images through the variations in local tip-surface capacitance. In non-contact techniques such as EFM and open-loop SSPM the topographic artifacts can be pinpointed by contrast behavior under reversed tip bias as discussed in Chapters 3 and 5. However, in contact regime, variations of contact area and tip surface capacitance related to the topographical features can alter the response amplitude on the very small scale (1-10 nm). This effect is expected to be small because topographic slopes are minimal at this length scale. Larger topographic features are not expected to influence PFM contrast. The possible exception can be PFM imaging of very steep surface features (e.g. micropatterned surfaces) when tip can contact the surface by lateral surface rather than tip apex as illustrated in Figure 6.18.

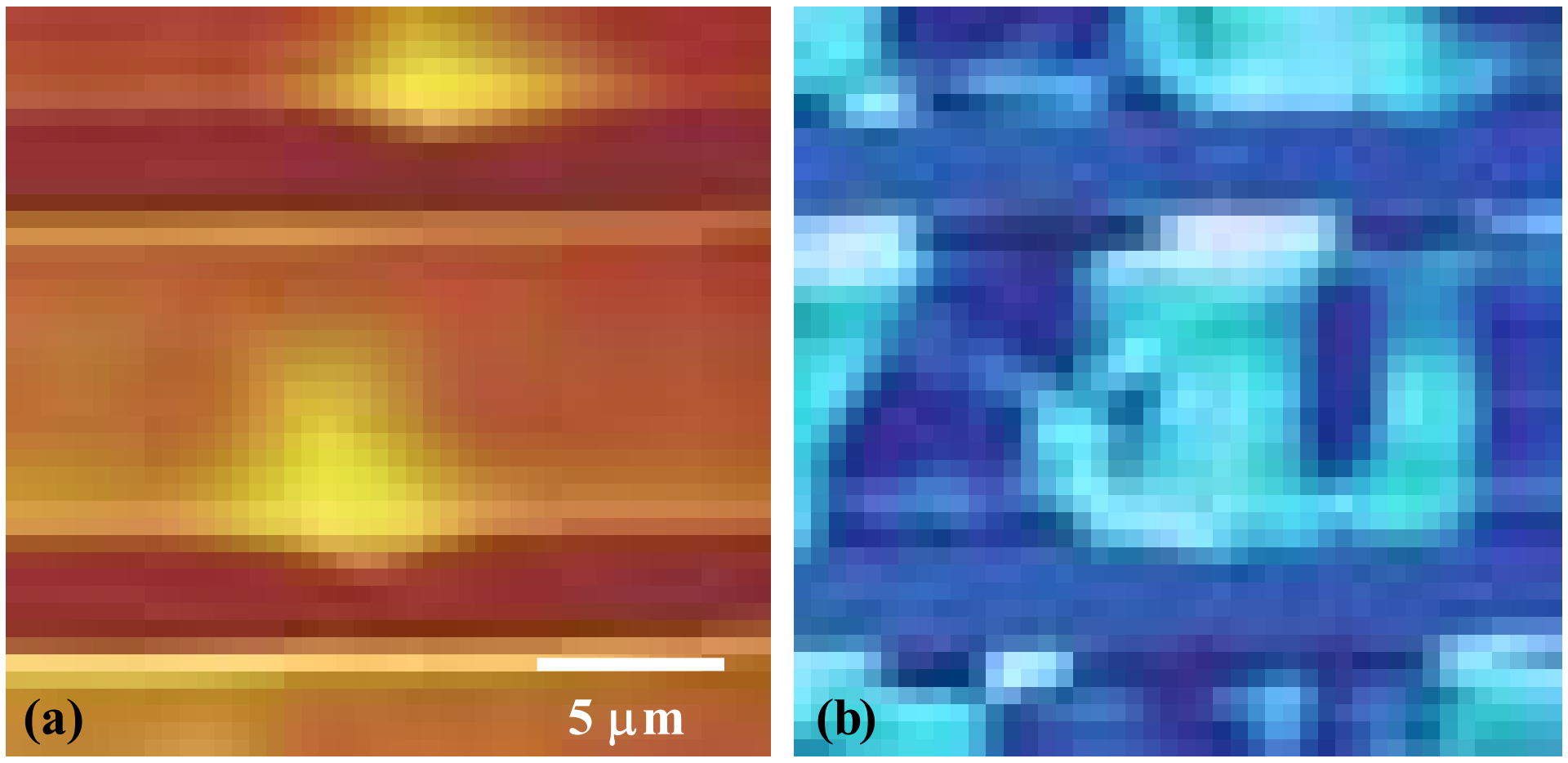

**Figure 6.18.** Surface topography (a) and piezoresponse images (b) of micropatterned PZT lines (samples are courtesy of I. Aksay, M. Ozenbas, Princeton University). Piezoresponse is higher in the regions with high slopes, where tip touches the surface by the side rather than apex. Vertical scale is 1 μm (a).

Similar effect is observed in conductive AFM imaging.[84] At the same time, it can be expected that presence of local inhomogeneities such as grain boundaries, crystallographic facets, etc. will also affect the local ferroelectric properties of material either through elastic and clamping effects or due to the modification of local domain structure. Therefore, topographic artifacts cannot be reliably distinguished from materials behavior in the general case. Nevertheless, experimental observations suggest that the topographic influence on PFM image is extremely small and even pronounced topographic features such as pores, etc do not contribute to PFM contrast.

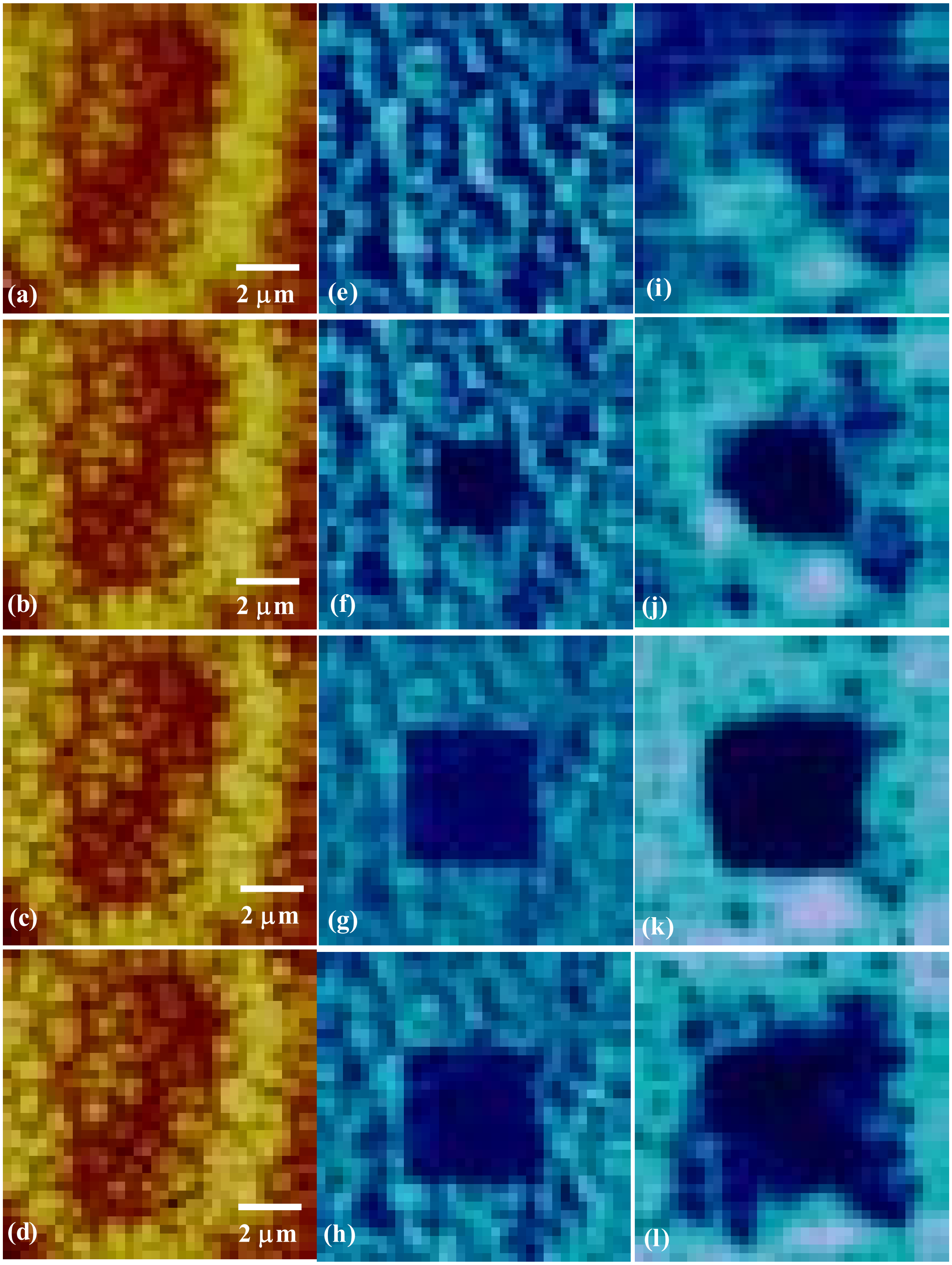

**Figure 6.19.** Surface topography (a-d), piezoresponse images (e-h) and surface potential (i-l) images of PZT surface in the initial state (a,e,i) and after sequential switching by negatively biased tip (-10V) in 2 μm (b,f,j), 4 μm(c,j,k) and 7 μm squares (d,h,l).

Another important effect in the PFM and especially in the hysteresis measurements and polarization switching is the possibility of surface charging by the tip. This effect was extensively studied by non-contact SPMs and several experimental studies on the magnitude of transferred charge and charge relaxation time have been reported.[40,42] Charge effects in the PFM are less studied. One of the experimental limitations in this case is that for the large tip-surface potential differences, i.e. under typical polarization switching conditions, the capacitive tip-surface and cantilever-surface forces can dominate over the Coulombic forces as discussed in Section 6.3. However, the presence of this effect can be established by sequential acquisition of SPM images in the non-contact and contact modes as illustrated on Figure 6.19. Note that initial piezoresponse and potential images exhibit distinctly different structure thus providing complimentary information on surface properties. Polarization switching on the central area 2x2 μm by the negatively biased tip results both in polarization switching and deposition of negative charge on the surface. The size of the charged region is larger (~300nm) on potential image due to the lateral spreading of the charge. Potential and piezoresponse features far from the switched region are unaltered. This effect can be reproduced for a larger scanned region of 4x4 μm. Finally, when the switching was attempted on a large area of 7x7 μm, polarization remained unswitched, but the deposition of negative charge is still observed. This effect can be attributed to both the faster relative tip velocity and deterioration of the tip-surface contact. Similar results were reported by Chen et al.[43,85] Relaxation of surface charges as opposed to remanence of switched polarization was presented by Ahn.[33] Correlation between potential and piezoresponse images of multiphase ceramics was studied by Borisevich et al.[32] It has been shown that the information provided by PFM and SSPM is in general case complementary and reflects different properties of the surface as discussed in Chapter 2 and Chapter 5.

### 6.8.2. Dynamic Tip-surface Contrast Transfer

The effect of tip-surface contrast transfer can be estimated from the frequency dependence of the average response amplitude, domain contrast and cantilever oscillation phase. The average amplitude is defined as $2PR_{av} = PR\left(c^{+}\right) + PR\left(c^{-}\right)$, while relative contrast of the domains is $2PR_{rel} = PR\left(c^{+}\right) - PR\left(c^{-}\right)$, where $c^{+}$ and $c^{-}$ refer to the domains with opposite polarization orientation. As shown by Allegrini *et. al.*,[38] this frequency behavior can be rather complex and a number of well-formed resonances in the frequency dependence of cantilever response can be observed. Imaging in the vicinity of these resonances allows higher signal to noise ratio. However, the experimental approach employed by the authors suggests the possibility of intermittent tip-surface contact in these measurements. Such cantilever resonances were reported to contribute to the observed dynamic behavior of cantilever in the contact force-modulation mode and are being extensively studied.[86,87,88,89]

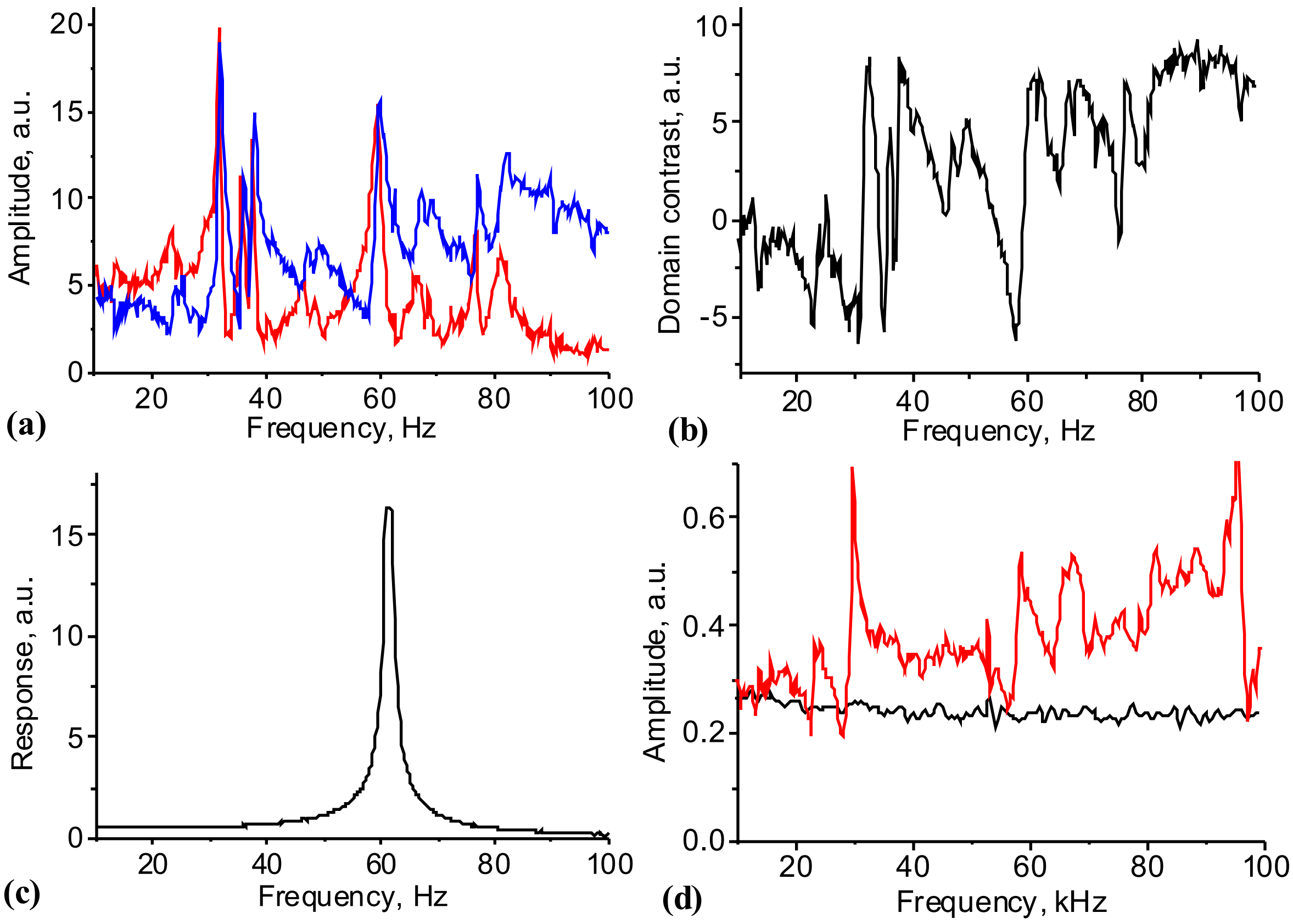

**Figure 6.20.** Frequency dependence of PFM amplitude on PZT for two adjacent PZT grains (a) and frequency dependence of piezoresponse contrast (b). Amplitude frequency dependence of a free cantilever oscillations (c). Frequency dependence of PFM signal on $BiFeO_3$ surface from tip modulation (red) and sample modulation (black).

In the contact regime, shown on Figure 6.20 is frequency dependence of response amplitude on two grains in polycrystalline PZT sample for tip modulation ($k$ = 1 N/m) using "bad" tip. Note that frequency dependence of response amplitude shows numerous resonances and relative response of the grains inverts as a function of frequency. This complex behavior can be attributed to scanner resonances similarly to intermittent contact mode imaging in fluids[90] or to intrinsic materials resonance. From the general considerations, resonant frequency of excited region below the tip can be estimated as $f_r \propto \sqrt{E^* / \rho l^2}$ , where $f_r$ is the resonant frequency, $\rho$ is material density and $l$ is the characteristic size of excited region.[91] For $l \sim 100$ nm resonant frequency is estimated as $f_r \sim 60$ GHz. Therefore, the non-linear dynamic behavior of PFM should be attributed solely to the probe dynamics. Interestingly, this effect can be significantly reduced by using sample modulation (i.e. bias is applied to the tip). Shown in Figure 6.20 is frequency dependence of PFM contrast on semiconductive (600 kOhm·m) $BiFeO_3$ surface for tip modulation and modulation applied to the sample. Note that in the latter case the frequency dependence of response amplitude is essentially constant. Similar frequency behavior is observed for the lateral PFM signal. These observations suggest that in some cases the dynamic properties of the cantilever significantly affect PFM signal. In such cases, quantitative measurements of piezoelectric properties are clearly impossible and weak frequency dependence of PFM amplitude with few or no resonances is a requirement quantitative and even qualitative experiment.

### 6.9. Simultaneous Acquisition of PFM and Potential Images

Electrostatic tip-surface interactions can be significantly affected by local surface charging. [33,40,42,43] Clearly, elucidating the charge effects in the PFM requires a reliable way to probe local piezoresponse and long-range electrostatic forces simultaneously. This is especially important for investigations of dynamic phenomena in which large time intervals between sequential PFM/SSPM images are unacceptable. Under equilibrium conditions, simultaneous acquisition of piezoresponse and potential images can facilitate the correlation between topographic, potential and piezoresponse features and analysis of surface properties. We have shown that simultaneous PFM and SSPM imaging can be implemented using the usual lift mode so that the topography and piezoresponse are acquired in contact and potential is collected on the interleave line (Appendix A).[92] Figure 6.21 illustrates several examples of simultaneous piezoresponse and potential imaging on $BaTiO_3$ and PZT. An open loop version of SSPM is used. For $BaTiO_3$ both SSPM and PFM features are related to the surface domain structure and therefore are closely correlated. For PZT the information provided by the two is complementary.

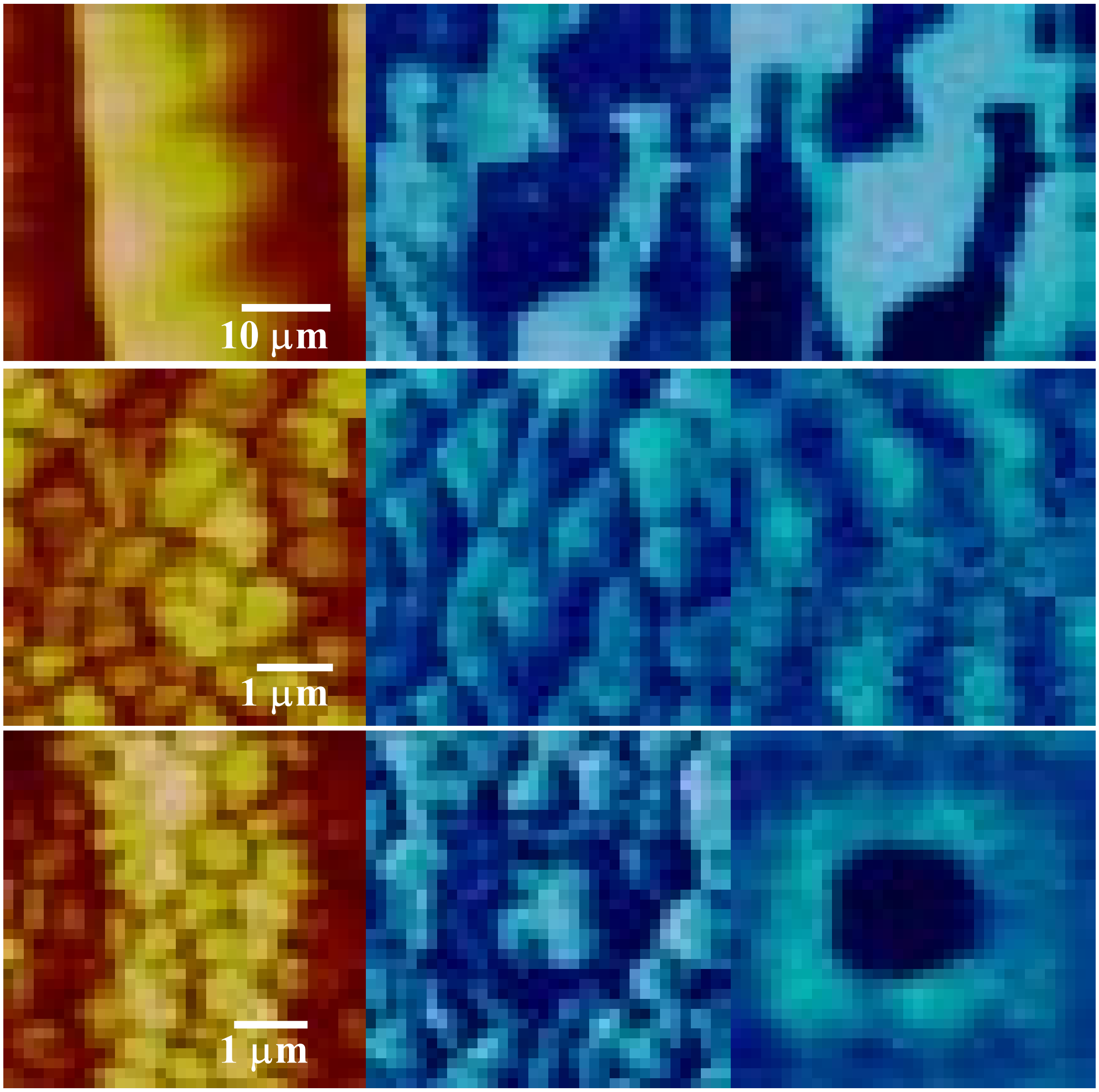

**Figure 6.21.** Surface topography (left), piezoresponse (central) and open-loop SSPM (right) images from *a-c* domains on the $BaTiO_3$ (100) surface (top), for a pristine PZT surface (middle) and for PZT after switching by 10 V at 2.5 μm and -10 V at 1 μm. Potential and piezoresponse images are obtained simultaneously.

However, after polarization switching the regions with deposited charge and reversed polarization are distinguished. This illustrates the approach to independently obtain information that allows capacitive vs. electromechanical interactions to be quantified.

## 6.10. Conclusions

Analytical models for electrostatic and electromechanical contrast in PFM have been developed. Image charge calculations are used to determine potential and field distributions in the tip-surface junction between a spherical tip and an anisotropic dielectric half plane. For high dielectric constant materials the surface potential directly below the tip is significantly smaller than the tip potential, implying the presence of an effective dielectric gap. The effect of the unscreened polarization charge during PFM is estimated and is shown to be negligible under ambient conditions for $BaTiO_3$. Within the electromechanical regime, strong (classical) and weak (field induced) indentation limits were distinguished. These solutions can be extended to domains of random orientation and to the analysis of stress effects in thin films by using renormalized effective electromechanical constants. Expressions for potential and field in the tip-surface junction and in the ferroelectric provide a framework for analyzing polarization switching phenomena and quantification of local hysteresis loops. The contributions of different electroelastic constants of the material to response amplitude were investigated and an almost linear correlation between piezoresponse and $d_{33}$ was illustrated for a series of PZT materials in the strong indentation regime. These solutions are represented by Contrast Mechanism Maps that elucidate the effect of experimental conditions on PFM. Based on these solutions the temperature dependence of piezoresponse on a $BaTiO_3$ surface was interpreted in terms of weak indentation/dielectric gap model, resolving apparent inconsistency between the divergence of $d_{33}$ at the Curie temperature and the experimental decay of the PFM signal with temperature.

Simple quantitative criterion for non-local cantilever-surface interactions in PFM is developed. The effective displacement due to cantilever buckling is inversely proportional to the spring constant of the cantilever. Depending on cantilever geometry, non-local interactions are small for cantilevers with spring constants $k_{eff} > 1$ N/m. This analysis can be used to introduce a non-local cantilever correction to local hysteresis loops obtained by PFM.

An approach for simultaneous acquisition of piezoresponse and surface potential image was developed. These data were shown to be complementary for the general case.

[65] One of the ways to ensure the quality of the contact is measurements of contact resistance on the similar conductive material, as discussed in Chapter 3. For example, PFM imaging on $BaTiO_3$ surface can be considered reliable if tip-surface current on the similarly prepared reduced $SrTiO_3$ surface is high; alternatively, zero tip-surface current implies high degree of contamination that limits both conductive and PFM imaging.

[71] This analysis does not take into account the effect of buckling oscillations of the cantilever as discussed in Section 6. However, the latter depends on the cantilever spring constant, rather than indentation force; therefore, the magnitudes of these effects can not be compared directly.

# 7. FERROELECTRIC LITHOGRAPHY: FROM DOMAIN ENGINEERING TO NANOSTRUCTURE FABRICATION

## 7.1. Introduction

To complement the results reported in Chapter 5, it can be expected that photochemical activity of ferroelectric titanates will strongly be affected by local polarization. A number of early results on polarization dependent catalytic activity of ferroelectrics were reported in the late sixties. It has also been known for a long time that rutile $TiO_2$ and perovskite titanates are extremely efficient catalysts in photooxidation and photoreduction processes. However, it was only recently that Rohrer *et. al.*[1] have discovered domain specific photoreduction of aqueous metal cations on polycrystalline $BaTiO_3$, this being the first example of domain selective photochemical process.

Immediately after this discovery, it was realized[1,2] that the PFM lithography on ferroelectric materials can be combined with metal photodeposition to fabricate metal and semiconductor nanostructures. The results here are directed to the practical implementation of ferroelectric lithography for the nanofabrication and analysis of the range of the systems (both substrate and metal) active in photodeposition processes. As SPM patterning is currently incompatible with large-scale fabrication, the opportunities for e-beam patterning ferroelectric domain structure are briefly discussed. Finally, the possible process flow for the device fabrication is presented. Many results presented here are obtained in the long standing collaboration with other members of Bonnell group, most notably T. Alvarez, R. Shao, E. Peng, Dr. X. Lei and Dr. J.H. Ferris, as reflected in the publication list. Extensive research in this field is currently underway and this chapter summarizes the results of proof-of-the-concept experiments.

## 7.2. Experimental Procedures

<u>Sample Preparation</u>

Polycrystalline barium titanate samples were prepared by sintering commercial $BaTiO_3$ powder (Aldrich). Powder was ball-milled and pressed into pellets. Pellets were annealed for 12h at 1400°C in air. Samples were cut by a diamond saw and exposed surfaces were polished with SiC media down to 1 μm grit size and alumina slurry down to 50 nm size. To relieve polishing damage, pellets were thermally etched at 1200°C for 12 h in air. Etching results in the formation of the grooves at the grain boundaries and surface faceting, which provide topographic contrast that can be used as markers in the AFM experiments. The annealing is also crucial to relieve surface damage associated with polishing.

A number of $BaTiO_3$ epitaxial films from several groups (Max Plank Institute, University of Michigan) were studied in the experiments. Unfortunately, PFM imaging did not reveal switchable polarization in either of the films and therefore no deposition experiments were attempted.

PZT thin films were prepared by sol-gel method on Pt/Si substrate (S. Dunn, Cranfield University, England). The thickness of the films was ~200 nm and characteristic grain size 50-100 nm.

<u>Piezoresponse Force Microscopy</u>

Contact-mode AFM and PFM was performed on a commercial instrument (Digital Instruments Dimension 3000 NS-III). To perform piezoresponse measurements, the AFM was additionally equipped with a function generator and lock-in amplifier (DS340, SRS 830, Stanford Research Systems). Pt coated tips ($l \approx 125$ μm, resonant frequency ~ 350 kHz) (Micromasch NSCS12 $W_2C$) and conductive diamond coated tips (DDESP, Digital Instruments) were used for these measurements. To perform polarization switching in $BaTiO_3$ polycrystalline samples, the microscope was equipped by PS310 high voltage power supply (Stanford Research Systems). To protect the electronic system, the electrical connections between the microscope and the tip were severed. A wire was connected from the function generator to the tip using a custom-build sample holder. This set-up allowed high voltages (up to 150 $V_{dc}$) to be applied. Internal microscope signals limits the switching voltage to 12 $V_{dc}$. The modulation amplitude in the PFM imaging was 6 $V_{pp}$. Using larger modulation amplitudes results in polarization reversal in the switched regions.

To perform local polarization switching, the ac tip bias was discontinued and a dc voltage was applied to the tip. To perform local patterning, a function generator output was controlled through a GPIB card using homebuilt control software.[3] After scanning a selected region, the tip was ac biased and scanned over larger region thus allowing the switched domain to be read. Polarization in PZT samples can be switched by voltages as low as 10 $V_{dc}$. Switching in $BaTiO_3$ ceramics requires high voltages (> 100 $V_{dc}$).

Photodeposition

To perform photodepositon, $BaTiO_3$ samples were placed in the 0.01 M $AgNO_3$ solution and irradiated by Xe UV lamp for 10 s at 100W. Variation in exposition time allowed particle size and density to be controlled. To study the wavelength dependence of deposition process, the experimental setup was additionally equipped by monochromator. After deposition, samples were washed and dried by airflow. Deposition conditions are specific for individual cations and substrates, e.g. Pd/$BaTiO_3$ or Ag/PZT requires longer exposure time (~ 30 min from 0.01M $PdCl_2$ and 0.01 M $AgNO_3$ solution respectively).

## 7.3. Ferroelectric Lithography on Polycrystalline $BaTiO_3$

The first experiments were aimed towards optimizing the properties of $BaTiO_3$ ceramics and reaction conditions. Photochemical reactivity and the structural properties of photodeposited metal are strongly dependent on semiconducting properties of $BaTiO_3$. Weakly *n*-doped/intrinsic ceramic material develops noticeable deposition layer in ~ 1-10 min and deposited layer is comprised from metal particles weakly or intermediately bound to the surface. Heavily donor doped $BaTiO_3$ (commercial PTCR sample) was extremely active under the irradiation and active reaction accompanied by the formation of hydrogen and metallic silver was observed. Similar behavior was observed for semiconducting ZnO. Metal was extremely weakly bound to the surface and was floating in the solution. Formation of large silver crystallites bound to the surface was also observed. Finally, unintentionally *p*-doped (iron impurity) materials exhibited weak or no photochemical reactivity.

Photochemical activity of $BaTiO_3$ was extremely sensitive to surface condition. Generally, as polished samples did not developed domain specific deposition patterns and thermal etching at 1200°C was necessary to achieve desired reactivity. Etching by HCl yields inferior results partially ascribed to surface contamination with chloride ions and subsequent formation of photoactive AgCl.

The wavelength dependence of deposition rate determined a threshold wavelength corresponding to the band gap of $BaTiO_3$ ($E_g$ = 3.1 eV). The threshold value for the PZT is higher (4.1 eV).

Based on the results of preliminary experiments, the optimal deposition was achieved on undoped thermally etched $BaTiO_3$ under white radiation. The use of a monochromator significantly reduces the overall intensity of radiation resulting in long (> 30 min) deposition times. The details of individual steps in domain imaging and engineering are presented below.

### 7.3.1. Domain Imaging and Photodeposition

Piezoresponse force microscopy is used to determine the domain structure of the sample before the photodeposition. Surface topography and piezoresponse force microscopy (PFM) images of $BaTiO_3$ surface are shown in Figure 7.1a,b. While no distinct topographic features are seen in Figure 7.1a, PFM image clearly reveals a complicated lamellar domain pattern with domain size of order of ~200-300 nm. Note that topographic defect (pore) results only in a minor alteration of domain structure and does not impair PFM contrast.

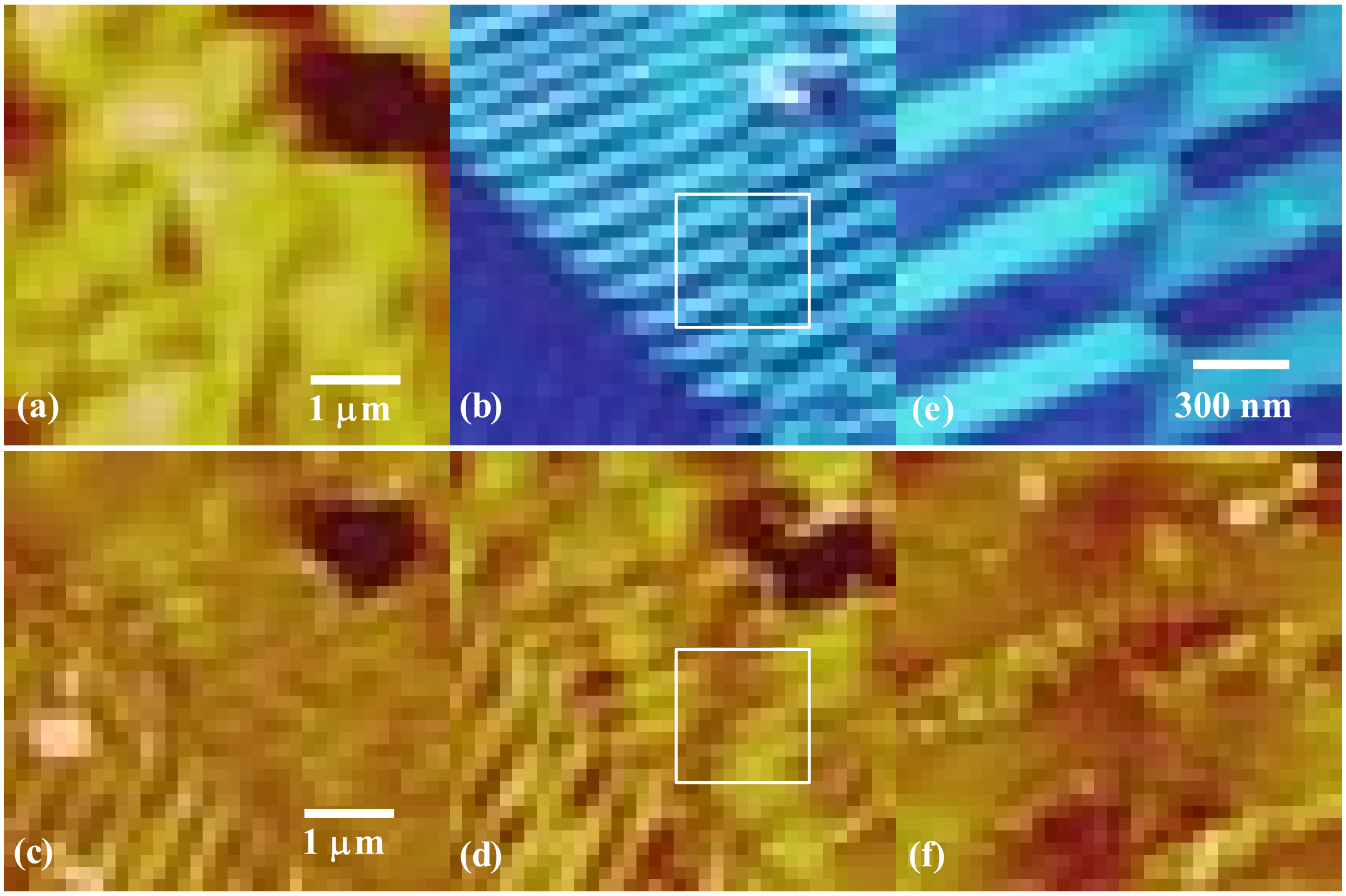

**Figure 7.1.** Local contact mode topography (a) and piezoresponse image (b) of $BaTiO_3$ surface prior to the deposition. Topography after silver (c) and palladium (d) deposition. (e) and (f) are small scans corresponding to the boxed areas on (b) and (d).

After the domain characterization, the sample was placed in a 0.01 M $AgNO_3$ solution and irradiated by Xe lamp for ~1 min. The photodeposited metal pattern is clearly seen in Figure 7.1c. Imaging in this case is difficult due to the small particle size and weak bonding between the individual silver particles and the surface. During AFM tapping mode imaging both pick-up of the particles by the tip and displacement of the particles is possible with corresponding reduction in image quality (Figure 7.1f). After imaging, silver particles were mechanically removed and palladium was deposited on the surface, Figure 7.1d. Note that the polarization distribution on the pristine surface and deposition patterns of silver and palladium are identical. It was shown that the polarization distribution on the surface does not change during deposition process. The reactivity of the ferroelectric surface is not limited by the degree to which the reaction has proceeded and removal and deposition steps can be repeated several times.

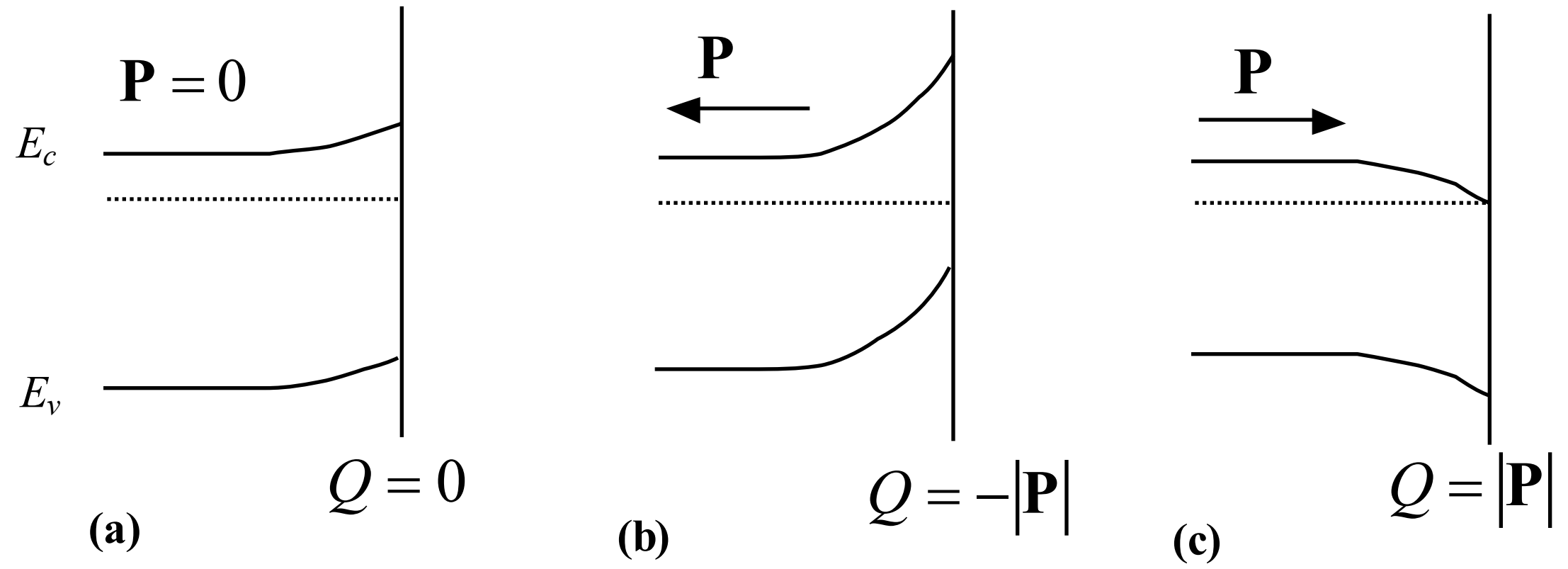

**Figure 7.2.** Schematic diagram of band bending in the paraelectric perovskite above the Curie temperature (a) and in the ferroelectric perovskite in the $c^-$ (b) and $c^+$ (c) domain regions.

The mechanism for domain selective photoreduction is closely related to intrinsic screening on ferroelectric surfaces as illustrated in Figure 7.2. In the absence of vacancy or step edge defects, transition metal oxide surfaces have a low density of surface states in the gap between the conduction band, formed predominantly from the *d* states of the transition metal, and valence band, formed predominantly from oxygen *p* states. In regions with negative polarization ($c^-$ domains), the effective surface charge becomes more negative and, therefore, upward band bending occurs. In the regions with positive polarization ($c^+$ domains), surface charge is positive, with associated downward band bending. Irradiation with super band gap light results in the formation of an electron-hole pair. In ambient, the space charge field results in separation of the electron-hole pair and charge accumulation on the surface, i.e. the photovoltage effect. However, on a surface immersed in a cationic solution the electrons can reduce the metal cations preventing charge accumulation at the surface. Reduction is expected on positive domains, while oxidation is expected at negative domains. Thus, particles form preferentially at $c^+$ domains.

To get some insight into the relationship between grain orientation and photodeposition rate, a number of grains were studied. Several examples of silver deposition pattern on samples with different grain orientations are shown in Figure 7.3. The domain structure of the first grain is formed by the lamellar regions with almost zero piezoresponse interspaced by the regions with well-defined dark and bright regions. This domain structure as well as the absence of surface faceting indicates that the grain orientation is close to (100); the domain structure is thus lamellar *a*-*c* domain structure, *c* domains having both $c^+$ and $c^-$ orientations (comp. Figure 5.3d). The domain structure of the second grain clearly has some *a*-*c* character (large features), but now the response in the "*a*" domain regions is not completely zero, indicating the deviation from (100) orientation. The topographic structure of this grain is formed by elongated steps. The combination of topographic and piezoresponse imaging suggests that the grain has an orientation close to (*hk*0); the facets in this case are due to the faceting into (100) surfaces. The topography of the third grain exhibits triangular pyramidal features indicating that the grain has a general orientation (*hkl*). The piezoresponse image in this case is considerably more complex exhibiting herringbone patterns; the domains thus can not be indexed as *a* or *c*. Note that domain structures in adjacent grains are closely correlated, i.e. there exist strong intergrain coupling of polarization. This can be expected for the minimization of electrostatic energy at the grain boundaries.

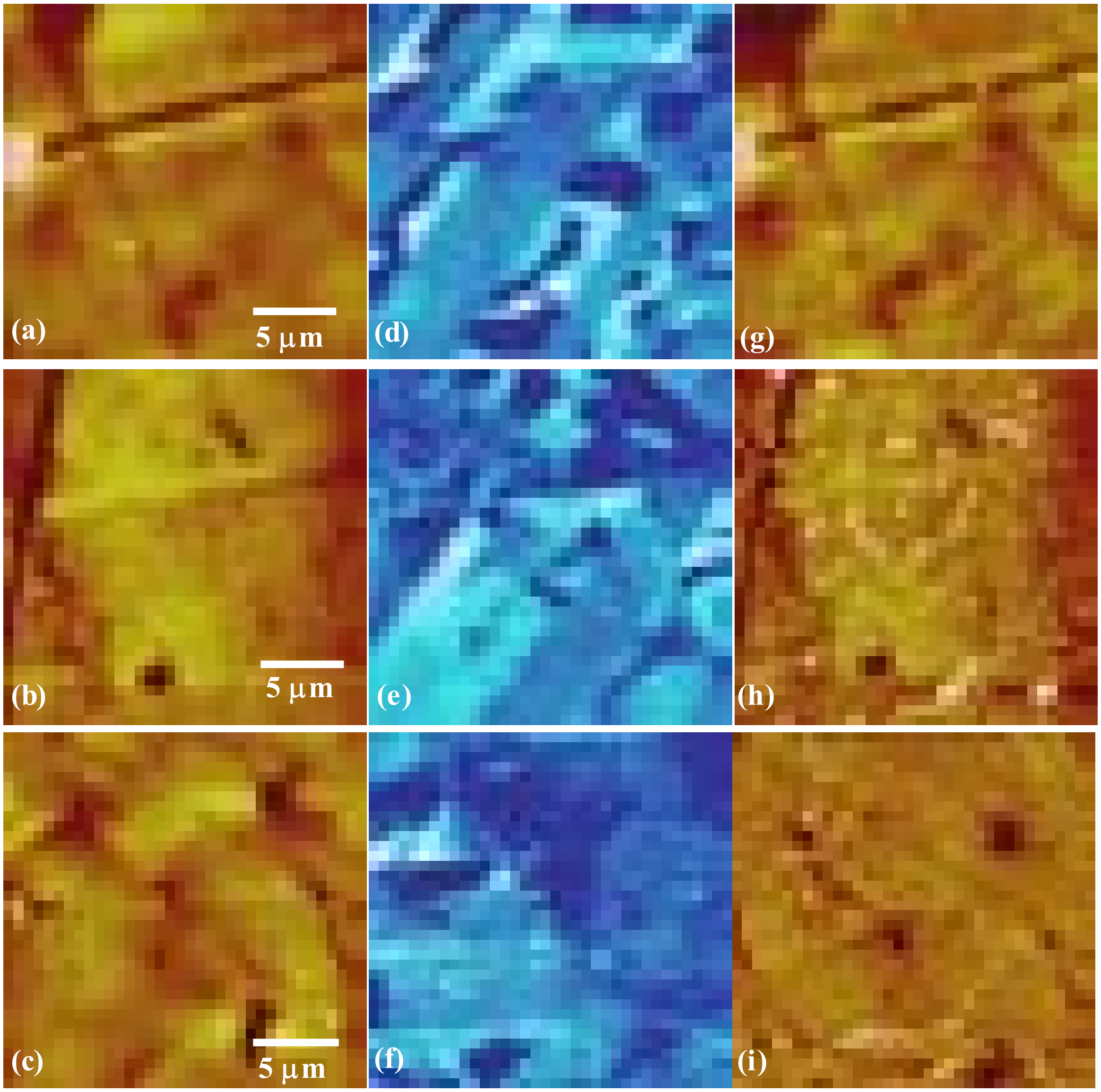

**Figure 7.3.** Local contact mode topography (a,b,c) and piezoresponse (d,e,f) images of $BaTiO_3$ surface prior to the deposition. Topography after silver deposition (g,h,i). Note the qualitative agreement between metal deposition pattern and PFM image.

The deposition patterns for all three grains are shown in Figure 7.3. It can be clearly seen that there is a one-to-one correspondence between the domain orientation and the deposition pattern. Unlike the single crystal, in the general case it is impossible to unambiguously relate the photochemical activity of the domain and piezoresponse image (note well defined photodeposition on "*a*" domains in Figure 7.2g). Quantifying the relationship between PFM contrast and local photochemical activity is challenging since it is determined as a product of at least three dependencies (chemical activity in paraelectric phase on orientation; polarization charge on orientation; measured piezoresponse on orientation) that have maxima at different orientations. The analysis of orientation dependence of domain related reactivity requires extensive statistical analysis of deposition data combined with orientation SEM imaging.

The additional complications to the orientation dependence studies of photochemical activity arise due to the marked tendency for the surface domain structure reconstructions. Stresses due to the processing and topographical features due to faceting on the high index surfaces promote the formation of nanoscale (~10-100 nm) near surface domains different from the dominant bulk domain structure (comp. to Figure 5.4 for single crystal $BaTiO_3$). Careful inspection of Figure 7.3 shows that the deposition pattern is defined by both large-scale and nanoscale domain structures as resolved by PFM.

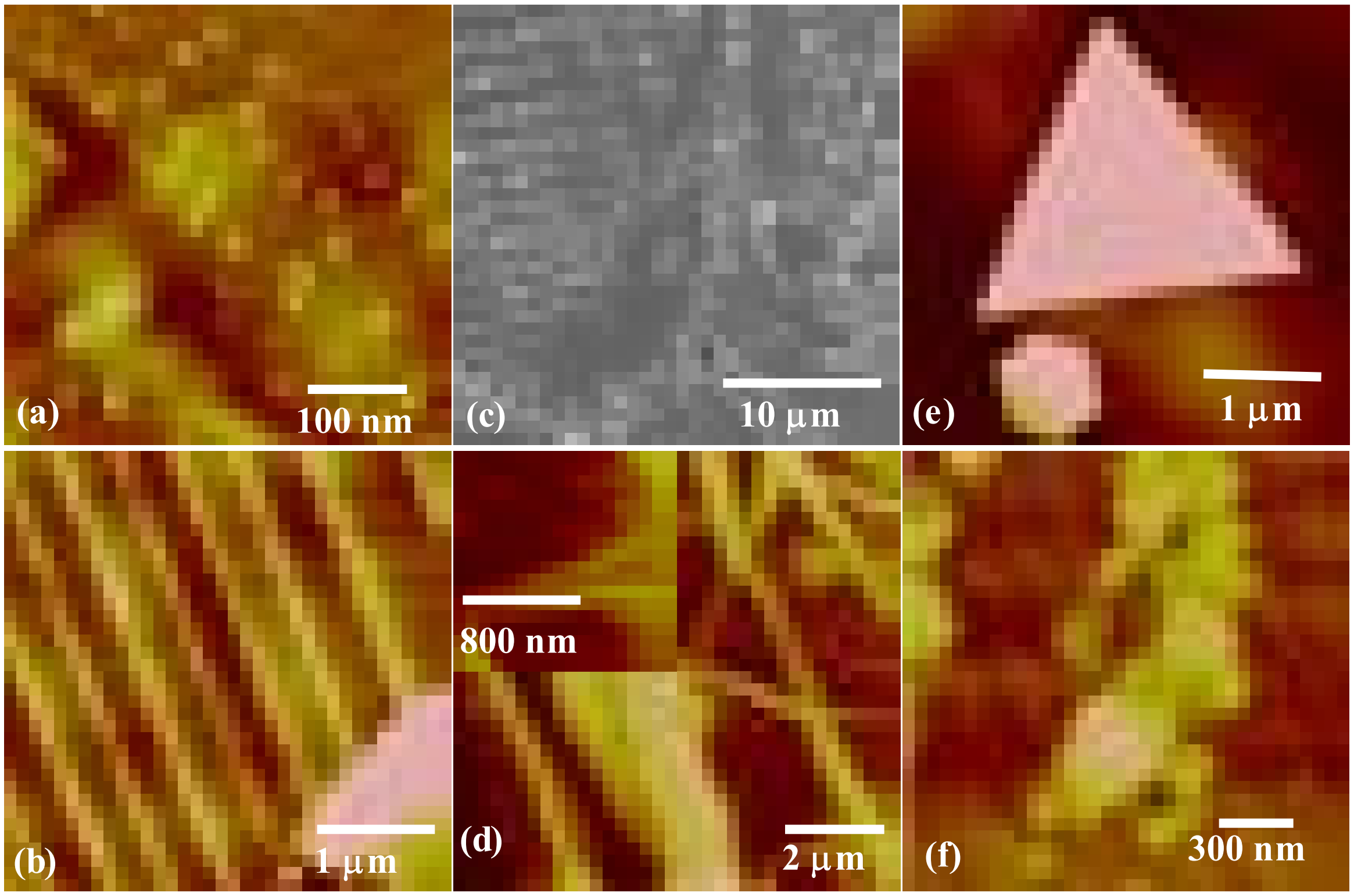

**Figure 7.4.** Surface topography (a,b) of Ag nanoparticles at the early stages of photo reduction of aqueous $AgNO_3$. On the larger length scales, particles are assembled in the lamellar arrays corresponding to the underlying positive ferroelectric domains (b). SEM image of crystalline Ag aggregates formed at longer reaction times (c). AFM image of silver nanowires formed on single crystal surfaces (d). AFM images of larger Au particles from photo reduction of $HAuCl_4$ (image by T. Alvarez) (e). Palladium deposition pattern (image by T. Alvarez) (f).

Some of the morphologies for metal photodeposition are illustrated in Figure 7.4. On the initial stage of growth (~3 sec) the silver deposited in the form of nanometer (3-10 nm) sized particles (Figure 7.4a). For longer exposure times (~1 min), the particles form lines on the corresponding domains (Figure 7.4b). The prolonged irradiation (30 min) results in the growth of silver and large (~400 nm) crystallites are formed as shown in Figure 7.4c. At this stage, the shape of the crystallites is well defined and the development of crystallographic planes can be seen. Deposition on single crystal $BaTiO_3$ with a relatively inactive surface results in the formation of peculiar meandering wires as illustrated in Figure 7.4d. Photodeposition of a number of other metals was also attempted. We have found that both silver and palladium grow domain specifically, while gold and rhodium tend to reduce non-selectively. Interesting crystallite morphologies were observed for gold in the form of triangular and hexagonal crystals in Figure 7.4e. Finally, deposition of Pd is similar to that of Ag; well-defined palladium crystallites are shown in Figure 7.4f. From these observations, the conditions for photodeposition are that the redox potential of the cation in the solution is larger than that of $H_2/H^+$ at a given pH. At the same time, for too large a potential the deposition is not selective (as for Au). Clearly, the pH cannot exceed the stability range for $BaTiO_3$ surface. We believe that the optimal control over photodeposition process can be achieved by the judicious choice of complexating agent for the cation and solvent and further research is under way.

### 7.3.2. Domain Engineering in Polycrystalline $BaTiO_3$

The purpose of this experiment is to establish the validity of a lithographic process based on controlled photodeposition on a ferroelectric material with an engineered domain structure. Here, we attempted local poling of polycrystalline $BaTiO_3$ ceramics with an AFM tip.

The results of a successful switching process are illustrated in Figure 7.5. Surface topography and piezoresponse images of a pristine $BaTiO_3$ surface are shown in Figure 7.5a,b. Note the presence of large domains associated with the grains and small *a-c* domain pattern. After the imaging, a bias of -100 V was applied to the tip while scanning over a 6x6 μm region. After switching, a large negative feature in PFM image is shown in Figure 7.5c. This switched domain structure was metastable and tended to disappear with time if the imaging ac bias was high (> 5 $V_{pp}$). The evolution of domain structure after consecutive scans is illustrated in the Figure 7.6.

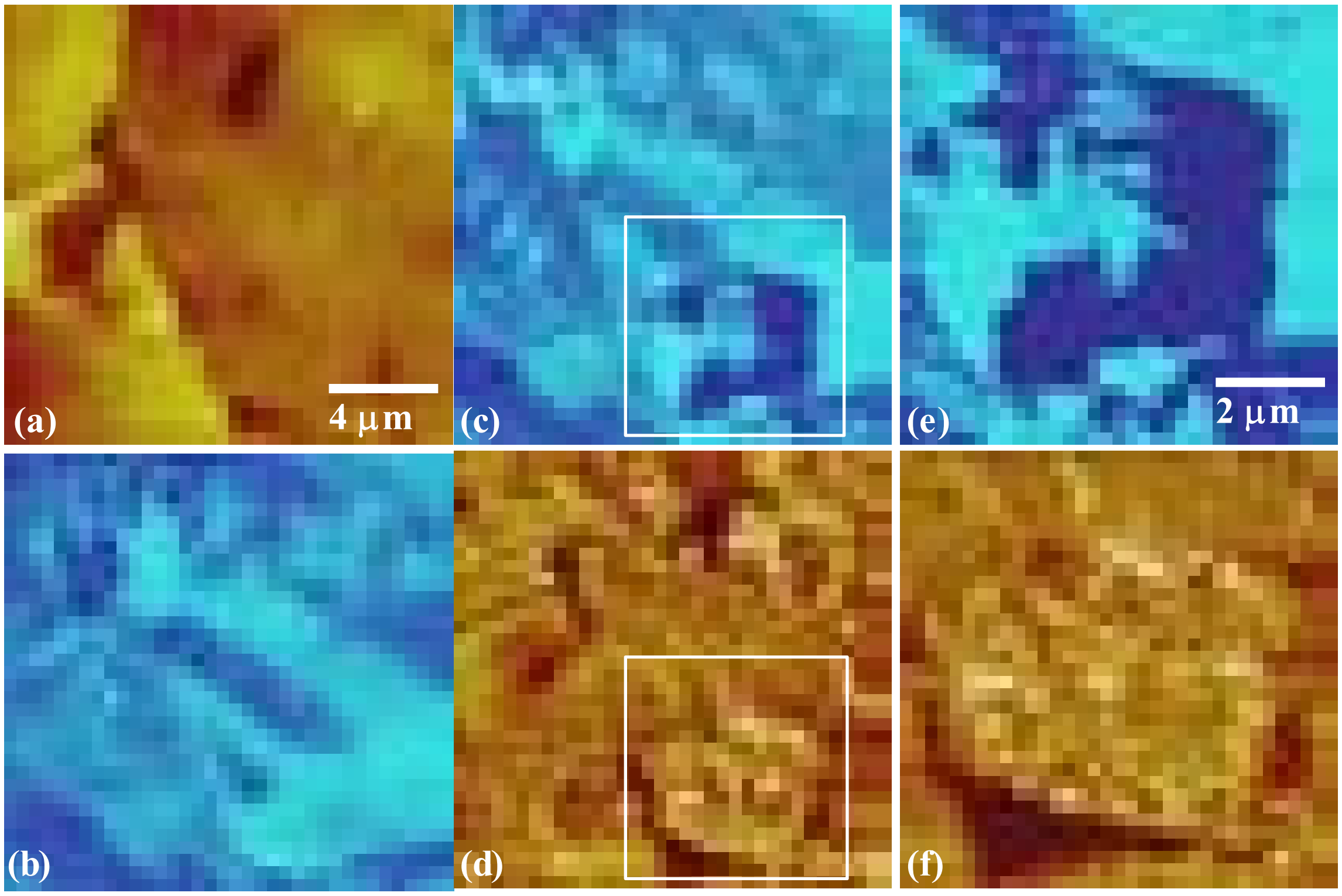

**Figure 7.5.** Surface topography (a) and piezoresponse image (b) of the pristine $BaTiO_3$ surface. PFM image of $BaTiO_3$ surface after local polarization switching by -100 $V_{dc}$ in the 6 μm region (c). Surface topography after silver photodeposition (d). (e) and (f) are small scans corresponding to the boxed areas on (c) and (d).

Note that the lamellar domains in the near surface regions remained unaltered (horizontal streaks in the switched region). Immediately after the switching, the sample was placed in the silver nitrate solution and the deposition was performed. The resulting deposition pattern is shown in Figure 7.5d in which heavier deposition is evident in the switched region. At the same time, due to the presence of large amount of *a-c* domains (stripes) the deposition is not localized in the switched region. These results couldn't be ascribed to the negative charge injection into ceramic material with subsequent reduction of silver. First, no deposition was observed on the regions after applied high voltages (~ -150 V) if the switching did not occur. Second, typical amount of deposited material (> 100 nm) would require unrealistically high initial injected charge densities (~ $5 \cdot 10^3$ C/m$^2$ corresponding to ~2000 electrons per unit cell).

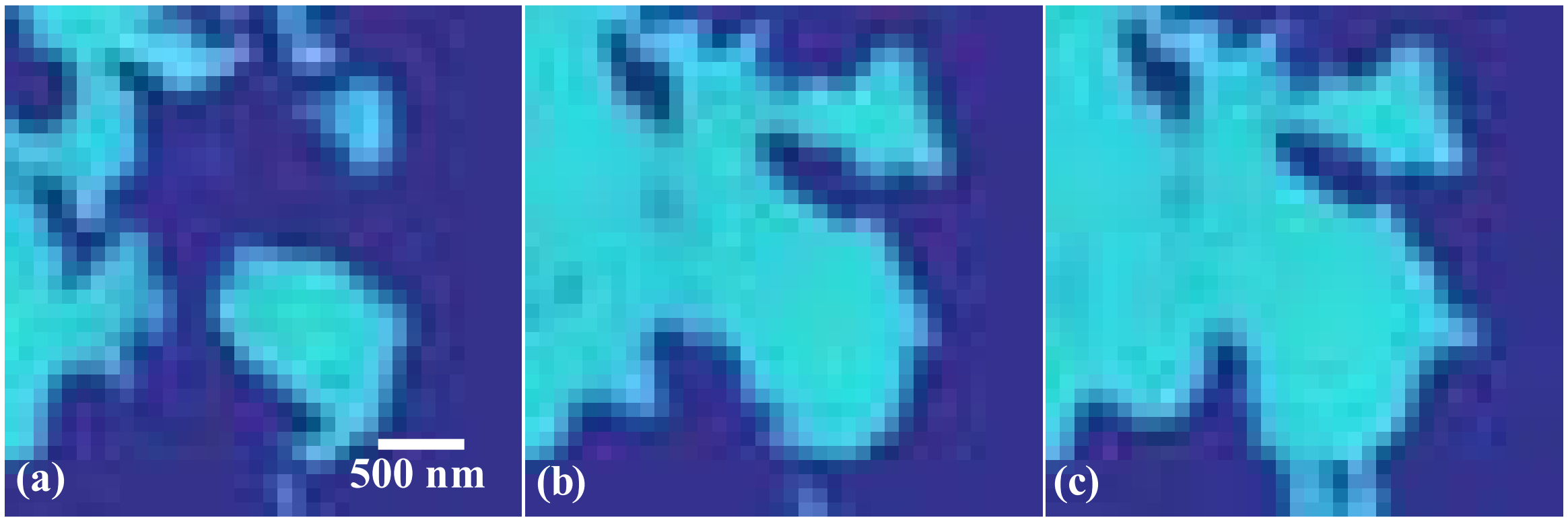

**Figure 7.6.** Polarization instability in switched region in Fig. 6.5. (a) is acquired immediately after switching by –100 V, (b) is the second scan, (c) is the third scan. It was shown latter that polarization reversal can be minimized by smaller driving voltages during PFM imaging.

Despite this result, local switching of polycrystalline $BaTiO_3$ is shown to be extremely difficult and unreliable. Poling requires extremely high tip voltages (~ 100-150 $V_{dc}$). Even for these high voltages (and inherent danger of such experiments), only two attempts out of ~ 10 were successful. To rationalize these observations, one can speculate that the poling voltage is determined by the substrate coercive field and the geometry of the system. In thin film samples on conductive substrates, the application of relatively small voltages (5-10 V) to the tip results in a large fields (~$10^6$-$10^8$ V/m) sufficient for local polarization reversal (provided that tip-surface contact is good, as discussed in Chapters 3 and 6). In contrast, polarization switching in bulk materials is considerably more difficult due to the effect of uncompensated charge on the lower surface of the growing domain[4] and high dielectric constant resulting in poor tip-surface contact (Chapter 6). Switching is also hindered by surface *a-c* surface domain structure reconstruction and associated elastic stress fields that effectively pin 90° domain walls. Therefore, even though bulk *c* domains can be oriented by poling at high voltages, surface *a* domains will remain and result in an uncontrollable deposition pattern as seen on Figure 7.5d. Finally, the poling voltage is clearly dependent on crystallographic grain orientation. While for (100) surfaces the poling results in a single variant of domain orientation and polarization vector orients in (100) direction (with a few exceptions for strongly anisotropic materials[5]), for grain with e.g. (111) orientation three domains orientation will be equivalent and more complex types of domain structures in the poled region can result.

To summarize, the results on $BaTiO_3$ prove that local polarization controlled with subsequent metal photodeposition can indeed be used for directed creation of metallic structures. $BaTiO_3$ ceramics and thin films, while being very active in the photodeposition process, are ill suited for AFM patterning. Therefore, further progress can be achieved by the search of alternative substrates.

### 7.4. Ferroelectric Lithography on PZT Thin Films

The difficulties in polarization switching in $BaTiO_3$ can be traced to the small number of possible domain orientations (six equivalent (100) axes) and relatively high *a*/*c* ratio of the unit cell that leads to the enhanced domain wall pinning. Lead-zirconate titanate (PZT) based materials, especially in the vicinity of morphotropic phase boundary, possess multiple equivalent polarization orientation (8 or 12) greatly facilitating polarization reversal processes. High quality epitaxial and oriented PZT films are widely used for many applications and thus are available commercially. Additionally, due to the close crystallographic and electronic structure similarity between $BaTiO_3$ and PZT it can be conjectured that domain specific photochemical activity will be characteristic for this material as well. However, until now no information on the photochemical reactions on PZT was available.

Preliminary experiments have shown that the color of PZT films irradiated by UV light in the silver nitrate solution indeed changes, indicating the deposition of the metal. The required deposition times for PZT are longer than those for $BaTiO_3$, reflecting the difference in the band gaps for these materials ($E_g$ = 3.1 eV for $BaTiO_3$ and 4.1 eV for PZT). The active region of the spectrum is narrower for the PZT.

Surface topography and piezoresponse image of PZT films before poling are shown in Figure 7.7a,b. The inset on the topographic image shows that the film consists of 50-100 nm size grains. The inset on the piezoresponse image verifies that ferroelectric domains are present in the film and domain size is comparable with the grain size. Careful inspection of PFM images illustrates that most small grains (~50 nm) are in the single domain state, while larger grains can contain multiple domains. Unlike $BaTiO_3$ crystals, domains are small and do not form ordered patterns; therefore, comparison of domain structure before the photodeposition and metal deposition pattern is all but impossible. To avoid this problem, we have fabricated relatively large-scale lines by intermittent application of +10 V and -10 V to the tip. The resulting domain pattern is shown in Figure 7.7c exhibiting "random" polarization orientations with domain size of 50-100 nm on the edges of the image and line regions with positive or negative polarization orientation in the central part of the image. Patterned samples were placed in the silver nitrate solution and irradiated by the UV lamp for 30 min. Resulting metal patterns were clearly visible in the optical microscope; the corresponding topographic image is shown in Figure 7.7d. Careful inspection of the lines has shown that they are formed by small (10-50 nm) silver particles. The total amount of deposited material corresponds to ~100 nm layer (comp. to the thickness of the PZT film ~ 200 nm). Note the one-to-one correspondence between the polarization pattern on the PFM image and photodeposited silver pattern. Deposition occurs exclusively on the domains written by the negative voltage, i.e. positive domains. Very little silver particle density was observed on the negative domains. The lateral size of the smallest feature on the image is ~700 nm. The lateral size of the features that can be written on PZT surface is limited by the grain size of the film and was found to be 50-100 nm.

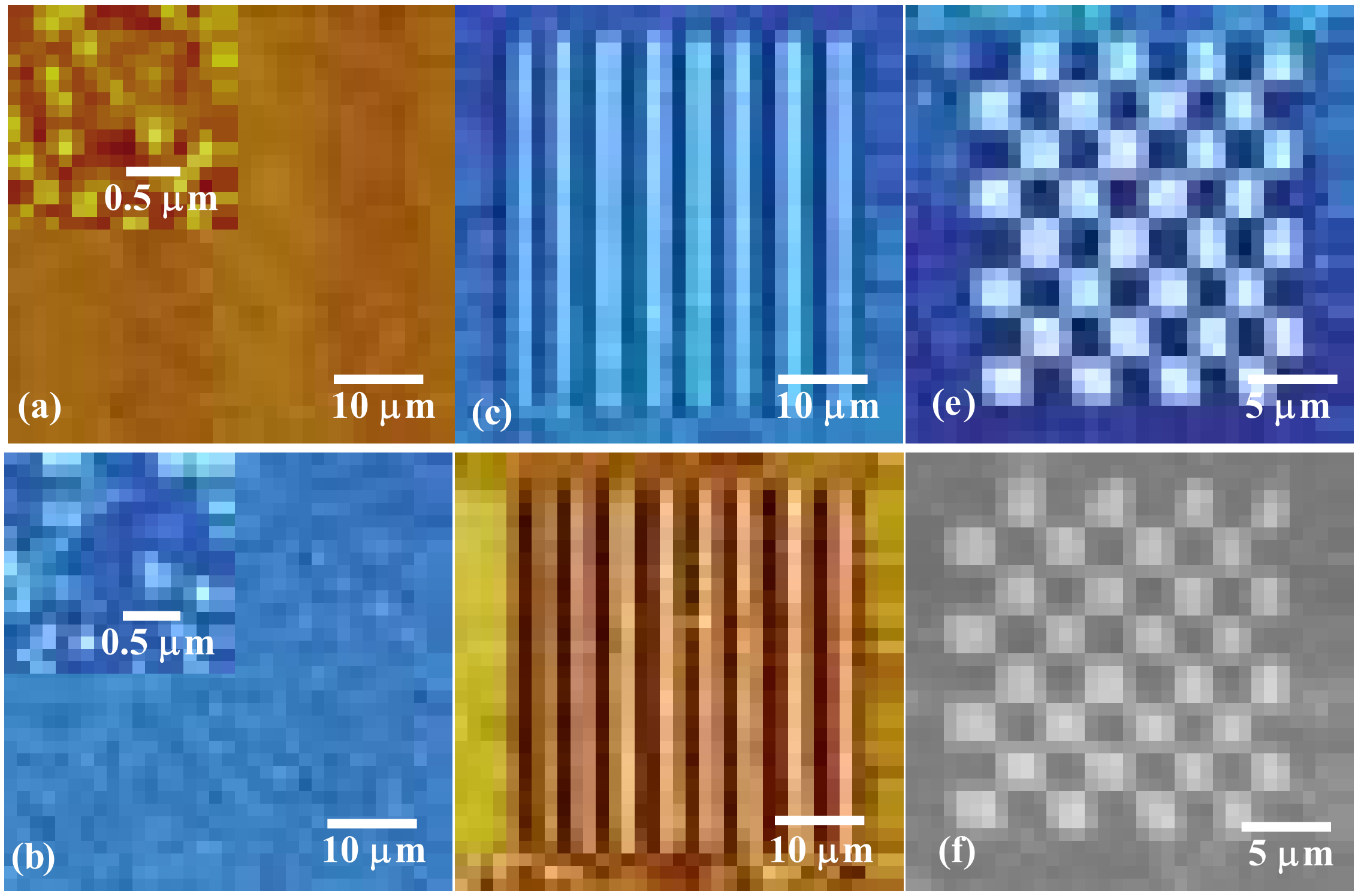

**Figure 7.7.** Surface topography (a) and piezoresponse image (b) of PZT thin film. The inset shows that the PFM contrast is not random but is due to the small (~50-100 nm) ferroelectric domains associated with grains. PFM image (c) of lines patterned with alternating +10 and -10 $V_{dc}$. Surface topography (d) after deposition of Ag nanoparticles. Piezoresponse image of checkerboard domain structure fabricated using lithographic system (e) and SEM image of corresponding silver photo deposition pattern (f).

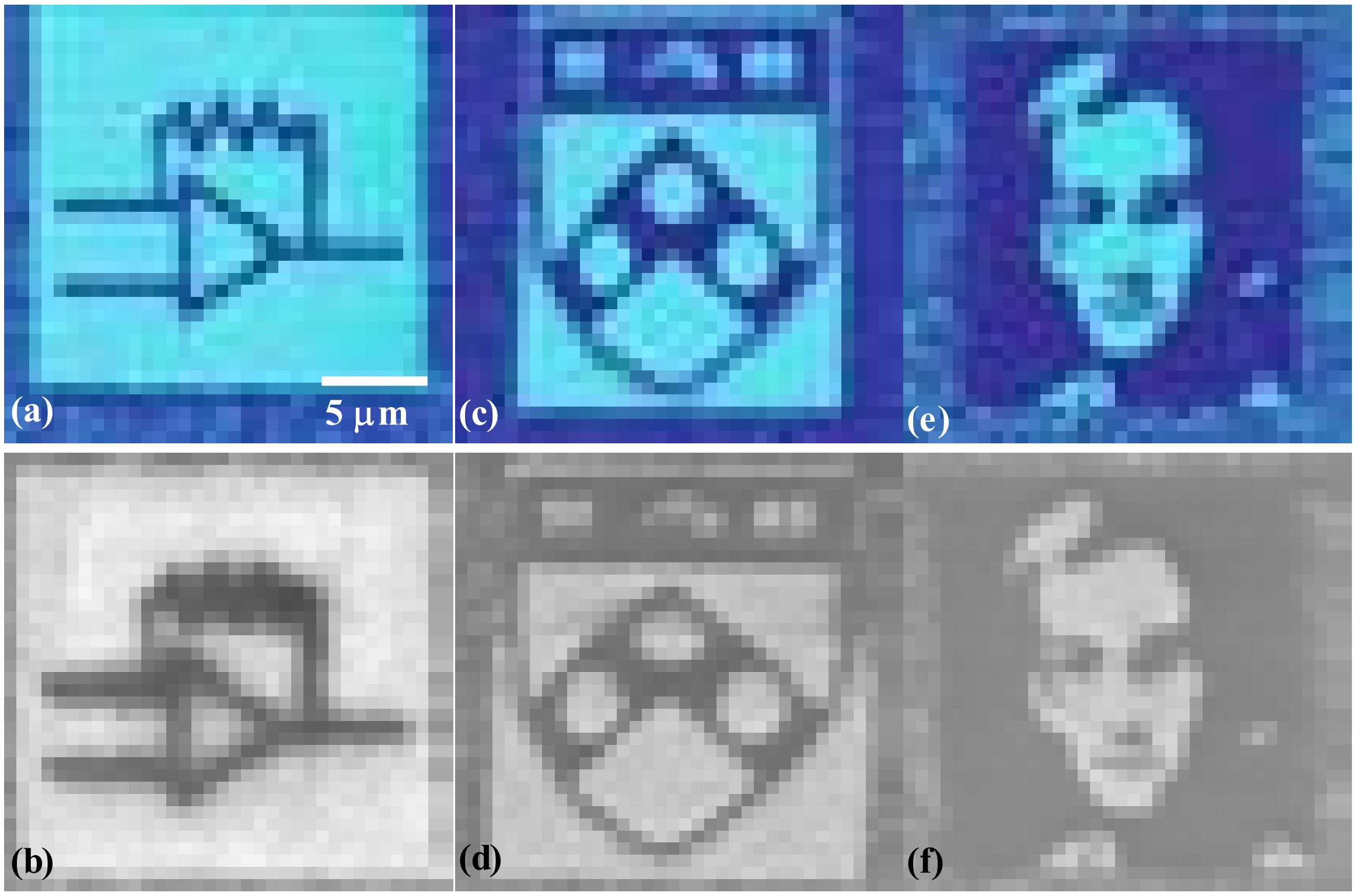

**Figure 7.8.** Piezoresponse images (a,c,e) patterned by lithographic system and SEM micrographs (b,d,f) of photodeposited silver. Note one to one correspondence between PFM image and silver deposition pattern. The images represent "nanocircuit" (a,b), Penn logo (c,d), thesis author (e,f). Image size is 15 μm.

To extend this approach towards complex structure fabrication, we have built in-house ferroelectric lithography system. Briefly, the input file (TIFF image) is digitized with controlled thresholding level resulting in binary data array. This array is used to control the voltage output of function generator through the GPIB interface. The voltage generator output is applied to the AFM tip. Thus, during the scanning the tip voltage is altered in a controlled manner, resulting in the image transfer on ferroelectric surface. A number of images produced by this approach are shown in Figure 7.8. The image size in all cases is 15x15 μm and is limited by the grain size of PZT thin film (for small image sizes, the grain effect becomes important, for larger image sizes, there are unswitched regions between the lines) and mechanical stability of the tip (for large scan sizes, large tip velocity enhances wear).

To summarize, we have discovered that PZT is a second ferroelectric material active in photodeposition process. This behavior can be projected to other ferroelectric titanates and possibly ferroelectric semiconductors in general, however, this remains to be proven. In PZT films a large number of polarization orientations, small dielectric constant, weak coupling between the grains and small number of non-180° domain walls facilitate the local poling. At the same time, the photodeposition process is selective and constrained to positively poled regions only; deposition level on negatively poled domains is extremely small. Therefore, local domain patterning and photodeposition can be performed selectively with high quality of image transfer. The minimal feature size achieved so far is limited by the grain size of the film and is ~ 100 nm. At the same time, particle size can be significantly smaller (~3-10 nm). Further improvement in resolution of these techniques will be achieved with PZT epitaxial thin films on conductive substrates and is expected to be at least 10 nm.

### 7.5. Domain Patterning by E-beam

As illustrated in the previous section, combination of ferroelectric patterning and metal photodeposition, further referred to as ferroelectric lithography, can be used for the fabrication of the nanometer scale structures. However, while nanofabrication with probe tips is useful in fundamental studies, it is not readily amenable to large-scale applications. At the same time, it was reported that ferroelectric domain structure in materials such as single crystal $LiNbO_3$ could be controlled by the e-beam.[6,7] Due to the close similarity between PZT and $LiNbO_3$ it can be conjectured that e-beam patterning can be applied to the former material as well.

To confirm this conjecture, PZT thin films were exposed to the e-beam in the SEM (JEOL 6400, ~30x40 μm area, 20 keV, 10 min). The exposed region was then imaged using PFM. Figure 7.9 clearly illustrates that e-beam exposure results in the reorientation of polarization. The sign of polarization reversal depends on the exposure time – 10 min exposure has led to both significant carbon deposition and formation of positively poled regions, whereas short (1-3 sec) exposures resulted in the formation of negatively poled regions. Surface potential imaging immediately after exposure has shown that there is no significant charge accumulation in the exposed regions (not shown), which may account for PFM contrast. The polarization orientations are confirmed by reversing the domains with a locally applied electric field and comparing the associated piezoelectric response. It is found that polarization in negatively switched regions can be back switched by relatively small tip voltages (~10 V, comparable to pristine film) whereas in the positively switched region carbon deposition resulted in increased (30 V) switching voltages.

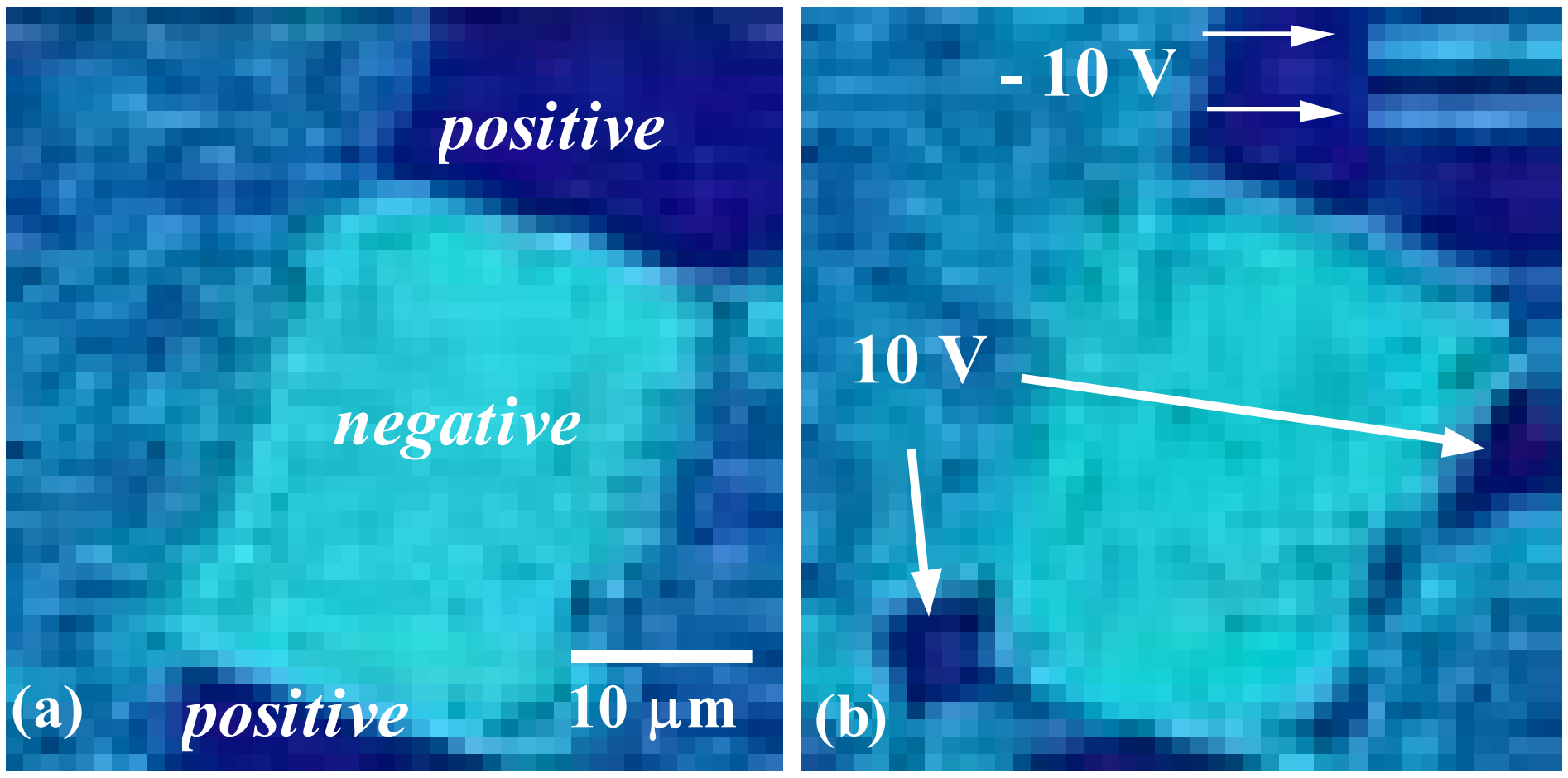

**Figure 7.9.** Piezoresponse image of PZT surface exposed to an e-beam (a). Polarization switching can be confirmed by reversing the polarization orientation by applying an electric field (b).

The origins of e-beam induced switching are straightforward. Primary electrons from the beam injected into the surface interact with the solid causing secondary electrons (core electrons, valence electrons, Auger electrons) to be emitted. If the ratio of emitted electrons to incident electrons is <1, the surface will be negatively charged; if > 1 the surface will be positively charged. The distribution of primary and emitted electrons is determined by penetration depths of the beam (1-5 μm)[8] and the escape depth of the secondary electrons (1-3 nm). The charge is also compensated by the band bending and carrier influx from the bulk of the materials. Beam induced charge distribution is clearly inhomogeneous and results in a local electric field that may cause realignment of atomic polarization. The formation of e-beam induced dipole is used for local domain imaging in Surface Acoustic Microscopy (SAM)[9] of ferroelectric materials and the mechanism of SAM imaging was extensively studied by X.X. Li. However, the primary object in these studies was polycrystalline $BaTiO_3$, in which polarization switching is hindered; no beam switching was reported.

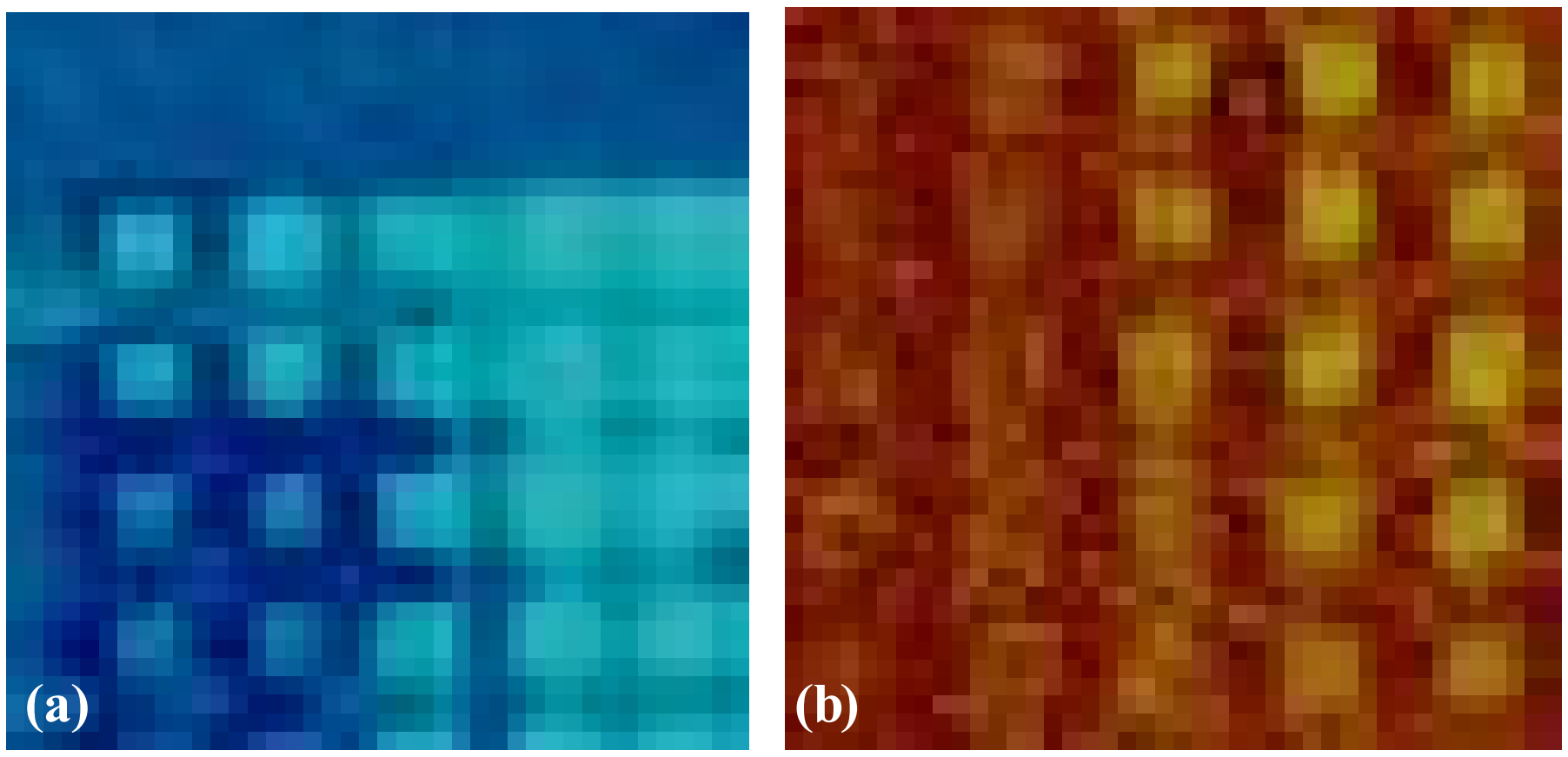

**Figure 7.10.** (a) Piezoresponse image of e-beam switched pattern for different doses and (b) topographic image photodeposited silver pattern (courtesy of J.H. Ferris and T. Alvarez).

In the case illustrated in Figure 7.9, the domains in a PZT thin film are oriented negatively by beam energies between 200 eV and 10 keV, which produce a positive surface charge. The opposite orientation is produced in regions where a thin (<1 nm) carbon film deposits, which has the opposite primary to secondary electron ratio and, consequently, the opposite charge. It is the combination of high excitation cross-section and small escape depth that facilitates development of a sufficiently large field to reorient domains. The energy dependence of secondary electron yield for oxides is such that positive poling is expected at energies >20 kV even in the absence of carbon deposition.[10]

This approach can be further developed by use of the standard e-beam lithographic tool for the local poling of the ferroelectric substrate. Figure 7.10a illustrates an array of square regions irradiated at different doses (500 $\mu C/m^2$ for the bottom left to 5.184 $mC/m^2$ in the top right). Figure 7.10b shows the corresponding deposition pattern. Note that the amount of deposited material strongly depends on the dose.

## 7.6. Device Fabrication

After the preliminary results on the controlled deposition of silver on PZT surface with engineered domain structure were achieved, we attempted the integration of this process with conventional semiconductor technology. At the first stage, we attempted the fabrication of the silver metal line between macroscopic contacts. The crucial step in this fabrication is to prevent the random silver deposition on the PZT surface areas unaffected by the tip. This can be achieved by a) fabricating the PZT islands on the non-conductive non-photoactive substrate, b) negative poling of the PZT surface prior to the SPM or e-beam patterning and c) creation of the protective dielectric layer that leaves open PZT mesas small enough to be controlled by the tip.

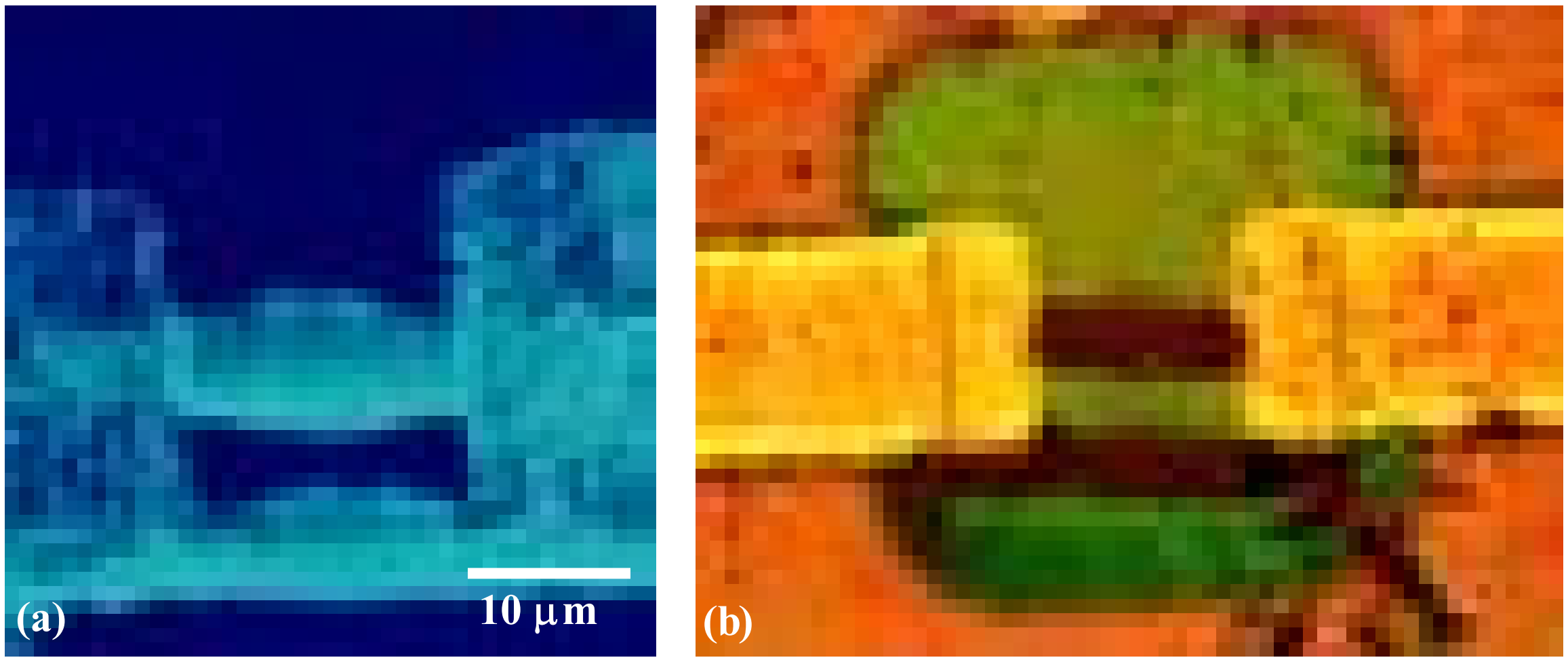

**Figure 7.11.** Piezoresponse phase image (a) of positively switched lines on the negative background. (b) Optical micrograph after silver deposition.

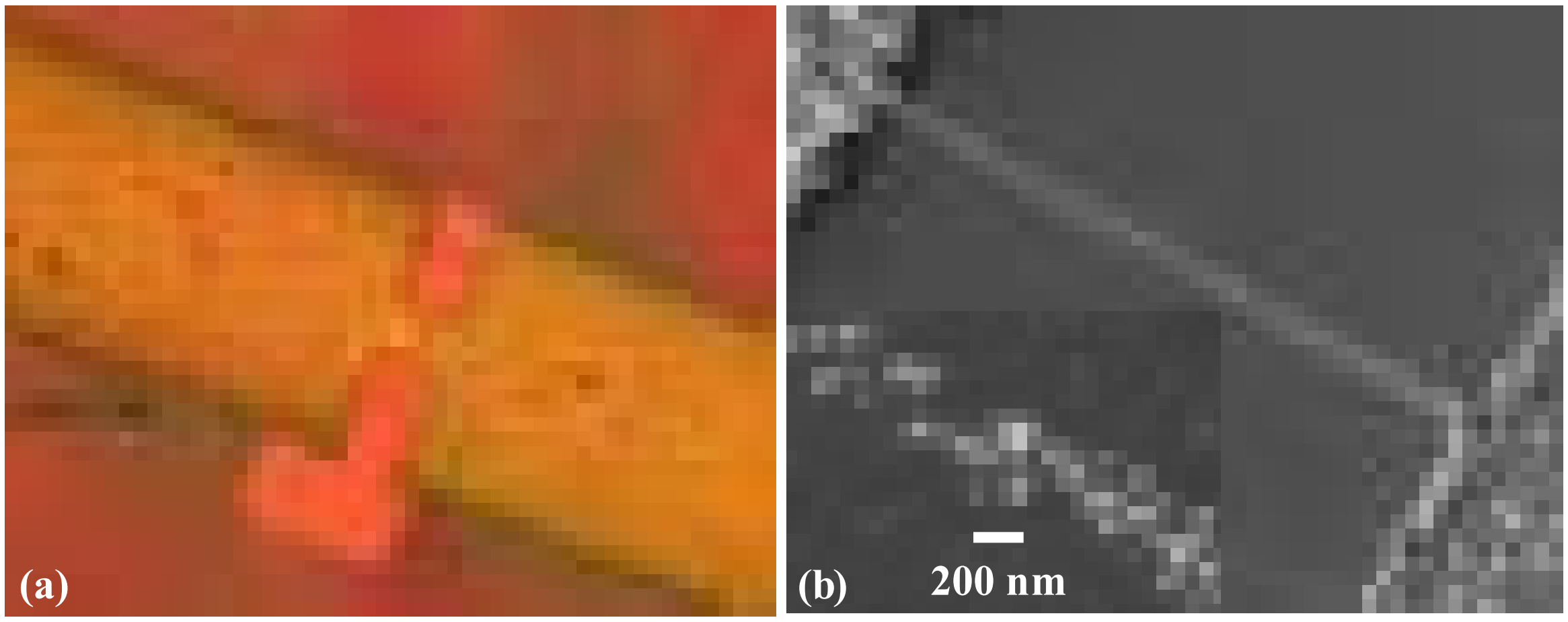

**Figure 7.12.** Optical micrograph (a) of the silver line fabricated between two gold electrodes. (b) SEM image of the same region. The inset shows that the 400 nm wide line is formed by the individual silver nanoclusters of 10-70 nm size (courtesy of R. Shao, X. Lei and J. Ferris).

At this point, we have actively pursued second and third approach. In the latter case, the overbaked resist layer was used to fabricate the dielectric structure. Second resist was used to fabricate gold leads using lift-off technique. It turned out that in this approach resists tend to contaminate the PZT layer, thus precluding the local PFM poling

After several trial and error steps, it was found that this effect can be alleviated by oxygen plasma etch with subsequent HCl etch to remove the organic contaminant and restore the active PZT surface. The double line structure fabricated by this approach is shown in Figure 7.11. This approach was found to be very sensitive to the resist dielectric properties (the surface PZT surface becomes contaminated again in time) and in the future alternative dielectric layers ($Al_2O_3$ or $SiO_2$) are required.

In an alternate approach, gold contacts were directly deposited on the PZT surface using the stencil mask, thus avoiding resist contamination. The line obtained by this approach is shown in Figure 7.12. Note the clear optical contrast between the regions with positive, negative and random poling state. In the future, the particle deposition far from the switched area will be controlled by either corona poling or use microcontact electroded stamp. On the SEM image the line is comprised of 20-50 nm individual Ag particles. The size of the particles and interparticle spacing can be controlled by the choice of deposition conditions, making this approach promising for the future nanodevices fabrication.

### 7.7. Conclusions and Perspectives

Ferroelectric lithography studies on polycrystalline $BaTiO_3$ surfaces illustrate that polarization patterning with subsequent metal photodeposition can be used for directed creation of metallic structures. $BaTiO_3$, while being very active in the photodeposition process, is not suitable for AFM patterning due to the high switching field required. To avoid this problem, lead zirconate titanate (PZT) is suggested as an alternative and its photochemical activity is demonstrated. In PZT films a large number of polarization orientations, small dielectric constant, weak coupling between the grains and small number of non-180° domain walls facilitate the local poling. The photodeposition process is selective and constrained to positively poled regions only; deposition level on negatively poled domains is extremely small. Local domain patterning and photodeposition can be performed selectively with high quality of image transfer. The minimal feature size achieved so far is limited by the grain size of the film. Further improvement in resolution of these techniques will be achieved with PZT epitaxial thin films on conductive substrates and is expected to be at least 10 nm. It is projected that this behavior is common to ferroelectric titanates and possibly other ferroelectric semiconductors.

These results clearly illustrate the potential of controlled polarization switching with subsequent metal photodeposition for the creation for metal meso- and nanoscale structures. Polarization and deposition steps can be repeated thus allowing fabrication of nanostructures comprised of several deposited materials on ferroelectric substrate. It is important to note that this mechanism of directed assembly differs fundamentally from those that utilize local electrostatic attraction to assemble nanostructures onto templates of patterned charge.[11] In the latter case, local charge can be used to locally deposit charged particles from colloidal solution. However, on the most surfaces in ambient the charge deposition is limited to sizes ~100 nm or larger. In these cases, the positions of the charges are not pinned; therefore, the pattern is susceptible to diffusion. Local polarization switching allows creation and manipulation of nanodomains down to 10 nm.[12] On a ferroelectric substrate, local surface charge is due to atomic polarization and therefore is stable. More importantly, since the reaction mechanism involves controlling the surface electronic structure, the reaction product is not limited by the amount of local charge. This approach can be combined with existing silicon technology.

# 8. CONCLUSIONS

The conclusions derived in this thesis can be divided into three categories: those related to the technique development and quantification, grain boundary phenoena and ferroelectric surface behavior.

## 8.1. Techniques

1.SSPM under applied lateral bias provides spatially resolved potential images not affected by surface adsorption. This technique provides a quantitative tool for spatially resolved resistance measurements with high (100-300 nm) spatial resolution (compare to ~ 4 μm for the best available probe-based techniques). The sources of error in local (< 300 nm) potential measurements are traced to the non-local cantilever contribution to the probe-surface interactions and non-ideality of the feedback loop.

2. For the quasi one-dimensional systems such as electroactive interfaces, it is possible to obtain the I-V characteristic of the interface from the local SSPM measurements under applied bias. The contribution of stray resistances in the circuit can be accounted for by the variation of circuit termination resistors.

3. A novel SPM technique, further referred to as Scanning Impedance Microscopy, is developed for the quantitative imaging of ac transport. SIM imaging at low frequencies visualizes the resistive barriers at the interfaces, whereas SIM imaging at high frequencies visualizes capacitive barriers at the interfaces. The crossover between the regimes occurs at frequencies equal or higher than intrinsic relaxation frequency of the interfaces, 1/RC. Unlike many SPM techniques, SIM yields quantitative information consistent with macroscopic impedance spectroscopy measurements.

4. Combination of SIM and SSPM allows local bias at the interface and interface capacitance to be determined simultaneously avoiding the contributions of contacts and bulk impedances. Ability to obtain **local** *I-V* and *C-V* characteristics of interfaces with lateral resolution of ~50 nm is extremely valuable for semiconductor device characterization and imaging of operational nanoelectronic circuits.

5. The tip related artifacts on the nanoscale device imaging is illustrated on an example of carbon nanotube circuit. In turn, by measuring the force signal over biased carbon nanotube, tip surface transfer function is determined, thus rendering carbon nanotubes as efficient standards for electrostatic tip characterization.

6. An approach for current based frequency resolved measurements (AFM-assisted impedance spectroscopy) is developed and the structure of tip-surface equivalent circuit is determined. The stray probe-surface capacitance is measured directly to be $C_s \sim 0.5$ pF. This capacitance imposes the limit on the measurable tip-surface resistances as $R < 1/(\omega C_s)$, practically limiting imaging to $R < 10^{12}$ Ohm.

7. To describe image formation mechanism in the piezoresponse force microscopy it is necessary to include local electromechanical, local electrostatic and non-local electrostatic contributions. Non-local contribution due to buckling cantilever oscillations is inversely proportional to the cantilever spring constant and becomes negligible for stiff cantilevers ($k >> 1$N/m).

8. The results of electrostatic image charge calculations give the upper boundary of local electrostatic response as ~5 pm/V and indicate the existence of intrinsic dielectric gap between metallic tip and ferroelectric surface.

9. Electromechanical piezoelectric response is heavily impaired by tip-surface contact. Depending on the contact, strong (classical), contact limited and weak (field induced) indentation limits are distinguished. Typical response in the classical regime is ~100 pm/V and is comparable with corresponding $d_{33}$. In the weak indentation regime, response is ~10 pm/V and depends on tip geometry. The contribution of different electroelastic constants of the material to response amplitude is investigated and an almost linear correlation between piezoresponse and $d_{33}$ is found.

10. It is possible to obtain the quantitative information on materials properties from PFM imaging in the strong indentation regime. In the weak indentation regime, the detailed knowledge of tip geometry is required (which is usually not available). No information can be obtained in the electrostatic regime.

11. The guidelines for quantitative PFM imaging and spectroscopy are set forth in the form of Contrast Mechanism Maps that elucidate the effect of experimental conditions such as indenttion force and tip radius of curvature. Quantitative imaging is possible using relatively high indentation forces (~100 nN) and blunt tips (~50-100 nm), while high resolution (~5 nm) imaging requires sharp probes.

12. PFM contrast is observable even for the relatively conductive materials (e.g. $BiFeO_3$ or doped $BaTiO_3$), for which classical current-based ferroelectric measurements are impossible due to the high sample conductivity, thus allowing it to be the reliable tool for establishing the ferroelectric ordering in semiconductors.

## 8.2. Transport at Electroactive Interfaces

1. Frequency dependence of zero-bias interface resistance and capacitance in $\Sigma 5$ Nb-doped $SrTiO_3$ bicrystal is determined by SPM and variable temperature impedance and I-V measurements. Experimental data show that the dielectric constant is lowered in the vicinity of $SrTiO_3$ grain boundaries. Corresponding theory was developed and the dielectric nonlinearity was shown to be in agreement with reference data.
2. The dominant transport mechanism in $SrTiO_3$ at room temperature is diffusion. It is shown that for lower temperatures mobility increases and the transport mechanism switches to thermionic emission over the barrier. At high interface biases conductance is limited by space charge limited current, thus partially explaining low non-linearity coefficient typical for $SrTiO_3$ varistors.
3. In polycrystalline ZnO, SSPM on the grounded surface indicates the presence of the second phase inclusions. Imaging under applied dc bias illustrated the presence of a large number of rectifying interfaces, indicating that symmetric I-V characteristic of polycrystalline samples represent the average properties. Potential gradients within individual grains allow the current distribution in the sample to be determined.
4. In polycrystalline $BaTiO_3$ with a positive temperature coefficient of resistance (PTCR), the electric activity of the interfaces and ferroelectric activity of the grains is mapped using variable temperature SSPM and PFM. The formation of resistive grain boundary barriers was observed below the nominal transition temperature, while piezoresponse activity was observed in the PTCR region. These results indicate the gradual nature of the transition, which is a direct consequence of large dispersion of grain boundary properties.
5. In polycrystalline $BiFeO_3$ ceramics it is shown that the grain boundaries, rather than ferroelectric domain walls, control ac conductivity.

**8.3. Polarization-related Phenomena on Ferroelectric Surfaces**

1. A combination of topographic imaging and SSPM/PFM measurements is required for unambiguous reconstruction of domain structure on the well defined ferroelectric surfaces.

2. Quantitative analysis of voltage and distance SSPM and EFM data on $BaTiO_3$ (100) surface was used to determine that the polarization charge is almost completely screened on $BaTiO_3$ (100) surface in air.

3. Spontaneous polarization changes instantly with temperature, while relaxation time for screening charge is relatively large. Variation of temperature results in the formation of uncompensated charge that slowly relaxes with time. The sign of domain potential is determined by screening charge rather than polarization bound charge; i.e. $c^+$ domains are negative and $c^-$ domains are positive on SSPM image.

4. The dynamics of phase transition and magnitude of potential suggest that screening is due to surface adsorbates. Temperature programmed desorption data suggest that the screening is due to the dissociative water adsorption.

5. Kinetics and thermodynamics of adsorption can be determined from VT SSPM measurements. The enthalpy for this process is $\Delta H_{ads}$ = 164.6 kJ/mole, the entropy is $\Delta S_{ads}$ = -126.6 J/mole K. Relaxation of screening charges is slow; characteristic times are ~10 min, corresponding activation energy is ~4 kJ/mole.

7. Piezoresponse measurements on $BaTiO_3$ are found to be strongly affected by the dielectric tip-surface gap. This effect severely reduces the measured response amplitude and governs the temperature dependence of the PFM signal. This effect is not nearly as important for PZT as it is for $BaTiO_3$ due to the smaller dielectric constant of the latter.

8. Ferroelectric field effects are used for the fabrication of metal nanostructures by domain-selective metal photodeposition on the engineered domain structures. The process flow for the device fabrication using ferroelectric lithography is developed.